\documentclass{article}
\usepackage{arxiv}

\usepackage[dvipsnames]{xcolor} 
\usepackage{amsmath}
\usepackage[labelfont={bf}]{caption}
\usepackage{amssymb}
\allowdisplaybreaks

\usepackage[most]{tcolorbox}
\usetikzlibrary{patterns}
\pgfdeclarepatternformonly{mystrikeout}{\pgfqpoint{-1pt}{-1pt}}{\pgfqpoint{11pt}{11pt}}{\pgfqpoint{10pt}{10pt}}%
{
  \pgfsetlinewidth{0.4pt}
  \pgfpathmoveto{\pgfqpoint{0pt}{0pt}}
  \pgfpathlineto{\pgfqpoint{10.1pt}{10.1pt}}
  \pgfusepath{stroke}
}
\newtcolorbox{tcbstrikeout}{breakable,
 enhanced jigsaw,
 opacityback=0,
 parbox=false,
 boxrule=0mm,
 top=0mm,bottom=0pt,left=0pt,right=0pt,
 boxsep=0pt,
 frame hidden,
 finish={\fill[pattern=mystrikeout] (frame.north west) rectangle (frame.south east);}
}

\usepackage{ifthen}
\usepackage{xparse}

\usepackage{xspace}
\usepackage{scalerel} 
\newcommand*{\paral}{\stretchrel*{\parallel}{\perp}}

\newcommand{\Borel}[1]{\mathcal{B} \left( #1 \right)}
\newcommand{\ellone}{$\mathcal{L}_1$\xspace}
\newcommand{\ellonedrac}{$\mathcal{L}_1$-DRAC\xspace}
\newcommand{\br}[1]{\left( #1 \right)} 
\newcommand{\sbr}[1]{\left[ #1 \right]} 
\newcommand{\cbr}[1]{\left\{ #1 \right\}} 
\newcommand{\indicator}[2]{\mathbbm{1}_{#1}\left( #2 \right)} 
\newcommand{\Lip}[1]{\mathbbm{1}_{\text{Lip}}\left( #1 \right)} 

\newcommand{\norm}[1]{\left\| #1 \right\|}
\newcommand{\Trace}[1]{\xspace\text{Tr}\left[#1\right]}

\newcommand{\Probability}[1]{\mathbb{P}\left( #1 \right)}

\newcommand{\pWass}[3]{\mathsf{W}_{#1}\left(#2,#3\right)}

\newcommand{\Pas}{$\mathbb{P}$-a.s.\xspace}

\newcommand{\absolute}[1]{\left| #1 \right|}

\newcommand{\Wt}[1]{W_{#1}}
\newcommand{\Wstart}[1]{W^\star_{#1}}
\newcommand{\Wrt}[1]{\widehat{W}_{#1}}
\newcommand{\Qt}[1]{Q_{#1}}

\newcommand{\Xt}[1]{X_{#1}}

\newcommand{\Yt}[1]{Y_{#1}}

\newcommand{\Xstart}[1]{X^\star_{#1}}

\newcommand{\Xrt}[1]{X^r_{#1}}

\newcommand{\Xhatt}[1]{\hat{X}_{#1}}
\newcommand{\Xtildet}[1]{\tilde{X}_{#1}}

\newcommand{\Zt}[1]{Z_{#1}}

\newcommand{\Wfilt}[1]{\mathfrak{W}_{#1}}
\newcommand{\Wstarfilt}[1]{\mathfrak{W}^\star_{#1}}

\newcommand{\Xdist}[1]{\mathbb{X}_{#1}}
\newcommand{\Zdist}[1]{\mathbb{Z}_{#1}}

\newcommand{\Xrdist}[1]{\mathbb{X}^r_{#1}}

\newcommand{\Xstardist}[1]{\mathbb{X}^\star_{#1}}

\newcommand{\Ydist}[1]{\mathbb{Y}_{#1}}

\newcommand{\Ut}[1]{U_{#1}}
\newcommand{\Urt}[1]{U^r_{#1}}

\newcommand{\ULt}[1]{U_{\mathcal{L}_1, #1}}
\newcommand{\expo}[1]{e^{ #1 }}

\NewDocumentCommand{\FL}{o o}{ 
  \mathcal{F}_{\mathcal{L}_1} \IfValueT{#1}{\left(#1\right)} \IfValueT{#2}{\left(#2\right)}
}

\NewDocumentCommand{\Filter}{o o}{
  \mathcal{F}_\Boldomega \IfValueT{#1}{\left( #1 \right)} \IfValueT{#2}{(#2)} 
} 

\NewDocumentCommand{\FilterW}{o o}{
  \mathcal{F}_{\mathcal{N},\Boldomega} \IfValueT{#1}{\left( #1 \right)} \IfValueT{#2}{(#2)} 
}

\NewDocumentCommand{\AdaptationLaw}{o o o}{
  \mathcal{F}_{\BoldTs} \IfValueT{#1}{\left( #1, \IfValueT{#2}{#2} \right)} \IfValueT{#3}{(#3)} 
}
\NewDocumentCommand{\AdaptationLawParal}{o o}{
  \mathcal{F}^{\paral}_{\BoldTs} \IfValueT{#1}{\left( #1 \right)} \IfValueT{#2}{(#2)} 
}

\NewDocumentCommand{\Predictor}{o o}{
  \mathcal{F}_{\lambda_s} \IfValueT{#1}{\left( #1 \right)} \IfValueT{#2}{(#2)} 
}

\NewDocumentCommand{\ReferenceInput}{o o}{
  \mathcal{F}_r \IfValueT{#1}{\left( #1 \right)} \IfValueT{#2}{(#2)} 
}

\NewDocumentCommand{\ControlError}{o}{
  \widehat{\mathcal{F}} \IfValueT{#1}{\left( #1 \right)}  
}

\NewDocumentCommand{\ControlErrorTilde}{o}{
  \widetilde{\mathcal{F}} \IfValueT{#1}{\left( #1 \right)}  
}
\NewDocumentCommand{\ControlErrorTildeDrift}{o}{
  \widetilde{\mathcal{F}}_\mu \IfValueT{#1}{\left( #1 \right)}  
}

\NewDocumentCommand{\ControlErrorTildeDiffusion}{o}{
  \widetilde{\mathcal{F}}_\sigma \IfValueT{#1}{\left( #1 \right)}  
}

\NewDocumentCommand{\ControlErrorHatDrift}{o}{
  \widehat{\mathcal{F}}_\mu \IfValueT{#1}{\left( #1 \right)}  
}

\NewDocumentCommand{\ControlErrorHatDiffusion}{o}{
  \widehat{\mathcal{F}}_\sigma \IfValueT{#1}{\left( #1 \right)}  
}

\NewDocumentCommand{\ControlErrorDrift}{o}{
  \widetilde{\Gamma}_\mu \IfValueT{#1}{\left( #1 \right)}  
}

\NewDocumentCommand{\ControlErrorDiffusion}{o}{
  \widetilde{\Gamma}_\sigma \IfValueT{#1}{\left( #1 \right)}  
}

\NewDocumentCommand{\ControlErrorIntegrand}{o}{
  \widetilde{\Gamma} \IfValueT{#1}{\left( #1 \right)}  
}

\NewDocumentCommand{\ControlErrorLastStep}{o o}{
  \IfValueT{#1}{\breve{\mathcal{F}}_{#1}} \IfValueT{#2}{\left( #2 \right)}  
}

\NewDocumentCommand{\ControlErrorDriftFinal}{o o}{
  \IfValueT{#1}{\widetilde{\mathcal{G}}_{\mu_{#1}}} \IfValueT{#2}{\left( #2 \right)}  
}

\NewDocumentCommand{\ControlErrorDiffusionFinalAlt}{o o}{
  \IfValueT{#1}{\widetilde{\mathcal{G}}_{\sigma_{#1}}} \IfValueT{#2}{\left( #2 \right)}  
}

\NewDocumentCommand{\ControlErrorDiffusionFinal}{o}{
  \widetilde{\mathcal{G}}_{\sigma} \IfValueT{#1}{\left( #1 \right)}  
}

\newcommand{\istar}[1]{i^\star\left(#1\right)}

\newcommand{\RefRho}{ {\color{Aquamarine}   \bm{\varrho_r ( } } \sfp {\color{Aquamarine}   \bm{)} }   }
\newcommand{\RefRhoFunction}[1]{ {\color{Aquamarine}   \bm{\varrho_r} \left( {\color{black} #1} \right)  }   }

\newcommand{\AdaptRho}{ {\color{Magenta}   \bm{\varrho_a ( } } \sfp {\color{Magenta}   \bm{)} }   }
\newcommand{\AdaptRhoFunction}[1]{ {\color{Magenta}   \bm{\varrho_a} \left( {\color{black} #1} \right)  }   }

\newcommand{\TotalRho}{ {\color{NavyBlue}   \bm{\varrho ( } } \sfp {\color{NavyBlue}   \bm{)} }   }
\newcommand{\TotalRhoFunction}[1]{ {\color{NavyBlue}   \bm{\varrho} \left( {\color{black} #1} \right)  }   }
\newcommand{\TotalRhoUUB}[1]{ {\color{NavyBlue}   \bm{\hat{\varrho}} \left( {\color{black} #1} \right) }  }

\newcommand{\RhoPrime}{ {\color{NavyBlue}   \bm{\varrho' ( } } \sfp {\color{NavyBlue}   \bm{)} }   }
\newcommand{\RhoDPrime}{ {\color{NavyBlue}   \bm{\varrho'' ( } } \sfp {\color{NavyBlue}   \bm{)} }   }

\newcommand{\RefRhoUUB}[1]{ {\color{Aquamarine}   \bm{\hat{\varrho}_r} \left( {\color{black} #1} \right) }  }

\newcommand{\pwrfourth}[1]{\left(#1\right)^\frac{1}{4}}

\newcommand{\LpRefError}[1]{ {\color{ForestGreen}   \widetilde{\mathcal{E}}_{r,#1}(2\sfp) } }
\newcommand{\LpRef}[1]{ {\color{ForestGreen}   \mathcal{E}_{r,#1}(2\sfp) } }

\newcommand{\LpRefErrorPrime}[2]{ {\color{ForestGreen}   \widetilde{\mathcal{E}}'_{r,#1}(#2,2\sfp) } }
\newcommand{\LpRefPrime}[2]{ {\color{ForestGreen}   \mathcal{E}'_{r,#1}(#2,2\sfp) } }

\newcommand{\LTwoRefErrorPrime}[1]{ {\color{ForestGreen}   \widetilde{\mathcal{E}}'_{r,#1}(t,2) } }

\newcommand{\LpTrueError}{ {\color{ForestGreen}   \widetilde{\mathcal{E}}(2\sfp) } }
\newcommand{\LpTrue}{ {\color{ForestGreen}   \mathcal{E}(2\sfp) } }
\newcommand{\LpTrueRef}{ {\color{ForestGreen}   \mathcal{E}_r(2\sfp) } }

\newcommand{\LpTrueErrorPrime}[1]{ {\color{ForestGreen}   \widetilde{\mathcal{E}}'(#1,2\sfp) } }
\newcommand{\LTwoTrueErrorPrime}[1]{ {\color{ForestGreen}   \widetilde{\mathcal{E}}'(#1,2) } }

\newcommand{\Boldomega}{ {\color{Magenta}   \bm{\omega}}  }
\newcommand{\BoldomegaRoot}{ {\color{Magenta}   \bm{\sqrt{\omega}}}  }

\newcommand{\BoldTs}{ {\color{NavyBlue} \bm{T_s}  }  }
\newcommand{\BoldTsroot}{ {\color{NavyBlue} \bm{\sqrt{T_s}}  }  }

\newcommand{\Jmu}[1]{J_\mu \left( #1 \right)}
\newcommand{\Jsigma}[1]{J_\sigma \left( #1 \right)}
\newcommand{\JNmu}[1]{J_{N,\mu} \left( #1 \right)}
\newcommand{\JNsigma}[1]{J_{N,\sigma} \left( #1 \right)}

\newcommand{\ito}{It\^{o}\xspace}

\newcommand{\Lmu}[1]{\Lambda_\mu \left( #1 \right)}
\newcommand{\LmuAll}[1]{\Lambda_\mu^{\cbr{\cdot,\paral,\perp}} \left( #1 \right)}

\newcommand{\Lsigma}[1]{\Lambda_\sigma \left( #1 \right)}
\newcommand{\Lparamu}[1]{\Lambda_\mu^{\paral} \left( #1 \right)}

\newcommand{\Lperpmu}[1]{\Lambda_\mu^\perp \left( #1 \right)}
\newcommand{\Lparasigma}[1]{\Lambda_\sigma^{\paral} \left( #1 \right)}
\newcommand{\Lperpsigma}[1]{\Lambda_\sigma^\perp \left( #1 \right)}

\newcommand{\LmuError}[2]{\widetilde{\Lambda}_{\mu_{#1}} \left( #2 \right)}
\newcommand{\LparamuError}[2]{\widetilde{\Lambda}^{\paral}_{\mu_{#1}} \left( #2 \right)}
\newcommand{\LmuErrorAll}[2]{\widetilde{\Lambda}^{\cbr{\cdot,\paral}}_{\mu_{#1}} \left( #2 \right)}

\newcommand{\LZmu}[1]{\mathring{\Lambda}_\mu \left( #1 \right)}
\newcommand{\LZparamu}[1]{\mathring{\Lambda}^{\paral}_\mu \left( #1 \right)}
\newcommand{\LZperpmu}[1]{\mathring{\Lambda}^{\perp}_\mu \left( #1 \right)}
\newcommand{\LZmuAll}[1]{\mathring{\Lambda}^{ \cbr{\cdot,\paral,\perp}  }_\mu \left( #1 \right)}

\newcommand{\FZsigma}[1]{\mathring{F}_\sigma \left( #1 \right)}
\newcommand{\FNZsigma}[1]{\mathring{F}_{N,\sigma} \left( #1 \right)}
\newcommand{\FNZmu}[1]{\mathring{F}_{N,\mu} \left( #1 \right)}

\newcommand{\FZparasigma}[1]{\mathring{F}^{\paral}_\sigma \left( #1 \right)}
\newcommand{\FZperpsigma}[1]{\mathring{F}^{\perp}_\sigma \left( #1 \right)}
\newcommand{\FZsigmaAll}[1]{\mathring{F}^{ \cbr{\cdot,\paral,\perp}  }_\sigma \left( #1 \right)}

\newcommand{\fZ}[1]{\mathring{f}\left( #1 \right)}

\NewDocumentCommand{\ReferenceInputZ}{o o}{
  \mathring{\mathcal{F}}_r \IfValueT{#1}{\left( #1 \right)} \IfValueT{#2}{(#2)} 
}

\newcommand{\SigmaError}[2]{ \widetilde{\Sigma}_{#1} \left( #2 \right)}
\newcommand{\SigmaparaError}[2]{ \widetilde{\Sigma}^{\paral}_{#1} \left( #2 \right)}
\newcommand{\SigmaErrorAll}[2]{ \widetilde{\Sigma}^{\cbr{\cdot,\paral}}_{#1} \left( #2 \right)}

\newcommand{\Lhat}[1]{\hat{\Lambda}\left( #1\right)}

\newcommand{\Lparahat}[1]{\hat{\Lambda}^{\paral}\left( #1\right)}

\newcommand{\ppara}[1]{p^{\paral}\left( #1  \right)}
\newcommand{\pperp}[1]{p^{\perp}\left( #1  \right)}

\newcommand{\Fmu}[1]{F_\mu \left( #1 \right)}
\newcommand{\Fsigma}[1]{F_\sigma \left( #1 \right)}
\newcommand{\FsigmaAll}[1]{F_\sigma^{\cbr{\cdot,\paral,\perp}} \left( #1 \right)}
\newcommand{\Fbarmu}[1]{\bar{F}_\mu \left( #1 \right)}
\newcommand{\Fbarsigma}[1]{\bar{F}_\sigma \left( #1 \right)}

\newcommand{\Fparasigma}[1]{F^{\paral}_\sigma \left( #1 \right)}

\newcommand{\Fperpsigma}[1]{F^{\perp}_\sigma \left( #1 \right)}

\newcommand{\LsigmaAll}[1]{\Lambda_\sigma^{\cbr{\cdot,\paral}} \left( #1 \right)}
\newcommand{\psigmaAll}[1]{p^{\cbr{\cdot,\paral}} \left( #1 \right)}

\newcommand{\FsigmaError}[2]{\widetilde{F}_{\sigma_{#1}} \left( #2 \right)}
\newcommand{\FparasigmaError}[2]{\widetilde{F}^{\paral}_{\sigma_{#1}} \left( #2 \right)}
\newcommand{\FsigmaErrorAll}[2]{\widetilde{F}^{\cbr{\cdot,\paral}}_{\sigma_{#1}} \left( #2 \right)}

\newcommand{\Grmu}[1]{G_\mu \left( #1 \right)}
\newcommand{\Grsigma}[1]{G_\sigma \left( #1 \right)}
\newcommand{\GrNmu}[1]{G_{N,\mu} \left( #1 \right)}
\newcommand{\GrNsigma}[1]{G_{N,\sigma} \left( #1 \right)}
\newcommand{\Hrsigma}[1]{H_\sigma \left( #1 \right)}
\newcommand{\HrNsigma}[1]{H_{N,\sigma} \left( #1 \right)}

\newcommand{\Ksigma}[1]{K_\sigma \left( #1 \right)}
\newcommand{\KNsigma}[1]{K_{N,\sigma} \left( #1 \right)}

\newcommand{\FNmu}[1]{F_{N,\mu} \left( #1 \right)}
\newcommand{\FNsigma}[1]{F_{N,\sigma} \left( #1 \right)}
\newcommand{\FbarNmu}[1]{\bar{F}_{N,\mu} \left( #1 \right)}
\newcommand{\FbarNsigma}[1]{\bar{F}_{N,\sigma} \left( #1 \right)}

\newcommand{\continuous}[2]{\mathcal{C} \left([0, #1]; \mathbb{R}^{#2}  \right)   }

\newcommand{\ELaw}[2]{\mathbb{E}_{#1}\left[ #2 \right]}

\newcommand{\QuadVar}[1]{\Bigl \langle #1 \Bigr \rangle}
\newcommand{\CrossVar}[2]{\Bigl \langle #1,#2 \Bigr \rangle}

\newcommand{\Frobenius}[1]{\norm{#1}_F}

\newcommand{\LpLaw}[3]{\norm{#3}_{L_{#1}}^{#2}}
\newcommand{\pLpLaw}[2]{\norm{#2}_{L_\mathsf{p}}^{#1}}

\newcommand{\tstar}{t^\star}

\newcommand{\that}{\hat{t} }

\newcommand{\sfp}{{\mathsf{p}}}
\newcommand{\sfq}{\mathsf{q} }

\newcommand{\Half}{\frac{1}{2}}
\newcommand{\Quarter}{\frac{1}{4}}

\usepackage{tikz}

\makeatletter
\def\widebreve{\mathpalette\wide@breve}
\def\wide@breve#1#2{\sbox\z@{$#1#2$}%
     \mathop{\vbox{\m@th\ialign{##\crcr
\kern0.08em\brevefill#1{0.8\wd\z@}\crcr\noalign{\nointerlineskip}%
                    $\hss#1#2\hss$\crcr}}}\limits}
\def\brevefill#1#2{$\m@th\sbox\tw@{$#1($}%
  \hss\resizebox{#2}{\wd\tw@}{\rotatebox[origin=c]{90}{\upshape(}}\hss$}
\makeatletter

\usepackage{stmaryrd}
\usepackage{leftindex}
\usepackage{cancel}
\usepackage{soul}
\usepackage{bbm} 
\usepackage{dsfont}  
\usepackage{romannum} 
\usepackage{mathtools}
\usepackage{wrapfig}
\usepackage{centernot}
\usepackage{subfigure}
\usepackage{enumitem} 
\usepackage[bb=boondox]{mathalfa} 
\usepackage{lipsum}
\usepackage{bm} 
\usepackage{upgreek}

\makeatletter
\newsavebox{\@brx}
\newcommand{\llangle}[1][]{\savebox{\@brx}{\(\m@th{#1\langle}\)}%
  \mathopen{\copy\@brx\mkern2mu\kern-0.9\wd\@brx\usebox{\@brx}}}
\newcommand{\rrangle}[1][]{\savebox{\@brx}{\(\m@th{#1\rangle}\)}%
  \mathclose{\copy\@brx\mkern2mu\kern-0.9\wd\@brx\usebox{\@brx}}}
\makeatother

\usepackage{amsthm}
\usepackage{latexsym}

\usepackage{booktabs}
\usepackage{multirow}
\usepackage{diagbox}

\newtheoremstyle{coloreditalicsbold}{}{}{\it}{}{\color{blue}\bfseries}{}{ }{}
\newtheoremstyle{ForRemarks}{}{}{}{}{\color{blue}\bfseries}{}{ }{}
\newtheoremstyle{altcoloreditalicsbold}{}{}{\it}{}{\color{red}\bfseries}{}{ }{}

\newtheoremstyle{coloreditalics}{}{}{\it}{}{\color{blue}}{}{ }{}

\theoremstyle{coloreditalics}
\newtheorem{remark}{Remark}[section]

\theoremstyle{coloreditalicsbold}

\theoremstyle{coloreditalicsbold}
\newtheorem{assumption}{Assumption}

\theoremstyle{coloreditalicsbold}
\newtheorem{definition}{Definition}

\theoremstyle{coloreditalicsbold}
\newtheorem{theorem}{Theorem}[section]

\theoremstyle{coloreditalicsbold}
\newtheorem{lemma}{Lemma}[section]

\theoremstyle{coloreditalicsbold}
\newtheorem{proposition}{Proposition}[section]

\theoremstyle{coloreditalicsbold}
\newtheorem{corollary}{Corollary}[section]

\theoremstyle{altcoloreditalicsbold}

\usepackage[nonewpage]{imakeidx} 
\makeindex[columns=2, title= {List of Constants\label{idx:ListofConstants}}, options= -s DRAC_idx.ist]

\usepackage{stackengine}
\usepackage{graphicx}
\usepackage{tikz}
\stackMath
\makeatletter
\newcommand{\ostar}{\mathbin{\mathpalette\make@circled\star}}
\newcommand{\make@circled}[2]{%
  \ooalign{$\m@th#1\smallbigcirc{#1}$\cr\hidewidth$\m@th#1#2$\hidewidth\cr}%
}
\newcommand{\smallbigcirc}[1]{%
  \vcenter{\hbox{\scalebox{0.77778}{$\m@th#1\bigcirc$}}}%
}
\makeatother

\newcommand{\RefNorm}[1]{\mathcal{V}\left( #1 \right)}
\newcommand{\RefDiffNorm}[1]{ \partial \mathcal{V} \left( #1 \right)}
\newcommand{\RefDiffNormPoint}[1]{ \partial_i \mathcal{V} \left( #1 \right)}

\newcommand{\ZNorm}[1]{\mathcal{V}\left( #1 \right)}
\newcommand{\ZDiffNorm}[1]{ \partial \mathcal{V} \left( #1 \right)}

\usepackage[disable]{todonotes}

\usepackage{hyperref} 
\hypersetup{
    colorlinks=true,
    linkcolor=red,
    citecolor=green,
    filecolor=magenta,      
    urlcolor=cyan,
    breaklinks=true
    }

\begin{document}
\pagenumbering{arabic}

\title{\ellonedrac: Distributionally Robust Adaptive Control\\ \emph{Global Results}}

\author{ Aditya Gahlawat \\
    Mechanical Science and Engineering\\
	University of Illinois at Urbana-Champaign\\
	Urbana, IL 61801 \\
	\texttt{gahlawat@illinois.edu} \\
	\And
	Sambhu H. Karumanchi \\
    Mechanical Science and Engineering\\
	University of Illinois at Urbana-Champaign\\
	Urbana, IL 61801 \\
	\texttt{shk9@illinois.edu} \\
    \And
	\And
	Naira Hovakimyan \\
    Mechanical Science and Engineering\\
	University of Illinois at Urbana-Champaign\\
	Urbana, IL 61801 \\
	\texttt{nhovakim@illinois.edu} \\
}

\date{}

\renewcommand{\headeright}{}
\renewcommand{\undertitle}{}



\maketitle

\begin{abstract}
    Data-driven machine learning methodologies have attracted considerable attention for the control and estimation of dynamical systems. 
    However, such implementations suffer from a lack of predictability and robustness. 
    Thus, adoption of data-driven tools has been minimal for safety-aware applications despite their impressive empirical results. 
    While classical tools like robust adaptive control can ensure predictable performance, their consolidation with data-driven methods remains a challenge and, when attempted, leads to conservative results. 
    The difficulty of consolidation stems from the inherently different `spaces' that robust control and data-driven methods occupy. 
    Data-driven methods suffer from the \emph{distribution-shift} problem, which current robust adaptive controllers can only tackle if using over-simplified learning models and unverifiable assumptions. 
    In this paper, we present \emph{\ellone distributionally robust adaptive control (\ellonedrac)}:  a control methodology for uncertain stochastic processes that guarantees robustness certificates in terms of uniform (finite-time) and maximal distributional deviation. 
    We leverage the \ellone adaptive control methodology to ensure the existence of Wasserstein \emph{ambiguity} set around a nominal distribution, which is guaranteed to contain the true distribution. 
    The uniform ambiguity set produces an \emph{ambiguity tube} of distributions centered on the nominal temporally-varying nominal distribution. 
    The designed controller generates the ambiguity tube in response to both \emph{epistemic} (model uncertainties) and \emph{aleatoric} (inherent randomness and disturbances) uncertainties. 
    
\end{abstract}

\section{Introduction}\label{sec:introduction}

Consider a spectrum of control methodologies with classical tools like robust adaptive control~\cite{ioannou1996robust,lavretsky2012robust} on one end and data-driven approaches relying on data, computation, and deep-learning on the other end~\cite{kuutti2020survey, shi2019neural, arulkumaran2017deep, arulkumaran2017deep, lenz2015deepmpc}. 
Robust adaptive control was born out of a need for certifiable robustness margins and has had a long and rich history of use with safety-critical applications such as aviation~\cite{lavretsky2012robust}. 
On the other end, while data-driven methods have displayed impressive empirical performance and can often learn control policies purely from data (e.g., model-free reinforcement learning~\cite{bertsekas2019reinforcement}), such methods lack robustness guarantees and safety guarantees~\cite{moos2022robust}. 

Ideally, one would like to consolidate the robustness guarantees of the classic methodologies with the empirical performance of the data-driven methodologies. 
However, such a consolidation is far from a straightforward exercise and, indeed, is a significant hurdle in the adoption of data-driven controllers/methods for safety-critical systems. 
In an attempt to provide safety guarantees for systems operating with data-driven learned components in the loop, recent attempts have focused on considering a robust worst-case analysis and bounded disturbances, which lead to overly conservative results, see e.g.~\cite{mitchell2005time, herbert2017fastrack, singh2019robust} and references therein. 
Instead, one can study average-case (or high-probability) stochastic safety guarantees to alleviate the conservativeness since then one analyzes the distributions and their associated statistical properties instead of purely their supports. Of course, the study of dynamical systems subject to stochastic perturbations brings forth further challenges in their analysis, and thus, a majority of existing work relies on relatively simple models that allow for amenable statistical properties like, e.g., linear systems that preserve the Gaussian nature of perturbations under integration or assumptions that are difficult to verify~\cite{borkar2009stochastic, mao1994exponential}. 

 In addition to safety guarantees with reduced conservatism, another appealing feature of systems operating under stochastic perturbations is their representation as time-evolving distributions (measures on probability spaces). 
 For example, under moderate regularity conditions, the probability density functions associated with the transition probabilities of solutions to certain stochastic differential equations (SDEs) exist and evolve as per the Kolmogorov forward equation (also known as the Fokker-Planck equation)~\cite[Chp.~2]{kloeden1992stochastic},~\cite{oksendal2013stochastic},~\cite[Chp.~8]{oksendal2013stochastic}. 
 Of course, along with the distributional representation of systems with stochastic perturbations, we still retain their representation in terms of trajectories (sample paths) as in deterministic systems. 
The distributional representation of stochastic systems is beneficial for using data-driven learned systems and control policies. 
Rooted in statistical learning theory~\cite{vapnik1999overview}, one can use concepts from generalization theory like (empirical) risk minimization~\cite{mohri2018foundations} to train models that achieve low-error on average over the training data distribution. 
Therefore, while stochastic systems can be represented distributionally, data-driven models are trained over distributions, and this commonality can be exploited to consolidate safe and robust control with the use of data-driven models.  
 It should be noted that deterministic evolution modeled by ordinary/partial differential equations also admit a distributional representation via the degenerate dirac measures concentrated on the system states. 
 Deterministic evolution is therefore not a viable candidate to represent general measures since it has only the single distributional representative.  
 In contrast, even the relatively simple class of SDEs with state-independent diffusion vector fields can represent general distributions as evidenced by their usage in generative modelling~\cite{songscore}.

While the distributional nature shared by stochastic systems and data-driven learned models is helpful, one needs to consider the issues that learned models contend with since these will affect the downstream task of their use in the control of systems. 
A central assumption in the learning-based method is that the training and testing datasets (testing refers to the actual system implementation) are samples from the same distribution~\cite{bishop2006pattern}. 
The assumption of the same training and testing (implementation) data distributions allows one to use generalization theory~\cite{mohri2018foundations} to obtain results like the one in~\cite{boffi2021learning} where the authors were able to provide the validity of learned stability certificates to new trajectories initialized from unseen initial conditions, but from the \emph{same} distribution over initial conditions that generated the training data. 
However, as one may expect, the assumption of encountering the same distribution of scenarios that the learned models are trained on will seldom hold in real-life applications.
Thus, a significant issue that learned systems have to contend with is the \emph{distribution shift problem}: the real-life scenario wherein a learned model has to provide predictions and actions in response to an input from a distribution that is different from the distribution it was trained on~\cite{quinonero2008dataset, sugiyama2012machine}. 
Indeed, distribution shift offers a significant hurdle in using learned models and policies in safety-critical systems, see~\cite{wiles2021fine, filos2020can} and references therein. Distribution shifts can occur, for example, when deploying a controller learned in a simulator. 
The controller is learned purely based on the quality (accuracy) of the simulator and its representational capabilities to offer sufficiently rich and realistic scenarios to the learning agent during training. 
The distribution shift problem that a simulator-based learned controller encounters in its deployment on a real system can be broadly classified under the sim2real transfer problem~\cite{peng2018sim, du2021auto, ho2021retinagan}. 
Distribution shifts can also occur due to different distributions of initial conditions and the shift between the control policy that generates the training data and the optimized learned policy deployed on the system. 
While far from trivial to address, researchers have made significant progress in tackling the distribution shifts due to change between data logging and learned policy, as is common in imitation learning (IL), using robustness properties of systems~\cite{pfrommer2022tasil}. 

Perhaps the most significant source of distribution shift is uncertainties in system dynamics models, also known as \emph{epistemic uncertainties}. 
Note that epistemic uncertainties contribute to the reality gap in the sim2real transfer problem since their effect manifests as the lack of the simulator in representing reality. 
Furthermore, if one considers stochastic systems subject to random perturbations (\emph{aleatoric uncertainties}), the effects of epistemic uncertainties can be worsened due to the contribution of the aleatoric uncertainties. 
Of course, as we mentioned above, classical control tools like robust adaptive methodologies were explicitly developed to counter the effects of such uncertainties in a predictable and guaranteed manner. 
However, the often unrealistic need for explicitly parameterized uncertainties and bounded and deterministic sets to which uncertainties belong causes one to discard the distributional nature of learned models to use the classical tools. 
As we mentioned previously, in doing so, we obtain conservative results and also lose the distributional representation of learned models and rely solely on bounded and deterministic representations, which contain far less valuable and actionable information.    
Therefore, we would like to \ul{develop a general control methodology for stochastic systems --- subject to both epistemic and aleatoric uncertainties ---  to produce certificates of robustness that are distributional in nature}.
Such approaches generalize the notion of robustness since the certificates are defined, not on the underlying state space, but instead on the set of probability measures on the state spaces.
We posit that the development of the stated methodology can facilitate seamless and non-conservative integration of robust control with data-driven learned models that can be verified to produce predictable and, hence, safe behavior. 
In other words, instead of restricting the capabilities of data-driven learned models to make them amenable to consolidation with classical control tools, we aim to raise the abstractions of robust-adaptive control tools so that such tools are conducive to working with learned models by design.

\begin{figure}[t]
    \centering
    \subfigure[]{\includegraphics[width=0.45\textwidth]{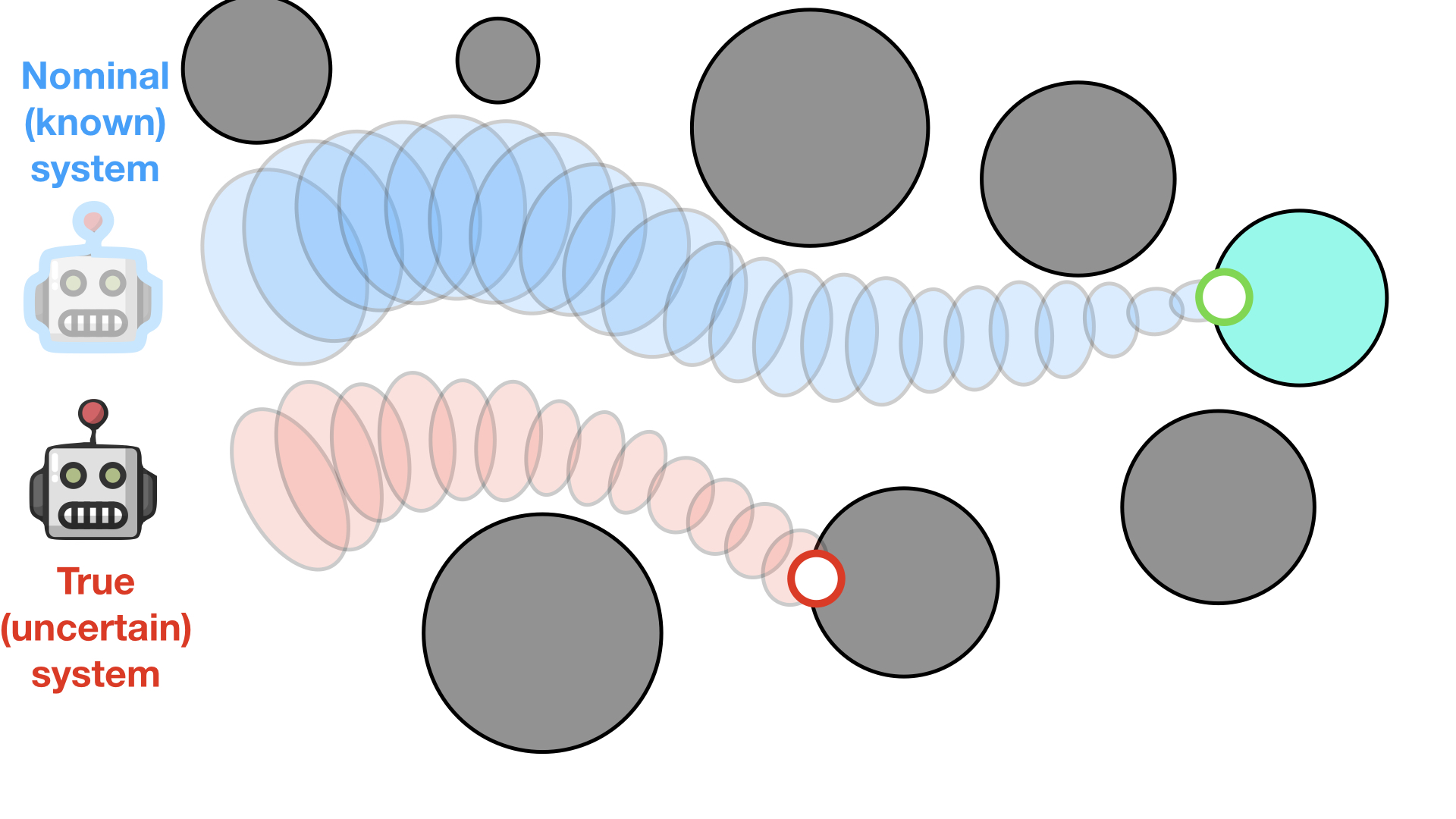}} 
    \subfigure[]{\includegraphics[width=0.45\textwidth]{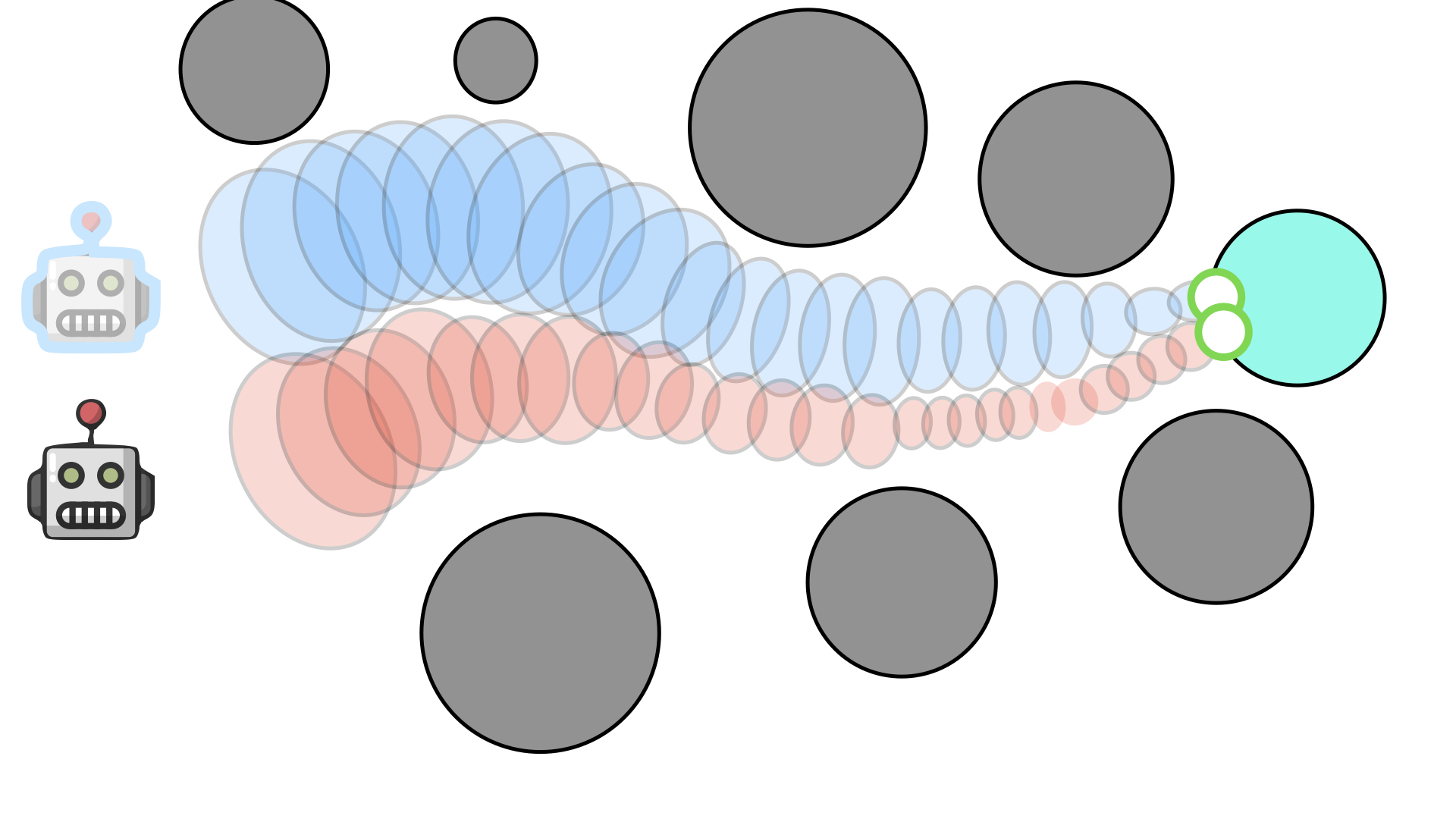}} 
    \subfigure[]{\includegraphics[width=0.80\textwidth]{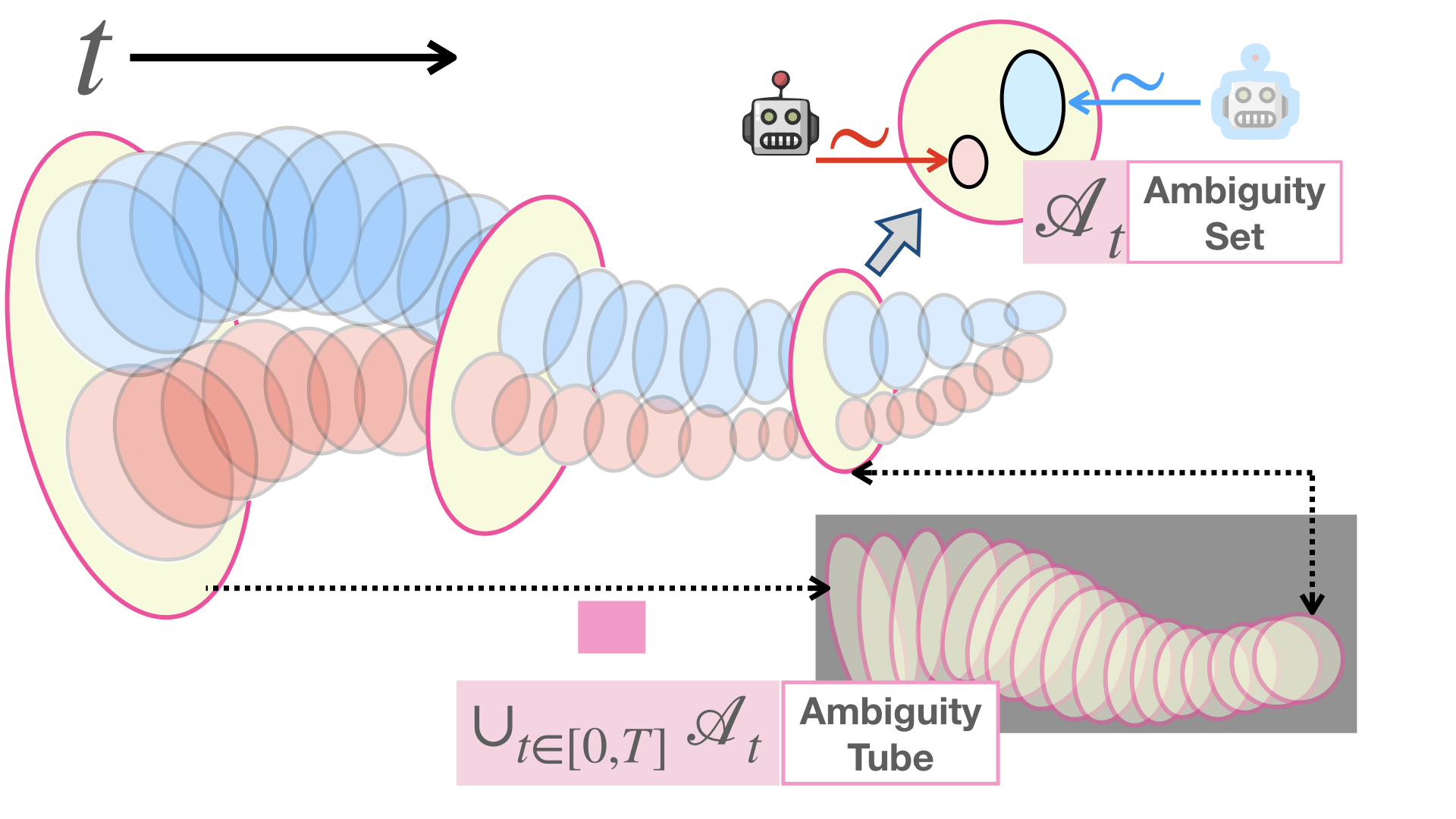}\label{fig:MainIllustration:Ambiguity}}
    \caption{Consider the problem of safely navigating an uncertain stochastic system to a goal set (green circle), avoiding unsafe subsets of the state-space (grey circles).  One constructs a control policy for the uncertainty-free nominal (known) system (the robot with faded colors and a blue outline) as it represents the best knowledge available for the true (uncertain) system (solid-colored robot). (a) While the control policy successfully guides the nominal (known) system to the goal set, as we illustrate with temporal state distributions in light blue, applying the same policy to the true (uncertain) system leads to unquantifiable and undesirable behaviors due to the presence of uncertainties (illustrated with light-red temporal state-distributions).  (b) Thus, one attempts to design an additional feedback policy to handle uncertainties such that the original policy can still guide the true (uncertain) system predictably and safely.  (c) We provide one such approach, the \ellonedrac~control, for the design of robust adaptive feedback such that we are assured of having the \emph{a priori} deviations between the nominal (known) system's state distribution and the true (uncertain) system's state distribution. These guarantees are in the form of \emph{ambiguity sets} ($\mathcal{A}_t$) within which both the nominal and true state-distributions are guaranteed to lie, $\forall t \geq 0$. Due to the uniform (finite-time) guarantees on the system's distributional transients, one can extrude such ambiguity tubes in time to obtain \emph{ambiguity tubes} ($\cup_{t \in [0, T]} \mathcal{A}_t$, for any $T \in (0, \infty)$), which enables safe predictive planning. }
    \label{fig:MainIllustration}
\end{figure}

We present \ellone-distributionally robust adaptive control (\ellonedrac): a robust adaptive methodology to control uncertain stochastic processes whose evolution is governed by stochastic differential equations (SDEs). 
We design \ellonedrac ~such that we can quantify and control the `distance' between the state distributions of the known (nominal) system and the true (uncertain) system. We use the Wasserstein norm~\cite[Chp.~6]{villani2009optimal} as a metric on the space of Borel measure to quantify the distance between the nominal and true distributions. 
\ul{We use \emph{ambiguity sets}\footnote{
    Sets of distributions (Borel measures on vector spaces over the field of reals).} defined via the Wasserstein distance between the nominal and true state distributions as the robustness certificate against distribution shifts due to epistemic and aleatoric uncertainties}. 
We develop \ellonedrac~using the architecture of \ellone-adaptive control~\cite{hovakimyan2010ℒ1}, a robust adaptive methodology that decouples estimation from control and provides transient guarantees in response to uncertainties. 
The \ellone-adaptive control has been successfully implemented on NASA's AirStar 5.5\% subscale generic transport aircraft model \cite{gregory2010flight}, Calspan's Learjet \cite{ackerman2017evaluation}, and uncrewed aerial vehicles  \cite{kaminer2010path,jafarnejadsani2017optimized}. 
A high-level illustration of the goals and capabilities of \ellonedrac~are illustrated in Fig.~\ref{fig:MainIllustration}.

\subsection{Prior art}\label{subsec:PriorArt} We now discuss the existing results in the literature that provide results for problems similar to the one we consider in this manuscript.  
%
  
\textbf{\emph{(\romannum{1})~Uniform (finite-time) guarantees:}} We begin by discussing results that provide uniform (finite-time) guarantees for controlled stochastic systems. The authors in~\cite{pflug1986stochastic} and~\cite{benveniste1982measure} consider asymptotic reference tracking for discrete-time stochastic systems. However, the safe operation of (stochastic and uncertain) systems requires uniform (finite-time) guarantees instead of asymptotic guarantees. The closed-loop system's behavior should remain predictable $\forall t \geq 0$, not just when $t \rightarrow 0$. Results on finite-time (uniform) guarantees for stochastic systems in discrete-time can be found in~\cite{kumar2019non}, and for continuous-time in~\cite{pham2009contraction} and~\cite{mazumdar2020high}. Using a different analysis, the authors in~\cite{mazumdar2020high} were able to provide bounds on higher-order moments leading to tighter tracking error bounds. Furthermore, the results in~\cite{benveniste1982measure} and~\cite{lakshminarayanan2018linear} require linearity of the systems under consideration.   
 

\textbf{\emph{(\romannum{2})~Control-theoretic approaches:}}
Control Lyapunov function (CLF)-based approaches~\cite{sontag1983lyapunov, khalil2002nonlinear} are the most well-known and applied methods for controlling perturbed deterministic systems. Such methods thus also led to the development of Lyapunov-based approaches for stochastic systems~\cite{deng2001stabilization, liu2008output, deng1997stochastic, deng1999output} wherein notions like that of globally asymptotically stable in probability are used~\cite[Chp.~2]{kushner1967stochastic},~\cite[Chp.~5]{khasminskii2011stochastic}. Asymptotic stability in probability  is a property of the trivial solution (similar to the deterministic counterpart) and hence applies only to systems whose noise vector field vanishes at zero~\cite[Sec.~5.1]{khasminskii2011stochastic}. The absence of such constraints prevents the stabilization of the trivial solution and thus requires robust approaches, e.g., see~\cite[Sec.~4]{deng2001stabilization}. Still, however, being Lyapunov-based approaches, as in the references mentioned above, one attempts to compute a stochastic Lyapunov function which depends explicitly on the noise (diffusion) vector field~\cite[Lem.~2.1]{kushner1967stochastic},~\cite[Thm.~5.3]{khasminskii2011stochastic}. Such requirements exacerbate the already challenging problem of synthesizing (control) Lyapunov functions. 

An approach that avoids the synthesis of stochastic Lyapunov functions is to consider stochastic systems whose robust stability can be determined owing to the stability of the deterministic system counterpart (noiseless stochastic system). For example, the authors in~\cite{mazumdar2020high} derive the stability of a stochastic system using a Lyapunov function for the deterministic counterpart system. However, this approach requires synthesizing a CLF for the deterministic system, which is still no trivial task for nonlinear systems. As an alternative, contraction theory-based solutions offer a computationally tractable convex formulation for searching CLFs for nonlinear systems~\cite {lohmiller1998contraction}. In fact, contraction theory offers necessary \emph{and} sufficient characterization for the stability of trajectories of nonlinear systems~\cite{angeli2002lyapunov}. A few examples of contraction theory-based solutions for control and estimation of nonlinear deterministic systems can be found in, e.g.~\cite{singh2017robust, manchester2017control} and in~\cite{dani2014observer, pham2009contraction} for stochastic systems. The tutorial paper~\cite{tsukamoto2021contraction} provides a comprehensive and exhaustive discussion on using contraction theory to control and estimate nonlinear systems. Using the robustness of the contraction-based controllers, the authors in~\cite{tsukamoto2020robust, tsukamoto2021contraction} prove the robustness of the stochastic counterpart systems. However, as the authors in~\cite{mazumdar2020high} observe, the transfer of the stability properties of the deterministic systems to their stochastic counterparts only holds under certain conditions~\cite{oksendal2013stochastic}. This fact holds for the robust controllers in~\cite{tsukamoto2020robust, tsukamoto2021contraction}, which apply only to stochastic systems whose noise (diffusion) vector fields are uniformly spatially and temporally bounded. The authors in~\cite{tsukamoto2020neural} developed a learning-based synthesis of contraction theory-based control for nonlinear stochastic systems; however, similar to~\cite{tsukamoto2020robust}, the robustness guarantees only hold for systems with bounded noise (diffusion) vector fields.   

In addition to the references provided above, a further few examples of adaptive control for uncertain nonlinear stochastic systems can be found in, e.g.~\cite{min2019globally} where the authors consider a strict-feedback system with time-delays and utilize a stochastic Lyapunov function to show global asymptotic stability in probability, and~\cite{ji2006adaptive}, wherein the authors consider output-feedback tracking for a system with linearly parameterized uncertainties and bounded noise (diffusion) vector field. 
Using deep-learning methods, the authors in~\cite{tsukamoto2021learning} (see also~\cite[Sec.~8]{tsukamoto2021contraction}) provide adaptive control schemes with robustness guarantees for affine and multiplicatively-separable parametric uncertainties and uniformly bounded disturbances. 
The assumption on bounded disturbances precludes the use of very general classes of random processes like the L\'{e}vy processes with Brownian motion as an example with continuous paths~\cite{karatzas1991brownian}.
Similarly, the assumption on structured parametric uncertainties can also be too restrictive to verify. 
Another example of learning-based adaptive control synthesis can be found in~\cite{richards2021adaptive}, where the authors pose the synthesis of adaptive controllers as a meta-learning problem. 
The authors used model ensembling to counter epistemic uncertainties and displayed robustness to distribution shifts. However, this was achieved empirically, thus preventing its use for \emph{a priori} predictive planning. 
Furthermore, the distribution shift was shown for distributions with compact supports thus unable to harness the generality of unbounded support of L\'{e}vy processes like the Brownian motion.  
The result in~\cite{sun2013performance} also uses \ellone-adaptive control, albeit for linear systems, with additive Brownian motion and for linearly parametrized uncertainties affecting solely the drift vector field.

\textbf{\emph{(\romannum{3})~Learning-based control:}} One can find recent results on handling epistemic uncertainties and the associated distribution shift under the umbrella term of learning-based control. For example, the authors in~\cite{kang2022lyapunov} consider discrete-time deterministic systems and combine density models with Lyapunov functions. Hence, the learned agent remains in, or close to, the training distribution during implementation (testing). While the work in~\cite{kang2022lyapunov} and ours share the same goal, the approach taken by the authors is primarily a learning-based data-driven approach where the system is controlled to remain near the training data distribution, thus avoiding high-uncertainty subsets of the state space. 
Reachability analysis for uncertain systems is a popular and effective approach for the safe control of uncertain systems, e.g.~\cite{seo2022real} wherein the authors use the Hamilton-Jacobi reachability analysis for systems under unknown but bounded disturbances. Similarly, the authors in~\cite{knuth2023statistical} analyze goal reachability for a class of uncertain stochastic systems under unknown but bounded stochastic disturbances. As we stated before, using bounded stochastic disturbances disregards beneficial statistical properties of the disturbances and uncertainties, and hence the state distributions. Thus, one must then rely solely on considering the supports of the distributions leading to conservative analyses. While the assumption of bounded disturbances simplifies the analysis relatively since, e.g., as in~\cite{knuth2023statistical}, the system state is differentiable in time even in the presence of stochastic perturbations, such assumptions exclude the use of models driven by L\'{e}vy processes like Brownian motion~\cite{applebaum2009levy}.
Reachability analysis of systems perturbed by noise with non-compactly supported distributions is a challenging prospect as evidenced by recent results in~\cite{hewing2018stochastic, hewing2019scenario, fiacchini2021probabilistic} wherein the authors consider linear systems with additive random disturbances that are possibly correlated, and~\cite{vinod2021stochastic} wherein the authors consider both linear and nonlinear discrete-time systems with the vector fields  assumed to be known. 
The work in~\cite{sivaramakrishnan2023stochastic} considers reachability analysis for nonlinear discrete-time stochastic systems in terms of probability measures corresponding to the state distributions. 
Finally, we refer the interested reader to~\cite{liu2024safety} and~\cite{jafarpour2025probabilistic} where the authors develop non-conservative and computationally efficient approaches for reachability of known continuous-time stochastic systems driven by Brownian motion.   

\textbf{\emph{(\romannum{4})~Distributionally robust control:}} Distributionally robust optimization (DRO) is a sub-field of mathematical optimization that considers obtaining extrema of distributionally ambiguous (uncertain) cost functions, see e.g.~\cite{lin2022distributionally, rahimian2019distributionally, mohajerin2018data} and references therein.  
Therefore, it stands to reason that DRO can assist in achieving the same goals for systems with distributional uncertainties, similar to how one synthesizes robust controllers against systems with compactly supported uncertainties ($\in$ compact sets). Indeed, several recent results use DRO for distributionally robust control (DRC). The results in~\cite{coulson2019regularized, dixit2022distributionally} consider disturbances with unknown distributions in discrete time; however, the dynamics are assumed to be linear with the disturbances affecting the systems additively. Under the same assumptions of linear dynamics and additive disturbances, the authors in~\cite{renganathan2020towards} provided a distributionally robust integration of perception, planning, and control. While the authors in~\cite{zhong2022distributionally} consider DRC for nonlinear systems, the noise is assumed to perturb the system additively. 
The authors in~\cite{yang2020wasserstein, hakobyan2023distributionally} consider nonlinear systems with disturbances of unknown distributions also affecting the system in a nonlinear fashion; however, the authors assume the availability of a (finite) number of samples from the actual disturbance distribution along with the accurate knowledge of the dynamics.
The authors in~\cite{hakobyan2022wasserstein} recently demonstrated the use of DRC for partially observable, albeit linear systems. 
When samples from the true distributions are unavailable, assumptions on the existence of an ambiguity set of distributions is assumed as in~\cite{schuurmans2020learning}. 
Instead of considering systems with distributionally uncertain disturbances,  the DRO formulation can also be used for the case of uncertain environments, see e.g.~\cite{farid2022task, hakobyan2021toward, ren2022distributionally, hakobyan2021wasserstein}.

\subsection{Contributions}\label{subsec:Contributions}
\begin{wrapfigure}[17]{R}{0.5\textwidth}
    \centering
    \includegraphics[width=0.45\textwidth]{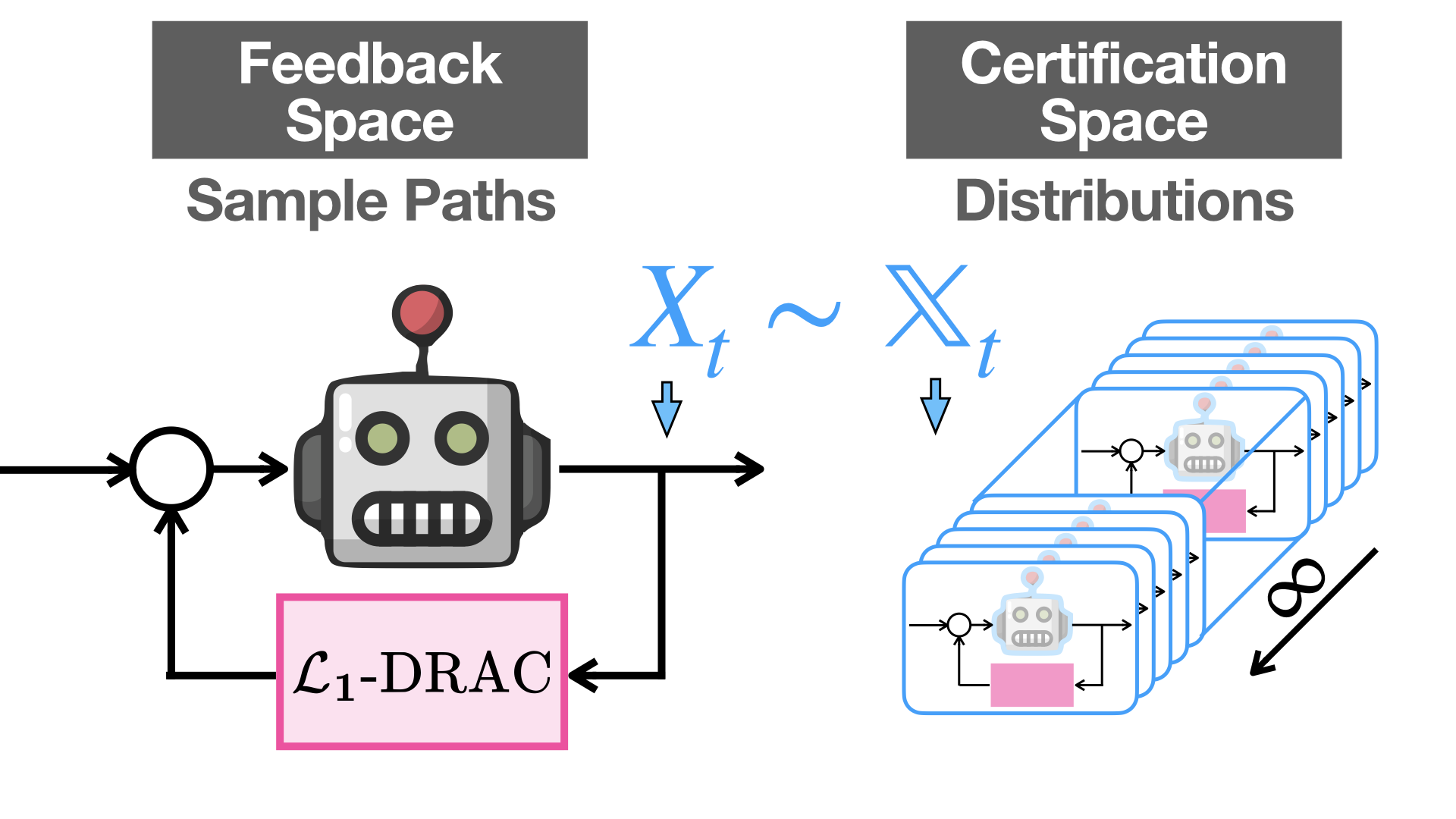}
    \caption{The \ellonedrac controller is designed for a standard implementation and operates under state feedback (the state is denoted by $\Xt{t}$).  
        However, \ellonedrac provides certificates on the {state distribution} $\Xdist{t}$ ($\Xt{t} \sim \Xdist{t}$).  }
    \label{fig:feedback_certificate_spaces}
\end{wrapfigure}
The following are the contributions and the features of the \ellonedrac~control compared to existing results above in Sec.~\ref{subsec:PriorArt}.

\textbf{\emph{(\romannum{1})}}  The design of \ellonedrac~control~ensures the existence of \textbf{\emph{a priori} computable uniform (finite-time) guarantees} for the closed-loop system in terms of \textbf{maximal deviation between the distributions} (probability measures) of the true (uncertain) stochastic system and its nominal (known) version. 
The uniform bounds of distributional deviation are in the form of \textbf{Wasserstein metric} that define the \textbf{guaranteed ambiguity sets (and tubes)}, see Fig.~\ref{fig:MainIllustration:Ambiguity}. 
Moreover, \ellonedrac's guarantees mirror those of the nominal (known) version; \ellonedrac~can ensure the existence of \emph{a priori} Wasserstein ambiguity sets (tubes) \textbf{up to any order identical to the nominal (known) system}, thus leading to, e.g., tighter tracking errors. 
Unlike the existing results we mention in Sec.~\ref{subsec:PriorArt}, we provide finite-time guarantees for \textbf{uncertain nonlinear stochastic systems}, thus avoiding the assumption of accurate knowledge of the system dynamics or their structure, e.g., linear dynamics and additive disturbances. 


\textbf{\emph{(\romannum{2})}} The \ellonedrac~control relies only on the stability of the deterministic counterpart of the known (nominal) subsystem as in~\cite{mazumdar2020high, tsukamoto2020neural, tsukamoto2020robust}. 
However, when compared to the existing control-theoretic approaches, \ellonedrac~is applicable to \textbf{nonlinear systems with (globally) unbounded and state-dependent diffusion and drift uncertainties} which leads to \textbf{instability of the true (uncertain) system} despite the stability of the nominal (known) system. 
The \textbf{uncertainties are \emph{not} required to have any parametric structures} and require mild assumptions on their growth but not their global boundedness. 
Despite the unbounded and non-parametric uncertainties, \ellonedrac~is applicable to systems driven by Brownian motion.


\textbf{\emph{(\romannum{3})}} Unlike the existing learning-based approaches, the design of \ellonedrac~\textbf{does not use data-driven learning} to produce guarantees of uniform and maximal distribution shift and instead relies on an \textbf{adaptive mechanism} to compensate for the uncertainties. While the nominal (known) system and a controller to stabilize it might be learned from data, the distributional robustness that the \ellonedrac~controller guarantees does not require learning. 
Since the \ellonedrac~control ensures  uniform guarantees of maximal distribution-shift, we consider \textbf{\ellonedrac~as an approach where one uses control to enable safe use of data-driven learning}, as opposed to using learning to enable safe control. 

A significant benefit of using \ellonedrac~control is the \textbf{ease of ensuring the safety of uncertain nonlinear stochastic systems} when compared to the existing state-of-the-art. 
For example, \ellonedrac~control can provide a \textbf{computationally inexpensive (optimization-free)} approach to compute \textbf{reachable sets or collision-free plans for safe operation of uncertain nonlinear stochastic systems}. 
Indeed, the \emph{a priori} existence of ambiguity sets that \ellonedrac~control ensures (see Fig.~\ref{fig:MainIllustration}) provides a reachable set of state-distributions (probability measures) for any $t \geq 0$. 
Thus, one only needs to analyze the reachability of the \emph{nominal (known) system}, and the ambiguity set guarantees ``snap-on'' to guarantee reachability for the true (uncertain) system without any additional computation and optimization. 

\textbf{\emph{(\romannum{4})}} The results on distributionally robust control (DRC) we reference above consider accurate dynamics but are perturbed by disturbances of unknown distributions, which implies uncertain distributions over states despite the accurate knowledge of the dynamics. 
Thus, one may assume the accurate dynamics to correspond to the known (nominal) version of the true (uncertain) systems. 
At the same time, the unknown distribution of the disturbance subsumes the effects of epistemic uncertainties and noise. 
We instead consider \textbf{uncertain nonlinear systems driven by Brownian motion}. 
Even though the distribution of Brownian motion is well-known, e.g., normally distributed and independent increments~\cite{durrett1996stochastic, karatzas1991brownian}, the presence of state-dependent uncertainties in \emph{both} drift and diffusion terms imply that the \textbf{random perturbations affecting the systems   possess unknown distributions, are correlated, appear multiplicatively, can affect the control channel, and thus lead to uncertain state distributions}.
Therefore, while the DRC results and our work consider different systems regarding the knowledge of the dynamics and the distribution of disturbance/noise, both lead to uncertain state distributions.

However, unlike the state-of-the-art for DRC, \textbf{\ellonedrac~does not require \emph{a priori} existence or knowledge of ambiguity sets for the true (uncertain) state distributions, nor the \emph{a priori} availability of (finite) samples from the uncertain distributions}. 
Indeed, the existence and knowledge of ambiguity sets are not guaranteed for the uncertain systems we consider. 
We do not make the unverifiable assumption of \emph{a priori} availability of samples from the true (uncertain) state distributions because the \textbf{\ellonedrac~controller} uses state-feedback, and hence, it \textbf{receives exactly \emph{one} sample from the time-varying uncertain state distribution at each point in time}. 
We emphasize the point that \ellonedrac retains a standard controller implementation operating under state-feedback while offering robustness guarantees in the space of distributions. 
Thus, unlike a majority of approaches where the control law's feedback space and the certification space are identical (e.g. feedback $\in$ state space $\mathbb{R}^n$ and invariant sets also $\in \,\mathbb{R}^n$), the \ellonedrac is designed for the case when the feedback space (state space) $\neq$ certification space (probability measures on the state space), see Fig.~\ref{fig:feedback_certificate_spaces}. 
Under slightly stronger conditions than the ones required for well-posedness for nonlinear SDEs, we show 1) the \textbf{existence of the ambiguity sets (tubes) for the true (uncertain) distributions}, 2) the existence of the \textbf{\emph{a priori} computable radius of ambiguity sets over temporal state measures}, and 3) a certain amount of control on the radius of the ambiguity sets \textbf{allowing us to exercise a degree of control, under reasonable tradeoffs, on limiting the distribution shift}.

\subsection{Article Organization}

We present the systems under consideration in Sec.~\ref{subsec:Systems}, followed by the required assumptions, their justifications, and consequences in Sec.~\ref{subsec:Assumptions}.  
The problem statement is stated in Sec.~\ref{subsec:ProblemStatement}.
Section~\ref{sec:L1DRAC:defn} presents the core results on the design and analysis of the \ellonedrac law. 
Section~\ref{subsec:L1DRAC:Architecture} provides the architecture of \ellonedrac, and Sec.~\ref{subsec:L1DRAC:Parameters} provides the design choice of the pertinent control parameters. 
The performance analysis of \ellonedrac is provided in Sec.~\ref{sec:L1DRAC:Analysis}, and the associated discussion is presented in Sec.~\ref{sec:Discussion}.
The manuscript is concluded with Sec.~\ref{sec:ConclusionFuture}. 

A majority of derivations are presented in the appendix: Appendix~\ref{app:Definitions} collects all the pertinent definitions as a quick reference, Appendix~\ref{app:TechnicalResults} contains the supporting technical results, and Appendices~\ref{app:ReferenceProcess} and~\ref{app:TrueProcess} provide derivations for the analysis in Sec.~\ref{subsec:Analysis:ReferenceProcess} and~\ref{subsec:Analysis:TrueUncertainProcess}, respectively. 

\subsection{Notation}\label{sec:Notation}

We define $a \wedge b \doteq \min\cbr{a,b}$, for any $a,b \in \mathbb{R}$, and $\indicator{x \in A}{x}$ is the indicator function of set $A$ evaluated at $x$.
We denote by $\mathbb{R}_{> 0}$ ($\mathbb{R}_{\geq 0}$) the set of positive (non-negative) reals. 
For any $\alpha \in \mathbb{N}$, we denote by $\mathbb{N}_{\geq \alpha}$ the set of naturals $\geq \alpha$.
The set of linear operators $\mathbb{R}^n \rightarrow \mathbb{R}^m$, $n,m \in \mathbb{N}$, are denoted by $\mathbb{R}^{n \times m}$, while $\mathbb{S}^n$ denotes self-adjoint operators  $\mathbb{R}^n \rightarrow \mathbb{R}^n$.    
We denote by $\mathbb{S}^n_{\succ 0}$ ($\mathbb{S}^n_{\succeq 0}$) the set of positive (semi)definite symmetric maps on $\mathbb{R}^n$.
Unless explicitly stated, $\norm{\cdot}$ denotes the Euclidean 2-norm on $\mathbb{R}^n$, $n \in \mathbb{N}$.
On the space $\mathbb{R}^{n \times m}$, $n,m \in \mathbb{N}$, $\norm{\cdot}$ denotes the induced 2-norm, while we denote the Frobenius norm on the same space by $\Frobenius{\cdot}$.
The set of all continuous maps from some $\mathcal{X}$ to $\mathcal{Y}$ is denoted by  $\mathcal{C}\br{\mathcal{X};\mathcal{Y}}$. 
Similarly, $\mathcal{C}^k \br{\mathcal{X};\mathcal{Y}}$, $k \in \cbr{\mathbb{N},\infty}$, denotes $k$-times continuously differentiable maps.   
For any $f \in \mathcal{C}^2\left(\mathbb{R}^n;\mathbb{R} \right)$
\begin{align*}
    \nabla f(x) \doteq \begin{bmatrix} \partial_{x_1}f(x) & \cdots \partial_{x_n}f(x)  \end{bmatrix}^\top \in \mathbb{R}^n, 
    \\
    \nabla^2 f(x) \doteq \nabla \cdot \left( \nabla f \right)^\top = \begin{bmatrix} \nabla \partial_{x_1}f(x) & \cdots \nabla \partial_{x_n}f(x)  \end{bmatrix} \in \mathbb{S}^{n}.
\end{align*}
Similarly, 
\begin{align*} 
    \nabla g(x) \doteq \begin{bmatrix} \nabla g_1(x) & \cdots & \nabla g_m(x)   \end{bmatrix} \in \mathbb{R}^{n \times m},
    \quad 
    g \in \mathcal{C}^1\left(\mathbb{R}^n;\mathbb{R}^m \right).
\end{align*}

For $T,p \in (0,\infty)$ and any finite-dimensional space $\mathcal{X}$, we define
\begin{align*}
    \mathfrak{L}_{p}\br{[0,T];\mathcal{X}}
    =
    \left\{
        z:[0,T] \rightarrow \mathcal{X}
        ~|~
        \norm{z}_{\mathfrak{L}_{p}} = 
        \left( \int_0^T \norm{z(\nu)}_{\mathcal{X}}^p d\nu \right)^\frac{1}{p} < \infty
    \right\},
\end{align*}
and we say $z \in \mathfrak{L}_{p}^{loc}\br{\mathcal{X}}$ if $z \in \mathfrak{L}_{p}\br{[0,T];\mathcal{X}}$ for every $T>0$.  

Let $\br{\Omega, \mathcal{F}, \mathbb{P}}$ be a complete probability space that we consider as the underlying space throughout the manuscript.
For a filtration $\mathcal{F}_t$ on the probability space above, we define\footnote{Identical definitions can be found in, e.g., $\mathcal{L}_{ad}$ in~\cite[Notation~5.1.1]{kuo2006introduction} and $\mathcal{L}^p$ in~\cite{mao2007stochastic}. } 
\begin{align*}
    L_{p}\left( [0,T];\mathcal{X}~|~\mathcal{F}_t \right)
    =
    \left\{ 
        \vphantom{\frac{1}{2}}
        \mathfrak{F}_t\text{-adapted }
        z(t,\cdot): \Omega \rightarrow \mathcal{X}
        ~|~
        z(\cdot,\upomega) \in \mathfrak{L}_{p}\br{[0,T];\mathcal{X} }
        ,
        \quad 
        \upomega \in \Omega
    \right\},
\end{align*} 
and $z \in L_{p}^{loc}\left(\mathcal{X}~|~\mathcal{F}_t \right)$ if $z \in L_{p}\left( [0,T];\mathcal{X}~|~\mathcal{F}_t \right)$ at every $T > 0$ assuming the filtration exists. 
We similarly define 
\begin{align*}
    \begin{multlined}[][0.8\linewidth]
        \mathcal{M}_{p}\left( [0,T];\mathcal{X}~|~\mathcal{F}_t \right)
        =
        \left\{
            z
            \in 
            L_{p}\left( [0,T];\mathcal{X}~|~\mathcal{F}_t \right)
            ~|~
            \ELaw{}{ \norm{z}_{\mathfrak{L}_p}^p }^\frac{1}{p}
            = 
            \left(
                \int_\Omega 
                    \norm{z(\cdot,\upomega)}_{\mathfrak{L}_p}^p
                \Probability{d\upomega}
            \right)^\frac{1}{p}
            < \infty
        \right\},
    \end{multlined} 
\end{align*}
and $\mathcal{M}_{p}^{loc}\left(\mathcal{X}~|~\mathcal{F}_t \right)$ is defined analogously to $L_{p}^{loc}\left(\mathcal{X}~|~\mathcal{F}_t \right)$.
With an abuse of notation, for any $\mathfrak{F}_t$-adapted $z \in \mathcal{C}\left([0,T];\mathcal{X}\right)$ we write 
\begin{align*}
    \norm{z(t)}_{L_p}
    =
    \ELaw{}{ \norm{z}_{\mathcal{X}}^p }^\frac{1}{p}
    = 
    \left(
        \int_\Omega 
            \norm{z(\cdot,\upomega)}_{\mathcal{X}}^p
        \Probability{d\upomega}
    \right)^\frac{1}{p}
    < \infty, \quad t \in [0,T].
\end{align*}
Note that $\norm{\cdot}_{L_p}$ is \textbf{not a norm} on the space $L_p$ defined above since the integrand in $\norm{\cdot}_{L_p}$ uses the finite-dimensional norm $\norm{\cdot}_{\mathcal{X}}$, and not the infinite-dimensional $\norm{\cdot}_{{\mathfrak{L}_p}}$.
\begin{remark}
    While it is standard in the literature to represent the sample points $\in \Omega$ by $\omega$, we use the upright version $\upomega \in \Omega$ instead.
    This allows us to use the standard version $\Boldomega$ to represent the control parameter of filter bandwidth (yet to be defined). 
    Moreover, as is standard, we suppress the argument $\upomega \in \Omega$ for functions with domain $\Omega$.
    Thus, from this point onward and unless explicitly stated, $\Boldomega$ denotes a control parameter, and not a sample point of $\Omega$. 
\end{remark}
The Borel $\sigma$-algebra over a topological space $\mathcal{Y}$ is denoted by $\Borel{\mathcal{Y}}$, and $\sigma (z)$ denotes the $\sigma$-algebra generated by any measurable $z:\br{\Omega,\mathcal{F}} \rightarrow \br{\mathcal{Y},\Borel{\mathcal{Y}}} $.
The notation $\mathcal{F}_1 \vee \mathcal{F}_2$ denotes the join of $\sigma$-algebras $\mathcal{F}_1$ and $\mathcal{F}_2$.

The Wasserstein metric of order $p \in [0,\infty)$ between probability measures $\varsigma_1$ and $\varsigma_2$ on a Polish space (complete and separable metric space~\cite[Defn.~18.1]{fristedt2013modern}) $\left(\mathcal{Z}, d_{\mathcal{Z}}\right)$ is given as follows~\cite[Defn.~6.1]{villani2009optimal}:
\begin{align*}
    \pWass{p}{\varsigma_1}{\varsigma_2}
    =
    \left(
        \inf_{\pi \in \Pi(\varsigma_1,\varsigma_2)}
        \int_{\mathcal{Z}}
            d_{\mathcal{Z}}\left(z_1,z_2\right)^p
        \pi\left(dz_1,dz_2\right)
    \right)^\frac{1}{p},
\end{align*}
where $d_{\mathcal{Z}}$ is the distance function on $\mathcal{Z}$, and $\Pi(\varsigma_1,\varsigma_2)$ is the set of all couplings between measures $\varsigma_1$ and $\varsigma_2$.

\section{Systems, Assumptions, and Problem Statement }\label{sec:SystemsAssumptionsProblemStatement}

We begin with the definitions of the stochastic processes under consideration.

\subsection{The Systems}\label{subsec:Systems}

The following defines the known and unknown drift and diffusion vector fields. 
\begin{definition}[Vector Fields]\label{def:VectorFields}
    Consider the \textbf{known functions} $f:\mathbb{R}_{\geq 0} \times \mathbb{R}^n \rightarrow \mathbb{R}^n$, $g: \mathbb{R}_{\geq 0} \rightarrow  \mathbb{R}^{n \times m}$, and $p:\mathbb{R}_{\geq 0} \times \mathbb{R}^n \rightarrow \mathbb{R}^{n \times d}$, for $n,m,d \in \mathbb{N}$. 
    Consider also the \textbf{unknown functions} $\Lambda_\mu :\mathbb{R}_{\geq 0} \times \mathbb{R}^n \rightarrow \mathbb{R}^n$ and $\Lambda_\sigma :\mathbb{R}_{\geq 0} \times \mathbb{R}^n \rightarrow \mathbb{R}^{n \times d}$.
    For any $(a,b,,t) \in \mathbb{R}^n \times \mathbb{R}^m \times \mathbb{R}_{\geq 0}$, we denote by
    \begin{align}\label{eqn:TrueVectorFields}
            \Fmu{t,a,b} \doteq  f(t,a) + g(t) b + \Lmu{t,a}  \in \mathbb{R}^n, \quad
            \Fsigma{t,a} \doteq   p(t,a) + \Lsigma{t,a}  \in \mathbb{R}^{n \times d}
    \end{align}
    the \textbf{true (uncertain) drift and diffusion vector fields}, respectively.
    Similarly, for any $(a,t) \in \mathbb{R}^n \times \mathbb{R}_{\geq 0}$, we denote by
    \begin{align}\label{eqn:NominalVectorFields}
            \Fbarmu{t,a} \doteq  f(t,a)  \in \mathbb{R}^n, \quad
            \Fbarsigma{t,a} \doteq   p(t,a) \in \mathbb{R}^{n \times d} 
    \end{align}
    the \textbf{nominal (known) drift and diffusion vector fields}, respectively.
    Note that we have the following decompositions
    \begin{align}\label{eqn:VectorFields:Decomposition}
        \Fmu{t,a,b} = \Fbarmu{t,a} + g(t) b + \Lmu{t,a}\in \mathbb{R}^n,
        \quad 
        \Fsigma{t,a} = \Fbarsigma{t,a} + \Lsigma{t,a}  \in \mathbb{R}^{n \times d}.
    \end{align}
\end{definition}

We now use the vector fields in Definition~\ref{def:VectorFields} to define the true and nominal processes that we study with respect to each other in this manuscript.
The notations in the following mostly follow the convention in~\cite{oksendal2013stochastic}.
The underlying complete probability space is $\br{\Omega, \mathcal{F}, \mathbb{P}}$ that we introduced in Sec.~\ref{sec:Notation}.
\begin{definition}[Processes]\label{def:TrueandNominalProcesses}
    We denote by $\Wt{t}$ and $\Wstart{t}$ any two \textbf{independent} $\mathbb{P}$-Brownian motions. 
    The filtrations generated by $\Wt{t}$ and $\Wstart{t}$ are denoted by $\Wfilt{t}$ and $\Wstarfilt{t}$, respectively, and  we define $\Wfilt{\infty} = \sigma \br{ \cup_{t \geq} \Wfilt{t}}$ and $\Wstarfilt{\infty} = \sigma \br{ \cup_{t \geq} \Wstarfilt{t}}$.
    Let $x_{0} \sim \xi_0$ and $x_0^\star \sim \xi_0^\star$ be two $\mathbb{R}^n$-valued random variables that are respectively independent of $\Wfilt{\infty}$ and $\Wstarfilt{\infty}$, where $\xi_0$ and $\xi_0^\star$ are probability measures on $\Borel{\mathbb{R}^n}$.
    We further define $\Wfilt{0,t} = \sigma\br{\xi_0} \vee \Wfilt{t}$ and $\Wstarfilt{0,t} = \sigma\br{\xi_0^\star} \vee \Wstarfilt{t}$. 

    For any $T \in (0,\infty)$, we say that $\Xt{}, \Xstart{} \in \continuous{T}{n}$ are the \textbf{true (uncertain) and nominal (known) processes}, respectively, if they are respectively adapted to the filtrations $\Wfilt{0,t}$ and $\Wstarfilt{0,t}$, and are the unique strong solutions~\cite[Defn.~5.2.1]{karatzas1991brownian} to the following \ito stochastic differential equations (SDEs), for all $t \in [0,T]$:  
    \begin{subequations}\label{eqn:AllSDE}
        \begin{align}
            d\Xt{t} = \Fmu{t,\Xt{t},\Ut{t}}dt + \Fsigma{t,\Xt{t}}d\Wt{t}, \quad \Xt{0} = x_0 \sim \xi_0~(\text{\Pas}), \label{eqn:TrueUncertainSDE}
            \\
            d\Xstart{t} = \Fbarmu{t,\Xstart{t}}dt + \Fbarsigma{t,\Xstart{t}}d\Wstart{t}, \quad \Xstart{0} = x_0^\star \sim \xi^\star_0~(\text{\Pas}), \label{eqn:NominalKnownSDE}
        \end{align}
    \end{subequations}
    where the vector fields $F_{\cbr{\mu,\sigma}}$ and $\bar{F}_{\cbr{\mu,\sigma}}$, are presented in Definition~\ref{def:VectorFields}, and $\Ut{t}$ is some yet-to-be-defined process.
    We refer to $\Ut{t}$ as the \textbf{feedback process}.\footnote{We will drop the quantifier \Pas from here on.} 
    
    We define the \textbf{true law} $\Xdist{t}$ as 
    \begin{align*}
        \Xdist{t}\br{B} \doteq \Probability{\Xt{t}^{-1}\br{B}},  
        \quad 
        (t,B) \in [0,T] \times \Borel{\mathbb{R}^n}.
    \end{align*} 
    The \textbf{nominal law} $\Xstardist{t}$ is defined analogously for $\Xstart{t}$. 
    Thus, $\Xt{t} \sim \Xdist{t}$ and $\Xstart{t} \sim \Xstardist{t}$.

\end{definition}

The independence of the Brownian motions $\Wt{t}$ and $\Wstart{t}$ follows from the fact that distinct real-world systems will be affected by distinct realizations of the random noise.
We will later discuss the consequences on the main results of the manuscript. 
See~\cite{pham2009contraction} for a further discussion.

\begin{remark}\label{rem:Laws}
    Since $\Xt{} \in \continuous{T}{n}$, instead of considering $\Xt{t}$ as a pointwise-in-time random variable  on $ \mathbb{R}^n$, one may consider the trajectory (sample path) $\Xt{[0,T]} \doteq \left(\Xt{t}\right)_{0 \leq t \leq T}$ as a random variable on $ \continuous{T}{n}$. 
    This then motivates the definition of the (path) law $\Xdist{[0,T]}$, $\Xt{[0,T]} \sim \Xdist{[0,T]}$, as follows~\cite[Sec.~2.4, p.~60]{karatzas1991brownian}: 
    \begin{align*}
        \Xdist{[0,T]}\br{B}
        =
        \Probability{\Xt{[0,T]}^{-1}\br{B}},  
        \quad 
        (t,B) \in [0,T] \times \Borel{\continuous{T}{n}}.
    \end{align*}
    Along with the separability of strong solutions~\cite[Thm.~5.2.1]{ash2014topics}, the Kolmogorov extension theorem~\cite[Thm.~2.2]{karatzas1991brownian} implies that the finite-dimensional cylinder sets form a determining class for the probability measures on $\Borel{\continuous{T}{n}}$~\cite[Thm~2.0]{balakrishnan2012stochastic}.
    Consequently, the path law can be represented alternatively as follows~\cite[Sec.~7.1]{oksendal2013stochastic}:
    \begin{align*}
        \Xdist{t_1\cdots t_k}\br{B_1 \times  \cdots \times B_k}
        =
        \Probability{\Xt{t_1}^{-1}\br{B_1},\dots,\Xt{t_k}^{-1}\br{B_k}},
    \end{align*}
    for all choice of temporal instances $0 < t_1 < \ldots < t_k \leq T$, and all $B_1, \ldots, B_k \in \Borel{\mathbb{R}^n}$~\cite[Sec.2.1,~p.~12]{balakrishnan2012stochastic}.
    We thus see that the law $\Xdist{t}$, $t \in [0,T]$, in Definition~\ref{def:TrueandNominalProcesses} can be interpreted as the probability measure defined on a cylinder set with $k=1$.
\end{remark}

\begingroup
\endgroup

\subsection{Assumptions}\label{subsec:Assumptions}

We begin with the assumptions for the nominal (known) system in~\eqref{eqn:NominalKnownSDE}.
\begin{assumption}[Nominal (Known) System]\label{assmp:KnownFunctions}
    The known drift function $f(t,a)$ in Definition~\ref{def:VectorFields} is locally Lipschitz in $a \in \mathbb{R}^n$, uniformly in $t \in \mathbb{R}_{\geq 0}$, and there exists a known $\Delta_f  \in \mathbb{R}_{>0}$ such that
    \begin{align*}
        \norm{f(t,a)}^2 \leq \Delta_f^2 \left(1 + \norm{a}^2 \right),
        \quad 
        \forall (t,a) \in \mathbb{R}_{\ge 0} \times \mathbb{R}^n.
    \end{align*} 
    Furthermore, the known diffusion function $p(t,a)$ in Definition~\ref{def:VectorFields} is uniformly bounded and globally Lipschitz in its arguments, and there exist known $\Delta_p, \hat{L}_p, L_p  \in \mathbb{R}_{>0}$ such that 
    \begin{align*}
        \norm{p(t,a)}_F \leq \Delta_p, 
        \, \Frobenius{p(t,a) - p(t',a')} \leq \hat{L}_p \absolute{t - t'} + L_p \norm{a - a'}^\frac{1}{2},
        \quad 
        \forall (t,t',a,a') \in \mathbb{R}_{\ge 0} \times \mathbb{R}_{\ge 0} \times \mathbb{R}^n \times \mathbb{R}^n. 
    \end{align*}
 
    The input operator $g: \mathbb{R}_{\geq 0} \rightarrow \mathbb{R}^{n \times m}$ has full column rank $\forall t \in \mathbb{R}_{\geq 0}$ and satisfies 
    \begin{align*}
        g \in \mathcal{C}^1([0,\infty);\mathbb{R}^{n \times m}), \quad \norm{g(t)}_F \leq \Delta_g, \quad \norm{\dot{g}(t)}_F \leq \Delta_{\dot{g}}, \forall t \in \mathbb{R}_{\geq 0},
    \end{align*}
    where $\Delta_g,\Delta_{\dot{g}} \in \mathbb{R}_{>0}$ are assumed to be known.
    Since $g(t)$ is full rank, we can construct a $g^\perp: \mathbb{R}_{\geq 0} \rightarrow \mathbb{R}^{n \times n-m}$ such that $\text{\emph{Im}} \, g(t)^\perp = \text{\emph{ker}} \, g(t)^\top$ and $\text{\emph{rank}}\br{\bar{g}(t)}=n$, $\forall t \in \mathbb{R}_{\geq 0}$ where 
    \begin{align*}
        \bar{g}(t) \doteq \begin{bmatrix} g(t) & g(t)^\perp  \end{bmatrix} \in \mathbb{R}^{n \times n}.
    \end{align*}
    We assume that $g^\perp(t) \in \mathbb{R}^{n \times n-m}$ is uniformly Lipschitz continuous in $t \in \mathbb{R}_{\geq 0}$, and $\norm{g(t)^\perp}_F \leq \Delta_g^\perp$, $\forall t \in \mathbb{R}_{\geq 0}$, where $\Delta_g^\perp \in \mathbb{R}_{>0}$ is assumed to be known.
    We further assume there exists a known scalar $\Delta_\Theta \in \mathbb{R}_{>0}$ such that
    \begin{align*}
        \Frobenius{\Theta_{ad}(t)} \leq \Delta_\Theta, \quad \forall t \in \mathbb{R}_{\geq 0}, 
        \text{ where }
        \Theta_{ad}(t) \doteq \begin{bmatrix}\mathbb{I}_m & 0_{m,n-m}  \end{bmatrix} \bar{g}(t)^{-1} \in \mathbb{R}^{m \times n}.
    \end{align*}  
    
    Finally, the initial condition $x^\star_0 \sim \xi^\star_0$ satisfies $\LpLaw{2\mathsf{p}^\star}{}{x^\star_0} < \infty$, for some $\mathsf{p}^\star \in \mathbb{N}_{\geq 1}$.  
\end{assumption}
\begin{remark}\label{rem:KnownFunctionsAssumption}
    %
    We assume the uniform boundedness of the known diffusion term $p(t,x)$, instead of linear growth, since it is a sufficient condition to transfer the stability of the deterministic counterpart of~\eqref{eqn:NominalKnownSDE} ($\Wstart{t} \equiv 0$) to the stochastic version via robustness arguments, see e.g.,~\cite{pham2009contraction, tsukamoto2021contraction, tsukamoto2020robust}. 
    However, as we will see below (see Assumption~\ref{assmp:UnknownFunctions} and the subsequent comments in Remark~\ref{assmp:UnknownFunctions}), we do not place any such assumptions on the diffusion term of the true (uncertain) system in~\eqref{eqn:TrueUncertainSDE}. 
    Therefore, the stability of~\eqref{eqn:TrueUncertainSDE} is not guaranteed by the stability of its deterministic counterpart. 
\end{remark}
\begin{remark}
    The assumption on the initial condition implies bounded second moments as a minimum. 
    Along with the initial condition's independence from $\Wstarfilt{\infty}$ (see Defn.~\ref{def:TrueandNominalProcesses}), this assumption is standard for the well-posedness of SDEs, see e.g.~[Thm.~5.2.1]\cite{oksendal2013stochastic}.
\end{remark}

We now place assumptions on the stability of the deterministic counterpart ($\Wstart{t} \equiv 0$) of the nominal (known) system in~\eqref{eqn:NominalKnownSDE}.  
We need the following definition of the the \emph{induced logarithmic norm} of $A \in \mathbb{R}^{n \times n}$:
    \begin{align*}
        \mu\left(A\right) = \lim_{h \rightarrow 0^+} \frac{\norm{\mathbb{I}_n + hA} - 1}{h}.        
    \end{align*}
The induced logarithmic norm (\emph{log norm} for brevity) is guaranteed to exist and can be interpreted as the derivative of $\expo{At}$ in the direction of $A$ and evaluated at $\mathbb{I}_n$.
Refer to~\cite[Sec.~2.3]{bulloContractionBook} for further details.  
\begin{assumption}[Nominal (Known) System Stability]\label{assmp:ILF}
    There exists a known $\lambda \in \mathbb{R}_{>0}$ such that 
    \begin{align}\label{eqn:ILFConditions} 
        \mu\left(
           \nabla_x \Fbarmu{t,x}  
        \right)
        \leq -\lambda, \quad \forall (t,x) \in \mathbb{R}_{\geq 0} \times \mathbb{R}^n,
    \end{align}
    where $\bar{F}_\mu$ is defined in~\eqref{eqn:NominalVectorFields}.
\end{assumption}
The condition in~\eqref{eqn:ILFConditions} allows one to use contraction theory to conclude the \emph{incremental exponential stability (IES)} of the deterministic system $\dot{x} = \Fbarmu{t,x}$~\cite[Thm.~3.9]{bulloContractionBook}.
The following lemma provides a vital consequence of the above.
\begin{lemma}\label{lem:ILFConditions:Consequence}[\cite{davydov2022non}]
    Given the deterministic system $\dot{x} = \Fbarmu{t,x}$, the statement in~\eqref{eqn:ILFConditions} is equivalent to  
    \begin{equation}\label{eqn:ILFConditions:Consequence}
        \left(x-y\right)^\top 
        \left( \Fbarmu{t,x} - \Fbarmu{t,y}  \right)
        \leq      
        -\lambda
        \norm{x-y}^2
        , 
        \quad 
        \forall (t,x,y) \in \mathbb{R}_{\geq 0} \times \mathbb{R}^n \times \mathbb{R}^n.
    \end{equation}
\end{lemma}
The use of the matrix measure (log norm) is one of the many approaches for characterizing the contractive property of the vector field $\Fbarmu{t,x}$~\cite{sontag2010contractive,aminzare2014contraction,davydov2022non}. 
Additional approaches include: incremental Lyapunov Functions~\cite{angeli2002lyapunov,angeli2009further}, Finsler-Lyapunov functions~\cite{forni2013differential,wu2021further}; and Riemannian metric conditions~\cite{lohmiller1998contraction, simpson2014contraction}. 
From a synthesis perspective, the concept of a \emph{control contraction metric (CCM)}, a Riemannian metric, has shown immense utility for controlled systems, e.g.,~\cite{manchester2017control,singh2019robust,tsukamoto2021contraction}.

\begingroup
\endgroup

Since we aim to establish the stability of the true (uncertain) process~\eqref{eqn:TrueUncertainSDE} with respect to the nominal (known) process~\eqref{eqn:NominalKnownSDE}, it is a reasonable requirement that the latter 
    satisfies certain properties that represent the desired behavior that we wish to endow upon the uncertain system.  
We state these in the next assumption on the statistical nature of~\eqref{eqn:NominalKnownSDE}.
\begin{assumption}\label{assmp:NominalSystem:FiniteMomentsWasserstein}
    There exists a known $\Delta_\star \in \mathbb{R}_{>0}$, such that, for any $T \in (0,\infty)$, the strong solution $\Xstart{t}$ of the nominal (known) process~\eqref{eqn:NominalKnownSDE} exists and satisfies
    \begin{align*}
        \LpLaw{2 \mathsf{p}^\star}{}{\Xstart{t} } \leq \Delta_\star,
        \quad \forall t \in [0,T], 
    \end{align*}
    where $\mathsf{p}^\star \in \mathbb{N}_{\geq 1}$ is stated in Assumption~\ref{assmp:KnownFunctions}.
\end{assumption}

\begin{remark}
    %
   
    The well-posedness of~\eqref{eqn:NominalKnownSDE} is straightforward to establish under general conditions, see e.g.~\cite[Definition~5.2.1]{karatzas1991brownian},~\cite[Sec.~4.5]{kloeden1992stochastic}, and~\cite[Sec.~5.2]{oksendal2013stochastic}. 
\end{remark}

Next, we state the assumptions that we place on the true (uncertain) system in~\eqref{eqn:TrueUncertainSDE}.
\begin{assumption}[True (Uncertain) System]\label{assmp:UnknownFunctions}
    The unknown functions $\Lambda_\mu$ and $\Lambda_\sigma$, presented in Definition~\ref{def:VectorFields}, satisfy 
    \begin{align*}
        \norm{\Lmu{t,a}}^2 \leq \Delta_\mu^2 \br{1 + \norm{a}^2}, \quad 
        \norm{\Lsigma{t,a}}_F^2 \leq \Delta_\sigma^2 \br{1 + \norm{a}^2}^\frac{1}{2},~\forall (t,a) \in \mathbb{R}_{\geq 0} \times \mathbb{R}^n, 
    \end{align*}
    where $\Delta_{\mu}, \Delta_{\sigma} \in \mathbb{R}_{>0}$ are known. Furthermore, 
    define $\Lambda_\mu^{\paral}: \mathbb{R}_{\geq 0} \times \mathbb{R}^n \rightarrow \mathbb{R}^m$, $\Lambda_\mu^\perp: \mathbb{R}_{\geq 0} \times \mathbb{R}^n \rightarrow \mathbb{R}^{n-m}$, $\Lambda_\sigma^{\paral}: \mathbb{R}_{\geq 0} \times \mathbb{R}^n \rightarrow \mathbb{R}^{m \times d}$, and $\Lambda_\sigma^\perp: \mathbb{R}_{\geq 0} \times \mathbb{R}^n \rightarrow \mathbb{R}^{n-m \times d}$ as
    \begin{align}\label{eqn:UnknownFunctions:Decomposition}
        \begin{bmatrix} \Lparamu{t,a} \\ \Lperpmu{t,a} \end{bmatrix}
        = \begin{bmatrix} g(t) & g(t)^\perp \end{bmatrix}^{-1} \Lmu{t,a}, \quad 
        \begin{bmatrix} \Lparasigma{t,a} \\ \Lperpsigma{t,a} \end{bmatrix}
        = \begin{bmatrix} g(t) & g(t)^\perp \end{bmatrix}^{-1} \Lsigma{t,a},~\forall (t,a) \in \mathbb{R}_{\geq 0} \times \mathbb{R}^n,
    \end{align}
    where the input operator $g$ is presented in Definition~\ref{def:VectorFields}, and $g^\perp$ is defined in Assumption~\ref{assmp:KnownFunctions}.
    As a consequence of the growth bounds, the functions above satisfy
    \begin{align*}
        \norm{\Lambda_\mu^{\cbr{\paral,\perp}}(t,a)}^2 
        \leq 
        \br{\Delta_\mu^{\cbr{\paral,\perp}}}^2 \br{1 + \norm{a}^2}, \quad 
        \norm{\Lambda_\sigma^{\cbr{\paral,\perp}}(t,a)}_F^2 
        \leq 
        \br{\Delta_\sigma^{\cbr{\paral,\perp}}}^2 \br{1 + \norm{a}^2}^\frac{1}{2},~\forall (t,a) \in \mathbb{R}_{\geq 0} \times \mathbb{R}^n,
    \end{align*}
    for known $\Delta_\mu^{\paral},\, \Delta_\mu^\perp,\, \Delta_\sigma^{\paral},\, \Delta_\sigma^\perp \in \mathbb{R}_{>0}$.
 \end{assumption}
\begin{remark}\label{rem:UnknownFunctions}
    The growth bound on drift uncertainty $\Lambda_\mu$ is the standard general linear growth condition. 
    For diffusion uncertainty $\Lambda_\sigma$, our analysis can compensate for uncertainties growing sub-linearly; we do not constrain $\Lambda_\sigma$ to be uniformly bounded. 
    An implication of this is that the stability of the nominal (known) system in~\eqref{eqn:NominalKnownSDE} cannot be extended to the uncertain (true) system in~\eqref{eqn:TrueUncertainSDE} via robustness arguments as done in~\cite{tsukamoto2020robust, tsukamoto2021contraction}. 
    The \ellonedrac~control can, however, accommodate unbounded drift uncertainties satisfying the sub-linear growth condition.
\end{remark}
Similar to the decomposition of $\Lambda_\mu$ and $\Lambda_\sigma$ in the last assumption, we decompose the known diffusion term $p$ as presented below. 
\begin{assumption}\label{assmp:knownDiffusion:Decomposition}
    For the known diffusion term $p$ in Definition~\ref{def:VectorFields}, we define 
    $p^{\paral}: \mathbb{R}_{\geq 0} \times \mathbb{R}^n \rightarrow \mathbb{R}^{m \times d}$, and $p^\perp: \mathbb{R}_{\geq 0} \times \mathbb{R}^n \rightarrow \mathbb{R}^{n-m \times d}$ as
    \begin{align}\label{eqn:knownDiffusion:Decomposition:p} 
        \begin{bmatrix} \ppara{t,a} \\ \pperp{t,a} \end{bmatrix}
        = \begin{bmatrix} g(t) & g(t)^\perp \end{bmatrix}^{-1} p(t,a),~\forall (t,a) \in \mathbb{R}_{\geq 0} \times \mathbb{R}^n,
    \end{align}
    where, we assume that $p^{\paral}$ and $p^\perp$ are Lipschitz continuous, locally in $a \in \mathbb{R}^n$ and uniformly in $t \in \mathbb{R}_{\geq 0}$.
    
    Due to the uniform boundedness and H\"{o}lder continuity of $p$ in Assumption~\ref{assmp:KnownFunctions}, we assume that 
    there exist known $\Delta_p^{\paral}, \Delta_p^\perp, L^{\paral}_p, \hat{L}^{\paral}_p, L^{\perp}_p, \hat{L}^{\perp}_p \in \mathbb{R}_{>0}$ such that
    \begin{align*}
        \norm{\ppara{t,a}}_F \leq \Delta_p^{\paral}, \quad \norm{\pperp{t,a}}_F \leq \Delta_p^\perp,
        \quad
        \Frobenius{\ppara{t,a} - \ppara{t',a'}}  
        \leq     
        \hat{L}^{\paral}_p \absolute{t - t'} + L^{\paral}_p \norm{a - a'}^\frac{1}{2},
        \\
        \Frobenius{\pperp{t,a} - \pperp{t',a'}}  
        \leq     
        \hat{L}^{\perp}_p \absolute{t - t'} + L^{\perp}_p \norm{a - a'}^\frac{1}{2},
    \end{align*}
    for all $(t,t',a,a') \in \mathbb{R}_{\geq 0} \times \mathbb{R}_{\geq 0} \times \mathbb{R}^n \times \mathbb{R}^n$.
 \end{assumption}
 For notational simplicity, and in light of Assumptions~\ref{assmp:UnknownFunctions}-\ref{assmp:knownDiffusion:Decomposition}, we define the following:
 \begin{definition}\label{def:Diffusion:Decomposed}
    We define functions $F^{\paral}_\sigma: \mathbb{R}_{\geq 0} \times \mathbb{R}^n \rightarrow \mathbb{R}^{m \times d}$ and $F^{\perp}_\sigma: \mathbb{R}_{\geq 0} \times \mathbb{R}^n \rightarrow \mathbb{R}^{n-m \times d}$ as follows:
    \begin{align*}
        \Fparasigma{t,a} \doteq \ppara{t,a} + \Lparasigma{t,a}, \quad \Fperpsigma{t,a} \doteq \pperp{t,a} + \Lperpsigma{t,a},~\forall (t,a) \in \mathbb{R}_{\geq 0} \times \mathbb{R}^n.
    \end{align*}  
 \end{definition}
The next assumption characterizes the continuity of the uncertain vector fields. Note that these are standard for well-posedness, and can be relaxed for local results. 
\begin{assumption}\label{assmp:LipschitzContinuity}
    The drift uncertainty $\Lambda_\mu: \mathbb{R}_{\geq 0} \times \mathbb{R}^n \rightarrow \mathbb{R}^n$, and its decomposed components $\Lambda_\mu^{\paral}: \mathbb{R}_{\geq 0} \times \mathbb{R}^n \rightarrow \mathbb{R}^m$ and $\Lambda_\mu^{\perp}: \mathbb{R}_{\geq 0} \times \mathbb{R}^n \rightarrow \mathbb{R}^{n-m}$, are globally Lipschitz continuous in their arguments, and there exist known positive scalars $L_\mu, \hat{L}_\mu, L^{\paral}_\mu, \hat{L}^{\paral}_\mu, L^{\perp}_\mu, \hat{L}^{\perp}_\mu$ such that       
    \begin{align*}
        \norm{
            \LmuAll{t,a}
            -
            \LmuAll{t',a'}
        } 
        \leq     
        \hat{L}_\mu^{\cbr{\cdot,\paral,\perp}} \absolute{t - t'} + L_\mu^{\cbr{\cdot,\paral,\perp}} \norm{a - a'}, 
    \end{align*}    
    for all $(t,t',a,a') \in \mathbb{R}_{\geq 0} \times \mathbb{R}_{\geq 0} \times \mathbb{R}^n \times \mathbb{R}^n$.

    The diffusion uncertainty $\Lambda_\sigma: \mathbb{R}_{\geq 0} \times \mathbb{R}^n \rightarrow \mathbb{R}^{n \times d}$, and its decomposed components $\Lambda_\sigma^{\paral}: \mathbb{R}_{\geq 0} \times \mathbb{R}^n \rightarrow \mathbb{R}^{m \times d}$ and $\Lambda_\sigma^{\perp}: \mathbb{R}_{\geq 0} \times \mathbb{R}^n \rightarrow \mathbb{R}^{n-m \times d}$, are globally Lipschitz continuous in their temporal variable, and H\"{o}lder continuous with exponent $1/2$ in their state-dependent arguments, and there exist known positive scalars $L_\sigma, \hat{L}_\sigma, L^{\paral}_\sigma, \hat{L}^{\paral}_\sigma, L^{\perp}_\sigma, \hat{L}^{\perp}_\sigma$ such that  
    \begin{align*}
        \Frobenius{
            \Lambda_\sigma^{\cbr{\cdot,\paral,\perp}}(t,a) 
            - 
            \Lambda_\sigma^{\cbr{\cdot,\paral,\perp}}(t',a')
        } 
        \leq     
        \hat{L}_\sigma^{\cbr{\cdot,\paral,\perp}} \absolute{t - t'} + L_\sigma^{\cbr{\cdot,\paral,\perp}} \norm{a - a'}^\frac{1}{2}, 
    \end{align*}    
    for all $(t,t',a,a') \in \mathbb{R}_{\geq 0} \times \mathbb{R}_{\geq 0} \times \mathbb{R}^n \times \mathbb{R}^n$.
\end{assumption}
\begin{remark}
    The global Lipschitz continuity in Assumption~\ref{assmp:LipschitzContinuity} implies the linear growth of the drift uncertainty $\Lambda_\mu$ in its state-dependent argument, as stated in Assumption~\ref{assmp:UnknownFunctions}.
    However, the diffusion uncertainty $\Lambda_\sigma$ is assumed to satisfy the stronger condition of H\"{o}lder continuity with exponent $1/2$ in its state-dependent argument, which is a sufficient condition for sub-linear growth, as stated in Assumption~\ref{assmp:UnknownFunctions}.
\end{remark}

The following assumption is regarding the stability of the deterministic counterpart of the true (uncertain) system in~\eqref{eqn:TrueUncertainSDE} in response to the unmatched drift uncertainty.
\begin{assumption}\label{assmp:InternalStability}
    The constant $\lambda \in \mathbb{R}_{>0}$ in Assumption~\ref{assmp:ILF} satisfies  
    \begin{equation*}
        \lambda 
        > 
        \Delta_g^{\perp} \cdot \max \cbr{ \Delta_\mu^{\perp}, L_\mu^{\perp} }
        ,
    \end{equation*}
    where the bounds $\Delta_g^\perp,\Delta_\mu^\perp, L_\mu^{\perp} \in \mathbb{R}_{>0}$ are presented in Assumptions~\ref{assmp:KnownFunctions},~\ref{assmp:UnknownFunctions}, and~\ref{assmp:LipschitzContinuity}, respectively.
\end{assumption}
\begin{remark}
    The condition in Assumption~\ref{assmp:InternalStability} ensures that the true (uncertain) system in~\eqref{eqn:TrueUncertainSDE} in the absence of diffusion terms ($F_\sigma \equiv 0$) and the matched drift uncertainty $\Lambda_\mu^{\paral} \equiv 0$, i.e.,deterministic counterpart of the true (uncertain) system with only the unmatched uncertainty $\Lambda_\mu^{\perp}$, maintains its incremental exponential stability.
    The assumption is required since, by definition, the control action cannot ``reach'' the unmatched uncertainties.
    This assumption is similar to Assumption~\ref{assmp:ILF} in that it pertains only to the deterministic counterparts of the \ito SDEs that we consider.  
\end{remark}

Lastly, we can obtain stronger results if instead of the local Lipschitz continuity of the known drift vector field in Assumption~\ref{assmp:KnownFunctions}, we assume the global Lipschitz continuity as below.  
\begin{assumption}[Stronger Regularity of Known Drift Vector Field]\label{assmp:KnownFunctions:GlobalLip}
    There exist known $L_f, \hat{L}_f \in \mathbb{R}_{>0}$ such that the known drift $f(t,a)$ in Definition~\ref{def:VectorFields} satisfy: 
    \begin{align*}
        \norm{f(t,a) - f(t',a')} 
        \leq     
        \hat{L}_f \absolute{t - t'} + L_f \norm{a - a'}, 
        \quad  
        \forall 
        (t,t',a,a') \in \mathbb{R}_{\geq 0} \times \mathbb{R}_{\geq 0} \times \mathbb{R}^n \times \mathbb{R}^n.
    \end{align*}
    
\end{assumption}

\subsection{Problem Statement}\label{subsec:ProblemStatement}

\begin{figure}[t]
    \centering
    \includegraphics[width=0.85\textwidth]{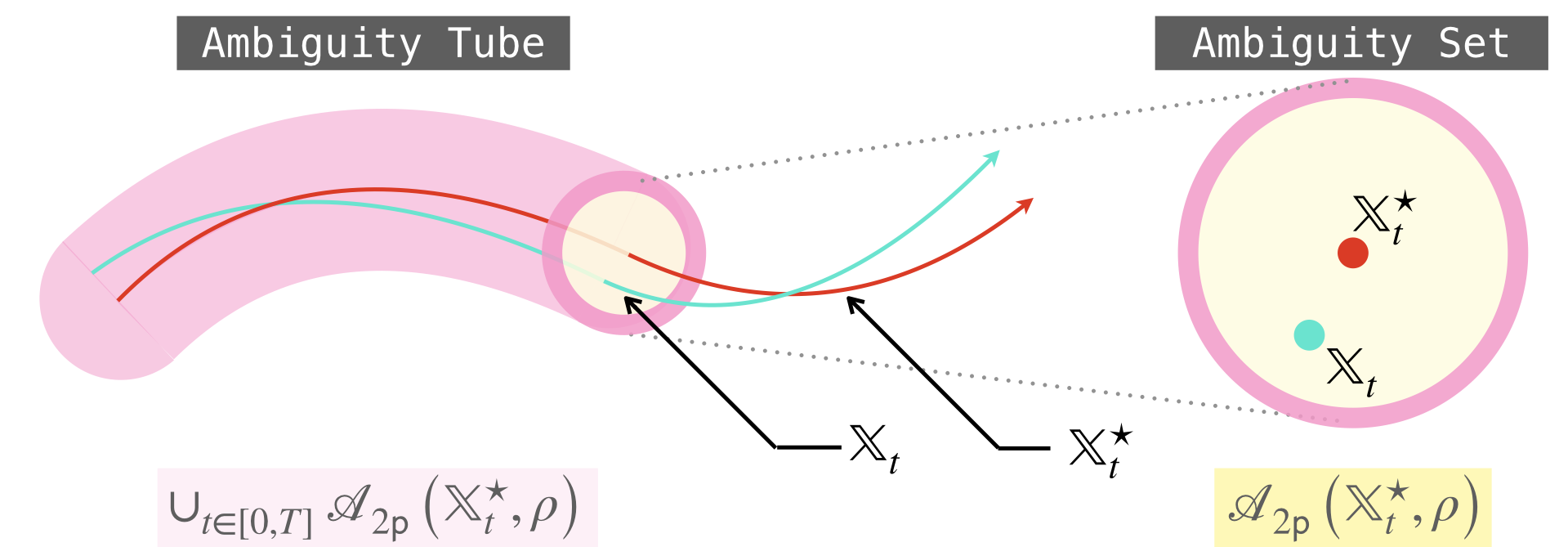}
    \caption{The uniform ambiguity tube (left) and ambiguity set (right) of laws defined in~\eqref{eqn:Bound:Law:Uniform} as sets of radius $\rho$ around the nominal (known) distribution $\Xstardist{t}$.   
    The \ellonedrac feedback is designed to ensure that the true (uncertain) distribution $\Xdist{t}$ remains inside the ambiguity set for all $t \in [0,T]$.
    }
    \label{fig:AmbiguitySetTube}
\end{figure}

Consider the \textbf{closed-loop true (uncertain) \ito~SDE}, for any $T \in (0,\infty)$: 
\begin{equation}\label{eqn:TrueUncertainClosedLoopSDE}
    d\Xt{t} = \Fmu{t,\Xt{t},\ULt{t}}dt + \Fsigma{t,\Xt{t}}d\Wt{t}, \quad \Xt{0} = x_0 \sim \xi_0,\forall t \in [0,T],
\end{equation}
that is induced by closing the feedback loop on the true (uncertain) \ito~SDE in~\eqref{eqn:TrueUncertainSDE} with the \textbf{feedback process $\Ut{t} = \ULt{t}$}, defined as
\begin{align}\label{eqn:L1FeedbackOperator}
    \ULt{t} \doteq \FL[\Xt{}][t], \quad  
    \FL :\continuous{T}{n} \rightarrow \continuous{T}{m}, \quad \forall t \in [0,T], 
    \quad  
\end{align}  
where $\FL$ is the \textbf{\ellonedrac feedback operator}. 
We wish to synthesize the \ellonedrac feedback operator $\FL$ such that the following conditions are satisfied:
\begin{enumerate}[wide, labelindent=0pt, topsep=0pt, label={\color{blue} \textit{\textbf{Condition \arabic*:}}}]
    \item The true uncertain process $\Xt{t} \sim \Xdist{t}$ exists and is a unique strong solution of the closed-loop true (uncertain) \ito~SDE in~\eqref{eqn:TrueUncertainClosedLoopSDE} for any $T \in (0,\infty)$. 
    \item There exists an a priori known $\rho \in \mathbb{R}_{>0}$ such that
    \begin{align}\label{eqn:Bound:Law:Uniform}
        \Xdist{t} \in \mathcal{A}_{2\sfp}\left(\Xstardist{t}, \rho \right)
        \doteq 
        \cbr{
            \text{Probability measures }\nu \text{ on }\Borel{\mathbb{R}^n}~|~\pWass{2\sfp}{\Xdist{t}^\star}{\nu} < \rho
        },  \quad \forall t \in [0,T],
    \end{align} 
    where $\Xstardist{t}$ is the law of the nominal (known) processes~\eqref{eqn:NominalKnownSDE}, and $\sfp \in \cbr{1,\dots,\sfp^\star}$ for $\mathsf{p}^\star \in \mathbb{N}_{\geq 1}$ introduced in Assumptions~\ref{assmp:KnownFunctions} and~\ref{assmp:NominalSystem:FiniteMomentsWasserstein}. 
    We refer to $\mathcal{A}_{2\sfp}\left(\Xstardist{t}, \rho \right)$ as the \textbf{uniform ambiguity set of laws}.
    Similarly, we refer to $ \cup_{t \in [0,T]} \mathcal{A}_{2\sfp}\left(\Xstardist{t}, \rho \right)$ as the \textbf{uniform ambiguity tube of laws}.
    The uniform ambiguity set and the uniform ambiguity tube are illustrated in Fig.~\ref{fig:AmbiguitySetTube}.  
    \item Furthermore, there exists an a priori known $\mathbb{R}_{>0} \ni \hat{\rho}\br{\that} < \rho$, $\that \in (0,T)$, such that 
    \begin{align}\label{eqn:Bound:Law:UniformlyUltimate}
        \Xdist{t} \in \mathcal{A}_{2\sfp}\left(\Xstardist{t}, \hat{\rho}\br{\that}  \right) \subset \mathcal{A}_{2\sfp}\left(\Xstardist{t}, \rho \right), \quad \forall t \in \sbr{\that,T}.
    \end{align}
    Analogous to the previous nomenclature, we refer to $\mathcal{A}_{2\sfp}\left(\Xstardist{t}, \hat{\rho}\br{\that}  \right)$ and $\cup_{t \in [0,T]} \mathcal{A}_{2\sfp}\left(\Xstardist{t}, \hat{\rho}\br{\that}  \right)$ as the \textbf{uniform ultimate ambiguity set of laws} and \textbf{uniform ultimate ambiguity tube of laws}, respectively. 
\end{enumerate}  

\begin{remark}
    The ambiguity tube (union of ambiguity sets of laws over $\Borel{\mathbb{R}^n}$) is not the same as ambiguity set of laws on $\Borel{\continuous{T}{n}}$.
    The ambiguity sets we consider are defined using the Wasserstein metric on the Polish space $\mathbb{R}^n$ under the Euclidean metric. 
    Laws on the trajectories (paths) are instead defined on the Polish space $\continuous{T}{n}$ equipped with the $\sup$-metric or some other appropriately chosen metrics~\cite[Sec.~18.1]{fristedt2013modern}. 
    We have previously discussed the difference between the underlying spaces for the laws in Remark.~\ref{rem:Laws}. 
\end{remark}

\section{$\mathcal{L}_1$-DRAC~Control: Design}\label{sec:L1DRAC:defn}

In this section we define and analyze the closed loop true (uncertain) system in~\eqref{eqn:TrueUncertainClosedLoopSDE} with the \ellonedrac~feedback control. 
We begin with the the definition of \ellonedrac~feedback operator $\FL$ that we introduced in~\eqref{eqn:L1FeedbackOperator}.

\subsection{ Architecture and Definition}\label{subsec:L1DRAC:Architecture}

\begin{figure}[t]
    \centering
    \includegraphics[width=0.50\textwidth]{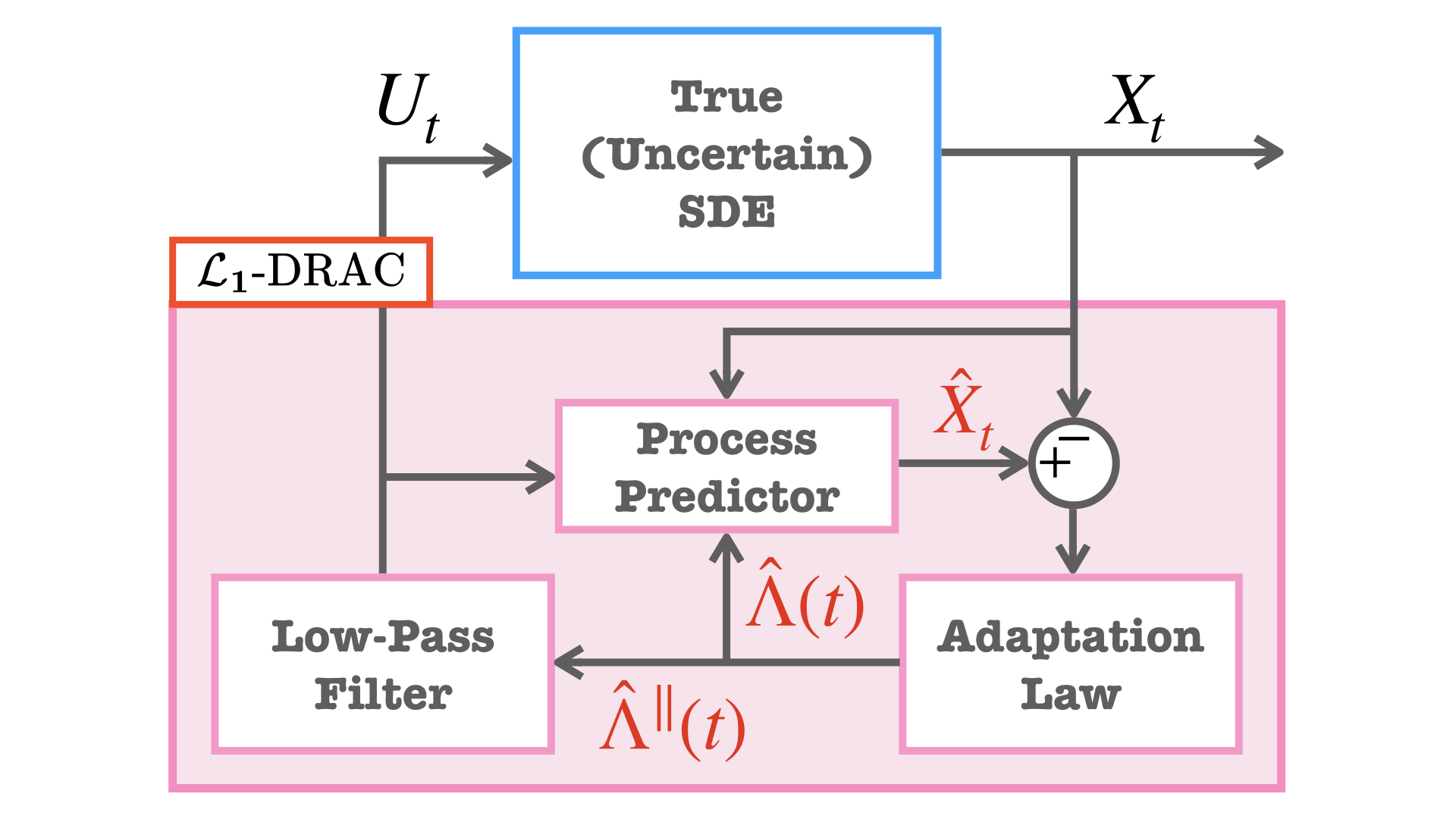}
    \caption{The architecture of the \ellonedrac~controller. The controller has three components: a \emph{process predictor} with output $\hat{X}_t$, an \emph{adaptation law}, and a \emph{low pass filter}.}
    \label{fig:L1DRAC}
\end{figure}

The design of $\FL$ is based on the \ellone-adaptive control methodology~\cite{hovakimyan2010ℒ1}. 
We will design the \ellonedrac~feedback operator $\mathcal{F}_{\mathcal{L}_1}$ such that the closed-loop true (uncertain) It\^{o} SDE in~\eqref{eqn:TrueUncertainClosedLoopSDE}  satisfies the conditions we set forth in Sec.~\ref{subsec:ProblemStatement}. 
Following the architecture of \ellone-adaptive control~\cite{hovakimyan2010ℒ1}, the \ellonedrac~feedback operator $\FL$ consists of a process predictor, an adaptation law, and a low-pass filter as illustrated in Fig.~\ref{fig:L1DRAC}. 

We define the \ellonedrac~feedback operator $\FL:\mathcal{C}([0,T]:\mathbb{R}^n) \rightarrow \mathcal{C}([0,T]:\mathbb{R}^m)$ as follows: 
\begin{equation}\label{eqn:L1DRAC:Definition:FeedbackOperator}
    \FL[y] \doteq \Filter \circ \AdaptationLaw \circ \Predictor[y], \quad y \in \mathcal{C}([0,T]:\mathbb{R}^n),  
\end{equation}
which we can alternatively represent as
\begin{align}\label{eqn:L1DRAC:Definition:FeedbackOperator:Alt}
    \FL[y] 
    = 
    \Filter[\hat{\Lambda}^{\paral}],
    \quad
    \hat{\Lambda}^{\paral} = \AdaptationLawParal[\hat{\Lambda}],
    \quad 
    \hat{\Lambda} = \AdaptationLaw[\hat{y}][y],
    \quad 
    \hat{y} = \Predictor[y]
    ,
\end{align} 
where
\begin{subequations}\label{eqn:L1DRAC:Definition:FeedbackOperator:Components}
    \begin{align}
        \Filter[\hat{\Lambda}^{\paral}][t] 
        \doteq  
        - \Boldomega \int_0^t \expo{-\Boldomega(t-\nu)}\Lparahat{\nu}d\nu,  \quad \text{\emph{(Low-pass filter)}}
        \label{eqn:L1DRAC:Definition:FeedbackOperator:Filter}
        \\
        \begin{aligned}
            \Lparahat{t} 
            =&
            \AdaptationLawParal[\hat{\Lambda}][t]
            = 
            \sum_{i=0}^{\lfloor \frac{t}{\BoldTs} \rfloor}  
            \Theta_{ad}(i\BoldTs) \Lhat{t}
            \indicator{[i\BoldTs,(i+1)\BoldTs)}{t}
            , 
            \\
            \Lhat{t} 
            =&
            \AdaptationLaw[\hat{y}][y][t]
            =  
            0_n \indicator{[0,\BoldTs)}{t} 
            \\
            &+
            \lambda_s \br{1 - e^{\lambda_s \BoldTs}}^{-1} 
            \sum_{i=1}^{\lfloor \frac{t}{\BoldTs} \rfloor}    
            \br{\hat{y}\br{i\BoldTs} - y\br{i\BoldTs}}
            \indicator{[i\BoldTs,(i+1)\BoldTs)}{t}, 
        \end{aligned} 
        \quad \text{\emph{(Adaptation Law)}}
        \label{eqn:L1DRAC:Definition:FeedbackOperator:AdaptationLaw}
        \\
        \begin{aligned}
            \hat{y}(t) 
            =&
            \Predictor[y][t] \Rightarrow \text{ solution to the integral equation: } 
            \\
            \hat{y}(t)=& x_0 +  
            \int_0^t \br{-\lambda_s \mathbb{I}_n \left(\hat{y}(\nu) - y(\nu) \right)+ f(\nu,y(\nu)) +  g(\nu)\FL[y][\nu] + \Lhat{\nu}} d\nu, 
        \end{aligned}
        \label{eqn:L1DRAC:Definition:FeedbackOperator:Predictor}
        \quad \text{\emph{(Process Predictor)}}
    \end{align}
\end{subequations}
for $t \in [0,T]$, where $\Boldomega, \BoldTs, \lambda_s \in \mathbb{R}_{>0}$ are the control parameters.
The parameters $\Boldomega$ and $\BoldTs$ and are referred to as the \textbf{\emph{filter bandwidth}} and the \textbf{\emph{sampling period}}, respectively.  
Additionally, $ \Theta_{ad}(t) = \begin{bmatrix}\mathbb{I}_m & 0_{m,n-m}  \end{bmatrix} \bar{g}(t)^{-1} \in \mathbb{R}^{m \times n}$, where $\bar{g}(t) = \begin{bmatrix} g(t) & g(t)^\perp  \end{bmatrix} \in \mathbb{R}^{n \times n}$, and here $g^\perp$ is defined in Assumption~\ref{assmp:UnknownFunctions}.
Finally, $x_0 \sim \xi_0$ in~\eqref{eqn:L1DRAC:Definition:FeedbackOperator:Predictor} is the initial condition of the true (uncertain) process~\eqref{eqn:TrueUncertainSDE}.
By initializing the process predictor with the initial condition of the true (uncertain) process, we simplify the analysis, however, this is not a strict requirement. 
See~\cite{L1book}\todo{Citation} for a more general treatment of the process predictor initialization for deterministic systems.  
 

\subsection{ \ellonedrac Parameters }\label{subsec:L1DRAC:Parameters}


We now present the choice of the control parameters $\Boldomega \in \mathbb{R}_{>0}$ and $\BoldTs \in \mathbb{R}_{>0}$, the bandwidth for the low-pass filter in~\eqref{eqn:L1DRAC:Definition:FeedbackOperator:Filter} and the sampling period for the adaptation law in~\eqref{eqn:L1DRAC:Definition:FeedbackOperator:AdaptationLaw}, respectively. 
Suppose that the assumptions in Sec.~\ref{subsec:Assumptions} hold.
For an arbitrarily chosen $\epsilon_r, \epsilon_a \in \mathbb{R}_{>0}$, and any $\sfp \in \cbr{1,\dots,\sfp^\star}$, select $\RefRho, \AdaptRho \in \mathbb{R}_{>0}$ such that
\begin{subequations}\label{eqn:Definitions:Total:Rho}
    \begin{align}
        \left(
            1
            -
            \frac{\Delta_g^\perp \Delta_\mu^\perp}{\lambda}
        \right)
        \RefRho^2
        \geq
        \alpha\left(\Xrt{0},\Xstart{0}\right)_{2\sfp}^2
        + 
        \Gamma_r\left(\RefRho,\sfp,\Boldomega\right)
        +
        \epsilon_r
        ,    
        \label{eqn:Definitions:Reference:Rho}
        \\
        \left(1 - \frac{\Delta_g^\perp L_\mu^\perp}{\lambda} \right) 
        \AdaptRho^2 
        \geq 
        \Gamma_a\left(\AdaptRho,\sfp,\Boldomega\right)
        +
        \epsilon_a
        ,    
        \label{eqn:Definitions:True:Rho}
    \end{align}
\end{subequations}
and define 
\begin{align}\label{eqn:Definitions:Eventual:Rho}
    \TotalRho = \RefRho + \AdaptRho,
\end{align}
where 
\begin{subequations}
    \begin{align}
        &\alpha\left(\Xrt{0},\Xstart{0}\right)_{2\sfp}
        \in       
        \left\{
            \norm{\Xrt{0}-\Xstart{0}}_{L_{2\sfp}},
            \pWass{2\sfp}{\xi_0}{\xi^\star_0}
        \right\}, 
        \\
        &\begin{aligned}[b]
            \Gamma_r\left(\RefRho,\sfp,\Boldomega\right)
            =&
            \frac{\Delta^r_{\circ_1}}{2\lambda}
            + 
            \frac{
                \Boldomega
                \Delta^r_{\circ_4}(\sfp)
                }{\absolute{2\lambda - \Boldomega}}
            +
            \left(
                \frac{\Delta^r_{\circledcirc_1}}{2\lambda}
                +
                \frac{
                    \Boldomega
                    \Delta^r_{\circledcirc_4}(\sfp)
                    }{\absolute{2\lambda - \Boldomega}}
            \right)
            \left(\RefRho+\Delta_\star\right)^\frac{1}{2}
            \\
            &+
            \left(
                \frac{\Delta^r_{\odot_1}}{2\lambda}
                +
                \frac{
                    \Boldomega
                    \Delta^r_{\odot_8}(\sfp)
                    }{\absolute{2\lambda - \Boldomega}}
            \right)
            \left(\RefRho+\Delta_\star\right)
            \\
            &+
            \left(
                \frac{\Delta^r_{\odot_2}}{2\lambda}
                +
                \frac{\Delta^r_{\odot_3}(\sfp)}{2\sqrt{\lambda}}
                +
                \frac{\Delta^r_{\circledast_1}}{2\lambda}
                \Delta_\star
            \right)
            \RefRho
            +
                \frac{\Delta^r_{\otimes_1}(\sfp)}{2\sqrt{\lambda}}
            \left(\RefRho+\Delta_\star\right)^\frac{3}{2}
            , 
        \end{aligned}    
        \label{eqn:Definitions:Reference:Gamma}
        \\
        &\Gamma_a\left(\AdaptRho,\sfp,\Boldomega\right)
        =
        \left(
            \frac{\Delta_{\odot_1}}{2 \lambda}  
            +  
            \frac{ \Boldomega \Delta_{\odot_4}(\sfp) }{  \absolute{2\lambda - \Boldomega}}
            +
            \frac{\Delta_{\otimes_1}(\sfp)}{2\sqrt{\lambda}}
            \left(\AdaptRho\right)^\frac{1}{2} 
        \right)
        \AdaptRho 
        ,
        \label{eqn:Definitions:True:Gamma}
    \end{align}
\end{subequations}
and where the definitions of the constants $\Delta^r_i(\sfp),\Delta_i(\sfp)$, $i \in \cbr{ \circ, \circledcirc,\odot, \otimes, \circledast   }$, are provided in Appendix~\ref{app:Definitions}.

\begin{remark}[Existence of $\RefRho$ and $\AdaptRho$]\label{rem:Reference:Design:Feasibility}    
   The condition in Assumption~\ref{assmp:InternalStability} implies  $\Delta_g^{\perp} \Delta_\mu^{\perp}/\lambda \in (0,1)$.
    It is also straightforward to see that 
    \begin{align*}
        \Gamma_r\left(\RefRho,\sfp,\Boldomega\right) \in \mathcal{O}\left(\RefRho^\frac{3}{2}\right), \text{ as } \RefRho \rightarrow \infty,
        \text{ and }
        \Gamma_r\left(\RefRho,\sfp,\Boldomega\right) \in \mathcal{O}\left(1\right), \text{ as } \Boldomega \rightarrow \infty.
    \end{align*} 
    It then follows 
    \begin{align*}
        \left(
            1
            -
            \frac{\Delta_g^\perp \Delta_\mu^\perp}{\lambda}
        \right)
        \RefRho^2
        -
        \Gamma_r\left(\RefRho,\sfp,\Boldomega\right)
         \in \mathcal{O}\left(\RefRho^2 \right), \text{ as } \RefRho \rightarrow \infty, 
    \end{align*}  
    and 
    \begin{align}\label{eqn:Reference:Remark:asymptotics}
        \left(
            1
            -
            \frac{\Delta_g^\perp \Delta_\mu^\perp}{\lambda}
        \right)
        \RefRho^2
        -
        \Gamma_r\left(\RefRho,\sfp,\Boldomega\right)
            \in \mathcal{O}\left(1\right), \text{ as } \Boldomega \rightarrow \infty.
    \end{align}  
    Therefore, one can always choose a $\RefRho \in \mathbb{R}_{>0}$ that satisfies~\eqref{eqn:Definitions:Reference:Rho}. 
    Importantly, the choice of $\RefRho \in \mathbb{R}_{>0}$ remains \textbf{independent} of the value of $\Boldomega \in \mathbb{R}_{>0}$.
    Similar arguments apply for the existence of $\AdaptRho$ satisfying~\eqref{eqn:Definitions:True:Rho}.
\end{remark}
Since $\Delta_g^{\perp} \Delta_\mu^{\perp}/\lambda \in (0,1)$ and $\Gamma_r\left(\RefRho,\sfp,\Boldomega\right) \geq 0$, $\forall \cbr{\RefRho,\sfp,\Boldomega} \in \mathbb{R}_{>0} \times \cbr{1,\dots,\sfp^\star} \times \mathbb{R}_{>0} $, a consequence of~\eqref{eqn:Definitions:Reference:Rho} is that
\begin{align}\label{eqn:Reference:Remark:RefRho:Lowerbound}
    \RefRho
    > 
    \alpha\left(\Xrt{0},\Xstart{0}\right)_{2\sfp}
    , 
    \quad 
    \alpha\left(\Xrt{0},\Xstart{0}\right)_{2\sfp}
    \in       
    \left\{
        \norm{\Xrt{0}-\Xstart{0}}_{L_{2\sfp}},
        \pWass{2\sfp}{\xi_0}{\xi^\star_0}
    \right\}.
\end{align}

\subsubsection{ Filter Bandwidth }\label{subsubsec:Design:Bandwidth}

Given $\RefRho$ and $\AdaptRho$ that satisfy~\eqref{eqn:Definitions:Total:Rho}, the bandwidth $\Boldomega \in \mathbb{R}_{>0}$ of the filter $\Filter$ in~\eqref{eqn:L1DRAC:Definition:FeedbackOperator:Filter} is chosen so as to verify the following conditions:
\begin{subequations}\label{eqn:Definitions:Total:BandwithCondition}
    \begin{align}
        \frac{1}{\absolute{2\lambda - \Boldomega}}
        \Theta_r\left(\RefRho,\sfp,\Boldomega\right)
        <
        \left(
            1
            -
            \frac{\Delta_g^\perp \Delta_\mu^\perp}{\lambda}
        \right)
        \RefRho^2
        - 
        \alpha\left(\Xrt{0},\Xstart{0}\right)_{2\sfp}^2
        -
        \Gamma_r\left(\RefRho,\sfp,\Boldomega\right) 
        \label{eqn:Definitions:Reference:BandwithCondition}
        ,
        \\
        \frac{ 1 }{  \absolute{2\lambda - \Boldomega}}
        \Theta_a\left(\AdaptRho,\RefRho,\sfp,\Boldomega\right)
        <
        \left(1 - \frac{\Delta_g^\perp L_\mu^\perp}{\lambda} \right) 
        \AdaptRho^2
        -
        \Gamma_a\left(\AdaptRho,\sfp,\Boldomega\right),  
        \label{eqn:Definitions:True:BandwithCondition}
    \end{align}    
\end{subequations}
where 
\begin{multline}\label{eqn:Definitions:Reference:Theta}
    \begin{aligned}
        \Theta_r\left(\RefRho,\sfp,\Boldomega\right)
        =&
        \Delta^r_{\circ_2}(\sfp)
        +
        \BoldomegaRoot 
        \Delta^r_{\circ_3}(\sfp)
        +
        \left(
            \Delta^r_{\circledcirc_2}(\sfp) 
            +
            \BoldomegaRoot
            \Delta^r_{\circledcirc_3}(\sfp)
        \right)
        \left(\RefRho+\Delta_\star\right)^\frac{1}{2}
        \\    
        &+
        \left(
                \Delta^r_{\odot_4}(\sfp)
                +
                \BoldomegaRoot
                \Delta^r_{\odot_6}(\sfp)
        \right)
        \left(\RefRho+\Delta_\star\right)
        +
        \left(
                \Delta^r_{\odot_5}(\sfp)
                +
                \BoldomegaRoot
                \Delta^r_{\odot_7}(\sfp)
        \right)
        \RefRho
        \\
        &+
        \left[
            \left(
                \Delta^r_{\otimes_2}(\sfp) 
            +
            \BoldomegaRoot
            \Delta^r_{\otimes_4}(\sfp)
            \right)
            \left(\RefRho+\Delta_\star\right) 
            +
            \left(
                \Delta^r_{\otimes_3}(\sfp)
                +    
                \BoldomegaRoot
            \Delta^r_{\otimes_5}(\sfp)
            \right)
            \RefRho
        \right]
        \left(\RefRho+\Delta_\star\right)^\frac{1}{2} 
    \end{aligned}
        \\
        +
        \left[
            \Delta^r_{\circledast_2}(\sfp)
            \left(\RefRho+\Delta_\star\right)
            +
            \Delta^r_{\circledast_3}(\sfp)
            \RefRho
        \right]
        \left(\RefRho+\Delta_\star\right),
\end{multline}
and  
\begin{multline}\label{eqn:Definitions:True:Theta}
    \Theta_a\left(\AdaptRho,\RefRho,\sfp,\Boldomega\right)
        =  
        \left[
            \sqrt{\Boldomega}
            \Delta_{\circledcirc_1}(\sfp)
            +
            \sqrt{\Boldomega}
            \Delta_{\otimes_3}(\sfp)
            \RhoDPrime 
            +
            \left(
                \Delta_{\otimes_2}(\sfp)
                +
                \sqrt{\Boldomega} 
                \Delta_{\otimes_4}(\sfp)
            \right)
            \AdaptRho
            \vphantom{\left(\AdaptRho\right)^\frac{1}{2}}
        \right] 
        \left(\AdaptRho\right)^\frac{1}{2}
        \\
        +  
        \left[
            \Delta_{\odot_2}(\sfp) 
            + 
            \sqrt{\Boldomega}
            \Delta_{\odot_3}(\sfp)
            +
            \Delta_{\circledast_2}
            \RhoDPrime
            + 
            \Delta_{\circledast_3}
            \AdaptRho
            \vphantom{\left(\AdaptRho\right)^\frac{1}{2}}
        \right]
        \AdaptRho,
\end{multline}
with $\RhoDPrime =  \AdaptRho + 2\left(\RefRho+\Delta_\star\right)$.

\begin{remark}[Feasibility of~\eqref{eqn:Definitions:Total:BandwithCondition}]\label{rem:Reference:Bandwidth:Feasibility}    
    It is easily shown that 
    \begin{align*}
        \left\{ 
            \frac{1}{\absolute{2\lambda - \Boldomega}} 
            \Theta_r\left(\RefRho,\sfp,\Boldomega\right),
            \,
            \frac{1}{\absolute{2\lambda - \Boldomega}} 
            \Theta_a\left(\AdaptRho,\RefRho,\sfp,\Boldomega\right)
        \right\}
        \in 
        \mathcal{O}\left(\frac{1}{\sqrt{\Boldomega}}\right), \text{ as } \Boldomega \rightarrow \infty.
    \end{align*}
    Moreover,~\eqref{eqn:Definitions:Reference:Rho} and~\eqref{eqn:Reference:Remark:asymptotics} imply that 
    \begin{align*}
        \left(
            1
            -
            \frac{\Delta_g^\perp \Delta_\mu^\perp}{\lambda}
        \right)
        \RefRho^2
        - 
        \alpha\left(\Xrt{0},\Xstart{0}\right)_{2\sfp}^2
        -
        \Gamma_r\left(\RefRho,\sfp,\Boldomega\right)
        \geq    
        \epsilon_r > 0, 
    \end{align*}
    independent of $\Boldomega \in \mathbb{R}_{>0}$, where $\alpha\left(\Xrt{0},\Xstart{0}\right)_{2\sfp}
        \in       
        \left\{
            \norm{\Xrt{0}-\Xstart{0}}_{L_{2\sfp}},
            \pWass{2\sfp}{\xi_0}{\xi^\star_0}
        \right\}$.
    As a consequence, one can always choose a filter bandwidth such that 
    \begin{align*}
        0
        <
        \frac{1}{\absolute{2\lambda - \Boldomega}}
        \Theta_r\left(\RefRho,\sfp,\Boldomega\right)
        <
        \left(
            1
            -
            \frac{\Delta_g^\perp \Delta_\mu^\perp}{\lambda}
        \right)
        \RefRho^2
        - 
        \alpha\left(\Xrt{0},\Xstart{0}\right)_{2\sfp}^2
        -
        \Gamma_r\left(\RefRho,\sfp,\Boldomega\right), 
    \end{align*}
    thus proving the feasibility of~\eqref{eqn:Definitions:Reference:BandwithCondition}.
    The feasibility of~\eqref{eqn:Definitions:True:BandwithCondition} follows similarly.  
\end{remark}

\subsubsection{ Sampling Period }\label{subsubsec:Design:SamplingPeriod}

Given $\RefRho$ and $\AdaptRho$ that satisfy~\eqref{eqn:Definitions:Total:Rho}, and filter bandwidth $\Boldomega$ that satisfies~\eqref{eqn:Definitions:Total:BandwithCondition}, the sampling period $\BoldTs \in \mathbb{R}_{>0}$ of the adaptation law $\AdaptationLaw$ in~\eqref{eqn:L1DRAC:Definition:FeedbackOperator:AdaptationLaw} is chosen such that 
\begin{multline}\label{eqn:Definitions:Total:SamplingPeriodCondition}
    \sum_{i=1}^3
    \Upsilon_{a_i}\left(\RhoPrime,\AdaptRho; \sfp,\Boldomega,\BoldTs\right)
    \leq 
    \left(1 - \frac{\Delta_g^\perp L_\mu^\perp}{\lambda} \right) 
    \AdaptRho^2
    - 
    \Gamma_a\left(\AdaptRho,\sfp,\Boldomega\right)
    \\
    -
    \frac{ 1 }{  \absolute{2\lambda - \Boldomega}}
    \Theta_a\left(\AdaptRho,\RefRho,\sfp,\Boldomega\right) 
    ,
\end{multline}
where $\RhoPrime =  \AdaptRho + \RefRho+\Delta_\star$, and we have defined the following for any $\xi_1,\xi_2 \in \mathbb{R}_{\geq 0}$:
\begin{subequations}\label{eqn:Definitions:Total:SamplingPeriodCondition:MasterUpsilons}
    \begin{align}
        \Upsilon_{a_1}\left(\xi_1,\xi_2; \sfp,\Boldomega,\BoldTs\right)
        =
        \widetilde{\Upsilon}^{-}\left(\xi_1; \sfp,\Boldomega,\BoldTs\right)
        \xi_2 
        +
        \frac{ \Boldomega }{ \lambda \absolute{2\lambda - \Boldomega}}  
        \mathring{\Upsilon}\left(\xi_2; \sfp,\Boldomega\right)
        \Upsilon^{-}\left(\xi_1; \sfp,\Boldomega,\BoldTs\right),
        \\
        \Upsilon_{a_2}\left(\xi_1,\xi_2; \sfp,\Boldomega,\BoldTs\right)
            =
            \frac{1}{\lambda}
            \left(
                \Delta_g\xi_2
                +
                \frac{\sqrt{\Boldomega}}{\absolute{2\lambda - \Boldomega}}
                \mathring{\Upsilon}\left(\xi_2;\sfp,\Boldomega\right)
            \right)
            \left(
                \Upsilon'_1\left(\xi_1; \sfp,\Boldomega,\BoldTs\right)
                + 
                \sqrt{\Boldomega}
                \Upsilon'_2\left(\xi_1; \sfp,\Boldomega,\BoldTs\right)
            \right),
        \\
        \Upsilon_{a_3}\left(\xi_1,\xi_2; \sfp,\Boldomega,\BoldTs\right)
        =
        \frac{1}{\absolute{2\lambda - \Boldomega}}
        \left(
            2
            \Delta_g
            \xi_2
            +  
            \frac{\sqrt{\Boldomega}}{\lambda }
            \mathring{\Upsilon}\left(\xi_2; \sfp,\Boldomega\right)
        \right)
        \Upsilon'_3\left(\xi_1; \sfp,\Boldomega,\BoldTs\right),
    \end{align}
\end{subequations}
and where the maps $\Upsilon'_{\cbr{1,2,3}}$, $\widetilde{\Upsilon}^-$, $\Upsilon^-$, and $\mathring{\Upsilon}$  are defined in~\eqref{eqn:app:Constants:True:Maps:UpsilonPrime} and~\eqref{eqn:app:Constants:True:Maps:UpsilonTilde} in Section~\ref{subsubsec:app:Definitions:True:PCA}.

\begin{remark}[Feasibility of~\eqref{eqn:Definitions:Total:SamplingPeriodCondition}]
    As a consequence of~\eqref{eqn:prop:Appendix:Constants:Maps:Properties:Limit} in Proposition~\ref{prop:Appendix:Constants:Maps:Properties}, one sees that 
    \begin{align*}
        \lim_{\BoldTs \downarrow 0}
        \sum_{i=1}^3
        \Upsilon_{a_i}\left(\RhoPrime,\AdaptRho; \sfp,\Boldomega,\BoldTs\right)
        = 0,
    \end{align*}
    for any chosen $\left(\RefRho,\AdaptRho,\Boldomega,\sfp\right) \in \mathbb{R}_{>0} \times \mathbb{R}_{>0} \times \mathbb{R}_{>0} \times \mathbb{N}_{\geq 1}$.
    Furthermore,~\eqref{eqn:Definitions:True:Rho} and~\eqref{eqn:Definitions:True:BandwithCondition} imply that 
    \begin{align*}
        \left(1 - \frac{\Delta_g^\perp L_\mu^\perp}{\lambda} \right) 
        \AdaptRho^2
        - 
        \Gamma_a\left(\AdaptRho,\sfp,\Boldomega\right)
        -
        \frac{ 1 }{  \absolute{2\lambda - \Boldomega}}
        \Theta_a\left(\AdaptRho,\RefRho,\sfp,\Boldomega\right) 
        >
        0
        .
    \end{align*}
    It then follows from the above two facts that one can always choose a sampling period $\BoldTs \in \mathbb{R}_{>0}$ that renders~\eqref{eqn:Definitions:Total:SamplingPeriodCondition} true. 
     
\end{remark}
\section{$\mathcal{L}_1$-DRAC~Control: Analysis}\label{sec:L1DRAC:Analysis}

We now present the analysis to obtain the performance bounds for the \ellonedrac~true (uncertain) It\^{o} SDE in~\eqref{eqn:TrueUncertainClosedLoopSDE}.
We follow a two-step approach as used for deterministic \ellone adaptive control analysis, see e.g.~\cite{hovakimyan2010ℒ1,lakshmanan2020safe}.
First, in Sec.~\ref{subsec:Analysis:ReferenceProcess} we introduce an intermediate and non-realizable process that we call the \emph{reference process} and obtain the performance bounds between it and the nominal (known) process. 
Then, in Sec.~\ref{subsec:Analysis:TrueUncertainProcess} we analyze the performance of the \ellonedrac~true (uncertain) process relative to the reference process.
The last step leads us to the relative performance bounds between the true (uncertain) process and the nominal (known) process via the triangle inequality of the Wasserstein metric.  

\subsection{Performance Analysis: Reference Process}\label{subsec:Analysis:ReferenceProcess}


We begin with the definition of the reference process. 
\begin{definition}[Reference Process]\label{def:ReferenceProcess}
    We say that $\Xrt{t}$, $t \in [0,T]$, for any $T \in (0,\infty)$, is the \textbf{reference process}, if $\Xrt{t}$ is a unique strong solution to the following \textbf{reference It\^{o} SDE}:
    \begin{equation}\label{eqn:ReferenceSDE}
        d\Xrt{t} = \Fmu{t,\Xrt{t},\Urt{t}}dt + \Fsigma{t,\Xrt{t}}d\Wt{t}, \quad \Xrt{0} = x_0 \sim \xi_{0},~\forall t \in [0,T],
    \end{equation}
    where the initial condition $x_0 \sim \xi_0$ and the driving Brownian motion $\Wt{t}$ are identical to those for the true (uncertain) process in~\eqref{eqn:TrueUncertainClosedLoopSDE} (and~\eqref{eqn:TrueUncertainSDE}).
    The \textbf{reference feedback process} $\Urt{t}$ is defined via the \textbf{reference feedback operator} $\ReferenceInput$ as follows:
    \begin{align}\label{eqn:ReferenceFeedbackOperatorProcess}
        \Urt{t} = \ReferenceInput[\Xrt{}][t],
        \quad   
        \ReferenceInput[\Xrt{}]
        \doteq 
        \Filter[\Lparamu{\cdot,\Xrt{}}] + \FilterW[\Fparasigma{\cdot,\Xrt{}}, \Wt{}], \quad t \in [0,T],
    \end{align}
    where the operator $\Filter$ is the low-pass filter defined in~\eqref{eqn:L1DRAC:Definition:FeedbackOperator:Filter}, the vector field $F_\sigma^{\paral}$ is introduced in Definition~\ref{def:Diffusion:Decomposed}, and 
    \begin{align}\label{eqn:ReferenceFeedbackOperatorProcess:BrownianFilter}
        \FilterW[\Fparasigma{\cdot,\Xrt{}}, \Wt{}][t] 
        \doteq  
        - \Boldomega \int_0^t \expo{-\Boldomega(t-\nu)}\Fparasigma{\nu,\Xrt{\nu}}d\Wt{\nu}.
    \end{align}
    Moreover, the functions $\Lparamu{t,\Xrt{t}} \in \mathbb{R}^m$ and $\Lparasigma{t,\Xrt{t}} \in \mathbb{R}^{m \times d}$ are defined in Assumption~\ref{assmp:UnknownFunctions}, and $\ppara{t,\Xrt{t}} \in \mathbb{R}^{m \times d}$ is defined in Assumption~\ref{assmp:knownDiffusion:Decomposition}.
   
    Finally, similar to $\Xdist{t}$ and $\Xstardist{t}$ in Definition~\ref{def:TrueandNominalProcesses}, we denote by $\Xrdist{t}$ the \textbf{reference law (path/trajectory probability measures)} induced by the process $\Xrt{t}$ on $\Borel{\mathbb{R}^n}$, i.e., $\Xrt{t} \sim \Xrdist{t}$.
\end{definition}
 The reference process is obtained by closing the loop of the true (uncertain) process in~\eqref{eqn:TrueUncertainSDE} with the feedback process $\Urt{t}$ that is composed of the filtered matched drift uncertainty $\Filter[\Lparamu{\cdot,\Xrt{}}][t]$ and the filtered \textbf{totality} of the matched diffusion vector field $\FilterW[\ppara{\cdot,\Xrt{}}+\Lparasigma{\cdot,\Xrt{}}, \Wt{}][t]$ expressed as an \ito integral with respect to the driving true Brownian motion. 
Thus, the reference process is \textbf{non-realizable} since, by definition, we do not have knowledge of the epistemic uncertainties $\Lambda_\mu$ and $\Lambda_\sigma$, and the aleatoric uncertainty $\Wt{}$. 
The reference process represents the \textbf{best achievable performance} since it quantifies, as a function of the low-pass filter bandwidth $\Boldomega$, how the system operates under the non-realizable assumption of perfect knowledge of the uncertainties. 
\begin{remark}
    In addition to the epistemic and aleatoric uncertainties $\Lambda_\mu$, $\Lambda_\sigma$, and $\Wt{}$, the reference feedback process $\Urt{t}$ also includes the \emph{known} diffusion term $p\br{\cdot,\Xrt{}}$ in its definition.
    Therefore, the reference feedback process $\Urt{t}$ further attempts to remove the effects of the known diffusion term $p\br{t,\Xrt{t}}$ from the reference process $\Xrt{t}$. 
    The reason for the inclusion of the known diffusion term $p$ is that due to the state-multiplicative nature of the term $p\br{t,\Xrt{t}}d\Wt{t}$, one cannot in general disambiguate the effects of the known term $p\br{t,\Xrt{t}}$ from the total uncertainty $\int_0^t \Lmu{\nu,\Xrt{\nu}}d\nu + \int_0^t \br{p\br{\nu,\Xrt{\nu}} + \Lsigma{\nu,\Xrt{\nu}} }d\Wt{\nu}$.
\end{remark}
For the performance analysis of the reference process with respect to the nominal (known) process, we define the following:
\begin{definition}[Joint Known (Nominal)-Reference Process]\label{def:Reference:JointProcess}
    We say that $\Yt{t} \in \mathbb{R}^{2n}$, $t \in [0,T]$, for any $T \in (0,\infty)$, is the \textbf{joint known(nominal)-reference process}, if it is a unique strong solution of the following \textbf{joint known(nominal)-reference It\^{o} SDE} on $\br{\Omega, \mathcal{F}, \Wfilt{0,t} \times \Wstarfilt{0,t},  \mathbb{P}}$ (see Definition~\ref{def:TrueandNominalProcesses} for the filtrations):
    \begin{align}\label{eqn:Reference:JointProcess}
        d\Yt{t}
        &= \Grmu{t,\Yt{t}}dt
        +
        \Grsigma{t,\Yt{t}}d\Wrt{t}, \quad t \in [0,T], 
        \quad 
        \Yt{0} = y_0 \sim \xi_0 \times \xi^\star_0, 
        \quad 
        \Yt{t} \sim \Ydist{t},
    \end{align}
    where
    \begin{align*}
        y_0 \doteq \begin{bmatrix} x_0 \\ x^\star_0 \end{bmatrix} \in \mathbb{R}^{2n}, 
        \quad 
        \Yt{t} \doteq \begin{bmatrix} \Xrt{t} \\ \Xstart{t} \end{bmatrix} \in \mathbb{R}^{2n}, 
        \quad 
        \Wrt{t} \doteq \begin{bmatrix} \Wt{t} \\  \Wstart{t}   \end{bmatrix} \in \mathbb{R}^{2d}, 
        \\
        \Grmu{t, \Yt{t}} \doteq \begin{bmatrix} \Fmu{t,\Xrt{t},\Urt{t}} \\ \Fbarmu{t,\Xstart{t}}     \end{bmatrix} \in \mathbb{R}^{2n}, \quad 
        \Grsigma{t, \Yt{t}} \doteq \begin{bmatrix} \Fsigma{t,\Xrt{t}} & 0_{n,d} \\ 
        0_{n,d} & \Fbarsigma{t,\Xstart{t}} \end{bmatrix} \in \mathbb{R}^{2n \times 2d},
    \end{align*}
    and where $\xi_0$ and $\xi^\star_0$ are the initial condition distributions and $\Ydist{t}$ denotes the law of $\br{\Xrt{t},\Xstart{t}}$ on $\Borel{\mathbb{R}^{2n}}$.
\end{definition}

We obtain the desired results using the Khasminskii-type theorem~\cite[Thm.~3.2]{mao2007stochastic}.
For this purpose, we require a truncated version of~\eqref{eqn:Reference:JointProcess} to build local solutions and extend them to the global solution using a limiting procedure. 
\begin{definition}[Truncated Joint Known (Nominal)-Reference Process]\label{def:Reference:TruncatedJointProcess}
    We first define 
    \begin{equation}\label{eqn:OpenBoundedSet}
        U_N \doteq \left\{ a \in \mathbb{R}^{2n}~:~\norm{a} < N \right\} \subset \subset \mathbb{R}^{2n}, \quad \forall N \in \mathbb{R}_{>0},  
    \end{equation}
    where $\subset \subset \mathbb{R}^{2n}$ denotes compact containment in $\mathbb{R}^{2n}$~\cite[Sec.~A.2]{evans2022partial}.
    Next, we define the 
    \textbf{truncated joint known(nominal)-reference It\^{o} SDE} as
    \begin{equation}\label{eqn:Reference:TruncatedJointProcess}
        d\Yt{N,t}
        = \GrNmu{t,\Yt{N,t}}dt + \GrNsigma{t,\Yt{N,t}}d\Wrt{t}, \quad \Yt{N,0} = \Yt{0},
    \end{equation}
    where, the process and the drift and diffusion vector fields are defined as
    \begin{align*}
    \Yt{N,t} \doteq \begin{bmatrix} \Xrt{N,t} \\ \Xstart{N,t} \end{bmatrix}, 
    \quad 
    \GrNmu{t, a} \br{ \GrNsigma{t,a} } = &
        \begin{cases}
            \Grmu{t,a} \br{ \Grsigma{t,a} }, \quad &~\norm{a} \leq N
            \\
            0_{2n} \br{ 0_{2n,2d} }, \quad &~\norm{a} \geq 2N
        \end{cases},
        \quad \forall (a,t) \in \mathbb{R}^{2n} \times [0,T],
    \end{align*}
    for any $\GrNmu{t,a}$ and $\GrNsigma{t,a}$ that are \textbf{uniformly} bounded and Lipschitz continuous for all $a \in \mathbb{R}^{2n}$ and $t \in \mathbb{R}_{\geq 0}$.
    Similar to $\Yt{t}$, we refer to $\Yt{N,t} \in \mathbb{R}^{2n}$ as the \textbf{truncated joint known(nominal)-reference process} if it is a unique strong solution of~\eqref{eqn:Reference:TruncatedJointProcess}.     
\end{definition}
An example of explicit construction of functions of the form $G_{N,\cbr{\mu,\sigma}}$ can be found in~\cite[p.~191]{durrett1996stochastic}.

With the setup complete, we start the analysis of the reference process by first establishing the existence and uniqueness of strong solutions of the truncated joint process. 
\begin{proposition}[Well-Posedness of~\eqref{eqn:Reference:TruncatedJointProcess}]\label{prop:Reference:TruncatedWellPosedness}
    If Assumptions~\ref{assmp:KnownFunctions} and~\ref{assmp:UnknownFunctions} hold true, then for any $N \in \mathbb{R}_{>0}$, $\Yt{N,t}$ is a unique strong solution of~\eqref{eqn:Reference:TruncatedJointProcess}, $\forall t \in [0,T]$, for any $T \in (0,\infty)$ and is a strong Markov process $\forall t \in \mathbb{R}_{\geq 0}$.
    Furthermore, define\footnote{The stopping time $\tau_N$ takes on the value of either the finite horizon $T$, or the first exit time of $\Yt{N,t}$ from set $U_N$.}
    \begin{equation}\label{eqn:Reference:FirstExitTime}
        \tau_N \doteq T \wedge \inf \cbr{t \in \mathbb{R}_{\geq 0}~:~\Yt{N,t} \notin U_N},         
    \end{equation}
    where the open and bounded set $U_N$ is defined in~\eqref{eqn:OpenBoundedSet}, for an arbitrary $N \in \mathbb{R}_{>0}$.
    Then, $\Yt{N,t} \in \mathcal{M}_2\left([0,\tau_N],\mathbb{R}^{2n}~|~\Wfilt{0,t} \times \Wstarfilt{0,t} \right)$ uniquely solves~\eqref{eqn:Reference:JointProcess}, in the strong sense, for all $t \in [0, \tau_N]$.
\end{proposition}
\begin{proof}
    See Appendix~\ref{app:ReferenceProcess}.
\end{proof}
Using the well-posedness of the truncated process, we can now define the transition probability for the process as (\cite[Sec.~3.1]{khasminskii2011stochastic},~\cite[Sec.~2.9]{mao2007stochastic}):
\begin{equation}\label{eqn:Reference:TransitionProbability}
    \Ydist{N}\left(\Yt{N,s},s;B,t\right)
    \doteq  
    \Probability{\Yt{N,t} \in B~|~\Yt{N,s}}, 
    \quad 
    \forall 0 \leq s \leq t \leq T, \quad B \in \Borel{\mathbb{R}^{2n}}.
\end{equation} 

Next, we derive bounds on the \emph{uniform in time} moments between the truncated versions of the reference and known (nominal) processes. 
\begin{lemma}\label{lem:Reference:dV}
    Let the assumptions in Sec.~\ref{subsec:Assumptions} hold.
    For an arbitrary $N \in \mathbb{R}_{>0}$, let the stopping time $\tau_N$ be as in~\eqref{eqn:Reference:FirstExitTime} for the truncated joint process in Definition~\ref{def:Reference:TruncatedJointProcess}.  
    For any constant $t^\star \in \mathbb{R}_{>0}$ define\footnote{The stopping time $\tau^\star$ is either the constant $t^\star$, the finite horizon $T$, or the first exit time of $\Yt{N,t}$ from the set $U_N$. The variable $\tau(t)$ is identical to $t$ up to $\tau^\star$, and zero afterwards.} 
    \begin{align}\label{eqn:Reference:StoppingTimes}
        \tau^\star = t^\star \wedge \tau_N,
        \quad
        \tau(t) = t \wedge \tau^\star.
    \end{align}
    Then, with $\RefNorm{\Yt{N}} \doteq \norm{\Xrt{N} - \Xstart{N}}^2$, the truncated joint process $\Yt{N,t} = \br{\Xrt{N,t},\Xstart{N,t}}$ in Definition~\ref{def:Reference:TruncatedJointProcess} satisfies:
    \begin{multline}\label{eqn:lem:Reference:dV:Supremums:UB:p=1}
        \ELaw{y_0}{\RefNorm{\Yt{N,\tau(t)}}}
        \leq
        \ELaw{y_0}{
            \expo{-2\lambda \tau(t)} 
        }
        \RefNorm{y_0}
        +
        \ELaw{y_0}{
            \expo{-2\lambda \tau(t)}
            \Xi^r\br{\tau(t),\Yt{N}}
        }
        \\
        +
        \ELaw{y_0}{
            \expo{-(2\lambda+\Boldomega) \tau(t)}
            \Xi^r_{\mathcal{U}}\br{\tau(t),\Yt{N};\Boldomega}
        }
        , 
        \quad \forall t \in \mathbb{R}_{\geq 0},
    \end{multline}
    and
    \begin{multline}\label{eqn:lem:Reference:dV:Supremums:UB}
        \pLpLaw{y_0}{
            \RefNorm{\Yt{N,\tau(t)}} 
        }
        \leq
        \pLpLaw{y_0}{
            \expo{-2\lambda \tau(t)} 
        }
        \RefNorm{y_0}
        +
        \pLpLaw{y_0}{
            \expo{-2\lambda \tau(t)}
            \Xi^r\br{\tau(t),\Yt{N}}
        }
        \\
        +
        \pLpLaw{y_0}{
            \expo{-(2\lambda+\Boldomega) \tau(t)}
            \Xi^r_{\mathcal{U}}\br{\tau(t),\Yt{N};\Boldomega}
        }
        , 
        \quad \forall (t,\sfp) \in \mathbb{R}_{\geq 0} \times \mathbb{N}_{\geq 2},
    \end{multline}    
    where, for any Borel measurable function $a:\mathbb{R}^{2n} \rightarrow \mathbb{R}$, 
    \begin{align}\label{eqn:lem:Reference:dV:CondExp:Definition}
        \ELaw{y_0}{
            a\br{\Yt{N,t}} 
        }
        \doteq
        \ELaw{}{
            a\br{\Yt{N,t}}~|~\Yt{N,0} = y_0 
        } 
        =
        \int_{\mathbb{R}^{2n}} a\br{y} \Ydist{N}\left(y_0,0;dy,t\right)
        ,
    \end{align}
    and where $\Ydist{N}$ is defined in~\eqref{eqn:Reference:TransitionProbability}.
    Similarly, $\pLpLaw{y_0}{a(t)}$ is defined using the probability measure $\Ydist{N}\left(y_0,0;\cdot,t\right)$. 
    Furthermore, 
    \begin{align}\label{eqn:lem:Reference:dV:Xi:Functions}
        \begin{aligned}
            \Xi^r\br{\tau(t),\Yt{N}}
            =
            \int_0^{\tau(t)}  \expo{ 2 \lambda  \nu }
            \left(
                \phi^r_{\mu}\br{\nu,\Yt{N,\nu}}d\nu
                +   
                \phi^r_{\sigma_\star}\br{\nu,\Yt{N,\nu}} d\Wstart{\nu}
                +
                \phi^r_{\sigma}\br{\nu,\Yt{N,\nu}}d\Wt{\nu}
            \right)
            ,
            \\
            \Xi^r_{\mathcal{U}}\br{\tau(t),\Yt{N};\Boldomega}
            =
            \int_0^{\tau(t)} 
            \left( 
                \mathcal{U}^r_{\mu}\br{\tau(t),\nu,\Yt{N};\Boldomega}d\nu
                +
                \mathcal{U}^r_{\sigma}\br{\tau(t),\nu,\Yt{N};\Boldomega}d\Wt{\nu}
            \right),
        \end{aligned}
    \end{align}
    where 
    \begin{align}\label{eqn:lem:Reference:dV:phi:Functions}
        \begin{aligned}
            \phi^r_{\mu}\br{\nu,\Yt{N,\nu}}
            =
            \RefDiffNorm{\Yt{N}}^\top 
            g(\nu)^\perp\Lperpmu{\nu,\Xrt{N,\nu}} 
            +
            \Frobenius{\Fsigma{\nu,\Xrt{N,\nu}}}^2
            +
            \Frobenius{\Fbarsigma{\nu,\Xstart{N,\nu}}}^2,
            \\
            \phi^r_{\sigma_\star}\br{\nu,\Yt{N,\nu}} 
            = 
            -\RefDiffNorm{\Yt{N,\nu}}^\top 
            \Fbarsigma{\nu,\Xstart{\nu}},
            \quad 
            \phi^r_{\sigma}\br{\nu,\Yt{N,\nu}} 
            = 
            \RefDiffNorm{\Yt{N,\nu}}^\top  
            g(\nu)^\perp \Fperpsigma{\nu,\Xrt{N,t}},
            \\
            \mathcal{U}^r_{\mu}\br{\tau(t),\nu,\Yt{N};\Boldomega}
            =
            \psi^r(\tau(t),\nu,\Yt{N})
            \Lparamu{\nu, \Xrt{N,\nu}}
            ,
            \quad 
            \mathcal{U}^r_{\sigma}\br{\tau(t),\nu,\Yt{N};\Boldomega}
            =
            \psi^r(\tau(t),\nu,\Yt{N})
            \Fparasigma{\nu,\Xrt{N,\nu}},
        \end{aligned}
    \end{align}
    and 
    \begin{multline}\label{eqn:lem:Reference:dV:psi:Functions}
        \psi^r(\tau(t),\nu,\Yt{N})
        =
        \frac{\Boldomega}{2\lambda - \Boldomega}
        \left(
            \expo{\Boldomega(\tau(t)+\nu)}
            \mathcal{P}^r\br{\tau(t),\nu}  
            -
            \expo{ (2\lambda \tau(t)+\Boldomega \nu)}
            \RefDiffNorm{\Yt{N,\tau(t)}}^\top g(\tau(t))
        \right)
        \\
        +
        \frac{2\lambda}{2\lambda - \Boldomega}
        \expo{(\Boldomega \tau(t) + 2 \lambda \nu)} 
        \RefDiffNorm{\Yt{N,\nu}}^\top g(\nu)
        \in \mathbb{R}^{1 \times m}.
    \end{multline}
    In the expressions above, we have defined 
    \begin{align*}
        \RefDiffNorm{\Yt{N}} = \partial_{\Xrt{N} - \Xstart{N}}\RefNorm{\Yt{N}} = 2 \left(\Xrt{N} - \Xstart{N}\right) \in \mathbb{R}^n,
        \quad  
        H_\sigma = G_\sigma G_\sigma^\top \in \mathbb{S}^{2n},
    \end{align*} 
    and 
    \begin{align}\label{eqn:Reference:ddV}
        \mathcal{P}^r\br{\tau(t),\nu}
        =
        \int_\nu^{\tau(t)}
            e^{ (2\lambda - \Boldomega) \beta }  d_\beta\sbr{\RefDiffNorm{\Yt{N,\beta}}^\top g(\beta)}
        \in \mathbb{R}^{1 \times m}, \quad 0 \leq \nu \leq \tau(t),
    \end{align}
    where $d_\beta \sbr{\cdot}$ denotes the stochastic differential with respect to $\beta$.

\end{lemma}
\begin{proof}
    We assume w.l.o.g. that $\mathbb{P}\left(y_0 \in U_N\right)=1$, since the current and all consecutive proofs can be extended for the general case by following the approach in~\cite[Thm.~8.3.1]{gikhman1969random}.  

    Since $\tstar$ is a constant, the fact that $\tau_N$ is a stopping time implies that $\tau^\star$ is a stopping time as well~\cite[Sec.~6.1]{evans2012introduction}. 
    Additionally, from Proposition~\ref{prop:Reference:TruncatedWellPosedness}, we know that $\Yt{N,t}$ is a unique strong solution of~\eqref{eqn:Reference:TruncatedJointProcess}, for all $t \in [0,T]$. 
    Consequently, the assumptions on the regularity of the vector fields in Sec.~\ref{subsec:Assumptions} imply that the vector fields $G_{N,\mu}$ and $G_{N,\sigma}$, that define the truncated process~\eqref{eqn:Reference:TruncatedJointProcess}, are uniformly bounded and globally Lipschitz continuous on $\mathbb{R}^{2n}$.
    It is thus straightforward to show that $\Yt{N,t}$ is a strong Markov process by invoking~\cite[Thm.~2.9.3]{mao2007stochastic}.
    Hence, the process $\Yt{N,\tau(t)}$ obtained by stopping the process $\Yt{N,t}$ at either the first (random) instant it leaves the set $U_N$ ( or when $t = t^\star$) is a Markov process and is well-defined~\cite[Lem.~3.2]{khasminskii2011stochastic}. 
    We may thus apply the \ito lemma~\cite[Thm.~1.6.4]{mao2007stochastic} to $e^{2\lambda \tau(t)} \RefNorm{\Yt{N,\tau(t)}}$ using the dynamics in~\eqref{eqn:Reference:TruncatedJointProcess} to obtain 
    \begin{multline*}
        e^{2\lambda \tau(t)} \RefNorm{\Yt{N,\tau(t)}}  
        \\
        = \RefNorm{y_0}
        +
        2 \lambda
        \int_0^{\tau(t)}  
            e^{2 \lambda \nu} \RefNorm{\Yt{N,\nu}} 
        d \nu 
        +
        \int_0^{\tau(t)}  
            e^{2 \lambda \nu} \nabla \RefNorm{\Yt{N,\nu}}^\top \GrNsigma{\nu,\Yt{N,\nu}} 
        d \Wrt{\nu} \notag 
        \\
        + \int_0^{\tau(t)}  
            e^{2 \lambda \nu} 
            \left( 
                \nabla \RefNorm{\Yt{N,\nu}}^\top \GrNmu{\nu, \Yt{N,\nu}}
                +
                \frac{1}{2} \Trace{ \HrNsigma{\nu, \Yt{N,\nu}} \nabla^2 \RefNorm{\Yt{N,\nu}}  }
            \right)
        d\nu, 
    \end{multline*}
    for all $t \in \mathbb{R}_{\geq 0}$, where $\HrNsigma{\nu, \Yt{N,\nu}} = \GrNsigma{\nu,\Yt{N,\nu}} \GrNsigma{\nu,\Yt{N,\nu}}^\top$.     
    Note that since $[0,\tau^\star] \subseteq [0,\tau_N]$, Proposition~\ref{prop:Reference:TruncatedWellPosedness} allows us to conclude that $\Yt{N,t}$ is also unique strong solution of~\eqref{eqn:Reference:JointProcess}, for all $t \in [0,\tau^\star]$.
    Thus, we may replace $G_{N,\mu}$, $G_{N,\sigma}$, and $H_{N,\sigma}$ with $G_{\mu}$, $G_{\sigma}$, and $H_{\sigma}$, respectively, in the above inequality and obtain
    \begin{multline*}
        e^{2\lambda \tau(t)} \RefNorm{\Yt{N,\tau(t)}}  
        = 
        \RefNorm{y_0}
        +
        2 \lambda
        \int_0^{\tau(t)}  
            e^{2 \lambda \nu} \RefNorm{\Yt{N,\nu}} 
        d \nu 
        +
        \int_0^{\tau(t)}  
            e^{2 \lambda \nu} \nabla \RefNorm{\Yt{N,\nu}}^\top \Grsigma{\nu,\Yt{N,\nu}} 
        d \Wrt{\nu} \notag 
        \\
        + \int_0^{\tau(t)}  e^{2 \lambda \nu} \left( 
            \nabla \RefNorm{\Yt{N,\nu}}^\top \Grmu{\nu, \Yt{N,\nu}}
            +
            \frac{1}{2} \Trace{ \Hrsigma{\nu, \Yt{N,\nu}} \nabla^2 \RefNorm{\Yt{N,\nu}}  }
        \right)d\nu, 
    \end{multline*}
    for all $t \in \mathbb{R}_{\geq 0}$, where $\Hrsigma{\nu, \Yt{N,\nu}} = \Grsigma{\nu,\Yt{N,\nu}} \Grsigma{\nu,\Yt{N,\nu}}^\top$.
    Substituting the expressions in~\eqref{eqn:Appendix:ReferenceProcess:prop:dV:Bound}, Proposition~\ref{prop:Appendix:ReferenceProcess:dV}, for the last two terms on the right hand side of the above expression and re-arranging terms leads to 
    \begin{multline}\label{lem:Reference:dV:1}
        e^{2\lambda \tau(t)} \RefNorm{\Yt{N,\tau(t)}}  
        \leq 
        \RefNorm{y_0}
        +
        \int_0^{\tau(t)}  
            e^{2 \lambda \nu}
            \left( 
                \phi^r_{\mu}\br{\nu,\Yt{N,\nu}}
                d\nu
                +
                \phi^r_{\sigma_\star}\br{\nu,\Yt{N,\nu}} 
                d\Wstart{\nu}
                + 
                \phi^r_{\sigma}\br{\nu,\Yt{N,\nu}}
                d\Wt{\nu}
            \right)   
        \\
        + 
        \int_0^{\tau(t)}  
            e^{2 \lambda \nu}
            \left( 
                \left[
                    \phi^r_{U}\br{\nu,\Yt{N,\nu}}
                    +
                    \phi^r_{\mu^{\paral}}\br{\nu,\Yt{N,\nu}}
                \right]
                d\nu
                + 
                \phi^r_{\sigma^{\paral}}\br{\nu,\Yt{N,\nu}}
                d\Wt{\nu}
            \right)
        , 
    \end{multline}
    for all $t \in \mathbb{R}_{\geq 0}$, where
    \begin{align*}
        \phi^r_{U}\br{\nu,\Yt{N,\nu}}
        =
        \RefDiffNorm{\Yt{N,\nu}}^\top
        g(\nu) 
        \Urt{\nu},
        \quad 
        \phi^r_{\mu^{\paral}}\br{\nu,\Yt{N,\nu}}
        =
        \RefDiffNorm{\Yt{N,\nu}}^\top
        g(\nu) 
        \Lparamu{\nu,\Xrt{N,\nu}},
        \\
        \phi^r_{\sigma^{\paral}}\br{\nu,\Yt{N,\nu}}
        =
        \RefDiffNorm{\Yt{N,\nu}}^\top
        g(\nu) \Fparasigma{\nu,\Xrt{N,\nu}}.
    \end{align*}
    Next, we use Proposition~\ref{prop:Appendix:ReferenceProcess:dV:U} to obtain the following expression 
    \begin{multline}\label{lem:Reference:dV:2}
        \int_0^{\tau(t)}  \expo{ 2\lambda \nu } 
        \left(
            \left[\phi^r_{U}\br{\nu,\Yt{N,\nu}} + \phi^r_{\mu^{\paral}}\br{\nu,\Yt{N,\nu}}\right]d\nu
            +
            \phi^r_{\sigma^{\paral}}\br{\nu,\Yt{N,\nu}}d\Wt{\nu}
        \right)
        \\
        =
        \int_0^{\tau(t)}  \expo{ 2\lambda \nu} 
        \left(
            \phi^r_{\mu^{\paral}}\br{\nu,\Yt{N,\nu}} + \phi^r_{U_\mu}\br{\nu,\Yt{N,\nu};\Boldomega} 
        \right)
        d\nu
        +
        \int_0^{\tau(t)}  \expo{ 2 \lambda \nu } 
        \left(
            \phi^r_{\sigma^{\paral}}\br{\nu,\Yt{N,\nu}} + \phi^r_{U_\sigma}\br{\nu,\Yt{N,\nu};\Boldomega}
        \right)
        d\Wt{\nu}
        \\
        +
        \int_0^{\tau(t)}
        \left(  
            \hat{\mathcal{U}}^r_{\mu}\br{\tau(t),\nu,\Yt{N};\Boldomega}d\nu 
            + 
            \hat{\mathcal{U}}^r_{\sigma}\br{\tau(t),\nu,\Yt{N};\Boldomega}d\Wt{\nu}
        \right),
        \quad \forall t \in \mathbb{R}_{\geq 0},
    \end{multline} 
    where 
    \begin{align*}
        \begin{multlined}[b][0.9\linewidth]
            \hat{\mathcal{U}}^r_{\mu}\br{\tau(t),\nu,\Yt{N};\Boldomega}
            \\
            =
            \expo{-\Boldomega \tau(t)}
            \frac{\Boldomega}{2\lambda - \Boldomega}
            \left(
                \expo{\Boldomega \tau(t)}  
                \mathcal{P}^r\br{\tau(t),\nu}
                -
                \expo{ 2\lambda \tau(t)}
                \RefDiffNorm{\Yt{N,\tau(t)}}^\top
                g(\tau(t))
            \right)
            \expo{\Boldomega \nu}
            \Lparamu{\nu, \Xrt{N,\nu}}
            ,
        \end{multlined}
        \\
        \begin{multlined}[b][0.9\linewidth]
            \hat{\mathcal{U}}^r_{\sigma}\br{\tau(t),\nu,\Yt{N};\Boldomega}
            \\
            =
            \expo{-\Boldomega \tau(t)}
            \frac{\Boldomega}{2\lambda - \Boldomega}
            \left( 
                \expo{\Boldomega \tau(t)} 
                \mathcal{P}^r\br{\tau(t),\nu}
                -
                \expo{2\lambda \tau(t)}
                \RefDiffNorm{\Yt{N,\tau(t)}}^\top g(\tau(t))
            \right)
            \expo{\Boldomega \nu}
            \Fparasigma{\nu,\Xrt{N,\nu}}
            ,
        \end{multlined}
        \\
        \phi^r_{U_\mu}\br{\nu,\Yt{N,\nu};\Boldomega}
        =
        \frac{\Boldomega}{2\lambda - \Boldomega}
        \RefDiffNorm{\Yt{N,\nu}}^\top g(\nu) 
        \Lparamu{\nu, \Xrt{N,\nu}},
        \\
        \phi^r_{U_\sigma}\br{\nu,\Yt{N,\nu};\Boldomega}
        =
        \frac{\Boldomega}{2\lambda - \Boldomega}
        \RefDiffNorm{\Yt{N,\nu}}^\top g(\nu)
        \Fparasigma{\nu,\Xrt{N,\nu}} 
        .
    \end{align*}
    Using the definitions of $\phi^r_{\mu^{\paral}}$ and $\phi^r_{\sigma^{\paral}}$ in~\eqref{lem:Reference:dV:1}, we obtain    
    \begin{align*}
        \phi^r_{\mu^{\paral}}\br{\nu,\Yt{N,\nu}} + \phi^r_{U_\mu}\br{\nu,\Yt{N,\nu};\Boldomega}
        =
        \frac{2\lambda}{2\lambda - \Boldomega}
        \RefDiffNorm{\Yt{N,\nu}}^\top g(\nu) 
        \Lparamu{\nu, \Xrt{N,\nu}},
        \\
        \phi^r_{\sigma^{\paral}}\br{\nu,\Yt{N,\nu}} + \phi^r_{U_\sigma}\br{\nu,\Yt{N,\nu};\Boldomega}
        =
        \frac{2\lambda}{2\lambda - \Boldomega}
        \RefDiffNorm{\Yt{N,\nu}}^\top g(\nu) 
        \Fparasigma{\nu, \Xrt{N,\nu}}.
    \end{align*}
    Substituting into~\eqref{lem:Reference:dV:2} thus produces
    \begin{multline*}
        \int_0^{\tau(t)}  \expo{ 2\lambda \nu } 
        \left(
            \left[\phi^r_{U}\br{\nu,\Yt{N,\nu}} + \phi^r_{\mu^{\paral}}\br{\nu,\Yt{N,\nu}}\right]d\nu
            +
            \phi^r_{\sigma^{\paral}}\br{\nu,\Yt{N,\nu}}d\Wt{\nu}
        \right)
        \\
        =
        \expo{-\Boldomega \tau(t)}
        \int_0^{\tau(t)}
        \left(  
            \mathcal{U}^r_{\mu}\br{\tau(t),\nu,\Yt{N};\Boldomega}d\nu 
            + 
            \mathcal{U}^r_{\sigma}\br{\tau(t),\nu,\Yt{N};\Boldomega}d\Wt{\nu}
        \right),
        \quad \forall t \in \mathbb{R}_{\geq 0},
    \end{multline*} 
    where $\mathcal{U}^r_{\mu}$ and $\mathcal{U}^r_{\sigma}$ are defined in~\eqref{eqn:lem:Reference:dV:phi:Functions}.
    Substituting the above expression for the last integral on the right hand side of~\eqref{lem:Reference:dV:1} and using the definitions of $\Xi^r$ and $\Xi^r_{\mathcal{U}}$ in~\eqref{eqn:lem:Reference:dV:Xi:Functions} yields the following bound: 
    \begin{equation*}
        \expo{2\lambda \tau(t)} 
        \RefNorm{\Yt{N,\tau(t)}} 
        \leq 
        \RefNorm{y_0}
        +
        \Xi^r\br{\tau(t),\Yt{N}}
        +
        \expo{-\Boldomega \tau(t)}
        \Xi^r_{\mathcal{U}}\br{\tau(t),\Yt{N};\Boldomega}
        , 
        \quad \forall t \in \mathbb{R}_{\geq 0},
    \end{equation*}
    which in turn produces
    \begin{equation}\label{lem:Reference:dV:3} 
        \RefNorm{\Yt{N,\tau(t)}} 
        \leq
        \expo{-2\lambda \tau(t)} 
        \RefNorm{y_0}
        +
        \expo{-2\lambda \tau(t)}
        \Xi^r\br{\tau(t),\Yt{N}}
        +
        \expo{-(2\lambda+\Boldomega) \tau(t)}
        \Xi^r_{\mathcal{U}}\br{\tau(t),\Yt{N};\Boldomega}
        , 
        \quad \forall t \in \mathbb{R}_{\geq 0}.
    \end{equation}
    Then, using the linearity of the expectation operator, we obtain 
    \begin{multline*}
        \ELaw{y_0}{\RefNorm{\Yt{N,\tau(t)}}}
        \leq
        \ELaw{y_0}{
            \expo{-2\lambda \tau(t)} 
        }
        \RefNorm{y_0}
        +
        \ELaw{y_0}{
            \expo{-2\lambda \tau(t)}
            \Xi^r\br{\tau(t),\Yt{N}}
        }
        \\
        +
        \ELaw{y_0}{
            \expo{-(2\lambda+\Boldomega) \tau(t)}
            \Xi^r_{\mathcal{U}}\br{\tau(t),\Yt{N};\Boldomega}
        }
        , 
        \quad \forall t \in \mathbb{R}_{\geq 0},
    \end{multline*}
    which establishes~\eqref{eqn:lem:Reference:dV:Supremums:UB:p=1}.
    Finally, applying the Minkowski's inequality to~\eqref{lem:Reference:dV:3} for $\sfp \in \mathbb{N}_{\geq 2}$ yields the bound in~\eqref{eqn:lem:Reference:dV:Supremums:UB}, thus concluding the proof.

\end{proof}

The next result establishes moment bounds for the truncated system up to the first exit time $\tau_N$.
\begin{lemma}\label{lem:Reference:MomentBounds:FirstExit}
    Let the assumptions in Sec.~\ref{subsec:Assumptions} hold and let $\RefRho \in \mathbb{R}_{>0}$, $\sfp \in \cbr{1,\dots,\sfp^\star}$, be as defined in~\eqref{eqn:Definitions:Reference:Rho} with $\alpha\left(\Xrt{0},\Xstart{0}\right)_{2\sfp} = \norm{\Xrt{0}-\Xstart{0}}_{L_{2\sfp}}$, where $\sfp^\star$ is defined in Assumption~\ref{assmp:NominalSystem:FiniteMomentsWasserstein}.
    If the filter bandwidth condition in~\eqref{eqn:Definitions:Reference:BandwithCondition} holds  with $\alpha\left(\Xrt{0},\Xstart{0}\right)_{2\sfp} = \norm{\Xrt{0}-\Xstart{0}}_{L_{2\sfp}}$, then the truncated joint process $\Yt{N,t} = \br{\Xrt{N,t},\Xstart{N,t}}$ in Definition~\ref{def:Reference:TruncatedJointProcess}, satisfies 
    \begin{align}\label{eqn:lem:Reference:FirstExit:UB}
        \LpLaw{2\sfp}{}{\Xrt{N,t}-\Xstart{N,t}} < \RefRho, \quad \forall t \in [0,\tau_N],
    \end{align}
    where the stopping time $\tau_N$ is as in~\eqref{eqn:Reference:FirstExitTime} for an arbitrary $N \in \mathbb{R}_{>0}$.
\end{lemma}
\begin{proof}
    We prove~\eqref{eqn:lem:Reference:FirstExit:UB} by contradiction.
    The $t$-continuity of the strong solutions $\Xrt{N,t}$ and $\Xstart{N,t}$ on $[0,\tau_N]$ implies that $\LpLaw{2\sfp}{}{\Xrt{N,t}-\Xstart{N,t}}$ is $t$-continuous as well.
    Moreover, as we presented in~\eqref{eqn:Reference:Remark:RefRho:Lowerbound}, the definition of $\RefRho \in \mathbb{R}_{>0}$ in~\eqref{eqn:Definitions:Reference:Rho} implies that 
    \begin{align}\label{eqn:Reference:Scratch:Statement:P}
        \RefRho 
        >  
        \LpLaw{2\sfp}{}{\Xrt{N,0}-\Xstart{N,0}}.
    \end{align} 
    Then,~\eqref{eqn:Reference:Scratch:Statement:P} and the deterministic nature of $t \mapsto \LpLaw{2\sfp}{}{\Xrt{N,t}-\Xstart{N,t}}$ allow us to formulate the negation of~\eqref{eqn:lem:Reference:FirstExit:UB} as 
    \begin{align}\label{eqn:Reference:Scratch:Statement:Not:Q}
        \lnot~\text{\eqref{eqn:lem:Reference:FirstExit:UB}}~: 
        ~
        \exists~~(deterministic)~~\tstar\in [0,\tau_N] \text{ such that }
        \begin{cases}
            \LpLaw{2\sfp}{}{\Xrt{N,t}-\Xstart{N,t}} < \RefRho, \quad & \forall t \in [0,\tstar)
            \\ 
            \LpLaw{2\sfp}{}{\Xrt{N,\tstar}-\Xstart{N,\tstar}} = \RefRho.
        \end{cases} 
    \end{align} 

    Now, let $\pi_0$ be the probability measure on $\Borel{\mathbb{R}^{2n}}$ induced by the initial condition $\Yt{N,0} = \Yt{0} = \br{\Xrt{0},\Xstart{0}} = \br{x_0,x_0^\star} = y_0$ of the truncated joint process.  
    Since the initial vector $\Yt{N,0}=\Yt{0}$ is independent of the Brownian motion $\Wrt{t}$, the strong solution $\Yt{N,\tau(t)}$ is adapted to the filtration generated by $Y_0$ and $\Wrt{t}$~\cite[Thm.~5.2.1]{oksendal2013stochastic}.
    Let us denote by $\Ydist{N}(\cdot,t)$ the law of the process $\Yt{N,t}$ under $\Yt{N,0} \sim \pi_0$.
    Hence, $\Ydist{N}(\cdot,t)$ is a probability measure on $\mathbb{R}^{2n}$ given as
    \begin{align}\label{eqn:Reference:Scratch:TotalLaw}
        \Ydist{N}(B,t)
        =
        \Probability{\Yt{N,t} \in B}
        = 
        \int_{\mathbb{R}^{2n}}
            \Ydist{N}\left(y_0,0;B,t\right)
        \pi_0(dy_0), 
        \quad 
        \forall B \in \Borel{\mathbb{R}^{2n}},
    \end{align} 
    where the integrand is the transition probability defined in~\eqref{eqn:Reference:TransitionProbability}.
    Then,  
    \begin{align*}
        \ELaw{}{ \RefNorm{\Yt{N,\tau(t)}} }
        =
        \int_{\mathbb{R}^{2n}}
            \RefNorm{y}    
        \Ydist{N}(dy,\tau(t))
        =&
        \int_{\mathbb{R}^{2n}}
        \int_{\mathbb{R}^{2n}}
            \RefNorm{y}
            \Ydist{N}\left(y_0,0;dy,\tau(t)\right)
        \pi_0(dy_0)
        \notag 
        \\
        =&
        \int_{\mathbb{R}^{2n}}
            \ELaw{y_0}{
                \RefNorm{\Yt{N,\tau(t)}}
                }
        \pi_0(dy_0),
        \quad  
        t \in \mathbb{R}_{\geq 0},
    \end{align*} 
    where the last expression follows from~\eqref{eqn:lem:Reference:dV:CondExp:Definition}, and $\tau(t)$ is defined in~\eqref{eqn:Reference:StoppingTimes}.
    Since we are assuming that $\tstar < \tau_N$, we have from~\eqref{eqn:Reference:StoppingTimes} that $\tau^\star = \tstar$ and thus $\tau(t) = t \wedge \tstar \in [0,\tstar]$, for all $t \in \mathbb{R}_{\geq 0}$.
    Hence, we can write the expression above as 
    \begin{align}\label{eqn:Reference:Scratch:TotalExpectation:p=1}
        \ELaw{}{
            \RefNorm{\Yt{N,t}}
            }
        =
        \int_{\mathbb{R}^{2n}}
            \ELaw{y_0}{
                \RefNorm{\Yt{N,\tau(t)}}
                }
        \pi_0(dy_0),
        \quad  
        t \in [0,\tstar].
    \end{align}
    Similarly, we can write
    \begin{align}\label{eqn:Reference:Scratch:TotalExpectation:p>=2}
        \LpLaw{\sfp}{}{\RefNorm{\Yt{N,t}}}
        =
        \ELaw{}{\RefNorm{\Yt{N,t}}^\sfp}^\frac{1}{\sfp}
        =&
        \left(
            \int_{\mathbb{R}^{2n}}
                \ELaw{y_0}{\RefNorm{\Yt{N,\tau(t)}}^\sfp}
            \pi_0(dy_0)
        \right)^\frac{1}{\sfp}
        \notag 
        \\ 
        =&
        \left(
            \int_{\mathbb{R}^{2n}}
            \left(\LpLaw{\sfp}{y_0}{\RefNorm{\Yt{N,\tau(t)}}}\right)^\sfp
            \pi_0(dy_0)
        \right)^\frac{1}{\sfp}
        ,
        \quad  
        t \in [0,\tstar].
    \end{align}  
    As stated above $\tau(t) \in [0,\tstar]$, for all $t \in \mathbb{R}_{\geq 0}$.
    Thus, we may use Lemma~\ref{lem:Reference:dV} by substituting~\eqref{eqn:lem:Reference:dV:Supremums:UB:p=1} into~\eqref{eqn:Reference:Scratch:TotalExpectation:p=1}, which produces 
    \begin{multline}\label{eqn:Reference:Scratch:TotalExpectation:Bound:p=1:1}
        \ELaw{}{\RefNorm{\Yt{N,t}}}
        \leq 
        \int_{\mathbb{R}^{2n}}
            \RefNorm{y_0}
        \pi_0(dy_0)
        + 
        \int_{\mathbb{R}^{2n}}
            \ELaw{y_0}{
                \expo{-2\lambda \tau(t)}
                \Xi^r\br{\tau(t),\Yt{N}}
            }
        \pi_0(dy_0)
        \\ 
        + 
        \int_{\mathbb{R}^{2n}}
            \ELaw{y_0}{
                \expo{-(2\lambda+\Boldomega) \tau(t)}
                \Xi^r_{\mathcal{U}}\br{\tau(t),\Yt{N};\Boldomega}
            }
        \pi_0(dy_0)
        ,
        \quad  
        \forall t \in [0,\tstar],
    \end{multline}
    where we have used the bound $\expo{-2\lambda \tau(t)} \leq 1$.
    Now, recall the definition $\RefNorm{\Yt{N}} \doteq \norm{\Xrt{N} - \Xstart{N}}^2$ from the statement of Lemma~\ref{lem:Reference:dV}. Hence, one sees that
    \begin{align*}
        \int_{\mathbb{R}^{2n}}
            \RefNorm{y_0}
        \pi_0(dy_0)
        =&     
        \int_{\mathbb{R}^{2n}}
            \norm{x_0 - x^\star_0}^2
        \pi_0(dy_0)
        \\
        =&
        \int_{\mathbb{R}^{2n}}
            \norm{x_0 - x^\star_0}^2
        \pi_0(dx_0,dx_0^\star)
        =
        \ELaw{}{\norm{\Xrt{0}-\Xstart{0}}^2}
        = 
        \norm{\Xrt{0}-\Xstart{0}}_{L_2}^2,
    \end{align*} 
    which, upon substituting into~\eqref{eqn:Reference:Scratch:TotalExpectation:Bound:p=1:1} produces  
    \begin{multline}\label{eqn:Reference:Scratch:TotalExpectation:Bound:p=1:2}
        \ELaw{}{\RefNorm{\Yt{N,t}}}
        \leq 
        \norm{\Xrt{0}-\Xstart{0}}_{L_2}^2
        + 
        \int_{\mathbb{R}^{2n}}
            \ELaw{y_0}{
                \expo{-2\lambda \tau(t)}
                \Xi^r\br{\tau(t),\Yt{N}}
            }
        \pi_0(dy_0)
        \\ 
        + 
        \int_{\mathbb{R}^{2n}}
            \ELaw{y_0}{
                \expo{-(2\lambda+\Boldomega) \tau(t)}
                \Xi^r_{\mathcal{U}}\br{\tau(t),\Yt{N};\Boldomega}
            }
        \pi_0(dy_0)
        ,
        \quad  
        \forall t \in [0,\tstar].
    \end{multline}
    Similarly, substituting~\eqref{eqn:lem:Reference:dV:Supremums:UB} into~\eqref{eqn:Reference:Scratch:TotalExpectation:p>=2} leads to
    \begin{multline}\label{eqn:Reference:Scratch:TotalExpectation:p>=2:1}
        \LpLaw{\sfp}{}{\RefNorm{\Yt{N,t}}}
        \leq
        \left(
            \int_{\mathbb{R}^{2n}}
                \RefNorm{y_0}^\sfp
            \pi_0(dy_0)
        \right)^\frac{1}{\sfp}
        \\ 
        + 
        \left(
            \int_{\mathbb{R}^{2n}}
            \left(
                \pLpLaw{y_0}{
                    \expo{-2\lambda \tau(t)}
                    \Xi^r\br{\tau(t),\Yt{N}}
                }
                +
                \pLpLaw{y_0}{
                    \expo{-(2\lambda+\Boldomega) \tau(t)}
                    \Xi^r_{\mathcal{U}}\br{\tau(t),\Yt{N};\Boldomega}
                }
            \right)^\sfp
            \pi_0(dy_0)
        \right)^\frac{1}{\sfp}
        ,
    \end{multline}
    for all $(t,\sfp) \in [0,\tstar] \times \cbr{2,\dots,\sfp^\star}$
    , where, as above, we have used the bound $\expo{-2\lambda \tau(t)} \leq 1$, along with the Minkowski's inequality. 
    Once again, it follows from the definition of $\RefNorm{\Yt{N}}$ that
    \begin{align*}
        \left(
            \int_{\mathbb{R}^{2n}}
                \RefNorm{y_0}^\sfp
            \pi_0(dy_0)
        \right)^\frac{1}{\sfp}
        =&     
        \left(
            \int_{\mathbb{R}^{2n}}
                \norm{x_0 - x^\star_0}^{2\sfp}
            \pi_0(dy_0)
        \right)^\frac{1}{\sfp}
        \\
        =&
        \left(
            \int_{\mathbb{R}^{2n}}
                \norm{x_0 - x^\star_0}^{2\sfp}
            \pi_0(dx_0,dx_0^\star)
        \right)^\frac{1}{\sfp}
        = 
        \norm{\Xrt{0}-\Xstart{0}}_{L_{2\sfp}}^2.
    \end{align*} 
    Substituting the above into~\eqref{eqn:Reference:Scratch:TotalExpectation:p>=2:1} leads to 
    \begin{multline}\label{eqn:Reference:Scratch:TotalExpectation:p>=2:2}
        \LpLaw{\sfp}{}{\RefNorm{\Yt{N,t}}}
        \leq
        \norm{\Xrt{0}-\Xstart{0}}_{L_{2\sfp}}^2
        \\ 
        + 
        \left(
            \int_{\mathbb{R}^{2n}}
            \left(
                \pLpLaw{y_0}{
                    \expo{-2\lambda \tau(t)}
                    \Xi^r\br{\tau(t),\Yt{N}}
                }
                +
                \pLpLaw{y_0}{
                    \expo{-(2\lambda+\Boldomega) \tau(t)}
                    \Xi^r_{\mathcal{U}}\br{\tau(t),\Yt{N};\Boldomega}
                }
            \right)^\sfp
            \pi_0(dy_0)
        \right)^\frac{1}{\sfp}
        ,
    \end{multline}
    for all $(t,\sfp) \in [0,\tstar] \times \cbr{2,\dots,\sfp^\star}$.
    Now, let us define 
    \begin{subequations}\label{eqn:Reference:Scratch:ReDefn}
        \begin{align}
            \LpRefErrorPrime{y_0}{t}
            \doteq
            \LpLaw{2 \sfp}{y_0}{\Xrt{N,t} - \Xstart{N,t}}
            , 
            \quad
            \LpRefPrime{y_0}{t}
            \doteq
            \LpLaw{2 \sfp}{y_0}{\Xrt{N,t}}
            , 
            \label{eqn:Reference:Scratch:ReDefn:A}
            \\ 
            \LpRefError{y_0}
            \doteq
            \sup_{\nu \in [0,\tstar]}
            \LpLaw{2 \sfp}{y_0}{\Xrt{N,\nu} - \Xstart{N,\nu}}, 
            \quad 
            \LpRef{y_0}
            \doteq
            \sup_{\nu \in [0,\tstar]}
            \LpLaw{2 \sfp}{y_0}{\Xrt{N,\nu}},
            \label{eqn:Reference:Scratch:ReDefn:B}
        \end{align}
    \end{subequations}
    for $\sfp \in \cbr{1,\dots,\sfp^\star}$.
    Then, we can write 
    \begin{align*}
        \ELaw{}{\RefNorm{\Yt{N,t}}}
        =
        \int_{\mathbb{R}^{2n}}
            \ELaw{y_0}{\RefNorm{\Yt{N,t}}}
        \pi_0(dy_0)
        =&      
        \int_{\mathbb{R}^{2n}}
            \LTwoRefErrorPrime{y_0}^2
        \pi_0(dy_0)
        \\
        =&
        \ELaw{\pi_0}{
            \LTwoRefErrorPrime{y_0}^2
            } 
        =
        \LpLaw{1}{\pi_0}{\LTwoRefErrorPrime{y_0}^2}
        ,
        \quad  
        t \in [0,\tstar],
    \end{align*}
    and 
    \begin{align*}
        \LpLaw{\sfp}{}{\RefNorm{\Yt{N,t}}}
        =&
        \left(
            \int_{\mathbb{R}^{2n}}
                \left(\LpLaw{\sfp}{y_0}{\RefNorm{\Yt{N,t}}}\right)^\sfp
            \pi_0(dy_0)
        \right)^\frac{1}{\sfp}
        \\
        =& 
        \left(
            \int_{\mathbb{R}^{2n}}
                \LpRefErrorPrime{y_0}{t}^{2\sfp}
            \pi_0(dy_0)
        \right)^\frac{1}{\sfp} 
        = 
        \LpLaw{\sfp}{\pi_0}{\LpRefErrorPrime{y_0}{t}^2}
        ,
        \quad  
        t \in [0,\tstar].
    \end{align*}
    Hence, we can write~\eqref{eqn:Reference:Scratch:TotalExpectation:Bound:p=1:2} and~\eqref{eqn:Reference:Scratch:TotalExpectation:p>=2:2} as
    \begin{multline}\label{eqn:Reference:Scratch:TotalExpectation:Bound:p=1:3}
        \LpLaw{1}{\pi_0}{\LTwoRefErrorPrime{y_0}^2}
        \leq 
        \norm{\Xrt{0}-\Xstart{0}}_{L_2}^2
        + 
        \int_{\mathbb{R}^{2n}}
            \ELaw{y_0}{
                \expo{-2\lambda \tau(t)}
                \Xi^r\br{\tau(t),\Yt{N}}
            }
        \pi_0(dy_0)
        \\ 
        + 
        \int_{\mathbb{R}^{2n}}
            \ELaw{y_0}{
                \expo{-(2\lambda+\Boldomega) \tau(t)}
                \Xi^r_{\mathcal{U}}\br{\tau(t),\Yt{N};\Boldomega}
            }
        \pi_0(dy_0)
        ,
        \quad  
        \forall t \in [0,\tstar],
    \end{multline} 
    and 
    \begin{multline}\label{eqn:Reference:Scratch:TotalExpectation:p>=2:3}
        \LpLaw{\sfp}{\pi_0}{\LpRefErrorPrime{y_0}{t}^2}
        \leq
        \norm{\Xrt{0}-\Xstart{0}}_{L_{2\sfp}}^2
        \\ 
        + 
        \left(
            \int_{\mathbb{R}^{2n}}
            \left(
                \pLpLaw{y_0}{
                    \expo{-2\lambda \tau(t)}
                    \Xi^r\br{\tau(t),\Yt{N}}
                }
                +
                \pLpLaw{y_0}{
                    \expo{-(2\lambda+\Boldomega) \tau(t)}
                    \Xi^r_{\mathcal{U}}\br{\tau(t),\Yt{N};\Boldomega}
                }
            \right)^\sfp
            \pi_0(dy_0)
        \right)^\frac{1}{\sfp}
        ,
    \end{multline}
    for all $(t,\sfp) \in [0,\tstar] \times \cbr{2,\dots,\sfp^\star}$.
    Since $\tau^\star = \tstar$ by assumption, we can substitute the bounds derived in Lemma~\ref{lem:Appendix:ReferenceProcess:Final:Bound} into the above inequalities and get
    \begin{align}\label{eqn:Reference:Scratch:TotalExpectation:Bound:p=1:4}
        \LpLaw{1}{\pi_0}{\LTwoRefErrorPrime{y_0}^2}
        \leq 
        \norm{\Xrt{0}-\Xstart{0}}_{L_2}^2
        + 
        \int_{\mathbb{R}^{2n}}
            \Phi^r(1,\Boldomega)
        \pi_0(dy_0)
        ,
        \quad  
        \forall t \in [0,\tstar],
    \end{align} 
    and
    \begin{align}\label{eqn:Reference:Scratch:TotalExpectation:p>=2:4}
        \LpLaw{\sfp}{\pi_0}{\LpRefErrorPrime{y_0}{t}^2}
        \leq
        \norm{\Xrt{0}-\Xstart{0}}_{L_{2\sfp}}^2
        + 
        \left(
            \int_{\mathbb{R}^{2n}}
                \Phi^r(\sfp,\Boldomega)^\sfp
            \pi_0(dy_0)
        \right)^\frac{1}{\sfp}
        ,
        \quad \forall (t,\sfp) \in [0,\tstar] \times \cbr{2,\dots,\sfp^\star},
    \end{align}
    where we have defined
    \begin{multline*}
        \Phi^r(\sfp,\Boldomega)
        = 
        \Delta^r_\circ(\sfp, \Boldomega)
        +
        \widebreve{\Delta}^r_{\circledcirc}(\sfp, \Boldomega)
        \left(\LpRef{y_0}\right)^\frac{1}{2}
        +
        \widebreve{\Delta}^r_{\odot}(\sfp, \Boldomega)
        \LpRef{y_0}
        +
        \Delta^r_\odot(\sfp, \Boldomega)
        \LpRefError{y_0}
        \\
        +
        \left(
            \widebreve{\Delta}^r_{\otimes}(\sfp,\Boldomega)
            \LpRef{y_0}
            +
            \Delta^r_{\otimes}(\sfp,\Boldomega)
            \LpRefError{y_0}
        \right)
        \left(\LpRef{y_0}\right)^\frac{1}{2}
        \\
        +
        \left(
        \widebreve{\Delta}^r_{\circledast}(\sfp,\Boldomega)
            \LpRef{y_0}
            +
            \Delta^r_{\circledast}(\sfp,\Boldomega)
            \LpRefError{y_0}
        \right)
        \left(\LpRef{y_0}\right),
        \quad \sfp \in \cbr{1,\dots,\sfp^\star}.
    \end{multline*} 
    The two bounds above can be combined into the following single bound over $\sfp \in \cbr{1,\dots,\sfp^\star}$:
    \begin{align}\label{eqn:Reference:Scratch:TotalExpectation:1:Pre}
        \LpLaw{\sfp}{\pi_0}{\LpRefErrorPrime{y_0}{t}^2}
        \leq
        \norm{\Xrt{0}-\Xstart{0}}_{L_{2\sfp}}^2
        + 
        \left(
            \int_{\mathbb{R}^{2n}}
                \Phi^r(\sfp,\Boldomega)^\sfp
            \pi_0(dy_0)
        \right)^\frac{1}{\sfp}
        ,
        \quad \forall (t,\sfp) \in [0,\tstar] \times \cbr{1,\dots,\sfp^\star}.
    \end{align}
    It then follows from the Minkowski's and the Cauchy-Schwarz inequalities that 
    \begin{multline}\label{eqn:Reference:Scratch:TotalExpectation:Phi:1}
        \left(
            \int_{\mathbb{R}^{2n}}
                \Phi^r(\sfp,\Boldomega)^\sfp
            \pi_0(dy_0)
        \right)^\frac{1}{\sfp}
        \\
        \leq
        \Delta^r_\circ(\sfp, \Boldomega)
        +
        \widebreve{\Delta}^r_{\circledcirc}(\sfp, \Boldomega)
        \LpLaw{\sfp}{\pi_0}{ \LpRef{y_0}^\frac{1}{2} }
        +
        \widebreve{\Delta}^r_{\odot}(\sfp, \Boldomega)
        \LpLaw{\sfp}{\pi_0}{\LpRef{y_0}}
        +
        \Delta^r_\odot(\sfp, \Boldomega)
        \LpLaw{\sfp}{\pi_0}{\LpRefError{y_0}}
        \\
        +
        \widebreve{\Delta}^r_{\otimes}(\sfp,\Boldomega)
        \LpLaw{\sfp}{\pi_0}{\LpRef{y_0}^\frac{3}{2}} 
        +
        \Delta^r_{\otimes}(\sfp,\Boldomega)
        \LpLaw{2\sfp}{\pi_0}{ \LpRefError{y_0}}
        \LpLaw{2\sfp}{\pi_0}{ \LpRef{y_0}^\frac{1}{2}}
        \\
        +
        \widebreve{\Delta}^r_{\circledast}(\sfp,\Boldomega)
        \LpLaw{\sfp}{\pi_0}{\LpRef{y_0}^2}
        +
        \Delta^r_{\circledast}(\sfp,\Boldomega)
        \LpLaw{2\sfp}{\pi_0}{\LpRefError{y_0}} 
        \LpLaw{2\sfp}{\pi_0}{\LpRef{y_0}} 
        ,
    \end{multline}
    for all $(t,\sfp) \in [0,\tstar] \times \cbr{1,\dots,\sfp^\star}$.
    Now, we have the following inequalities due to the Cauchy Schwarz (C-S) and Jensen's (J) inequalities:
    \begin{subequations}
        \begin{align}
            \LpLaw{\sfp}{\pi_0}{ \LpRef{y_0}^\frac{1}{2} }
            \overset{\text{(C-S)}}{\leq}
            \LpLaw{2\sfp}{\pi_0}{ \LpRef{y_0}^\frac{1}{2} }
            \overset{\text{(J)}}{\leq}
            \left(\LpLaw{2\sfp}{\pi_0}{ \LpRef{y_0} }\right)^\frac{1}{2},
            \quad 
            \LpLaw{\sfp}{\pi_0}{\LpRef{y_0}}
            \overset{\text{(C-S)}}{\leq}
            \LpLaw{2\sfp}{\pi_0}{\LpRef{y_0}}, 
            \\
            \LpLaw{\sfp}{\pi_0}{\LpRef{y_0}^\frac{3}{2}}
            \overset{\text{(C-S)}}{\leq}
            \LpLaw{2\sfp}{\pi_0}{\LpRef{y_0}}
            \LpLaw{2\sfp}{\pi_0}{\LpRef{y_0}^\frac{1}{2}}
            \overset{\text{(J)}}{\leq}
            \left(\LpLaw{2\sfp}{\pi_0}{\LpRef{y_0}}\right)^\frac{3}{2},
            \\ 
            \LpLaw{\sfp}{\pi_0}{\LpRef{y_0}^2}
            =
            \LpLaw{\sfp}{\pi_0}{\LpRef{y_0} \LpRef{y_0}}
            \overset{\text{(C-S)}}{\leq}
            \left(\LpLaw{2\sfp}{\pi_0}{\LpRef{y_0}}\right)^2, 
            \\ 
            \LpLaw{\sfp}{\pi_0}{\LpRefError{y_0}}
            \overset{\text{(C-S)}}{\leq}
            \LpLaw{2\sfp}{\pi_0}{\LpRefError{y_0}}.
        \end{align}
    \end{subequations}
    Substituting these bounds into~\eqref{eqn:Reference:Scratch:TotalExpectation:Phi:1} produces 
    \begin{multline}\label{eqn:Reference:Scratch:TotalExpectation:Phi:2}
        \left(
            \int_{\mathbb{R}^{2n}}
                \Phi^r(\sfp,\Boldomega)^\sfp
            \pi_0(dy_0)
        \right)^\frac{1}{\sfp}
        \\
        \leq
        \Delta^r_\circ(\sfp, \Boldomega)
        +
        \widebreve{\Delta}^r_{\circledcirc}(\sfp, \Boldomega)
        \left(\LpLaw{2\sfp}{\pi_0}{ \LpRef{y_0} }\right)^\frac{1}{2}
        +
        \widebreve{\Delta}^r_{\odot}(\sfp, \Boldomega)
        \LpLaw{2\sfp}{\pi_0}{\LpRef{y_0}}
        +
        \Delta^r_\odot(\sfp, \Boldomega)
        \LpLaw{2\sfp}{\pi_0}{\LpRefError{y_0}}
        \\
        +
        \left(
            \widebreve{\Delta}^r_{\otimes}(\sfp,\Boldomega)
            \LpLaw{2\sfp}{\pi_0}{\LpRef{y_0}} 
            +
            \Delta^r_{\otimes}(\sfp,\Boldomega)
            \LpLaw{2\sfp}{\pi_0}{ \LpRefError{y_0}}
        \right)
        \left(\LpLaw{2\sfp}{\pi_0}{ \LpRef{y_0} }\right)^\frac{1}{2}
        \\
        +
        \left(
            \widebreve{\Delta}^r_{\circledast}(\sfp,\Boldomega)
            \LpLaw{2\sfp}{\pi_0}{\LpRef{y_0}}
            +
            \Delta^r_{\circledast}(\sfp,\Boldomega)
            \LpLaw{2\sfp}{\pi_0}{\LpRefError{y_0}}  
        \right)
        \LpLaw{2\sfp}{\pi_0}{\LpRef{y_0}}
        ,
    \end{multline}
    for all $(t,\sfp) \in [0,\tstar] \times \cbr{1,\dots,\sfp^\star}$.

    Next, using the continuity of $\nu \mapsto \LpLaw{2\sfp}{y_0}{\Xrt{N,\nu} - \Xstart{N,\nu}  }$ over the closed interval $[0,\tstar]$, we may assume the existence of some $t' \in [0,\tstar]$, such that 
    \begin{align*}
        \sup_{\nu \in [0,\tstar]} \LpLaw{2\sfp}{y_0}{\Xrt{N,\nu} - \Xstart{N,\nu}  } 
        =
        \LpLaw{2\sfp}{y_0}{\Xrt{N,t'} - \Xstart{N,t'}  }.
    \end{align*}
    Therefore, using the definition of $\LpRefError{y_0}$ in~\eqref{eqn:Reference:Scratch:ReDefn:B}, we see that
    \begin{align*}
        \LpLaw{2\sfp}{\pi_0}{\LpRefError{y_0}}
        =&
        \left(
            \int_{\mathbb{R}^{2n}}
                \left( 
                    \sup_{\nu \in [0,\tstar]}
                    \LpLaw{2\sfp}{y_0}{\Xrt{N,\nu} - \Xstart{N,\nu}  } 
                \right)^{2\sfp}
            \pi_0(dy_0)
        \right)^\frac{1}{2\sfp}
        \\ 
        =&
        \left(
            \int_{\mathbb{R}^{2n}}
                \left( 
                    \LpLaw{2\sfp}{y_0}{\Xrt{N,t'} - \Xstart{t'}  } 
                \right)^{2\sfp}
            \pi_0(dy_0)
        \right)^\frac{1}{2\sfp}
        = 
        \left(
            \int_{\mathbb{R}^{2n}}
                    \ELaw{y_0}{ \norm{\Xrt{N,t'} - \Xstart{t'}}^{2\sfp}  } 
            \pi_0(dy_0)
        \right)^\frac{1}{2\sfp}
        .
    \end{align*}
    We can further develop the expression as follows 
    \begin{align}
        \LpLaw{2\sfp}{\pi_0}{\LpRefError{y_0}}
        =& 
        \left(
            \int_{\mathbb{R}^{2n}}
                    \ELaw{y_0}{ \norm{\Xrt{N,t'} - \Xstart{t'}}^{2\sfp}  } 
            \pi_0(dy_0)
        \right)^\frac{1}{2\sfp}
        \notag 
        \\
        =& 
        \left(
            \int_{\mathbb{R}^{2n}}
                    \ELaw{}{ \norm{\Xrt{N,t'} - \Xstart{N,t'}}^{2\sfp}~|~\Yt{N,0}=y_0  } 
            \pi_0(dy_0)
        \right)^\frac{1}{2\sfp}
        =
        \LpLaw{2\sfp}{}{\Xrt{N,t'}-\Xstart{N,t'}}
        , 
    \end{align}
    for some $t' \in [0,\tstar]$, and thus 
    \begin{align}\label{eqn:Reference:Scratch:TotalExpectation:RefRho}
        \LpLaw{2\sfp}{\pi_0}{\LpRefError{y_0}}
        \leq
        \sup_{\nu \in [0,\tstar]} 
        \LpLaw{2\sfp}{}{\Xrt{N,\nu}-\Xstart{N,\nu}}.
    \end{align}
    Similarly, using the definition of $\LpRef{y_0}$ in~\eqref{eqn:Reference:Scratch:ReDefn:B} and the Minkowski's inequality, we conclude that there exists some $t'' \in [0,\tstar]$ such that
    \begin{align*}
        \LpLaw{2\sfp}{\pi_0}{\LpRef{y_0}}
        =
        \LpLaw{2\sfp}{}{\Xrt{N,t''}}
        \leq     
        \LpLaw{2\sfp}{}{\Xrt{N,t''}-\Xstart{N,t''}}
        +
        \LpLaw{2\sfp}{}{\Xstart{N,t''}},
        \quad t'' \in [0,\tstar],
    \end{align*}
    which further implies
    \begin{align}\label{eqn:Reference:Scratch:TotalExpectation:RefRhoError}
        \LpLaw{2\sfp}{\pi_0}{\LpRef{y_0}}
        \leq 
        \sup_{\nu \in [0,\tstar]}    
        \LpLaw{2\sfp}{}{\Xrt{N,\nu}-\Xstart{N,\nu}}
        +
        \sup_{\nu \in [0,\tstar]}
        \LpLaw{2\sfp}{}{\Xstart{N,\nu}}.
    \end{align} 
   It then follows from~\eqref{eqn:Reference:Scratch:TotalExpectation:RefRho}~-~\eqref{eqn:Reference:Scratch:TotalExpectation:RefRhoError}, the contradiction statement~\eqref{eqn:Reference:Scratch:Statement:Not:Q}, and Assumption~\ref{assmp:NominalSystem:FiniteMomentsWasserstein}
    \begin{align*}
    \LpLaw{2\sfp}{\pi_0}{\LpRefError{y_0}}
    \leq
    \sup_{\nu \in [0,\tstar]} 
    \LpLaw{2\sfp}{}{\Xrt{N,\nu}-\Xstart{N,\nu}}
    \leq \RefRho, 
    \\
    \LpLaw{2\sfp}{\pi_0}{\LpRef{y_0}}
    \leq 
    \sup_{\nu \in [0,\tstar]}    
    \LpLaw{2\sfp}{}{\Xrt{N,\nu}-\Xstart{N,\nu}}
    +
    \sup_{\nu \in [0,\tstar]}
    \LpLaw{2\sfp}{}{\Xstart{N,\nu}}
    \leq     
    \RefRho+\Delta_\star,
    \end{align*}
    for $\sfp \in \cbr{1,\dots,\sfp^\star}$.
    Substituting the above into~\eqref{eqn:Reference:Scratch:TotalExpectation:Phi:2}, and substituting the resulting bound into~\eqref{eqn:Reference:Scratch:TotalExpectation:1:Pre}, we get:
    \begin{align*}
        \LpLaw{\sfp}{\pi_0}{\LpRefErrorPrime{y_0}{t}^2}
        \leq&
        \norm{\Xrt{0}-\Xstart{0}}_{L_{2\sfp}}^2
        + 
        \Delta^r_\circ(\sfp, \Boldomega)
        +
        \widebreve{\Delta}^r_{\circledcirc}(\sfp, \Boldomega)
        \left(\RefRho+\Delta_\star\right)^\frac{1}{2}
        \\
        &+
        \widebreve{\Delta}^r_{\odot}(\sfp, \Boldomega)
        \left(\RefRho+\Delta_\star\right)
        +
        \Delta^r_\odot(\sfp, \Boldomega)
        \RefRho
        \\
        &+
        \left(
            \widebreve{\Delta}^r_{\otimes}(\sfp,\Boldomega)
            \left(\RefRho+\Delta_\star\right) 
            +
            \Delta^r_{\otimes}(\sfp,\Boldomega)
            \RefRho
        \right)
        \left(\RefRho+\Delta_\star\right)^\frac{1}{2}
        \\
        &+
        \left(
            \widebreve{\Delta}^r_{\circledast}(\sfp,\Boldomega)
            \left(\RefRho+\Delta_\star\right)
            +
            \Delta^r_{\circledast}(\sfp,\Boldomega)
            \RefRho  
        \right)
        \left(\RefRho+\Delta_\star\right)
        ,
    \end{align*}
    for all $t \in [0,\tstar]$.
    Substituting the definition of constants $\Delta^r_i(\sfp,\Boldomega)$, $i \in \cbr{ \circ, \circledcirc,\odot, \otimes, \circledast   }$, from Lemma~\ref{lem:Appendix:ReferenceProcess:Final:Bound} into the above inequality and re-arranging the terms leads to 
    \begin{align}\label{eqn:Reference:Scratch:TotalExpectation:1}
            \LpLaw{\sfp}{\pi_0}{\LpRefErrorPrime{y_0}{t}^2}
            -
            \frac{\Delta_g^\perp \Delta_\mu^\perp}{\lambda}
            \RefRho^2
        \leq
        \norm{\Xrt{0}-\Xstart{0}}_{L_{2\sfp}}^2
        + 
        \Gamma_r\left(\RefRho,\sfp,\Boldomega\right)
        +
        \frac{1}{\absolute{2\lambda - \Boldomega}}
        \Theta_r\left(\RefRho,\sfp,\Boldomega\right)
        ,
    \end{align}
    for all $t \in [0,\tstar]$, where $\Gamma_r\left(\RefRho,\sfp,\Boldomega\right)$ and $\Theta_r\left(\RefRho,\sfp,\Boldomega\right)$ are defined in~\eqref{eqn:Definitions:Reference:Gamma} and~\eqref{eqn:Definitions:Reference:Theta}, respectively.
    Now, using the definition of  $\LpRefErrorPrime{y_0}{t}$ from~\eqref{eqn:Reference:Scratch:ReDefn:A}, we see that 
    \begin{align*}
        \LpLaw{\sfp}{\pi_0}{\LpRefErrorPrime{y_0}{t}^2}
        =
        \left(
            \int_{\mathbb{R}^{2n}}
                \left( 
                    \LpLaw{2\sfp}{y_0}{\Xrt{N,t} - \Xstart{N,t}  } 
                \right)^{2\sfp}
            \pi_0(dy_0)
        \right)^\frac{1}{\sfp}
        =&
        \left(
            \int_{\mathbb{R}^{2n}}
                    \ELaw{y_0}{ \norm{\Xrt{N,t} - \Xstart{N,t}}^{2\sfp}  } 
            \pi_0(dy_0)
        \right)^\frac{1}{\sfp}
        \\
        =&
        \left(
            \int_{\mathbb{R}^{2n}}
                    \ELaw{y_0}{ \norm{\Xrt{N,t} - \Xstart{N,t}}^{2\sfp}  } 
            \pi_0(dy_0)
        \right)^\frac{2}{2\sfp}
        , 
    \end{align*}
    which implies that 
    \begin{align*}
        \LpLaw{\sfp}{\pi_0}{\LpRefErrorPrime{y_0}{t}^2}
        =
        \left(
            \int_{\mathbb{R}^{2n}}
                    \ELaw{}{ \norm{\Xrt{N,t} - \Xstart{N,t}}^{2\sfp}~|~\Yt{N,0}=y_0  } 
            \pi_0(dy_0)
        \right)^\frac{2}{2\sfp}
        = 
        \norm{\Xrt{N,t}-\Xstart{N,t}}_{L_{2\sfp}}^2
        .
    \end{align*}
    We can thus write~\eqref{eqn:Reference:Scratch:TotalExpectation:1} as  
    \begin{multline}\label{eqn:Reference:Scratch:TotalExpectation:2}
            \norm{\Xrt{N,t}-\Xstart{N,t}}_{L_{2\sfp}}^2
            -
            \frac{\Delta_g^\perp \Delta_\mu^\perp}{\lambda}
            \RefRho^2
        \\
        \leq
        \norm{\Xrt{0}-\Xstart{0}}_{L_{2\sfp}}^2
        + 
        \Gamma_r\left(\RefRho,\sfp,\Boldomega\right)
        +
        \frac{1}{\absolute{2\lambda - \Boldomega}}
        \Theta_r\left(\RefRho,\sfp,\Boldomega\right)
        ,
        \quad 
        \forall t \in [0,\tstar].
    \end{multline}
    Now, at the temporal instance $t = \tstar$, we have from the contradiction statement~\eqref{eqn:Reference:Scratch:Statement:Not:Q} that 
    \begin{align*}
        \norm{\Xrt{N,\tstar}-\Xstart{N,\tstar}}_{L_{2\sfp}}^2
        = 
        \RefRho^2.
    \end{align*}
    Hence, at $t = \tstar$, the inequality in~\eqref{eqn:Reference:Scratch:TotalExpectation:2} reduces to
    \begin{align*}
        \left(
            1
            -
            \frac{\Delta_g^\perp \Delta_\mu^\perp}{\lambda}
        \right)
        \RefRho^2
        \leq
        \norm{\Xrt{0}-\Xstart{0}}_{L_{2\sfp}}^2
        + 
        \Gamma_r\left(\RefRho,\sfp,\Boldomega\right)
        +
        \frac{1}{\absolute{2\lambda - \Boldomega}}
        \Theta_r\left(\RefRho,\sfp,\Boldomega\right)
        ,
    \end{align*}
    which implies that  
    \begin{align}\label{eqn:Reference:Scratch:TotalExpectation:4}
        \frac{1}{\absolute{2\lambda - \Boldomega}}
        \Theta_r\left(\RefRho,\sfp,\Boldomega\right)
        \geq 
        \left(
            1
            -
            \frac{\Delta_g^\perp \Delta_\mu^\perp}{\lambda}
        \right)
        \RefRho^2
        - 
        \norm{\Xrt{0}-\Xstart{0}}_{L_{2\sfp}}^2
        - 
        \Gamma_r\left(\RefRho,\sfp,\Boldomega\right)
        .
    \end{align}
    
    However, one sees from~\eqref{eqn:Definitions:Reference:BandwithCondition}, with $\alpha\left(\Xrt{0},\Xstart{0}\right)_{2\sfp} = \norm{\Xrt{0}-\Xstart{0}}_{L_{2\sfp}}$, that the choice of $\Boldomega$ ensures that
    \begin{align*}
        \frac{1}{\absolute{2\lambda - \Boldomega}}
        \Theta_r\left(\RefRho,\sfp,\Boldomega\right)
        <
        \left(
            1
            -
            \frac{\Delta_g^\perp \Delta_\mu^\perp}{\lambda}
        \right)
        \RefRho^2
        - 
        \norm{\Xrt{0}-\Xstart{0}}_{L_{2\sfp}}^2
        -
        \Gamma_r\left(\RefRho,\sfp,\Boldomega\right),
    \end{align*} 
    which contradicts the conclusion above in~\eqref{eqn:Reference:Scratch:TotalExpectation:4} that follows from the postulation in~\eqref{eqn:Reference:Scratch:Statement:Not:Q}. 
    Therefore, the desired result in~\eqref{eqn:lem:Reference:FirstExit:UB} is proved since its negation~\eqref{eqn:Reference:Scratch:Statement:Not:Q} leads to a contradiction.

\end{proof}

Next, we extend the strong solutions of the truncated process to the original joint process in Definition~\ref{def:Reference:JointProcess}. 
\begin{lemma}\label{lem:Reference:WellPosedness}

    Let the assumptions in Sec.~\ref{subsec:Assumptions} hold.
    Additionally, for $\alpha\left(\Xrt{0},\Xstart{0}\right)_{2\sfp} = \norm{\Xrt{0}-\Xstart{0}}_{L_{2\sfp}}$ and $\sfp \in \cbr{1,\dots,\sfp^\star}$, let $\RefRho \in \mathbb{R}_{>0}$ be as defined in~\eqref{eqn:Definitions:Reference:Rho} and let filter bandwidth condition in~\eqref{eqn:Definitions:Reference:BandwithCondition} hold, where $\sfp^\star$ is defined in Assumption~\ref{assmp:NominalSystem:FiniteMomentsWasserstein}.
   Then the joint process in Definition~\ref{def:Reference:JointProcess} admits a unique strong solution $\Yt{t} = \br{\Xrt{t},\Xstart{t}}$ for all $T \in (0,\infty)$.
    Furthermore, the strong solution satisfies   
    \begin{align}\label{eqn:lem:Reference:WellPosedness:Bound}
        \LpLaw{2\sfp}{}{\Xrt{t}-\Xstart{t}} < \RefRho, \quad \forall t \in [0,T],
    \end{align}
    for any $T \in (0,\infty)$.
\end{lemma}
\begin{proof}
    We prove the existence of a unique strong solution to the reference process in Definition~\ref{def:ReferenceProcess} by following the approach in~\cite[Thm.~3.5]{khasminskii2011stochastic},~\cite[Thm.~5.3.1]{durrett1996stochastic}.
    Then, along with well-posedness of the known (nominal) system in Assumption~\ref{assmp:NominalSystem:FiniteMomentsWasserstein}, we  will be able to conclude the well-posedness of the joint known (nominal)-reference process. 
    Let us define 
    \begin{align*}
        t_N \doteq \inf \cbr{t \in \mathbb{R}_{\geq 0}~:~\Yt{N,t} \notin U_N},         
    \end{align*}
    where the open and bounded set $U_N$ is defined in~\eqref{eqn:OpenBoundedSet}, for an arbitrary $N \in \mathbb{R}_{>0}$.
    Then, we can write the definition of $\tau_N$ in~\eqref{eqn:Reference:FirstExitTime} as 
    \begin{align*}
        \tau_N \doteq T \wedge t_N,         
    \end{align*}
    where $T \in (0,\infty)$ is the horizon, and $t_N$ is the first exit time of the truncated joint process $\Yt{N,t} = \br{\Xrt{N,t},\Xstart{N,t}}$ from the set $U_N$.
    
    Now, Lemma~\ref{lem:Reference:MomentBounds:FirstExit} and Jensen's inequality allow us to conclude
    \begin{align*}
        \LpLaw{2}{}{\Xrt{N,t}-\Xstart{N,t}}
        \leq
        \LpLaw{2\sfp}{}{\Xrt{N,t}-\Xstart{N,t}} < \RefRho, \quad \forall t \in [0,T \wedge t_N],
    \end{align*}
    which can be further expressed as 
    \begin{align}\label{eqn:lem:Reference:Global:FirstExit:A}
        \ELaw{}{ \norm{\Xrt{N,t}-\Xstart{N,t}}^2 } < \RefRho^2, \quad \forall t \in [0,T \wedge t_N].
    \end{align}
    Furthermore, this bound and Assumption~\ref{assmp:NominalSystem:FiniteMomentsWasserstein} allow us to conclude 
    \begin{align}\label{eqn:lem:Reference:Global:FirstExit:B}
        \ELaw{}{\norm{\Xrt{N,t}}^2} 
        \overset{(\star)}{\leq}       
        2\ELaw{}{\norm{\Xrt{N,t}-\Xstart{N,t}}^2}
        + 
        2\ELaw{}{\norm{\Xstart{N,t}}^2}
        \leq     
        2\left( \RefRho^2 + \Delta_\star^2 \right), 
        \quad \forall  t \in [0,T \wedge t_N], 
    \end{align}
    where $(\star)$ is due to the triangle inequality followed by~\cite[Prop.~3.1.10~(\romannum 3)]{athreya2006measure}. 

    In order to establish the well-posedness of the reference process over the interval $[0,T]$ it suffices to show that
    \begin{align*}
        \lim_{N \rightarrow \infty }
        \Probability{t_N \leq T} = 0, \quad \forall T \in (0,\infty).
    \end{align*}
    Using the definition of the expectation operator, we derive the estimate 
    \begin{align*}
        \ELaw{}{\norm{\Xrt{N,t_N \wedge T}}^2}
        =&
        \int_\Omega \norm{\Xrt{N,t_N \wedge T}}^2 d\mathbb{P}
        \\
        =&
        \int_{\cbr{t_N \leq T}} \norm{\Xrt{N,t_N}}^2 d\mathbb{P}
        + 
        \int_{\cbr{t_N > T}} \norm{\Xrt{N,T}}^2 d\mathbb{P}
        \geq     
        \int_{\cbr{t_N \leq T}} \norm{\Xrt{N,t_N}}^2 d\mathbb{P}.
    \end{align*}
    We similarly obtain
    \begin{align*}
        \ELaw{}{\norm{\Xrt{N,t_N \wedge T} - \Xstart{N,t_N \wedge T}}^2}
        \geq     
        \int_{\cbr{t_N \leq T}} \norm{\Xrt{N,t_N}-\Xstart{N,t_N}}^2 d\mathbb{P}.
    \end{align*}
    Combining the above two inequalities leads to 
    \begin{multline*}
        \int_{\cbr{t_N \leq T}} \norm{\Xrt{N,t_N}}^2 d\mathbb{P}
        +
        \int_{\cbr{t_N \leq T}} \norm{\Xrt{N,t_N}-\Xstart{N,t_N}}^2 d\mathbb{P}
        \\
        \leq 
        \ELaw{}{\norm{\Xrt{N,t_N \wedge T}}^2}
        +
        \ELaw{}{\norm{\Xrt{N,t_N \wedge T} - \Xstart{N,t_N \wedge T}}^2}, 
        \quad
        \forall T \in (0,\infty).
    \end{multline*}
   It then follows from the bounds in~\eqref{eqn:lem:Reference:Global:FirstExit:A} and~\eqref{eqn:lem:Reference:Global:FirstExit:B}
    \begin{multline}\label{eqn:lem:Reference:Global:Existence:1}
        \int_{\cbr{t_N \leq T}} \norm{\Xrt{N,t_N}}^2 d\mathbb{P}
        +
        \int_{\cbr{t_N \leq T}} \norm{\Xrt{N,t_N}-\Xstart{N,t_N}}^2 d\mathbb{P}
        \\
        \leq 
        \ELaw{}{\norm{\Xrt{N,t_N \wedge T}}^2}
        +
        \ELaw{}{\norm{\Xrt{N,t_N \wedge T} - \Xstart{N,t_N \wedge T}}^2}
        \leq     
        3\RefRho^2 + 2\Delta_\star^2
        , 
        \quad 
        \forall T \in (0,\infty)
        .
    \end{multline}
    Since $t_N$ is the first exit time of the process $\Yt{N,t}$ from the set $U_N \subset \subset \mathbb{R}^{2n}$ defined in~\eqref{eqn:OpenBoundedSet}, the stopped process $\Yt{N,t_N}$ satisfies 
    \begin{align*}
        \Yt{N,t_N} = \br{\Xrt{N,t_N},\Xstart{N,t_N}} \in \partial U_N,
        \quad   
        \partial U_N
        = 
        \cbr{
            z = (z_1,z_2),~
            z_1,z_2 \in \mathbb{R}^{n}~|~
            \norm{z_1}^2 + \norm{z_2}^2 = N^2
        }.
    \end{align*}
    Therefore, we deduce from~\eqref{eqn:lem:Reference:Global:Existence:1} that 
    \begin{multline}\label{eqn:lem:Reference:Global:Existence:2}
        \inf_{(z_1,z_2) \in \partial U_N}
        \left(    
            \norm{z_1}^{2} 
            +
            \norm{z_1 - z_2}^{2} 
        \right)
        \Probability{t_N \leq T}
        \\
        \leq 
        \ELaw{}{\norm{\Xrt{N,t_N \wedge T}}^2}
        +
        \ELaw{}{\norm{\Xrt{N,t_N \wedge T} - \Xstart{N,t_N \wedge T}}^2}
        \leq     
        3\RefRho^2 + 2\Delta_\star^2
        , 
        \quad 
        \forall T \in (0,\infty)
        .
    \end{multline}
    Using the method of Lagrange multipliers, one can show that 
    \begin{align*}
        \inf_{(z_1,z_2) \in \partial U_N}
        \left(    
            \norm{z_1}^{2} 
            +
            \norm{z_1 - z_2}^{2} 
        \right)
        =
        \frac{1}{2}
        \min  
        \left\{
            3 + \sqrt{5}
            ,
            3 - \sqrt{5}
        \right\} 
        N^2
        \geq 
        \frac{N^2}{3}.
    \end{align*}
    \begingroup
    \endgroup
    Hence,~\eqref{eqn:lem:Reference:Global:Existence:2} can be developed into 
    \begin{align*}
        \frac{N^2}{3}
        \Probability{t_N \leq T}
        \leq
        \inf_{(z_1,z_2) \in \partial U_N}
        \left(    
            \norm{z_1}^{2} 
            +
            \norm{z_1 - z_2}^{2} 
        \right)
        \Probability{t_N \leq T}
        \leq    
        3\RefRho^2 + 2\Delta_\star^2
        , 
        \quad 
        \forall T \in (0,\infty)
        ,
    \end{align*}
    which further implies that 
    \begin{align*}
        \Probability{t_N \leq T}
        \leq    
        \frac{3}{N^2}
        \left(3\RefRho^2 + 2\Delta_\star^2\right)
        , 
        \quad 
        \forall T \in (0,\infty)
        .
    \end{align*}
    Consequently, 
    \begin{align}\label{eqn:lem:Reference:Global:Existence:Limit}
        \lim_{N \rightarrow \infty}
        \Probability{t_N \leq T}
        =
        0
        , 
        \quad 
        \forall T \in (0,\infty),
    \end{align}
    which establishes the well-posedness of the reference process. 
    We thus define $\Yt{t} = \br{\Xrt{t},\Xstart{t}}$, where $\Xrt{t} \doteq \lim_{N \rightarrow \infty}\Xrt{\tau_N(t)}$, and $\tau_N(t) = t \wedge \tau_N$. 
    Therefore, it follows from Fatou's lemma~\cite[Thm.~2.3.7]{athreya2006measure} and Lemma~\ref{lem:Reference:MomentBounds:FirstExit} 
    \begin{align*}
        \LpLaw{2\sfp}{}{\Xrt{t}-\Xstart{t}} 
        \leq 
        \liminf_{N \rightarrow \infty} \LpLaw{2\sfp}{}{\Xrt{\tau_N(t)}-\Xstart{\tau_N(t)}}
        < 
        \RefRho, 
        \quad 
        \forall t \in [0,T],
    \end{align*}
    for any $T \in (0,\infty)$, which establishes the desired result in~\eqref{eqn:lem:Reference:WellPosedness:Bound}.

\end{proof}

We now formulate uniform ultimate bounds (UUBs) in the following proposition. 
\begin{proposition}\label{prop:Reference:UUB}
     Let the assumptions in Sec.~\ref{subsec:Assumptions} hold and let $\RefRho \in \mathbb{R}_{>0}$, $\sfp \in \cbr{1,\dots,\sfp^\star}$, be as defined in~\eqref{eqn:Definitions:Reference:Rho} with $\alpha\left(\Xrt{0},\Xstart{0}\right)_{2\sfp} = \norm{\Xrt{0}-\Xstart{0}}_{L_{2\sfp}}$, where $\sfp^\star$ is defined in Assumption~\ref{assmp:NominalSystem:FiniteMomentsWasserstein}.
    If the filter bandwidth condition in~\eqref{eqn:Definitions:Reference:BandwithCondition} holds with $\alpha\left(\Xrt{0},\Xstart{0}\right)_{2\sfp} = \norm{\Xrt{0}-\Xstart{0}}_{L_{2\sfp}}$, then for any deterministic $\that \in (0,T)$,  the joint process $\Yt{t} = \br{\Xrt{t},\Xstart{t}}$ in Definition~\ref{def:Reference:JointProcess} satisfies
    \begin{align}\label{eqn:prop:Reference:UUB:Main}
        \LpLaw{2\sfp}{}{\Xrt{t}-\Xstart{t}} 
        \leq
        \RefRhoUUB{\sfp,\Boldomega,\that} < \RefRho, 
        \quad 
        \forall t \in [\that,T],
    \end{align} 
    where 
    \begin{multline}\label{eqn:Definitions:Reference:VarTheta}
        \RefRhoUUB{\sfp,\Boldomega,\that}
        \\
        =
        \left(
            \expo{-2\lambda \that}
            \norm{\Xrt{0}-\Xstart{0}}_{L_{2\sfp}}^2
            +
            \frac{\Delta_g^\perp \Delta_\mu^\perp}{\lambda}
            \RefRho^2
            + 
            \Gamma_r\left(\RefRho,\sfp,\Boldomega\right)
            +
            \frac{1}{\absolute{2\lambda - \Boldomega}}
            \Theta_r\left(\RefRho,\sfp,\Boldomega\right)
        \right)^{\frac{1}{2}},
    \end{multline}
    and where the pertinent definitions are provided in Sec~\ref{subsec:L1DRAC:Parameters}.

\end{proposition}
\begin{proof}
    Owing to Lemma~\ref{lem:Reference:WellPosedness}, the results of Lemma~\ref{lem:Reference:dV} hold for the stopping time $T(t) \doteq t \wedge T$.
    That is, for any deterministic $\that \in (0,T)$, the inequalities~\eqref{eqn:lem:Reference:dV:Supremums:UB:p=1} and~\eqref{eqn:lem:Reference:dV:Supremums:UB} can be written as 
    \begin{multline}\label{eqn:prop:Reference:UUB:p=1}
        \ELaw{y_0}{\RefNorm{\Yt{T(t)}}}
        \leq
        \expo{-2\lambda \that}
        \RefNorm{y_0}
        +
        \ELaw{y_0}{
            \expo{-2\lambda T(t)}
            \Xi^r\br{T(t),\Yt{}}
        }
        \\
        +
        \ELaw{y_0}{
            \expo{-(2\lambda+\Boldomega) T(t)}
            \Xi^r_{\mathcal{U}}\br{T(t),\Yt{};\Boldomega}
        }
        , 
        \quad \forall t \in [\that,\infty),
    \end{multline}
    and
    \begin{multline}\label{eqn:prop:Reference:UUB:p>=2}
        \pLpLaw{y_0}{
            \RefNorm{\Yt{T(t)}} 
        }
        \leq
        \expo{-2\lambda \that}
        \RefNorm{y_0}
        +
        \pLpLaw{y_0}{
            \expo{-2\lambda T(t)}
            \Xi^r\br{T(t),\Yt{}}
        }
        \\
        +
        \pLpLaw{y_0}{
            \expo{-(2\lambda+\Boldomega) T(t)}
            \Xi^r_{\mathcal{U}}\br{T(t),\Yt{};\Boldomega}
        }
        , 
        \quad \forall (t,\sfp) \in [\that,\infty) \times \mathbb{N}_{\geq 2},
    \end{multline}   
    where we have further used Lemma~\ref{lem:Reference:WellPosedness} to replace the truncated joint process $\Yt{N,t} = \br{\Xrt{N,t},\Xstart{N,t}}$ wth the unique strong solution of the joint process $\Yt{t} = \br{\Xrt{t},\Xstart{t}}$.
    Next, as in the proof of Lemma~\ref{lem:Reference:MomentBounds:FirstExit}, by integrating~\eqref{eqn:prop:Reference:UUB:p=1} and~\eqref{eqn:prop:Reference:UUB:p>=2} with respect to the joint distribution  $\pi_0$ of the initial state $\Yt{0}$, we obtain the following bounds that are analogous to~\eqref{eqn:Reference:Scratch:TotalExpectation:Bound:p=1:2} and~\eqref{eqn:Reference:Scratch:TotalExpectation:p>=2:2}: 
    \begin{multline}\label{eqn:prop:Reference:UUB:p=1:2}
        \ELaw{}{\RefNorm{\Yt{t}}}
        \leq 
        \alpha_2
        \expo{-2\lambda \that}
        \ELaw{}{\norm{\Xrt{0}-\Xstart{0}}^2}
        + 
        \int_{\mathbb{R}^{2n}}
            \ELaw{y_0}{
                \expo{-2\lambda T(t)}
                \Xi^r\br{T(t),\Yt{}}
            }
        \pi_0(dy_0)
        \\ 
        + 
        \int_{\mathbb{R}^{2n}}
            \ELaw{y_0}{
                \expo{-(2\lambda+\Boldomega) T(t)}
                \Xi^r_{\mathcal{U}}\br{T(t),\Yt{};\Boldomega}
            }
        \pi_0(dy_0)
        ,
        \quad  
        \forall t \in [\that,T],
    \end{multline}
    and
    \begin{multline}\label{eqn:prop:Reference:UUB:p>=2:2}
        \LpLaw{\sfp}{}{\RefNorm{\Yt{t}}}
        \leq
        \alpha_2
        \expo{-2\lambda \that}
        \norm{\Xrt{0}-\Xstart{0}}_{L_{2\sfp}}^2
        \\ 
        + 
        \left(
            \int_{\mathbb{R}^{2n}}
            \left(
                \pLpLaw{y_0}{
                    \expo{-2\lambda T(t)}
                    \Xi^r\br{T(t),\Yt{}}
                }
                +
                \pLpLaw{y_0}{
                    \expo{-(2\lambda+\Boldomega) T(t)}
                    \Xi^r_{\mathcal{U}}\br{T(t),\Yt{};\Boldomega}
                }
            \right)^\sfp
            \pi_0(dy_0)
        \right)^\frac{1}{\sfp}
        ,
    \end{multline}
    for all $(t,\sfp) \in [\that,T] \times \cbr{2,\dots,\sfp^\star}$.
    Next, for the particular $\sfp \in \cbr{1,\dots,\sfp^\star}$ used in the definitions and conditions in Section~\ref{subsec:L1DRAC:Parameters}, we have the following from Lemma~\ref{lem:Reference:WellPosedness} and Assumption~\ref{assmp:NominalSystem:FiniteMomentsWasserstein}:
    \begin{align}\label{eqn:Reference:Scratch:TotalExpectation:SupBounds}
        \sup_{\nu \in \sbr{\that,T}} 
        \LpLaw{2\sfp}{}{\Xrt{\nu}-\Xstart{\nu}}
        \leq \RefRho, 
        \quad 
        \sup_{\nu \in \sbr{\that,T}}    
        \LpLaw{2\sfp}{}{\Xrt{\nu}-\Xstart{\nu}}
        +
        \sup_{\nu \in \sbr{\that,T}}
        \LpLaw{2\sfp}{}{\Xstart{\nu}}
        \leq     
        \RefRho+\Delta_\star.
    \end{align}
    Using~\eqref{eqn:prop:Reference:UUB:p=1:2},~\eqref{eqn:prop:Reference:UUB:p>=2:2}, and~\eqref{eqn:Reference:Scratch:TotalExpectation:SupBounds}, we following the analysis in the proof of Lemma~\ref{lem:Reference:MomentBounds:FirstExit} to obtain the following that is analogous to~\eqref{eqn:Reference:Scratch:TotalExpectation:2}{}
    \begin{align}\label{eqn:prop:Reference:UUB:Total:Final}
        &\norm{\Xrt{t}-\Xstart{t}}_{L_{2\sfp}}^2
        \notag 
        \\
        &\leq
        \expo{-2\lambda \that}
        \norm{\Xrt{0}-\Xstart{0}}_{L_{2\sfp}}^2
        +
        \frac{\Delta_g^\perp \Delta_\mu^\perp}{\lambda}
        \RefRho^2
        + 
        \Gamma_r\left(\RefRho,\sfp,\Boldomega\right)
        +
        \frac{1}{\absolute{2\lambda - \Boldomega}}
        \Theta_r\left(\RefRho,\sfp,\Boldomega\right)
        \doteq   
        \RefRhoUUB{\sfp,\Boldomega,\that}^2
        ,
    \end{align}
    for all $t \in \sbr{\that,T}$.
    Using the filter bandwidth condition in~\eqref{eqn:Definitions:Reference:BandwithCondition}  with $\alpha\left(\Xrt{0},\Xstart{0}\right)_{2\sfp} = \norm{\Xrt{0}-\Xstart{0}}_{L_{2\sfp}}$, we see that
    \begin{align*}
        \RefRhoUUB{\sfp,\Boldomega,\that}^2
        =
        \expo{-2\lambda \that}
        \norm{\Xrt{0}-\Xstart{0}}_{L_{2\sfp}}^2
        +
        \frac{\Delta_g^\perp \Delta_\mu^\perp}{\lambda}
        \RefRho^2
        + 
        \Gamma_r\left(\RefRho,\sfp,\Boldomega\right)
        +
        \frac{1}{\absolute{2\lambda - \Boldomega}}
        \Theta_r\left(\RefRho,\sfp,\Boldomega\right)
        \\
        <
        \RefRho^2
        -
        \left(1 - \expo{-2\lambda \that} \right)  
        \norm{\Xrt{0}-\Xstart{0}}_{L_{2\sfp}}^2
        \\
        < 
        \RefRho^2
        ,
    \end{align*}
    for all $\that \in (0,T)$.
    Coupled to the fact in~\eqref{eqn:Reference:Remark:RefRho:Lowerbound} that $\RefRho > \norm{\Xrt{0}-\Xstart{0}}_{L_{2\sfp}}$, we conclude that 
    \begin{align*}
        \RefRhoUUB{\sfp,\Boldomega,\that} \in (0,\RefRho), \quad \forall \that \in (0,T),
    \end{align*} 
    which, along with~\eqref{eqn:prop:Reference:UUB:Total:Final} concludes the proof.

\end{proof}

We now state the main result of the section. 
\begin{lemma}\label{lem:Reference:Main}
     Let the assumptions in Sec.~\ref{subsec:Assumptions} hold and let $\RefRho \in \mathbb{R}_{>0}$, $\sfp \in \cbr{1,\dots,\sfp^\star}$, be as defined in~\eqref{eqn:Definitions:Reference:Rho} with $\alpha\left(\Xrt{0},\Xstart{0}\right)_{2\sfp} = \pWass{2\sfp}{\xi_0}{\xi^\star_0}$, where $\sfp^\star$ is defined in Assumption~\ref{assmp:NominalSystem:FiniteMomentsWasserstein}.
    If the filter bandwidth condition in~\eqref{eqn:Definitions:Reference:BandwithCondition} holds with $\alpha\left(\Xrt{0},\Xstart{0}\right)_{2\sfp}=\pWass{2\sfp}{\xi_0}{\xi^\star_0}$, then
    \begin{subequations}\label{eqn:thm:Reference:Wassersetin:Main}
        \begin{align}
            \pWass{2\sfp}{\Xrdist{t}}{\Xdist{t}^\star}
            <  
            \RefRho
            ,
            \quad 
            \forall t \in [0,T]
            ,
            \quad 
            &\text{\emph{(Distributional UB)}}
            \label{eqn:thm:Reference:Wassersetin:Main:UB}
            \\
            \pWass{2\sfp}{\Xrdist{t}}{\Xdist{t}^\star} 
            \leq
            \RefRhoUUB{\sfp,\Boldomega,\that} 
            < 
            \RefRho, 
            \quad 
            \forall t \in [\that,T]
            ,
            \quad 
            &\text{\emph{(Distributional UUB)}}
            \label{eqn:thm:Reference:Wassersetin:Main:UUB}
        \end{align}
    \end{subequations} 
    where $\that \in (0,T)$ is any deterministic constant, and 
    \begin{align}\label{eqn:Definitions:Reference:VarUpsilonUUB}
        \RefRhoUUB{\sfp,\Boldomega,\that}
        =
        \left(
            \expo{-2\lambda \that}
            \pWass{2\sfp}{\xi_0}{\xi_0^\star}^2
            +
            \frac{\Delta_g^\perp \Delta_\mu^\perp}{\lambda}
            \RefRho^2
            + 
            \Gamma_r\left(\RefRho,\sfp,\Boldomega\right)
            +
            \frac{1}{\absolute{2\lambda - \Boldomega}}
            \Theta_r\left(\RefRho,\sfp,\Boldomega\right)
        \right)^{\frac{1}{2}},
    \end{align}
    and where the pertinent definitions are provided in Sec~\ref{subsec:L1DRAC:Parameters}.

    Furthermore, if the hypotheses stated above hold for $\alpha\left(\Xrt{0},\Xstart{0}\right)_{2\sfp} = \norm{\Xrt{0}-\Xstart{0}}_{L_{2\sfp}}$, then for any $\delta \in (0,1)$ such that $\log \sqrt{1/\delta} = \sfp$, 
    \begin{subequations}\label{eqn:thm:Reference:Markov:Main}
        \begin{align}
            \Probability{\norm{\Xrt{t}-\Xstart{t}} \geq \kappa_r(\delta)} 
            < 
            \delta
            , 
            \quad 
            \forall t \in [0,T]
            ,
            \quad 
            &\text{\emph{(Pathwise UB)}}
            \label{eqn:thm:Reference:Markov:Main:UB}
            \\
            \Probability{\norm{\Xrt{t}-\Xstart{t}} \geq \hat{\kappa}_r(\delta,\Boldomega,\that)} 
            < 
            \delta
            , 
            \quad 
            \forall t \in [\that,T]
            ,
            \quad 
            &\text{\emph{(Pathwise UUB)}}
            \label{eqn:thm:Reference:Markov:Main:UUB}
        \end{align}       
    \end{subequations} 
    where $\kappa_r(\delta) = e \cdot \RefRhoFunction{\log \sqrt{1/\delta}}$ and $\hat{\kappa}_r(\delta,\Boldomega,\that) = e \cdot \RefRhoUUB{\log \sqrt{1/\delta},\Boldomega,\that}$. 

\end{lemma}
\begin{proof}
    Recall that we proved Lemma~\ref{lem:Reference:MomentBounds:FirstExit} for moments with respect to the law $\Ydist{N}(\cdot,t):\Borel{\mathbb{R}^{2n}} \rightarrow [0,1]$ that resulted under the arbitrary initial coupling $\Yt{0} = \Yt{N,0} \sim \pi_0$ (see~\eqref{eqn:Reference:Scratch:TotalLaw}).  
    In order to prove~\eqref{eqn:thm:Reference:Wassersetin:Main}, we will therefore appeal to the proof of Lemma~\ref{lem:Reference:MomentBounds:FirstExit} but under the law that results from the joint process initialized under the \emph{optimal} coupling. 
    Let us denote by $\pi^\star_0$ the optimal coupling between the probability measures $\xi_0$ and $\xi^\star_0$ that are the respective distributions of the initial states $x_0$ and $x^\star_0$~\cite[Chp.~1]{villani2009optimal}.   
    From Lemma~\ref{lem:Reference:WellPosedness}, the unique strong solution $\Yt{t}$ of the joint process exists, and is adapted to the filtration generated by $Y_0$ and $\Wrt{t}$~\cite[Thm.~5.2.1]{oksendal2013stochastic} (initial vector $\Yt{0}$ is independent of the Brownian motion $\Wrt{t}$).
    Let us denote by $\Ydist{\pi^\star_0}(\cdot,t)$ the law of the process $\Yt{t}$ under $\Yt{0} \sim \pi^\star_0$.
    Hence, $\Ydist{\pi^\star_0}(\cdot,t)$ is a probability measure on $\mathbb{R}^{2n}$ given as
    \begin{align}\label{eqn:them:Reference:Wasserstein:TotalLaw}
        \Ydist{\pi^\star_0}(B,t)
        =
        \Probability{\Yt{t} \in B}
        = 
        \int_{\mathbb{R}^{2n}}
            \Ydist{}\left(y_0,0;B,t\right)
        \pi^\star_0(dy_0), 
        \quad 
        \forall B \in \Borel{\mathbb{R}^{2n}},
    \end{align} 
    where $\Ydist{}\left(y_0,0;B,t\right)$ denotes the transition probability for the process $\Yt{t}$, initialized from $y_0$~\cite[Sec.~3.1]{khasminskii2011stochastic}.
    Then, for $\RefNorm{\Yt{t}} = \norm{\Xrt{t} - \Xstart{t}}^2$ we can define  
    \begin{align}\label{eqn:thm:Reference:Wasserstein:Divergence}
        \ELaw{\pi^\star_0}{\RefNorm{\Yt{t}}^\sfp}
        =
        \int_{\mathbb{R}^{2n}}
            \RefNorm{y}^\sfp
        \Ydist{\pi^\star_0}(dy,t)
        =&
        \int_{\mathbb{R}^{2n}}
        \int_{\mathbb{R}^{2n}}
            \RefNorm{y}^\sfp
            \Ydist{}\left(y_0,0;dy,t\right)
        \pi^\star_0(dy_0)
        \notag 
        \\
        =&
        \int_{\mathbb{R}^{2n}}
            \ELaw{}{ \RefNorm{\Yt{t}}^\sfp ~|~\Yt{0}=y_0}
        \pi^\star_0(dy_0)
        =
        \int_{\mathbb{R}^{2n}}
            \ELaw{y_0}{ \RefNorm{\Yt{t}}^\sfp}
        \pi^\star_0(dy_0)
        ,
    \end{align}
    for $t \in [0,T]$. 
    Note that if we were to replace the optimal initial coupling $\pi^\star_0$ with an arbitrary coupling $\pi_0$, then~\eqref{eqn:thm:Reference:Wasserstein:Divergence} yields~\eqref{eqn:Reference:Scratch:TotalExpectation:p=1} and~\eqref{eqn:Reference:Scratch:TotalExpectation:p>=2}.
    Since $\pi^\star_0$ is the optimal coupling between the initial measures $\xi_0$ and $\xi^\star_0$, it follows from the definition of the Wasserstein distance that 
    \begin{align*}
        \int_{\mathbb{R}^{2n}}
            \RefNorm{y_0}^\sfp
        \pi^\star_0(dy_0)
        = 
        \int_{\mathbb{R}^{2n}}
            \norm{x_0 - x^\star_0}^{2\sfp}
        \pi^\star_0(dx_0,dx_0^\star)
        =
        \pWass{2\sfp}{\xi_0}{\xi^\star_0}^{2\sfp}
        .
    \end{align*}
    We can thus follow the analysis in Lemma~\ref{lem:Reference:MomentBounds:FirstExit} for the law $\Ydist{\pi^\star_0}(\cdot,t)$ of the joint process $\Yt{t}$ up to the stopping time $T$ and obtain 
    \begin{align}\label{eqn:thm:Reference:Wasserstein:UB:1}
        \LpLaw{2\sfp}{\pi^\star_0}{\Xrt{t}-\Xstart{t}}
        =
        \ELaw{\pi^\star_0}{\norm{\Xrt{t}-\Xstart{t}}^{2\sfp}}^\frac{1}{2\sfp}
        < 
        \RefRho, \quad \forall t \in [0,T],
    \end{align}
    where $\RefRho$ is defined in~\eqref{eqn:Definitions:Reference:Rho} with $\alpha\left(\Xrt{0},\Xstart{0}\right)_{2\sfp}=\pWass{2\sfp}{\xi_0}{\xi^\star_0}$. 
    We now follow the approach in~\cite{bouvrie2019wasserstein} and, using~\eqref{eqn:them:Reference:Wasserstein:TotalLaw}, define the measure $\pi_t = \Ydist{\pi^\star_0}(\cdot,t)$. 
    It then follows from~\eqref{eqn:thm:Reference:Wasserstein:Divergence}
    \begin{align*}
        \ELaw{\pi^\star_0}{\norm{\Xrt{t}-\Xstart{t}}^{2\sfp}}
        =
        \ELaw{\pi^\star_0}{\RefNorm{\Yt{t}}^\sfp}
        =
        \int_{\mathbb{R}^{2n}}
            \RefNorm{y}^\sfp
        \Ydist{\pi^\star_0}(dy,t)
        =
        \int_{\mathbb{R}^{2n}}
            \RefNorm{y}^\sfp
        \pi_t(dy).
    \end{align*}   
    Since $\pi_t$ is not necessarily the optimal coupling between the state distributions $\Xrdist{t}$ and $\Xdist{t}^\star$, we reason that 
    \begin{align*}
        \pWass{2\sfp}{\Xrdist{t}}{\Xdist{t}^\star}^{2\sfp}
        =
        \int_{\mathbb{R}^{2n}}
            \RefNorm{y}^\sfp
        \pi^\star_t(dy)
        \leq      
        \int_{\mathbb{R}^{2n}}
            \RefNorm{y}^\sfp
        \pi_t(dy)
        =
        \ELaw{\pi^\star_0}{\norm{\Xrt{t}-\Xstart{t}}^{2\sfp}}, \quad \forall t \in [0,T],
    \end{align*}
    where $\pi^\star_t$ denotes the optimal coupling between the state distributions $\Xrdist{t}$ and $\Xdist{t}^\star$. 
    Substituting the above into~\eqref{eqn:thm:Reference:Wasserstein:UB:1} then establishes~\eqref{eqn:thm:Reference:Wassersetin:Main:UB}.
     The bound in~\eqref{eqn:thm:Reference:Wassersetin:Main:UUB} follows \emph{mutatis mutandis} from the analysis above and the proof of Proposition~\ref{prop:Reference:UUB}. 

    Next, for any $\kappa \in \mathbb{R}_{>0}$, we can use the Markov's inequality~\cite[Cor.~3.1.2]{athreya2006measure} and Lemma~\ref{lem:Reference:WellPosedness} to obtain
    \begin{align*}
        \Probability{\norm{\Xrt{t}-\Xstart{t}} \geq \kappa} \leq \frac{\ELaw{}{ \norm{\Xrt{t}-\Xstart{t}}^{2\sfp} }}{ \kappa^{2\sfp} }
        < 
        \left(\frac{ \RefRho }{ \kappa }\right)^{2\sfp}
        , 
        \quad 
        \forall (t,\sfp) \in [0,T] \times [1,\sfp^\star].
    \end{align*}
    By setting $\sfp = \log \sqrt{\frac{1}{\delta}}$ and $\kappa = e \cdot \RefRho$, we obtain
    \begin{align*}
        \Probability{\norm{\Xrt{t}-\Xstart{t}} \geq \kappa_r(\delta)} 
        < 
        \delta
        , 
        \quad 
        \forall t \in [0,T],
    \end{align*} 
    thus establishing~\eqref{eqn:thm:Reference:Markov:Main:UB}. 
    The proof of~\eqref{eqn:thm:Reference:Markov:Main:UUB} follows \emph{mutatis mutandis} by using the uniform ultimate bound from Proposition~\ref{prop:Reference:UUB}.
    
\end{proof}
\begingroup

\endgroup


\subsection{Performance Analysis: True (Uncertain) Process}\label{subsec:Analysis:TrueUncertainProcess}


We now consider the true (uncertain) system in~\eqref{eqn:TrueUncertainSDE} operating under the \ellonedrac feedback law $\FL$ defined in~\eqref{eqn:L1DRAC:Definition:FeedbackOperator}, Sec.~\ref{subsec:L1DRAC:Architecture}.
\begin{definition}[True (Uncertain) \ellonedrac Closed-loop Process]\label{def:True:TrueL1Process}
    We say that $\Xt{t}$, $t \in [0,T]$, for any $T \in (0,\infty)$, is the \textbf{true (uncertain) \ellonedrac process}, if $\Xt{t}$ is a unique strong solution to the true (uncertain) \ito SDE in~\eqref{eqn:TrueUncertainSDE} under the \ellonedrac feedback law $\FL$ in~\eqref{eqn:L1DRAC:Definition:FeedbackOperator}: 
    \begin{align}\label{eqn:True:SDE}
        d\Xt{t} = \Fmu{t,\Xt{t},\ULt{t}}dt + \Fsigma{t,\Xt{t}}d\Wt{t}, \quad \Xt{0} = x_0 \sim \xi_0~(\text{\Pas}), 
    \end{align}
    where 
    \begin{equation}\label{eqn:True:L1DRACOperator:Initial}
        \ULt{t} \doteq \FL[\Xt{}][t].
    \end{equation}
    The \ellonedrac~feedback operator $\FL:\mathcal{C}([0,T]:\mathbb{R}^n) \rightarrow \mathcal{C}([0,T]:\mathbb{R}^m)$ is defined in~\eqref{eqn:L1DRAC:Definition:FeedbackOperator}, which we restate next. 
    Using the definition in~\eqref{eqn:L1DRAC:Definition:FeedbackOperator}  
    \begin{align}\label{eqn:True:L1DRACOperator}
        \FL[\Xt{}] = \Filter \circ \AdaptationLaw \circ \Predictor[\Xt{}],
    \end{align}
    we write 
    \begin{align}\label{eqn:True:L1DRACInput}
        U_{\mathcal{L}_1}
        =
        \FL[X]
        =
        \Filter[\hat{\Lambda}^{\paral}],
        \quad
        \hat{\Lambda}^{\paral} = \AdaptationLawParal[\hat{\Lambda}],
        \quad 
        \hat{\Lambda} = \AdaptationLaw[\Xhatt{}][\Xt{}],
        \quad 
        \Xhatt{} = \Predictor[\Xt{}].
    \end{align}
    Using~\eqref{eqn:L1DRAC:Definition:FeedbackOperator:Filter}, 
    we see that the input $\ULt{t}$ is defined as the output of the \textbf{low-pass filter}:
    \begin{align}\label{eqn:True:Filter}
        \ULt{t} 
        =
        \Filter[\hat{\Lambda}^{\paral}][t]  
        =  
        - \Boldomega \int_0^t \expo{-\Boldomega(t-\nu)}\Lparahat{\nu}d\nu,  
    \end{align}
    where $\Boldomega \in \mathbb{R}_{>0}$ is the \textbf{filter bandwidth}.
    The \textbf{adaptive estimates} $\hat{\Lambda}^{\paral}$ and $\hat{\Lambda}$ are obtained via the \textbf{adaptation law} operator $\AdaptationLaw \left( \AdaptationLawParal \right)$ in~\eqref{eqn:L1DRAC:Definition:FeedbackOperator:AdaptationLaw} as follows:
    \begin{align}\label{eqn:True:AdaptationLaw}
        \begin{aligned}
            \Lparahat{t} 
            =&
            \AdaptationLawParal[\hat{\Lambda}][t]
            = 
            \sum_{i=0}^{\lfloor \frac{t}{\BoldTs} \rfloor}  
            \Theta_{ad}(i\BoldTs) \Lhat{t}
            \indicator{[i\BoldTs,(i+1)\BoldTs)}{t}
            ,  
            \\
            \Lhat{t} 
            =&
            \AdaptationLaw[\Xhatt{t}][\Xt{}][t]
            \\
            =&  
            0_n \indicator{[0,\BoldTs)}{t} 
            +
            \lambda_s \br{1 - e^{\lambda_s \BoldTs}}^{-1} 
            \sum_{i=1}^{\lfloor \frac{t}{\BoldTs} \rfloor}    
            \Xtildet{i\BoldTs}
            \indicator{[i\BoldTs,(i+1)\BoldTs)}{t},
            \quad 
            \Xtildet{i\BoldTs} \doteq \Xhatt{i\BoldTs} - \Xt{i\BoldTs},
        \end{aligned}
    \end{align} 
    where $\BoldTs \in \mathbb{R}_{>0}$ is the \textbf{sampling period} and $ \Theta_{ad}(t) = \begin{bmatrix}\mathbb{I}_m & 0_{m,n-m}  \end{bmatrix} \bar{g}(t)^{-1} \in \mathbb{R}^{m \times n}$, with $\bar{g}(t) = \begin{bmatrix} g(t) & g(t)^\perp  \end{bmatrix} \in \mathbb{R}^{n \times n}$ for $g^\perp$ defined in Assumption~\ref{assmp:UnknownFunctions}.
    The parameter $\lambda_s \in \mathbb{R}_{>0}$ contributes to the \textbf{prediction process} $\Xhatt{t}$ as a parameter to the operator $\Predictor$ in~\eqref{eqn:L1DRAC:Definition:FeedbackOperator:Predictor} which induces the \textbf{process predictor} as follows: 
    \begin{align}\label{eqn:True:ProcessPredictor}
        \Xhatt{t} =& \Predictor[\Xt{}][t] 
        \notag 
        \\
        &\Rightarrow
        \Xhatt{t} 
        =
        x_0
        +
        \int_0^t 
            \left(-\lambda_s \mathbb{I}_n \Xtildet{\nu}+ f(\nu,\Xt{\nu}) +  g(\nu) \FL[\Xt{}][\nu] + \Lhat{\nu}\right) d\nu,
        \quad 
        \Xtildet{t} \doteq \Xhatt{t} - \Xt{t}.
    \end{align} 
    We collectively refer to $\cbr{\Boldomega,\BoldTs,\lambda_s}$ as the control parameters. 

    Finally, in Definition~\ref{def:TrueandNominalProcesses}, $\Xdist{t}$ denotes the \textbf{true law } induced by the process $\Xt{t}$ on $\Borel{\mathbb{R}^n}$, i.e., $\Xt{t} \sim \Xdist{t}$.

\end{definition}

Our goal in this section is to establish the performance of the \ellonedrac closed-loop true (uncertain) process in~\eqref{eqn:True:SDE} relative to the reference process in Definition~\ref{def:ReferenceProcess}.
For this purpose, we provide the following:
\begin{definition}[True (Uncertain)-Reference Error Process]\label{def:True:JointProcess}
    Consider the true process $\Xt{t}$ and the reference process $\Xrt{t} \sim \Xrdist{t}$ in Definitions~\ref{def:True:TrueL1Process} and~\ref{def:ReferenceProcess}, respectively.
    We say that $\Zt{t}$, $t \in [0,T]$, for any $T \in (0,\infty)$, is the \textbf{true (uncertain)-reference joint process}, if it is a unique strong solution of the following \textbf{true (uncertain)-reference joint \ito SDE} on $\br{\Omega, \mathcal{F}, \Wfilt{0,t},  \mathbb{P}}$:
    \begin{align}\label{eqn:True:JointProcess}
        d\Zt{t}
        &= 
        \Jmu{t,\Zt{t}}dt
        +
        \Jsigma{t,\Zt{t}}d\Wt{t}, \quad t \in [0,T], 
        \quad 
        \Zt{0} = 0,
        \quad 
        \Zt{t} \sim \Zdist{t},
    \end{align}
    where 
    \begin{align*}
        \Zt{t} \doteq \Xt{t} - \Xrt{t} \in \mathbb{R}^{n},
        \quad 
        \Jmu{t, \Zt{t}} \doteq \Fmu{t,\Xt{t},\ULt{t}} - \Fmu{t,\Xrt{t},\Urt{t}}  \in \mathbb{R}^{n}, 
        \\ 
        \Jsigma{t, \Zt{t}} \doteq \Fsigma{t,\Xt{t}} - \Fsigma{t,\Xrt{t}} \in \mathbb{R}^{n \times d}.  
    \end{align*}
\end{definition}
\begin{remark}
    The definition of the joint process above between the true (uncertain) process and the reference process is \textbf{not} a direct analogue to the joint known (nominal)-reference process in Definition~\ref{def:Reference:JointProcess}. 
    The reason is that the true (uncertain) process and the reference process are driven by the identical Brownian motion $\Wt{t}$ and share the same initial condition $x_0 \sim \xi_0$.
    This is in contrast to the joint known (nominal)-reference process in Definition~\ref{def:Reference:JointProcess}, where the known (nominal) process and the reference process are driven by \textbf{independent Brownian motions} $\Wstart{t}$ and $\Wt{t}$ and have \textbf{distinct initial data} $x_0 \sim \xi_0$ and $x^\star_0 \sim \xi^\star_0$, respectively.
    Recall from our discussion after Definition~\ref{def:ReferenceProcess} that the reference process is \textbf{not-realizable} and represents the best achievable performance, and thus we designed the reference process to be driven by the identical Brownian motion and share the same initial condition as the true (uncertain) process.
\end{remark}

Similar to the truncated joint known (nominal)-reference process in Definition~\ref{def:Reference:TruncatedJointProcess}, we require the following:
\begin{definition}[Truncated Joint True (Uncertain)-Reference Process]\label{def:True:TruncatedJointProcess}
    Let
    \begin{equation}\label{eqn:True:OpenBoundedSet}
        \widetilde{U}_N \doteq \left\{ a \in \mathbb{R}^{n}~:~\norm{a} < N \right\} \subset \subset \mathbb{R}^{n}, \quad \forall N \in \mathbb{R}_{>0}.  
    \end{equation}
    Next, we define the \textbf{truncated joint true (uncertain)-reference \ito SDE} as
    \begin{equation}\label{eqn:True:TruncatedJointProcess}
        d\Zt{N,t}
        = \JNmu{t,\Zt{N,t}}dt + \JNsigma{t,\Zt{N,t}}d\Wt{t}, \quad \Zt{N,0} = \Zt{0} = 0,
    \end{equation}
    where, the process and the drift and diffusion vector fields are defined as
    \begin{align*}
    \Zt{N,t} \doteq \Xt{N,t} - \Xrt{N,t} \in \mathbb{R}^{n}, 
    \quad 
    \JNmu{t, a} \br{ \JNsigma{t,a} } = &
        \begin{cases}
            \Jmu{t,a} \br{ \Jsigma{t,a} }, \quad &~\norm{a} \leq N
            \\
            0_{n} \br{ 0_{n,d} }, \quad &~\norm{a} \geq 2N
        \end{cases},
    \end{align*}
    for all $(a,t) \in \mathbb{R}^{n} \times [0,T]$, where $\JNmu{t,a}$ and $\JNsigma{t,a}$ are any \textbf{uniformly} Lipschitz continuous maps for all $a \in \mathbb{R}^{n}$ and $t \in \mathbb{R}_{\geq 0}$.
    Similar to $\Zt{t}$, we refer to $\Zt{N,t} \in \mathbb{R}^{n}$ as the \textbf{truncated joint known(nominal)-reference process} if it is a unique strong solution of~\eqref{eqn:True:TruncatedJointProcess}.     
\end{definition}

We begin by establishing the uniqueness and existence of strong solutions for the truncated joint true (uncertain)-reference process.  
\begin{proposition}[Well-Posedness of~\eqref{eqn:True:TruncatedJointProcess}]\label{prop:True:TruncatedWellPosedness}
    If Assumptions~\ref{assmp:KnownFunctions} and~\ref{assmp:UnknownFunctions} hold, then for any $N \in \mathbb{R}_{>0}$, $\Zt{N,t}$ is a unique strong solution of~\eqref{eqn:True:TruncatedJointProcess}, $\forall t \in [0,T]$, for any $T \in (0,\infty)$ and is a strong Markov process $\forall t \in \mathbb{R}_{\geq 0}$.
    Furthermore, define\footnote{The stopping time $\tau_N$ takes on the value of either the finite horizon $T$, or the first exit time of $\Zt{N,t}$ from set $\widetilde{U}_N$.}
    \begin{equation}\label{eqn:True:FirstExitTime}
        \tau_N \doteq T \wedge \inf\cbr{t \in [0,T]~:~\Zt{N,t} \notin \widetilde{U}_N},        
    \end{equation}
    where the open and bounded set $\widetilde{U}_N$ is defined in~\eqref{eqn:True:OpenBoundedSet}, for an arbitrary $N \in \mathbb{R}_{>0}$.
    Then, $\Zt{N,t}$ uniquely solves~\eqref{eqn:True:JointProcess}, in the strong sense, for all $t \in [0, \tau_N]$.
\end{proposition}
\begin{proof}
    See Appendix~\ref{app:TrueProcess}.
\end{proof}
\begin{remark}
    Recall that $\tau_N$ in~\eqref{eqn:Reference:FirstExitTime}, Proposition~\ref{prop:Reference:TruncatedWellPosedness}, denotes the first exit time of the process $\Yt{N,t}$ from the set $U_N \subset \subset \mathbb{R}^{2n}$.
    Similarly, with an abuse of notation and from this point onward, $\tau_N$ in~\eqref{eqn:True:FirstExitTime} denotes the first exit time of the process $\Zt{N,t}$ from the set $\widetilde{U}_N \subset \subset \mathbb{R}^{n}$. 
\end{remark}

Using the well-posedness of the truncated process, we can now define the following transition probability:
\begin{equation}\label{eqn:True:TransitionProbability}
    \Zdist{N}\left(\Zt{N,s},s;B,t\right)
    \doteq  
    \Probability{\Zt{N,t} \in B~|~\Zt{N,s}}, 
    \quad 
    \forall 0 < s \leq t \leq T, \quad B \in \Borel{\mathbb{R}^{n}}
    .
\end{equation} 
With an abuse of notation, let us also denote the law of the process $\Zt{N,t}$ by $\Zdist{N}$.
Thus, $\Zdist{N}(\cdot,t)$ is a probability measure on $\mathbb{R}^n$ given as 
\begin{align*}
    \Zdist{N}(B,t) = \Probability{\Zt{N,t} \in B} 
    = 
    \int_{\mathbb{R}^n}
        \Zdist{N}\left(z_0,0;B,t\right)
    \mathbb{P}(dz_0),
    \quad 
    \forall B \in \Borel{\mathbb{R}^n},
\end{align*}
where $\mathbb{P}(z_0)$ denotes the distribution of the initial condition $\Zt{N,0}$.
Hence, for any Borel measurable function $a:\mathbb{R}^{n} \rightarrow \mathbb{R}$, we have
\begin{align*}
    \ELaw{}{
        a\br{\Zt{N,t}} 
    }
    = 
    \int_{\mathbb{R}^n} a(z) \Zdist{N}(dz,t)
    =&
    \int_{\mathbb{R}^n} 
    \int_{\mathbb{R}^n}
        a(z) \Zdist{N}\left(z_0,0;dz,t\right)
    \mathbb{P}(dz_0)
    \\
    =&
    \int_{\mathbb{R}^n} 
        \ELaw{}{a\br{\Zt{N,t}}~|~\Zt{N,0}=z_0 }
    \mathbb{P}(dz_0)
    .
\end{align*}
However, since $\Probability{\Zt{N,0}=0}=1$, the previous expression can be written as 
\begin{align}\label{eqn:lem:True:dV:CondExp:Definition}
    \ELaw{}{
        a\br{\Zt{N,t}} 
    }
    =
    \int_{\mathbb{R}^n} 
        \ELaw{}{a\br{\Zt{N,t}}~|~\Zt{N,0}=z_0 }
    \delta_0(dz_0)
    =&
    \ELaw{}{a\br{\Zt{N,t}}~|~\Zt{N,0}=0 }
    \notag 
    \\
    =& 
    \int_{\mathbb{R}^n}
        a(z) 
    \Zdist{N}\left(0,0;dz,t\right)
    ,
\end{align}
where $\delta_0$ is the Dirac measure supported on $0 \in \mathbb{R}^n$.
Hence, the following analysis uses the total expectation as it is identical to the expectation conditioned on the zero initial condition. 

Next, we study the \emph{uniform in time} bounds on the moments of the truncated joint true (uncertain)-reference process.
In order to streamline the derivations in this section, we define the following:
\begin{subequations}\label{def:True:ErrorFunctions}
    \begin{align}
        \fZ{\nu,\Zt{N,\nu}}
        = 
        f\left(\nu,\Xt{N,\nu}\right) 
        - 
        f\left(\nu,\Xrt{N,\nu}\right), 
        \\
        \LZmuAll{\nu,\Zt{N,\nu}}
        =
        \LmuAll{\nu,\Xt{N,\nu}}
        - 
        \LmuAll{\nu,\Xrt{N,\nu}}, 
        \\
        \FZsigmaAll{\nu,\Zt{N,\nu}}
        =
        \FsigmaAll{\nu,\Xt{N,\nu}}
        - 
        \FsigmaAll{\nu,\Xrt{N,\nu}}. 
        \\ 
        \ReferenceInputZ[\Zt{N}][\nu] 
        =
        \ReferenceInput[\Xt{N}][\nu] 
        -
        \ReferenceInput[\Xrt{N}][\nu].
    \end{align} 
\end{subequations}
\begin{lemma}\label{lem:True:dV} 
    Let the assumptions in Sec.~\ref{subsec:Assumptions} hold.
    For an arbitrary $N \in \mathbb{R}_{>0}$, let the stopping time $\tau_N$ be as in~\eqref{eqn:Reference:FirstExitTime} for the truncated joint process in Definition~\ref{def:Reference:TruncatedJointProcess}.  
    For any constant $t^\star \in \mathbb{R}_{>0}$ define\footnote{The stopping time $\tau^\star$ is either the constant $t^\star$, the finite horizon $T$, or the first exit time of $\Zt{N,t}$ from the set $\widetilde{U}_N$. The variable $\tau(t)$ is identical to $t$ up to $\tau^\star$, and zero afterwards.}  
    \begin{align}\label{eqn:True:StoppingTimes}
        \tau^\star = t^\star \wedge \tau_N,
        \quad
        \tau(t) = t \wedge \tau^\star.
    \end{align}
    Then, with $\ZNorm{\Zt{N,t}} \doteq \norm{\Zt{N,t}}^2 = \norm{\Xt{N,t} - \Xrt{N,t}}^2$, the truncated joint process $\Zt{N,t}$ in Definition~\ref{def:True:TruncatedJointProcess} satisfies:
    \begin{multline}\label{eqn:lem:True:dV:Supremums:UB:p=1}
        \ELaw{}{\ZNorm{\Zt{N,\tau(t)}}}
        \leq
        \ELaw{}{
            e^{-2\lambda \tau(t)}  
            \Xi\left(\tau(t),\Zt{N} \right)
        }
        +
        \ELaw{}{
            \expo{-(2\lambda+\Boldomega) \tau(t)}
            \Xi_{\mathcal{U}}\left(\tau(t),\Zt{N} \right)
        }
        \\
        +
        \ELaw{}{
            e^{-2\lambda \tau(t)}
            \widetilde{\Xi}_{\mathcal{U}}\left(\tau(t),\Zt{N} \right)
        }
        ,
        \quad  
        \forall t \in \mathbb{R}_{\geq 0},
    \end{multline}
    and
    \begin{multline}\label{eqn:lem:True:dV:Supremums:UB}
            \LpLaw{\sfp}{}{\ZNorm{\Zt{N,\tau(t)}}}
        \leq
            \LpLaw{\sfp}{}{
            \expo{-2\lambda \tau(t)}
            \Xi\left(\tau(t),\Zt{N} \right)
            }
        +
            \LpLaw{\sfp}{}{
            \expo{-(2\lambda+\Boldomega) \tau(t)}
            \Xi_{\mathcal{U}}\left(\tau(t),\Zt{N} \right)
            }
        \\
        +
            \LpLaw{\sfp}{}{
            \expo{-2\lambda \tau(t)}
            \widetilde{\Xi}_{\mathcal{U}}\left(\tau(t),\Zt{N} \right)
            }
        ,
        \quad \forall (t,\sfp) \in \mathbb{R}_{\geq 0} \times \mathbb{N}_{\geq 2},
    \end{multline}
    where, for any Borel measurable function $a:\mathbb{R}^{n} \rightarrow \mathbb{R}$, $\ELaw{}{a\br{\Zt{N,t}}}$ is defined in~\eqref{eqn:lem:True:dV:CondExp:Definition}, and $\pLpLaw{}{a(t)}$ follows analogously.
    The above bounds are defined using the following: 
    \begin{subequations}\label{eqn:lem:True:dV:Xi:Functions}
        \begin{align}
                \Xi\left(\tau(t),\Zt{N} \right)
                =
                \int_0^{\tau(t)}  
                    \expo{2 \lambda \nu} 
                    \left( \phi_\mu\left(\nu,\Zt{N,\nu}\right)d\nu  + \phi_\sigma\left(\nu,\Zt{N,\nu}\right)d\Wt{\nu} \right),
                \label{eqn:lem:True:dV:Xi:Functions:A}
                \\
                \Xi_{\mathcal{U}}\left(\tau(t),\Zt{N} \right)
                =
                \int_0^{\tau(t)}
                \left(  
                    \mathcal{U}_{\mu}\br{\tau(t),\nu,\Zt{N};\Boldomega}d\nu 
                    + 
                    \mathcal{U}_{\sigma}\br{\tau(t),\nu,\Zt{N};\Boldomega}d\Wt{\nu}
                \right),
                \label{eqn:lem:True:dV:Xi:Functions:B}
                \\ 
                \widetilde{\Xi}_{\mathcal{U}}\left(\tau(t),\Zt{N} \right)
                =
                \int_0^{\tau(t)}  
                    \expo{2 \lambda \nu} 
                    \widetilde{\mathcal{U}}\left(\nu,\Zt{N,\nu}\right)d\nu
                ,
                \label{eqn:lem:True:dV:Xi:Functions:C}
        \end{align} 
    \end{subequations}
    where 
    \begin{subequations}\label{eqn:lem:True:dV:phi:Functions}
        \begin{align}
            \phi_\mu\left(\nu,\Zt{N,\nu}\right)
            =
            \ZDiffNorm{\Zt{N,\nu}}^\top
            g(\nu)^\perp
            \LZperpmu{\nu,\Zt{N,\nu}}
            + 
            \Frobenius{\FZsigma{\nu,\Zt{N,\nu}}}^2,
            \\
            \phi_\sigma\left(\nu,\Zt{N,\nu}\right)
            =
            \ZDiffNorm{\Zt{N,\nu}}^\top
            g(\nu)^\perp
            \FZperpsigma{\nu,\Zt{N,\nu}},  
            \\
            \widetilde{\mathcal{U}}\left(\nu,\Zt{N,\nu}\right)
            =
            \ZDiffNorm{\Zt{N,\nu}}^\top
            g(\nu)
            \left(\FL - \ReferenceInput\right)(\Xt{N})(\nu), 
            \label{eqn:lem:True:dV:U:Functions}
        \end{align}
    \end{subequations}
    and
    \begin{align}\label{eqn:lem:True:dV:phi:Functions:2}
        \mathcal{U}_{\mu}\br{\tau(t),\nu,\Zt{N};\Boldomega}
        =
        \psi(\tau(t),\nu,\Zt{N})
        \LZparamu{\nu,\Zt{N,\nu}}
        ,
        \quad 
        \mathcal{U}_{\sigma}\br{\tau(t),\nu,\Zt{N};\Boldomega}
        =
        \psi(\tau(t),\nu,\Zt{N})
        \FZparasigma{\nu,\Zt{N,\nu}}
        .
    \end{align}
    In the above, along with $\ZDiffNorm{\Zt{N,\nu}} = 2\left(\Xt{N,\nu} - \Xrt{N,\nu}\right)$ and the definitions in~\eqref{def:True:ErrorFunctions}, we have set 
    \begin{multline}\label{eqn:lem:True:dV:psi:Functions}
        \psi(\tau(t),\nu,\Zt{N})
        =
        \frac{\Boldomega}{2\lambda - \Boldomega}
        \left(
            \expo{\Boldomega(\tau(t)+\nu)}
            \mathcal{P}\br{\tau(t),\nu}  
            -
            \expo{ (2\lambda \tau(t)+\Boldomega \nu)}
            \ZDiffNorm{\Zt{N,\tau(t)}}^\top g(\tau(t))
        \right)
        \\
        +
        \frac{2\lambda}{2\lambda - \Boldomega}
        \expo{(\Boldomega \tau(t) + 2 \lambda \nu)} 
        \ZDiffNorm{\Zt{N,\nu}}^\top g(\nu)  
        \in \mathbb{R}^{1 \times m},
    \end{multline}
    and
    \begin{align}\label{eqn:True:ddV}
        \mathcal{P}\br{\tau(t),\nu} 
        =
        \int_\nu^{\tau(t)}
            e^{ (2\lambda - \Boldomega) \beta }  
            d_\beta
            \left[
                \ZDiffNorm{\Zt{N,\beta}}^\top g(\beta)
            \right] 
        \in \mathbb{R}^{1 \times m}
        , \quad 0 \leq \nu \leq \tau(t),
    \end{align}
    where $d_\beta \sbr{\cdot}$ denotes the stochastic differential with respect to $\beta$.

\end{lemma}
\begin{proof}

    Since $\Zt{N,0}=0$, we have that $\Probability{\Zt{N,0} \in \widetilde{U}_N}=1$. 
    We therefore apply the \ito lemma~\cite[Thm.~1.6.4]{mao2007stochastic} to $e^{2\lambda \tau(t)} \ZNorm{\Zt{N,t}}$, using the dynamics in~\eqref{eqn:True:TruncatedJointProcess}, to obtain  
    \begin{multline*}
        e^{2\lambda \tau(t)} \ZNorm{\Zt{N,\tau(t)}}
        = 
        2 \lambda
        \int_0^{\tau(t)}  e^{2 \lambda \nu} \ZNorm{\Zt{N,\nu}} d \nu 
        +
        \int_0^{\tau(t)}  e^{2 \lambda \nu} \nabla \ZNorm{\Zt{N,\nu}}^\top \JNsigma{\nu,\Zt{N,\nu}} d \Wt{\nu}  
        \\
        + \int_0^{\tau(t)}  e^{2 \lambda \nu} \left( 
            \nabla \ZNorm{\Zt{N,\nu}}^\top
            \JNmu{\nu, \Zt{N,\nu}}
            +
            \frac{1}{2} \Trace{ \KNsigma{\nu, \Zt{N,\nu}}  \nabla^2 \ZNorm{\Zt{N,\nu}}}
        \right)d\nu, 
    \end{multline*}
    for all $t \in \mathbb{R}_{\geq 0}$, where $\KNsigma{\nu, \Yt{N,\nu}} = \JNsigma{\nu,\Yt{N,\nu}} \JNsigma{\nu,\Yt{N,\nu}}^\top \in \mathbb{S}^n$. 
    Next, from Proposition~\ref{prop:True:TruncatedWellPosedness}, $\Zt{N,t}$ is also unique strong solution of~\eqref{eqn:True:JointProcess}, for all $t \in [0,\tau^\star]$ because $[0,\tau^\star] \subseteq [0,\tau_N]$.
    Therefore, we may replace $J_{N,\mu}$, $J_{N,\sigma}$, and $K_{N,\sigma}$ with $J_{\mu}$, $J_{\sigma}$, and $K_{\sigma}$, respectively, in the above inequality and obtain 
    \begin{multline*}
        e^{2\lambda \tau(t)} \ZNorm{\Zt{N,\tau(t)}}
        = 
        2 \lambda
        \int_0^{\tau(t)}  e^{2 \lambda \nu} \ZNorm{\Zt{N,\nu}} d \nu 
        +
        \int_0^{\tau(t)}  
            e^{2 \lambda \nu} 
            \nabla \ZNorm{\Zt{N,\nu}}^\top \Jsigma{\nu,\Zt{N,\nu}} 
        d \Wt{\nu}  
        \\
        + 
        \int_0^{\tau(t)}  e^{2 \lambda \nu} 
        \left(  
            \nabla \ZNorm{\Zt{N,\nu}}^\top
            \Jmu{\nu, \Zt{N,\nu}}
            +
            \frac{1}{2} \Trace{ \Ksigma{\nu, \Zt{N,\nu}}  \nabla^2 \ZNorm{\Zt{N,\nu}}}
        \right)
        d\nu, 
    \end{multline*}
    for all $t \in \mathbb{R}_{\geq 0}$, where $\Ksigma{\nu, \Zt{N,\nu}} = \Jsigma{\nu,\Zt{N,\nu}} \Jsigma{\nu,\Zt{N,\nu}}^\top \in \mathbb{S}^n$. 
    Substituting the expressions in~\eqref{eqn:Appendix:TrueProcess:prop:dV:Bound:Total}, Proposition~\ref{prop:Appendix:TrueProcess:dV}, for the last two terms on the right hand side leads to 
    \begin{multline}\label{lem:True:dV:1}
        e^{2\lambda \tau(t)} \ZNorm{\Zt{N,\tau(t)}}
        \leq
        \int_0^{\tau(t)}  
            e^{2 \lambda \nu}
            \left( 
                \left[  
                    \phi_\mu\left(\nu,\Zt{N,\nu}\right)
                    +
                    \widetilde{\mathcal{U}}\left(\nu,\Zt{N,\nu}\right)
                \right]
                d\nu
                +
                \phi_\sigma\left(\nu,\Zt{N,\nu}\right)
                d\Wt{\nu}      
            \right)
        \\
        +
        \int_0^{\tau(t)}  
            e^{2 \lambda \nu}
            \left( 
                \left[  
                    \phi_{\mu^{\paral}}\left(\nu,\Zt{N,\nu}\right)
                    +
                    \phi_{U}\left(\nu,\Zt{N,\nu}\right)
                \right]
                d\nu
                +
                    \phi_{\sigma^{\paral}}\left(\nu,\Zt{N,\nu}\right)
                d\Wt{\nu}      
            \right),  
    \end{multline}
    for all $t \in \mathbb{R}_{\geq 0}$, where 
    \begin{align*}
        \phi_{\mu^{\paral}}\left(\nu,\Zt{N,\nu}\right)
        =
        \ZDiffNorm{\Zt{N,\nu}}^\top
        g(\nu)
        \LZparamu{\nu,\Zt{N,\nu}},
        \quad 
        \phi_{\sigma^{\paral}}\left(\nu,\Zt{N,\nu}\right)
        =
        \ZDiffNorm{\Zt{N,\nu}}^\top 
        g(\nu) 
        \FZparasigma{\nu,\Zt{N,\nu}},
        \\
        \phi_{U}\left(\nu,\Zt{N,\nu}\right)
        =
        \ZDiffNorm{\Zt{N,\nu}}^\top
        g(\nu) 
        \ReferenceInputZ[\Zt{N}][\nu].
    \end{align*}
    Next, we use Proposition~\ref{prop:Appendix:TrueProcess:dV:U} to obtain the following expression 
    \begin{multline}\label{lem:True:dV:2}
            \int_0^{\tau(t)}  
            \expo{2 \lambda \nu} 
            \left(\left[\phi_U\left(\nu,\Zt{N,\nu}\right)+\phi_{\mu^{\paral}}\left(\nu,\Zt{N,\nu}\right)\right]d\nu + \phi_{\sigma^{\paral}}\left(\nu,\Zt{N,\nu}\right)d\Wt{\nu} \right)
        \\
        =
        \int_0^{\tau(t)}  \expo{ 2\lambda \nu} 
        \left(
            \phi_{\mu^{\paral}}\br{\nu,\Zt{N,\nu}} + \phi_{U_\mu}\br{\nu,\Zt{N,\nu};\Boldomega} 
        \right)
        d\nu
        +
        \int_0^{\tau(t)}  \expo{ 2 \lambda \nu } 
        \left(
            \phi_{\sigma^{\paral}}\br{\nu,\Zt{N,\nu}} + \phi_{U_\sigma}\br{\nu,\Zt{N,\nu};\Boldomega}
        \right)
        d\Wt{\nu}
        \\
        +
        \int_0^{\tau(t)}
        \left(  
            \hat{\mathcal{U}}_{\mu}\br{\tau(t),\nu,\Zt{N};\Boldomega}d\nu 
            + 
            \hat{\mathcal{U}}_{\sigma}\br{\tau(t),\nu,\Zt{N};\Boldomega}d\Wt{\nu}
        \right),
        \quad \forall t \in \mathbb{R}_{\geq 0},
    \end{multline} 
    where 
    \begin{align*}
        \begin{multlined}[b][0.9\linewidth]
            \hat{\mathcal{U}}_{\mu}\br{\tau(t),\nu,\Zt{N};\Boldomega}
            \\
            =
            \expo{-\Boldomega \tau(t)}  
            \frac{\Boldomega}{2\lambda - \Boldomega}
            \left( 
                \expo{\Boldomega \tau(t)}
                \mathcal{P}\br{\tau(t),\nu}
                - 
                e^{ 2\lambda  \tau(t) }  
                \ZDiffNorm{\Zt{N,\tau(t)}}^\top g(\tau(t))
            \right) 
            \expo{ \Boldomega \nu} 
            \LZparamu{\nu,\Zt{N,\nu}}
            ,
        \end{multlined}
        \\
        \begin{multlined}[b][0.9\linewidth]
            \hat{\mathcal{U}}_{\sigma}\br{\tau(t),\nu,\Zt{N};\Boldomega}
            \\
            =
            \expo{-\Boldomega \tau(t)}  
            \frac{\Boldomega}{2\lambda - \Boldomega}
            \left( 
                \expo{\Boldomega \tau(t)}
                \mathcal{P}\br{\tau(t),\nu}
                - 
                e^{ 2\lambda  \tau(t) } 
                \ZDiffNorm{\Zt{N,\tau(t)}}^\top g(\tau(t)) 
            \right)   
            \expo{ \Boldomega \nu}  
            \FZparasigma{\nu,\Zt{N,\nu}}
            ,
        \end{multlined}
        \\
        \phi_{U_\mu}\br{\nu,\Zt{N,\nu};\Boldomega}
        =
        \frac{\Boldomega}{2\lambda - \Boldomega}
        \ZDiffNorm{\Zt{N,\nu}}^\top g(\nu)
        \LZparamu{\nu,\Zt{N,\nu}}
        ,
        \\
        \phi_{U_\sigma}\br{\nu,\Zt{N,\nu};\Boldomega}
        = 
        \frac{\Boldomega}{2\lambda - \Boldomega}
        \ZDiffNorm{\Zt{N,\nu}}^\top g(\nu) 
        \FZparasigma{\nu,\Zt{N,\nu}} 
        .
    \end{align*}
     Using the definitions of $\phi_{\mu^{\paral}}$ and $\phi_{\sigma^{\paral}}$ in~\eqref{lem:True:dV:1}, we obtain    
    \begin{align*}
        \phi_{\mu^{\paral}}\br{\nu,\Zt{N,\nu}} + \phi_{U_\mu}\br{\nu,\Zt{N,\nu};\Boldomega}
        =
        \frac{2\lambda}{2\lambda - \Boldomega}
        \ZDiffNorm{\Zt{N,\nu}}^\top g(\nu)
        \LZparamu{\nu,\Zt{N,\nu}}
        ,
        \\
        \phi_{\sigma^{\paral}}\br{\nu,\Zt{N,\nu}} + \phi_{U_\sigma}\br{\nu,\Zt{N,\nu};\Boldomega}
        =
        \frac{2\lambda}{2\lambda - \Boldomega}
        \ZDiffNorm{\Zt{N,\nu}}^\top g(\nu) 
        \FZparasigma{\nu,\Zt{N,\nu}} 
        .
    \end{align*}
    Substituting into~\eqref{lem:True:dV:2} thus produces
    \begin{multline*}
        \int_0^{\tau(t)}  
            \expo{2 \lambda \nu} 
            \left(\left[\phi_U\left(\nu,\Zt{N,\nu}\right)+\phi_{\mu^{\paral}}\left(\nu,\Zt{N,\nu}\right)\right]d\nu + \phi_{\sigma^{\paral}}\left(\nu,\Zt{N,\nu}\right)d\Wt{\nu} \right)
        \\
        = 
        \expo{-\Boldomega \tau(t)}
        \int_0^{\tau(t)}
        \left(  
            \mathcal{U}_{\mu}\br{\tau(t),\nu,\Zt{N};\Boldomega}d\nu 
            + 
            \mathcal{U}_{\sigma}\br{\tau(t),\nu,\Zt{N};\Boldomega}d\Wt{\nu}
        \right),
        \quad \forall t \in \mathbb{R}_{\geq 0},
    \end{multline*} 
    where $\mathcal{U}_{\mu}$ and $\mathcal{U}_{\sigma}$ are defined in~\eqref{eqn:lem:True:dV:phi:Functions:2}.
    Substituting the above expression for the last integral on the right hand side of~\eqref{lem:True:dV:1} and using the definitions of $\Xi$, $\Xi_{\mathcal{U}}$, and $\widetilde{\Xi}_{\mathcal{U}}$ in~\eqref{eqn:lem:True:dV:Xi:Functions} yields the following bound: 
    \begin{align*}
        e^{2\lambda \tau(t)} \ZNorm{\Zt{N,\tau(t)}}
        \leq
        \Xi\left(\tau(t),\Zt{N} \right)
        +
        \expo{-\Boldomega \tau(t)}
        \Xi_{\mathcal{U}}\left(\tau(t),\Zt{N} \right)
        +
        \widetilde{\Xi}_{\mathcal{U}}\left(\tau(t),\Zt{N} \right)
        ,
        \quad \forall t \in \mathbb{R}_{\geq 0},
    \end{align*}
    which implies
    \begin{align}\label{lem:True:dV:3}
        \ZNorm{\Zt{N,\tau(t)}}
        \leq
        \expo{-2\lambda \tau(t)}
        \Xi\left(\tau(t),\Zt{N} \right)
        +
        \expo{-(2\lambda+\Boldomega) \tau(t)}
        \Xi_{\mathcal{U}}\left(\tau(t),\Zt{N} \right)
        +
        \expo{-2\lambda \tau(t)}
        \widetilde{\Xi}_{\mathcal{U}}\left(\tau(t),\Zt{N} \right)
        ,
        \quad \forall t \in \mathbb{R}_{\geq 0}.
    \end{align}
    Then, using the linearity of the expectation operator, we obtain
    \begin{multline*}
        \ELaw{}{\ZNorm{\Zt{N,\tau(t)}}}
        \leq
        \ELaw{}{
            e^{-2\lambda \tau(t)}  
            \Xi\left(\tau(t),\Zt{N} \right)
        }
        +
        \ELaw{}{
            \expo{-(2\lambda+\Boldomega) \tau(t)}
            \Xi_{\mathcal{U}}\left(\tau(t),\Zt{N} \right)
        }
        \\
        +
        \ELaw{}{
            e^{-2\lambda \tau(t)}
            \widetilde{\Xi}_{\mathcal{U}}\left(\tau(t),\Zt{N} \right)
        }
        ,
        \quad  
        \forall t \in \mathbb{R}_{\geq 0},
    \end{multline*}
    which establishes~\eqref{eqn:lem:True:dV:Supremums:UB:p=1}.
    Finally, applying  Minkowski's inequality to~\eqref{lem:True:dV:3} for $\sfp \in \mathbb{N}_{\geq 2}$ yields the bound in~\eqref{eqn:lem:True:dV:Supremums:UB}, thus concluding the proof.
    \begingroup
    \endgroup
\end{proof}

We next derive bounds on the conditional moments of the truncated joint true (uncertain)-reference process up to the stopping time $\tau_N$. 
\begin{lemma}\label{lem:True:MomentBounds:FirstExit}
    Let the assumptions in Sec.~\ref{subsec:Assumptions} hold and let $\AdaptRho \in \mathbb{R}_{>0}$, $\sfp \in \cbr{1,\dots,\sfp^\star}$, be as defined in~\eqref{eqn:Definitions:True:Rho}, where $\sfp^\star$ is defined in Assumption~\ref{assmp:NominalSystem:FiniteMomentsWasserstein}.
    Further suppose that the filter bandwidth and sampling period conditions in~\eqref{eqn:Definitions:Total:BandwithCondition} and~\eqref{eqn:Definitions:Total:SamplingPeriodCondition}, respectively, hold. 
    Then the truncated joint process $\Zt{N,t} = \Xt{N,t} - \Xrt{N,t}$ in Definition~\ref{def:True:TruncatedJointProcess}, satisfies 
    \begin{align}\label{eqn:lem:True:FirstExit:UB}
        \LpLaw{2\sfp}{}{\Xt{N,t} - \Xrt{N,t}} < \AdaptRho, \quad \forall t \in [0,\tau_N],
    \end{align}
    where the stopping time $\tau_N$ is as in~\eqref{eqn:True:FirstExitTime} for arbitrary $N \in \mathbb{R}_{>0}$.
\end{lemma}
\begin{proof}
    We prove~\eqref{eqn:lem:True:FirstExit:UB} by contradiction.
    The $t$-continuity of the strong solutions $\Xt{N,t}$ and $\Xrt{N,t}$ on $[0,\tau_N]$ implies that $\LpLaw{2\sfp}{}{\Xt{N,t}-\Xrt{N,t}}$ is $t$-continuous as well.
    We have from Definitions~\ref{def:True:JointProcess}~-~\ref{def:True:TruncatedJointProcess} that 
    \begin{align*} 
        \norm{\Zt{N,0}}
        =
        \norm{\Xt{N,0}-\Xrt{N,0}}
        = 0.
    \end{align*}  
    Thus, since $\AdaptRho > 0$, the deterministic nature and continuity of $t \mapsto \LpLaw{2\sfp}{}{\Xt{N,t}-\Xrt{N,t}}$ allow us to formulate the negation of~\eqref{eqn:lem:True:FirstExit:UB} as 
    \begin{align}\label{eqn:True:Scratch:Statement:Not:Q}
        \lnot~\text{\eqref{eqn:lem:True:FirstExit:UB}}~: 
        ~
        \exists~~(deterministic)~~\tstar\in [0,\tau_N] \text{ such that }
        \begin{cases}
            \LpLaw{2\sfp}{}{\Xt{N,t}-\Xrt{N,t}} < \AdaptRho, \quad & \forall t \in [0,\tstar),
            \\ 
            \LpLaw{2\sfp}{}{\Xt{N,\tstar}-\Xrt{N,\tstar}} = \AdaptRho.
        \end{cases} 
    \end{align}

    Now, $\tau(t) = (t \wedge \tstar) \in [0,\tstar]$, for all $t \in \mathbb{R}_{\geq 0}$ because $\tau^\star = \tstar$ in~\eqref{eqn:True:StoppingTimes} since we are assuming that $\tstar < \tau_N$.
    Thus, we may write the bounds~\eqref{eqn:lem:True:dV:Supremums:UB:p=1} and~\eqref{eqn:lem:True:dV:Supremums:UB} in Lemma~\ref{lem:True:dV} as 
    \begin{multline}\label{eqn:True:Scratch:TotalExpectation:Bound:p=1:1}
        \ELaw{}{\ZNorm{\Zt{N,t}}}
        \leq
        \ELaw{}{
            e^{-2\lambda \tau(t)}  
            \Xi\left(\tau(t),\Zt{N} \right)
        }
        +
        \ELaw{}{
            \expo{-(2\lambda+\Boldomega) \tau(t)}
            \Xi_{\mathcal{U}}\left(\tau(t),\Zt{N} \right)
        }
        \\
        +
        \ELaw{}{
            e^{-2\lambda \tau(t)}
            \widetilde{\Xi}_{\mathcal{U}}\left(\tau(t),\Zt{N} \right)
        }
        ,
        \quad  
        \forall t \in [0,\tstar],
    \end{multline}
    and 
    \begin{multline}\label{eqn:True:Scratch:TotalExpectation:p>=2:1}
            \LpLaw{\sfp}{}{\ZNorm{\Zt{N,t}}}
        \leq
            \LpLaw{\sfp}{}{
            \expo{-2\lambda \tau(t)}
            \Xi\left(\tau(t),\Zt{N} \right)
            }
        +
            \LpLaw{\sfp}{}{
            \expo{-(2\lambda+\Boldomega) \tau(t)}
            \Xi_{\mathcal{U}}\left(\tau(t),\Zt{N} \right)
            }
        \\
        +
            \LpLaw{\sfp}{}{
            \expo{-2\lambda \tau(t)}
            \widetilde{\Xi}_{\mathcal{U}}\left(\tau(t),\Zt{N} \right)
            }
        ,
    \end{multline}
    for all $(t,\sfp) \in [0,\tstar] \times \cbr{2,\dots,\sfp^\star}$.
    Now, let us define 
    \begin{subequations}\label{eqn:True:Scratch:ReDefn}
        \begin{align}
            \LpTrueErrorPrime{t}
            \doteq
            \LpLaw{2 \sfp}{}{\Xt{N,t} - \Xrt{N,t}}
            , 
            \label{eqn:True:Scratch:ReDefn:A}
            \\ 
            \LpTrueError
            \doteq
            \sup_{\nu \in [0,\tstar]}
            \LpLaw{2 \sfp}{}{\Xt{N,\nu} - \Xrt{N,\nu}}, 
            \quad 
            \LpTrue
            \doteq
            \sup_{\nu \in [0,\tstar]}
            \LpLaw{2 \sfp}{}{\Xt{N,\nu}},
            \quad 
            \LpTrueRef
            \doteq
            \sup_{\nu \in [0,\tstar]}
            \LpLaw{2 \sfp}{}{\Xrt{N,\nu}},
            \label{eqn:True:Scratch:ReDefn:B}
        \end{align}
    \end{subequations}
    for $\sfp \in \cbr{1,\dots,\sfp^\star}$.
    Then, using the definition $\ZNorm{\Zt{N,t}} \doteq \norm{\Zt{N,t}}^2 = \norm{\Xt{N,t} - \Xrt{N,t}}^2$ in Lemma~\ref{lem:True:dV}, along with the probability measure $\Zdist{N}\left(0,0;dz,t\right)$ defined in~\eqref{eqn:lem:True:dV:CondExp:Definition}, we see that 
    \begin{align*}
        \LpLaw{\sfp}{}{\ZNorm{\Zt{N,t}}}
        =
        \left(
            \int_{\mathbb{R}^n}
                \norm{\Zt{N,t}}^{2\sfp}
            \Zdist{N}\left(0,0;dz,t\right)
        \right)^\frac{2}{2\sfp}
        =
        \left(\LpLaw{2\sfp}{}{\Zt{N,t}}\right)^2
        =&
        \left(\LpLaw{2\sfp}{}{\Xt{N,t} - \Xrt{N,t}}\right)^2
        =
        \left(\LpTrueErrorPrime{t}\right)^2.
    \end{align*}
    Hence, we can write~\eqref{eqn:True:Scratch:TotalExpectation:Bound:p=1:1} and~\eqref{eqn:True:Scratch:TotalExpectation:p>=2:1} as  
    \begin{multline*}
        \left(\LTwoTrueErrorPrime{t}\right)^2
        \leq
        \ELaw{}{
            e^{-2\lambda \tau(t)}  
            \Xi\left(\tau(t),\Zt{N} \right)
        }
        +
        \ELaw{}{
            \expo{-(2\lambda+\Boldomega) \tau(t)}
            \Xi_{\mathcal{U}}\left(\tau(t),\Zt{N} \right)
        }
        \\
        +
        \ELaw{}{
            e^{-2\lambda \tau(t)}
            \widetilde{\Xi}_{\mathcal{U}}\left(\tau(t),\Zt{N} \right)
        }
        ,
        \quad  
        \forall t \in [0,\tstar],
    \end{multline*}
    and 
    \begin{align*}
        \left(\LpTrueErrorPrime{t}\right)^2
        \leq
            \LpLaw{\sfp}{}{
            \expo{-2\lambda \tau(t)}
            \Xi\left(\tau(t),\Zt{N} \right)
            }
        +
            \LpLaw{\sfp}{}{
            \expo{-(2\lambda+\Boldomega) \tau(t)}
            \Xi_{\mathcal{U}}\left(\tau(t),\Zt{N} \right)
            }
        +
            \LpLaw{\sfp}{}{
            \expo{-2\lambda \tau(t)}
            \widetilde{\Xi}_{\mathcal{U}}\left(\tau(t),\Zt{N} \right)
            }
        ,
    \end{align*}
    for all $(t,\sfp) \in [0,\tstar] \times \cbr{2,\dots,\sfp^\star}$.
    Since $\tau^\star = \tstar$ by assumption, we can substitute the bounds derived in Lemma~\ref{lem:Appendix:TrueProcess:Final:Bound} into the above inequalities and get
    \begin{align*}
        \left(\LTwoTrueErrorPrime{t}\right)^2 \leq \Phi\left(1,\Boldomega,\BoldTs\right), 
        \quad 
        \left(\LpTrueErrorPrime{t}\right)^2 \leq \Phi\left(\sfp,\Boldomega,\BoldTs\right),
        \quad \forall (t,\sfp) \in [0,\tstar] \times \cbr{2,\dots,\sfp^\star},
    \end{align*}
    where we have defined the following for $\sfp \in \cbr{1,\dots,\sfp^\star}$:
    \begin{align}\label{eqn:True:Scratch:TotalExpectation:Phi:1}
        \Phi\left(\sfp,\Boldomega,\BoldTs\right)
        =&
        \Delta_{\circledcirc}(\sfp,\Boldomega)
        \left(\LpTrueError\right)^\frac{1}{2}
        +
        \Delta_{\odot}(\sfp,\Boldomega)
        \LpTrueError
        \notag 
        \\ 
        &+
        \left(
            \widebreve{\Delta}_{\otimes}(\sfp,\Boldomega)
            \left( \LpTrue + \LpTrueRef \right)
            +
            \Delta_{\otimes}(\sfp,\Boldomega)
            \LpTrueError
        \right)
        \left(\LpTrueError\right)^\frac{1}{2}
        \notag 
        \\
        &+
        \left(
            \widebreve{\Delta}_\circledast(\Boldomega)
            \left( \LpTrue + \LpTrueRef \right)
            +
            \Delta_\circledast(\Boldomega)
            \LpTrueError
        \right)
        \LpTrueError
        +
        \sum_{i=1}^3
        \Upsilon_i\left(\LpTrue,\LpTrueError; \sfp,\Boldomega,\BoldTs\right).
    \end{align}
    The two bounds above can be combined into the following single bound over $\sfp \in \cbr{1,\dots,\sfp^\star}$:
    \begin{align}\label{eqn:True:Scratch:TotalExpectation:1}
        \left(\LpTrueErrorPrime{t}\right)^2 \leq \Phi\left(\sfp,\Boldomega,\BoldTs\right),
        \quad \forall (t,\sfp) \in [0,\tstar] \times \cbr{1,\dots,\sfp^\star}.
    \end{align}
    Next, as in the proof of Lemma~\ref{lem:Reference:MomentBounds:FirstExit}, the continuity of $\nu \mapsto \LpLaw{2\sfp}{}{\Xt{N,\nu} - \Xrt{N,\nu}  }$ over the closed interval $[0,\tstar]$ implies the existence of some $t' \in [0,\tstar]$, such that 
    \begin{align*}
        \LpTrueError
        \doteq
        \sup_{\nu \in [0,\tstar]} \LpLaw{2\sfp}{}{\Xt{N,\nu} - \Xrt{N,\nu}  } 
        =
        \LpLaw{2\sfp}{}{\Xt{N,t'} - \Xrt{N,t'}  }.
    \end{align*}
    Hence, the contradiction statement in~\eqref{eqn:True:Scratch:Statement:Not:Q} leads us to conclude that 
    \begin{align}\label{eqn:True:Scratch:TotalExpectation:RhoError:1}
        \LpTrueError
        \doteq
        \sup_{\nu \in [0,\tstar]} \LpLaw{2\sfp}{}{\Xt{N,\nu} - \Xrt{N,\nu}  } 
        \leq \AdaptRho,
    \end{align}
    for $\sfp \in \cbr{1,\dots,\sfp^\star}$.
    Similarly, the definitions of $\LpTrue$ and $\LpTrueRef$ in~\eqref{eqn:True:Scratch:ReDefn:B}, along with the contradiction statement in~\eqref{eqn:True:Scratch:Statement:Not:Q}, Assumption~\ref{assmp:NominalSystem:FiniteMomentsWasserstein}, and the result in Lemma~\ref{lem:Reference:WellPosedness} imply 
    \begin{subequations}\label{eqn:True:Scratch:TotalExpectation:RhoError:2}
        \begin{align}
            \LpTrueRef
            \doteq
            \sup_{\nu \in [0,\tstar]}
            \LpLaw{2 \sfp}{}{\Xrt{N,\nu}}
            = 
            \sup_{\nu \in [0,\tstar]}
            \LpLaw{2 \sfp}{}{\Xrt{\nu}}
            \leq     
            \sup_{\nu \in [0,\tstar]}
            \LpLaw{2 \sfp}{}{\Xrt{\nu} - \Xstart{\nu}}
            +
            \sup_{\nu \in [0,\tstar]}
            \LpLaw{2 \sfp}{}{\Xstart{\nu}}
            \notag 
            \\
            < 
            \RefRho + \Delta_\star,
            \\
            \LpTrue
            \doteq
            \sup_{\nu \in [0,\tstar]}
            \LpLaw{2 \sfp}{}{\Xt{N,\nu}}
            \leq 
            \sup_{\nu \in [0,\tstar]}
            \LpLaw{2 \sfp}{}{\Xt{N,\nu} - \Xrt{N,\nu}}
            +
            \sup_{\nu \in [0,\tstar]}
            \LpLaw{2 \sfp}{}{\Xrt{N,\nu}}
            < 
            \AdaptRho + \RefRho + \Delta_\star.
        \end{align}
    \end{subequations}
    Substituting~\eqref{eqn:True:Scratch:TotalExpectation:RhoError:1}~-~\eqref{eqn:True:Scratch:TotalExpectation:RhoError:2} into~\eqref{eqn:True:Scratch:TotalExpectation:Phi:1}, and substituting the resulting bound into~\eqref{eqn:True:Scratch:TotalExpectation:1} gives:
    \begin{multline}\label{eqn:True:Scratch:TotalExpectation:2}
        \left(\LpTrueErrorPrime{t}\right)^2 
        < 
        \Delta_{\circledcirc}(\sfp,\Boldomega)
        \left(\AdaptRho\right)^\frac{1}{2}
        +
        \Delta_{\odot}(\sfp,\Boldomega)
        \AdaptRho
        +
        \left(
            \widebreve{\Delta}_{\otimes}(\sfp,\Boldomega)
            \RhoDPrime 
            +
            \Delta_{\otimes}(\sfp,\Boldomega)
            \AdaptRho
        \right)
        \left(\AdaptRho\right)^\frac{1}{2} 
        \\
        +
        \left(
            \widebreve{\Delta}_\circledast(\Boldomega)
            \RhoDPrime
            +
            \Delta_\circledast(\Boldomega)
            \AdaptRho
        \right)
        \AdaptRho
        +
        \sum_{i=1}^3
        \Upsilon_i\left(\RhoPrime,\AdaptRho; \sfp,\Boldomega,\BoldTs\right),
    \end{multline}
    for all $t \in [0,\tstar]$, where 
    \begin{align*}
        \RhoPrime =  \AdaptRho + \RefRho+\Delta_\star,
        \quad
        \RhoDPrime =  \AdaptRho + 2\left(\RefRho+\Delta_\star\right).      
    \end{align*}
    Furthermore, we have used the following due to the definitions in~\eqref{eqn:Definitions:Total:SamplingPeriodCondition:MasterUpsilons}, Section~\ref{subsubsec:Design:SamplingPeriod}, and the monotonicity property~\eqref{eqn:prop:Appendix:Constants:Maps:Properties:Limit} in Proposition~\ref{prop:Appendix:Constants:Maps:Properties}:
    \begin{align*}
        \Upsilon_i\left(\LpTrue,\LpTrueError; \sfp,\Boldomega,\BoldTs\right)
        \leq 
        \Upsilon_i\left(\RhoPrime,\AdaptRho; \sfp,\Boldomega,\BoldTs\right), \quad \forall i \in \cbr{1,2,3}.
    \end{align*}
    Now, using the definition of $\LpTrueErrorPrime{\cdot}$ in~\eqref{eqn:True:Scratch:ReDefn:A} and the contradiction statement in~\eqref{eqn:True:Scratch:Statement:Not:Q}, one sees that $\LpTrueErrorPrime{\tstar} = \AdaptRho$ at the temporal instance $t = \tstar$,
    Hence, at the temporal instance $t=\tstar$, the inequality in~\eqref{eqn:True:Scratch:TotalExpectation:2} reduces to
    \begin{multline*}
        \AdaptRho^2 
        < 
        \Delta_{\circledcirc}(\sfp,\Boldomega)
        \left(\AdaptRho\right)^\frac{1}{2}
        +
        \Delta_{\odot}(\sfp,\Boldomega)
        \AdaptRho
        +
        \left(
            \widebreve{\Delta}_{\otimes}(\sfp,\Boldomega)
            \RhoDPrime 
            +
            \Delta_{\otimes}(\sfp,\Boldomega)
            \AdaptRho
        \right)
        \left(\AdaptRho\right)^\frac{1}{2} 
        \\
        +
        \left(
            \widebreve{\Delta}_\circledast(\Boldomega)
            \RhoDPrime
            +
            \Delta_\circledast(\Boldomega)
            \AdaptRho
        \right)
        \AdaptRho
        +
        \sum_{i=1}^3
        \Upsilon_i\left(\RhoPrime,\AdaptRho; \sfp,\Boldomega,\BoldTs\right).
    \end{multline*}
    Substituting the definition of constants $\Delta_i(\sfp,\Boldomega)$, $i \in \cbr{ \circledcirc,\odot, \otimes, \circledast   }$ from Lemma~\ref{lem:Appendix:TrueProcess:Final:Bound} into the above inequality and re-arranging the terms leads to
    \begin{align*}
        \left(1 - \frac{\Delta_g^\perp L_\mu^\perp}{\lambda} \right) 
        \AdaptRho^2 
        < 
        \Gamma_a\left(\AdaptRho,\sfp,\Boldomega\right)
        +&
        \frac{ 1 }{  \absolute{2\lambda - \Boldomega}}
        \Theta_a\left(\AdaptRho,\RefRho,\sfp,\Boldomega\right)
        \\
        +&
        \sum_{i=1}^3
        \Upsilon_{a_i}\left(\RhoPrime,\AdaptRho; \sfp,\Boldomega,\BoldTs\right)
        ,
    \end{align*}
    which further implies that
    \begin{multline}\label{eqn:True:Scratch:TotalExpectation:3}
        \sum_{i=1}^3
        \Upsilon_{a_i}\left(\RhoPrime,\AdaptRho; \sfp,\Boldomega,\BoldTs\right)
        >
        \left(1 - \frac{\Delta_g^\perp L_\mu^\perp}{\lambda} \right) 
        \AdaptRho^2
        -
        \Gamma_a\left(\AdaptRho,\sfp,\Boldomega\right)
        \\
        -
        \frac{ 1 }{  \absolute{2\lambda - \Boldomega}}
        \Theta_a\left(\AdaptRho,\RefRho,\sfp,\Boldomega\right) 
        .
    \end{multline}
    However, one sees from~\eqref{eqn:Definitions:Total:SamplingPeriodCondition} that the choice of $\BoldTs$ ensures that
    \begin{multline*}
        \sum_{i=1}^3
        \Upsilon_{a_i}\left(\RhoPrime,\AdaptRho; \sfp,\Boldomega,\BoldTs\right)
        \leq 
        \left(1 - \frac{\Delta_g^\perp L_\mu^\perp}{\lambda} \right) 
        \AdaptRho^2
        - 
        \Gamma_a\left(\AdaptRho,\sfp,\Boldomega\right)
        \\
        -
        \frac{ 1 }{  \absolute{2\lambda - \Boldomega}}
        \Theta_a\left(\AdaptRho,\RefRho,\sfp,\Boldomega\right) 
        ,
    \end{multline*}
    which contradicts the conclusion above in~\eqref{eqn:True:Scratch:TotalExpectation:3} that follows from the postulation in~\eqref{eqn:True:Scratch:Statement:Not:Q}. 
    Therefore, the desired result in~\eqref{eqn:lem:True:FirstExit:UB} is proved since its negation~\eqref{eqn:True:Scratch:Statement:Not:Q} leads to a contradiction.

\end{proof}

Next, we extend the strong solutions of the truncated process $\Zt{N,t}$ to the original joint process $\Zt{t}$ in Definition~\ref{def:True:JointProcess}. 
\begin{lemma}\label{lem:True:WellPosedness}
    If the hypotheses of Lemmas~\ref{lem:Reference:WellPosedness} and~\ref{lem:True:MomentBounds:FirstExit} hold, then the joint process in Definition~\ref{def:True:JointProcess} admits a unique strong solution $\Zt{t} = \Xt{t} - \Xrt{t}$, for all $T \in (0,\infty)$.
    Furthermore,    
    \begin{align}\label{eqn:lem:True:WellPosedness:Bound}
        \LpLaw{2\sfp}{}{\Xt{t}-\Xrt{t}} < \AdaptRho, \quad \forall t \in [0,T],
    \end{align}
    for any $T \in (0,\infty)$.
\end{lemma}
\begin{proof}
    We omit the details of the proof since it follows identically to Lemma~\ref{lem:Reference:WellPosedness} with $\Yt{N} = \br{\Xrt{N},\Xstart{N}} \in \mathbb{R}^{2n}$ and $U_N \subset \subset \mathbb{R}^{2n}$ (see~\eqref{eqn:OpenBoundedSet}), replaced by $\Zt{N,t} = \Xt{N,t} - \Xrt{N,t} \in \mathbb{R}^n$ and $\widetilde{U}_N \subset \subset \mathbb{R}^n$ (see~\eqref{eqn:True:OpenBoundedSet}), respectively.  
\end{proof}

The next result follows from Lemma~\ref{lem:Reference:Main} \emph{mutatis mutandis}.
\begin{lemma}\label{lem:True:Main}
    Let the hypotheses of Lemma~\ref{lem:True:MomentBounds:FirstExit} hold. 
    Then, if in addition, the hypotheses of Lemma~\ref{lem:Reference:WellPosedness} holds for $\alpha\left(\Xrt{0},\Xstart{0}\right)_{2\sfp}=\pWass{2\sfp}{\xi_0}{\xi^\star_0}$, then 
    \begin{align}\label{eqn:thm:True:Wassersetin:Main:UB}
        \pWass{2\sfp}{\Xdist{t}}{\Xrdist{t}}
        <  
        \AdaptRho
        ,
        \quad 
        \forall t \in [0,T]
        .
    \end{align}

    If instead the hypotheses of Lemma~\ref{lem:Reference:WellPosedness} holds for $\alpha\left(\Xrt{0},\Xstart{0}\right)_{2\sfp} = \norm{\Xrt{0}-\Xstart{0}}_{L_{2\sfp}}$, then for any $\delta \in (0,1)$ such that $\log \sqrt{1/\delta} = \sfp$, 
    \begin{align}\label{eqn:thm:True:Markov:Main:UB}
        \Probability{\norm{\Xt{t}-\Xrt{t}} \geq \kappa_a(\delta)} 
        < 
        \delta
        , 
        \quad 
        \forall t \in [0,T]
        ,
    \end{align}       
    where $\kappa_a(\delta) = e \cdot \AdaptRhoFunction{\log \sqrt{1/\delta}}$. 

\end{lemma}

We can now state the main result of the manuscript which follows directly from Lemmas~\ref{lem:Reference:Main} and~\ref{lem:True:Main}. 
\begin{theorem}\label{thm:main}
    Suppose that the assumptions in Sec.~\ref{subsec:Assumptions} hold, and let $\RefRho, \AdaptRho \in \mathbb{R}_{>0}$, $\sfp \in \cbr{1,\dots,\sfp^\star}$, be as defined in~\eqref{eqn:Definitions:Total:Rho} with $\alpha\left(\Xrt{0},\Xstart{0}\right)_{2\sfp} = \pWass{2\sfp}{\xi_0}{\xi^\star_0}$.
    Additionally, let the filter bandwidth and sampling period conditions in~\eqref{eqn:Definitions:Total:BandwithCondition} and~\eqref{eqn:Definitions:Total:SamplingPeriodCondition}, respectively, hold.   
    Then, 
    \begin{subequations}\label{eqn:thm:Total:Wassersetin:Main}
        \begin{align}
            \pWass{2\sfp}{\Xdist{t}}{\Xdist{t}^\star}
            <  
            \TotalRho 
            ,
            \quad 
            \forall t \in [0,T]
            ,
            \quad 
            &\text{\emph{(Distributional UB)}}
            \label{eqn:thm:Total:Wassersetin:Main:UB}
            \\
            \pWass{2\sfp}{\Xrdist{t}}{\Xdist{t}^\star} 
            \leq
            \TotalRhoUUB{\sfp,\Boldomega,\that} 
            < 
            \TotalRho, 
            \quad 
            \forall t \in [\that,T]
            ,
            \quad 
            &\text{\emph{(Distributional UUB)}}
            \label{eqn:thm:Total:Wassersetin:Main:UUB}
        \end{align}
    \end{subequations} 
    where $\that \in (0,T)$ is any deterministic constant, and as given in~\eqref{eqn:Definitions:Eventual:Rho}, $\TotalRho = \RefRho + \AdaptRho$ .
    Moreover,  
    \begin{align*}
        \TotalRhoUUB{\sfp,\Boldomega,\that}
        =
        \RefRhoUUB{\sfp,\Boldomega,\that} + \AdaptRho,
    \end{align*}
    where $\RefRhoUUB{\sfp,\Boldomega,\that}$ is defined in statement of Lemma~\ref{lem:Reference:Main}.

    If instead the above hypotheses are satisfied for $\alpha\left(\Xrt{0},\Xstart{0}\right)_{2\sfp} = \norm{\Xrt{0}-\Xstart{0}}_{L_{2\sfp}}$, then for any $\delta \in (0,1)$ such that $\log \sqrt{1/\delta} = \sfp$, 
    \begin{subequations}\label{eqn:thm:Total:Markov:Main}
        \begin{align}
            \Probability{\norm{\Xt{t}-\Xstart{t}} \geq \kappa(\delta)} 
            < 
            \delta
            , 
            \quad 
            \forall t \in [0,T]
            ,
            \quad 
            &\text{\emph{(Pathwise UB)}}
            \label{eqn:thm:Total:Markov:Main:UB}
            \\
            \Probability{\norm{\Xt{t}-\Xstart{t}} \geq \hat{\kappa}(\delta,\Boldomega,\that)} 
            < 
            \delta
            , 
            \quad 
            \forall t \in [\that,T]
            ,
            \quad 
            &\text{\emph{(Pathwise UUB)}}
            \label{eqn:thm:Total:Markov:Main:UUB}
        \end{align}       
    \end{subequations} 
    where $\kappa(\delta) = e \cdot \TotalRhoFunction{\log \sqrt{1/\delta}}$ and $\hat{\kappa}(\delta,\Boldomega,\that) = e \cdot \TotalRhoUUB{\log \sqrt{1/\delta},\Boldomega,\that}$. 
\end{theorem}
\begin{proof}
    Since the Wasserstein metrics satisfy the axioms of distance~\cite{villani2009optimal}, we get 
    \begin{align*}
        \pWass{2\sfp}{\Xdist{t}}{\Xdist{t}^\star} 
        \leq
        \pWass{2\sfp}{\Xdist{t}}{\Xrdist{t}}
        + 
        \pWass{2\sfp}{\Xrdist{t}}{\Xdist{t}^\star}
        , \quad \forall t \in [0,T].
    \end{align*}
    The bounds in~\eqref{eqn:thm:Total:Wassersetin:Main} then follow from~\eqref{eqn:thm:Reference:Wassersetin:Main} and~\eqref{eqn:thm:True:Wassersetin:Main:UB} in Lemmas~\ref{lem:Reference:Main} and~\ref{lem:True:Main}, respectively.
    
    The bounds in~\eqref{eqn:thm:Total:Markov:Main} follow similarly from~\eqref{eqn:thm:Reference:Markov:Main} and~\eqref{eqn:thm:True:Markov:Main:UB} in Lemmas~\ref{lem:Reference:Main} and~\ref{lem:True:Main}, respectively, and the following due to the Minkowski's inequality:
    \begin{align*}
        \LpLaw{2\sfp}{}{\Xt{t}-\Xstart{t}}
        \leq      
        \LpLaw{2\sfp}{}{\Xt{t}-\Xrt{t}}
        + 
        \LpLaw{2\sfp}{}{\Xrt{t}-\Xstart{t}}, 
        \quad 
        \forall t \in [0,T].
    \end{align*}
\end{proof}


\subsection{Discussion}\label{sec:Discussion}

A few critical points merit a discussion.

\ul{\emph{Tighter High-probability Guarantees:}}
Recall the definition $\TotalRho = \RefRho + \AdaptRho$ from~\eqref{eqn:Definitions:Eventual:Rho}, where $\RefRho$ and $\AdaptRho$ are defined in~\eqref{eqn:Definitions:Total:Rho}.
The terms $\RefRho$ and $\AdaptRho$ depend on the moment $\sfp \in \cbr{1,\dots,\sfp^\star}$, through the constants presented in Appendix~\ref{subsec:app:Definitions:Reference}. 
It is easy to establish that  $\cbr{\RefRho,\AdaptRho} \in \mathcal{O}\br{\sfp^2}$, as $\sfp \rightarrow \infty$. 
It then follows that the pathwise uniform bound $\kappa(\delta)$ in~\eqref{eqn:thm:Total:Markov:Main:UB}, Thm.~\ref{thm:main}, satisfies  $\kappa(\delta) \in \mathcal{O}\left( \log^2 \sqrt{1 / \delta}  \right)$, as $\delta \downarrow 0$.
It can be similarly shown that the pathwise uniform ultimate bound $\hat{\kappa}(\delta,\Boldomega,\that)$ in~\eqref{eqn:thm:Total:Markov:Main:UUB} is also of the same order in $\delta \in (0,1)$.

The order $2$ \emph{polylog} (polylogarithmic) dependence of the bounds $\kappa$ and $\hat{\kappa}$ on the probability of violation $\delta \in (0,1)$ implies much tighter bounds for small $\delta$ when compared to bounds resulting from using up to only the first two moments.  
For example, if the \ellonedrac controller could only guarantee bounds up to $\sfp=1$ (moments up to $2\sfp$), then $\kappa \in \mathcal{O}\left(\sqrt{1 / \delta}\right)$, as $\delta \downarrow 0$, which leads to substantially large bounds for small $\delta$. 
As an example, for $\delta = 10^{-10}$, $\log^2 \sqrt{1 / \delta} = 132.55$, while $\sqrt{1 / \delta} = 10^{5}$.

\ul{\emph{Generalization of the Deterministic Design and Performance:}} 
As discussed in Remark~\ref{rem:Reference:Bandwidth:Feasibility}, the filter bandwidth conditions~\eqref{eqn:Definitions:Total:BandwithCondition} are $\mathcal{O}\left(1 / \sqrt{\Boldomega}\right)$, as  $\Boldomega \rightarrow \infty$.  
For the purposes of this discussion, we may interpret the order of $1/\Boldomega$ on which the bandwidth functions depend as the `strength' of the controller's ability. 
In the absence of the diffusion vector fields, however, one sees that
\begin{align*}
    \left\{
    \frac{1}{\absolute{2\lambda - \Boldomega}} \Theta_r\left(\RefRho,\sfp,\Boldomega\right)
    ,
    \,
    \frac{ 1 }{  \absolute{2\lambda - \Boldomega}} \Theta_a\left(\AdaptRho,\RefRho,\sfp,\Boldomega\right)
    \right\}
    \in 
    \mathcal{O}\left(\frac{1}{\Boldomega}\right),
\end{align*} 
as $\Boldomega \rightarrow \infty$, which represents a `stronger' ability of the controller.  
Indeed, the $\mathcal{O}\left(1/\Boldomega\right)$ dependence of the bandwidth conditions is consistent with our previous results on \ellone adaptive controller design for deterministic systems, see e.g.~\cite{lakshmanan2020safe, wang2012l1}.
It can be similarly shown that the functions $\Upsilon_{a_i}$, $i \in \cbr{1,2,3}$, that define the sampling period condition in~\eqref{eqn:Definitions:Total:SamplingPeriodCondition} satisfy 
\begin{align*}
    \sum_{i=1}^3
    \Upsilon_{a_i}\left(\RhoPrime,\AdaptRho; \sfp,\Boldomega,\BoldTs\right) 
    \in \mathcal{O}\br{\sqrt{\BoldTs}},\text{ as } \BoldTs \downarrow 0. 
\end{align*}
Instead, in the absence of the diffusion terms, the function above belongs to $\mathcal{O}\br{\BoldTs}$, as $\BoldTs \downarrow 0$, which is similar to piecewise constant adaptation laws utilized in the \ellone adaptive controller for deterministic systems, see e.g.~\cite{zhao2024robust, zhao2020adaptive}.
Finally, in the absence of diffusion terms, and unmatched uncertainties, i.e., deterministic system subject to only matched uncertainties, we recover definitions of the bounds $\RefRho$ and $\AdaptRho$ that are consistent with our previous results on deterministic systems in~\cite{lakshmanan2020safe}.

These observations lead to the conclusion that the \ellonedrac control generalizes \ellone adaptive control design for deterministic systems, in both design requirements and performance guarantees.  

\section{Conclusion}\label{sec:ConclusionFuture}

We provide \ellonedrac methodology that allows the synthesis of controllers that can guarantee distributionally robust control on uncertain nonlinear processes. 
The \ellonedrac is based on the \ellone adaptive control architecture.
The robustness guarantees ensured by the \ellonedrac feedback are in the form of Wasserstein ambiguity sets centered on nominal state distributions with \emph{a priori} known radius.
The class of uncertain nonlinear systems for which the \ellonedrac controller is applicable consists of systems with unbounded epistemic uncertainties in both drift and diffusion: a significantly larger class of systems than currently considered for similar synthesis problems in the literature.


\section{Acknowledgements}

The authors are thankful to Evangelos Theodorou from GaTech for many useful discussions in the early stages of this development. The authors gratefully acknowledge funding support by AFOSR Grant FA9550-21-1-0411, the National Aeronautics and Space Administration (NASA) under Grants 80NSSC22M0070 and 80NSSC20M0229, and by the National Science Foundation (NSF) under Grants CMMI 2135925 and  IIS 2331878.

\bibliographystyle{IEEEtran}
\bibliography{SectionsFinal/references_DRAC}

\renewcommand{\thesection}{\Alph{section}}
\renewcommand{\theequation}{\Alph{section}.\arabic{equation}}
\renewcommand{\theHsection}{\Alph{section}}
\renewcommand{\theHequation}{\Alph{section}.\arabic{equation}}
\setcounter{section}{0}
\setcounter{equation}{0}
\section*{\LARGE Appendices}


\section{Constants}\label{app:Definitions}


In this section we collect cumbersome constants that we use throughout the manuscript. 
The subsequent definitions use the  constants $\Delta_{\br{\cdot}}$ defined in Assumptions~\ref{assmp:KnownFunctions}-\ref{assmp:KnownFunctions:GlobalLip} in Sec.~\ref{subsec:Assumptions}.
We first define a few constants that are used globally in the manuscript:
\begin{subequations}\label{eqn:app:Global:Constants}
    \begin{align}
        \mathfrak{p}(\sfp) = \left(\sfp \frac{\sfp - 1}{2}\right)^\frac{1}{2}, 
        \quad
        \mathfrak{p}'(\sfp)= \left(\sfp \frac{2\sfp - 1}{2}\right)^\frac{1}{2},
        \quad
        \mathfrak{p}''(\sfp)= \left(\sfp \left(4\sfp - 1 \right) \right)^\frac{1}{2}, 
        \label{eqn:app:Constants:FrakP}
        \\
        \Lip{f}
        \doteq
        \begin{cases}
            1, 
            &
            \text{if Assumption~\ref{assmp:KnownFunctions:GlobalLip} holds}
            \\
            0,
            &
            \text{otherwise}
        \end{cases}
        .
        \label{eqn:app:Constants:GlobalLipschitz}
    \end{align}  
\end{subequations}

Next, we provide the definitions of constants and maps bespoke to the analysis of the reference process in Sec.~\ref{subsec:Analysis:ReferenceProcess} and Appendix~\ref{app:ReferenceProcess}.


\subsection{Reference Process Analysis}\label{subsec:app:Definitions:Reference}

We begin with the definition of constants $\widehat{\Delta}^r_1$, $\widehat{\Delta}^r_2$, $\widehat{\Delta}^r_3$, and $\widehat{\Delta}^r_4$:
\begin{subequations}\label{eqn:app:Constants:Ref:DelHat}
    \begin{align}
        \widehat{\Delta}^r_1(\sfp)
        =
        \Delta_g
        \left(
            \frac{ 1 }{ \sqrt{\lambda} }
            \left[
                \Delta_f \left(2 + \Delta_\star \right)
                \left(1 - \Lip{f}\right)
                +
                \Delta_\mu
            \right] 
            + 
            \mathfrak{p}(\sfp)
            \left(
                2\Delta_p + \Delta_\sigma
            \right)
        \right),
        \\
        \widehat{\Delta}^r_2(\sfp)
        =
        \Delta_g
        \mathfrak{p}(\sfp)
        \Delta_\sigma,
        \quad
        \widehat{\Delta}^r_3
        =
        \frac{ 1 }{ \sqrt{\lambda} }
        \Delta_g
        \left( 
            \Delta_f\left(1 - \Lip{f}\right) 
            + 
            \Delta_\mu 
        \right),
        \quad
        \widehat{\Delta}^r_4
        =
        \frac{ 1}{ \sqrt{\lambda} }
        \left( 
            \Delta_g
            L_f
            \Lip{f}
            + 
            \Delta_{\dot{g}}
        \right),
    \end{align}
\end{subequations}
where $\Lip{f}$ is defined above in~\eqref{eqn:app:Constants:GlobalLipschitz}.

Next, we define the constants $\Delta^r_\circ$, $\Delta^r_\circledcirc$, $\Delta^r_\odot$, $\Delta^r_\otimes$, $\Delta^r_\circledast$. 
In addition to the constants $\Delta_{\br{\cdot}}$ and $\lambda$ defined in Assumptions~\ref{assmp:KnownFunctions}-\ref{assmp:knownDiffusion:Decomposition} in Sec.~\ref{subsec:Assumptions}, the following also use the constants in~\eqref{eqn:app:Constants:Ref:DelHat} above. 
We first begin with $\Delta^r_{\circ_i}$, $i \in \cbr{1,\dots,4}$, that are defined as follows:
\begin{subequations}\label{eqn:app:Constants:Ref:DelCirc}
    \begin{align}
        \Delta^r_{\circ_1}
        =
        \Delta_p^2
        +
        \left(\Delta_p + \Delta_\sigma\right)^2,
        \quad
        \Delta^r_{\circ_2}(\sfp)
        =
        \frac{\Delta_\mu^{\paral}}{\sqrt{\lambda}}
        \left(
            \widehat{\Delta}^r_1(\sfp) 
            +
            \frac{ \Delta_g^2 \Delta_\mu^{\paral} }{\sqrt{\lambda}}
        \right),
        \\
        \Delta^r_{\circ_3}(\sfp)
        =
        \frac{\mathfrak{p}'(\sfp)}{\sqrt{\lambda}}
        \left(\Delta_p^{\paral} +  \Delta_\sigma^{\paral}\right) 
        \left(
            \widehat{\Delta}^r_1(\sfp) 
            +
            \frac{2 \Delta_g^2 \Delta_\mu^{\paral} }{ \sqrt{\lambda} }
        \right)
        ,
        \\
        \Delta^r_{\circ_4}(\sfp)
        = 
        \left(\Delta_p^{\paral} +  \Delta_\sigma^{\paral}\right) 
        \frac{\Delta_g}{\lambda}
        \left(
            \mathfrak{p}'(\sfp)^2
            \Delta_g 
            \left(
                \Delta_p^{\paral} +  \Delta_\sigma^{\paral} 
            \right)
            + 
            \sqrt{m}
            \left(
                2\Delta_p + \Delta_\sigma
            \right)
        \right)
        .
    \end{align}
\end{subequations}
Next, we define the constants $\Delta^r_{\circledcirc_i}$, $i \in \cbr{1,\dots,4}$, as follows:
\begin{subequations}\label{eqn:app:Constants:Ref:DelCircledCirc}
    \begin{align}
        \Delta^r_{\circledcirc_1}
        =
        2
        \Delta_\sigma
        \left(\Delta_p + \Delta_\sigma \right),
        \quad
        \Delta^r_{\circledcirc_2}(\sfp)
        =
        \frac{\Delta_\mu^{\paral} }{\sqrt{\lambda}}
        \widehat{\Delta}^r_2(\sfp),
        \\
        \Delta^r_{\circledcirc_3}(\sfp)
        =
        \frac{\mathfrak{p}'(\sfp)}{\sqrt{\lambda} }
        \left(
            \left(
                \Delta_p^{\paral}
                +
                \Delta_\sigma^{\paral}    
            \right) 
            \widehat{\Delta}^r_2(\sfp)
            +
            \Delta_\sigma^{\paral}
            \left(
                \widehat{\Delta}^r_1(\sfp) 
                +
                \frac{2 \Delta_g^2 \Delta_\mu^{\paral} }{\sqrt{\lambda}}
            \right)
        \right)
        ,
        \\
        \Delta^r_{\circledcirc_4}(\sfp)
        =
        \frac{\Delta_g}{\lambda}
        \left(
            \vphantom{\frac{\Delta_g}{\lambda}}
            \left(\Delta_p^{\paral} +  \Delta_\sigma^{\paral}\right) 
            \left(
                2
                \mathfrak{p}'(\sfp)^2 
                \Delta_g 
                \Delta_\sigma^{\paral}
                +
                \sqrt{m}
                \Delta_\sigma 
            \right)
            +
            \sqrt{m}
            \Delta_\sigma^{\paral}
            \left(
                2\Delta_p + \Delta_\sigma
            \right)
        \right)            
        .
    \end{align}
\end{subequations}
The constants $\Delta^r_{\odot_i}$, $i \in \cbr{1,\dots,8}$, are defined as: 
\begin{subequations}\label{eqn:app:Constants:Ref:DelOdot}
    \begin{align}
        \Delta^r_{\odot_1}
        =
        \Delta_\sigma^2
        ,
        \quad
        \Delta^r_{\odot_2}
        =
        2 
        \Delta_g^\perp
        \Delta_\mu^\perp
        ,
        \quad
        \Delta^r_{\odot_3}(\sfp)
        =
        2 
        \mathfrak{p}(\sfp)
        \left(
            \Delta_g^\perp
            \left(\Delta_p^{\perp} +  \Delta_\sigma^{\perp}\right) 
            + 
            \Delta_p
        \right)
        ,
        \\
        \Delta^r_{\odot_4}(\sfp)
        =
        \frac{\Delta_\mu^{\paral}}{\sqrt{\lambda}}
        \left(
            \widehat{\Delta}^r_1(\sfp) 
            + 
            \widehat{\Delta}^r_3
            +
            \frac{2 \Delta_g^2 \Delta_\mu^{\paral} }{\sqrt{\lambda}} 
        \right)
        , 
        \quad
        \Delta^r_{\odot_5}(\sfp)
        =
        \Delta_\mu^{\paral}
        \left(
            \frac{ \widehat{\Delta}^r_4 }{\sqrt{\lambda}}
            + 
            4
            \Delta_g  
        \right)
        +
        2
        \sqrt{\lambda} 
        \Delta_g
        \mathfrak{p}(\sfp)
        \left(\Delta_p^{\paral} +  \Delta_\sigma^{\paral}\right),
        \\
        \Delta^r_{\odot_6}(\sfp)
        =
        \frac{\mathfrak{p}'(\sfp)}{\sqrt{\lambda}}
        \left(
            \left(\Delta_p^{\paral} +  \Delta_\sigma^{\paral}\right) 
            \left( 
                \widehat{\Delta}^r_3 
                +
                \frac{ 2 \Delta_g^2 \Delta_\mu^{\paral} }{ \sqrt{\lambda} }
            \right)
            +
            \Delta_\sigma^{\paral}
            \widehat{\Delta}^r_2(\sfp) 
        \right) ,
        \\
        \Delta^r_{\odot_7}(\sfp)
        =
        \mathfrak{p}'(\sfp)
        \left(\Delta_p^{\paral} +  \Delta_\sigma^{\paral}\right) 
        \left(
            \frac{ \widehat{\Delta}^r_4 }{\sqrt{\lambda}}
            + 
            2
            \Delta_g  
        \right)
        ,
        \quad
        \Delta^r_{\odot_8}(\sfp)
        =
        \Delta_\sigma^{\paral}
        \frac{\Delta_g}{\lambda}
        \left(
            \mathfrak{p}'(\sfp)^2
            \Delta_g 
            \Delta_\sigma^{\paral}
            +
            \sqrt{m}
            \Delta_\sigma 
        \right),
    \end{align}
\end{subequations}
while the constants $\Delta^r_{\otimes_i}$, $i \in \cbr{1,\dots,5}$, are defined as follows:
\begin{subequations}\label{eqn:app:Constants:Ref:DelOtimes}
    \begin{align}
        \Delta^r_{\otimes_1}(\sfp)
        =
        2
        \mathfrak{p}(\sfp)  
        \Delta_g^\perp
        \Delta_\sigma^{\perp},
        \quad
        \Delta^r_{\otimes_2}(\sfp)
        = 
        \Delta_\mu^{\paral}
        \frac{ \widehat{\Delta}^r_2(\sfp) }{\sqrt{\lambda}},
        \quad
        \Delta^r_{\otimes_3}(\sfp)
        = 
        2
        \mathfrak{p}(\sfp)
        \sqrt{\lambda} 
        \Delta_g
        \Delta_\sigma^{\paral} ,
        \\
        \cbr{\Delta^r_{\otimes_4}(\sfp),\Delta^r_{\otimes_5}(\sfp)}
        =
        \mathfrak{p}'(\sfp)
        \Delta_\sigma^{\paral}
        \odot
        \cbr{
            \frac{1}{\sqrt{\lambda}}
            \left( 
                \widehat{\Delta}^r_3 
                +
                \frac{ 2 \Delta_g^2 \Delta_\mu^{\paral} }{ \sqrt{\lambda} }
            \right)
            ,
            \left(
                \frac{ \widehat{\Delta}^r_4 }{\sqrt{\lambda}}
                + 
                2
                \Delta_g  
            \right)
        }.
    \end{align}
\end{subequations}
Finally, we define the constants $\Delta^r_{\circledast_i}$, $i \in \cbr{1, \dots,3}$, as follows: 
\begin{subequations}\label{eqn:app:Constants:Ref:DelCircledAst}
    \begin{align}
        \Delta^r_{\circledast_1}
        &= 
        2 
        \Delta_g^\perp
        \Delta_\mu^\perp
        ,
        \quad
        \cbr{\Delta^r_{\circledast_2},\Delta^r_{\circledast_3}}
        =
        \Delta_\mu^{\paral}
        \odot
        \cbr{
            \frac{1}{\sqrt{\lambda}}
            \left( 
                \widehat{\Delta}^r_3 
                +
                \frac{ \Delta_g^2 \Delta_\mu^{\paral} }{ \sqrt{\lambda} }
            \right)
            ,
                \frac{ \widehat{\Delta}^r_4 }{\sqrt{\lambda}}
                + 
                4
                \Delta_g  
        }
        .
    \end{align}       
\end{subequations}

\subsection{True Process Analysis}\label{subsec:app:Definitions:True}

Similar to the definitions in Sec.~\ref{subsec:app:Definitions:Reference} above, we collect constants used in the analysis of the true process, starting with $\widehat{\Delta}_i$, $i \in \cbr{1,\dots,5}$:
    \begin{subequations}\label{eqn:app:Constants:True:DelHat}
        \begin{align}
            \widehat{\Delta}_1(\sfp)
            =
            \frac{ 2 }{  \sqrt{\lambda} }
            \Delta_{g}
            \Delta_f
            \left(1-\Lip{f}\right)
            ,
            \\
            \widehat{\Delta}_2(\sfp)
            =
            \Delta_{g}
            \mathfrak{p}(\sfp)
            \left(L_p  + L_\sigma\right),
            \quad 
            \widehat{\Delta}_3
            = 
            \frac{ 1 }{\sqrt{\lambda} }
                \Delta_{g}
                \Delta_f
                \left(1-\Lip{f}\right),
            \\
            \widehat{\Delta}_4
            =
            \frac{ 1 }{\sqrt{\lambda} }
                \left(
                    \Delta_{g}
                    \left(L_\mu + L_f \Lip{f}\right)
                    +
                    \Delta_{\dot{g}}
                \right),
            \quad
            \widehat{\Delta}_5
            =
            \sqrt{m}
            \Delta_{g}
            \left(L_p  + L_\sigma\right).
        \end{align}
    \end{subequations}

    
    Next, we define $\Delta_{\circledcirc_1} $ and $ \Delta_{\odot_i}$, $i \in \cbr{1,\dots,4}$ as follows:
    \begin{align}\label{eqn:app:Constants:True:DelCircledCirc}
        \Delta_{\circledcirc_1}(\sfp) 
        =
        \frac{ 1  }{\sqrt{\lambda }}
        \mathfrak{p}'(\sfp)
            \left( L_p^{\paral}  + L_\sigma^{\paral} \right)
            \widehat{\Delta}_1(\sfp)
        ,
    \end{align}
    and
    \begin{subequations}\label{eqn:app:Constants:True:DelOdot}
        \begin{align}
            \Delta_{\odot_1}
            =
            \left( L_p  + L_\sigma \right)^2
            ,
            \quad
            \Delta_{\odot_2}(\sfp) 
            =
            \frac{ 1  }{\sqrt{\lambda }}
            L_\mu^{\paral}
            \widehat{\Delta}_1(\sfp),
            \quad
            \Delta_{\odot_3}(\sfp) 
            =
            \frac{ 1  }{\sqrt{\lambda }}
            \mathfrak{p}'(\sfp)\left( L_p^{\paral}  + L_\sigma^{\paral} \right)
            \widehat{\Delta}_2(\sfp),
            \\
            \Delta_{\odot_4}(\sfp)
            =
            \frac{ 1 }{\lambda}
            \left(L_p^{\paral} + L_\sigma^{\paral}\right)
            \left( 
                \widehat{\Delta}_6
                +
                \Delta_g^2
                \mathfrak{p}'(\sfp)^2 
                \left( 
                    L_p^{\paral} + L_\sigma^{\paral} 
                \right)
            \right)
            .
        \end{align}
    \end{subequations}

    Finally, the constants $\Delta_{\otimes_i}$, $i \in \cbr{1,\dots,4}$, and $\Delta_{\circledast_j}$, $j \in \cbr{1,2,3}$, are defined as
    \begin{subequations}\label{eqn:app:Constants:True:DelOtimes}
        \begin{align}
            \Delta_{\otimes_1}(\sfp)
            =
            2 
            \Delta_g^\perp
            \mathfrak{p}(\sfp)
            \left( L_p^{\perp} + L_\sigma^{\perp} \right)
            ,
            \quad
            \Delta_{\otimes_2}(\sfp) 
            =
            2
            \sqrt{\lambda}
            \Delta_g
            \mathfrak{p}(\sfp)
            \left( 
                L_p^{\paral}  + L_\sigma^{\paral} 
            \right)
            + 
            L_\mu^{\paral}
            \frac{ 1  }{\sqrt{\lambda }}
            \widehat{\Delta}_2(\sfp),
            \\
            \Delta_{\otimes_3}(\sfp) 
            =
            \frac{ 1  }{\sqrt{\lambda }}
            \mathfrak{p}'(\sfp)\left( L_p^{\paral}  + L_\sigma^{\paral} \right)
            \widehat{\Delta}_3,
            \\
            \Delta_{\otimes_4}(\sfp) 
            =
            \mathfrak{p}'(\sfp)\left( L_p^{\paral}  + L_\sigma^{\paral} \right)
            \left(
                \frac{ 1  }{\sqrt{\lambda }}
                \widehat{\Delta}_4
                +
                2
                \Delta_g
                \left(
                    1  
                    + 
                    \frac{\Delta_g}{\lambda}  
                    L_\mu^{\paral} 
                \right)
            \right),
        \end{align}        
    \end{subequations}
    and
    \begin{align}\label{eqn:app:Constants:True:DelCircledAst}
        \Delta_{\circledast_1}
        =
        2  
        \Delta_g^\perp
        L_\mu^\perp
        ,
        \quad
        \Delta_{\circledast_2}
        =
        \frac{ 1  }{\sqrt{\lambda }}
        L_\mu^{\paral}
        \widehat{\Delta}_3,
        \quad
        \Delta_{\circledast_3}
        =
        L_\mu^{\paral}
        \left(
            \frac{ 1  }{\sqrt{\lambda }}
            \widehat{\Delta}_4
            +
            \Delta_g
            \left(
                4  
                + 
                \frac{\Delta_g}{\lambda}  
                L_\mu^{\paral} 
            \right)
        \right).
    \end{align}

\subsubsection{Piecewise Constant Adaptation Law}\label{subsubsec:app:Definitions:True:PCA}

    In the following we define constants and maps related to the piecewise constant adaptation law in~\eqref{eqn:L1DRAC:Definition:FeedbackOperator:AdaptationLaw} ( and~\eqref{eqn:True:AdaptationLaw}). 
    Note the explicit dependence of the following entities on the control parameters $\Boldomega, \BoldTs, \lambda_s \in \mathbb{R}_{>0}$ (see~\eqref{eqn:L1DRAC:Definition:FeedbackOperator}~-~\eqref{eqn:L1DRAC:Definition:FeedbackOperator:Components}).

    We begin with the definitions of $\gamma_i$, $\gamma'_i$, $i \in \cbr{1,2}$, and $\gamma''$:
    \begin{subequations}\label{eqn:app:Functions:True:gammaTs:Numbered}
        \begin{align}
            \gamma_1\left(\BoldTs\right)
            =
            \left(
                \lambda_s
                \frac{ \expo{ \lambda_s \BoldTs } + 1 }{ \expo{ \lambda_s \BoldTs } - 1 }
            \right)^\Half
            , 
            \quad
            \gamma_1'(\Boldomega,\BoldTs) 
            = 
            \left(1 - \expo{-2\lambda \BoldTs}\right) 
            \left( 1 - \expo{-\Boldomega \BoldTs}\right)
            \\
            \gamma_2(\Boldomega,\BoldTs)
            =
            \max\left\{ \expo{(\Boldomega - \lambda_s)\BoldTs},1 \right\} 
            \frac{\lambda_s}{\Boldomega} 
            \frac{\expo{\Boldomega\BoldTs} - 1}{\expo{\lambda_s \BoldTs} - 1},
            \\
            \gamma_2'(\Boldomega,\BoldTs)
            =
            \max 
            \left\{ 
                \absolute{
                    1
                    -
                    \expo{\left(\Boldomega - \lambda_s\right)\BoldTs}
                    \frac{\lambda_s}{\Boldomega} 
                    \frac{\expo{\Boldomega\BoldTs} - 1}{\expo{\lambda_s \BoldTs} - 1}
                }
                ,
                \absolute{
                    1
                    -
                    \frac{\lambda_s}{\Boldomega} 
                    \frac{\expo{\Boldomega\BoldTs} - 1}{\expo{\lambda_s \BoldTs} - 1}
                }
            \right\}
            ,  
            \\ 
            \gamma''\left(\sfp,\Boldomega,\BoldTs\right)
            =
            \mathfrak{p}'(\sfp)
            \gamma_2'(\Boldomega,\BoldTs)
            + 
            \mathfrak{p}''(\sfp)
            \left(1 - \expo{ -2\Boldomega \BoldTs } \right)^\frac{1}{2}
            \left(
                2    
                +   
                \gamma_2(\Boldomega,\BoldTs) 
            \right)
            .
            \label{eqn:app:Functions:True:gammaTs:Dblprime}
        \end{align}    
    \end{subequations}

    Similar to the above, we define
    \begin{subequations}\label{eqn:app:Functions:True:gammaTs:MuSigma}
        \begin{align} 
            \gamma_{\mu}\left(\Boldomega,\BoldTs\right)
            =
            L^{\paral}_{\mu}
            + 
            L_{\mu}
            \gamma_2(\Boldomega,\BoldTs)
            \Delta_\Theta, 
            \quad 
            \widehat{\gamma}_{\mu}\left(\Boldomega,\BoldTs\right)
            =
            \hat{L}^{\paral}_{\mu}
            + 
            \hat{L}_{\mu}
            \gamma_2(\Boldomega,\BoldTs)
            \Delta_\Theta
            ,
            \\ 
            \gamma_{\sigma}\left(\Boldomega,\BoldTs\right)
            =
            L^{\paral}_p
            + 
            L^{\paral}_\sigma
            +
            \gamma_2(\Boldomega,\BoldTs)
            \Delta_\Theta
            \left(
                L_p
                + 
                L_\sigma
            \right)
            , 
            \\
            \widehat{\gamma}_{\sigma}\left(\Boldomega,\BoldTs\right)
            =
            \hat{L}^{\paral}_p 
            + 
            \hat{L}^{\paral}_\sigma 
            +  
            \gamma_2(\Boldomega,\BoldTs)
            \Delta_\Theta
            \left( 
                \hat{L}_p 
                + 
                \hat{L}_\sigma
            \right)
            .
        \end{align}    
    \end{subequations}

    Next, the constants $\bar{\Delta}_i$, $i\in \cbr{1,\dots,3}$, are defined as follows:
    \begin{subequations}\label{eqn:app:Constants:True:DelBar}
        \begin{align}
            \begin{aligned}[b]
                \bar{\Delta}_1\left(\BoldTs,\sfp\right)
                =
                2
                \BoldTs 
                \left[
                    \Delta_f
                    + 
                    \left(
                        1
                        + 
                        \Delta_g 
                        \Delta_\Theta
                        e^{-\lambda_s \BoldTs}
                    \right)
                    \Delta_\mu 
                    +
                    \mathfrak{p}'(\sfp)
                    \Delta_g 
                    \Delta_\Theta
                    e^{-\lambda_s \BoldTs}
                    \gamma_1\left(\BoldTs\right)
                    \left(\Delta_p + \Delta_\sigma\right)
                \right]
                \\
                +
                2
                \sqrt{2}
                \BoldTsroot
                \mathfrak{p}'(\sfp)  
                \left(\Delta_p + \Delta_\sigma\right) 
                ,
            \end{aligned}
            \\ 
            \bar{\Delta}_2\left(\BoldTs,\sfp\right)
            =
            2
            \mathfrak{p}'(\sfp)  
            \Delta_\sigma
            \left(
                \sqrt{2}
                \BoldTsroot
                + 
                \BoldTs
                \Delta_g 
                \Delta_\Theta
                e^{-\lambda_s \BoldTs}
                \gamma_1\left(\BoldTs\right)
            \right), 
            \\ 
            \bar{\Delta}_3\left(\BoldTs\right)
            =
            2
            \BoldTs
            \left(
                \Delta_f  
                + 
                \left(
                    1
                    + 
                    \Delta_g 
                    \Delta_\Theta
                    e^{-\lambda_s \BoldTs}
                \right)
                \Delta_\mu    
            \right).
        \end{align} 
    \end{subequations}

    The final set of constants $\Delta_{\mu_i}$ and $\Delta_{\sigma_i}$, $i \in \cbr{1,\dots,3}$, are defined as follows: 
    \begin{subequations}\label{eqn:app:Constants:True:DelMu}
        \begin{align}
            {\Delta}_{\mu_1}\left(\sfp,\Boldomega,\BoldTs\right)
            =
            \BoldTs
            \widehat{\gamma}_\mu\left(\Boldomega,\BoldTs\right)
            +
            \bar{\Delta}_1\left(\BoldTs,\sfp\right)
            \gamma_\mu\left(\Boldomega,\BoldTs\right)
            +
            \Delta^{\paral}_\mu
            \left(1 - \expo{-\lambda_s \BoldTs}\right)
            ,
            \\
            {\Delta}_{\mu_2}\left(\sfp,\Boldomega,\BoldTs\right)
            =
            \bar{\Delta}_2\left(\BoldTs,\sfp\right)
            \gamma_\mu\left(\Boldomega,\BoldTs\right)
            , 
            \\
            {\Delta}_{\mu_3}\left(\sfp,\Boldomega,\BoldTs\right)
            =
            \BoldTs
            \gamma_\mu\left(\Boldomega,\BoldTs\right)
            \bar{\Delta}_3\left(\BoldTs\right) 
            + 
            \Delta^{\paral}_\mu
            \left(1 - \expo{-\lambda_s \BoldTs}\right),
        \end{align}
    \end{subequations}
    and 
    \begin{subequations}\label{eqn:app:Constants:True:DelSigma}
        \begin{align}
            \begin{multlined}[b][0.75\linewidth]
                {\Delta}_{\sigma_1}\left(\sfp,\Boldomega,\BoldTs\right)
                =
                \left[
                    \left(
                        L^{\paral}_p
                        + 
                        L^{\paral}_\sigma
                    \right)
                    \left(\bar{\Delta}_1\left(\BoldTs,\sfp\right)\right)^\frac{1}{2} 
                    +  
                    \BoldTs
                    \left( 
                        \hat{L}^{\paral}_p 
                        + 
                        \hat{L}^{\paral}_\sigma
                    \right)
                \right]
                \left(1 + \gamma_2(\Boldomega,\BoldTs)\right)
                \\
                +
                    \gamma_{\sigma}\left(\Boldomega,\BoldTs\right)
                    \left(\bar{\Delta}_1\left(\BoldTs,\sfp\right)\right)^\frac{1}{2}
                    +  
                    \BoldTs
                    \widehat{\gamma}_{\sigma}\left(\Boldomega,\BoldTs\right) 
                ,
            \end{multlined}
            \\
            {\Delta}_{\cbr{\sigma_2,\sigma_3}}\left(\sfp,\Boldomega,\BoldTs\right)
            =
            \left( \bar{\Delta}_{\cbr{2,3}}\left(\BoldTs,\sfp\right) \right)^\frac{1}{2}
            \left[
                \gamma_{\sigma}\left(\Boldomega,\BoldTs\right) 
                + 
                \left(1 + \gamma_2(\Boldomega,\BoldTs)\right)
                \left(
                    L^{\paral}_p
                    + 
                    L^{\paral}_\sigma
                \right) 
            \right].
        \end{align}
    \end{subequations}

    Next, we define the mappings $\Upsilon'_k\left(\cdot; \sfp,\Boldomega,\BoldTs\right):\mathbb{R}_{\geq 0} \rightarrow \mathbb{R}_{\geq 0}$, $k \in \cbr{1,2,3}$, as follows
    \begin{subequations}\label{eqn:app:Constants:True:Maps:UpsilonPrime}
        \begin{align}
            &\begin{multlined}[b][0.9\linewidth]
                \Upsilon'_1\left(\xi; \sfp,\Boldomega,\BoldTs\right)
                =
                \Delta_{\mu_1}\left(\sfp,\Boldomega,\BoldTs\right)
                + 
                \sqrt{\Boldomega}
                \mathfrak{p}'(\sfp)
                {\Delta}_{\sigma_1}\left(\sfp,\Boldomega,\BoldTs\right)
                + 
                \sqrt{\Boldomega}
                \mathfrak{p}'(\sfp)
                {\Delta}_{\sigma_2}\left(\sfp,\Boldomega,\BoldTs\right)
                \left(\xi\right)^\frac{1}{4}
                \\
                +
                \left(
                    {\Delta}_{\mu_2}\left(\sfp,\Boldomega,\BoldTs\right)
                    +   
                    \sqrt{\Boldomega}
                    \mathfrak{p}'(\sfp)
                    {\Delta}_{\sigma_3}\left(\sfp,\Boldomega,\BoldTs\right) 
                \right)
                \left(\xi\right)^\frac{1}{2}
                + 
                {\Delta}_{\mu_3}\left(\sfp,\Boldomega,\BoldTs\right)
                \xi
                ,
            \end{multlined}
            \\
            &\Upsilon'_2\left(\xi; \sfp,\Boldomega,\BoldTs\right)
            =
            \gamma''\left(\sfp,\Boldomega,\BoldTs\right)
            \left(
                \Delta^{\paral}_p 
                +
                \Delta^{\paral}_\sigma
                \left( 
                    1 
                    + 
                    \left(\xi\right)^\frac{1}{2}
                \right)
            \right),
            \\
            &\Upsilon'_3\left(\xi; \sfp,\Boldomega,\BoldTs\right)
            =
            \Delta^{\paral}_\mu
            \left(\expo{\Boldomega \BoldTs} - 1\right)
            \left(
                1
                + 
                \xi
            \right),
        \end{align}
    \end{subequations}
    for any $\xi \in \mathbb{R}_{\geq 0}$.
    We similarly define the mappings $\left\{ \widetilde{\Upsilon}^{-}\left(\cdot; \sfp,\Boldomega,\BoldTs\right), \Upsilon^{-}\left(\cdot; \sfp,\Boldomega,\BoldTs\right) \right\}:\mathbb{R}_{\geq 0} \rightarrow \mathbb{R}_{\geq 0}$ and $\mathring{\Upsilon}\left(\cdot; \sfp,\Boldomega\right):\mathbb{R}_{\geq 0} \rightarrow \mathbb{R}_{\geq 0}$ as follows
    \begin{subequations}\label{eqn:app:Constants:True:Maps:UpsilonTilde}
        \begin{align}
            &\begin{multlined}[b][0.8\linewidth]
                \widetilde{\Upsilon}^{-}\left(\xi; \sfp,\Boldomega,\BoldTs\right)
                =
                \frac{1}{\lambda}
                \left(1 - \expo{-2\lambda \BoldTs}\right)
                \Delta_g
                \\ 
                \times
                \left(
                    \sqrt{\Boldomega}
                    \mathfrak{p}'(\sfp)
                    \left( 1 - \expo{-2\Boldomega \BoldTs}  \right)^\frac{1}{2} 
                    \left(
                        \Delta^{\paral}_p 
                        +
                        \Delta^{\paral}_\sigma
                        \left(
                            1 
                            + 
                            \left(\xi\right)^\frac{1}{2}
                        \right)        
                    \right)
                    +
                    \Delta^{\paral}_\mu
                    \left(
                        1 
                        + 
                        \xi
                    \right)
                \right),
            \end{multlined}
            \\
            &\Upsilon^{-}\left(\xi; \sfp,\Boldomega,\BoldTs\right)
            =
            \gamma_1'(\Boldomega,\BoldTs)
            \left( 
                2 
                \mathfrak{p}'(\sfp)
                \left( \Delta^{\paral}_p + \Delta^{\paral}_\sigma \left(1 + \left(\xi\right)^\frac{1}{2} \right) \right)
                +
                \frac{ \Delta^{\paral}_\mu }{\sqrt{\Boldomega}}  
                \left( 1 +  \xi \right)
            \right),
            \\
            &\mathring{\Upsilon}\left(\xi; \sfp,\Boldomega\right)
            =
            \Delta_g^2
            \left(
                \mathfrak{p}'(\sfp)
                \left( 
                    L_p^{\paral} + L_\sigma^{\paral} 
                \right)
                \left(\xi\right)^\frac{1}{2}
                + 
                \frac{1}{\BoldomegaRoot}
                L_\mu^{\paral}
                \xi 
            \right),
        \end{align}
    \end{subequations}
    for any $\xi \in \mathbb{R}_{\geq 0}$.
    We conclude the section by proving a few key but straightforward properties of the maps defined above.
    \begin{proposition}\label{prop:Appendix:Constants:Maps:Properties} 
        The mapping $\xi \mapsto \check{\Upsilon}\left(\xi; \sfp,\Boldomega,\BoldTs\right)$, $\check{\Upsilon} \in \cbr{ \Upsilon'_1,\Upsilon'_2,\Upsilon'_3,\widetilde{\Upsilon}^-,\Upsilon^-,\mathring{\Upsilon} }$, is non-decreasing over $\mathbb{R}_{\geq 0}$, i.e. 
        \begin{align}\label{eqn:prop:Appendix:Constants:Maps:Properties:NonDecreasing}
            \check{\Upsilon}\left(\xi_1; \sfp,\Boldomega,\BoldTs\right)
            \leq     
            \check{\Upsilon}\left(\xi_2; \sfp,\Boldomega,\BoldTs\right), 
            \quad 
            \forall \br{\xi_1,\xi_2,\sfp,\Boldomega,\BoldTs} \in \mathbb{R}_{\geq 0} \times \mathbb{R}_{\geq 0} \times \mathbb{N} \times \mathbb{R}_{>0} \times \mathbb{R}_{>0}, \quad  
            \xi_1 \leq \xi_2.
        \end{align}
        Furthermore, the following limits are satisfied for any fixed $\cbr{\xi,\sfp,\Boldomega} \in \mathbb{R}_{\geq 0} \times \mathbb{N} \times \mathbb{R}_{>0}$:
        \begin{align}\label{eqn:prop:Appendix:Constants:Maps:Properties:Limit}
            \lim_{\BoldTs \downarrow 0 }
            \left\{
                \widetilde{\Upsilon}^{-}
                ,
                \Upsilon^{-}
                ,
                \Upsilon'_1
                ,
                \Upsilon'_2
                ,
                \Upsilon'_3
            \right\}
            \left(\xi; \sfp,\Boldomega,\BoldTs\right)
            =
            0.
        \end{align}

    \end{proposition}
    \begin{proof}
        The non-decreasing property in~\eqref{eqn:prop:Appendix:Constants:Maps:Properties:NonDecreasing} is trivial to prove.  
        In order to prove~\eqref{eqn:prop:Appendix:Constants:Maps:Properties:Limit}, we begin by considering the terms in~\eqref{eqn:app:Functions:True:gammaTs:Numbered} for any fixed $\lambda_s \in \mathbb{R}_{>0}$.
        We obtain the following limits in a straightforward manner:  
        \begin{subequations}\label{eqn:app:Functions:True:gammaTs:Numbered:Limits}
            \begin{align}
                \lim_{\BoldTs \downarrow 0 }
                \left(\BoldTs \gamma_1(\BoldTs)\right)
                =
                \sqrt{\lambda_s}
                \lim_{\BoldTs \downarrow 0}
                \left(\expo{ \lambda_s \BoldTs } + 1\right)^\Half
                \lim_{\BoldTs \downarrow 0}
                \frac{ \BoldTs  }{ \left(\expo{ \lambda_s \BoldTs } - 1\right)^\Half }
                = 0,
                 \label{eqn:prop:Appendix:Constants:Limit:Tsgamma1}
                \\
                \lim_{\BoldTs \downarrow 0 }
                \gamma_1'(\Boldomega,\BoldTs) 
                = 
                \lim_{\BoldTs \downarrow 0 }
                \left(1 - \expo{-2\lambda \BoldTs}\right)
                \lim_{\BoldTs \downarrow 0 } 
                \left( 1 - \expo{-\Boldomega \BoldTs}\right)
                =
                0,
                \\
                \lim_{\BoldTs \downarrow 0 }
                \gamma_2(\Boldomega,\BoldTs)
                =
                \frac{\lambda_s}{\Boldomega}
                \left(
                    \lim_{\BoldTs \downarrow 0 }
                    \max\left\{ \expo{(\Boldomega - \lambda_s)\BoldTs},1 \right\} 
                \right)
                \left(
                    \lim_{\BoldTs \downarrow 0 } 
                    \frac{\expo{\Boldomega\BoldTs} - 1}{\expo{\lambda_s \BoldTs} - 1}
                \right)
                =
                1,
                \label{eqn:prop:Appendix:Constants::Limit:gamma2}
                \\
                \lim_{\BoldTs \downarrow 0 }
                \gamma_2'(\Boldomega,\BoldTs)
                =
                \max 
                \left\{ 
                    \lim_{\BoldTs \downarrow 0 }
                    \absolute{
                        1
                        -
                        \expo{\left(\Boldomega - \lambda_s\right)\BoldTs}
                        \frac{\lambda_s}{\Boldomega} 
                        \frac{\expo{\Boldomega\BoldTs} - 1}{\expo{\lambda_s \BoldTs} - 1}
                    }
                    ,
                    \lim_{\BoldTs \downarrow 0 }
                    \absolute{
                        1
                        -
                        \frac{\lambda_s}{\Boldomega} 
                        \frac{\expo{\Boldomega\BoldTs} - 1}{\expo{\lambda_s \BoldTs} - 1}
                    }
                \right\}
                =0, 
                \\ 
                \lim_{\BoldTs \downarrow 0 }
                \gamma''\left(\sfp,\Boldomega,\BoldTs\right)
                =
                \mathfrak{p}'(\sfp)
                \lim_{\BoldTs \downarrow 0 }
                \gamma_2'(\Boldomega,\BoldTs)
                + 
                \mathfrak{p}''(\sfp)
                \lim_{\BoldTs \downarrow 0 }
                \left(1 - \expo{ -2\Boldomega \BoldTs } \right)^\frac{1}{2}
                \lim_{\BoldTs \downarrow 0 }
                \left(
                    2    
                    +   
                    \gamma_2(\Boldomega,\BoldTs) 
                \right)
                =
                0
                .
            \end{align}
        \end{subequations}
        Using the limit in~\eqref{eqn:prop:Appendix:Constants::Limit:gamma2}, we obtain the following for the terms in~\eqref{eqn:app:Functions:True:gammaTs:MuSigma}:
        \begin{subequations}\label{eqn:app:Functions:True:gammaTs:MuSigma:Limits}
            \begin{align} 
                \lim_{\BoldTs \downarrow 0 }
                \gamma_{\mu}\left(\Boldomega,\BoldTs\right)
                =
                L^{\paral}_{\mu}
                + 
                L_{\mu}
                \Delta_\Theta, 
                \quad 
                \lim_{\BoldTs \downarrow 0 }
                \widehat{\gamma}_{\mu}\left(\Boldomega,\BoldTs\right)
                =
                \hat{L}^{\paral}_{\mu}
                + 
                \hat{L}_{\mu}
                \Delta_\Theta
                ,
                \\
                \lim_{\BoldTs \downarrow 0 } 
                \gamma_{\sigma}\left(\Boldomega,\BoldTs\right)
                =
                L^{\paral}_p
                + 
                L^{\paral}_\sigma
                +
                \Delta_\Theta
                \left(
                    L_p
                    + 
                    L_\sigma
                \right)
                ,
                \quad
                \lim_{\BoldTs \downarrow 0 }
                \widehat{\gamma}_{\sigma}\left(\Boldomega,\BoldTs\right)
                =
                \hat{L}^{\paral}_p 
                + 
                \hat{L}^{\paral}_\sigma 
                +  
                \Delta_\Theta
                \left( 
                    \hat{L}_p 
                    + 
                    \hat{L}_\sigma
                \right)
                . 
            \end{align}    
        \end{subequations}
        Similarly, using the limit in~\eqref{eqn:prop:Appendix:Constants:Limit:Tsgamma1}, we obtain the following for the terms in~\eqref{eqn:app:Constants:True:DelBar}:
        \begin{align}\label{eqn:app:Constants:True:DelBar:Limits}
            \lim_{\BoldTs \downarrow 0 }
            \left\{
                \bar{\Delta}_1\left(\BoldTs,\sfp\right), \bar{\Delta}_2\left(\BoldTs,\sfp\right), \bar{\Delta}_3\left(\BoldTs \right)
            \right\}
            = 0.
        \end{align}
        We now use~\eqref{eqn:app:Functions:True:gammaTs:Numbered:Limits},~\eqref{eqn:app:Functions:True:gammaTs:MuSigma:Limits}, and~\eqref{eqn:app:Constants:True:DelBar:Limits} to conclude the following for the terms defined in~\eqref{eqn:app:Constants:True:DelMu} and~\eqref{eqn:app:Constants:True:DelSigma}: 
        \begin{align}\label{eqn:app:Constants:True:DelMuSigma:Limits}
            \lim_{\BoldTs \downarrow 0 }
            {\Delta}_{\mu_i}\left(\sfp,\Boldomega,\BoldTs\right)
            =0,
            \quad   
            \lim_{\BoldTs \downarrow 0 }
            {\Delta}_{\sigma_i}\left(\sfp,\Boldomega,\BoldTs\right)
            =0
            , 
            \quad  
            i \in \cbr{1,2,3}
            .
        \end{align}
        The claim in~\eqref{eqn:prop:Appendix:Constants:Maps:Properties:Limit} then follows by applying the above limits to the mappings in~\eqref{eqn:app:Constants:True:Maps:UpsilonPrime} and~\eqref{eqn:app:Constants:True:Maps:UpsilonTilde}.

    \end{proof}

\begingroup

\endgroup


\setcounter{equation}{0}
\section{Technical Results}\label{app:TechnicalResults}

\begin{remark}
    The results that we present in this appendix consider functions belonging to $\mathcal{M}_{2}^{loc}\left(\mathcal{X}~|~\mathcal{F}_t \right)$, for some finite-dimensional space $\mathcal{X}$ and filtration $\mathcal{F}_t$, $t \in [0,T]$. 
    However, it is straightforward to see that the results hold for the more general space $\mathcal{M}_{2}\left([0,T];\mathcal{X}~|~\mathcal{F}_t \right)$.
    In fact, Appendices~\ref{app:ReferenceProcess} and~\ref{app:TrueProcess} invoke the following results since they also hold over $\mathcal{M}_{2}\left([0,\tau \wedge T];\mathcal{X}~|~\mathcal{F}_t \right)$, where $\tau$ is any stopping-time.
\end{remark}

We begin with the following result that is a multidimensional version of the stochastic integration by parts formula in~\cite[Thm.~1.6.5]{mao2007stochastic}.
\begin{lemma}\label{lem:TechnicalResults:Fubini-ish}
    Consider any $L \in \mathcal{M}_{2}^{loc}\br{\mathbb{R}^{l \times l'}|\mathfrak{F}_t}$ and $S \in \mathcal{M}_{2}^{loc}\br{\mathbb{R}^{l' \times n_q}|\mathfrak{F}_t}$, for some filtration $\mathfrak{F}_t$ on $\br{\Omega, \mathcal{F}, \mathbb{P}}$.
    Then, for any $n_q$-dimensional $\mathfrak{F}_t$-adapted Brownian motion $Q_t$, we have that
    \begin{subequations}
        \begin{align}
            \int_0^t \int_0^\nu L(\nu) S(\beta) dQ_\beta d \nu 
            =
            \int_0^t \int_\nu^t L(\beta) S(\nu) d\beta dQ_\nu \in \mathbb{R}^{l}, 
            \label{eqn:lem:TechnicalResults:Fubini-ish:Identity}
            \\
            \int_0^t \int_\nu^t L(\nu) S(\beta) dQ_\beta d \nu 
            =
            \int_0^t \int_0^\nu L(\beta) S(\nu) d\beta dQ_\nu \in \mathbb{R}^{l}, 
            \label{eqn:lem:TechnicalResults:Fubini-ish:Identity:Alt}
        \end{align}
    \end{subequations} 
    for all $t \in [0,T]$.
\end{lemma}
\begin{proof}
    We provide a proof for~\eqref{eqn:lem:TechnicalResults:Fubini-ish:Identity} only since~\eqref{eqn:lem:TechnicalResults:Fubini-ish:Identity:Alt} can be established \emph{mutatis mutandis}.
    We begin by defining 
    \begin{align}\label{eqn:lem:TechnicalResults:Fubini-ish:G:Definition}
        \hat{L}(\varsigma) \doteq \int_0^\varsigma L(\beta) d\beta \in \mathbb{R}^{l \times l'}, \quad \varsigma \in \mathbb{R}_{\geq 0}.    
    \end{align}
    As a consequence of the above definition, we have that 
    \begin{align}\label{eqn:lem:TechnicalResults:Fubini-ish:G:Expression}
        \int_\nu^t L(\beta) d\beta = \int_0^t L(\beta) d\beta - \int_0^\nu L(\beta) d\beta \doteq \hat{L}(t) - \hat{L}(\nu) \in \mathbb{R}^{l \times l'}, \quad 0 \leq \nu \leq t \leq T.
    \end{align}
    Using the expression above, we may write the right hand side of~\eqref{eqn:lem:TechnicalResults:Fubini-ish:Identity} as 
    \begin{align*}
        \int_0^t \int_\nu^t L(\beta) S(\nu) d\beta dQ_\nu
        =
        \int_0^t \left(\int_\nu^t L(\beta)d\beta\right) S(\nu)  dQ_\nu
        =
        \int_0^t \left( \hat{L}(t) - \hat{L}(\nu)  \right)  S(\nu)  dQ_\nu 
        \in
        \mathbb{R}^l, 
    \end{align*}
    which can further be re-written as 
    \begin{multline}\label{eqn:lem:TechnicalResults:Fubini-ish:RHS:Pre}
        \begin{aligned}
            &\int_0^t \int_\nu^t L(\beta) S(\nu) d\beta dQ_\nu
            \\
            &
            = 
            \int_0^t 
            \begin{bmatrix} 
                \left(\hat{L}_1(t) - \hat{L}_1(\nu)\right)S(\nu)dQ_\nu 
                & \cdots &  
                \left(\hat{L}_l(t) - \hat{L}_l(\nu)\right)S(\nu)dQ_\nu 
            \end{bmatrix}^\top
        \end{aligned}
        \\
        = 
        \sum_{i=1}^{l'}
        \int_0^t 
        \begin{bmatrix} 
            \left(\hat{L}_{1,i}(t) - \hat{L}_{1,i}(\nu)\right)S_i(\nu)dQ_\nu 
            & \cdots &  
            \left(\hat{L}_{l,i}(t) - \hat{L}_{l,i}(\nu)\right)S_i(\nu)dQ_\nu 
        \end{bmatrix}^\top    
        \in
        \mathbb{R}^l, 
    \end{multline}
    where $\hat{L}_{1,\cdots,l} \in \mathbb{R}^{1 \times l'}$ and $S_{1,\cdots,l'} \in \mathbb{R}^{1 \times n_q}$ are the rows of $\hat{L}(\cdot) \in \mathbb{R}^{l \times l'}$ defined in~\eqref{eqn:lem:TechnicalResults:Fubini-ish:G:Definition} and $S(\cdot) \in \mathbb{R}^{l' \times n_q}$, respectively.
    Next, we define the following scalar \ito process:
    \begin{align}\label{eqn:lem:TechnicalResults:Fubini-ish:Z:Process}
        \hat{S}_i(t) = \int_0^t S_i(\varsigma) dQ_\varsigma, \quad d\hat{S}_i(t) = S_{i}(t) dQ_t, 
        \quad t \in [0,T],
        \quad \hat{S}_i(0)=0,  
        \quad i \in \cbr{1,\dots,l'},
    \end{align}	
    using which we derive the following expression: 
    \begin{multline}\label{eqn:lem:TechnicalResults:Fubini-ish:R:1}
        \int_0^t 
        \left(\hat{L}_{j,i}(t) - \hat{L}_{j,i}(\nu)\right)S_i(\nu)dQ_\nu
        =
        \int_0^t 
        \left(\hat{L}_{j,i}(t) - \hat{L}_{j,i}(\nu)\right) d \hat{S}_i(\nu)
        \\
        \begin{multlined}[b][0.5\linewidth]
            =
            \hat{L}_{j,i}(t) \int_0^t d\hat{S}_i(\nu)
            -
            \int_0^t \hat{L}_{j,i}(\nu) d\hat{S}_i(\nu)
            \\
            =
            \hat{L}_{j,i}(t)\hat{S}_i(t)
            -
            \int_0^t \hat{L}_{j,i}(\nu) d\hat{S}_i(\nu),
        \end{multlined} 
    \end{multline}
    for $(j,i,t) \in \cbr{1,\dots,l} \times \cbr{1,\dots,l'} \times [0,T]$.
    It is clear that $\hat{L}_{j,i}(t)$ is $\mathfrak{F}_t$-adapted due its definition in~\eqref{eqn:lem:TechnicalResults:Fubini-ish:G:Definition} and the assumed adaptedness of $L$.
    Moreover, we may write $\hat{L}_{j,i}(t) = \hat{L}_{j,i}^+(t) - \hat{L}_{j,i}^-(t)$, where
    \begin{align*}
        \hat{L}_{j,i}^+(t)
        = 
        \int_0^t \max \left\{ 0, L_{j,i}(\varsigma) \right\}d\varsigma, 
        \quad 
        \hat{L}_{j,i}^-(t)
        = 
		\int_0^t \max \left\{ -L_{j,i}(\varsigma),0 \right\} d\varsigma.
    \end{align*}
    Since both $\hat{L}_{j,i}^+(t)$ and $\hat{L}_{j,i}^-(t)$ are increasing processes, we conclude that $\hat{L}_{j,i}(t)$ is a finite variation process~\cite[Sec.~1.3]{mao2007stochastic}.
    Therefore we may use the integration by parts formula~\cite[Thm.~1.6.5]{mao2007stochastic} to obtain 
    \begin{align*}
        \int_0^t \hat{L}_{j,i}(\nu) d\hat{S}_i(\nu)
        =
        \hat{L}_{j,i}(t) \hat{S}_i(t)
        -
        \int_0^t  \hat{S}_i(\nu) d\hat{L}_{j,i}(\nu)
        , 
    \end{align*}
    where the last integral is a Lebesgue-Stieltjes integral.
    Substituting the above equality into~\eqref{eqn:lem:TechnicalResults:Fubini-ish:R:1}, we obtain
    \begin{align}\label{eqn:lem:TechnicalResults:Fubini-ish:R:2}
        \int_0^t 
        \left(\hat{L}_{j,i}(t) - \hat{L}_{j,i}(\nu)\right)S_i(\nu)dQ_\nu
        =
        \hat{L}_{j,i}(t)\hat{S}_i(t)
        -
        \int_0^t \hat{L}_{j,i}(\nu) d\hat{S}_i(\nu)
        =
        \int_0^t  \hat{S}_i(\nu) d\hat{L}_{j,i}(\nu)
        ,
    \end{align}
    for $ (j,i,t) \in \cbr{1,\dots,l} \times \cbr{1,\dots,l'} \times [0,T]$.
    Then, once again appealing to the decomposition $\hat{L}_{j,i}(t) = \hat{L}_{j,i}^+(t) - \hat{L}_{j,i}^-(t)$, and using the continuity of $L$ and the definition of $\hat{L}$ in~\eqref{eqn:lem:TechnicalResults:Fubini-ish:G:Definition}, we apply the fundamental theorem for Lebesgue-Stieltjes integrals~\cite[Thm.~7.7.1]{burk2007garden} to~\eqref{eqn:lem:TechnicalResults:Fubini-ish:R:2} and obtain  
    \begin{align*}
        \int_0^t 
        \left(\hat{L}_{j,i}(t) - \hat{L}_{j,i}(\nu)\right)S_i(\nu)dQ_\nu
        =
        \int_0^t  \hat{S}_i(\nu) d\hat{L}_{j,i}(\nu)
        =
        \int_0^t  \hat{S}_i(\nu) L_{j,i}(\nu)d\nu
        ,
    \end{align*} 
    for $(j,i,t) \in \cbr{1,\dots,l} \times \cbr{1,\dots,l'} \times [0,T]$.
    The proof of~\eqref{eqn:lem:TechnicalResults:Fubini-ish:Identity} is then concluded by substituting the above expression into~\eqref{eqn:lem:TechnicalResults:Fubini-ish:RHS:Pre} 
    to obtain 
    \begin{multline*}
        \begin{aligned}
            &\int_0^t \int_\nu^t L(\beta) S(\nu) d\beta dQ_\nu
            \\
            &
            = 
            \sum_{i=1}^{l'}
            \int_0^t 
            \begin{bmatrix} 
                \left(\hat{L}_{1,i}(t) - \hat{L}_{1,i}(\nu)\right)S_i(\nu)dQ_\nu 
                & \cdots &  
                \left(\hat{L}_{l,i}(t) - \hat{L}_{l,i}(\nu)\right)S_i(\nu)dQ_\nu 
            \end{bmatrix}^\top
            \\
            &
            = 
            \sum_{i=1}^{l'}
            \int_0^t 
            \begin{bmatrix} 
                \hat{S}_i(\nu) L_{1,i}(\nu)d\nu 
                & \cdots &  
                \hat{S}_i(\nu) L_{l,i}(\nu)d\nu 
            \end{bmatrix}^\top
        \end{aligned}
        \\
        \overset{(\star)}{=}
        \sum_{i=1}^{l'}
        \int_0^t
        \int_0^\nu  
        \begin{bmatrix} 
            L_{1,i}(\nu) S_i(\beta) dQ_\beta d\nu 
            & \cdots &  
            L_{l,i}(\nu) S_i(\beta) dQ_\beta d\nu 
        \end{bmatrix}^\top
        = 
        \int_0^t
        \int_0^\nu  
        L(\nu) S(\beta) dQ_\beta d\nu  
        , 
    \end{multline*}
    where the equality $(\star)$ is obtained by invoking the definition of the scalar process $\hat{S}_i(t)$ in~\eqref{eqn:lem:TechnicalResults:Fubini-ish:Z:Process}. 
\end{proof}

The next result is a straightforward moment bound for the Lebesgue integral of pertinent stochastic processes.
\begin{proposition}\label{prop:TechnicalResults:LebesgueMoment}
    Consider a complete filtered probability space $\br{\Omega, \mathcal{F}, \mathbb{P}}$ with  filtration $\mathfrak{F}_t$, and let $S \in \mathcal{M}_{2}^{loc}\br{\mathbb{R}^{n_s}|\mathfrak{F}_t}$ satisfy 
    \begin{align*}
        \ELaw{}{\int_0^T \norm{S(\nu)}^{\sfp}d\nu } < \infty, \quad \forall \sfp \geq 1.
    \end{align*}
    Additionally, for any constants $\theta \in \mathbb{R}_{> 0}$ and $\xi \in [0,T]$, define 
    \begin{equation*}
        N(t)
        = 
        \int_\xi^t 
            \expo{\theta \nu} \norm{S(\nu)}
        d\nu \in \mathbb{R}, \quad t \in [\xi,T].  
    \end{equation*}
    Then,  
    \begin{equation}\label{eqn:prop:TechnicalResults:LebesgueMoment:Bound:Final}
        \LpLaw{\sfp}{}{N(t)} 
        \leq 
        \frac{\expo{\theta t} - \expo{\theta \xi}}{\theta} 
        \sup_{\nu \in [\xi,t]}\LpLaw{\sfp}{}{S(\nu)}
        ,
        \quad \forall (t,\sfp) \in [\xi,T] \times \mathbb{N}_{\geq 1}
        .
    \end{equation} 
\end{proposition}
\begin{proof}
    For $\sfp=1$, we use the Fubini's theorem to conclude that 
    \begin{multline*}
        \ELaw{}{N(t)}
        =
        \int_\xi^t 
            \expo{\theta \nu} \ELaw{}{\norm{S(\nu)}}
        d\nu
        \leq
        \left( 
        \int_\xi^t 
            \expo{\theta \nu} 
        d\nu
        \right)
        \sup_{\nu \in [\xi,t]}
        \ELaw{}{\norm{S(\nu)}}
        =
        \frac{\expo{\theta t} - \expo{\theta \xi}}{\theta}
        \sup_{\nu \in [\xi,t]}
        \ELaw{}{\norm{S(\nu)}}
        ,
    \end{multline*}
    for all $t \in [\xi,T]$, thus establishing~\eqref{eqn:prop:TechnicalResults:LebesgueMoment:Bound:Final}.
    %
    %
    For $\sfp \in \mathbb{N}_{\geq 2}$, we obtain the following chain of inequalities:
    \begin{align*}
        \LpLaw{\sfp}{}{N(t)} 
        =     
        \ELaw{}{
            \left(
                \int_\xi^{t} 
                \expo{\theta \nu} \norm{S(\nu)} 
                d\nu 
            \right)^\sfp
        }^\frac{1}{\sfp}
        \overset{(\star)}{\leq}
        \int_\xi^{t} 
            \expo{\theta \nu} \ELaw{}{\norm{S(\nu)}^\sfp}^\frac{1}{\sfp} 
        d\nu 
        \leq
        \left(
            \int_\xi^{t} 
            \expo{\theta \nu}  
            d\nu 
        \right)
        \sup_{\nu \in [\xi,t]}\ELaw{}{\norm{S(\nu)}^\sfp}^\frac{1}{\sfp}
        ,
    \end{align*}
    for all $t \in [\xi,T]$, where $(\star)$ follows from the Minkowski’s integral inequality~\cite[Thm.~202]{hardy1952inequalities}. 
   The bound in~\eqref{eqn:prop:TechnicalResults:LebesgueMoment:Bound:Final}, $\forall \sfp \in \mathbb{N}_{\geq 2}$, then follows by solving the integral.

\end{proof}

The following is a simple corollary of the last result. 
\begin{corollary}\label{cor:TechnicalResults:LebesgueMoment}
    Let the hypotheses of Proposition~\ref{prop:TechnicalResults:LebesgueMoment} hold.
    Then, the process 
    \begin{equation*}
        N(t)
        = 
        \int_\xi^t 
            \norm{S(\nu)}
        d\nu \in \mathbb{R}, \quad t \in [\xi,T],  
    \end{equation*}
    satisfies
    \begin{equation}\label{eqn:cor:TechnicalResults:LebesgueMoment:Bound:Final} 
        \LpLaw{\sfp}{}{N(t)} 
        \leq
        (t-\xi)
        \sup_{\nu \in [\xi,t]}
        \LpLaw{\sfp}{}{S(\nu)}
        ,
        \quad \forall (t,\sfp) \in [\xi,T] \times \mathbb{N}_{\geq 1}
        .
    \end{equation} 
    
\end{corollary}
\begin{proof}
    The proof follows by noting that $N(t)$ in this corollary can be obtained from $N(t)$ in Proposition~\ref{prop:TechnicalResults:LebesgueMoment} by setting $\theta = 0$.
    
\end{proof}

Next, we apply Proposition~\ref{prop:TechnicalResults:LebesgueMoment} to obtain a moment bound for nested Lebesgue integrals. 
\begin{proposition}\label{prop:TechnicalResults:LebesgueNestedMoment}
    Consider a complete filtered probability space $\br{\Omega, \mathcal{F}, \mathbb{P}}$ with  filtration $\mathfrak{F}_t$, and let $S_1 \in \mathcal{M}_{2}^{loc}\br{\mathbb{R}^{n_1}|\mathfrak{F}_t}$ and $S_2 \in \mathcal{M}_{2}^{loc}\br{\mathbb{R}^{n_{2}}|\mathfrak{F}_t}$ satisfy 
    \begin{align*}
        \ELaw{}{\int_0^T \left[\norm{S_1(\nu)}^{2\sfp} + \norm{S_2(\nu)}^{2\sfp}\right] d\nu } < \infty, \quad \forall \sfp \geq 1.
    \end{align*}
    Additionally, for any constants $\theta_1 \in \mathbb{R}_{> 0}$, $\theta_2 \in \mathbb{R}_{> 0}$,  and $\xi \in [0,T]$, define 
    \begin{equation*}
        N(t)
        = 
        \int_\xi^t 
            \expo{(\theta_1 - \theta_2) \nu} 
            \norm{S_1(\nu)}
            \left(
                \int_0^\nu
                    \expo{\theta_2 \beta} \norm{S_2(\beta)}
                d\beta
            \right)
        d\nu \in \mathbb{R}, \quad t \in [\xi,T].  
    \end{equation*}
    Then,  
    \begin{equation}\label{eqn:prop:TechnicalResults:LebesgueNestedMoment:Bound:Final} 
        \LpLaw{\sfp}{}{N(t)} 
        \leq 
        \frac{ \expo{ \theta_1 t } - \expo{ \theta_1 \xi } }{ \theta_1 \theta_2 }
        \sup_{\nu \in [\xi,t]}  
        \LpLaw{2\sfp}{}{S_1(\nu)}
        \sup_{\nu \in [0,t]} 
        \LpLaw{2\sfp}{}{S_2(\nu)}
        ,
        \quad \forall (t,\sfp) \in [\xi,T] \times \mathbb{N}_{\geq 1}
        .
    \end{equation} 
\end{proposition}
\begin{proof}
    We begin with the case $\sfp=1$, for which we apply  Fubini's theorem, followed by the Cauchy-Schwarz inequality to obtain  
    \begin{align*}
        \ELaw{}{\absolute{N(t)}} 
        =& 
        \int_\xi^t 
        \expo{(\theta_1 - \theta_2) \nu} 
        \ELaw{}
        {
            \norm{S_1(\nu)}
            \left(
                \int_0^\nu
                    \expo{\theta_2 \beta} \norm{S_2(\beta)}
                d\beta
            \right)
        }
        d\nu
        \\
        \leq & 
        \int_\xi^t 
        \expo{(\theta_1 - \theta_2) \nu} 
        \ELaw{}{\norm{S_1(\nu)}^2}^\Half
        \ELaw{}
        {
            \left(
                \int_0^\nu
                    \expo{\theta_2 \beta} \norm{S_2(\beta)}
                d\beta
            \right)^2
        }^\Half
        d\nu
        , 
        \quad \forall t \in [\xi,T]
        , 
    \end{align*}
    where we have used the fact that, by definition, $N(t) \geq 0$, $\forall t \geq \xi$.
    Applying Proposition~\ref{prop:TechnicalResults:LebesgueMoment} to the second inner integral with $\cbr{\theta,\xi,n_s,S} = \cbr{\theta_2,0,n_2,S_2}$ leads to 
    \begin{align*}
        \ELaw{}{\absolute{N(t)}} 
        \leq & 
        \frac{1}{\theta_2}
        \int_\xi^t 
        \left(
            \expo{\theta_1 \nu} 
            \ELaw{}{\norm{S_1(\nu)}^2}^\Half
            \sup_{\beta \in [0,\nu]}
            \ELaw{}{\norm{S_2(\beta)}^2}^\Half
        \right)
        d\nu
        \\ 
        \leq & 
        \frac{1}{\theta_2}
        \left(
            \int_\xi^t 
                \expo{\theta_1 \nu} 
            d\nu
        \right)
        \sup_{\nu \in [\xi,t]}
        \ELaw{}{\norm{S_1(\nu)}^2}^\Half
        \sup_{\nu \in [0,t]}
        \ELaw{}{\norm{S_2(\nu)}^2}^\Half
        , 
        \quad \forall t \in [\xi,T]
        ,
    \end{align*}
    where in the first inequality we have used the bound $1-\expo{-\theta_2 \nu} \leq 1$, $\nu \in [\xi,t]$.
    Solving the integral on the right hand side then leads to~\eqref{eqn:prop:TechnicalResults:LebesgueNestedMoment:Bound:Final} for $\sfp = 1$. 

    Next, we consider the $L_\mathsf{p}$ norm of $N(t)$ for $\sfp \in \mathbb{N}_{\geq 2}$ given by
    \begin{align*}
        \LpLaw{\sfp}{}{N(t)} 
        =     
        \ELaw{}
        {
            \left(
                \int_\xi^{t} 
                \expo{(\theta_1 - \theta_2) \nu} 
                \norm{S_1(\nu)}
                \left(
                    \int_0^\nu
                        \expo{\theta_2 \beta} \norm{S_2(\beta)}
                    d\beta
                \right)
                d\nu 
            \right)^\sfp
        }^\frac{1}{\sfp}
        ,
        \quad \forall t \in [\xi,T]
        . 
    \end{align*}
    From Minkowski’s integral inequality~\cite[Thm.~202]{hardy1952inequalities}, we get
    \begin{align*}
        \LpLaw{\sfp}{}{N(t)} 
        =&
        \,
        \ELaw{}
        {
            \left(
                \int_\xi^{t} 
                \expo{(\theta_1 - \theta_2) \nu} 
                \norm{S_1(\nu)}
                \left(
                    \int_0^\nu
                        \expo{\theta_2 \beta} \norm{S_2(\beta)}
                    d\beta
                \right)
                d\nu 
            \right)^\sfp
        }^\frac{1}{\sfp}
        \notag 
        \\
        \leq
        & 
        \int_\xi^{t} 
        \expo{\left(\theta_1   - \theta_2 \right) \nu }
        \ELaw{}
        {
            \norm{S_1(\nu)}^\sfp
            \left( \int_0^\nu \expo{\theta_2 \beta} \norm{S_2(\beta)} d\beta   \right)^\sfp
        }^\frac{1}{\sfp}
        d\nu
        ,
    \end{align*}
    for all $t \in [\xi,T]$.
    Subsequently, applying the Cauchy-Schwarz inequality then leads to 
    \begin{align}\label{eqn:prop:TechnicalResults:LebesgueNestedMoment:Bound:A:Final}
        \LpLaw{\sfp}{}{N(t)} 
        \leq
        \int_\xi^{t} 
        \expo{\left(\theta_1   - \theta_2 \right) \nu }
        \ELaw{}{\norm{S_1(\nu)}^{2\sfp}}^\frac{1}{2\sfp}
        \ELaw{}
        {
            \left( \int_0^\nu \expo{\theta_2 \beta} \norm{S_2(\beta)} d\beta   \right)^{2\sfp}
        }^\frac{1}{2\sfp}
        d\nu
        ,
        \quad \forall t \in [\xi,T]
        .
    \end{align}
    Next, using Proposition~\ref{prop:TechnicalResults:LebesgueMoment} for $\cbr{\theta,\xi,n_s,S} = \cbr{\theta_2,0,n_2,S_2}$, we obtain
    \begin{multline*}
        \ELaw{}{
            \left( \int_0^\nu \expo{\theta_2 \beta} \norm{S_2(\beta)} d\beta   \right)^{2\sfp}
        }^\frac{1}{2\sfp}
        \leq     
        \left( \frac{ \expo{\theta_2\nu} - 1 }{\theta_2} \right)
        \sup_{\beta \in [0,\nu]} 
        \ELaw{}{\norm{S_2(\beta)}^{2\sfp}}^\frac{1}{2\sfp}
        \\
        \leq     
        \frac{ \expo{\theta_2 \nu} }{\theta_2}
        \sup_{\beta \in [0,\nu]} 
        \ELaw{}{\norm{S_2(\beta)}^{2\sfp}}^\frac{1}{2\sfp}
        , 
        \quad \forall \nu \in [\xi,t], 
    \end{multline*}
    where we have bounded $\expo{\theta_2\nu} - 1 $ by $\expo{\theta_2\nu}$.
    Substituting the above bound in~\eqref{eqn:prop:TechnicalResults:LebesgueNestedMoment:Bound:A:Final} yields
    \begin{align*}
        \LpLaw{\sfp}{}{N(t)} 
        \leq
        &
        \frac{ 1 }{\theta_2 }
        \int_\xi^{t}
        \expo{ \theta_1 \nu }
        \ELaw{}{\norm{S_1(\nu)}^{2\sfp}}^\frac{1}{2\sfp}
        \sup_{\beta \in [0,\nu]} 
        \ELaw{}{\norm{S_2(\beta)}^{2\sfp}}^\frac{1}{2\sfp}
        d\nu 
        \\
        \leq
        &
        \frac{ 1 }{\theta_2 }
        \left(
            \int_\xi^{t} 
                \expo{\theta_1 \nu }
            d\nu
        \right)^\frac{1}{\sfp}
        \sup_{\nu \in [\xi,t]}  
        \ELaw{}{\norm{S_1(\nu)}^{2\sfp}}^\frac{1}{2\sfp}
        \sup_{\nu \in [0,t]} 
        \ELaw{}{\norm{S_2(\nu)}^{2\sfp}}^\frac{1}{2\sfp}
        , 
        \\
        \leq
        &
        \frac{ \expo{ \theta_1 t } - \expo{ \theta_1 \xi } }{ \theta_1 \theta_2 }
        \sup_{\nu \in [\xi,t]}  
        \ELaw{}{\norm{S_1(\nu)}^{2\sfp}}^\frac{1}{2\sfp}
        \sup_{\nu \in [0,t]} 
        \ELaw{}{\norm{S_2(\nu)}^{2\sfp}}^\frac{1}{2\sfp}
        , 
        \quad \forall t \in [\xi,T],
    \end{align*}
    which establishes~\eqref{eqn:prop:TechnicalResults:LebesgueNestedMoment:Bound:Final} for all $\sfp \in \mathbb{N}_{\geq 2}$.

\end{proof}

The following result provides moment bound for a pertinent class of \ito integrals. 
\begin{lemma}\label{lemma:TechnicalResults:MartingaleMoment}
    Consider a complete filtered probability space $\br{\Omega, \mathcal{F}, \mathbb{P}}$ with  filtration $\mathfrak{F}_t$, and let $L \in \mathcal{M}_{2}^{loc}\br{\mathbb{R}^{n_l \times n_q}|\mathfrak{F}_t}$ satisfy 
    \begin{align*}
        \ELaw{}{\int_0^T \Frobenius{L(\nu)}^{\sfp}d\nu } < \infty, \quad \forall \sfp \geq 2.
    \end{align*}
    Additionally, for any constants $\theta \in \mathbb{R}_{> 0}$ and $\xi \in [0,T]$, and an $\mathfrak{F}_t$-adapted Brownian motion $Q_t \in \mathbb{R}^{n_q}$, define 
    \begin{equation*}
        N(t)
        = 
        \int_\xi^t 
            \expo{\theta \nu} L(\nu)
        d\Qt{\nu} \in \mathbb{R}^{n_l}, \quad t \in [\xi,T].
    \end{equation*}
    Then, 
    \begin{align}\label{eqn:lemma:TechnicalResults:MartingaleMoment:Main}
        \LpLaw{\sfp}{}{N(t)}
        \leq&
        \left(\sfp \frac{\sfp-1}{2}\right)^\Half  
        \left(\frac{\expo{ 2\theta t } - \expo{ 2\theta \xi }}{2\theta}\right)^\Half
        \sup_{\nu \in [\xi,t]}
        \LpLaw{\sfp}{}{L(\nu)}
        ,
        \quad \forall (t,\sfp) \in [\xi,T] \times \mathbb{N}_{\geq 2}
        .
    \end{align} 
    
\end{lemma}
\begin{proof}

    We begin with the case when $\sfp=2$. 
    Since $\xi \in [0,T]$ is a constant, and hence a stopping time, we use \ito isometry~\cite[Thm.~1.5.21]{mao2007stochastic} to obtain 
    \begin{align*}
        \mathbb{E}\left[\norm{N(t)}^{2}\right] 
        =
        \ELaw{}{\norm{\int_\xi^{t} e^{\theta \nu} L(\nu)dQ_\nu}^{2}}
        =
        \ELaw{}{\int_\xi^{t} e^{2\theta \nu} \Frobenius{L(\nu)}^2 d\nu}
        .
    \end{align*}
    It then follows from Fubini's theorem that  
    \begin{align*}
        \mathbb{E}\left[\norm{N(t)}^{2}\right] 
        \leq 
        \int_\xi^{t} e^{2\theta \nu} \ELaw{}{\Frobenius{L(\nu)}^2} d\nu
        \leq& 
        \left(\int_\xi^{t} e^{2\theta \nu}  d\nu\right)
        \sup_{\nu \in [\xi,t]}
        \ELaw{}{\Frobenius{L(\nu)}^2}
        \\
        =&
        \frac{\expo{2\theta t} - \expo{2\theta \xi}}{2\theta}
        \sup_{\nu \in [\xi,t]}
        \ELaw{}{\Frobenius{L(\nu)}^2}
        , 
        \quad \forall t \in [\xi,T]
        ,
    \end{align*} 
    which establishes~\eqref{eqn:lemma:TechnicalResults:MartingaleMoment:Main} for $\sfp=2$.

    Next, let $\sfp \in \mathbb{N}_{> 2}$. 
    Using \ito's lemma~\cite[Thm.~1.6.4]{mao2007stochastic} in a similar fashion to the proof of~\cite[Thm.~1.7.1]{mao2007stochastic}, we get
    \begin{align}\label{eqn:lemma:TechnicalResults:MartingaleMoment:1}
        \ELaw{}{\norm{N(t)}^{\sfp}} 
        = 
        \frac{\sfp}{2} 
        \ELaw{}
        {
            \int_\xi^t \left( \norm{N(\nu)}^{\sfp-2} \expo{2\theta\nu} \Frobenius{L(\nu)}^2
            +
            (\sfp-2)\norm{N(\nu)}^{\sfp-4} \expo{2\theta\nu} \norm{N(\nu)^\top L(\nu)}^2  \right)d\nu
        }
        ,
    \end{align}
    for $t \in [\xi,T]$.
    Thus, we obtain the following bound using the submultiplicativity of norms: 
    \begin{align*}
        \ELaw{}{\norm{N(t)}^{\sfp}}
        \leq&
        \sfp \frac{\sfp-1}{2} 
        \ELaw{}{\int_\xi^t \norm{N(\nu)}^{\sfp-2} \expo{2\theta \nu}\Frobenius{L(\nu)}^2 d\nu }
        \\
        =&
        \sfp \frac{\sfp-1}{2} 
        \ELaw{}{\int_\xi^t 
        \expo{ 2\theta \nu (\sfp-2)/\sfp } \norm{N(\nu)}^{\sfp-2} \expo{ 4\theta \nu /\sfp } \Frobenius{L(\nu)}^2 d\nu }
        ,
        \quad 
        \forall t \in [\xi,T]
        ,
    \end{align*}   
    where we obtain the last expression by writing $\expo{ 2\theta \nu} = \expo{ 2\theta \nu \left[(\sfp-2)/\sfp + 2/\sfp\right] } = \expo{ 2\theta \nu (\sfp-2)/\sfp }\expo{ 4\theta \nu /\sfp }$.
    Using H\"{o}lder's inequality with conjugates $\sfp / (\sfp - 2)$ and $\sfp/2$, followed by Fubini's theorem, leads to  
    \begin{align*}
        \ELaw{}{\norm{N(t)}^{\sfp}}
        \leq&
        \sfp \frac{\sfp-1}{2}  
        \ELaw{}{\int_\xi^t \expo{ 2\theta \nu } \norm{N(\nu)}^{ \sfp }  d\nu }^{\frac{\sfp-2}{\sfp}}
        \ELaw{}{\int_\xi^t \expo{ 2\theta \nu } \Frobenius{L(\nu)}^{\sfp} d\nu }^\frac{2}{\sfp}
        \\
        =& 
        \sfp \frac{\sfp-1}{2} 
        \left(\int_\xi^t \expo{ 2\theta \nu } \ELaw{}{\norm{N(\nu)}^{\sfp }}  d\nu \right)^{\frac{\sfp-2}{\sfp}}
        \left(\int_\xi^t \expo{ 2\theta \nu } \ELaw{}{\Frobenius{L(\nu)}^{\sfp}} d\nu \right)^\frac{2}{\sfp}
        ,
        \quad \forall t \in [\xi,T]. 
    \end{align*}
    Note from~\eqref{eqn:lemma:TechnicalResults:MartingaleMoment:1} that $\ELaw{}{\norm{N(t)}^{\sfp}}$ is non-decreasing in $t$.
    It then follows that
    \begin{align*}
        \ELaw{}{\norm{N(t)}^{\sfp}}
        \leq& 
        \sfp \frac{\sfp-1}{2}  
        \left(\int_\xi^t \expo{ 2\theta \nu } d\nu \right)^{\frac{\sfp-2}{\sfp}}
        \ELaw{}{\norm{N(t)}^{\sfp }}^{\frac{\sfp-2}{\sfp}}
        \left(\int_\xi^t \expo{ 2\theta \nu } \ELaw{}{\Frobenius{L(\nu)}^{\sfp}} d\nu \right)^\frac{2}{\sfp}
        \\
        \leq& 
        \sfp \frac{\sfp-1}{2}  
        \left(\int_\xi^t \expo{ 2\theta \nu } d\nu \right)^{\frac{\sfp-2}{\sfp}}
        \ELaw{}{\norm{N(t)}^{\sfp }}^{\frac{\sfp-2}{\sfp}}
        \left(\int_\xi^t \expo{ 2\theta \nu }  d\nu \right)^\frac{2}{\sfp}
        \sup_{\nu \in [\xi,t]}
        \ELaw{}{\Frobenius{L(\nu)}^{\sfp}}^\frac{2}{\sfp}
        \\
        =& 
        \sfp \frac{\sfp-1}{2}  
        \ELaw{}{\norm{N(t)}^{\sfp }}^{\frac{\sfp-2}{\sfp}}
        \left(\int_\xi^t \expo{ 2\theta \nu }  d\nu \right)
        \sup_{\nu \in [\xi,t]}
        \ELaw{}{\Frobenius{L(\nu)}^{\sfp}}^\frac{2}{\sfp}
        , 
        \quad \forall t \in [\xi,T]
    ,
    \end{align*}
    and thus
    \begin{align*}
        \ELaw{}{\norm{N(t)}^{\sfp}}^\frac{2}{\sfp}
        \leq & 
        \sfp \frac{\sfp-1}{2}  
        \left(\int_\xi^t \expo{ 2\theta \nu }  d\nu \right)
        \sup_{\nu \in [\xi,t]}
        \ELaw{}{\Frobenius{L(\nu)}^{\sfp}}^\frac{2}{\sfp}
        \\
        =&
        \sfp \frac{\sfp-1}{2}  
        \frac{\expo{ 2\theta t } - \expo{ 2\theta \xi }}{2\theta}
        \sup_{\nu \in [\xi,t]}
        \ELaw{}{\Frobenius{L(\nu)}^{\sfp}}^\frac{2}{\sfp}
        , 
        \quad \forall t \in [\xi,T]
    ,
    \end{align*}
    which proves~\eqref{eqn:lemma:TechnicalResults:MartingaleMoment:Main}, $\forall \sfp \in \mathbb{N}_{> 2}$.

\end{proof}

Next, we provide a bound on a particular Lebesgue integral with an \ito integral in the integrand.
\begin{lemma}\label{lemma:TechnicalResults:NestedSupMoment}
    Consider a complete filtered probability space $\br{\Omega, \mathcal{F}, \mathbb{P}}$ with  filtration $\mathfrak{F}_t$, and let $S \in \mathcal{M}_{2}^{loc}\br{\mathbb{R}^{n_s}|\mathfrak{F}_t}$ and $L \in \mathcal{M}_{2}^{loc}\br{\mathbb{R}^{n_l \times n_q}|\mathfrak{F}_t}$ satisfy 
    \begin{align*}
        \ELaw{}{\int_0^T \left( \norm{S(\nu)}^{2\sfp} + \Frobenius{L(\nu)}^{2\sfp} \right)d\nu } < \infty, \quad \forall \sfp \geq 1.
    \end{align*}
    Additionally, for any $\theta_1,\,\theta_2 \in \mathbb{R}_{> 0}$,  $\xi \in \mathbb{R}_{\geq 0}$, and an $\mathfrak{F}_t$-adapted Brownian motion $Q_t \in \mathbb{R}^{n_q}$ define 
    \begin{equation*}
        N(t)
        = 
        \int_{\xi}^t \expo{(\theta_1 - \theta_2) \nu} \norm{S(\nu)}
            \norm{\int_0^\nu \expo{\theta_2 \beta} \indicator{[\zeta_1,\zeta_2]}{\beta} L(\beta)dQ_\beta}
        d\nu \in \mathbb{R}, \quad t \in [\xi,T],  
    \end{equation*}
    where and $0 \leq \zeta_1 \leq \zeta_2 \leq t$ are any deterministic variables.
    Then, 
    \begin{align}\label{eqn:lem:TechnicalResults:NestedSupMoment:Bound:Final}
        \LpLaw{\sfp}{}{N(t)} 
        \leq  
        \left(
            \sfp \frac{2\sfp -1}{2} 
        \right)^\frac{1}{2}
        \frac{ \expo{\theta_1 t} - \expo{\theta_1 (\xi \vee \zeta_1)} }{\theta_1 \sqrt{\theta_2} }
        \left( 1 - \expo{-2\theta_2(\zeta_2 - \zeta_1)}  \right)^\frac{1}{2}
        \sup_{\nu \in [\xi,t]}
        \LpLaw{2\sfp}{}{S(\nu)}
        \sup_{\nu \in [\zeta_1,\zeta_2]} 
        \LpLaw{2\sfp}{}{L(\nu)}
        , 
    \end{align}
    for all $(t,\sfp) \in [\xi,T] \times \mathbb{N}_{\geq 1}$. 
\end{lemma}
\begin{proof}
    We provide the proof only for the case when $\sfp \in \mathbb{N}_{\geq 2}$, since the proof for $\sfp=1$ is established \emph{mutatis mutandis}.
    The presence of the indicator function in the inner integrand allows us to write
    \begin{multline*}
        N(t)
        = 
        \int_{\xi}^t 
            \expo{(\theta_1 - \theta_2) \nu} \norm{S(\nu)}
            \norm{ \int_0^\nu \expo{\theta_2 \beta} \indicator{[\zeta_1,\zeta_2]}{\beta} L(\beta)dQ_\beta }
        d\nu 
        \\
        = 
        \int_{\xi \vee \zeta_1}^t 
            \expo{(\theta_1 - \theta_2) \nu} \norm{S(\nu)}
            \norm{ \int_{\zeta_1}^{\nu \wedge \zeta_2} \expo{\theta_2 \beta}  L(\beta)dQ_\beta }
        d\nu 
        ,
        \quad t \in [\xi,T].  
    \end{multline*}
    Now, consider the $L_{\mathsf{p}}$ norm of $N(t)$. 
    By following the same line of reasoning that led to~\eqref{eqn:prop:TechnicalResults:LebesgueNestedMoment:Bound:A:Final} in the proof of Proposition~\ref{prop:TechnicalResults:LebesgueNestedMoment}, we obtain:
    \begin{align*}
        \LpLaw{\sfp}{}{N(t)} 
        \leq
        \int_{\xi \vee \zeta_1}^{t} 
            \expo{ \left(\theta_1   - \theta_2 \right) \nu }
            \ELaw{}{\norm{S(\nu)}^{2\sfp}}^\frac{1}{2\sfp} 
            \ELaw{}
            {
                \norm{\int_{\zeta_1}^{\nu \wedge \zeta_2} \expo{\theta_2 \beta} L(\beta) d\Qt{\beta}   }^{2\sfp}
            }^\frac{1}{2\sfp} 
        d\nu
        ,
        \quad t \in [\xi,T]
        .
    \end{align*}
    Next, using Lemma~\ref{lemma:TechnicalResults:MartingaleMoment} for $\cbr{\theta,\xi,t} = \cbr{\theta_2,\zeta_1,\nu \wedge \zeta_2}$ leads to 
    \begin{align*}
        \ELaw{}{
        \norm{ 
            \int_{\zeta_1}^{\nu \wedge \zeta_2} \expo{\theta_2 \beta} L(\beta) d\Qt{\beta}   }^{2\sfp}
        }^\frac{1}{2\sfp} 
        \leq& 
        \left(\sfp \frac{2\sfp -1}{2\theta_2} \right)^\frac{1}{2}
        \left( \expo{ 2\theta_2 (\nu \wedge \zeta_2)} - \expo{ 2\theta_2 \zeta_1 }  \right)^\frac{1}{2}
        \sup_{\beta \in [\zeta_1,\nu \wedge \zeta_2]} 
        \ELaw{}{\norm{L(\beta)}^{2\sfp}}^\frac{1}{2\sfp} 
        \\
        \leq& 
        \left(\sfp \frac{2\sfp -1}{2\theta_2} \right)^\frac{1}{2}
        \left( \expo{ 2\theta_2 (\nu \wedge \zeta_2)} - \expo{ 2\theta_2 \zeta_1 }  \right)^\frac{1}{2}
        \sup_{\beta \in [\zeta_1,\zeta_2]} 
        \ELaw{}{\norm{L(\beta)}^{2\sfp}}^\frac{1}{2\sfp} 
        ,
    \end{align*}
    for all $\nu \in [\xi,t]$, where the last inequality follows from the fact that $[\zeta_1,\nu \wedge \zeta_2] \subseteq [\zeta_1,\zeta_2]$, $\forall \nu \in [\xi,t]$.
    Therefore, 
    \begin{multline}\label{eqn:lem:TechnicalResults:NestedSupMoment:Bound:B:1}
        \LpLaw{\sfp}{}{N(t)} 
        \leq
        \int_{\xi \vee \zeta_1}^{t} 
            \expo{ \left(\theta_1   - \theta_2 \right) \nu }
            \ELaw{}{\norm{S(\nu)}^{2\sfp}}^\frac{1}{2\sfp} 
            \ELaw{}
            {
                \norm{\int_{\zeta_1}^{\nu \wedge \zeta_2} \expo{\theta_2 \beta} L(\beta) d\Qt{\beta}   }^{2\sfp}
            }^\frac{1}{2\sfp} 
        d\nu
        \\ 
        \leq
        \left(\sfp \frac{2\sfp -1}{2\theta_2} \right)^\frac{1}{2}
        \left(
        \int_{\xi \vee \zeta_1}^{t} 
            \expo{ \left(\theta_1   - \theta_2 \right) \nu }
            \left( \expo{ 2\theta_2 (\nu \wedge \zeta_2)} - \expo{ 2\theta_2 \zeta_1 }  \right)^\frac{1}{2}
        d\nu 
        \right)
        \\
        \times 
        \sup_{\nu \in [\xi,t]}
        \ELaw{}{\norm{S(\nu)}^{2\sfp}}^\frac{1}{2\sfp} 
        \sup_{\nu \in [\zeta_1,\zeta_2]} 
        \ELaw{}{\norm{L(\nu)}^{2\sfp}}^\frac{1}{2\sfp} 
        ,
    \end{multline}
    for all $t \in [\xi,T]$, where we have used the fact that $[\xi \vee \zeta_1,t] \subseteq [\xi,t]$.
    The definition $(\nu \wedge \zeta_2) \doteq \min(\nu,\zeta_2)$ implies that $(\nu \wedge \zeta_2) \leq \nu$ and $(\nu \wedge \zeta_2) \leq \zeta_2$. 
    Hence,
    \begin{align*}
        \left( \expo{2\theta_2(\nu \wedge \zeta_2)} - \expo{2\theta_2\zeta_1}  \right)^\frac{1}{2}
        =
        \expo{\theta_2 (\nu \wedge \zeta_2)}
        \left( 1 - \expo{-2\theta_2((\nu \wedge \zeta_2) - \zeta_1)}  \right)^\frac{1}{2}
        \leq     
        \expo{\theta_2  \nu }
        \left( 1 - \expo{-2\theta_2(\zeta_2 - \zeta_1)}  \right)^\frac{1}{2}
        .
    \end{align*}
    This bound allows us to further develop the following: 
    \begin{align*}
        \int_{\xi \vee \zeta_1}^{t}
            \expo{ \left(\theta_1   - \theta_2 \right) \nu }
            \left( \expo{2\theta_2(\nu \wedge \zeta_2)} - \expo{2\theta_2\zeta_1}  \right)^\frac{1}{2} 
        d\nu
        \leq
        \left(
            \int_{\xi \vee \zeta_1}^{t}
                \expo{ \theta_1 \nu }
            d\nu
        \right) 
        \left( 1 - \expo{-2\theta_2(\zeta_2 - \zeta_1)}  \right)^\frac{1}{2}
        .
    \end{align*}
    Substituting the above bound in~\eqref{eqn:lem:TechnicalResults:NestedSupMoment:Bound:B:1} yields
    \begin{multline}\label{eqn:lem:TechnicalResults:NestedSupMoment:Bound:B:Final}
       \LpLaw{\sfp}{}{N(t)} 
        \leq
        \left(\sfp \frac{2\sfp -1}{2\theta_2} \right)^\frac{1}{2}
        \left(
            \int_{\xi \vee \zeta_1}^{t}
            \expo{ \theta_1 \nu }
            d\nu
        \right)
        \left( 1 - \expo{-2\theta_2(\zeta_2 - \zeta_1)}  \right)^\frac{1}{2}
        \\
        \times 
        \sup_{\nu \in [\xi,t]}
        \ELaw{}{\norm{S(\nu)}^{2\sfp}}^\frac{1}{2\sfp} 
        \sup_{\nu \in [\zeta_1,\zeta_2]} 
        \ELaw{}{\norm{L(\nu)}^{2\sfp}}^\frac{1}{2\sfp} 
        ,
    \end{multline}
    for all $t \in [\xi,T]$.
    Solving the integral on the right hand side  establishes~\eqref{eqn:lem:TechnicalResults:NestedSupMoment:Bound:Final} for $\sfp \in \mathbb{N}_{\geq 2}$, thus completing the proof. 

\end{proof}

The following is a corollary of the last result. 
\begin{corollary}\label{cor:TechnicalResults:NestedSupMoment}
    Consider a complete filtered probability space $\br{\Omega, \mathcal{F}, \mathbb{P}}$ with  filtration $\mathfrak{F}_t$, and let $S \in \mathcal{M}_{2}^{loc}\br{\mathbb{R}^{n_s}|\mathfrak{F}_t}$ and $L \in \mathcal{M}_{2}^{loc}\br{\mathbb{R}^{n_l \times n_q}|\mathfrak{F}_t}$ satisfy 
    \begin{align*}
        \ELaw{}{\int_0^T \left( \norm{S(\nu)}^{2\sfp} + \Frobenius{L(\nu)}^{4\sfp} \right)d\nu } < \infty, \quad \forall \sfp \geq 1.
    \end{align*}
    Additionally, for any $\theta_1,\,\theta_2 \in \mathbb{R}_{> 0}$,  $\xi \in \mathbb{R}_{\geq 0}$, $\BoldTs \in \mathbb{R}_{>0}$, and an $\mathfrak{F}_t$-adapted Brownian motion $Q_t \in \mathbb{R}^{n_q}$ define 
    \begin{align*}
        N_{\cbr{a,b}}(t)
        = 
        \int_{\BoldTs}^t 
            \expo{(\theta_1 - \theta_2) \nu} 
            \norm{S(\nu)}
            \norm{
                R_{\cbr{a,b}}(\nu)
            }
        d\nu \in \mathbb{R}, \quad t \in [\BoldTs,T]
        ,  
    \end{align*}
    where  
    \begin{align*}
        R_a(\nu)    
        =
        L(\nu)
        \varsigma_1(\nu)
        \int_{(\istar{\nu}-1)\BoldTs}^{\istar{\nu}\BoldTs} 
            \expo{\theta_2 \beta}
            \varsigma_2(\beta) 
        dQ_\beta, 
        \quad  
        R_b(\nu)    
        =
        L(\nu)
        \int_{\istar{\nu}\BoldTs}^\nu 
            \expo{\theta_2 \beta} 
        dQ_\beta,
    \end{align*}
    and where $\istar{\nu} \doteq \max\cbr{i \in \mathbb{N}_{\geq 1} \, : \, i\BoldTs \leq \nu} = \left\lfloor \nu / \BoldTs \right\rfloor$.
    Furthermore, $\varsigma_{\cbr{1,2}} \in \mathcal{C} \left( \mathbb{R}_{\geq 0},\mathbb{R}\right) $ are any deterministic functions. 
    Then, for all $(t,\sfp) \in [\BoldTs,T] \times \mathbb{N}_{\geq 1}$:
    \begin{multline}\label{eqn:cor:TechnicalResults:NestedSupMoment:Na:Bound:Final}
        \LpLaw{\sfp}{}{N_a(t)} 
        \leq 
        \left(
            \sfp  (4\sfp-1)
        \right)^\frac{1}{2}  
        \frac{ \expo{ \theta_1 t } - \expo{ \theta_1 \BoldTs } }{ \theta_1 \sqrt{\theta_2} }
        \left( 1 - \expo{ -2\theta_2 \BoldTs } \right)^\frac{1}{2}
        \\ 
        \times
        \sup_{\nu \in [\BoldTs,t]}
        \left(  
            \LpLaw{2\sfp}{}{S(\nu)}
            \LpLaw{4\sfp}{}{L(\nu)}
            \absolute{\varsigma_1(\nu)}
            \sup_{\beta \in [(\istar{\nu}-1)\BoldTs,\istar{\nu}\BoldTs]}
            \absolute{\varsigma_2(\beta)}
        \right)
        , 
    \end{multline}
    and 
    \begin{align}\label{eqn:cor:TechnicalResults:NestedSupMoment:Nb:Bound:Final}
        \LpLaw{\sfp}{}{N_b(t)} 
        \leq 
        \left(
            \sfp (4\sfp-1)
        \right)^\frac{1}{2}  
        \frac{ \expo{ \theta_1 t } - \expo{ \theta_1 \BoldTs } }{ \theta_1 \sqrt{\theta_2} }
        \left( 1 - \expo{ -2\theta_2 \BoldTs } \right)^\frac{1}{2}
        \sup_{\nu \in [\BoldTs,t]}  
            \LpLaw{2\sfp}{}{S(\nu)}
        \sup_{\nu \in [\BoldTs,t]}
            \LpLaw{4\sfp}{}{L(\nu)}
        .
    \end{align} 
    
\end{corollary}
\begin{proof}
     We provide the proof only for the case when $\sfp \in \mathbb{N}_{\geq 2}$. 
     As in the proof of Lemma~\ref{lemma:TechnicalResults:NestedSupMoment}, we have 
    \begin{align}\label{eqn:cor:TechnicalResults:NestedSupMoment:Bound:Partial}
        \LpLaw{\sfp}{}{N_{\cbr{a,b}}(t)} 
        \leq 
        \int_{\BoldTs}^{t} 
            \expo{ \left(\theta_1   - \theta_2 \right) \nu }
            \ELaw{}{\norm{S(\nu)}^{2\sfp}}^\frac{1}{2\sfp}
            \ELaw{}{
                \norm{ R_{\cbr{a,b}}(\nu) }^{2\sfp}
            }^\frac{1}{2\sfp}
        d\nu
        , 
    \end{align}
    for all $(t,\sfp) \in [\BoldTs,T] \times \mathbb{N}_{\geq 2}$.
    Now, it follows from the Cauchy-Schwarz inequality that
    \begin{align*}
        \LpLaw{2\sfp}{}{R_a(\nu) }
        =& 
        \LpLaw{2\sfp}{}{
            L(\nu)
            \varsigma_1(\nu)
            \int_{(\istar{\nu}-1)\BoldTs}^{\istar{\nu}\BoldTs} 
                \expo{\theta_2 \beta}
                \varsigma_2(\beta) 
            dQ_\beta 
        }
        \\
        \leq& 
        \LpLaw{4\sfp}{}{ L(\nu) }
        \absolute{\varsigma_1(\nu)}
        \LpLaw{4\sfp}{}{ 
                \int_{(\istar{\nu}-1)\BoldTs}^{\istar{\nu}\BoldTs} 
                    \expo{\theta_2 \beta}
                    \varsigma_2(\beta) 
                dQ_\beta 
        }
        ,
        \quad  
        \forall \nu \in [\BoldTs,t],
    \end{align*}
    where we have used the fact that $\varsigma_1$ is a deterministic function.
    Using Lemma~\ref{lemma:TechnicalResults:MartingaleMoment} to bound the \ito integral term, we get 
    \begin{align}\label{eqn:cor:TechnicalResults:NestedSupMoment:Bound:B:a:Final}
        \LpLaw{2\sfp}{}{R_a(\nu) }
        &\leq
        \left(\sfp (4\sfp-1)\right)^\frac{1}{2}  
        \expo{ \theta_2  \istar{\nu} \BoldTs }
        \left(\frac{1 - \expo{ -2\theta_2 \BoldTs }}{\theta_2}\right)^\frac{1}{2}
        \LpLaw{4\sfp}{}{L(\nu)}
        \absolute{\varsigma_1(\nu)}
        \sup_{\beta \in [(\istar{\nu}-1)\BoldTs,\istar{\nu}\BoldTs]}
        \absolute{\varsigma_2(\beta)}
        \notag 
        \\
        &\leq
        \left(\sfp (4\sfp-1)\right)^\frac{1}{2}  
        \expo{ \theta_2 \nu  }
        \left(\frac{1 - \expo{ -2\theta_2 \BoldTs }}{\theta_2}\right)^\frac{1}{2}
        \LpLaw{4\sfp}{}{L(\nu)}
        \absolute{\varsigma_1(\nu)}
        \sup_{\beta \in [(\istar{\nu}-1)\BoldTs,\istar{\nu}\BoldTs]}
        \absolute{\varsigma_2(\beta)}
        , 
    \end{align}  
    for all $(t,\sfp) \in [\BoldTs,T] \times  \mathbb{N}_{\geq 2}$, where we have used the fact that $\varsigma_2$ is a deterministic function. Further, the second inequality follows from $\istar{\nu} \BoldTs \leq \nu$, $\forall \nu$.
    Similarly, using the Cauchy-Schwarz inequality and Lemma~\ref{lemma:TechnicalResults:MartingaleMoment}: 
    \begin{align}\label{eqn:cor:TechnicalResults:NestedSupMoment:Bound:B:b:Final}
        \LpLaw{2\sfp}{}{R_b(\nu) }
        \leq&
        \left(\sfp (4\sfp-1)\right)^\frac{1}{2}  
        \expo{ \theta_2 \nu }
        \left(\frac{1 - \expo{ 2\theta_2 (\istar{\nu}\BoldTs - \nu) }}{\theta_2}\right)^\frac{1}{2}
        \LpLaw{4\sfp}{}{L(\nu)}
        \notag 
        \\ 
        \leq&
        \left(\sfp (4\sfp-1)\right)^\frac{1}{2} 
        \expo{ \theta_2 \nu } 
        \left(\frac{1 - \expo{ -2\theta_2 \BoldTs }}{\theta_2}\right)^\frac{1}{2}
        \LpLaw{4\sfp}{}{L(\nu)}
        ,
        \quad 
        \forall (t,\sfp) \in [\BoldTs,T] \times \mathbb{N}_{\geq 2},
    \end{align}
    where we obtain the second inequality by using the fact that $\nu - \istar{\nu}\BoldTs \leq \BoldTs$, $\forall \nu$.
    Substituting~\eqref{eqn:cor:TechnicalResults:NestedSupMoment:Bound:B:a:Final} into~\eqref{eqn:cor:TechnicalResults:NestedSupMoment:Bound:Partial} yields
    \begin{multline*}
        \LpLaw{\sfp}{}{N_a(t)} 
        \leq 
        \left(\sfp (4\sfp-1)\right)^\frac{1}{2}  
        \left(
            \int_{\BoldTs}^{t} 
                \expo{ \theta_1 \nu }
            d\nu
        \right)
        \left(\frac{1 - \expo{ -2\theta_2 \BoldTs }}{\theta_2}\right)^\frac{1}{2}
        \\ 
        \times
        \sup_{\nu \in [\BoldTs,t]}
        \left[  
            \LpLaw{2\sfp}{}{S(\nu)}
            \LpLaw{4\sfp}{}{L(\nu)}
            \absolute{\varsigma_1(\nu)}
            \sup_{\beta \in [(\istar{\nu}-1)\BoldTs,\istar{\nu}\BoldTs]}
            \absolute{\varsigma_2(\beta)}
        \right]
        , 
    \end{multline*}
    for all $(t,\sfp) \in [\BoldTs,T] \times \mathbb{N}_{\geq 2}$.
    Solving the integral on the right hand side then leads to~\eqref{eqn:cor:TechnicalResults:NestedSupMoment:Na:Bound:Final}. 
    Substituting~\eqref{eqn:cor:TechnicalResults:NestedSupMoment:Bound:B:b:Final} into~\eqref{eqn:cor:TechnicalResults:NestedSupMoment:Bound:Partial} and performing the same manipulations as above, we obtain~\eqref{eqn:cor:TechnicalResults:NestedSupMoment:Nb:Bound:Final}.

\end{proof}

Next, we provide a converse result to Lemma~\ref{lemma:TechnicalResults:NestedSupMoment} in which we derive a bound on a particular \ito integral with a Lebesgue integral in the integrand.
\begin{lemma}\label{lemma:TechnicalResults:ConverseNestedSupMoment}
    Given any $S \in \mathcal{M}_{2}^{loc}\br{\mathbb{R}^{n_o}|\mathfrak{F}_t}$, $L \in \mathcal{M}_{2}^{loc}\br{\mathbb{R}^{n_o \times n_q}|\mathfrak{F}_t}$, $\mathfrak{F}_t$-adapted Brownian motion $Q_t \in \mathbb{R}^{n_q}$, and strictly positive constants $\theta_1,\,\theta_2 \in \mathbb{R}_{>0}$, define
    \begin{equation}\label{eqn:lemma:TechnicalResults:ConverseNestedSupMoment:N}
        N(t)
        = 
        \int_0^t 
        \expo{(\theta_1 - \theta_2) \nu}
        \left( \int_0^\nu \expo{\theta_2 \beta} S(\beta)d\beta   \right)^\top
         L(\nu)dQ_\nu
        \in \mathbb{R}, \quad t \in [0,T].  
    \end{equation}
    If 
    \begin{align*}
        \ELaw{}{\int_0^T \left( \norm{S(\nu)}^{2\sfp} + \Frobenius{L(\nu)}^{2\sfp} \right)d\nu } < \infty, \quad \forall \sfp \geq 2
        ,
    \end{align*}
    then
    \begin{align}\label{eqn:lemma:TechnicalResults:ConverseNestedSupMoment:Bound:Final}
        \LpLaw{\sfp}{}{N(t)} 
        \leq
        \left(
            \sfp (\sfp - 1)
        \right)^\frac{1}{2} 
        \frac{ \expo{ \theta_1 t }}{ 2 \sqrt{\theta_1} \theta_2}
        \sup_{\nu \in [0,t]} 
        \LpLaw{2\sfp}{}{S(\nu)}
        \sup_{\nu \in [0,t]}
        \LpLaw{2\sfp}{}{L(\nu)}
        , 
        \quad 
        \forall (t,\sfp) \in [0,T] \times \mathbb{N}_{\geq 2}. 
    \end{align}
\end{lemma}
\begin{proof}
    The proof for $\sfp = 2$ follows by using ~\cite[Thm.~1.5.21]{mao2007stochastic} and following the proof for $\sfp > 2$ \emph{mutatis mutandis}, which we present below.
    We begin by noting that the inner Lebesgue integral is absolute continuous on $[0,T]$ by fundamental theorem for the Lebesgue integral~\cite[Thm.~6.4.1]{burk2007garden}, and $S \in \mathcal{M}_{2}^{loc}\br{\mathbb{R}^{n_o}|\mathfrak{F}_t}$ implies that
    \begin{align*}
        \int_0^t \expo{\theta_2 \beta} S(\beta)d\beta \in \mathcal{M}_{2}^{loc}\br{\mathbb{R}^{n_o}|\mathfrak{F}_t}.
    \end{align*}
    It follows then from~\cite[Thm.~5.5.2]{kuo2006introduction} that $N(t)$ is a continuous local martingale with respect to the filtration $\mathfrak{F}_t$ since $L \in \mathcal{M}_{2}^{loc}\br{\mathbb{R}^{n_o \times n_q}|\mathfrak{F}_t}$.
    Then, applying \ito's lemma~\cite[Thm.~1.6.4]{mao2007stochastic} and using~\cite[Thm.~1.5.21]{mao2007stochastic} produces 
    \begin{multline}\label{eqn:lemma:TechnicalResults:ConverseNestedSupMoment:IncreasingN}
        \absolute{N(t)}^\sfp 
        = 
        \frac{\sfp}{2} 
        \int_0^t 
            \expo{2(\theta_1 - \theta_2) \nu} 
            \absolute{N(\nu)}^{\sfp-2}
            \norm{ L(\nu)^\top \left( \int_0^\nu \expo{\theta_2 \beta} S(\beta)d\beta   \right)
            }^2
        d\nu
        \\ 
        + 
        \frac{\sfp}{2}(\sfp - 2) 
        \int_0^t 
            \expo{2(\theta_1 - \theta_2) \nu} 
            \absolute{N(\nu)}^{\sfp-4}
            \norm{N(\nu) L(\nu)^\top \left( \int_0^\nu \expo{\theta_2 \beta} S(\beta)d\beta   \right) }^2
        d\nu
        ,
    \end{multline}
    and therefore 
    \begin{align*}
        \absolute{N(t)}^\sfp 
        \leq  
        \frac{\sfp}{2}(\sfp - 1) 
        \int_0^t 
            \expo{2(\theta_1 - \theta_2) \nu} 
            \absolute{N(\nu)}^{\sfp-2}
            \Frobenius{L(\nu)}^2
            \norm{\int_0^\nu \expo{\theta_2 \beta} S(\beta)d\beta }^2
        d\nu
        ,
        \quad \forall t \in [0,T].
    \end{align*}
    Using this bound, we obtain 
    \begin{align*}
        \ELaw{}{\absolute{N(t)}^\sfp} 
        =& 
        \frac{\sfp}{2}(\sfp - 1) 
        \ELaw{}{
            \int_0^t 
                \expo{2(\theta_1 - \theta_2) \nu} 
                \absolute{N(\nu)}^{\sfp-2}
                \Frobenius{L(\nu)}^2
                \norm{\int_0^\nu \expo{\theta_2 \beta} S(\beta)d\beta }^2
            d\nu
        }
        \notag 
        \\
        \overset{(\romannum{1})}{=}
        & 
        \frac{\sfp}{2}(\sfp - 1) 
        \ELaw{}{
        \int_0^{t} 
        \left(
            \expo{ 2\theta_1 (\sfp-2) \nu /\sfp   }
            \absolute{N(\nu)}^{\sfp-2} 
        \right)
        \left(
            \expo{ 2\left(2\theta_1 /\sfp  - \theta_2 \right) \nu }
            \Frobenius{L(\nu)}^2
            \norm{\int_0^\nu \expo{\theta_2 \beta} S(\beta)d\beta }^2
        \right)
        d\nu
        }
        \notag 
        \\
        \overset{(\romannum{2})}{\leq}
        &
        \frac{\sfp}{2}(\sfp - 1) 
        \ELaw{}{
            \int_0^{t} 
                \expo{ 2 \theta_1  \nu }
                \absolute{N(\nu)}^\sfp  
            d\nu
        }^\frac{\sfp-2}{\sfp}
        \ELaw{}{
            \int_0^{t} 
                \expo{ \left(2\theta_1  - \theta_2 \sfp \right) \nu }
                \Frobenius{L(\nu)}^\sfp
                \norm{\int_0^\nu \expo{\theta_2 \beta} S(\beta)d\beta }^\sfp
            d\nu
        }^\frac{2}{\sfp}, 
    \end{align*}
    for all $t \in [0,T]$, where we obtain $(\romannum{1})$ by writing 
    \begin{align*}
        \expo{ 2(\theta_1 - \theta_2) \nu }
        =
        \expo{ 2\left[\theta_1 \left( \frac{\sfp-2}{\sfp} + \frac{2}{\sfp} \right) - \theta_2 \right] \nu }
        =
        \expo{ 2\theta_1 (\sfp-2) \nu /\sfp   }
        \expo{ 2\left(2\theta_1 /\sfp  - \theta_2 \right) \nu },
    \end{align*} 
    and $(\romannum{2})$ follows from H\"{o}lder's inequality with conjugates $\sfp / (\sfp - 2)$ and $\sfp / 2$.
    Applying Fubini's theorem, followed by the Cauchy-Schwarz inequality then leads to 
    \begin{multline}\label{eqn:lemma:TechnicalResults:ConverseNestedSupMoment:Bound:1}
        \begin{aligned}
            \ELaw{}{\absolute{N(t)}^\sfp} 
            \leq & 
            \frac{\sfp}{2}(\sfp - 1) 
            \left(
                \int_0^{t} 
                    \expo{ 2 \theta_1  \nu }
                    \ELaw{}{\absolute{N(\nu)}^\sfp}  
                d\nu
            \right)^\frac{\sfp-2}{\sfp}
            \left(
                \int_0^{t} 
                    \expo{ \left(2\theta_1  - \theta_2 \sfp \right) \nu }
                    \ELaw{}{
                        \Frobenius{L(\nu)}^\sfp
                        \norm{\int_0^\nu \expo{\theta_2 \beta} S(\beta)d\beta }^\sfp
                    }
                d\nu
            \right)^\frac{2}{\sfp}
            \\
            \leq & 
            \frac{\sfp}{2}(\sfp - 1) 
            \left(
                \int_0^{t} 
                    \expo{ 2 \theta_1  \nu }
                    \ELaw{}{\absolute{N(\nu)}^\sfp}  
                d\nu
            \right)^\frac{\sfp-2}{\sfp}
        \end{aligned}
        \\
        \times 
        \left(
            \int_0^{t} 
                \expo{ \left(2\theta_1  - \theta_2 \sfp \right) \nu }
                \ELaw{}{
                    \Frobenius{L(\nu)}^{2\sfp}
                }^\frac{1}{2}
                \ELaw{}{
                    \norm{\int_0^\nu \expo{\theta_2 \beta} S(\beta)d\beta }^{2\sfp}
                }^\frac{1}{2}
            d\nu
        \right)^\frac{2}{\sfp}
        , 
    \end{multline}
    for all $t \in [0,T]$.

    Using~\eqref{eqn:lemma:TechnicalResults:ConverseNestedSupMoment:IncreasingN}, one sees that $\ELaw{}{\absolute{N(t)}^\sfp}$ is non-decreasing over $t \in [0,T]$, and thus 
    \begin{align}\label{eqn:lemma:TechnicalResults:ConverseNestedSupMoment:Bound:A:Final}
        \left(
            \int_0^{t} 
                \expo{ 2 \theta_1  \nu }
                \ELaw{}{\absolute{N(\nu)}^\sfp}  
            d\nu
        \right)^\frac{\sfp-2}{\sfp}
        \leq     
        \left(
            \int_0^{t} 
                \expo{ 2 \theta_1  \nu }  
            d\nu
        \right)^\frac{\sfp-2}{\sfp}
        \ELaw{}{\absolute{N(t)}^\sfp}^\frac{\sfp-2}{\sfp} 
        , 
        \quad \forall t \in [0,T].
    \end{align}
    
    Next, we use Proposition~\ref{prop:TechnicalResults:LebesgueMoment} for $\cbr{\theta,\xi,n_s} = \cbr{\theta_2,0,n_o}$ and obtain  
    \begin{align*}
        \ELaw{}{
            \norm{\int_0^\nu \expo{\theta_2 \beta} S(\beta)d\beta }^{2\sfp}
        }^\frac{1}{2}
        \leq     
        \ELaw{}{
            \left(\int_0^\nu \expo{\theta_2 \beta} \norm{S(\beta)}d\beta \right)^{2\sfp}
        }^\frac{1}{2}
        \leq&     
        \left( \frac{ \expo{\theta_2\nu} - 1 }{\theta_2} \right)^\sfp
        \sup_{\beta \in [0,\nu]} 
        \ELaw{}{\norm{S(\beta)}^{2\sfp}}^\Half
        \\
        \leq&     
        \frac{ \expo{\theta_2 \sfp \nu} }{\theta_2^\sfp }
        \sup_{\beta \in [0,\nu]} 
        \ELaw{}{\norm{S(\beta)}^{2\sfp}}^\Half
        , 
        \quad \forall \nu \in [0,t], 
    \end{align*}
    where we have bounded $\expo{\theta_2\nu} - 1 $ by $\expo{\theta_2\nu}$.
    Therefore, 
    \begin{multline}\label{eqn:lemma:TechnicalResults:ConverseNestedSupMoment:Bound:B:Final}
        \left(
            \int_0^{t} 
                \expo{ \left(2\theta_1  - \theta_2 \sfp \right) \nu }
                \ELaw{}{
                    \Frobenius{L(\nu)}^{2\sfp}
                }^\frac{1}{2}
                \ELaw{}{
                    \norm{\int_0^\nu \expo{\theta_2 \beta} S(\beta)d\beta }^{2\sfp}
                }^\frac{1}{2}
            d\nu
        \right)^\frac{2}{\sfp}
        \\
        \leq 
        \left( \frac{1}{\theta_2} \right)^2
        \left(
            \int_0^{t} 
                \expo{ 2\theta_1   \nu }
                \ELaw{}{
                    \Frobenius{L(\nu)}^{2\sfp}
                }^\frac{1}{2}
                \sup_{\beta \in [0,\nu]} 
                \ELaw{}{\norm{S(\beta)}^{2\sfp}}^\Half
            d\nu
        \right)^\frac{2}{\sfp}
        \\
        \leq 
        \left( \frac{1}{\theta_2} \right)^2
        \left(
            \int_0^{t} 
                \expo{ 2\theta_1   \nu }
            d\nu
        \right)^\frac{2}{\sfp}
        \sup_{\nu \in [0,t]}
        \ELaw{}{\Frobenius{L(\nu)}^{2\sfp}}^\frac{1}{\sfp}
        \sup_{\nu \in [0,t]} 
        \ELaw{}{\norm{S(\nu)}^{2\sfp}}^\frac{1}{\sfp}
        ,
    \end{multline}
    for all $t \in [0,T]$.
    Substituting~\eqref{eqn:lemma:TechnicalResults:ConverseNestedSupMoment:Bound:A:Final} and~\eqref{eqn:lemma:TechnicalResults:ConverseNestedSupMoment:Bound:B:Final} into~\eqref{eqn:lemma:TechnicalResults:ConverseNestedSupMoment:Bound:1} yields
    \begin{align*}
        \ELaw{}{\absolute{N(t)}^\sfp} 
        \leq&   
        \ELaw{}{\absolute{N(t)}^\sfp}^\frac{\sfp-2}{\sfp}
        \sfp \frac{(\sfp - 1)}{2} 
        \frac{1}{\theta_2^2}
        \left(
            \int_0^{t} 
                \expo{ 2\theta_1   \nu }
            d\nu
        \right)
        \sup_{\nu \in [0,t]}
        \ELaw{}{\Frobenius{L(\nu)}^{2\sfp}}^\frac{1}{\sfp}
        \sup_{\nu \in [0,t]} 
        \ELaw{}{\norm{S(\nu)}^{2\sfp}}^\frac{1}{\sfp}
        \\ 
        \leq&   
        \ELaw{}{\absolute{N(t)}^\sfp}^\frac{\sfp-2}{\sfp}
        \sfp \frac{(\sfp - 1)}{4} 
        \frac{ \expo{ 2\theta_1 t }}{\theta_1 \theta_2^2}
        \sup_{\nu \in [0,t]}
        \ELaw{}{\Frobenius{L(\nu)}^{2\sfp}}^\frac{1}{\sfp}
        \sup_{\nu \in [0,t]} 
        \ELaw{}{\norm{S(\nu)}^{2\sfp}}^\frac{1}{\sfp}
        , 
    \end{align*}
    for all $t \in [0,T]$, where we have bounded $ \expo{ 2\theta_1   t } - 1$ by $\expo{ 2\theta_1   t }$.
    This bound then implies that 
    \begin{align*}
        \ELaw{}{\absolute{N(t)}^\sfp}^\frac{2}{\sfp} 
        \leq&   
        \sfp  \frac{(\sfp - 1)}{4} 
        \frac{ \expo{ 2\theta_1 t }}{\theta_1 \theta_2^2}
        \sup_{\nu \in [0,t]}
        \ELaw{}{\Frobenius{L(\nu)}^{2\sfp}}^\frac{1}{\sfp}
        \sup_{\nu \in [0,t]} 
        \ELaw{}{\norm{S(\nu)}^{2\sfp}}^\frac{1}{\sfp}
        , 
        \quad 
        \forall t \in [0,T]
        ,
    \end{align*}
    thus establishing~\eqref{eqn:lemma:TechnicalResults:ConverseNestedSupMoment:Bound:Final} for $\sfp > 2$.


\end{proof}

We need the following corollary to the last result for Lemma~\ref{lemma:TechnicalResults:NestedSupMoment:Ito}.
\begin{corollary}\label{cor:TechnicalResults:ConverseNestedSupMoment}
    Consider a complete filtered probability space $\br{\Omega, \mathcal{F}, \mathbb{P}}$ with  filtration $\mathfrak{F}_t$, and let $L_1,\, L_2 \in \mathcal{M}_{2}^{loc}\br{\mathbb{R}^{n_l \times n_q}|\mathfrak{F}_t}$ satisfy 
    \begin{align*}
        \ELaw{}{\int_0^T \Frobenius{L_1(\nu) + L_2(\nu)}^{2\sfp}d\nu } < \infty, \quad \forall \sfp \geq 2.
    \end{align*}
    Additionally, for any strictly positive constants $\theta_1,\,\theta_2 \in \mathbb{R}_{>0}$ and an $\mathfrak{F}_t$-adapted Brownian motion $Q_t \in \mathbb{R}^{n_q}$ define 
    \begin{align*}
        \breve{N}(t)
        =
        \int_0^t 
            \expo{(\theta_1 - \theta_2) \nu} \left( \int_0^\nu \expo{\theta_2 \beta} L_2(\beta) dQ_\beta \right)^\top L_1(\nu)
        dQ_\nu \in \mathbb{R}.
    \end{align*}
    Then,  
    \begin{align}\label{eqn:cor:TechnicalResults:ConverseNestedSupMoment:Bound:Final}
        \LpLaw{\sfp}{}{\breve{N}(t)} 
        \leq
        \frac{\sfp \sqrt{(\sfp - 1)(2\sfp - 1)}}{2 \sqrt{2}} 
        \frac{\expo{ \theta_1 t }}{\sqrt{\theta_1 \theta_2}}
        \sup_{\nu \in [0,t]} 
        \LpLaw{2\sfp}{}{L_1(\nu)}
        \sup_{\nu \in [0,t]}
        \LpLaw{2\sfp}{}{L_2(\nu)}
        , 
        \quad 
        \forall (t,\sfp) \in [0,T] \times \mathbb{N}_{\geq 2}. 
    \end{align}
\end{corollary}
\begin{proof}
    As in the proof of Lemma~\ref{lemma:TechnicalResults:ConverseNestedSupMoment}, we provide a proof for $\sfp > 2$ only since the proof for $\sfp = 2$ follows \emph{mutatis mutandis} by using~\cite[Thm.~1.5.21]{mao2007stochastic}. 
    For $\sfp > 2$, applying \ito's lemma~\cite[Thm.~1.6.4]{mao2007stochastic} and using~\cite[Thm.~1.5.21]{mao2007stochastic} produces 
    \begin{multline}\label{eqn:cor:TechnicalResults:ConverseNestedSupMoment:IncreasingN}
        \absolute{\breve{N}(t)}^\sfp 
        = 
        \frac{\sfp}{2} 
        \int_0^t 
            \expo{2(\theta_1 - \theta_2) \nu} 
            \absolute{\breve{N}(\nu)}^{\sfp-2}
            \norm{ L_1(\nu)^\top \left( \int_0^\nu \expo{\theta_2 \beta} L_2(\beta)d\Qt{\beta}   \right)
            }^2
        d\nu
        \\ 
        + 
        \frac{\sfp}{2}(\sfp - 2) 
        \int_0^t 
            \expo{2(\theta_1 - \theta_2) \nu} 
            \absolute{\breve{N}(\nu)}^{\sfp-4}
            \norm{\breve{N}(\nu) L_1(\nu)^\top \left( \int_0^\nu \expo{\theta_2 \beta} L_2(\beta)d\Qt{\beta}   \right) }^2
        d\nu
        ,
    \end{multline}
    and therefore 
    \begin{align*}
        \absolute{\breve{N}(t)}^\sfp 
        \leq  
        \frac{\sfp}{2}(\sfp - 1) 
        \int_0^t 
            \expo{2(\theta_1 - \theta_2) \nu} 
            \absolute{\breve{N}(\nu)}^{\sfp-2}
            \Frobenius{L_1(\nu)}^2
            \norm{\int_0^\nu \expo{\theta_2 \beta} L_2(\beta)d\Qt{\beta}}^2
        d\nu
        ,
        \quad \forall t \in [0,T].
    \end{align*}
    Following the identical analysis that led to~\eqref{eqn:lemma:TechnicalResults:ConverseNestedSupMoment:Bound:1} in the proof of Lemma~\ref{lemma:TechnicalResults:ConverseNestedSupMoment}, we obtain
    \begin{multline}\label{eqn:prop:TechnicalResults:ConverseNestedSupMoment:Bound:1}
        \ELaw{}{\absolute{\breve{N}(t)}^\sfp} 
        \leq 
        \frac{\sfp}{2}(\sfp - 1) 
        \left(
            \int_0^{t} 
                \expo{ 2 \theta_1  \nu }
                \ELaw{}{\absolute{\breve{N}(\nu)}^\sfp}  
            d\nu
        \right)^\frac{\sfp-2}{\sfp}
        \\
        \times 
        \left(
            \int_0^{t} 
                \expo{ \left(2\theta_1  - \theta_2 \sfp \right) \nu }
                \ELaw{}{
                    \Frobenius{L_1(\nu)}^{2\sfp}
                }^\frac{1}{2}
                \ELaw{}{
                    \norm{
                        \int_0^\nu \expo{\theta_2 \beta} L_2(\beta)d\Qt{\beta}
                    }^{2\sfp}
                }^\frac{1}{2}
            d\nu
        \right)^\frac{2}{\sfp}
        , 
    \end{multline}
    for all $t \in [0,T]$.
    Using~\eqref{eqn:cor:TechnicalResults:ConverseNestedSupMoment:IncreasingN}, one sees that $\ELaw{}{\absolute{\breve{N}(t)}^\sfp}$ is non-decreasing over $t \in [0,T]$, and thus 
    \begin{align}\label{eqn:cor:TechnicalResults:ConverseNestedSupMoment:Bound:A:Final}
        \left(
            \int_0^{t} 
                \expo{ 2 \theta_1  \nu }
                \ELaw{}{\absolute{\breve{N}(\nu)}^\sfp}  
            d\nu
        \right)^\frac{\sfp-2}{\sfp}
        \leq     
        \left(
            \int_0^{t} 
                \expo{ 2 \theta_1  \nu }  
            d\nu
        \right)^\frac{\sfp-2}{\sfp}
        \ELaw{}{\absolute{\breve{N}(t)}^\sfp}^\frac{\sfp-2}{\sfp} 
        , 
        \quad \forall t \in [0,T].
    \end{align}

    Next, we use Lemma~\ref{lemma:TechnicalResults:MartingaleMoment} for $\cbr{\theta,\xi} = \cbr{\theta_2,0}$ and obtain  
    \begin{align*}
        \ELaw{}{
            \norm{
                \int_0^\nu \expo{\theta_2 \beta} L_2(\beta)d\Qt{\beta}
            }^{2\sfp}
        }^\frac{1}{2}
        \leq&
        \left( \sfp \frac{2\sfp - 1 }{2} \right)^\frac{\sfp}{2}
        \left( \frac{\expo{2\theta_2 \nu}}{\theta_2} \right)^\frac{\sfp}{2}
        \sup_{\beta \in [0,\nu]} 
        \ELaw{}{\Frobenius{L_2(\beta)}^{2\sfp}}^\Half
        , 
        \quad \forall \nu \in [0,t], 
    \end{align*}
    where we have bounded $\expo{2\theta_2\nu} - 1 $ by $\expo{2\theta_2\nu}$.
    Therefore, 
    \begin{multline}\label{eqn:cor:TechnicalResults:ConverseNestedSupMoment:Bound:B:Final}
        \left(
            \int_0^{t} 
                \expo{ \left(2\theta_1  - \theta_2 \sfp \right) \nu }
                \ELaw{}{
                    \Frobenius{L_1(\nu)}^{2\sfp}
                }^\frac{1}{2}
                \ELaw{}{
                    \norm{
                        \int_0^\nu \expo{\theta_2 \beta} L_2(\beta)d\Qt{\beta}
                    }^{2\sfp}
                }^\frac{1}{2}
            d\nu
        \right)^\frac{2}{\sfp}
       \\ 
       \leq    
       \sfp \frac{2\sfp - 1 }{2 \theta_2}
       \left(
            \int_0^{t} 
                \expo{ 2\theta_1 \nu }
                \ELaw{}{
                    \Frobenius{L_1(\nu)}^{2\sfp}
                }^\frac{1}{2}
                \sup_{\beta \in [0,\nu]} 
                \ELaw{}{\Frobenius{L_2(\beta)}^{2\sfp}}^\Half
            d\nu
        \right)^\frac{2}{\sfp}
        \\ 
       \leq    
       \sfp \frac{2\sfp - 1 }{2 \theta_2}
       \left(
            \int_0^{t} 
                \expo{ 2\theta_1 \nu }
            d\nu
        \right)^\frac{2}{\sfp}
        \sup_{\nu \in [0,t]}
            \ELaw{}{\Frobenius{L_1(\nu)}^{2\sfp}}^\frac{1}{\sfp}
        \sup_{\nu \in [0,t]} 
            \ELaw{}{\Frobenius{L_2(\nu)}^{2\sfp}}^\frac{1}{\sfp}
        ,
    \end{multline}
    for all $t \in [0,T]$.
    Substituting~\eqref{eqn:cor:TechnicalResults:ConverseNestedSupMoment:Bound:A:Final} and~\eqref{eqn:cor:TechnicalResults:ConverseNestedSupMoment:Bound:B:Final} into~\eqref{eqn:prop:TechnicalResults:ConverseNestedSupMoment:Bound:1} yields
    \begin{align*}
        \ELaw{}{\absolute{\breve{N}(t)}^\sfp} 
        \leq& 
        \ELaw{}{\absolute{\breve{N}(t)}^\sfp}^\frac{\sfp-2}{\sfp} 
        \frac{\sfp^2 (\sfp - 1)(2\sfp - 1) }{4 \theta_2} 
       \left(
            \int_0^{t} 
                \expo{ 2\theta_1 \nu }
            d\nu
        \right)
        \sup_{\nu \in [0,t]}
            \ELaw{}{\Frobenius{L_1(\nu)}^{2\sfp}}^\frac{1}{\sfp}
        \sup_{\nu \in [0,t]} 
            \ELaw{}{\Frobenius{L_2(\nu)}^{2\sfp}}^\frac{1}{\sfp}
        \\ 
        \leq& 
        \ELaw{}{\absolute{\breve{N}(t)}^\sfp}^\frac{\sfp-2}{\sfp} 
        \frac{1}{8}
        \sfp^2 (\sfp - 1)(2\sfp - 1)  
        \frac{\expo{ 2\theta_1 t }}{\theta_1 \theta_2}
        \sup_{\nu \in [0,t]}
            \ELaw{}{\Frobenius{L_1(\nu)}^{2\sfp}}^\frac{1}{\sfp}
        \sup_{\nu \in [0,t]} 
            \ELaw{}{\Frobenius{L_2(\nu)}^{2\sfp}}^\frac{1}{\sfp}
        , 
    \end{align*}
    for all $t \in [0,T]$, where we have bounded $\expo{ 2\theta_1 t } - 1$ by $\expo{ 2\theta_1 t }$.
    This bound then implies that 
    \begin{align*}
        \ELaw{}{\absolute{\breve{N}(t)}^\sfp}^\frac{2}{\sfp} 
        \leq 
        \frac{1}{8}
        \sfp^2 (\sfp - 1)(2\sfp - 1)  
        \frac{\expo{ 2\theta_1 t }}{\theta_1 \theta_2}
        \sup_{\nu \in [0,t]}
            \ELaw{}{\Frobenius{L_1(\nu)}^{2\sfp}}^\frac{1}{\sfp}
        \sup_{\nu \in [0,t]} 
            \ELaw{}{\Frobenius{L_2(\nu)}^{2\sfp}}^\frac{1}{\sfp}
        , 
        \quad \forall t \in [0,T],
    \end{align*} 
    thus establishing~\eqref{eqn:cor:TechnicalResults:ConverseNestedSupMoment:Bound:Final} for $\sfp > 2$.


\end{proof}

The following result provides a bound on a class of nested \ito integrals.
\begin{lemma}\label{lemma:TechnicalResults:NestedSupMoment:Ito}
    Consider a complete filtered probability space $\br{\Omega, \mathcal{F}, \mathbb{P}}$ with  filtration $\mathfrak{F}_t$, and let $L_1,\, L_2 \in \mathcal{M}_{2}^{loc}\br{\mathbb{R}^{n_l \times n_q}|\mathfrak{F}_t}$ satisfy 
    \begin{align}\label{eqn:lemma:TechnicalResults:NestedSupMoment:Ito:SquareIntegrability}
        \ELaw{}{\int_0^T \Frobenius{L_1(\nu) + L_2(\nu)}^{2\sfp}d\nu } < \infty, \quad \forall \sfp \geq 2.
    \end{align}
    Additionally, for any strictly positive constants $\theta_1,\,\theta_2 \in \mathbb{R}_{>0}$ and an $\mathfrak{F}_t$-adapted Brownian motion $Q_t \in \mathbb{R}^{n_q}$ define 
    \begin{equation*}
        N(t)
        = 
        \int_0^t 
            \expo{\theta_2 \nu}
            \left(
                \int_\nu^t 
                    \expo{(\theta_1 - \theta_2) \beta} L_1(\beta)
                dQ_\beta 
            \right)^\top  
            L_2(\nu)
        dQ_\nu 
        \in \mathbb{R},
        \quad t \in [0,T].  
    \end{equation*}
    Then, 
    \begin{align}\label{eqn:lemma:TechnicalResults:NestedSupMoment:Ito:Bound:p=1:Final}
        \ELaw{}{N(t)}
        \leq& 
        \sqrt{n_l}
        \frac{\expo{\theta_1 t}}{\theta_1} 
        \sup_{\nu \in [0,t]}
        \ELaw{}{\Frobenius{L_1(\nu)L_2(\nu)^\top}} 
        , 
        \quad \forall t \in [0,T],
    \end{align}
    and 
    \begin{multline}\label{eqn:lemma:TechnicalResults:NestedSupMoment:Ito:Bound:p>=2:Final}
        \LpLaw{\sfp}{}{N(t)}
        \leq 
        \frac{1}{\sqrt{2}}
        \left(\sfp \frac{\sfp-1}{2}\right)^\frac{1}{2}
        \left(\sfp \frac{2\sfp-1}{2}\right)^\frac{1}{2}
        \frac{\expo{ \theta_1 t }}{\sqrt{\theta_1 \theta_2}}
        \sup_{\nu \in [0,t]} 
        \LpLaw{2\sfp}{}{L_1(\nu)}
        \sup_{\nu \in [0,t]}
        \LpLaw{2\sfp}{}{L_2(\nu)}
        \\
        +
        \sqrt{n_l}
        \frac{\expo{\theta_1 t}}{\theta_1} 
        \sup_{\nu \in [0,t]}
        \LpLaw{\sfp}{}{L_1(\nu)L_2(\nu)^\top}
        , 
    \end{multline}
    for all $(t,\sfp) \in [0,T] \times \mathbb{N}_{\geq 2}$.

\end{lemma}
\begin{proof}
    We begin by setting 
    \begin{align*}
        N_1(t) = \int_0^t  \expo{ (\theta_1 - \theta_2) \nu} L_1(\nu)dQ_\nu \in \mathbb{R}^{n_l} ,
        \quad 
        N_2(t) = \int_0^t  \expo{ \theta_2 \nu} L_2(\nu)dQ_\nu \in \mathbb{R}^{n_l}, \quad t \in [0,T].
    \end{align*}
    It then follows from the definition of $N(t)$ that 
    \begin{align*}
        N(t)
        = 
        \int_0^t 
            \expo{\theta_2 \nu}
            \left(
                \int_\nu^t 
                    \expo{(\theta_1-\theta_2) \beta} L_1(\beta)
                dQ_\beta 
            \right)^\top  
            L_2(\nu)
        dQ_\nu
        =&
        \int_0^t \left(\int_\nu^t dN_1(\beta)\right)^\top dN_2(\nu).
    \end{align*}
    Hence, 
    \begin{multline}\label{eqn:lemma:TechnicalResults:NestedSupMoment:Ito:Decomposition}
        N(t)
        =
        \int_0^t \left(\int_0^t dN_1(\beta)\right)^\top dN_2(\nu)
        -
        \int_0^t \left(\int_0^\nu dN_1(\beta)\right)^\top dN_2(\nu)
        \\ 
        =
        N_1(t)^\top N_2(t) - \int_0^t N_1(\nu)^\top dN_2(\nu), \quad t \in [0,T].
    \end{multline}
    Next, applying \ito's lemma to $N_1(t)^\top N_2(t)$ (or alternatively using the \ito product rule~\cite[Sec.~4.4.2]{evans2012introduction} applied element wise) produces 
    \begin{align*}
        N_1(t)^\top N_2(t) 
        = 
        \int_0^t N_1(\nu)^\top dN_2(\nu)
        +
        \int_0^t N_2(\nu)^\top dN_1(\nu)
        +
        \sum_{i=1}^{n_l}\CrossVar{N_{1,i}}{N_{2,i}}_t, \quad t \in [0,T],
    \end{align*} 
    where $N_{1,i},\, N_{2,i} \in \mathbb{R}$ denote the $i^{th}$ scalar-valued process, $i \in \cbr{1,\dots,n_l}$, of $N_1(t)$ and $N_2(t)$, respectively, and $\CrossVar{N_{1,i}}{N_{2,i}}_t$ denotes the cross-variation process between $N_{1,i}(t)$ and $N_{2,i}(t)$~\cite[Defn.~2.3.9]{durrett1996stochastic}.
    Substituting the above expression for $N_1(t)^\top N_2(t)$ into~\eqref{eqn:lemma:TechnicalResults:NestedSupMoment:Ito:Decomposition} produces
    \begin{align*}
        N(t)
        =&
        \int_0^t N_2(\nu)^\top dN_1(\nu)
        +
        \sum_{i=1}^{n_l}\CrossVar{N_{1,i}}{N_{2,i}}_t, \quad t \in [0,T].
    \end{align*}
    It then follows from the linearity of the expectation operator and the Minkowski's inequality that
    \begin{align}\label{eqn:lemma:TechnicalResults:NestedSupMoment:Ito:Moments:p=1:Decomposition}
        \ELaw{}{N(t)}
        =& 
        \ELaw{}{
            \int_0^t N_2(\nu)^\top dN_1(\nu)
        }
        +
        \ELaw{}{
            \sum_{i=1}^{n_l}\CrossVar{N_{1,i}}{N_{2,i}}_t
        }
        \notag 
        \\ 
        \leq & 
        \ELaw{}{
            \int_0^t N_2(\nu)^\top dN_1(\nu)
        }
        +
        \ELaw{}{
            \absolute{\sum_{i=1}^{n_l}\CrossVar{N_{1,i}}{N_{2,i}}_t}
        }
        , 
        \quad \forall t \in [0,T],
    \end{align}
    and 
    \begin{align}\label{eqn:lemma:TechnicalResults:NestedSupMoment:Ito:Moments:p>=2:Decomposition}
        \ELaw{}{\absolute{N(t)}^\sfp}^\frac{1}{\sfp}
        \leq 
        \ELaw{}{
            \absolute{\int_0^t N_2(\nu)^\top dN_1(\nu)}^\sfp
        }^\frac{1}{\sfp}
        +
        \ELaw{}{
            \absolute{\sum_{i=1}^{n_l}\CrossVar{N_{1,i}}{N_{2,i}}_t}^\sfp
        }^\frac{1}{\sfp}
        , 
        \quad \forall (t,\sfp) \in [0,T] \times \mathbb{N}_{\geq 2}
        .
    \end{align}

    Now, let
    \begin{align*}
        \breve{N}(t)
        \doteq 
        \int_0^t N_2(\nu)^\top dN_1(\nu)
        =
        \int_0^t 
            \expo{(\theta_1 - \theta_2) \nu} \left( \int_0^\nu \expo{\theta_2 \beta} L_2(\beta) dQ_\beta \right)^\top L_1(\nu)
        dQ_\nu \in \mathbb{R}.
    \end{align*}
    The $t$-continuity of $N_2(t)$~\cite[Thm.~3.2.5]{oksendal2013stochastic}, along with $L_1,\, L_2 \in \mathcal{M}_{2}^{loc}\br{\mathbb{R}^{n_l \times n_q}|\mathfrak{F}_t}$ imply that $\breve{N}(t)$ is a continuous local martingale~\cite[Thm.~5.5.2]{kuo2006introduction}.
    Hence, it follows from~\cite[Thm.~1.5.21]{mao2007stochastic} that 
    \begin{align}\label{eqn:lemma:TechnicalResults:NestedSupMoment:Ito:Moments:Bound:p=1:A}
        \ELaw{}{
            \breve{N}(t)
        }
        =
        \ELaw{}{
            \int_0^t N_2(\nu)^\top dN_1(\nu)
        }
        =0, \quad \forall t \in [0,T]
        .
    \end{align}
    Furthermore, from Corollary~\ref{cor:TechnicalResults:ConverseNestedSupMoment}, we have that  
    \begin{align}\label{eqn:lemma:TechnicalResults:NestedSupMoment:Ito:Moments:Bound:p>=2:A}
        \ELaw{}{\absolute{\breve{N}(t)}^\sfp}^\frac{1}{\sfp}
        =&
        \ELaw{}{
            \absolute{\int_0^t N_2(\nu)^\top dN_1(\nu)}^\sfp
        }^\frac{1}{\sfp}
        \notag 
        \\
        \leq&
        \frac{\sfp \sqrt{(\sfp - 1)(2\sfp - 1)}}{2 \sqrt{2}} 
        \frac{\expo{ \theta_1 t }}{\sqrt{\theta_1 \theta_2}}
        \sup_{\nu \in [0,t]} 
        \LpLaw{2\sfp}{}{L_1(\nu)}
        \sup_{\nu \in [0,t]}
        \LpLaw{2\sfp}{}{L_2(\nu)}
        , 
    \end{align}
    for all $(t,\sfp) \in [0,T] \times \mathbb{N}_{\geq 2}$.

    Next, it follows from the definition of the cross-variation process~\cite[Defn.~2.3.9]{durrett1996stochastic} that  
    \begin{align*}
        \sum_{i=1}^{n_l}\CrossVar{N_{1,i}}{N_{2,i}}_t
        =
        \Quarter 
        \sum_{i=1}^{n_l}
        \left(\QuadVar{N_{1,i} + N_{2,i}}_t - \QuadVar{N_{1,i} - N_{2,i}}_t    \right)
        =
        \sum_{i=1}^{n_l}
        \int_0^t \expo{\theta_1 \nu}L_{1,i}(\nu) L_{2,i}(\nu)^\top d\nu,
    \end{align*}
    where $L_{1,i},\, L_{2,i} \in \mathbb{R}^{1 \times n_q}$ denote the $i^{th}$ rows of $L_1(t)$ and $L_2(t)$, respectively. 
    Hence, 
    \begin{align*}
        \absolute{\sum_{i=1}^{n_l}\CrossVar{N_{1,i}}{N_{2,i}}_t}
        \leq 
        \int_0^t 
            \expo{\theta_1\nu}
            \absolute{
            \sum_{i=1}^{n_l} L_{1,i}(\nu) L_{2,i}(\nu)^\top} 
        d\nu
        \leq& 
        \sqrt{n_l}
        \int_0^t \expo{\theta_1 \nu}
        \left(\sum_{i=1}^{n_l} \absolute{ L_{1,i}(\nu) L_{2,i}(\nu)^\top}^2\right)^\Half d\nu
        \\
        =&
        \sqrt{n_l}
        \int_0^t \expo{\theta_1\nu}
        \Frobenius{L_{1}(\nu) L_{2}(\nu)^\top} d\nu,
    \end{align*}
    for all $t \in [0,T]$.
    Then, applying Proposition~\ref{prop:TechnicalResults:LebesgueMoment} for $\cbr{\theta,\xi,\norm{S(t)}} = \cbr{\theta_1,0,\Frobenius{L_1(t)L_2(t)^\top}}$ leads to 
    \begin{align}\label{eqn:lemma:TechnicalResults:NestedSupMoment:Ito:Moments:Bound:B}
        \ELaw{}{
            \absolute{\sum_{i=1}^m\CrossVar{N_{1,i}}{N_{2,i}}_t}^\sfp
        }^\frac{1}{\sfp}
        \leq 
        \sqrt{n_l}
        \frac{\expo{\theta_1 t}}{\theta_1} 
        \sup_{\nu \in [0,t]}
        \LpLaw{\sfp}{}{L_1(\nu)L_2(\nu)^\top}
        , 
        \quad 
        \forall (t,\sfp) \in [0,T] \times \mathbb{N}_{\geq 1}
        ,
    \end{align}
    where we have bounded $\expo{\theta_1 t} - 1$ by $\expo{\theta_1 t}$.
    Substituting~\eqref{eqn:lemma:TechnicalResults:NestedSupMoment:Ito:Moments:Bound:p=1:A},~\eqref{eqn:lemma:TechnicalResults:NestedSupMoment:Ito:Moments:Bound:p>=2:A}, and~\eqref{eqn:lemma:TechnicalResults:NestedSupMoment:Ito:Moments:Bound:B} into~\eqref{eqn:lemma:TechnicalResults:NestedSupMoment:Ito:Moments:p=1:Decomposition} and~\eqref{eqn:lemma:TechnicalResults:NestedSupMoment:Ito:Moments:p>=2:Decomposition} produces the desired result.  

\end{proof}

The final major result of this section provides a bound on a class of nested \ito and Lebesgue integrals. 
\begin{lemma}\label{cor:TechnicalResults:UTildeBound}
    Consider a complete filtered probability space $\br{\Omega, \mathcal{F}, \mathbb{P}}$ with  filtration $\mathfrak{F}_t$, and let $R \in \mathcal{M}_{2}^{loc}\br{\mathbb{R}^{n_1 \times n_2}|\mathfrak{F}_t}$, $S_1 \in \mathcal{M}_{2}^{loc}\br{\mathbb{R}^{n_1}|\mathfrak{F}_t}$, $S_2 \in \mathcal{M}_{2}^{loc}\br{\mathbb{R}^{n_2}|\mathfrak{F}_t}$, $L_1 \in \mathcal{M}_{2}^{loc}\br{\mathbb{R}^{n_1 \times n_q}|\mathfrak{F}_t}$, and $L_2 \in \mathcal{M}_{2}^{loc}\br{\mathbb{R}^{n_2 \times n_q}|\mathfrak{F}_t}$ satisfy 
    \begin{align*}
        \sum_{i=1}^2
        \ELaw{}{\int_0^T \left(\Frobenius{R(\nu)}^{2\sfp} + \norm{S_i(\nu)}^{2\sfp} + \Frobenius{L_i(\nu)}^{2\sfp} \right)d\nu } < \infty, \quad \forall \sfp \geq 1.
    \end{align*}

    Additionally, for any constants $\theta_1,\,\theta_2 \in \mathbb{R}_{> 0}$ and an $\mathfrak{F}_t$-adapted Brownian motion $Q_t \in \mathbb{R}^{n_q}$ define 
    \begin{equation}\label{eqn:cor:TechnicalResults:UBound:N}
        N(t)
        = 
        \int_{\xi}^t 
            \expo{2(\theta_1 - \theta_2) \nu} 
            \widebreve{R}_1(\nu)^\top R(\nu) \widebreve{R}_2(\nu) 
        d\nu
        \in \mathbb{R}, \quad 0 \leq \xi \leq t \leq T,  
    \end{equation}
    where $\xi \in \mathbb{R}_{\geq 0}$ is a constant, and
    \begin{align*}
        \widebreve{R}_1(t)
        =
        \int_{0}^{t} 
            e^{\theta_2 \beta}
            \left[ S_1(\beta)d\beta + L_1(\beta)dQ_\beta \right] 
        \in \mathbb{R}^{n_1}
        , 
        \quad
        \widebreve{R}_2(t) = \int_0^t e^{\theta_2 \beta} \left[ S_2(\beta)d\beta + L_2(\beta)dQ_\beta \right] \in \mathbb{R}^{n_2}. 
    \end{align*}

    If there exists a constant $\Delta_R \in \mathbb{R}_{>0}$ such that 
    \begin{align*}
        \Frobenius{R(t)} \leq \Delta_R, \quad \forall t \in [0,T],
    \end{align*}
    then, the following bound holds for all $(t,\sfp) \in [\xi,T] \times \mathbb{N}_{\geq 1}$:
    \begin{multline}\label{cor:TechnicalResults:UTildeBound:Bound:Final}
        \LpLaw{\sfp}{}{N(t)}
        \leq
        \Delta_R
        \frac{\expo{ 2\theta_1 t } - \expo{ 2\theta_1 \xi }}{2\theta_1}
        \prod_{i=1}^{2}
        \left(
            \frac{1-\expo{-\theta_2 t}}{\theta_2}
            \sup_{\nu \in [0,t]}
            \LpLaw{2\sfp}{}{S_i(\nu)}
        \right.
        \\
        \left.
            + 
            \left(\sfp \frac{2\sfp-1}{2}\right)^\Half  
            \left(\frac{1 - \expo{ -2\theta_2 t }}{\theta_2}\right)^\Half
            \sup_{\nu \in [0,t]}
            \LpLaw{2\sfp}{}{L_i(\nu)}
        \right)
        . 
    \end{multline}
\end{lemma}
\begin{proof}
    We begin with the case $\sfp = 1$ by observing that one has the straightforward bound
    \begin{align*}
        \absolute{N(t)}
        \leq 
        \Delta_R
        \int_{\xi}^t 
            \expo{2(\theta_1 - \theta_2) \nu} 
            \norm{\widebreve{R}_1(\nu)} \norm{\widebreve{R}_2(\nu)} 
        d\nu
        , \quad \forall t \in [\xi,T]  
        .
    \end{align*}
    Applying Fubini's theorem, followed by the Cauchy-Schwarz inequality produces 
    \begin{align}\label{eqn:cor:TechnicalResults:FinalBoss:Bound:p=1:1}
        \ELaw{}{\absolute{N(t)}}
        \leq 
        \Delta_R
        \int_{\xi}^t 
            \expo{2(\theta_1 - \theta_2) \nu} 
            \LpLaw{2}{}{\widebreve{R}_1(\nu)} 
            \LpLaw{2}{}{\widebreve{R}_2(\nu)} 
        d\nu
        , 
        \quad \forall t \in [\xi,T]  
        .
    \end{align}
    Using the Minkowski's inequality, one can establish that 
    \begin{align*}
        \LpLaw{2}{}{
            \widebreve{R}_i(\nu)
        }
        \leq
        \LpLaw{2}{}{
            \int_{0}^{\nu} 
            e^{\theta_2 \beta}
            \norm{S_i(\beta)d\beta} 
        }
        + 
        \LpLaw{2}{}{
            \int_{0}^{\nu} 
            e^{\theta_2 \beta}
            L_i(\beta)dQ_\beta 
        }
        ,
        \quad 
        \forall (\nu,i) \in [\xi,t] \times \cbr{1,2}.
    \end{align*}
    Using Proposition~\ref{prop:TechnicalResults:LebesgueMoment} and Lemma~\ref{lemma:TechnicalResults:MartingaleMoment}, we obtain
    \begin{align*}
        \LpLaw{2}{}{
            \widebreve{R}_i(\nu)
        }
        \leq
        \expo{\theta_2 \nu}
        \left(
            \frac{1 - \expo{-\theta_2 \nu}}{\theta_2} 
            \sup_{\beta \in [0,\nu]}\LpLaw{2}{}{S_i(\beta)}
            +  
            \left( \frac{1 - \expo{ -2\theta_2 \nu }}{2\theta_2}\right)^\Half
            \sup_{\beta \in [0,\nu]}
            \LpLaw{2}{}{L_i(\beta)}
        \right)
        ,
    \end{align*}
    for all $(\nu,i) \in [\xi,t] \times \cbr{1,2}$.
    Substituting the above inequality into~\eqref{eqn:cor:TechnicalResults:FinalBoss:Bound:p=1:1} leads to 
    \begin{align*}
        \ELaw{}{\absolute{N(t)}}
        \leq 
        \Delta_R
        \int_{\xi}^t 
            \expo{2\theta_1\nu}
            \prod_{i=1}^{2} 
            \left(
                \frac{1 - \expo{-\theta_2 \nu}}{\theta_2} 
                \sup_{\beta \in [0,\nu]}\LpLaw{2}{}{S_i(\beta)}
                +  
                \left( \frac{1 - \expo{ -2\theta_2 \nu }}{2\theta_2}\right)^\Half
                \sup_{\beta \in [0,\nu]}
                \LpLaw{2}{}{L_i(\beta)}
            \right)
        d\nu
        ,
    \end{align*}
    which can be further developed into
    \begin{align*}
        \ELaw{}{\absolute{N(t)}}
        \leq 
        \Delta_R
        \left(
        \int_{\xi}^t 
            \expo{2\theta_1\nu}
        d\nu
        \right)
        \prod_{i=1}^{2} 
        \left(
            \frac{1 - \expo{-\theta_2 t}}{\theta_2} 
            \sup_{\nu \in [0,t]}\LpLaw{2}{}{S_i(\nu)}
            +  
            \left( \frac{1 - \expo{ -2\theta_2 t }}{2\theta_2}\right)^\Half
            \sup_{\nu \in [0,t]}
            \LpLaw{2}{}{L_i(\nu)}
        \right)
        ,
    \end{align*}
    for all $t \in [\xi,T] $.
    Solving the integral on the right hand side of the above inequality produces the desired result for $\sfp = 1$.

    Next, consider $\sfp \in \mathbb{N}_{\geq 2}$, and define 
    \begin{align}\label{eqn:cor:TechnicalResults:FinalBoss:NBreveDef}
        \widebreve{N}(t)
        =
        \int_{\xi}^t 
            \expo{2(\theta_1 - \theta_2) \nu} 
            \norm{\widebreve{R}_1(\nu)}  
            \norm{\widebreve{R}_2(\nu)} 
        d\nu
        \in \mathbb{R}, \quad t \in [\xi,T].
    \end{align}
    Then, 
    \begin{align*}
        \absolute{N(t)}
        \leq  
        \int_{\xi}^t 
            \expo{2(\theta_1 - \theta_2) \nu} 
            \norm{\widebreve{R}_1(\nu)} \Frobenius{R(\nu)} \norm{\widebreve{R}_2(\nu)} 
        d\nu
        \leq&  
        \Delta_R
        \int_{\xi}^t 
            \expo{2(\theta_1 - \theta_2) \nu} 
            \norm{\widebreve{R}_1(\nu)}  \norm{\widebreve{R}_2(\nu)} 
        d\nu
        \\
        =&
        \Delta_R
        \widebreve{N}(t),
    \end{align*}
    and thus
    \begin{equation}\label{eqn:cor:TechnicalResults:FinalBoss:Bound:PreSet}
        \LpLaw{\sfp}{}{N(t)}
        \leq      
        \Delta_R
        \LpLaw{\sfp}{}{\widebreve{N}(t)}
        , \quad \forall  (t,\sfp) \in [\xi,T] \times \mathbb{N}_{\geq 1}.
    \end{equation}
    By following the same line of reasoning that led to~\eqref{eqn:prop:TechnicalResults:LebesgueNestedMoment:Bound:A:Final} in the proof of Proposition~\ref{prop:TechnicalResults:LebesgueNestedMoment}, we obtain:
    \begin{align}\label{eqn:cor:TechnicalResults:FinalBoss:Bound:1}
        \LpLaw{\sfp}{}{\widebreve{N}(t)}
        \leq 
        \int_\xi^{t} 
            \expo{ 2\left(\theta_1   - \theta_2 \right) \nu }
            \ELaw{}{\norm{\widebreve{R}_1(\nu)}^{2\sfp}}^\frac{1}{2\sfp}
            \ELaw{}{\norm{\widebreve{R}_2(\nu)}^{2\sfp}}^\frac{1}{2\sfp}
        d\nu
        , 
        \quad  
        \forall t \in [\xi,T].
    \end{align}
    Now, it follows from the definition of $\widebreve{R}_i$, $i \in \cbr{1,2}$, that
    \begin{align*}
        \ELaw{}{
            \norm{\widebreve{R}_i(\nu)}^{2\sfp}
        }^\frac{1}{2\sfp}
        \leq&  
        \ELaw{}{
            \left( 
                \norm{
                    \int_{0}^{\nu} 
                    e^{\theta_2 \beta}
                    S_i(\beta)d\beta 
                }
                + 
                \norm{
                    \int_{0}^{\nu} 
                    e^{\theta_2 \beta}
                    L_i(\beta)dQ_\beta 
                }
            \right)^{2\sfp}
        }^\frac{1}{2\sfp}
        \notag 
        \\ 
        \overset{(\star)}{\leq}&  
        \ELaw{}{
            \norm{
                \int_{0}^{\nu} 
                e^{\theta_2 \beta}
                S_i(\beta)d\beta 
            }^{2\sfp}
        }^\frac{1}{2\sfp}
        + 
        \ELaw{}{  
            \norm{
                \int_{0}^{\nu} 
                e^{\theta_2 \beta}
                L_i(\beta)dQ_\beta 
            }^{2\sfp}
        }^\frac{1}{2\sfp}
        \notag  
        \\ 
        \leq& 
        \ELaw{}{
            \norm{
                \int_{0}^{\nu} 
                e^{\theta_2 \beta}
                \norm{S_i(\beta)} 
                d\beta
            }^{2\sfp}
        }^\frac{1}{2\sfp}
        + 
        \ELaw{}{  
            \norm{
                \int_{0}^{\nu} 
                e^{\theta_2 \beta}
                L_i(\beta)dQ_\beta 
            }^{2\sfp}
        }^\frac{1}{2\sfp}
        ,
        \quad 
        \forall (\nu,i) \in [\xi,t] \times \cbr{1,2},
    \end{align*}
    where $(\star)$ is due to the Minkowski's inequality. 
    Using Proposition~\ref{prop:TechnicalResults:LebesgueMoment} and Lemma~\ref{lemma:TechnicalResults:MartingaleMoment} then leads to 
    \begin{align*}
        &\ELaw{}{
            \norm{\widebreve{R}_i(\nu)}^{2\sfp}
        }^\frac{1}{2\sfp}
        \\
        &\leq
        \expo{\theta_2\nu}
        \left(
            \frac{1-\expo{-\theta_2 \nu}}{\theta_2}
            \sup_{\beta \in [0,\nu]}
            \LpLaw{2\sfp}{}{S_i(\beta)}
            + 
            \left(\sfp \frac{2\sfp-1}{2}\right)^\Half  
            \left(\frac{1 - \expo{ -2\theta_2 \nu }}{\theta_2}\right)^\Half
            \sup_{\beta \in [0,\nu]}
            \LpLaw{2\sfp}{}{L_i(\beta)}
        \right)
        ,
    \end{align*}
    for all $(\nu,i) \in [\xi,t] \times \cbr{1,2}$.
    We thus conclude that  
    \begin{multline*}
        \ELaw{}{
            \norm{\widebreve{R}_1(\nu)}^{2\sfp}
        }^\frac{1}{2\sfp}
        \ELaw{}{
            \norm{\widebreve{R}_2(\nu)}^{2\sfp}
        }^\frac{1}{2\sfp}
        \\
        \leq      
        \expo{2\theta_2 \nu}
        \prod_{i=1}^{2}
        \left(
            \frac{1-\expo{-\theta_2 \nu}}{\theta_2}
            \sup_{\beta \in [0,\nu]}
            \LpLaw{2\sfp}{}{S_i(\beta)}
            + 
            \left(\sfp \frac{2\sfp-1}{2}\right)^\Half  
            \left(\frac{1 - \expo{ -2\theta_2 \nu }}{\theta_2}\right)^\Half
            \sup_{\beta \in [0,\nu]}
            \LpLaw{2\sfp}{}{L_i(\beta)}
        \right)
        ,
    \end{multline*}
    for all $\nu \in [\xi,t]$.
    Substituting the above bound in~\eqref{eqn:cor:TechnicalResults:FinalBoss:Bound:1} yields
    \begin{multline*}
       \LpLaw{\sfp}{}{\widebreve{N}(t)}
        \\ 
        \leq 
        \int_\xi^{t} 
            \expo{ 2\theta_1 \nu }
            \prod_{i=1}^{2}
            \left(
                \frac{1-\expo{-\theta_2 \nu}}{\theta_2}
                \sup_{\beta \in [0,\nu]}
                \LpLaw{2\sfp}{}{S_i(\beta)}
                + 
                \left(\sfp \frac{2\sfp-1}{2}\right)^\Half  
                \left(\frac{1 - \expo{ -2\theta_2 \nu }}{\theta_2}\right)^\Half
                \sup_{\beta \in [0,\nu]}
                \LpLaw{2\sfp}{}{L_i(\beta)}
            \right)
        d\nu
       ,
    \end{multline*}
    which further implies
    \begin{multline*}
        \LpLaw{\sfp}{}{\widebreve{N}(t)}
        \\
        \leq 
        \left(
            \int_\xi^{t} 
                \expo{ 2\theta_1 \nu }
            d\nu
        \right)
        \prod_{i=1}^{2}
        \left(
            \frac{1-\expo{-\theta_2 t}}{\theta_2}
            \sup_{\nu \in [0,t]}
            \LpLaw{2\sfp}{}{S_i(\nu)}
            + 
            \left(\sfp \frac{2\sfp-1}{2}\right)^\Half  
            \left(\frac{1 - \expo{ -2\theta_2 t }}{\theta_2}\right)^\Half
            \sup_{\nu \in [0,t]}
            \LpLaw{2\sfp}{}{L_i(\nu)}
        \right)
        , 
    \end{multline*}
    for all $t \in [\xi,T]$.
    Solving the integral on the right hand side and substituting in~\eqref{eqn:cor:TechnicalResults:FinalBoss:Bound:PreSet} then leads to
    \begin{align*}
        &\LpLaw{\sfp}{}{N(t)}
        \\
        &\leq  
        \Delta_R
        \frac{\expo{ 2\theta_1 t } - \expo{ 2\theta_1 \xi }}{2\theta_1}
        \prod_{i=1}^{2}
        \left(
            \frac{1-\expo{-\theta_2 t}}{\theta_2}
            \sup_{\nu \in [0,t]}
            \LpLaw{2\sfp}{}{S_i(\nu)}
            + 
            \left(\sfp \frac{2\sfp-1}{2}\right)^\Half  
            \left(\frac{1 - \expo{ -2\theta_2 t }}{\theta_2}\right)^\Half
            \sup_{\nu \in [0,t]}
            \LpLaw{2\sfp}{}{L_i(\nu)}
        \right)
        , 
    \end{align*}
    for all $t \in [\xi,T]$,    
    thus, establishing~\eqref{cor:TechnicalResults:UTildeBound:Bound:Final}, for $\sfp \in \mathbb{N}_{\geq 2} $.

   
\end{proof}

We conclude the appendix with a couple of corollaries, starting with a corollary of the last lemma. 
\begin{corollary}\label{cor:TechnicalResults:Alt}
    Consider a complete filtered probability space $\br{\Omega, \mathcal{F}, \mathbb{P}}$ with  filtration $\mathfrak{F}_t$, and let $R \in \mathcal{M}_{2}^{loc}\br{\mathbb{R}^{n_1 \times n_2}|\mathfrak{F}_t}$, $S_1 \in \mathcal{M}_{2}^{loc}\br{\mathbb{R}^{n_1}|\mathfrak{F}_t}$, $S_2 \in \mathcal{M}_{2}^{loc}\br{\mathbb{R}^{n_2}|\mathfrak{F}_t}$, $L_1 \in \mathcal{M}_{2}^{loc}\br{\mathbb{R}^{n_1 \times n_q}|\mathfrak{F}_t}$, and $L_2 \in \mathcal{M}_{2}^{loc}\br{\mathbb{R}^{n_2 \times n_q}|\mathfrak{F}_t}$ satisfy 
    \begin{align*}
        \sum_{i=1}^2
        \ELaw{}{\int_0^T \left(\Frobenius{R(\nu)}^{2\sfp} + \norm{S_i(\nu)}^{2\sfp} + \Frobenius{L_i(\nu)}^{2\sfp} \right)d\nu } < \infty, \quad \forall \sfp \geq 1.
    \end{align*}
    Additionally, for any $\theta_1,\,\theta_2 \in \mathbb{R}_{> 0}$,  $\xi \in \mathbb{R}_{\geq 0}$, and an $\mathfrak{F}_t$-adapted Brownian motion $Q_t \in \mathbb{R}^{n_q}$ define 
    \begin{equation*}
        N_{\cbr{a,b}}(t)
        = 
        \int_{\xi}^t 
            \expo{2(\theta_1 - \theta_2) \nu} 
            \widebreve{R}_{\cbr{a,b}}(\nu)^\top R(\nu) \widebreve{R}_2(\nu) 
        d\nu
        \in \mathbb{R}, \quad 0 \leq \xi \leq t \leq T,  
    \end{equation*}
    where 
    \begin{align*}
        \widebreve{R}_{a}(\xi)
        =
        \int_{0}^{\xi} 
            e^{\theta_2 \beta}
            S_1(\beta)
        d\beta
        \in \mathbb{R}^{n_1}
        ,
        \quad  
        \widebreve{R}_{b}(\nu)
        =
        \int_{0}^{\nu} 
            e^{\theta_2 \beta}
            \indicator{[\zeta_1,\zeta_2]}{\beta}
            L_1(\beta)
        dQ_\beta 
        \in \mathbb{R}^{n_1}
        , 
        \\
        \widebreve{R}_2(t) = \int_0^t e^{\theta_2 \beta} \left[ S_2(\beta)d\beta + L_2(\beta)dQ_\beta \right] \in \mathbb{R}^{n_2}.
    \end{align*}
    In the above, $0 \leq \zeta_1 \leq \zeta_2 < t$ are any deterministic variables. 
    Then, for all $(t,\sfp) \in [\xi,T] \times \mathbb{N}_{\geq 1}$
    \begin{multline}\label{cor:TechnicalResults:Alt:Na:Bound:Final}
        \ELaw{}{\absolute{N_a(t)}^\sfp}^\frac{1}{\sfp} 
        \leq 
        \frac{\Delta_R}{2}
        \frac{\expo{2\theta_1 t} - \expo{2\theta_1 \xi}}{\theta_1 \theta_2}
            \left(\expo{\theta_2 \xi}-1\right)
            \sup_{\nu \in [0,t]}
            \LpLaw{2\sfp}{}{S_1(\nu)}
        \\
        \times 
        \left(
            \frac{1-\expo{-\theta_2 t}}{\theta_2}
            \sup_{\nu \in [0,t]}
            \LpLaw{2\sfp}{}{S_2(\nu)}
            + 
            \left(\sfp \frac{2\sfp-1}{2}\right)^\Half  
            \left(\frac{1 - \expo{ -2\theta_2 t }}{\theta_2}\right)^\Half
            \sup_{\nu \in [0,t]}
            \LpLaw{2\sfp}{}{L_2(\nu)}
        \right)
        , 
    \end{multline}
    and
    \begin{multline}\label{cor:TechnicalResults:Alt:Nb:Bound:Final}
        \ELaw{}{\absolute{N_b(t)}^\sfp}^\frac{1}{\sfp}
        \leq 
        \frac{\Delta_R}{2}
        \left(
            \sfp \frac{2\sfp -1}{2} \right)^\frac{1}{2}
        \frac{\expo{ 2\theta_1 t } - \expo{ 2\theta_1 (\xi \vee \zeta_1) }}{\theta_1 \sqrt{\theta_2}}
        \left( 1 - \expo{-2\theta_2(\zeta_2 - \zeta_1)}  \right)^\frac{1}{2}
        \sup_{\nu \in [\zeta_1,\zeta_2]} 
        \ELaw{}{\norm{L_1(\nu)}^{2\sfp}}^\frac{1}{2\sfp}
        \\
        \times 
        \left(
            \frac{1-\expo{-\theta_2 t}}{\theta_2}
            \sup_{\nu \in [0,t]}
            \LpLaw{2\sfp}{}{S_2(\nu)}
            + 
            \left(\sfp \frac{2\sfp-1}{2}\right)^\Half  
            \left(\frac{1 - \expo{ -2\theta_2 t }}{\theta_2}\right)^\Half
            \sup_{\nu \in [0,t]}
            \LpLaw{2\sfp}{}{L_2(\nu)}
        \right)
        .
    \end{multline}
                                                                                                     
\end{corollary}
\begin{proof}
    We provide a proof for the case $\sfp \in \mathbb{N}_{\geq 2}$ only and begin by writing 
    \begin{equation*} 
        N(t) = N_a(t) + N_b(t), \quad t \in [\xi,T],
    \end{equation*}
    where 
    \begin{align*}
        N_{\cbr{a,b}}(t)
        = 
        \int_{\xi}^t 
            \expo{2(\theta_1 - \theta_2) \nu} 
            \widebreve{R}_{\cbr{a,b}}(\nu)^\top R(\nu) \widebreve{R}_2(\nu) 
        d\nu,
        \\
        \widebreve{R}_{a}(\xi)
        =
        \int_{0}^{\xi} 
            e^{\theta_2 \beta}
            S_1(\beta)
        d\beta
        ,
        \quad  
        \widebreve{R}_{b}(\nu)
        =
        \int_{0}^{\nu} 
            e^{\theta_2 \beta}
            \indicator{[\zeta_1,\zeta_2]}{\beta}
            L_1(\beta)
        dQ_\beta 
        .
    \end{align*}

    Now, following the proof of Lemma~\ref{cor:TechnicalResults:UTildeBound} with $\widebreve{R}_1$ replaced by $\widebreve{R}_a$ leads to 
    \begin{multline*}
        \begin{aligned}
            &\ELaw{}{\absolute{N_a(t)}^\sfp}^\frac{1}{\sfp}
            \\ 
            &\leq 
            \Delta_R
            \left(\int_\xi^t \expo{(2\theta_1 - \theta_2)\nu} d\nu\right)
            \left( 
                \frac{\expo{\theta_2 \xi}-1}{\theta_2}
                \sup_{\nu \in [0,\xi]}
                \LpLaw{2\sfp}{}{S_1(\nu)}
            \right)
        \end{aligned}
        \\
        \times 
        \left(
            \frac{1-\expo{-\theta_2 t}}{\theta_2}
            \sup_{\nu \in [0,t]}
            \LpLaw{2\sfp}{}{S_2(\nu)}
            + 
            \left(\sfp \frac{2\sfp-1}{2}\right)^\Half  
            \left(\frac{1 - \expo{ -2\theta_2 t }}{\theta_2}\right)^\Half
            \sup_{\nu \in [0,t]}
            \LpLaw{2\sfp}{}{L_2(\nu)}
        \right)
        , 
    \end{multline*}
    for all $t \in [\xi,T]$.
    Note that since
    \begin{align*}
        \int_\xi^t \expo{(2\theta_1 - \theta_2)\nu} d\nu
        \leq
        \int_\xi^t \expo{2\theta_1\nu} d\nu
        =
        \frac{\expo{2\theta_1 t} - \expo{2\theta_1 \xi}}{2\theta_1}
        ,
    \end{align*}
    we can write the previous bound as 
    \begin{multline*}
        \ELaw{}{\absolute{N_a(t)}^\sfp}^\frac{1}{\sfp}
        \leq 
        \frac{\Delta_R}{2}
        \frac{\expo{2\theta_1 t} - \expo{2\theta_1 \xi}}{\theta_1 \theta_2}
        \left(\expo{\theta_2 \xi}-1\right)
        \sup_{\nu \in [0,t]}
        \LpLaw{2\sfp}{}{S_1(\nu)}
        \\
        \times 
        \left(
            \frac{1-\expo{-\theta_2 t}}{\theta_2}
            \sup_{\nu \in [0,t]}
            \LpLaw{2\sfp}{}{S_2(\nu)}
            + 
            \left(\sfp \frac{2\sfp-1}{2}\right)^\Half  
            \left(\frac{1 - \expo{ -2\theta_2 t }}{\theta_2}\right)^\Half
            \sup_{\nu \in [0,t]}
            \LpLaw{2\sfp}{}{L_2(\nu)}
        \right)
        , 
    \end{multline*}
    for all $t \in [\xi,T]$, where have also used the bound $\sup_{\nu \in [0,\xi]}\LpLaw{2\sfp}{}{S_1(\nu)} \leq \sup_{\nu \in [0,t]}\LpLaw{2\sfp}{}{S_1(\nu)}$, thus proving~\eqref{cor:TechnicalResults:Alt:Na:Bound:Final}.

    Next, as at the beginning of the proof of  Lemma~\ref{lemma:TechnicalResults:NestedSupMoment}, we can write 
    \begin{align*}
        N_{b}(t)
        =&
        \int_{\xi}^t 
            \expo{2(\theta_1 - \theta_2) \nu} 
            \left(
                \int_{0}^{\nu} 
                    e^{\theta_2 \beta}
                    \indicator{[\zeta_1,\zeta_2]}{\beta}
                    L_1(\beta)
                dQ_\beta 
            \right)^\top R(\nu) \widebreve{R}_2(\nu) 
        d\nu
        \\
        =&
        \int_{{\xi \vee \zeta_1}}^t 
            \expo{2(\theta_1 - \theta_2) \nu} 
            \left(
                \int_{\zeta_1}^{\nu \wedge \zeta_2} 
                    e^{\theta_2 \beta}
                    L_1(\beta)
                dQ_\beta 
            \right)^\top R(\nu) \widebreve{R}_2(\nu) 
        d\nu,
        \quad 
        \forall t \in [\xi,T].
    \end{align*}
    Therefore, as in~\eqref{eqn:lem:TechnicalResults:NestedSupMoment:Bound:B:Final}, and using the bound on $\ELaw{}{\norm{\widebreve{R}_2(\nu)}^{2\sfp}}^\frac{1}{2\sfp}$ in the proof of Lemma~\ref{cor:TechnicalResults:UTildeBound}, we obtain
    \begin{multline*}
        \begin{aligned}
            &\ELaw{}{\absolute{N_b(t)}^\sfp}^\frac{1}{\sfp}
            \\ 
            &\leq 
            \Delta_R
            \left(
                \int_{\xi \vee \zeta_1}^{t} 
                    \expo{ 2\theta_1 \nu }
                d\nu
            \right)
            \left(\sfp \frac{2\sfp -1}{2\theta_2} \right)^\frac{1}{2}
            \left( 1 - \expo{-2\theta_2(\zeta_2 - \zeta_1)}  \right)^\frac{1}{2}
            \sup_{\nu \in [\zeta_1,\zeta_2]} 
            \ELaw{}{\norm{L_1(\nu)}^{2\sfp}}^\frac{1}{2\sfp}
        \end{aligned}
        \\
        \times 
        \left(
            \frac{1-\expo{-\theta_2 t}}{\theta_2}
            \sup_{\nu \in [0,t]}
            \LpLaw{2\sfp}{}{S_2(\nu)}
            + 
            \left(\sfp \frac{2\sfp-1}{2}\right)^\Half  
            \left(\frac{1 - \expo{ -2\theta_2 t }}{\theta_2}\right)^\Half
            \sup_{\nu \in [0,t]}
            \LpLaw{2\sfp}{}{L_2(\nu)}
        \right)
        , 
    \end{multline*}
    for all $t \in [\xi,T]$, which then leads to 
    \begin{multline*}
            \ELaw{}{\absolute{N_b(t)}^\sfp}^\frac{1}{\sfp}
            \leq 
            \frac{\Delta_R}{2}
            \left(\sfp \frac{2\sfp -1}{2} \right)^\frac{1}{2}
            \frac{\expo{ 2\theta_1 t } - \expo{ 2\theta_1 (\xi \vee \zeta_1) }}{\theta_1 \sqrt{\theta_2}}
            \left( 1 - \expo{-2\theta_2(\zeta_2 - \zeta_1)}  \right)^\frac{1}{2}
            \sup_{\nu \in [\zeta_1,\zeta_2]} 
            \ELaw{}{\norm{L_1(\nu)}^{2\sfp}}^\frac{1}{2\sfp}
        \\
        \times 
        \left(
            \frac{1-\expo{-\theta_2 t}}{\theta_2}
            \sup_{\nu \in [0,t]}
            \LpLaw{2\sfp}{}{S_2(\nu)}
            + 
            \left(\sfp \frac{2\sfp-1}{2}\right)^\Half  
            \left(\frac{1 - \expo{ -2\theta_2 t }}{\theta_2}\right)^\Half
            \sup_{\nu \in [0,t]}
            \LpLaw{2\sfp}{}{L_2(\nu)}
        \right)
        , 
    \end{multline*}
    for all $t \in [\xi,T]$, thus proving~\eqref{cor:TechnicalResults:Alt:Nb:Bound:Final}.

\end{proof}

The final result considers a particular process resulting from the piecewise nature of the adaptation law. 
\begin{corollary}\label{cor:TechnicalResults:PCA}
    Consider a complete filtered probability space $\br{\Omega, \mathcal{F}, \mathbb{P}}$ with  filtration $\mathfrak{F}_t$, and let $R \in \mathcal{M}_{2}^{loc}\br{\mathbb{R}^{n_1 \times n_2}|\mathfrak{F}_t}$, $S \in \mathcal{M}_{2}^{loc}\br{\mathbb{R}^{n_2}|\mathfrak{F}_t}$, $L_a,\, L_b \in \mathcal{M}_{2}^{loc}\br{\mathbb{R}^{n_1 \times n_q}|\mathfrak{F}_t}$,  and $L \in \mathcal{M}_{2}^{loc}\br{\mathbb{R}^{n_2 \times n_q}|\mathfrak{F}_t}$ satisfy 
    \begin{align*}
        \ELaw{}{\int_0^T 
        \left(\Frobenius{R(\nu)}^{2\sfp} + \norm{S(\nu)}^{2\sfp} + \Frobenius{L(\nu)}^{2\sfp} + \Frobenius{L_a(\nu)}^{4\sfp} + \Frobenius{L_b(\nu)}^{4\sfp}  \right)d\nu } < \infty, \quad \forall \sfp \geq 1.
    \end{align*}
    Additionally, for any $\theta_1,\,\theta_2 \in \mathbb{R}_{> 0}$,  $\BoldTs \in \mathbb{R}_{> 0}$, and an $\mathfrak{F}_t$-adapted Brownian motion $Q_t \in \mathbb{R}^{n_q}$ define 
    \begin{equation}\label{cor:TechnicalResults:PCA:N}
        N_{\cbr{a,b}}(t)
        = 
        \int_{\BoldTs}^t 
            \expo{2(\theta_1 - \theta_2) \nu} 
            \widebreve{R}_{\cbr{a,b}}(\nu)^\top R(\nu) \widebreve{R}(\nu) 
        d\nu
        \in \mathbb{R}, \quad t \in [\BoldTs,T],  
    \end{equation}
    where 
    \begin{align*}
        \widebreve{R}_{a}(\nu)
        =
        L_a(\nu)
        \varsigma_1(\nu)
        \int_{(\istar{\nu}-1)\BoldTs}^{\istar{\nu}\BoldTs} 
            e^{\theta_2 \beta}
            \varsigma_2(\beta)
        d\Qt{\beta}
        \in \mathbb{R}^{n_1}
        ,
        \quad  
        \widebreve{R}_{b}(\nu)
        =
        L_b(\nu)
        \int_{\istar{\nu}\BoldTs}^\nu 
            e^{\theta_2 \beta}
        d\Qt{\beta}
        \in \mathbb{R}^{n_1}
        , 
        \\
        \widebreve{R}(\nu) = \int_0^\nu e^{\theta_2 \beta} \left[ S(\beta)d\beta + L(\beta)dQ_\beta \right] \in \mathbb{R}^{n_2}
        ,
    \end{align*}
    and where $\istar{\nu} \doteq \max\cbr{i \in \mathbb{N}_{\geq 1} \, : \, i\BoldTs \leq \nu} = \left\lfloor \nu / \BoldTs \right\rfloor$.
    Furthermore, $\varsigma_{\cbr{1,2}} \in \mathcal{C} \left( \mathbb{R}_{\geq 0},\mathbb{R}\right) $ are any deterministic functions. 
    Then, for all $(t,\sfp) \in [\BoldTs,T] \times \mathbb{N}_{\geq 1}$:
    \begin{align}\label{cor:TechnicalResults:PCA:Na:Bound:Final}
        \LpLaw{\sfp}{}{N_a(t)} 
        \leq&  
        \frac{\Delta_R}{2}
        \left(
            \sfp (4\sfp-1) 
        \right)^\frac{1}{2}   
        \frac{\expo{ 2\theta_1  t } - \expo{ 2\theta_1  \BoldTs }}{\theta_1 \sqrt{\theta_2}} 
            \left(1 - \expo{ -2\theta_2 \BoldTs } \right)^\frac{1}{2}
        \notag 
        \\
        & \times 
        \sup_{\nu \in [\BoldTs,t]}
        \left(
            \LpLaw{4\sfp}{}{L_a(\nu)}
            \absolute{\varsigma_1(\nu)}
            \sup_{\beta \in [(\istar{\nu}-1)\BoldTs,\istar{\nu}\BoldTs]}
            \absolute{\varsigma_2(\beta)}
        \right)
        \notag 
        \\
        &\times     
        \left(
            \frac{1-\expo{-\theta_2 t}}{\theta_2}
            \sup_{\nu \in [0,t]}
            \LpLaw{2\sfp}{}{S(\nu)}
            + 
            \left(\sfp \frac{2\sfp-1}{2}\right)^\Half  
            \left(\frac{1 - \expo{ -2\theta_2 t }}{\theta_2}\right)^\Half
            \sup_{\nu \in [0,t]}
            \LpLaw{2\sfp}{}{L(\nu)}
        \right)
        , 
    \end{align}
    and 
    \begin{align}\label{cor:TechnicalResults:PCA:Nb:Bound:Final}
        \LpLaw{\sfp}{}{N_b(t)} 
        \leq&  
        \frac{\Delta_R}{2}
        \left(
            \sfp (4\sfp-1) \right)^\frac{1}{2}   
        \frac{\expo{ 2\theta_1  t } - \expo{ 2\theta_1  \BoldTs }}{\theta_1 \sqrt{\theta_2}} 
            \left(1 - \expo{ -2\theta_2 \BoldTs } \right)^\frac{1}{2}
            \sup_{\nu \in [\BoldTs,t]}
            \LpLaw{4\sfp}{}{L_b(\nu)}
        \notag 
        \\
        &\times     
        \left(
            \frac{1-\expo{-\theta_2 t}}{\theta_2}
            \sup_{\nu \in [0,t]}
            \LpLaw{2\sfp}{}{S(\nu)}
            + 
            \left(\sfp \frac{2\sfp-1}{2}\right)^\Half  
            \left(\frac{1 - \expo{ -2\theta_2 t }}{\theta_2}\right)^\Half
            \sup_{\nu \in [0,t]}
            \LpLaw{2\sfp}{}{L(\nu)}
        \right)
        .
    \end{align}

\end{corollary}
\begin{proof}
    We provide a proof for the case $\sfp \in \mathbb{N}_{\geq 2}$ only and begin by defining 
    \begin{align}\label{eqn:cor:TechnicalResults:PCA:NBreveDef}
        \widebreve{N}_{\cbr{a,b}}(t)
        =
        \int_{\BoldTs}^t 
            \expo{2(\theta_1 - \theta_2) \nu} 
            \norm{\widebreve{R}_{\cbr{a,b}}(\nu)}  
            \norm{\widebreve{R}(\nu)} 
        d\nu
        \in \mathbb{R}, \quad t \in [\BoldTs,T].
    \end{align}
    Then, 
    \begin{multline*}
        \absolute{N_{\cbr{a,b}}(t)}
        \leq  
        \int_{\BoldTs}^t 
            \expo{2(\theta_1 - \theta_2) \nu} 
            \norm{\widebreve{R}_{\cbr{a,b}}(\nu)} \Frobenius{R(\nu)} \norm{\widebreve{R}(\nu)} 
        d\nu
        \\
        \leq  
        \Delta_R
        \int_{\BoldTs}^t 
            \expo{2(\theta_1 - \theta_2) \nu} 
            \norm{\widebreve{R}_{\cbr{a,b}}(\nu)}  \norm{\widebreve{R}(\nu)} 
        d\nu
        =
        \Delta_R
        \widebreve{N}_{\cbr{a,b}}(t),
    \end{multline*}
    and thus
    \begin{equation}\label{eqn:cor:TechnicalResults:PCA:PreSet}
        \LpLaw{\sfp}{}{N_{\cbr{a,b}}(t)}
        \leq      
        \Delta_R
        \LpLaw{\sfp}{}{\widebreve{N}_{\cbr{a,b}}(t)}
        , \quad \forall  (t,\sfp) \in [\BoldTs,T] \times \mathbb{N}_{\geq 2}.
    \end{equation}

    Next, as in the proof of Lemma~\ref{cor:TechnicalResults:UTildeBound}, we have that 
    \begin{align}\label{eqn:cor:TechnicalResults:PCA:1}
        \ELaw{}{\absolute{\widebreve{N}_{\cbr{a,b}}(t)}^\sfp}^\frac{1}{\sfp} 
        \leq 
        \int_{\BoldTs}^{t} 
            \expo{ 2\left(\theta_1   - \theta_2  \right) \nu }
            \ELaw{}{\norm{\widebreve{R}_{\cbr{a,b}}(\nu)}^{2\sfp}}^\frac{1}{2\sfp} 
            \ELaw{}{\norm{\widebreve{R}(\nu)}^{2\sfp}}^\frac{1}{2\sfp} 
        d\nu
        , 
        \quad  
        \forall t \in [\BoldTs,T].
    \end{align}
    Similarly, as in the proof of Lemma~\ref{cor:TechnicalResults:UTildeBound}, we further have that
    \begin{multline}\label{eqn:cor:TechnicalResults:PCA:Bound:B:1:Final}
        \ELaw{}{
            \norm{\widebreve{R}(\nu)}^{2\sfp}
        }^\frac{1}{2 \sfp}
        \leq
        \expo{\theta_2  \nu}
        \left(
            \frac{1-\expo{-\theta_2 \nu}}{\theta_2}
            \sup_{\beta \in [0,\nu]}
            \LpLaw{2\sfp}{}{S(\beta)}
        \right. 
        \\
        \left.
            + 
            \left(\sfp \frac{2\sfp-1}{2}\right)^\Half  
            \left(\frac{1 - \expo{ -2\theta_2 \nu }}{\theta_2}\right)^\Half
            \sup_{\beta \in [0,\nu]}
            \LpLaw{2\sfp}{}{L(\beta)}
        \right)
        ,
        \quad 
        \forall \nu \in [\BoldTs,t]
        .
    \end{multline}
    Next, by the Cauchy-Schwarz inequality
    \begin{align*}
        \LpLaw{2\sfp}{}{\widebreve{R}_{a}(\nu)}
        \leq 
        \LpLaw{4\sfp}{}{L_a(\nu)}
        \absolute{\varsigma_1(\nu)}
        \LpLaw{4\sfp}{}{
            \int_{(\istar{\nu}-1)\BoldTs}^{\istar{\nu}\BoldTs} 
                e^{\theta_2 \beta}
                \varsigma_2(\beta)
            d\Qt{\beta}
        }
        , 
        \quad 
        \forall \nu \in [\BoldTs,t]
        ,
    \end{align*}
    where we have used the fact that $\varsigma_1$ is a deterministic function. 
    Using Lemma~\ref{lemma:TechnicalResults:MartingaleMoment} to bound the \ito integral term, we get 
    \begin{align}\label{eqn:cor:TechnicalResults:PCA:Bound:B:a:Final}
        &\LpLaw{2\sfp}{}{\widebreve{R}_{a}(\nu)}
        \notag 
        \\
        &\leq
        \left(\sfp (4\sfp-1)\right)^\frac{1}{2}  
        \expo{ \theta_2  \istar{\nu} \BoldTs } 
        \left(\frac{1 - \expo{ -2\theta_2 \BoldTs }}{\theta_2}\right)^\frac{1}{2}
        \LpLaw{4\sfp}{}{L_a(\nu)}
        \absolute{\varsigma_1(\nu)}
        \sup_{\beta \in [(\istar{\nu}-1)\BoldTs,\istar{\nu}\BoldTs]}
        \absolute{\varsigma_2(\beta)}
        \notag 
        \\
        &\leq
        \left(\sfp (4\sfp-1)\right)^\frac{1}{2}  
        \expo{ \theta_2 \nu}
        \left(\frac{1 - \expo{ -2\theta_2 \BoldTs }}{\theta_2}\right)^\frac{1}{2}
        \LpLaw{4\sfp}{}{L_a(\nu)}
        \absolute{\varsigma_1(\nu)}
        \sup_{\beta \in [(\istar{\nu}-1)\BoldTs,\istar{\nu}\BoldTs]}
        \absolute{\varsigma_2(\beta)}
        , 
    \end{align}  
    for all $(t,\sfp) \in [\BoldTs,T] \times \mathbb{N}_{\geq 2}$, where we have used the fact that $\varsigma_2$ is a deterministic function. Moreover, the second inequality follows from $\istar{\nu} \BoldTs \leq \nu$, $\forall \nu$.

    Similarly, using the Cauchy-Schwarz inequality and Lemma~\ref{lemma:TechnicalResults:MartingaleMoment}: 
    \begin{align}\label{eqn:cor:TechnicalResults:PCA:Bound:B:b:Final}
        \LpLaw{2\sfp}{}{\widebreve{R}_{b}(\nu)}
        \leq&
        \left(\sfp (4\sfp-1)\right)^\frac{1}{2}  
        \expo{ \theta_2 \nu }
        \left(\frac{1 - \expo{ 2\theta_2 (\istar{\nu}\BoldTs - \nu) }}{\theta_2}\right)^\frac{1}{2}
        \LpLaw{4\sfp}{}{L_b(\nu)}
        \notag 
        \\ 
        \leq&
        \left(\sfp (4\sfp-1)\right)^\frac{1}{2}  
        \expo{ \theta_2 \nu }
        \left(\frac{1 - \expo{ -2\theta_2 \BoldTs }}{\theta_2}\right)^\frac{1}{2}
        \LpLaw{4\sfp}{}{L_b(\nu)}
        ,
    \end{align}
    for all $(t,\sfp) \in [\BoldTs,T] \times\mathbb{N}_{\geq 2}$, where, we obtain the second inequality by using the fact that $\nu - \istar{\nu}\BoldTs \leq \BoldTs$, $\forall \nu$.
    Substituting~\eqref{eqn:cor:TechnicalResults:PCA:Bound:B:1:Final}, and~\eqref{eqn:cor:TechnicalResults:PCA:Bound:B:a:Final} into~\eqref{eqn:cor:TechnicalResults:PCA:1} leads to  
    \begin{align*}
        \LpLaw{\sfp}{}{\widebreve{N}_a(t)} 
        \leq&  
        \left(
            \int_{\BoldTs}^{t}  
                \expo{ 2\theta_1  \nu } 
            d\nu
        \right)    
        \left(\sfp (4\sfp-1) \right)^\frac{1}{2}  
        \left(\frac{1 - \expo{ -2\theta_2 \BoldTs }}{\theta_2}\right)^\frac{1}{2}
        \\
        &\times
        \sup_{\nu \in [\BoldTs,t]}
        \left(
            \LpLaw{4\sfp}{}{L_a(\nu)}
            \absolute{\varsigma_1(\nu)}
            \sup_{\beta \in [(\istar{\nu}-1)\BoldTs,\istar{\nu}\BoldTs]}
            \absolute{\varsigma_2(\beta)}
        \right)
        \\
        &\times     
        \left(
            \frac{1-\expo{-\theta_2 t}}{\theta_2}
            \sup_{\nu \in [0,t]}
            \LpLaw{2\sfp}{}{S(\nu)}
            + 
            \left(\sfp \frac{2\sfp-1}{2}\right)^\Half  
            \left(\frac{1 - \expo{ -2\theta_2 t }}{\theta_2}\right)^\Half
            \sup_{\nu \in [0,t]}
            \LpLaw{2\sfp}{}{L(\nu)}
        \right)
        , 
    \end{align*}
    for all $t \in [\BoldTs,T]$.
    Similarly, substituting~\eqref{eqn:cor:TechnicalResults:PCA:Bound:B:1:Final}, and~\eqref{eqn:cor:TechnicalResults:PCA:Bound:B:b:Final} into~\eqref{eqn:cor:TechnicalResults:PCA:1} leads to  
    \begin{align*}
        \LpLaw{\sfp}{}{\widebreve{N}_b(t)} 
        \leq&  
        \left(
            \int_{\BoldTs}^{t}  
                \expo{ 2\theta_1  \nu } 
            d\nu
        \right)    
        \left(\sfp (4\sfp-1) \right)^\frac{1}{2}  
        \left(\frac{1 - \expo{ -2\theta_2 \BoldTs }}{\theta_2}\right)^\frac{1}{2}
        \sup_{\nu \in [\BoldTs,t]}
            \LpLaw{4\sfp}{}{L_b(\nu)}
        \\
        &\times     
        \left(
            \frac{1-\expo{-\theta_2 t}}{\theta_2}
            \sup_{\nu \in [0,t]}
            \LpLaw{2\sfp}{}{S(\nu)}
            + 
            \left(\sfp \frac{2\sfp-1}{2}\right)^\Half  
            \left(\frac{1 - \expo{ -2\theta_2 t }}{\theta_2}\right)^\Half
            \sup_{\nu \in [0,t]}
            \LpLaw{2\sfp}{}{L(\nu)}
        \right)
        , 
    \end{align*}
    for all $t \in [\BoldTs,T]$.
    Then, by solving the first integrals on the right hand side of the bounds above and substituting into~\eqref{eqn:cor:TechnicalResults:PCA:PreSet} lead to~\eqref{cor:TechnicalResults:PCA:Na:Bound:Final} and~\eqref{cor:TechnicalResults:PCA:Nb:Bound:Final}, thus completing the proof.

\end{proof}
\setcounter{equation}{0}
\section{Reference Process}\label{app:ReferenceProcess}

We provide the proof of Proposition~\ref{prop:Reference:TruncatedWellPosedness} below.
\begin{proof}[Proof of Proposition~\ref{prop:Reference:TruncatedWellPosedness}]
    We consider the case $\Probability{\Yt{N,0} = \Yt{0}  \in U_N} = 1$ w.l.o.g., since the general case can be handled by following the approach in~\cite[Thm.~8.3.1]{gikhman1969random}.

    We begin by observing that establishing the well-posedness of $\Yt{N,t} = \begin{bmatrix} \br{\Xrt{N.t}}^\top & \br{\Xstart{N.t}}^\top \end{bmatrix}$ is equivalent to establishing the individual well-posedness of the following systems:
    \begin{subequations}
        \begin{align}
            d\Xstart{N,t} = \FbarNmu{t,\Xstart{N,t}}dt + \FbarNsigma{t,\Xstart{N,t}}d\Wstart{t}, \quad \Xstart{N,0} = x^\star_0 \sim \xi^\star_0, \label{eqn:Reference:TruncatedWellPosedness:System:1} 
            \\ 
            d\Xrt{N,t} = \FNmu{t,\Xrt{N,t},\Urt{t}}dt + \FNsigma{t,\Xrt{N,t}}d\Wt{t}, \quad \Xrt{N,0} = x_0 \sim \xi_0, \label{eqn:Reference:TruncatedWellPosedness:System:2}
        \end{align}
    \end{subequations}
    for $t \in [0,\tau_N \wedge T)$, where $\bar{F}_{N,\cbr{\mu,\sigma}}$ and $F_{N,\cbr{\mu,\sigma}}$ are defined analogously to $G_{N,\cbr{\mu,\sigma}}$ in~\eqref{eqn:Reference:TruncatedJointProcess}. 
    
    The well-posedness of~\eqref{eqn:Reference:TruncatedWellPosedness:System:1} is straightforward to establish using Definition~\ref{def:VectorFields}, Assumption~\ref{assmp:KnownFunctions}~\cite[Thm.~3.4]{khasminskii2011stochastic}, and due to $\bar{F}_{N,\cbr{\mu,\sigma}}(a) \equiv 0$, $\forall a \in \mathbb{R}^n$ with $\norm{a} \geq 2N$.  

    Now, consider any $z \in \mathcal{M}_2\br{[0,T],\mathbb{R}^n~|~\Wfilt{0,t}}$, for  any $t \in [0,T]$, and  define 
    \begin{align}\label{eqn:Reference:TruncatedWellPosedness:M}
        M(z(t)) = \int_0^t \FNmu{\nu,z(\nu),\Urt{\nu}}d\nu + \int_0^t \FNsigma{\nu,z(\nu)}d\Wt{\nu}, \quad t \in [0,T].
    \end{align}
    Let $f_N$, $\Lambda_{N,\mu}$, $p_N$, and $\Lambda_{N,\sigma}$ be the truncated versions of the functions $f$, $\Lambda_{\mu}$, $p$, and $\Lambda_{\sigma}$, respectively, and where the truncation is defined as in~\eqref{eqn:Reference:JointProcess}. 
    Then, we have that  
    \begin{align*}
        M(z(t)) = \int_0^t \br{ f_N\br{\nu, z(\nu)} + g(\nu) \Urt{\nu} +  \Lambda_{N,\mu}\br{\nu, z(\nu)}    }d\nu + \int_0^t \br{p_N(\nu, z(\nu)) + \Lambda_{N,\sigma} \br{\nu, z(\nu)} }d\Wt{\nu}, 
    \end{align*}   
    for $t \in [0,T]$, where, from~\eqref{eqn:ReferenceFeedbackOperatorProcess}, we have that 
    \begin{align*}
        \Urt{\nu} = \Filter[\Lambda_{N,\mu}^{\paral} \br{\cdot, z}][\nu] + \FilterW[p_N^{\paral}(\cdot,z) + \Lambda_{N,\sigma}^{\paral} \br{\cdot, z}, \Wt{}][\nu].
    \end{align*}
    Therefore, the previous expression can be expressed as 
    \begin{align}\label{eqn:Reference:TruncatedWellPosedness:M:1}
        M(z(t)) =& \int_0^t \br{ f_N\br{\nu,z(\nu)} +  \Lambda_{N,\mu} \br{\nu, z(\nu)}    }d\nu 
        + \int_0^t g(\nu)\Filter[\Lambda_{N,\mu}^{\paral} \br{\cdot,z}][\nu] d\nu \notag 
        \\ 
        &+ \int_0^t g(\nu) \FilterW[p_N^{\paral}(\cdot,z) + \Lambda_{N,\sigma}^{\paral} \br{\cdot,z}, \Wt{}][\nu]  d\nu 
        + \int_0^t \br{p_N(\nu,z(\nu)) + \Lambda_{N,\sigma} \br{\nu,z(\nu)} }d\Wt{\nu},
    \end{align}
    for $t \in [0,T]$.
    Using the definition of $\Filter$ in~\eqref{eqn:L1DRAC:Definition:FeedbackOperator:Filter}, we have that
    \begin{align*}
        \int_0^t g(\nu)\Filter[\Lambda_{N,\mu}^{\paral} \br{\cdot,z}][\nu] d\nu
        =&
        -\Boldomega \int_0^t \int_0^\nu g(\nu) \expo{-\Boldomega(\nu - \beta )} \Lambda_{N,\mu}^{\paral} \br{\beta,z(\beta)}d\beta  d\nu, \quad t \in [0,T].
    \end{align*}
    Changing the order of integration in the double Lebesgue integral produces
    \begin{align}\label{eqn:Reference:TruncatedWellPosedness:DriftControl}
        \int_0^t g(\nu)\Filter[\Lambda_{N,\mu}^{\paral} \br{\cdot,z}][\nu] d\nu
        =&
        -\Boldomega \int_0^t \br{ \int_\nu^t g(\beta) \expo{-\Boldomega \beta} d\beta } \expo{\Boldomega \nu } \Lambda_{N,\mu}^{\paral} \br{\nu,z(\nu)}  d\nu, \quad t \in [0,T].
    \end{align}
    Next, using the definition of $\FilterW$ in~\eqref{eqn:ReferenceFeedbackOperatorProcess}, we have that  
    \begin{align*}
        &\int_0^t g(\nu) \FilterW[p_N^{\paral}(\cdot,z) + \Lambda_{N,\sigma}^{\paral} \br{\cdot,z}, \Wt{}][\nu]  d\nu
        \\
        & \qquad \qquad  = -\Boldomega \int_0^t \int_0^\nu  g(\nu) \expo{-\Boldomega(\nu - \beta)} 
        \br{ p_N^{\paral}\br{\beta,z(\beta)}  + \Lambda_{N,\sigma}^{\paral} \br{\beta,z(\beta)} } d\Wt{\beta}  d\nu, \quad t \in [0,T].
    \end{align*}
    Applying Lemma~\ref{lem:TechnicalResults:Fubini-ish} to the above expression for 
    \begin{align*}
        L(\nu) = g(\nu) \expo{-\Boldomega \nu} \in \continuous{T}{n \times m}, 
        \\ 
        S(\beta) = \expo{\Boldomega \beta} 
        \br{ p_N^{\paral}\br{\beta,z(\beta)}  + \Lambda_{N,\sigma}^{\paral} \br{\beta,z(\beta)} } 
        \in \mathcal{M}\br{[0,T];\mathbb{R}^{m \times d}~|~\Wfilt{0,t}},
        Q_\beta = \Wt{\beta},
    \end{align*}
    produces 
    \begin{align}\label{eqn:Reference:TruncatedWellPosedness:DiffusionControl}
        &\int_0^t g(\nu) \FilterW[p_N^{\paral}(\cdot,z) + \Lambda_{N,\sigma}^{\paral} \br{\cdot,z}, \Wt{}][\nu] d\nu \notag 
        \\
        & \qquad \qquad  = -\Boldomega \int_0^t   
        \br{ \int_\nu^t g(\beta) \expo{-\Boldomega \beta} d\beta } \expo{\Boldomega \nu } 
        \br{ p_N^{\paral}\br{\nu,z(\nu)}  + \Lambda_{N,\sigma}^{\paral} \br{\nu,z(\nu)} } d\Wt{\nu}
        , \quad t \in [0,T].
    \end{align}
    Substituting~\eqref{eqn:Reference:TruncatedWellPosedness:DriftControl} and~\eqref{eqn:Reference:TruncatedWellPosedness:DiffusionControl} into~\eqref{eqn:Reference:TruncatedWellPosedness:M:1} yields
    \begin{align}\label{eqn:Reference:TruncatedWellPosedness:M:Final}
        M(z(t)) 
        =& \int_0^t \FNmu{\nu,z(\nu),\Urt{\nu}}d\nu + \int_0^t \FNsigma{\nu,z(\nu)}d\Wt{\nu} \notag 
        \\
        =& \int_0^t M_\mu\br{t,\nu,z(\nu)}d\nu + \int_0^t M_\sigma\br{t,\nu,z(\nu)}d\Wt{\nu}, \quad t \in [0,T],
    \end{align}
    where 
    \begin{align*}
        M_\mu\br{t,\nu,z(\nu)} = f_N\br{\nu,z(\nu)} +  \Lambda_{N,\mu} \br{\nu, z(\nu)}   
        -\Boldomega \br{ \int_\nu^t g(\beta) \expo{-\Boldomega \beta} d\beta } \expo{\Boldomega \nu } \Lambda_{N,\mu}^{\paral} \br{\nu,z(\nu)},
        \\
        M_\sigma\br{t,\nu,z(\nu)} = p_N(\nu,z(\nu)) + \Lambda_{N,\sigma} \br{\nu,z(\nu)} 
        -\Boldomega   
        \br{ \int_\nu^t g(\beta) \expo{-\Boldomega \beta} d\beta } \expo{\Boldomega \nu } 
        \br{ p_N^{\paral}\br{\nu,z(\nu)}  + \Lambda_{N,\sigma}^{\paral} \br{\nu,z(\nu)} },
    \end{align*}
    for $\nu \in [0,t]$, and $t \in [0,T]$.
    Using Assumption~\ref{assmp:KnownFunctions}, we have that 
    \begin{multline}\label{eqn:Reference:TruncatedWellPosedness:FilterBound}
        \norm{ \br{ \int_\nu^t g(\beta) \expo{-\Boldomega \beta} d\beta } \expo{\Boldomega \nu } }_F
        \\
        \leq 
        \int_\nu^t \norm{g(\beta)}_F \expo{-\Boldomega \beta} d\beta  \expo{\Boldomega \nu }
        \leq 
        \Delta_g \frac{1 - \expo{-\Boldomega(t - \nu)}}{\Boldomega} 
        \leq 
        \Delta_g \frac{1}{\Boldomega}, \quad \forall \nu \in [0,t],\,t \in [0,T].
    \end{multline}
    Thus, we conclude that 
    \begin{subequations}\label{eqn:Reference:TruncatedWellPosedness:M:Bounds}
        \begin{align}
            \norm{M_\mu\br{t,\nu,z(\nu)}} \leq  \norm{f_N\br{\nu,z(\nu)}} +  \norm{\Lambda_{N,\mu} \br{\nu, z(\nu)}}   
            + \Delta_g  \norm{ \Lambda_{N,\mu}^{\paral} \br{\nu,z(\nu)}},
            \\
            \norm{M_\sigma\br{t,\nu,z(\nu)}}_F = \norm{p_N(\nu,z(\nu))}_F + \norm{\Lambda_{N,\sigma} \br{\nu,z(\nu)}}_F 
            + \Delta_g  
            \norm{ p_N^{\paral}\br{\nu,z(\nu)}  + \Lambda_{N,\sigma}^{\paral} \br{\nu,z(\nu)} }_F,
        \end{align}
    \end{subequations}
    for all $\nu \in [0,t]$, and $t \in [0,T]$.

    Let us set $x(t) \equiv x_0$, and define the Picard iterates for~\eqref{eqn:Reference:TruncatedWellPosedness:System:2} as
    \begin{align*}
        x_k(t) = x_0 + \int_0^t \FNmu{\nu,x_{k-1}(\nu),\Urt{\nu}}d\nu + \int_0^t \FNsigma{\nu,x_{k-1}(\nu)}d\Wt{\nu}, \quad k \in \mathbb{N}, \quad t \in [0,T]. 
    \end{align*} 
    Then, by the definition of $M$ in~\eqref{eqn:Reference:TruncatedWellPosedness:M}, we have that
    \begin{align*}
        x_k(t) 
        = 
        x_0 + M\br{x_{k-1}(t)} 
        = 
        x_0 +  \int_0^t M_\mu\br{t,\nu,x_{k-1}(\nu)}d\nu + \int_0^t M_\sigma\br{t,\nu,x_{k-1}(\nu)}d\Wt{\nu}, \quad t \in [0,T]. 
    \end{align*}
    Since the truncated functions $f_N$, $\Lambda_{N,\mu}$, $p_N$, and $\Lambda_{N,\sigma}$ agree with their non-truncated counterparts on $[0,\tau_N]$, we have that over the interval $[0,T] \supseteq [0,\tau_N]$, the Assumptions~\ref{assmp:KnownFunctions} and~\ref{assmp:UnknownFunctions}, the truncation definition in~\eqref{eqn:Reference:TruncatedJointProcess},  
    along with the bounds in~\eqref{eqn:Reference:TruncatedWellPosedness:M:Bounds}, imply linear growth of the integrands in the Picard iterates above. 
    Therefore, as in the proof of~\cite[Thm.~2.3.1]{mao2007stochastic}, we claim the existence of solutions to~\eqref{eqn:Reference:TruncatedWellPosedness:System:2} on $[0,T]$.  

    Similarly, using the assumptions of local Lipschitz continuity of $f$, and the global Lipschitz continuity of $\Lambda_{\mu}$, $p$, and $\Lambda_{\sigma}$, we can use the same arguments as above to establish the uniqueness of solutions to~\eqref{eqn:Reference:TruncatedWellPosedness:System:2} on $[0,T]$ as in the proof of~\cite[Thm.~2.3.1]{mao2007stochastic}.

    Furthermore, as above, using the definition of the truncation in~\eqref{eqn:Reference:TruncatedJointProcess}, we can use~\eqref{eqn:Reference:TruncatedWellPosedness:M:Bounds} and show linear growth bounds and Lipschitz continuity for $F_{N,\cbr{\mu,\sigma}}$, globally over $\mathbb{R}^n$ and uniformly in $t \in \mathbb{R}_{\geq 0}$, thus implying the strong Markov property of the solutions by~\cite[Thm.~2.9.3]{mao2007stochastic}.

    Finally, since $G_{N,\cbr{\mu,\sigma}}\br{t, \cdot} = G_{\cbr{\mu,\sigma}}\br{t, \cdot}$ for all $t \in [0,\tau_N]$, we may invoke~\cite[Thm.~3.5]{khasminskii2011stochastic},~\cite[Thm.~5.2.9]{karatzas1991brownian} to conclude that $\Yt{N,t}$ is a unique solution to~\eqref{eqn:Reference:JointProcess} on $[0,\tau_N]$.
\end{proof}

The next two results help us to derive the stochastic differential of $\RefNorm{\Yt{N,t}}$ in the proof of Lemma~\ref{lem:Reference:dV}. 
\begin{proposition}\label{prop:Appendix:ReferenceProcess:dV}
    Let $\Yt{N,t}$ be the strong solution of~\eqref{eqn:Reference:TruncatedJointProcess}, and let $\tau(t)$ be the stopping time defined in~\eqref{eqn:Reference:StoppingTimes}, Lemma~\ref{lem:Reference:dV}. 
    Moreover, define 
    \begin{align*}
        \RefNorm{\Yt{N}} \doteq \norm{\Xrt{N} - \Xstart{N}}^2,
        \quad 
        \RefDiffNorm{\Yt{N}} 
        \doteq 
        \partial_{\Xrt{N} - \Xstart{N}}\RefNorm{\Yt{N}} 
        = 
        2 \left(\Xrt{N} - \Xstart{N}\right) \in \mathbb{R}^n.
    \end{align*}
    Then,
    \begin{subequations}\label{eqn:Appendix:ReferenceProcess:prop:dV:Bound}
        \begin{align}
            \begin{multlined}[b][0.9\textwidth]
                \int_0^{\tau(t)}  
                    e^{2 \lambda \nu} 
                    \left( 
                        \nabla \RefNorm{\Yt{N,\nu}}^\top \Grmu{\nu, \Yt{N,\nu}}
                        +
                        \frac{1}{2} \Trace{ \Hrsigma{\nu, \Yt{N,\nu}} \nabla^2 \RefNorm{\Yt{N,\nu}}  }
                    \right)
                d\nu
                \\
                \leq
                -2 \lambda
                \int_0^{\tau(t)}  
                    e^{2 \lambda \nu}    
                    \RefNorm{\Yt{N,\nu}}       
                d\nu
                + 
                \int_0^{\tau(t)}  
                    e^{2 \lambda \nu}  
                    \phi^r_{U}\br{\nu,\Yt{N,\nu}}
                d\nu
                \\
                + 
                \int_0^{\tau(t)}  
                    e^{2 \lambda \nu}  
                    \left(
                        \phi^r_{\mu}\br{\nu,\Yt{N,\nu}}
                        + 
                        \phi^r_{\mu^{\paral}}\br{\nu,\Yt{N,\nu}}
                    \right)
                d\nu
                ,
            \end{multlined}
            \label{eqn:Appendix:ReferenceProcess:prop:dV:Bound:A}
            \\
            \begin{multlined}[b][0.9\textwidth]
                \int_0^{\tau(t)}  
                    e^{2 \lambda \nu} 
                    \nabla \RefNorm{\Yt{N,\nu}}^\top \Grsigma{\nu,\Yt{N,\nu}} 
                d \Wrt{\nu}
                \\
                =
                \int_0^{\tau(t)}  
                    e^{2 \lambda \nu}
                    \left( 
                        \phi^r_{\sigma_\star}\br{\nu,\Yt{N,\nu}} 
                        d\Wstart{\nu}
                        +
                        \left[ 
                            \phi^r_{\sigma}\br{\nu,\Yt{N,\nu}}
                            + 
                            \phi^r_{\sigma^{\paral}}\br{\nu,\Yt{N,\nu}}
                        \right]
                        d\Wt{\nu}
                    \right)
                ,  
            \end{multlined}
            \label{eqn:Appendix:ReferenceProcess:prop:dV:Bound:B}
        \end{align}
    \end{subequations}
    for all $ t \in \mathbb{R}_{\geq 0}$, where $\Hrsigma{\nu, \Yt{N,\nu}} = \Grsigma{\nu,\Yt{N,\nu}} \Grsigma{\nu,\Yt{N,\nu}}^\top$, $\Grmu{\nu,\Yt{N,\nu}}$ and $\Grsigma{\nu,\Yt{N,\nu}}$ are defined in~\eqref{eqn:Reference:JointProcess}.
    Additionally, we have defined 
    \begin{align*} 
        \phi^r_{\mu^{\paral}}\br{\nu,\Yt{N,\nu}}
        =
        \RefDiffNorm{\Yt{N,\nu}}^\top
        g(\nu) 
        \Lparamu{\nu,\Xrt{N,\nu}},
        \\
        \phi^r_{\mu}\br{\nu,\Yt{N,\nu}}
        =
        \RefDiffNorm{\Yt{N}}^\top 
        g(\nu)^\perp\Lperpmu{\nu,\Xrt{N,\nu}} 
        +
        \Frobenius{\Fsigma{\nu,\Xrt{N,\nu}}}^2
        +
        \Frobenius{\Fbarsigma{\nu,\Xstart{N,\nu}}}^2,
        \\
        \phi^r_{U}\br{\nu,\Yt{N,\nu}}
        =
        \RefDiffNorm{\Yt{N}}^\top
        g(\nu) 
        \Urt{\nu},
    \end{align*}
    and 
    \begin{align*}
        \phi^r_{\sigma^{\paral}}\br{\nu,\Yt{N,\nu}}
        =
        \RefDiffNorm{\Yt{N}}^\top  
        g(\nu) \Fparasigma{\nu,\Xrt{N, \nu}},
        \quad 
        \phi^r_{\sigma_\star}\br{\nu,\Yt{N,\nu}} 
        = 
        -\RefDiffNorm{\Yt{N}}^\top 
        \Fbarsigma{\nu,\Xstart{N,\nu}},
        \\
        \phi^r_{\sigma}\br{\nu,\Yt{N,\nu}} 
        = 
        \RefDiffNorm{\Yt{N}}^\top  
        g(\nu)^\perp \Fperpsigma{\nu,\Xrt{N,\nu}}. 
    \end{align*}
\end{proposition}
\begin{proof}
    Using the definition of $G_\mu$ from~\eqref{eqn:Reference:JointProcess} and the decomposition of $F_\mu$ in~\eqref{eqn:VectorFields:Decomposition}, we have that 
    \begin{multline*}
        \nabla \RefNorm{\Yt{N,\nu}}^\top \Grmu{\nu, \Yt{N,\nu}}
        =
        2 \left(\Xrt{N,\nu} - \Xstart{N,\nu}\right)^\top 
        \left(
            \Fbarmu{\nu,\Xrt{N,\nu}} - \Fbarmu{\nu,\Xstart{N,\nu}} 
        \right)
        \\
        + 
        2 \left(\Xrt{N,\nu} - \Xstart{N,\nu}\right)^\top 
        \br{ g(\nu)\Urt{\nu}  + \Lmu{\nu,\Xrt{N,\nu}} }, 
        \quad \nu \in [0,\tau(t)]
        ,
    \end{multline*}
    which, due to Lemma~\ref{lem:ILFConditions:Consequence} reduces to 
    \begin{align}\label{eqn:Appendix:ReferenceProcess:prop:dV:Term1:A:1}
        \nabla \RefNorm{\Yt{N,\nu}}^\top \Grmu{\nu, \Yt{N,\nu}}
        \leq 
        -2 \lambda  
        \RefNorm{\Yt{N,\nu}}
        + 
        2 \left(\Xrt{N,\nu} - \Xstart{N,\nu}\right)^\top 
        \br{ g(\nu)\Urt{\nu}  + \Lmu{\nu,\Xrt{N,\nu}} }, 
    \end{align}
    for all $\nu \in [0,\tau(t)]$.
    We develop the expression further by using~\eqref{eqn:UnknownFunctions:Decomposition} in Assumption~\ref{assmp:UnknownFunctions} to write 
    \begin{align*}
        \Lmu{\nu,\Xrt{N,\nu}}
        =
        \begin{bmatrix} g(\nu) & g(\nu)^\perp \end{bmatrix}
        \begin{bmatrix} \Lparamu{\nu,\Xrt{N,\nu}} \\ \Lperpmu{\nu,\Xrt{N,\nu}} \end{bmatrix}
        = g(\nu)\Lparamu{\nu,\Xrt{N,\nu}} + g(\nu)^\perp\Lperpmu{\nu,\Xrt{N,\nu}}.
    \end{align*}
    Substituting into~\eqref{eqn:Appendix:ReferenceProcess:prop:dV:Term1:A:1} yields
    \begin{multline*}
        \nabla \RefNorm{\Yt{N,\nu}}^\top \Grmu{\nu, \Yt{N,\nu}}
        \leq 
        -2 \lambda  
        \RefNorm{\Yt{N,\nu}}
        + 
        2 \left(\Xrt{N,\nu} - \Xstart{N,\nu}\right)^\top 
        g(\nu)^\perp\Lperpmu{\nu,\Xrt{N,\nu}} 
        \\
        + 
        2 \left(\Xrt{N,\nu} - \Xstart{N,\nu}\right)^\top
        g(\nu) 
        \left(
            \Urt{\nu}  + \Lparamu{\nu,\Xrt{N,\nu}} 
        \right), 
        \quad \forall \nu \in [0,\tau(t)],
    \end{multline*}
    which then leads to~\eqref{eqn:Appendix:ReferenceProcess:prop:dV:Bound:A} by further noting that 
    \begin{align*}
        \frac{1}{2} \Trace{ \Hrsigma{\nu, \Yt{N,\nu}}  \nabla^2 \RefNorm{\Yt{N,\nu}}}
        =&
        \Frobenius{\begin{bmatrix} \Fsigma{\nu,\Xrt{N,\nu}} & -\Fbarsigma{\nu,\Xstart{N,\nu}} \end{bmatrix}}^2
        \\
        =&
        \Frobenius{\Fsigma{\nu,\Xrt{N,\nu}}}^2
        +
        \Frobenius{\Fbarsigma{\nu,\Xstart{N,\nu}}}^2.
    \end{align*} 

    Next, using the definition of $G_\sigma$ from~\eqref{eqn:Reference:JointProcess}, we have that 
    \begin{align*}
        &\int_0^{\tau(t)}  
            e^{2 \lambda \nu} 
            \nabla \RefNorm{\Yt{N,\nu}}^\top \Grsigma{\nu,\Yt{N,\nu}} 
        d \Wrt{\nu}
        \\
        &=
        2
        \int_0^{\tau(t)}  
            e^{2 \lambda \nu} 
            \left(\Xrt{N,\nu} - \Xstart{N,\nu}\right)^\top 
            \left( 
                \Fsigma{\nu,\Xrt{\nu}}d \Wt{\nu}
                - \Fbarsigma{\nu,\Xstart{\nu}}d \Wstart{\nu}
            \right) 
        \\
        &=
        -2
        \int_0^{\tau(t)}  
            e^{2 \lambda \nu} 
            \left(\Xrt{N,\nu} - \Xstart{N,\nu}\right)^\top 
                \Fbarsigma{\nu,\Xstart{\nu}}d \Wstart{\nu}
        \\
        & \quad 
        +
        2
        \int_0^{\tau(t)}  
            e^{2 \lambda \nu} 
            \left(\Xrt{N,\nu} - \Xstart{N,\nu}\right)^\top  
            \left( p\br{\nu,\Xrt{N,\nu}} + \Lsigma{\nu,\Xrt{N,\nu}}  \right)
        d\Wt{\nu}
        , 
         \quad  t \in \mathbb{R}_{\geq 0}, 
    \end{align*}
    where we have used the definition of $F_\sigma$ in~\eqref{eqn:TrueVectorFields}.
    Since~\eqref{eqn:UnknownFunctions:Decomposition} and~\eqref{eqn:knownDiffusion:Decomposition:p} in Assumptions~\ref{assmp:UnknownFunctions} and~\ref{assmp:knownDiffusion:Decomposition}, respectively, along with Definition~\ref{def:Diffusion:Decomposed} imply that  
    \begin{multline*}
        p\br{\nu,\Xrt{N,t}} + \Lsigma{\nu,\Xrt{N,t}}
        \\
        = 
        g(\nu)^\perp\pperp{\nu,\Xrt{N,t}} + g(\nu)^\perp\Lperpsigma{\nu,\Xrt{N,t}}
        + g(\nu)\ppara{\nu,\Xrt{N,t}} + g(\nu)\Lparasigma{\nu,\Xrt{N,t}}
        \\
        =
        g(\nu)^\perp \Fperpsigma{\nu,\Xrt{N,t}} + g(\nu) \Fparasigma{\nu,\Xrt{N,t}}
        , \quad \forall \nu \in [0,\tau(t)],
    \end{multline*}
    the previous integral equality can be re-written as 
    \begin{multline*}
        \int_0^{\tau(t)}  
            e^{2 \lambda \nu} 
            \nabla \RefNorm{\Yt{N,\nu}}^\top \Grsigma{\nu,\Yt{N,\nu}} 
        d \Wrt{\nu}
        \\
        =
        -2
        \int_0^{\tau(t)}  
            e^{2 \lambda \nu} 
            \left(\Xrt{N,\nu} - \Xstart{N,\nu}\right)^\top 
                \Fbarsigma{\nu,\Xstart{N,\nu}}d \Wstart{\nu}
        +
        2
        \int_0^{\tau(t)}  
            e^{2 \lambda \nu} 
            \left(\Xrt{N,\nu} - \Xstart{N,\nu}\right)^\top  
            g(\nu)^\perp \Fperpsigma{\nu,\Xrt{N,\nu}}
        d\Wt{\nu}
        \\
        +
        2
        \int_0^{\tau(t)}  
            e^{2 \lambda \nu} 
            \left(\Xrt{N,\nu} - \Xstart{N,\nu}\right)^\top  
            g(\nu) \Fparasigma{\nu,\Xrt{N,\nu}}
        d\Wt{\nu}
        , 
         \quad  t \in \mathbb{R}_{\geq 0}, 
    \end{multline*}
    thus establishing the expression in~\eqref{eqn:Appendix:ReferenceProcess:prop:dV:Bound:B}.

\end{proof}

In the subsequent proposition, we derive the expression for how the reference feedback process $\Urt{}$~\eqref{eqn:ReferenceFeedbackOperatorProcess} enters the truncated joint process $\Yt{N,t}$.
\begin{proposition}\label{prop:Appendix:ReferenceProcess:dV:U}
    Let $\Yt{N,t}$ be the strong solution of~\eqref{eqn:Reference:TruncatedJointProcess}, and let $\tau(t)$ be the stopping time defined in~\eqref{eqn:Reference:StoppingTimes}, Lemma~\ref{lem:Reference:dV}. 
    Then, for the term $\phi^r_U$ defined in the statement of Proposition~\ref{prop:Appendix:ReferenceProcess:dV}, we have that    
    \begin{multline}\label{eqn:Appendix:ReferenceProcess:prop:dV:U}
        \int_0^{\tau(t)}  \expo{2\lambda \nu } 
        \phi^r_{U}\br{\nu,\Yt{N,\nu}}
        d\nu
        = 
        \int_0^{\tau(t)} 
        \left(
            \hat{\mathcal{U}}^r_{\mu}\br{\tau(t),\nu,\Yt{N};\Boldomega}d\nu
            +
            \hat{\mathcal{U}}^r_{\sigma}\br{\tau(t),\nu,\Yt{N};\Boldomega}d\Wt{\nu}
        \right) 
        \\
        + 
        \int_0^{\tau(t)}
            \expo{2\lambda \nu}
            \left( 
                \phi^r_{U_\mu}\br{\nu,\Yt{N,\nu};\Boldomega}d\nu
                +
                \phi^r_{U_\sigma}\br{\nu,\Yt{N,\nu};\Boldomega}d\Wt{\nu}
            \right)
        ,
        \quad t \in \mathbb{R}_{\geq 0},
    \end{multline}
    where 
    \begin{align*}
        \begin{multlined}[b][0.9\linewidth]
            \hat{\mathcal{U}}^r_{\mu}\br{\tau(t),\nu,\Yt{N};\Boldomega}
            \\
            =
            \expo{-\Boldomega \tau(t)}
            \frac{\Boldomega}{2\lambda - \Boldomega}
            \left(
                \expo{\Boldomega \tau(t)}  
                \mathcal{P}^r\br{\tau(t),\nu}
                -
                \expo{ 2\lambda \tau(t)}
                \RefDiffNorm{\Yt{N,\tau(t)}}^\top
                g(\tau(t))
            \right)
            \expo{\Boldomega \nu}
            \Lparamu{\nu, \Xrt{N,\nu}}
            ,
        \end{multlined}
        \\
        \begin{multlined}[b][0.9\linewidth]
            \hat{\mathcal{U}}^r_{\sigma}\br{\tau(t),\nu,\Yt{N};\Boldomega}
            \\
            =
            \expo{-\Boldomega \tau(t)}
            \frac{\Boldomega}{2\lambda - \Boldomega}
            \left( 
                \expo{\Boldomega \tau(t)} 
                \mathcal{P}^r\br{\tau(t),\nu}
                -
                \expo{2\lambda \tau(t)}
                \RefDiffNorm{\Yt{N,\tau(t)}}^\top g(\tau(t))
            \right)
            \expo{\Boldomega \nu}
            \Fparasigma{\nu,\Xrt{N,\nu}}
            ,
        \end{multlined}
        \\
        \phi^r_{U_\mu}\br{\nu,\Yt{N,\nu};\Boldomega}
        =
        \frac{\Boldomega}{2\lambda - \Boldomega}
        \RefDiffNorm{\Yt{N,\nu}}^\top g(\nu) 
        \Lparamu{\nu, \Xrt{N,\nu}},
        \\
        \phi^r_{U_\sigma}\br{\nu,\Yt{N,\nu};\Boldomega}
        =
        \frac{\Boldomega}{2\lambda - \Boldomega}
        \RefDiffNorm{\Yt{N,\nu}}^\top g(\nu)
        \Fparasigma{\nu,\Xrt{N,\nu}} 
        ,
    \end{align*}
    and 
    \begin{align*}
        \mathcal{P}^r\br{\tau(t),\nu}
        =
        \int_\nu^{\tau(t)}
            e^{ (2\lambda - \Boldomega) \beta }  d_\beta\sbr{\RefDiffNorm{\Yt{N,\beta}}^\top g(\beta)}
        .
    \end{align*}
\end{proposition}
\begin{proof}
    Using the definition of $\phi^r_U$ in the statement of Proposition~\ref{prop:Appendix:ReferenceProcess:dV}, we have that 
    \begin{align*}
        &\int_0^{\tau(t)}  \expo{ 2\lambda \nu} 
        \phi^r_{U}\br{\nu,\Yt{N,\nu}}
        d\nu
        \\
        &
        =
        \int_0^{\tau(t)}  
            \expo{ 2\lambda \nu } 
            \RefDiffNorm{\Yt{N,\nu}}^\top
            g(\nu) 
            \Urt{\nu}
        d\nu
        \\
        &=
        2
        \int_0^{\tau(t)}  
            \expo{ 2\lambda \nu }  
            \RefDiffNorm{\Yt{N,\nu}}^\top
            g(\nu) 
            \left( 
                \Filter[\Lparamu{\cdot,\Xrt{}}][\nu] + \FilterW[\ppara{\cdot,\Xrt{}}+\Lparasigma{\cdot,\Xrt{}}, \Wt{}][\nu] 
            \right) 
        d\nu,
    \end{align*}
    for all $t \geq 0$, where we have incorporated the definition of $\Urt{}$ from~\eqref{eqn:ReferenceFeedbackOperatorProcess}.
    Using the definitions of $\Filter$ and $\FilterW[\cdot,\Wt{}]$ in~\eqref{eqn:L1DRAC:Definition:FeedbackOperator:Filter} and~\eqref{eqn:ReferenceFeedbackOperatorProcess:BrownianFilter}, respectively, we get  
    \begin{multline*}
            \int_0^{\tau(t)}  \expo{ 2\lambda \nu } 
            \phi^r_{U}\br{\nu,\Yt{N,\nu}}
            d\nu
            \\
            =
            \int_0^{\tau(t)} 
            \int_0^\nu  
                \left(-\Boldomega \expo{(2\lambda - \Boldomega)\nu}\right)  
                \RefDiffNorm{\Yt{N,\nu}}^\top 
                g(\nu) 
                \left(\expo{ \Boldomega \beta} \Lparamu{\beta, \Xrt{N,\beta}}d\beta \right) 
            d\nu
        \\
        + 
        \int_0^{\tau(t)} \int_0^\nu  
            \left(-\Boldomega \expo{(2\lambda - \Boldomega)\nu}\right)  
            \RefDiffNorm{\Yt{N,\nu}}^\top 
            g(\nu) 
            \left(
                \expo{ \Boldomega \beta} 
                \Fparasigma{\beta,\Xrt{N,\beta}}
                d\Wt{\beta}
            \right) 
        d\nu,
    \end{multline*}
    for all $t \in \mathbb{R}_{\geq 0}$.
    Changing the order of integration in the first integral on the right hand side, and applying Lemma~\ref{lem:TechnicalResults:Fubini-ish} to the second integral, we get:
    \begin{multline}\label{eqn:Appendix:ReferenceProcess:prop:dV:U:1}
        \begin{aligned}
            &\int_0^{\tau(t)}  \expo{ 2\lambda \nu } 
            \phi^r_{U}\br{\nu,\Yt{N,\nu}}
            d\nu
            \\
            &
            = 
            \int_0^{\tau(t)}
                \left(
                    - 
                    \int_\nu^{\tau(t)}  
                        \Boldomega \expo{(2\lambda - \Boldomega)\beta}
                        \RefDiffNorm{\Yt{N,\beta}}^\top 
                        g(\beta)
                    d\beta 
                \right)
                \expo{ \Boldomega \nu} \Lparamu{\nu, \Xrt{N,\nu}}
            d\nu 
        \end{aligned}
        \\
        + 
        \int_0^{\tau(t)}   
            \left(
                - 
                \int_\nu^{\tau(t)}  
                    \Boldomega \expo{(2\lambda - \Boldomega)\beta}
                    \RefDiffNorm{\Yt{N,\beta}}^\top
                    g(\beta)
                d\beta 
            \right) 
            \expo{ \Boldomega \nu}\Fparasigma{\nu,\Xrt{N,\nu}}
            d\Wt{\nu}
            ,
    \end{multline}
    for all $t \in \mathbb{R}_{\geq 0}$, where in the first integral, we switch between the variables $\beta$ and $\nu$ after changing the order of integration.
    Now, the chain rule implies that  
    \begin{align*}
        d_\beta \sbr{e^{ (2\lambda - \Boldomega) \beta }  \RefDiffNorm{\Yt{N,\beta}}^\top g(\beta)}
        = \br{\frac{d}{d\beta} e^{ (2\lambda - \Boldomega) \beta}} 
        \RefDiffNorm{\Yt{N,\beta}}^\top 
        g(\beta)
        + 
        e^{ (2\lambda - \Boldomega) \beta }  
        d_\beta\sbr{\RefDiffNorm{\Yt{N,\beta}}^\top g(\beta)},
    \end{align*}
    where $d_\beta\sbr{\cdot}$ in the last term is interpreted as the stochastic differential with respect to the variable $\beta$. 
    Multiplying both sides by $-\frac{\Boldomega}{2\lambda - \Boldomega}$ yields 
    \begin{multline*}
        -\frac{\Boldomega}{2\lambda - \Boldomega}
        d_\beta \sbr{e^{ (2\lambda - \Boldomega) \beta }  \RefDiffNorm{\Yt{N,\beta}}^\top g(\beta)}
        \\
        = 
        -\frac{\Boldomega}{2\lambda - \Boldomega}
        \br{\frac{d}{d\beta} e^{ (2\lambda - \Boldomega) \beta}} \RefDiffNorm{\Yt{N,\beta}}^\top g(\beta)
        -
        \frac{\Boldomega}{2\lambda - \Boldomega}
        e^{ (2\lambda - \Boldomega) \beta }  d_\beta\sbr{\RefDiffNorm{\Yt{N,\beta}}^\top g(\beta)}
        \\
        = 
        -\Boldomega e^{ (2\lambda - \Boldomega) \beta} \RefDiffNorm{\Yt{N,\beta}}^\top g(\beta)
        - 
        \frac{\Boldomega}{2\lambda - \Boldomega}
        e^{ (2\lambda - \Boldomega) \beta }  d_\beta\sbr{\RefDiffNorm{\Yt{N,\beta}}^\top g(\beta)},
    \end{multline*}
    and thus 
    \begin{multline*}
        -\Boldomega e^{ (2\lambda - \Boldomega) \beta} \RefDiffNorm{\Yt{N,\beta}}^\top g(\beta)
        =
        -\frac{\Boldomega}{2\lambda - \Boldomega}
        d_\beta \sbr{e^{ (2\lambda - \Boldomega) \beta }  \RefDiffNorm{\Yt{N,\beta}}^\top g(\beta)}
        \\
        + 
        \frac{\Boldomega}{2\lambda - \Boldomega}
        e^{ (2\lambda - \Boldomega) \beta }  d_\beta\sbr{\RefDiffNorm{\Yt{N,\beta}}^\top g(\beta)}
        .
    \end{multline*}
    Integrating over the interval $[\nu,\tau(t)]$ leads to  
    \begin{multline*}
        -
        \int_\nu^{\tau(t)}  
            \Boldomega e^{ (2\lambda - \Boldomega) \beta} 
            \RefDiffNorm{\Yt{N,\beta}}^\top g(\beta)
        d\beta
        \\
        =
        -
        \frac{\Boldomega}{2\lambda - \Boldomega}
        \expo{ (2\lambda - \Boldomega) \tau(t) }  \RefDiffNorm{\Yt{N,\tau(t)}}^\top g(\tau(t))
        +
        \frac{\Boldomega}{2\lambda - \Boldomega}
        \expo{ (2\lambda - \Boldomega) \nu }  \RefDiffNorm{\Yt{N,\nu}}^\top g(\nu)
        \\
        + 
        \frac{\Boldomega}{2\lambda - \Boldomega}
        \mathcal{P}^r\br{\tau(t),\nu}
        ,
    \end{multline*}
    where
    \begin{align*}
        \mathcal{P}^r\br{\tau(t),\nu}
        =
        \int_\nu^{\tau(t)}
            e^{ (2\lambda - \Boldomega) \beta }  d_\beta\sbr{\RefDiffNorm{\Yt{N,\beta}}^\top g(\beta)}
        .
    \end{align*}
    Substituting the expression above for the two identical inner integrals on the right hand side of~\eqref{eqn:Appendix:ReferenceProcess:prop:dV:U:1} produces~\eqref{eqn:Appendix:ReferenceProcess:prop:dV:U}, thus completing the proof upon re-arranging terms.

\end{proof}

The next result provides an alternative representation of $\mathcal{P}^r$ that is amenable to the analysis of the reference process.
\begin{proposition}\label{prop:Appendix:ReferenceProcess:ddV:Bound}
    Recall the expression for $\mathcal{P}^r\br{\tau(t),\nu}$ in~\eqref{eqn:Reference:ddV} in the statement of Lemma~\ref{lem:Reference:dV}, which we restate below:
    \begin{align}\label{eqn:lem:Appendix:ReferenceProcess:ddV:Expression:Main}
        \mathcal{P}^r\br{\tau(t),\nu} = 
        \int_\nu^{\tau(t)} 
            \expo{ (2\lambda - \Boldomega) \beta }  
            d_\beta \sbr{\RefDiffNorm{\Yt{N,\beta}}^\top g(\beta)} 
        \in \mathbb{R}^{1 \times m}, \quad 0 \leq \nu \leq \tau(t),
    \end{align}
    where $\tau(t)$ is defined in~\eqref{eqn:Reference:StoppingTimes}, and $\RefDiffNorm{\Yt{N,t}} \doteq \partial_{\Xrt{N} - \Xstart{N}}\RefNorm{\Yt{N}} = 2 \left(\Xrt{N} - \Xstart{N}\right) \in \mathbb{R}^n$.
    Then, $\mathcal{P}^r\br{\tau(t),\nu}$ admits the following representation:
    \begin{equation}\label{eqn:lem:Appendix:ReferenceProcess:ddV:Expression:Pr:Main}
        \mathcal{P}^r\br{\tau(t),\nu}
        =
        \mathcal{P}^r_\circ \br{\tau(t),\nu}
        +
        \mathcal{P}^r_{ad} \br{\tau(t),\nu}
        \in \mathbb{R}^{1 \times m},
        \quad 
        0 \leq \nu \leq \tau(t), ~t \in \mathbb{R}_{\geq 0},
    \end{equation}
    where 
    \begin{subequations}
        \begin{align}
            \mathcal{P}^r_{\circ}\br{\tau(t),\nu}
            =  
            \int_\nu^{\tau(t)} 
                \expo{ (2\lambda - \Boldomega) \beta } 
                \left(
                    \mathcal{P}^r_{\mu}(\beta) 
                    d\beta 
                    +
                    \mathcal{P}^r_{\sigma}(\beta)
                    d\Wt{\beta}
                    +
                    \mathcal{P}^r_{\sigma_\star}(\beta)
                    d\Wstart{\beta}
                \right)^\top
            \in \mathbb{R}^{1 \times m},
            \label{eqn:lem:Appendix:ReferenceProcess:ddV:Expression:Pr}
            \\
            \mathcal{P}^r_{ad}\br{\tau(t),\nu}
            = 
            \int_\nu^{\tau(t)} 
                \expo{ (2\lambda - \Boldomega) \beta } 
                \mathcal{P}^r_\mathcal{U}(\beta)^\top 
            d\beta \in \mathbb{R}^{1 \times m},
            \label{eqn:lem:Appendix:ReferenceProcess:ddV:Expression:PrU}
        \end{align}
    \end{subequations}
    and where $\mathcal{P}^r_\mathcal{U}(\beta),\,\mathcal{P}^r_{\mu}(\beta) \in \mathbb{R}^m$ and $\mathcal{P}^r_{\sigma}(\beta),\,\mathcal{P}^r_{\sigma_\star}(\beta) \in \mathbb{R}^{m \times d}$ are defined as
    \begin{align*}
        \mathcal{P}^r_{\mu}(\beta)
        =
        \dot{g}(\beta)^\top
        \RefDiffNorm{\Yt{N,\beta}}
        + 
        2
        g(\beta)^\top
        \left(
            \Fbarmu{\beta,\Xrt{N,\beta}}
            - 
            \Fbarmu{\beta,\Xstart{N,\beta}}    
            + 
            \Lmu{\beta,\Xrt{N,\beta}}  
        \right)
        ,
        \\
        \mathcal{P}^r_\mathcal{U}(\beta)
        =
        2 
        g(\beta)^\top
        g(\beta)
        \Urt{\beta}
        , 
        \\
        \mathcal{P}^r_{\sigma}(\beta)
        =
        2
        g(\beta)^\top 
        \Fsigma{\beta,\Xrt{N,\beta}}
        , 
        \quad 
        \mathcal{P}^r_{\sigma_\star}(\beta)
        =
        -
        2
        g(\beta)^\top
        \Fbarsigma{\beta,\Xstart{N,\beta}} 
        .
    \end{align*}

\end{proposition}
\begin{proof}
    We begin by writing $\RefDiffNorm{\Yt{N,\beta}}^\top g(\beta)$ as
    \begin{align}\label{eqn:lem:Appendix:ReferenceProcess:ddV:Kernel:Decomposition}
        \RefDiffNorm{\Yt{N,\beta}}^\top g(\beta)
        =&
        \begin{bmatrix} 
            \RefDiffNorm{\Yt{N,\beta}}^\top g_{\cdot,1}(\beta) & \cdots &  
            \RefDiffNorm{\Yt{N,\beta}}^\top g_{\cdot,m}(\beta)   
        \end{bmatrix}
        \notag  
        \\
        =&
        \begin{bmatrix} 
            \sum_{i=1}^n \RefDiffNormPoint{\Yt{N,\beta}} g_{i,1}(\beta) & \cdots &  
            \sum_{i=1}^n \RefDiffNormPoint{\Yt{N,\beta}} g_{i,m}(\beta)   
        \end{bmatrix} \in \mathbb{R}^{1 \times m},
    \end{align} 
    where $g_{\cdot, j}(\beta) \in \mathbb{R}^n$ is the $j$-th column of $g(\beta)$, and 
    \begin{align*}
        \RefDiffNormPoint{\Yt{N,\beta}} \doteq \frac{\partial}{\partial \left(\Xrt{N} - \Xstart{N}\right)_i}\RefNorm{\Yt{N,\beta}} = 2\left(\Xrt{N,\beta} - \Xstart{N,\beta}\right)_i  \in \mathbb{R}, \quad i \in \cbr{1,\dots,n}.
    \end{align*} 
    Applying \ito's lemma to $\RefDiffNormPoint{\Yt{N,\beta}}g_{i,j}(\beta) \in \mathbb{R}$, $(i,j) \in \cbr{1,\dots,n} \times \cbr{1,\dots,m}$, and using the truncated dynamics in~\eqref{eqn:Reference:TruncatedJointProcess}  we get
    \begin{multline}\label{eqn:lem:Appendix:ReferenceProcess:ddV:Ito:Initial}
        d_\beta \sbr{\RefDiffNormPoint{\Yt{N,\beta}}g_{i,j}(\beta)} 
        =
        \left( 
            \RefDiffNormPoint{\Yt{N,\beta}}\dot{g}_{i,j}(\beta)
            +
            \nabla \RefDiffNormPoint{\Yt{N,\beta}}^\top  \Grmu{\beta, \Yt{N,\beta}} 
            g_{i,j}(\beta)
        \right)
        d\beta
        \\
        + 
        \nabla \RefDiffNormPoint{\Yt{N,\beta}}^\top  \Grsigma{\beta, \Yt{N,\beta}} g_{i,j}(\beta) d\Wrt{\beta},
    \end{multline}
    where we have used the straightforward fact that $\nabla^2 \RefDiffNormPoint{\Yt{N,\beta}} = 0 \in \mathbb{S}^{2n}$.
    We have further replaced $G_{N,\mu}$ and $G_{N,\sigma}$ with $G_{\mu}$ and $G_{\sigma}$ because from Proposition~\ref{prop:Reference:TruncatedWellPosedness}, $Y_{N,\beta}$ is also a strong solution of the joint process~\eqref{eqn:Reference:JointProcess} for all $\beta \in [\nu,\tau(t)] \subseteq [0,\tau^\star] \subseteq [0,\tau_N]$.
    Observe from~\eqref{eqn:lem:Appendix:ReferenceProcess:ddV:Kernel:Decomposition}  that 
    \begin{align*}
        d_\beta\left[ \RefDiffNorm{\Yt{N,\beta}}^\top g_{\cdot, j}(\beta)\right]
        =
        \sum_{i=1}^n d_\beta\left[ \RefDiffNormPoint{\Yt{N,\beta}}g_{i,j}(\beta)\right] \in \mathbb{R}, \quad j \in \cbr{1,\dots,m}.
    \end{align*} 
    It therefore follows from the expression in~\eqref{eqn:lem:Appendix:ReferenceProcess:ddV:Ito:Initial} 
    \begin{multline*}
        d_\beta\left[ \RefDiffNorm{\Yt{N,\beta}}^\top g_{\cdot, j}(\beta)\right]
        =
        \sum_{i=1}^n
        \left( 
            \RefDiffNormPoint{\Yt{N,\beta}}\dot{g}_{i,j}(\beta)
            +
            \nabla \RefDiffNormPoint{\Yt{N,\beta}}^\top  \Grmu{\beta, \Yt{N,\beta}} 
            g_{i,j}(\beta)
        \right)
        d\beta
        \\
        + 
        \sum_{i=1}^n
        \nabla \RefDiffNormPoint{\Yt{N,\beta}}^\top  \Grsigma{\beta, \Yt{N,\beta}} g_{i,j}(\beta) d\Wrt{\beta} 
        ,
    \end{multline*}
    which can be further written as
    \begin{multline}\label{eqn:lem:Appendix:ReferenceProcess:ddV:Ito:Pre:1}
        d_\beta\left[ \RefDiffNorm{\Yt{N,\beta}}^\top g_{\cdot, j}(\beta)\right]
        =
        \left( 
            \RefDiffNorm{\Yt{N,\beta}}^\top
            \dot{g}_{\cdot,j}(\beta)
            +
            g_{\cdot,j}(\beta)^\top
            \nabla \RefDiffNorm{\Yt{N,\beta}}^\top  \Grmu{\beta, \Yt{N,\beta}} 
        \right)
        d\beta
        \\
        + 
        g_{\cdot,j}(\beta)^\top
        \nabla \RefDiffNorm{\Yt{N,\beta}}^\top  \Grsigma{\beta, \Yt{N,\beta}}  d\Wrt{\beta} 
        \in \mathbb{R}
        ,
    \end{multline} 
    for $j \in \cbr{1,\dots,m}$.
    Once again,~\eqref{eqn:lem:Appendix:ReferenceProcess:ddV:Kernel:Decomposition} implies that 
    \begin{align*}
        d_\beta \left[\RefDiffNorm{\Yt{N,\beta}}^\top g(\beta)\right]
        =
        \begin{bmatrix} 
            d_\beta \left[\RefDiffNorm{\Yt{N,\beta}}^\top g_{\cdot,1}(\beta)\right]  
            & \cdots &  
            d_\beta \left[\RefDiffNorm{\Yt{N,\beta}}^\top g_{\cdot,m}(\beta)\right]   
        \end{bmatrix}
        \in \mathbb{R}^{1 \times m}.
    \end{align*} 
    Thus, we use the expression in~\eqref{eqn:lem:Appendix:ReferenceProcess:ddV:Ito:Pre:1} to write 
    \begin{multline}\label{eqn:lem:Appendix:ReferenceProcess:ddV:Ito:1}
        d_\beta \left[\RefDiffNorm{\Yt{N,\beta}}^\top g(\beta)\right]
        =
        \left( 
            \RefDiffNorm{\Yt{N,\beta}}^\top
            \dot{g}(\beta)
            +
            \left(
                g(\beta)^\top
                \nabla \RefDiffNorm{\Yt{N,\beta}}^\top  \Grmu{\beta, \Yt{N,\beta}} 
            \right)^\top
        \right)
        d\beta
        \\
        + 
        \left(
            g(\beta)^\top
            \nabla \RefDiffNorm{\Yt{N,\beta}}^\top  \Grsigma{\beta, \Yt{N,\beta}}  d\Wrt{\beta}
        \right)^\top
        \in \mathbb{R}^{1 \times m}. 
    \end{multline}
    Now, using the definition of the gradient of vector valued functions in Sec.~\ref{sec:Notation}, we observe that
    \begin{align*}
        \nabla \RefDiffNorm{\Yt{N,\beta}}
        = 
        2
        \begin{bmatrix}
            \nabla \left(\Xrt{N,\beta} - \Xstart{N,\beta}\right)_1
            & \cdots & 
            \nabla \left(\Xrt{N,\beta} - \Xstart{N,\beta}\right)_n
        \end{bmatrix}
        =
        2 
        \begin{bmatrix} 
            \mathbb{I}_n \\ - \mathbb{I}_n
        \end{bmatrix} 
        \in \mathbb{R}^{2n \times n}.    
    \end{align*}
    Thus, it follows from the definitions of the vector fields in~\eqref{eqn:Reference:JointProcess} that 
    \begin{align*}
        \nabla \RefDiffNorm{\Yt{N,\beta}}^\top  \Grmu{\beta, \Yt{N,\beta}}
        =&
        2 
        \left(
            \Fmu{\beta,\Xrt{N,\beta},\Urt{\beta}} - \Fbarmu{\beta,\Xstart{N,\beta}}     
        \right)
        \\ 
        =& 
            2
            \Fbarmu{\beta,\Xrt{N,\beta}}  
            + 
            2
            \Lmu{\beta,\Xrt{N,\beta}}
            - 
            2
            \Fbarmu{\beta,\Xstart{N,\beta}}     
        + 
        2 g(\beta)\Urt{\beta}
        \in \mathbb{R}^n,
        \\
        \nabla \RefDiffNorm{\Yt{N,\beta}}^\top  \Grsigma{\beta, \Yt{N,\beta}}
        =& 
        2  
        \begin{bmatrix} 
            \Fsigma{\beta,\Xrt{N,\beta}} 
            & 
            -\Fbarsigma{\beta,\Xstart{N,\beta}} 
        \end{bmatrix}
        \in \mathbb{R}^{n \times 2d},
    \end{align*}
    where we have used the following decomposition from~\eqref{eqn:VectorFields:Decomposition} in Definition~\ref{def:VectorFields}: 
    \begin{align*}
        \Fmu{\beta,\Xrt{N,\beta},\Urt{\beta}}
        =
        \Fbarmu{\beta,\Xrt{N,\beta}} + g(\beta)\Urt{\beta} + \Lmu{\beta,\Xrt{N,\beta}}.
    \end{align*}
    Substituting the above identities into~\eqref{eqn:lem:Appendix:ReferenceProcess:ddV:Ito:1} produces 
    \begin{align*}
        d_\beta \left[\RefDiffNorm{\Yt{N,\beta}}^\top g(\beta)\right]
        = 
        \left[\mathcal{P}^r_{\mu}(\beta) + \mathcal{P}^r_\mathcal{U}(\beta) \right]^\top d\beta
        +
        \left[
            \mathcal{P}^r_{\sigma}(\beta)d\Wt{\beta}
            +
            \mathcal{P}^r_{\sigma_\star}(\beta)d\Wstart{\beta}
        \right]^\top
        \in \mathbb{R}^{1 \times m}.
    \end{align*}
    Then,~\eqref{eqn:lem:Appendix:ReferenceProcess:ddV:Expression:Pr:Main} is established by substituting the above into~\eqref{eqn:lem:Appendix:ReferenceProcess:ddV:Expression:Main}.

\end{proof}

The next result establishes the bounds for the pertinent entities in the last proposition. 
\begin{proposition}\label{prop:Appendix:ReferenceProcess:Pparts:Bound}
    If the stopping time $\tau^\star$, defined in~\eqref{eqn:Reference:StoppingTimes}, Lemma~\ref{lem:Reference:dV}, satisfies  $\tau^\star = \tstar$, then the functions $\mathcal{P}^r_{\mu}(t) \in \mathbb{R}^m$ and $\mathcal{P}^r_{\sigma}(t),\,\mathcal{P}^r_{\sigma_\star}(t) \in \mathbb{R}^{m \times d}$  defined in the statement of Proposition~\ref{prop:Appendix:ReferenceProcess:ddV:Bound} satisfy the following bounds for any $\sfp \in \mathbb{R}_{\geq 1}$:  
    \begin{subequations}\label{prop:Appendix:ReferenceProcess:Pparts:Bounds:Final}
        \begin{align}
            \sup_{\nu \in [0,\tstar]}
            \LpLaw{2\sfp}{y_0}{\mathcal{P}^r_{\mu}(\nu)}
            \leq
            2
            \Delta_g
            \Delta^r_{\mathcal{P}_\mu}(\sfp)
            +
            \Delta_{\dot{g}}
            \sup_{\nu \in [0,\tstar]}
            \LpLaw{2\sfp}{y_0}{\RefDiffNorm{\Yt{N,\nu}}}
            ,
            \label{prop:Appendix:ReferenceProcess:Pparts:Bound:Mu:Final} 
            \\
            \sum_{i \in \cbr{\sigma,\sigma_\star}}
            \sup_{\nu \in [0,\tstar]}
            \LpLaw{2\sfp}{y_0}{\mathcal{P}^r_{i}(\nu)}
            \leq 
            2
            \Delta_g
            \Delta^r_{\mathcal{P}_\sigma}(\sfp)
            ,
            \label{prop:Appendix:ReferenceProcess:Pparts:Bound:Sigma:Final}
        \end{align}
    \end{subequations}
    where 
    \begin{align*}
        \Delta^r_{\mathcal{P}_\mu}(\sfp)
        = 
        \sup_{\nu \in [0,\tstar]}
        \left(
            \LpLaw{2\sfp}{y_0}{\Fbarmu{\nu,\Xrt{N,\nu}}-\Fbarmu{\nu,\Xstart{N,\nu}}}
            + 
            \LpLaw{2\sfp}{y_0}{\Lmu{\nu,\Xrt{N,\nu}}}  
        \right),
        \\
        \Delta^r_{\mathcal{P}_\sigma}(\sfp)
        =
        \sup_{\nu \in [0,\tstar]}
            \left( 
                \LpLaw{2\sfp}{y_0}{\Fbarsigma{\nu,\Xrt{N,\nu}}}
                +
                \LpLaw{2\sfp}{y_0}{\Lsigma{\nu,\Xrt{N,\nu}}}
            \right)
            +
            \sup_{\nu \in [0,\tstar]}
            \LpLaw{2\sfp}{y_0}{\Fbarsigma{\nu,\Xstart{N,\nu}}}.
    \end{align*}

\end{proposition}
\begin{proof}
    We begin by using the definition of $\mathcal{P}^r_{\mu}$ in~\eqref{eqn:lem:Appendix:ReferenceProcess:ddV:Expression:Pr} to obtain
    \begin{align*}
        \norm{\mathcal{P}^r_{\mu}(\nu)}
        \leq
        \Delta_{\dot{g}}
        \norm{\RefDiffNorm{\Yt{N,\nu}}}
        + 
        2
        \Delta_g
        \left(
            \norm{\Fbarmu{\nu,\Xrt{N,\nu}}-\Fbarmu{\nu,\Xstart{N,\nu}}}
            + 
            \norm{\Lmu{\nu,\Xrt{N,\nu}}}  
        \right), 
    \end{align*}
    for all $\nu \in [0,\tstar]$, where we have used the bounds on $g(t)$ and $\dot{g}(t)$ in Assumption~\ref{assmp:KnownFunctions}.
    It then follows from the Minkowski's inequality that
    \begin{align*}
        \LpLaw{2\sfp}{y_0}{\mathcal{P}^r_{\mu}(\nu)}
        \leq
        \Delta_{\dot{g}}
        \LpLaw{2\sfp}{y_0}{\RefDiffNorm{\Yt{N,\nu}}}
        + 
        2
        \Delta_g
        \left(
            \LpLaw{2\sfp}{y_0}{\Fbarmu{\nu,\Xrt{N,\nu}}-\Fbarmu{\nu,\Xstart{N,\nu}}}
            + 
            \LpLaw{2\sfp}{y_0}{\Lmu{\nu,\Xrt{N,\nu}}}  
        \right), 
    \end{align*} 
    for all $\nu \in [0,\tstar]$, which in turn implies~\eqref{prop:Appendix:ReferenceProcess:Pparts:Bound:Mu:Final}.
    
    Next, using the definitions of $\mathcal{P}^r_\sigma$ and $\mathcal{P}^r_{\sigma_\star}$, we obtain 
    \begin{align*}
        \Frobenius{\mathcal{P}^r_{\sigma}(\nu)}
        \leq 
        2
        \Delta_g
        \Frobenius{\Fsigma{\nu,\Xrt{N,\nu}}},
        \quad 
        \Frobenius{\mathcal{P}^r_{\sigma_\star}(\nu)}
        \leq
        2
        \Delta_g
        \Frobenius{\Fbarsigma{\nu,\Xstart{N,\nu}}},
        \quad \forall \nu \in [0,\tstar],
    \end{align*}
    where we have used the bound on $g(t)$ in Assumption~\ref{assmp:KnownFunctions}.  
    It then follows from the decomposition $F_\sigma = \bar{F}_\sigma + \Lambda_\sigma$ in~\eqref{eqn:VectorFields:Decomposition} that
    \begin{multline*}
        \Frobenius{\mathcal{P}^r_{\sigma}(\nu)}
        \leq 
        2
        \Delta_g
        \left( 
            \Frobenius{\Fbarsigma{\nu,\Xrt{N,\nu}}}
            +
            \Frobenius{\Lsigma{\nu,\Xrt{N,\nu}}}
        \right),
        \\
        \Frobenius{\mathcal{P}^r_{\sigma_\star}(\nu)}
        \leq
        2
        \Delta_g
        \Frobenius{\Fbarsigma{\nu,\Xstart{N,\nu}}},
        \quad \forall \nu \in [0,\tstar].
    \end{multline*} 
    From the Minkowski's inequality, we get
    \begin{multline*}
        \LpLaw{2\sfp}{y_0}{\mathcal{P}^r_{\sigma}(\nu)}
        \leq 
        2
        \Delta_g
        \left( 
            \LpLaw{2\sfp}{y_0}{\Fbarsigma{\nu,\Xrt{N,\nu}}}
            +
            \LpLaw{2\sfp}{y_0}{\Lsigma{\nu,\Xrt{N,\nu}}}
        \right),
        \\
        \LpLaw{2\sfp}{y_0}{\mathcal{P}^r_{\sigma_\star}(\nu)}
        \leq
        2
        \Delta_g
        \LpLaw{2\sfp}{y_0}{\Fbarsigma{\nu,\Xstart{N,\nu}}},
        \quad \forall \nu \in [0,\tstar],
    \end{multline*} 
    which then leads to~\eqref{prop:Appendix:ReferenceProcess:Pparts:Bound:Sigma:Final} in a straightforward manner.

\end{proof}
Next, we provide a result that is essential to the proof of the main results of the section.
\begin{proposition}\label{prop:Appendix:ReferenceProcess:N:Bounds}
    Consider the following scalar processes:
    \begin{align}\label{eqn:prop:Appendix:ReferenceProcess:N:Definitions}
        \begin{aligned}
            N^r_1(t)
            =
            \int_0^{t}
                \expo{\Boldomega \nu}
                M_\mu(t,\nu)\Lparamu{\nu, \Xrt{N,\nu}}d\nu,
            \quad 
            N^r_2(t)
            =
            \int_0^{t}
                \expo{\Boldomega \nu}
                M_\mu(t,\nu)\Fparasigma{\nu,\Xrt{N,\nu}}d\Wt{\nu},
            \\
            N^r_3(t)
            =
            \int_0^{t}
                \expo{\Boldomega \nu}
                M_\sigma(t,\nu)\Lparamu{\nu, \Xrt{N,\nu}}d\nu,
            \quad 
            N^r_4(t)
            =
            \int_0^{t}
                \expo{\Boldomega \nu}
                M_\sigma(t,\nu)\Fparasigma{\nu,\Xrt{N,\nu}}d\Wt{\nu},
        \end{aligned}
    \end{align}
    where  
    \begin{align*}
        M_\mu(t,\nu)
        =
        \int_\nu^{t} 
            \expo{ (2\lambda - \Boldomega) \beta } \mathcal{P}^r_{\mu}(\beta)^\top
        d\beta, 
        \quad 
        M_\sigma(t,\nu)
        =
        \int_\nu^{t} 
            \expo{ (2\lambda - \Boldomega) \beta } 
            \left[
            \mathcal{P}^r_{\sigma}(\beta)d\Wt{\beta}
            +
            \mathcal{P}^r_{\sigma_\star}(\beta)d\Wstart{\beta}
            \right]^\top,
    \end{align*}
    for $0 \leq \nu \leq t \leq T$, and  where $\mathcal{P}^r_{\mu}(\beta) \in \mathbb{R}^m$ and $\mathcal{P}^r_{\sigma}(\beta),\,\mathcal{P}^r_{\sigma_\star}(\beta) \in \mathbb{R}^{m \times d}$ are defined in the statement of Proposition~\ref{prop:Appendix:ReferenceProcess:ddV:Bound}.
    If the stopping time $\tau^\star$, defined in~\eqref{eqn:Reference:StoppingTimes}, Lemma~\ref{lem:Reference:dV}, satisfies  $\tau^\star = \tstar$, then  we have the following bounds:
    \begin{multline}\label{eqn:prop:Appendix:ReferenceProcess:N:Bounds:p=1:Main}
        \sum_{i=1}^4 \absolute{\ELaw{y_0}{N^r_i(t)}}
        \leq
        \frac{ \expo{ 2\lambda t } }{ \sqrt{\lambda} \Boldomega }  
        \Delta^r_{\mathcal{P}_1}(1,\tstar)
        \sup_{\nu \in [0,\tstar]} 
            \LpLaw{2}{y_0}{\Lparamu{\nu, \Xrt{N,\nu}}}
        \\
        + 
        \left(
            \frac{\expo{ 2\lambda t }}{\sqrt{\lambda \Boldomega}}
            \Delta^r_{\mathcal{P}_2}(1,\tstar)
            +
            \frac{\expo{2\lambda t}}{\lambda} 
            \Delta^r_{\mathcal{P}_3}(1,\tstar)
        \right)
        \sup_{\nu \in [0,\tstar]}
                \LpLaw{2}{y_0}{\Fparasigma{\nu,\Xrt{N,\nu}}}
        ,
    \end{multline}
    for all $t \in [0,\tstar]$, and
    \begin{multline}\label{eqn:prop:Appendix:ReferenceProcess:N:Bounds:p>=2:Main}
        \sum_{i=1}^4 \pLpLaw{y_0}{N^r_i(t)}
        \leq 
        \frac{ \expo{ 2\lambda t } }{ \sqrt{\lambda} \Boldomega }
        \Delta^r_{\mathcal{P}_1}(\sfp,\tstar)
        \sup_{\nu \in [0,\tstar]} 
            \LpLaw{2\sfp}{y_0}{\Lparamu{\nu, \Xrt{N,\nu}}}
        \\
        +
        \left( 
            \frac{\expo{ 2\lambda t }}{\sqrt{\lambda \Boldomega }} 
            \Delta^r_{\mathcal{P}_2}(\sfp,\tstar)
            + 
            \frac{\expo{2\lambda t}}{\lambda}
            \Delta^r_{\mathcal{P}_3}(\sfp,\tstar)
        \right)
        \sup_{\nu \in [0,\tstar]}\LpLaw{2\sfp}{y_0}{\Fparasigma{\nu,\Xrt{N,\nu}}}
        ,
    \end{multline}
    for all $(t,\sfp) \in [0,\tstar] \times \mathbb{N}_{\geq 2}$, where 
    \begin{align*}
        \Delta^r_{\mathcal{P}_1}(\sfp,\tstar)
        = 
        \frac{ 1 }{ 2  \sqrt{\lambda} }
            \sup_{\nu \in [0,\tstar]}  
            \LpLaw{2\sfp}{y_0}{\mathcal{P}^r_{\mu}(\nu)}
        + 
        \frac{\mathfrak{p}(\sfp)}{2 }
        \sup_{\nu \in [0,\tstar]}
        \left(
            \LpLaw{2\sfp}{y_0}{\mathcal{P}^r_{\sigma}(\nu)}
            +
            \LpLaw{2\sfp}{y_0}{\mathcal{P}^r_{\sigma_\star}(\nu)} 
        \right)
            ,
        \\
            \Delta^r_{\mathcal{P}_2}(\sfp,\tstar)
            =
            \mathfrak{p}'(\sfp)
            \Delta^r_{\mathcal{P}_1}(\sfp,\tstar)
            ,
        \quad
        \Delta^r_{\mathcal{P}_3}(\sfp,\tstar)
        = 
        \frac{\sqrt{m}}{2}  
        \sup_{\nu \in [0,\tstar]}
        \left(
            \LpLaw{2\sfp}{y_0}{\mathcal{P}^r_{\sigma}(\nu)}
            +
            \LpLaw{2\sfp}{y_0}{\mathcal{P}^r_{\sigma_\star}(\nu)} 
        \right)
        ,
    \end{align*}
    and where the constants $\mathfrak{p}(\sfp)$ and $\mathfrak{p}'(\sfp)$ are defined in~\eqref{eqn:app:Constants:FrakP}. 
                          
\end{proposition}
\begin{proof}
    We begin with the process $N^r_1$ and use the definition of $M_\mu$ to obtain
    \begin{align*}
        N^r_1(t)
        =
        \int_0^{t}
            \expo{\Boldomega \nu}
            \int_\nu^{t} 
                \expo{ (2\lambda - \Boldomega) \beta } \left[\mathcal{P}^r_{\mu}(\beta)\right]^\top
            d\beta\Lparamu{\nu, \Xrt{N,\nu}}d\nu
        ,
        \quad t \in [0,t^\star].
    \end{align*}
    Changing the order of integration and switching the variables $\beta$ and $\nu$ leads to 
    \begin{align*}
        N^r_1(t)
        =
        \int_0^{t}
            \expo{ (2\lambda - \Boldomega) \nu }
            \left[\mathcal{P}^r_{\mu}(\nu)\right]^\top
            \left(
                \int_0^{\nu} 
                    \expo{\Boldomega \beta}
                    \Lparamu{\beta, \Xrt{N,\beta}}
                d\beta
            \right) 
        d\nu
        ,
        \quad t \in [0,t^\star].
    \end{align*}
    Hence,
    \begin{align*}
        \absolute{N^r_1(t)}
        \leq 
        \int_0^{t}
            \expo{ (2\lambda - \Boldomega) \nu }
            \norm{\mathcal{P}^r_{\mu}(\nu)}
            \left(
                \int_0^{\nu} 
                    \expo{\Boldomega \beta}
                    \norm{\Lparamu{\beta, \Xrt{N,\beta}}}
                d\beta
            \right) 
        d\nu
        ,
        \quad t \in [0,t^\star].
    \end{align*}
    Applying Proposition~\ref{prop:TechnicalResults:LebesgueNestedMoment} for $\cbr{\theta_1,\theta_2,\xi,S_1,S_2} = \cbr{2\lambda,\Boldomega, 0, \mathcal{P}^r_{\mu_i}, \Lambda_\mu^{\paral}}$ to the integral on the right hand side of the above inequality produces 
    \begin{align}\label{eqn:prop:Appendix:ReferenceProcess:N1:Bound:Final}
        \LpLaw{\sfp}{y_0}{N^r_1(t)}
        \leq&      
        \frac{ \expo{ 2\lambda t } }{ 2\lambda \Boldomega }
        \sup_{\nu \in [0,t]}  
        \LpLaw{2\sfp}{y_0}{\mathcal{P}^r_{\mu}(\nu)}
        \sup_{\nu \in [0,t]} 
        \LpLaw{2\sfp}{y_0}{\Lparamu{\nu, \Xrt{N,\nu}}}
        \notag 
        \\ 
        \leq&  
        \frac{ \expo{ 2\lambda t } }{ 2\lambda \Boldomega }
        \sup_{\nu \in [0,\tstar]}  
        \LpLaw{2\sfp}{y_0}{\mathcal{P}^r_{\mu}(\nu)}
        \sup_{\nu \in [0,\tstar]} 
        \LpLaw{2\sfp}{y_0}{\Lparamu{\nu, \Xrt{N,\nu}}}
        ,
        \quad \forall (t,\sfp) \in [0,\tstar] \times \mathbb{N}_{\geq 1}
        ,
    \end{align} 
    where we have bounded $\expo{ 2\lambda t } - 1$ by $\expo{ 2\lambda t }$.

    \begingroup
    \endgroup

    Next, from the definitions of $M_\mu(t,\nu)$ and $N^r_2$ we obtain
    \begin{multline*}
        N^r_2(t)
        =
        \int_0^{t}
            \expo{\Boldomega \nu}
            M_\mu(t,\nu)\Fparasigma{\nu,\Xrt{N,\nu}}d\Wt{\nu}
        \\
        =
        \int_0^{t}
            \expo{\Boldomega \nu}
            \int_\nu^{t} 
                \expo{ (2\lambda - \Boldomega) \beta } 
                \left[\mathcal{P}^r_{\mu}(\beta)\right]^\top
            d\beta
            \Fparasigma{\nu,\Xrt{N,\nu}}d\Wt{\nu}
        \\
        =
        \int_0^{t}
            \int_\nu^{t} 
                \expo{ (2\lambda - \Boldomega) \beta } 
                \left[\mathcal{P}^r_{\mu}(\beta)\right]^\top
            \expo{\Boldomega \nu}
            \begin{bmatrix}\Fparasigma{\nu,\Xrt{N,\nu}} & 0_{m,d} \end{bmatrix} d\beta d\Wrt{\nu},
    \end{multline*}
    where recall from Definition~\ref{def:Reference:JointProcess} that  $\Wrt{t} = \begin{bmatrix}\left(\Wt{t}\right)^\top & \left(\Wstart{t}\right)^\top \end{bmatrix}^\top \in \mathbb{R}^{2d}$. 
    Using the definition of $\mathcal{P}^r_{\mu}$ in Proposition~\ref{prop:Appendix:ReferenceProcess:ddV:Bound}, and the regularity assumptions in Sec.~\ref{subsec:Assumptions}, it is straightforward to show that $\mathcal{P}^r_{\mu} \in \mathcal{M}_{2}^{loc}\br{\mathbb{R}^{m}|\Wfilt{t} \times \Wstarfilt{t}}$ and $\begin{bmatrix}\Fparasigma{\nu,\Xrt{N,\nu}} & 0_{m,d} \end{bmatrix} \in \mathcal{M}_{2}^{loc}\br{\mathbb{R}^{m \times 2d}|\Wfilt{t} \times \Wstarfilt{t}}$.  
    Therefore, we may apply Lemma~\ref{lem:TechnicalResults:Fubini-ish} to the right hand side of the previous expression and obtain 
    \begin{align*}
        N^r_2(t)
        =
        \int_0^{t}
            \expo{ (2\lambda - \Boldomega) \nu } 
            \left[\mathcal{P}^r_{\mu}(\nu)\right]^\top
            \left(
                \int_0^\nu
                \expo{\Boldomega \beta} 
                \begin{bmatrix}\Fparasigma{\beta,\Xrt{N,\beta}} & 0_{m,d} \end{bmatrix} 
                d\Wrt{\beta}  
            \right)
        d\nu
        \in \mathbb{R}
        ,
    \end{align*}
    and thus 
    \begin{align*}
        \absolute{N^r_2(t)}
        \leq 
        \int_0^{t}
            \expo{ (2\lambda - \Boldomega) \nu } 
            \norm{\mathcal{P}^r_{\mu}(\nu)}
            \norm{
                \int_0^\nu
                \expo{\Boldomega \beta} 
                \begin{bmatrix}\Fparasigma{\beta,\Xrt{N,\beta}} & 0_{m,d} \end{bmatrix} 
                d\Wrt{\beta}  
            }
        d\nu
        .
    \end{align*}
    Since the process on the right hand side can be cast in the form of the process $N(t)$ in Lemma~\ref{lemma:TechnicalResults:NestedSupMoment} by setting 
    \begin{align*}
        Q_t = \Wrt{t} \in \mathbb{R}^{2d}\,(n_q=2d), 
        \quad 
        \mathfrak{F}_t = \Wfilt{t} \times \Wstarfilt{t},
        \quad 
        \theta_1 = 2\lambda, 
        \quad 
        \theta_2 = \Boldomega,
        \quad 
        \xi = 0, 
        \quad 
        \zeta_1 = 0, 
        \quad 
        \zeta_2 =t,
        \\
        S(t) = \mathcal{P}^r_{\mu}(t) \in \mathcal{M}_{2}^{loc}\br{\mathbb{R}^{m}|\Wfilt{t} \times \Wstarfilt{t}}\,(n_s=m), 
        \\ 
        L(t) = \begin{bmatrix}\Fparasigma{t,\Xrt{N,t}} & 0_{m,d} \end{bmatrix}
        \in
        \mathcal{M}_{2}^{loc}\br{\mathbb{R}^{m \times 2d}|\Wfilt{t} \times \Wstarfilt{t}}
        ,   
    \end{align*}
    we may apply Lemma~\ref{lemma:TechnicalResults:NestedSupMoment} to obtain the following bound:
    \begin{align}\label{eqn:prop:Appendix:ReferenceProcess:N2:Bound:Final}
        \LpLaw{\sfp}{y_0}{N^r_2(t)}
        \leq& 
        \left(\sfp \frac{ 2\sfp-1}{2} \right)^\Half
        \frac{\expo{ 2\lambda t }}{2\lambda \BoldomegaRoot}
        \sup_{\nu \in [0,t]}
        \LpLaw{2\sfp}{y_0}{\mathcal{P}^r_{\mu}(\nu)}
        \sup_{\nu \in [0,t]}
        \LpLaw{2\sfp}{y_0}{\Fparasigma{\nu,\Xrt{N,\nu}}}
        \notag 
        \\
        \leq& 
        \left(\sfp \frac{ 2\sfp-1}{2} \right)^\Half
        \frac{\expo{ 2\lambda t }}{2\lambda \BoldomegaRoot}
        \sup_{\nu \in [0,\tstar]}
        \LpLaw{2\sfp}{y_0}{\mathcal{P}^r_{\mu}(\nu)}
        \sup_{\nu \in [0,\tstar]}
        \LpLaw{2\sfp}{y_0}{\Fparasigma{\nu,\Xrt{N,\nu}}}
        ,
    \end{align} 
    for all $(t,\sfp) \in [0,\tstar] \times \mathbb{N}_{\geq 1}$.

    Next, we consider the process $N^r_3(t)$ given by 
    \begin{multline*}
        N^r_3(t)
        =
        \int_0^{t}
            \expo{\Boldomega \nu} M_\sigma(t,\nu)\Lparamu{\nu, \Xrt{N,\nu}}
        d\nu
        \\
        =
        \int_0^{t}
            \expo{\Boldomega \nu} 
            \left( 
                \int_\nu^{t} 
                    \expo{ (2\lambda - \Boldomega) \beta } 
                    \left[
                    \mathcal{P}^r_{\sigma}(\beta)d\Wt{\beta}
                    +
                    \mathcal{P}^r_{\sigma_\star}(\beta)d\Wstart{\beta}
                    \right]^\top
            \right)\Lparamu{\nu, \Xrt{N,\nu}}
        d\nu
        \\
        =
        \int_0^{t}
            \int_\nu^{t} 
                \expo{\Boldomega \nu}
                \Lparamu{\nu, \Xrt{N,\nu}}^\top  
                \expo{ (2\lambda - \Boldomega) \beta }
                \begin{bmatrix}
                    \mathcal{P}^r_{\sigma}(\beta) &  \mathcal{P}^r_{\sigma_\star}(\beta)
                \end{bmatrix} 
            d\Wrt{\beta}
        d\nu.
    \end{multline*}
    Using Lemma~\ref{lem:TechnicalResults:Fubini-ish}, one sees that 
    \begin{align*}
        N^r_3(t)
        =&
        \int_0^{t}
            \int_0^{\nu} 
                \expo{\Boldomega \beta}
                \Lparamu{\beta, \Xrt{N,\beta}}^\top  
                \expo{ (2\lambda - \Boldomega) \nu }
                \begin{bmatrix}
                    \mathcal{P}^r_{\sigma}(\nu) &  \mathcal{P}^r_{\sigma_\star}(\nu)
                \end{bmatrix} 
            d\beta
        d\Wrt{\nu}
        \\
        =&
        \int_0^{t}
            \expo{ (2\lambda - \Boldomega) \nu }
            \left( 
                \int_0^{\nu} 
                    \expo{\Boldomega \beta}
                    \Lparamu{\beta, \Xrt{N,\beta}}
                d\beta
            \right)^\top  
            \begin{bmatrix}
                \mathcal{P}^r_{\sigma}(\nu) &  \mathcal{P}^r_{\sigma_\star}(\nu)
            \end{bmatrix} 
        d\Wrt{\nu}.
    \end{align*}
    Note that $N^r_3(t)$ can be cast in the form of the process $N(t)$ in Lemma~\ref{lemma:TechnicalResults:ConverseNestedSupMoment} by setting 
    \begin{align*}
        Q_t = \Wrt{t} \in \mathbb{R}^{2d}\,(n_q=2d), 
        \quad 
        \mathfrak{F}_t = \Wfilt{t} \times \Wstarfilt{t},
        \quad 
        \theta_1 = 2\lambda, 
        \quad 
        \theta_2 = \Boldomega,
        \\
        S(t) = \Lparamu{t, \Xrt{N,t}} \in \mathcal{M}_{2}^{loc}\br{\mathbb{R}^{m}|\Wfilt{t} \times \Wstarfilt{t}}\,(n_o=m),
        \\ 
        L(t) = \begin{bmatrix}
                        \mathcal{P}^r_{\sigma}(t) &  \mathcal{P}^r_{\sigma_\star}(t)
                    \end{bmatrix}
        \in     
        \mathcal{M}_{2}^{loc}\br{\mathbb{R}^{m \times 2d}|\Wfilt{t} \times \Wstarfilt{t}}
        .   
    \end{align*}
    Therefore, we use Lemma~\ref{lemma:TechnicalResults:ConverseNestedSupMoment} and obtain 
    \begin{align}\label{eqn:prop:Appendix:ReferenceProcess:N3:Bound:1}
        \LpLaw{\sfp}{y_0}{N^r_3(t)}
        \leq&   
        \left(\sfp (\sfp - 1)\right)^\frac{1}{2} 
        \frac{ \expo{ 2\lambda t }}{ 2\sqrt{2\lambda} \Boldomega }
        \sup_{\nu \in [0,t]} 
        \LpLaw{2\sfp}{}{\Lparamu{\nu, \Xrt{N,\nu}}}
        \sup_{\nu \in [0,t]}
        \LpLaw{2\sfp}{}{
            \begin{bmatrix}
                \mathcal{P}^r_{\sigma}(\nu) &  \mathcal{P}^r_{\sigma_\star}(\nu)
            \end{bmatrix}
        } 
        \notag 
        \\
        \leq&   
        \left(\sfp (\sfp - 1)\right)^\frac{1}{2} 
        \frac{ \expo{ 2\lambda t }}{2\sqrt{2\lambda} \Boldomega }
        \sup_{\nu \in [0,\tstar]} 
        \LpLaw{2\sfp}{}{\Lparamu{\nu, \Xrt{N,\nu}}}
        \sup_{\nu \in [0,\tstar]}
        \LpLaw{2\sfp}{}{
            \begin{bmatrix}
                \mathcal{P}^r_{\sigma}(\nu) &  \mathcal{P}^r_{\sigma_\star}(\nu)
            \end{bmatrix}
        }
        , 
    \end{align}
    for all $(t,\sfp) \in [0,\tstar] \times  \mathbb{N}_{\geq 2}$.
    Since
    \begin{align*}
        \Frobenius{\begin{bmatrix}
            \mathcal{P}^r_{\sigma}(\nu) &  \mathcal{P}^r_{\sigma_\star}(\nu)
        \end{bmatrix}}
        =
        \left( 
            \Frobenius{\mathcal{P}^r_{\sigma}(\nu)}^2 + \Frobenius{\mathcal{P}^r_{\sigma_\star}(\nu)}^2
        \right)^\Half
        \leq 
        \Frobenius{\mathcal{P}^r_{\sigma}(\nu)} + \Frobenius{\mathcal{P}^r_{\sigma_\star}(\nu)}
        ,
        \quad \forall \nu \in [0,\tstar],
    \end{align*}
    we may use the Minkowski's inequality to obtain
    \begin{align}\label{eqn:prop:Appendix:ReferenceProcess:FrobeniusTrick}
        \LpLaw{2\sfp}{}{
            \begin{bmatrix}
                \mathcal{P}^r_{\sigma}(\nu) &  \mathcal{P}^r_{\sigma_\star}(\nu)
            \end{bmatrix}
        }
        \leq 
        \LpLaw{2\sfp}{}{\mathcal{P}^r_{\sigma}(\nu)} + \LpLaw{2\sfp}{}{\mathcal{P}^r_{\sigma_\star}(\nu)}
        ,
        \quad \forall \nu \in [0,\tstar].
    \end{align}
    Hence,~\eqref{eqn:prop:Appendix:ReferenceProcess:N3:Bound:1} can be further developed into 
    \begin{align}\label{eqn:prop:Appendix:ReferenceProcess:N3:Bound:p>=2:Final}
        \LpLaw{\sfp}{y_0}{N^r_3(t)}
        \leq  
        \left(\sfp (\sfp - 1)\right)^\frac{1}{2} 
        \frac{ \expo{ 2\lambda t }}{2\sqrt{2\lambda} \Boldomega }
        \sup_{\nu \in [0,\tstar]} 
        \LpLaw{2\sfp}{y_0}{\Lparamu{\nu, \Xrt{N,\nu}}}
        \sup_{\nu \in [0,\tstar]} 
        \left(
            \LpLaw{2\sfp}{y_0}{\mathcal{P}^r_{\sigma}(\nu)} 
            + 
            \LpLaw{2\sfp}{y_0}{\mathcal{P}^r_{\sigma_\star}(\nu)} 
        \right)
        , 
    \end{align}
    for all $ (t,\sfp) \in [0,\tstar] \times  \mathbb{N}_{\geq 2}$
    Additionally, using~\cite[Thm.~1.5.21]{mao2007stochastic}, one obtains 
    \begin{equation}\label{eqn:prop:Appendix:ReferenceProcess:N3:Bound:p=1:Final}
        \ELaw{y_0}{N^r_3(t)} = 0, \quad \forall t \in [0,\tstar].
    \end{equation}

    Finally, consider the process $N^r_4(t)$ given as 
    \begin{multline*}
        N^r_4(t)
        =
        \int_0^{t}
            \expo{\Boldomega \nu}
            M_\sigma(t,\nu)\Fparasigma{\nu,\Xrt{N,\nu}}
        d\Wt{\nu}
        \\
        =
        \int_0^{t}
            \expo{\Boldomega \nu}
            \int_\nu^{t} 
                \expo{ (2\lambda - \Boldomega) \beta } 
                \left[
                \mathcal{P}^r_{\sigma}(\beta)d\Wt{\beta}
                +
                \mathcal{P}^r_{\sigma_\star}(\beta)d\Wstart{\beta}
                \right]^\top
            \Fparasigma{\nu,\Xrt{N,\nu}}
        d\Wt{\nu}
        \\
        =
        \int_0^{t}
            \expo{\Boldomega \nu}
            \left(
                \int_\nu^{t} 
                    \expo{ (2\lambda - \Boldomega) \beta } 
                        \begin{bmatrix}
                            \mathcal{P}^r_{\sigma}(\beta) &  \mathcal{P}^r_{\sigma_\star}(\beta)
                        \end{bmatrix} 
                    d\Wrt{\beta}
            \right)^\top 
            \begin{bmatrix}\Fparasigma{\nu,\Xrt{N,\nu}} & 0_{m,d} \end{bmatrix} 
        d\Wrt{\nu},
    \end{multline*}
    which can be cast in the form of the process $N(t)$ in Lemma~\ref{lemma:TechnicalResults:NestedSupMoment:Ito} by choosing 
    \begin{align*}
        Q_t = \Wrt{t} \in \mathbb{R}^{2d}\,(n_q=2d), 
        \quad 
        \mathfrak{F}_t = \Wfilt{t} \times \Wstarfilt{t},
        \quad 
        \theta_1 = 2\lambda, 
        \quad 
        \theta_2 = \Boldomega,
        \\
        L_1(t) = \begin{bmatrix}
                        \mathcal{P}^r_{\sigma}(t) &  \mathcal{P}^r_{\sigma_\star}(t)
                    \end{bmatrix}
        \in     
        \mathcal{M}_{2}^{loc}\br{\mathbb{R}^{m \times 2d}|\Wfilt{t} \times \Wstarfilt{t}}\,(n_l = m),
        \\
        L_2(t) = \begin{bmatrix}\Fparasigma{t,\Xrt{N,t}} & 0_{m,d} \end{bmatrix}
        \in     
        \mathcal{M}_{2}^{loc}\br{\mathbb{R}^{m \times 2d}|\Wfilt{t} \times \Wstarfilt{t}}
        .   
    \end{align*}
    It then follows from Lemma~\ref{lemma:TechnicalResults:NestedSupMoment:Ito} that
    \begin{align}\label{eqn:prop:Appendix:ReferenceProcess:N4:Bound:p=1:1}
        \ELaw{}{N_4^r(t)}
        \leq& 
        \sqrt{m}
        \frac{\expo{2\lambda t}}{2\lambda} 
        \sup_{\nu \in [0,\tstar]}
        \ELaw{}{
            \Frobenius{
                \begin{bmatrix}
                    \mathcal{P}^r_{\sigma}(\nu) &  \mathcal{P}^r_{\sigma_\star}(\nu)
                \end{bmatrix}
                \Fparasigma{\nu,\Xrt{N,\nu}}^\top
            }
        } 
        , 
        \quad \forall t \in [0,\tstar],
    \end{align}
    and 
    \begin{multline}\label{eqn:prop:Appendix:ReferenceProcess:N4:Bound:p>=1:1}
        \LpLaw{\sfp}{}{N_4^r(t)}
        \leq 
        \frac{\sfp \sqrt{(\sfp - 1)(2\sfp - 1)}}{4} 
        \frac{\expo{ 2\lambda t }}{\sqrt{\lambda \Boldomega }}
        \sup_{\nu \in [0,\tstar]} 
        \LpLaw{2\sfp}{}{
            \begin{bmatrix}
                \mathcal{P}^r_{\sigma}(\nu) &  \mathcal{P}^r_{\sigma_\star}(\nu)
            \end{bmatrix}
        }
        \sup_{\nu \in [0,\tstar]}
        \LpLaw{2\sfp}{}{\Fparasigma{\nu,\Xrt{N,\nu}}}
        \\
        +
        \sqrt{m}
        \frac{\expo{2\lambda t}}{2\lambda} 
        \sup_{\nu \in [0,\tstar]}
        \LpLaw{\sfp}{}{
            \begin{bmatrix}
                \mathcal{P}^r_{\sigma}(\nu) &  \mathcal{P}^r_{\sigma_\star}(\nu)
            \end{bmatrix}
            \Fparasigma{\nu,\Xrt{N,\nu}}^\top
        }
        , 
    \end{multline}
    for all $(t,\sfp) \in [0,\tstar] \times \mathbb{N}_{\geq 2}$.
    Now, using the Cauchy-Schwarz inequality and~\eqref{eqn:prop:Appendix:ReferenceProcess:FrobeniusTrick}, one obtains  
    \begin{align*}
        \LpLaw{\sfq}{}{
            \begin{bmatrix}
                \mathcal{P}^r_{\sigma}(\nu) &  \mathcal{P}^r_{\sigma_\star}(\nu)
            \end{bmatrix}
            \Fparasigma{\nu,\Xrt{N,\nu}}^\top
        }
        \leq& 
        \LpLaw{2\sfq}{}{
            \begin{bmatrix}
                \mathcal{P}^r_{\sigma}(\nu) &  \mathcal{P}^r_{\sigma_\star}(\nu)
            \end{bmatrix}
        }
        \LpLaw{2\sfq}{}{
            \Fparasigma{\nu,\Xrt{N,\nu}}^\top
        }
        \\
        \leq& 
        \left(
            \LpLaw{2\sfq}{y_0}{\mathcal{P}^r_{\sigma}(\nu)} 
            + 
            \LpLaw{2\sfq}{y_0}{\mathcal{P}^r_{\sigma_\star}(\nu)} 
        \right)
        \LpLaw{2\sfq}{y_0}{
            \Fparasigma{\nu,\Xrt{N,\nu}}^\top
        }
        , 
    \end{align*}
    for all $(\nu,\sfq) \in [0,\tstar] \times \mathbb{N}_{\geq 1}$.
    Using this inequality for $\sfq =1$ and $\sfq = \sfp \in \mathbb{N}_{\geq 2}$, the bounds in~\eqref{eqn:prop:Appendix:ReferenceProcess:N4:Bound:p=1:1} and~\eqref{eqn:prop:Appendix:ReferenceProcess:N4:Bound:p>=1:1} can be further developed into
    \begin{align}\label{eqn:prop:Appendix:ReferenceProcess:N4:Bound:p=1:Final}
        \ELaw{y_0}{N_4^r(t)}
        \leq& 
        \sqrt{m}
        \frac{\expo{2\lambda t}}{2\lambda}  
        \sup_{\nu \in [0,\tstar]}
        \left(
            \LpLaw{2}{y_0}{\mathcal{P}^r_{\sigma}(\nu)} 
            + 
            \LpLaw{2}{y_0}{\mathcal{P}^r_{\sigma_\star}(\nu)} 
        \right)
        \sup_{\nu \in [0,\tstar]}
        \LpLaw{2}{y_0}{
            \Fparasigma{\nu,\Xrt{N,\nu}}^\top
        }
        , 
        \quad \forall t \in [0,\tstar],
    \end{align}
    and 
    \begin{multline}\label{eqn:prop:Appendix:ReferenceProcess:N4:Bound:p>=2:Final}
        \LpLaw{\sfp}{y_0}{N_4^r(t)}
        \leq 
        \left(
            \frac{\sfp \sqrt{(\sfp - 1)(2\sfp - 1)}}{4} 
            \frac{\expo{ 2\lambda t }}{\sqrt{\lambda \Boldomega }}
            +
            \sqrt{m}
            \frac{\expo{2\lambda t}}{2\lambda} 
        \right)
        \\
        \times 
        \sup_{\nu \in [0,\tstar]}
        \left(
            \LpLaw{2\sfp}{y_0}{\mathcal{P}^r_{\sigma}(\nu)} 
            + 
            \LpLaw{2\sfp}{y_0}{\mathcal{P}^r_{\sigma_\star}(\nu)} 
        \right)
        \sup_{\nu \in [0,\tstar]}
        \LpLaw{2\sfp}{y_0}{
            \Fparasigma{\nu,\Xrt{N,\nu}}^\top
        }
        , 
    \end{multline}
    for all $(t,\sfp) \in [0,\tstar] \times \mathbb{N}_{\geq 2}$.
    Then, the bound in~\eqref{eqn:prop:Appendix:ReferenceProcess:N:Bounds:p=1:Main} is established by adding together~\eqref{eqn:prop:Appendix:ReferenceProcess:N1:Bound:Final},~\eqref{eqn:prop:Appendix:ReferenceProcess:N2:Bound:Final},~\eqref{eqn:prop:Appendix:ReferenceProcess:N3:Bound:p=1:Final}, and~\eqref{eqn:prop:Appendix:ReferenceProcess:N4:Bound:p=1:Final}. 
    Similarly, the bounds in~\eqref{eqn:prop:Appendix:ReferenceProcess:N1:Bound:Final},~\eqref{eqn:prop:Appendix:ReferenceProcess:N2:Bound:Final},~\eqref{eqn:prop:Appendix:ReferenceProcess:N3:Bound:p>=2:Final}, and~\eqref{eqn:prop:Appendix:ReferenceProcess:N4:Bound:p>=2:Final}, lead to~\eqref{eqn:prop:Appendix:ReferenceProcess:N:Bounds:p>=2:Main}.

\end{proof}

Similar to the last proposition, the following result establishes the bound on the input to the reference system.
\begin{proposition}\label{prop:Appendix:ReferenceProcess:NU:Bounds}
    Consider the following scalar process:
    \begin{align}\label{eqn:prop:Appendix:ReferenceProcess:NU:Definitions}
        N^r_{\mathcal{U}}(t)
        =  
        \int_0^{t}
            \expo{\Boldomega \nu}
            M_\mathcal{U}(t,\nu)
            \left( 
                \Lparamu{\nu, \Xrt{N,\nu}}d\nu
                +
                \Fparasigma{\nu,\Xrt{N,\nu}}d\Wt{\nu}
            \right)
        ,
    \end{align}
    where  
    \begin{align*}
        M_\mathcal{U}(t,\nu)
        =
        \int_\nu^{t} 
            \expo{ (2\lambda - \Boldomega) \beta } \mathcal{P}^r_\mathcal{U}(\beta)^\top
        d\beta, 
    \end{align*}
    for $0 \leq \nu \leq t \leq T$, and  where $\mathcal{P}^r_{\mathcal{U}}(\beta) \in \mathbb{R}^m$ is  defined in the statement of Proposition~\ref{prop:Appendix:ReferenceProcess:ddV:Bound}.
    If the stopping time $\tau^\star$, defined in~\eqref{eqn:Reference:StoppingTimes}, Lemma~\ref{lem:Reference:dV}, satisfies  $\tau^\star = \tstar$, then
    \begin{align}\label{eqn:prop:Appendix:ReferenceProcess:NU:Bounds:Main}
        \LpLaw{\sfp}{y_0}{N^r_{\mathcal{U}}(t)}
        \leq
        \Delta_g^2 
        \frac{\expo{ 2\lambda t } }{\lambda}
        \left( 
            \frac{1}{\BoldomegaRoot}
            \sup_{\nu \in [0,\tstar]}
            \LpLaw{2\sfp}{y_0}{\Lparamu{\nu,\Xrt{N,\nu}}}
            + 
            \left(\sfp \frac{2\sfp-1}{2}\right)^\Half  
            \sup_{\nu \in [0,\tstar]}
            \LpLaw{2\sfp}{y_0}{\Fparasigma{\nu,\Xrt{N,\nu}}}
        \right)^2
        ,
    \end{align} 
   for all $(t,\sfp) \in [0,\tstar] \times \mathbb{N}_{\geq 1}$.
    
\end{proposition}
\begin{proof}
    We begin by decomposing the process $N^r_{\mathcal{U}}(t)$ as follows:
    \begin{equation}\label{eqn:prop:Appendix:ReferenceProcess:NU:Decomposed}
        N^r_{\mathcal{U}}(t)
        =  
        N^r_{\mathcal{U}_1}(t) + N^r_{\mathcal{U}_2}(t), 
    \end{equation}
    where 
    \begin{align*}
        N^r_{\mathcal{U}_1}(t)
        =
        \int_0^{t}
            \expo{\Boldomega \nu}
            M_\mathcal{U}(t,\nu)\Lparamu{\nu, \Xrt{N,\nu}}d\nu
        ,
        \quad 
        N^r_{\mathcal{U}_2}(t)
        =
        \int_0^{t}
            \expo{\Boldomega \nu}
            M_\mathcal{U}(t,\nu)\Fparasigma{\nu,\Xrt{N,\nu}}d\Wt{\nu}.
    \end{align*}

    Now, consider the process $N^r_{\mathcal{U}_1}(t)$, which, using the definition of $M_\mathcal{U}$ can be written as 
    \begin{multline*}
        N^r_{\mathcal{U}_1}(t)
        =
        \int_0^{t}
            \expo{\Boldomega \nu}M_\mathcal{U}(t,\nu)\Lparamu{\nu, \Xrt{N,\nu}}
        d\nu
        \\
        =
        \int_0^{t}
            \expo{\Boldomega \nu}
            \int_\nu^{t} 
                \expo{ (2\lambda - \Boldomega) \beta } \mathcal{P}^r_\mathcal{U}(\beta)^\top
            d\beta
            \Lparamu{\nu, \Xrt{N,\nu}}
        d\nu
        \\
        =
        \int_0^{t}
        \int_\nu^{t}
            \expo{\Boldomega \nu} 
                \expo{ (2\lambda - \Boldomega) \beta } \mathcal{P}^r_\mathcal{U}(\beta)^\top
            \Lparamu{\nu, \Xrt{N,\nu}}
            d\beta
        d\nu.
    \end{multline*}
    Since the integral above is a nested Lebesgue integral with $t$-continuous integrands, we may change the order of integration as follows: 
    \begin{multline}\label{eqn:prop:Appendix:ReferenceProcess:NU1:ReDefined}
        N^r_{\mathcal{U}_1}(t)
        =
        \int_0^{t}
        \int_0^{\beta}
            \expo{\Boldomega \nu} 
                \expo{ (2\lambda - \Boldomega) \beta } \mathcal{P}^r_\mathcal{U}(\beta)^\top
            \Lparamu{\nu, \Xrt{N,\nu}}
            d\nu
        d\beta
        \\
        =
        \int_0^{t}
            \expo{ (2\lambda - \Boldomega) \beta }
            \mathcal{P}^r_\mathcal{U}(\beta)^\top
            \left(
            \int_0^{\beta}
                \expo{\Boldomega \nu} 
                \Lparamu{\nu, \Xrt{N,\nu}}
            d\nu
            \right)
        d\beta
        \\
        =
        \int_0^{t}
            \expo{ (2\lambda - \Boldomega) \nu }
            \mathcal{P}^r_\mathcal{U}(\nu)^\top
            \left(
            \int_0^{\nu}
                \expo{\Boldomega \beta} 
                \Lparamu{\beta, \Xrt{N,\beta}}
            d\beta
            \right)
        d\nu,
    \end{multline}
    where in the last integral we have switched between the variables $\beta$ and $\nu$.

    Next, consider the process $N^r_{\mathcal{U}_2}(t)$, which, using the definition of $M_\mathcal{U}$ can be written as 
    \begin{multline*}
        N^r_{\mathcal{U}_2}(t)
        =
        \int_0^{t}
            \expo{\Boldomega \nu}
            M_\mathcal{U}(t,\nu)\Fparasigma{\nu,\Xrt{N,\nu}}
        d\Wt{\nu}
        \\
        =
        \int_0^{t}
        \int_\nu^{t} 
                \left(\expo{ (2\lambda - \Boldomega) \beta } \mathcal{P}^r_\mathcal{U}(\beta)^\top\right)
                \left(\expo{\Boldomega \nu} \Fparasigma{\nu,\Xrt{N,\nu}}\right)
            d\beta
        d\Wt{\nu}
        \\
        =
        \int_0^{t}
        \int_\nu^{t} 
                \left(\expo{ (2\lambda - \Boldomega) \beta } \mathcal{P}^r_\mathcal{U}(\beta)^\top\right)
                \left(\expo{\Boldomega \nu} \begin{bmatrix} \Fparasigma{\nu,\Xrt{N,\nu}} & 0_{m,d} \end{bmatrix}\right)
            d\beta
        d\Wrt{\nu},
    \end{multline*}
    where once again recall from Definition~\ref{def:Reference:JointProcess} that  $\Wrt{t} = \begin{bmatrix}\left(\Wt{t}\right)^\top & \left(\Wstart{t}\right)^\top \end{bmatrix}^\top \in \mathbb{R}^{2d}$.
    Using the definition of $P^r_\mathcal{U}$ in Proposition~\ref{prop:Appendix:ReferenceProcess:ddV:Bound}, and the regularity assumptions in Sec.~\ref{subsec:Assumptions}, it is straightforward to show that $P^r_\mathcal{U} \in \mathcal{M}_{2}^{loc}\br{\mathbb{R}^{m}|\Wfilt{t} \times \Wstarfilt{t}}$ and $\begin{bmatrix}\Fparasigma{\nu,\Xrt{N,\nu}} & 0_{m,d} \end{bmatrix} \in \mathcal{M}_{2}^{loc}\br{\mathbb{R}^{m \times 2d}|\Wfilt{t} \times \Wstarfilt{t}}$.  
    Therefore, we invoke Lemma~\ref{lem:TechnicalResults:Fubini-ish} to obtain 
    \begin{multline}\label{eqn:prop:Appendix:ReferenceProcess:NU2:ReDefined}
        N^r_{\mathcal{U}_2}(t)
        =
        \int_0^{t}
            \expo{\Boldomega \nu}
            M_\mathcal{U}(t,\nu)\Fparasigma{\nu,\Xrt{N,\nu}}
        d\Wt{\nu}
        \\
        =
        \int_0^{t}
            \left(\expo{ (2\lambda - \Boldomega) \nu } \mathcal{P}^r_\mathcal{U}(\nu)^\top\right)
            \int_0^\nu 
                \left(\expo{\Boldomega \beta} \begin{bmatrix} \Fparasigma{\beta,\Xrt{N,\beta}} & 0_{m,d} \end{bmatrix}\right)
            d\Wrt{\beta}
        d\nu
        \\
        =
        \int_0^{t}
            \expo{ (2\lambda - \Boldomega) \nu } 
            \mathcal{P}^r_\mathcal{U}(\nu)^\top
            \left( 
                \int_0^\nu 
                    \expo{\Boldomega \beta} \Fparasigma{\beta,\Xrt{N,\beta}}
                d\Wt{\beta}
            \right)
        d\nu.
    \end{multline}
    Substituting~\eqref{eqn:prop:Appendix:ReferenceProcess:NU1:ReDefined} and~\eqref{eqn:prop:Appendix:ReferenceProcess:NU2:ReDefined} into~\eqref{eqn:prop:Appendix:ReferenceProcess:NU:Decomposed} yields 
    \begin{align}\label{eqn:prop:Appendix:ReferenceProcess:NU:Redefined}
        N^r_{\mathcal{U}}(t)
        = 
        \int_0^{t}
            \expo{ (2\lambda - \Boldomega) \nu }
            \mathcal{P}^r_\mathcal{U}(\nu)^\top
            \left(
            \int_0^{\nu}
                \expo{\Boldomega \beta}
                \left[ 
                    \Lparamu{\beta, \Xrt{N,\beta}}d\beta
                    +
                    \Fparasigma{\beta,\Xrt{N,\beta}}
                    d\Wt{\beta}
                \right]
            \right)
        d\nu
        .
    \end{align}
    Substituting the definition of $\mathcal{P}^r_{\mathcal{U}}$ from Proposition~\ref{prop:Appendix:ReferenceProcess:ddV:Bound} into the above:
    \begin{align*}
        \mathcal{P}^r_\mathcal{U}(t)
        =&
        2 
        g(t)^\top
        g(t)
        \Urt{t}
        \\
        \overset{(\star)}{=}&
        2 
        g(t)^\top
        g(t)
        \left( 
            \Filter[\Lparamu{\cdot,\Xrt{N}}][t] 
            + \FilterW[\Fparasigma{\cdot,\Xrt{N}},\Wt{}][t]
        \right)
        \\
        =&
        2 
        g(t)^\top
        g(t)
        \left(
            -\Boldomega 
            \int_0^t \expo{-\Boldomega(t-\nu)}\Lparamu{\nu,\Xrt{N,\nu}}d\nu
            -\Boldomega
            \int_0^t \expo{-\Boldomega(t-\nu)}\Fparasigma{\nu,\Xrt{N,\nu}}d\Wt{\nu}
        \right)
        \\
        =&
        -2\Boldomega
        \expo{-\Boldomega t}
        g(t)^\top
        g(t)
        \left( 
            \int_0^t 
            \expo{\Boldomega \nu}
            \left[ 
                \Lparamu{\nu,\Xrt{N,\nu}}d\nu
                +
                \Fparasigma{\nu,\Xrt{N,\nu}}d\Wt{\nu}
            \right]
        \right),
    \end{align*}
    where $(\star)$ follows from the definition of $\Urt{t}$ in~\eqref{eqn:ReferenceFeedbackOperatorProcess} for the truncated process~\eqref{eqn:Reference:TruncatedJointProcess}.
    Substituting the above expression into~\eqref{eqn:prop:Appendix:ReferenceProcess:NU:Redefined} produces 
    \begin{multline}\label{eqn:prop:Appendix:ReferenceProcess:NU:Redefined:Final}
        N^r_{\mathcal{U}}(t)
        = 
        -2\Boldomega
        \int_0^{t}
            \expo{ 2(\lambda - \Boldomega) \nu }
            \left( 
                    \int_0^\nu 
                    \expo{\Boldomega \beta}
                    \left[ 
                            \Lparamu{\beta,\Xrt{N,\beta}}d\beta
                            +
                            \Fparasigma{\beta,\Xrt{N,\beta}}d\Wt{\beta}
                    \right]
            \right)^\top 
            g(\nu)^\top g(\nu)
        \\
        \times 
            \left(
            \int_0^{\nu}
                \expo{\Boldomega \beta}
                \left[ 
                    \Lparamu{\beta, \Xrt{N,\beta}}d\beta
                    +
                    \Fparasigma{\beta,\Xrt{N,\beta}}
                    d\Wt{\beta}
                \right]
            \right)
        d\nu
        .
    \end{multline}
    Observe that $N^r_\mathcal{U}(t)$ can be expressed as the process $N(t)$ in the statement of Lemma~\ref{cor:TechnicalResults:UTildeBound} by setting 
    \begin{align*}
        Q_t = \Wrt{t} \in \mathbb{R}^{2d}\,(n_q=2d), 
        \quad 
        \mathfrak{F}_t = \Wfilt{t} \times \Wstarfilt{t},
        \quad 
        \theta_1 = \lambda, 
        \quad 
        \theta_2 = \Boldomega,
        \quad 
        \xi =0
        ,
        \\
        R(t) = -2\Boldomega g(t)^\top g(t) \in \mathcal{M}_{2}^{loc}\br{\mathbb{S}^m|\Wfilt{t} \times \Wstarfilt{t}},
        \\
        S_1(t) = S_2(t) = \Lparamu{t,\Xrt{N,t}} \in \mathcal{M}_{2}^{loc}\br{\mathbb{R}^m|\Wfilt{t} \times \Wstarfilt{t}}, 
        \\
        L_1(t) = L_2(t) = \begin{bmatrix}\Fparasigma{t,\Xrt{N,t}} & 0_{m,d} \end{bmatrix}
        \in     
        \mathcal{M}_{2}^{loc}\br{\mathbb{R}^{m \times 2d}|\Wfilt{t} \times \Wstarfilt{t}}
        .   
    \end{align*}
    Furthermore, as a consequence of Assumption~\ref{assmp:KnownFunctions}, we may set $\Delta_R$ in the hypothesis of Lemma~\ref{cor:TechnicalResults:UTildeBound} as $\Delta_R = 2 \Delta_g^2  \Boldomega$.
    Hence, applying Lemma~\ref{cor:TechnicalResults:UTildeBound} to the process $N^r_{\mathcal{U}}(t)$ in~\eqref{eqn:prop:Appendix:ReferenceProcess:NU:Redefined:Final} leads to 
    \begin{multline*}
        \LpLaw{\sfp}{y_0}{N^r_{\mathcal{U}}(t)}
        \leq
        \Delta_g^2  \Boldomega
        \frac{\expo{ 2\lambda t } - 1}{\lambda}
        \left( 
            \frac{1-\expo{-\Boldomega t}}{\Boldomega}
            \sup_{\nu \in [0,\tstar]}
            \LpLaw{2\sfp}{y_0}{\Lparamu{\nu,\Xrt{N,\nu}}}
        \right.
        \\
        \left.
            + 
            \left(\sfp \frac{2\sfp-1}{2}\right)^\Half  
            \left(\frac{1 - \expo{ -2\Boldomega t }}{\Boldomega}\right)^\Half
            \sup_{\nu \in [0,\tstar]}
            \LpLaw{2\sfp}{y_0}{\Fparasigma{\nu,\Xrt{N,\nu}}}
        \right)^2
        ,
    \end{multline*}
   for all $(t,\sfp) \in [0,\tstar] \times \mathbb{N}_{\geq 1}$.
   Then, bounding $\expo{ 2\lambda t } - 1$ by $\expo{ 2\lambda t }$, and  $1-\expo{-\Boldomega t}$ and $1 - \expo{ -2\Boldomega t }$  by $1$ produces desired result in~\eqref{eqn:prop:Appendix:ReferenceProcess:NU:Bounds:Main}.


\end{proof}


The next lemma establishes the bound on $\Xi^r$.
\begin{lemma}\label{lem:Appendix:ReferenceProcess:Xir}
    If the stopping time $\tau^\star$, defined in~\eqref{eqn:Reference:StoppingTimes}, Lemma~\ref{lem:Reference:dV}, satisfies  $\tau^\star = \tstar$, then 
    \begin{subequations}\label{eqn:lem:Appendix:ReferenceProcess:Xir:Bound:Alt:Final}
        \begin{align}
            \ELaw{y_0}{
                \expo{-2\lambda t}
                \Xi^r\br{t,\Yt{N}}
            }
            \leq 
            \frac{\Delta_{\Xi_1}^r(1,\tstar)}{2\lambda} 
            , 
            \quad 
            \forall t \in [0,\tstar],
            \label{eqn:lem:Appendix:ReferenceProcess:Xir:Bound:Alt:Final:A}
            \\
            \pLpLaw{y_0}{
                \expo{-2\lambda t}
                \Xi^r\br{t,\Yt{N}}
            }
            \leq 
            \frac{\Delta_{\Xi_1}^r(\sfp,\tstar)}{2\lambda} 
            +
            \frac{\Delta_{\Xi_2}^r(\sfp,\tstar)}{2\sqrt{\lambda}} 
            ,
            \quad 
            \forall (t,\sfp) \in [0,\tstar] \times \mathbb{N}_{\geq 2},
            \label{eqn:lem:Appendix:ReferenceProcess:Xir:Bound:Alt:Final:B}
        \end{align}
    \end{subequations}
    where the term $\Xi^r\br{\tau(t),\Yt{N}}$ is defined in~\eqref{eqn:lem:Reference:dV:Xi:Functions}, Lemma~\ref{lem:Reference:dV}, and $\Delta_{\Xi_1}^r(\sfp,\tstar)$ and $\Delta_{\Xi_2}^r(\sfp,\tstar)$ are defined for $\sfp \in \mathbb{N}_{\geq 1}$ as follows: 
    \begin{align*}
        \begin{multlined}[][0.8\linewidth]
            \Delta_{\Xi_1}^r(\sfp,\tstar)
            =
            \sup_{\nu \in [0,\tstar]}
            \left(
                \Delta_g^\perp
                \LpLaw{2\sfp}{y_0}{
                    \RefDiffNorm{\Yt{N,\nu}} 
                } 
                \LpLaw{2\sfp}{y_0}{
                    \Lperpmu{\nu,\Xrt{N,\nu}}
                }
            \right.
            \\
            \left.  
                +
                \ELaw{y_0}{
                    \Fsigma{\nu,\Xrt{N,\nu}}^{2\sfp}
                }^\frac{1}{\sfp}
                +
                \ELaw{y_0}{
                    \Fbarsigma{\nu,\Xstart{N,\nu}}^{2\sfp}
                }^\frac{1}{\sfp}     
            \right)
            ,
        \end{multlined}
        \\
        \Delta_{\Xi_2}^r(\sfp,\tstar)
        =
        \mathfrak{p}(\sfp)
        \sup_{\nu \in [0,\tstar]}
        \left(
            \Delta_g^\perp
            \LpLaw{2\sfp}{y_0}{
                \Fperpsigma{\nu,\Xrt{N,\nu}}
            }
            + 
            \LpLaw{2\sfp}{y_0}{
                \Fbarsigma{\nu,\Xstart{N,\nu}}
            }
        \right)
        \sup_{\nu \in [0,\tstar]}
        \LpLaw{2\sfp}{y_0}{
            \RefDiffNorm{\Yt{N,\nu}}
        } 
        ,
    \end{align*}
    where the constant $\mathfrak{p}(\sfp)$ is defined in~\eqref{eqn:app:Constants:FrakP}.

\end{lemma}
\begin{proof}
    We begin with the definition of $\Xi^r$ in~\eqref{eqn:lem:Reference:dV:Xi:Functions} to write  
    \begin{multline}\label{eqn:lem:Appendix:ReferenceProcess:Xir:Bound:PreExpectation}
        \expo{-2\lambda t}
        \Xi^r\br{t,\Yt{N}}
        =
        \expo{-2\lambda t}
        \int_0^{t}  
            \expo{ 2 \lambda  \nu }\phi^r_{\mu}\br{\nu,\Yt{N,\nu}}
        d\nu
        \\
        +
        \expo{-2\lambda t}
        \int_0^{t}  
            \expo{ 2 \lambda  \nu }  
            \begin{bmatrix}
            \phi^r_{\sigma}\br{\nu,\Yt{N,\nu}}
            &
            \phi^r_{\sigma_\star}\br{\nu,\Yt{N,\nu}}
            \end{bmatrix}
        d\Wrt{\nu},
        \quad t \in [0,\tstar],
    \end{multline}
    where $\Wrt{\nu} \in \mathbb{R}^{2d}$ is defined in Definition~\ref{def:Reference:JointProcess}.  
    The regularity assumptions and the $t$-continuity of the strong solutions $\Xrt{N,t}$ and $\Xstart{N,t}$ imply that $\begin{bmatrix}\phi^r_{\sigma}\br{\nu,\Yt{N,\nu}} & \phi^r_{\sigma_\star}\br{\nu,\Yt{N,\nu}}\end{bmatrix} \in \mathcal{M}^{loc}_2\left(\mathbb{R}^{1 \times 2d}|\Wfilt{t} \times \Wstarfilt{t} \right)$.
    Hence, taking the expectation of~\eqref{eqn:lem:Appendix:ReferenceProcess:Xir:Bound:PreExpectation} and using the linearity of the expectation operator, along with~\cite[Thm.~1.5.21]{mao2007stochastic}, leads to  
    \begin{align}\label{eqn:lem:Appendix:ReferenceProcess:Xir:Bound:A:Initial}
        \ELaw{y_0}{
            \expo{-2\lambda t}
            \Xi^r\br{t,\Yt{N}}
        }
        =&
        \expo{-2\lambda t}
        \ELaw{y_0}{
            \int_0^{t}  
                \expo{ 2 \lambda  \nu }\phi^r_{\mu}\br{\nu,\Yt{N,\nu}}
            d\nu
        }
        \notag 
        \\
        \leq&     
        \expo{-2\lambda t}
        \ELaw{y_0}{
            \int_0^{t}  
                \expo{ 2 \lambda  \nu }\absolute{\phi^r_{\mu}\br{\nu,\Yt{N,\nu}}}
            d\nu
        }
        ,
        \quad t \in [0,\tstar].
    \end{align}
    In the above, we have used the fact that since $\tau^\star = \tstar$, and $\tstar$ is a constant, 
    \begin{align*} 
        \mathbb{E}\left[\expo{-2\lambda t}\right]
        =
        \expo{-2\lambda t}
        ,
        \, \forall t \in [0,\tstar].
    \end{align*} 
    Moreover,~\eqref{eqn:lem:Appendix:ReferenceProcess:Xir:Bound:PreExpectation} and the Minkowski's inequality imply that  
    \begin{multline}\label{eqn:lem:Appendix:ReferenceProcess:Xir:Bound:B:Initial}
        \pLpLaw{y_0}{
            \expo{-2\lambda t}
            \Xi^r\br{t,\Yt{N}}
        }
        \leq
        \expo{-2\lambda t}
        \pLpLaw{y_0}{
            \int_0^{t}  
                \expo{ 2 \lambda  \nu } \absolute{\phi^r_{\mu}\br{\nu,\Yt{N,\nu}}}
            d\nu
        }
        \\
        +
        \expo{-2\lambda t}
        \pLpLaw{y_0}{
            \int_0^{t}  
                \expo{ 2 \lambda  \nu }  
                \begin{bmatrix}
                \phi^r_{\sigma}\br{\nu,\Yt{N,\nu}}
                &
                \phi^r_{\sigma_\star}\br{\nu,\Yt{N,\nu}}
                \end{bmatrix}
            d\Wrt{\nu}
        }
        ,
    \end{multline}
    for all $(t,\sfp) \in [0,\tstar] \times \mathbb{N}_{\geq 2}$.
    
    Now, the regularity assumptions and the $t$-continuity of the strong solutions $\Xrt{N,t}$ and $\Xstart{N,t}$ over $[0,\tstar] \subseteq [0,\tau_N]$ imply that $\phi^r_{\mu}\br{\nu,\Yt{N,\nu}} \in \mathcal{M}^{loc}_2\left(\mathbb{R}|\Wfilt{t} \times \Wstarfilt{t} \right)$.
    Thus, we invoke Proposition~\ref{prop:TechnicalResults:LebesgueMoment} for $\cbr{\theta,\xi,n_s,S} = \cbr{2\lambda,0,1,\phi^r_{\mu}\br{\cdot,\Yt{N}}}$ to obtain
    \begin{align}\label{eqn:lem:Appendix:ReferenceProcess:Xir:Bound:Mu:Initial}
        \pLpLaw{y_0}{
            \int_0^{t}  
                \expo{ 2 \lambda  \nu }\absolute{\phi^r_{\mu}\br{\nu,\Yt{N,\nu}}}
            d\nu
        }
        \leq& 
        \frac{\expo{2\lambda t}}{2\lambda} 
        \sup_{\nu \in [0,t]}
        \pLpLaw{y_0}{\phi^r_{\mu}\br{\nu,\Yt{N,\nu}}}
        \leq 
        \frac{\expo{2\lambda t}}{2\lambda} 
        \sup_{\nu \in [0,\tstar]}
        \pLpLaw{y_0}{\phi^r_{\mu}\br{\nu,\Yt{N,\nu}}}
        , 
    \end{align}
    for all $ (t,\sfp) \in [0,\tstar] \times \mathbb{N}_{\geq 1}$, where we have bounded $\expo{2\lambda t} -1$ by $\expo{2\lambda t}$. 
    Now, we obtain the following from the definition of $\phi^r_{\mu}$ in~\eqref{eqn:lem:Reference:dV:phi:Functions}: 
    \begin{align*}
        \absolute{\phi^r_{\mu}\br{\nu,\Yt{N,\nu}}}
        \leq&
        \norm{\RefDiffNorm{\Yt{N, \nu}}} 
        \Frobenius{g(\nu)^\perp}
        \norm{\Lperpmu{\nu,\Xrt{N,\nu}}} 
        +
        \Frobenius{\Fsigma{\nu,\Xrt{N,\nu}}}^2
        +
        \Frobenius{\Fbarsigma{\nu,\Xstart{N,\nu}}}^2
        \\
        \leq&
        \Delta_g^\perp
        \norm{\RefDiffNorm{\Yt{N, \nu}}} 
        \norm{\Lperpmu{\nu,\Xrt{N,\nu}}} 
        +
        \Frobenius{\Fsigma{\nu,\Xrt{N,\nu}}}^2
        +
        \Frobenius{\Fbarsigma{\nu,\Xstart{N,\nu}}}^2
        ,
    \end{align*}
    where we have used the bound on $g(t)^\perp$ from Assumption~\ref{assmp:KnownFunctions}.
    Using the Minkowski's inequality followed by the Cauchy-Schwarz inequality, we obtain  
    \begin{multline*}
        \pLpLaw{y_0}{\phi^r_{\mu}\br{t,\Yt{N,t}}}
        \leq
        \Delta_g^\perp
        \LpLaw{2\sfp}{y_0}{
            \RefDiffNorm{\Yt{N,t}} 
        } 
        \LpLaw{2\sfp}{y_0}{
            \Lperpmu{\nu,\Xrt{N,t}}
        } 
        \\
        +
        \ELaw{y_0}{
            \Fsigma{\nu,\Xrt{N,t}}^{2\sfp}
        }^\frac{1}{\sfp}
        +
        \ELaw{y_0}{
            \Fbarsigma{\nu,\Xstart{N,t}}^{2\sfp}
        }^\frac{1}{\sfp}, 
        \quad 
            \forall (t,\sfp) \in [0,\tstar] \times \mathbb{N}_{\geq 1}.
    \end{multline*}
    Substituting the above bound into~\eqref{eqn:lem:Appendix:ReferenceProcess:Xir:Bound:Mu:Initial} produces
    \begin{multline}\label{eqn:lem:Appendix:ReferenceProcess:Xir:Bound:Mu:Final}
        \pLpLaw{y_0}{
            \int_0^{t}  
                \expo{ 2 \lambda  \nu }\absolute{\phi^r_{\mu}\br{\nu,\Yt{N,\nu}}}
            d\nu
        }
        \leq 
        \frac{\expo{2\lambda t}}{2\lambda}
        \sup_{\nu \in [0,\tstar]}
        \left(
            \Delta_g^\perp
            \LpLaw{2\sfp}{y_0}{
                \RefDiffNorm{\Yt{N,\nu}} 
            } 
            \LpLaw{2\sfp}{y_0}{
                \Lperpmu{\nu,\Xrt{N,\nu}}
            } 
        \right. 
        \\
        \left.
            +
            \ELaw{y_0}{
                \Fsigma{\nu,\Xrt{N,\nu}}^{2\sfp}
            }^\frac{1}{\sfp}
            +
            \ELaw{y_0}{
                \Fbarsigma{\nu,\Xstart{N,\nu}}^{2\sfp}
            }^\frac{1}{\sfp}     
        \right)
        , 
    \end{multline}
    for all $(t,\sfp) \in [0,\tstar] \times \mathbb{N}_{\geq 1}$.
    Next, using the definitions of $\phi^r_{\sigma_\star}$ and $\phi^r_{\sigma}$ in~\eqref{eqn:lem:Reference:dV:phi:Functions}, we obtain
    \begin{multline*}
        \int_0^{t}  
            \expo{ 2 \lambda  \nu }  
            \begin{bmatrix}
                \phi^r_{\sigma}\br{\nu,\Yt{N,\nu}}
                &
                \phi^r_{\sigma_\star}\br{\nu,\Yt{N,\nu}}
            \end{bmatrix}
        d\Wrt{\nu}
        \\
        =
        \int_0^{t}  
            \expo{ 2 \lambda  \nu }  
            \begin{bmatrix}
                \RefDiffNorm{\Yt{N,\nu}}^\top  
                g(\nu)^\perp \Fperpsigma{\nu,\Xrt{N,\color{brown}\nu}}
                &
                -\RefDiffNorm{\Yt{N,\nu}}^\top 
                \Fbarsigma{\nu,\Xstart{N,\nu}} 
            \end{bmatrix}
        d\Wrt{\nu}
        ,
    \end{multline*}
    where $\Wrt{\nu} \in \mathbb{R}^{2d}$ is defined in Definition~\ref{def:Reference:JointProcess}. 
    Recall from the discussion after~\eqref{eqn:lem:Appendix:ReferenceProcess:Xir:Bound:PreExpectation} that $\begin{bmatrix}\phi^r_{\sigma}\br{\nu,\Yt{N,\nu}} & \phi^r_{\sigma_\star}\br{\nu,\Yt{N,\nu}}\end{bmatrix} \in \mathcal{M}^{loc}_2\left(\mathbb{R}^{1 \times 2d}|\Wfilt{t} \times \Wstarfilt{t} \right)$.
    Thus, we may use Lemma~\ref{lemma:TechnicalResults:MartingaleMoment} with $\cbr{\theta,\xi,n_l,n_q,\Qt{}} = \cbr{2\lambda,0,1,2d, \Wrt{} }$, and 
    \begin{align*}
        L(t)
        =
        \begin{bmatrix}
            \RefDiffNorm{\Yt{N,t}}^\top  
            g(t)^\perp \Fperpsigma{t,\Xrt{N,t}}
            &
            -\RefDiffNorm{\Yt{N,t}}^\top 
            \Fbarsigma{\nu,\Xstart{N,t}} 
        \end{bmatrix}
        ,
    \end{align*}
    to obtain
    \begin{multline}\label{eqn:lem:Appendix:ReferenceProcess:Xir:Bound:Sigma:1}
        \pLpLaw{y_0}{
            \int_0^{t}  
                \expo{ 2 \lambda  \nu }  
                \begin{bmatrix}
                    \phi^r_{\sigma}\br{\nu,\Yt{N,\nu}}
                    &
                    \phi^r_{\sigma_\star}\br{\nu,\Yt{N,\nu}}
                \end{bmatrix}
            d\Wrt{\nu}
        }
        \\
        \leq 
        \left(\sfp \frac{\sfp-1}{2}\right)^\Half  
        \frac{\expo{ 2\lambda t }}{2\sqrt{\lambda}}
        \sup_{\nu \in [0,t]}
        \LpLaw{\sfp}{y_0}{
            \begin{bmatrix}
                \RefDiffNorm{\Yt{N,\nu}}^\top  
                g(t)^\perp \Fperpsigma{ \nu, \Xrt{N,\nu}}
                &
                -\RefDiffNorm{\Yt{N,\nu}}^\top 
                \Fbarsigma{\nu,\Xstart{N,\nu}} 
            \end{bmatrix}
        }
        \\
        \leq 
        \left(\sfp \frac{\sfp-1}{2}\right)^\Half  
        \frac{\expo{ 2\lambda t }}{2\sqrt{\lambda}}
        \sup_{\nu \in [0,\tstar]}
        \LpLaw{\sfp}{y_0}{
            \begin{bmatrix}
                \RefDiffNorm{\Yt{N,\nu}}^\top  
                g(t)^\perp \Fperpsigma{\nu,\Xrt{N,\nu}}
                &
                -\RefDiffNorm{\Yt{N,\nu}}^\top 
                \Fbarsigma{\nu,\Xstart{N,\nu}} 
            \end{bmatrix}
        }
        ,
    \end{multline}
    for all $ (t,\sfp) \in [0,\tstar] \times \mathbb{N}_{\geq 2}$, where we have bounded $\left(\expo{4\lambda t} -1\right)^\Half$ by $\expo{2\lambda t}$. 
    Next, observe that    
    \begin{multline*}
        \norm{
            \begin{bmatrix}
                \RefDiffNorm{\Yt{N,t}}^\top  
                g(t)^\perp \Fperpsigma{t,\Xrt{N,t}}
                &
                -\RefDiffNorm{\Yt{N,t}}^\top 
                \Fbarsigma{ t,\Xstart{N,t}} 
            \end{bmatrix}
        }
        \\
        =
        \left( 
            \norm{
                \RefDiffNorm{\Yt{N,t}}^\top  
                g(t)^\perp \Fperpsigma{t,\Xrt{N,t}}
            }^2 
            + 
            \norm{
                \RefDiffNorm{\Yt{N,t}}^\top 
                \Fbarsigma{t,\Xstart{N,t}}
            }^2
        \right)^\Half
        \\
        \leq 
        \left( 
            \norm{
                \RefDiffNorm{\Yt{N,t}}^\top  
                g(t)^\perp \Fperpsigma{t,\Xrt{N,t}}
            } 
            + 
            \norm{
                \RefDiffNorm{\Yt{N,t}}^\top 
                \Fbarsigma{t,\Xstart{N,t}}
            }
        \right),
    \end{multline*}
    and thus, an application of the Minkowski's inequality yields
    \begin{multline*}
        \LpLaw{\sfp}{y_0}{
            \begin{bmatrix}
                \RefDiffNorm{\Yt{N,t}}^\top  
                g(t)^\perp \Fperpsigma{t,\Xrt{N,t}}
                &
                -\RefDiffNorm{\Yt{N,t}}^\top 
                \Fbarsigma{t,\Xstart{N,t}} 
            \end{bmatrix}
        }
        \\
        \leq
        \left( 
            \LpLaw{\sfp}{y_0}{
                \RefDiffNorm{\Yt{N,t}}^\top  
                g(t)^\perp \Fperpsigma{t,\Xrt{N,t}}
            } 
            + 
            \LpLaw{\sfp}{y_0}{
                \RefDiffNorm{\Yt{N,t}}^\top 
                \Fbarsigma{t,\Xstart{N,t}}
            }
        \right)
        \\
        \leq 
        \left( 
            \Delta_g^\perp
            \LpLaw{2\sfp}{y_0}{
                \RefDiffNorm{\Yt{N,t}}
            } 
            \LpLaw{2\sfp}{y_0}{
                \Fperpsigma{t,\Xrt{N,t}}
            } 
            + 
            \LpLaw{2\sfp}{y_0}{
                \RefDiffNorm{\Yt{N,t}}
            }
            \LpLaw{2\sfp}{y_0}{
                \Fbarsigma{t,\Xstart{N,t}}
            }
        \right)
        ,
    \end{multline*}
    where we have used the bound on $g(t)^\perp$ from Assumption~\ref{assmp:KnownFunctions}, and the Cauchy-Schwarz inequality to obtain the last bound. 
    Substituting the above bound into~\eqref{eqn:lem:Appendix:ReferenceProcess:Xir:Bound:Sigma:1} produces
    \begin{multline}\label{eqn:lem:Appendix:ReferenceProcess:Xir:Bound:Sigma:Final}
        \pLpLaw{y_0}{
            \int_0^{t}  
                \expo{ 2 \lambda  \nu }  
                \begin{bmatrix}
                    \phi^r_{\sigma}\br{\nu,\Yt{N,\nu}}
                    &
                    \phi^r_{\sigma_\star}\br{\nu,\Yt{N,\nu}}
                \end{bmatrix}
            d\Wrt{\nu}
        }
        \\
        \leq 
        \left(\sfp \frac{\sfp-1}{2}\right)^\Half  
        \frac{\expo{ 2\lambda t }}{2\sqrt{\lambda}}
        \sup_{\nu \in [0,\tstar]}
        \left( 
            \Delta_g^\perp
            \LpLaw{2\sfp}{y_0}{
                \RefDiffNorm{\Yt{N,\nu}}
            } 
            \LpLaw{2\sfp}{y_0}{
                \Fperpsigma{\nu,\Xrt{N,\nu}}
            } 
        \right.
        \\
        \left.
            + 
            \LpLaw{2\sfp}{y_0}{
                \RefDiffNorm{\Yt{N,\nu}}
            }
            \LpLaw{2\sfp}{y_0}{
                \Fbarsigma{\nu,\Xstart{N, \nu}}
            }
        \right)
        ,
    \end{multline}
    for all $ (t,\sfp) \in [0,\tstar] \times \mathbb{N}_{\geq 2}$.
    Substituting~\eqref{eqn:lem:Appendix:ReferenceProcess:Xir:Bound:Mu:Final} into~\eqref{eqn:lem:Appendix:ReferenceProcess:Xir:Bound:A:Initial} leads to~\eqref{eqn:lem:Appendix:ReferenceProcess:Xir:Bound:Alt:Final:A}, and substituting~\eqref{eqn:lem:Appendix:ReferenceProcess:Xir:Bound:Mu:Final} and~\eqref{eqn:lem:Appendix:ReferenceProcess:Xir:Bound:Sigma:Final} into~\eqref{eqn:lem:Appendix:ReferenceProcess:Xir:Bound:B:Initial} leads to~\eqref{eqn:lem:Appendix:ReferenceProcess:Xir:Bound:Alt:Final:B}, thus completing the proof. 

    \begingroup
    \endgroup
\end{proof}

We next derive the bound on $\Xi^r_{\mathcal{U}}$ in the following lemma. 
\begin{lemma}\label{lem:Appendix:ReferenceProcess:XirU}
    If the stopping time $\tau^\star$, defined in~\eqref{eqn:Reference:StoppingTimes}, Lemma~\ref{lem:Reference:dV}, satisfies  $\tau^\star = \tstar$, then the term $\Xi^r_{\mathcal{U}}\br{\tau(t),\Yt{N};\Boldomega}$ defined in~\eqref{eqn:lem:Reference:dV:Xi:Functions}, Lemma~\ref{lem:Reference:dV}, satisfies the following bounds:
    \begin{subequations}\label{eqn:lem:Appendix:ReferenceProcess:XirU:Bound:Final}
        \begin{align}
            \begin{multlined}[b][0.9\linewidth]
                \ELaw{y_0}{
                    \expo{-(2\lambda+\Boldomega) \tau(t)}
                    \Xi^r_{\mathcal{U}}\br{\tau(t),\Yt{N};\Boldomega}
                }
                \\
                \leq
                \frac{ 1  }{\absolute{2\lambda - \Boldomega}}
                \left(
                    \Delta^r_{\mathcal{U}_1}(1,\tstar)
                    +
                    \BoldomegaRoot 
                    \Delta^r_{\mathcal{U}_2}(1,\tstar)
                    +
                    \Boldomega
                    \Delta^r_{\mathcal{U}_3}(1,\tstar)
                \right)
                ,
                \quad \forall t \in [0,\tstar]
                ,
            \end{multlined}
            \label{eqn:lem:Appendix:ReferenceProcess:XirU:Bound:A:Final}
            \\
            \begin{multlined}[b][0.9\linewidth]
                \pLpLaw{y_0}{
                    \expo{-(2\lambda+\Boldomega) \tau(t)}
                    \Xi^r_{\mathcal{U}}\br{\tau(t),\Yt{N};\Boldomega}
                }
                \\
                \leq
                \frac{ 1  }{\absolute{2\lambda - \Boldomega}}
                \left(
                    \Delta^r_{\mathcal{U}_1}(\sfp,\tstar)
                    +
                    \BoldomegaRoot 
                    \Delta^r_{\mathcal{U}_2}(\sfp,\tstar)
                    +
                    \Boldomega
                    \Delta^r_{\mathcal{U}_3}(\sfp,\tstar)
                \right)
                ,
                \quad \forall (t,\sfp) \in [0,\tstar] \times \mathbb{N}_{\geq 2}
                ,
            \end{multlined}
            \label{eqn:lem:Appendix:ReferenceProcess:XirU:Bound:B:Final}
        \end{align}
    \end{subequations}
    where 
    \begin{align*}
        \begin{multlined}[][0.9\linewidth]
            \Delta^r_{\mathcal{U}_1}(\sfp,\tstar)
            =
            \sqrt{\lambda} 
            \Delta_g
            \mathfrak{p}(\sfp)  
            \sup_{\nu \in [0,\tstar]}
            \LpLaw{2\sfp}{}{
                \RefDiffNorm{\Yt{N,\nu}} 
            }
            \sup_{\nu \in [0,\tstar]}
            \LpLaw{2\sfp}{}{
                \Fparasigma{\nu,\Xrt{N,\nu}}
            }
            \\
            \qquad \qquad
            +
            \left(
                \frac{\Delta^r_{\mathcal{P}_1}(\sfp,\tstar)}{\sqrt{\lambda}}
                + 
                2\Delta_g 
                \sup_{\nu \in [0,\tstar]}\LpLaw{2\sfp}{y_0}{\RefDiffNorm{\Yt{N,\nu}}}
                + 
                \frac{\Delta_g^2}{\lambda}
                \sup_{\nu \in [0,\tstar]}
                \LpLaw{2\sfp}{y_0}{\Lparamu{\nu,\Xrt{N,\nu}}}
            \right)
            \\
            \times 
            \sup_{\nu \in [0,\tstar]}
            \LpLaw{2\sfp}{y_0}{\Lparamu{\nu,\Xrt{N,\nu}}}
            ,
        \end{multlined}
        \\   
        \begin{multlined}[][0.9\linewidth] 
            \Delta^r_{\mathcal{U}_2}(\sfp,\tstar)
            =
            \left(
                \frac{\Delta^r_{\mathcal{P}_2}(\sfp,\tstar)}{\sqrt{\lambda}}
                + 
                \Delta_g
                \mathfrak{p}'(\sfp)
                \left[   
                    \sup_{\nu \in [0,\tstar]}\LpLaw{2\sfp}{y_0}{\RefDiffNorm{\Yt{N,\nu}}}
                    +
                    \frac{2 \Delta_g }{\lambda}  
                    \sup_{\nu \in [0,\tstar]}
                    \LpLaw{2\sfp}{y_0}{\Lparamu{\nu,\Xrt{N,\nu}}}
                \right]
            \right)
            \\
            \times
            \sup_{\nu \in [0,\tstar]}
                \LpLaw{2\sfp}{y_0}{\Fparasigma{\nu,\Xrt{N,\nu}}}
            ,
        \end{multlined}
        \\
        \begin{aligned}
            \Delta^r_{\mathcal{U}_3}(\sfp,\tstar)
            =
            \left(
                \frac{ \Delta^r_{\mathcal{P}_3}(\sfp,\tstar)}{\lambda}
                +
                \frac{ \Delta_g^2  \mathfrak{p}'(\sfp)^2 }{\lambda}
                    \sup_{\nu \in [0,\tstar]}
                    \LpLaw{2\sfp}{y_0}{\Fparasigma{\nu,\Xrt{N,\nu}}}
            \right)
            \sup_{\nu \in [0,\tstar]}
                \LpLaw{2\sfp}{y_0}{\Fparasigma{\nu,\Xrt{N,\nu}}}
            ,
        \end{aligned}
    \end{align*}
    and where the constants $\Delta^r_{\mathcal{P}_1}(\sfp,\tstar)$, $\Delta^r_{\mathcal{P}_2}(\sfp,\tstar)$ and $\Delta^r_{\mathcal{P}_3}(\sfp,\tstar)$ are defined in the statement of Proposition~\ref{prop:Appendix:ReferenceProcess:N:Bounds}. 
    Furthermore, $\mathfrak{p}(\sfp)$ and $\mathfrak{p}'(\sfp)$ are defined in~\eqref{eqn:app:Constants:FrakP}.
\end{lemma}
\begin{proof}
    We begin by recalling the definition of $\psi^r(\tau(t),\nu,\Yt{N})$ from~\eqref{eqn:lem:Reference:dV:psi:Functions} 
    \begin{multline*}
        \psi^r(\tau(t),\nu,\Yt{N})
        =
        \frac{\Boldomega}{2\lambda - \Boldomega}
        \left(
            \expo{\Boldomega(\tau(t)+\nu)}
            \mathcal{P}^r\br{\tau(t),\nu}  
            -
            \expo{ (2\lambda \tau(t)+\Boldomega \nu)}
            \RefDiffNorm{\Yt{N,\tau(t)}}^\top g(\tau(t))
        \right)
        \\
        +
        \frac{2\lambda}{2\lambda - \Boldomega}
        \expo{(\Boldomega \tau(t) + 2 \lambda \nu)} 
        \RefDiffNorm{\Yt{N,\nu}}^\top g(\nu),
    \end{multline*}
    which, upon using the decomposition~\eqref{eqn:lem:Appendix:ReferenceProcess:ddV:Expression:Pr:Main} in Proposition~\ref{prop:Appendix:ReferenceProcess:ddV:Bound}, can be re-written as
    \begin{multline*}
        \psi^r(\tau(t),\nu,\Yt{N})
        =
        \frac{\Boldomega}{2\lambda - \Boldomega}
        \left(
            \expo{\Boldomega(\tau(t)+\nu)}
            \mathcal{P}^r_\circ \br{\tau(t),\nu} 
            -
            \expo{ (2\lambda \tau(t)+\Boldomega \nu)}
            \RefDiffNorm{\Yt{N,\tau(t)}}^\top g(\tau(t))
        \right)
        \\
        +
        \frac{\Boldomega}{2\lambda - \Boldomega}
        \expo{\Boldomega(\tau(t)+\nu)}
        \mathcal{P}^r_{ad} \br{\tau(t),\nu} 
        +
        \frac{2\lambda}{2\lambda - \Boldomega}
        \expo{(\Boldomega \tau(t) + 2 \lambda \nu)} 
        \RefDiffNorm{\Yt{N,\nu}}^\top g(\nu),
    \end{multline*}
    We thus decompose $\psi^r(\tau(t),\nu,\Yt{N})$ in~\eqref{eqn:lem:Reference:dV:psi:Functions} as follows: 
    \begin{equation}\label{eqn:lem:Appendix:ReferenceProcess:XirU:psi_r:Decomposition}
        \psi^r(\tau(t),\nu,\Yt{N})
        = 
        \sum_{i=1}^3 \psi_i^r(\tau(t),\nu,\Yt{N})
        +
        \psi_{ad}^r(\tau(t),\nu,\Yt{N})
        \in \mathbb{R}^{1 \times m},
    \end{equation}
    where 
    \begin{align*}
        \psi_1^r(\tau(t),\nu,\Yt{N})
        =
        \frac{\Boldomega}{2\lambda - \Boldomega}
        \expo{\Boldomega(\tau(t)+\nu)}
        \mathcal{P}^r_\circ\br{\tau(t),\nu},  
        \\
        \psi_2^r(\tau(t),\nu,\Yt{N})
        =
        -
        \frac{\Boldomega}{2\lambda - \Boldomega}
        \expo{ (2\lambda \tau(t)+\Boldomega \nu)}
        \RefDiffNorm{\Yt{N,\tau(t)}}^\top g(\tau(t)),
        \\
        \psi_3^r(\tau(t),\nu,\Yt{N})
        =
        \frac{2\lambda}{2\lambda - \Boldomega}
        \expo{(\Boldomega \tau(t) + 2 \lambda \nu)} 
        \RefDiffNorm{\Yt{N,\nu}}^\top g(\nu),
    \end{align*}
    and 
    \begin{align*}
        \psi_{ad}^r(\tau(t),\nu,\Yt{N})
        =
        \frac{\Boldomega}{2\lambda - \Boldomega}
        \expo{\Boldomega(\tau(t)+\nu)}
        \mathcal{P}^r_{ad} \br{\tau(t),\nu}.
    \end{align*}

    Next, using the definitions of $\mathcal{U}^r_{\mu}\br{\tau(t),\nu,\Yt{N};\Boldomega}$ and $\mathcal{U}^r_{\sigma}\br{\tau(t),\nu,\Yt{N};\Boldomega}$ in~\eqref{eqn:lem:Reference:dV:phi:Functions}, the decomposition in~\eqref{eqn:lem:Appendix:ReferenceProcess:XirU:psi_r:Decomposition} produces the following expression:
    \begin{multline*}
        \mathcal{U}^r_{\mu}\br{\tau(t),\nu,\Yt{N};\Boldomega}
        =
        \left(\sum_{i=1}^3 \psi_i^r(\tau(t),\nu,\Yt{N}) + \psi_{ad}^r(\tau(t),\nu,\Yt{N})\right)
        \Lparamu{\nu, \Xrt{N,\nu}} \in \mathbb{R}
        ,
        \\
        \mathcal{U}^r_{\sigma}\br{\tau(t),\nu,\Yt{N};\Boldomega}
        =
        \left(\sum_{i=1}^3 \psi_i^r(\tau(t),\nu,\Yt{N}) + \psi_{ad}^r(\tau(t),\nu,\Yt{N})\right)
        \Fparasigma{\nu,\Xrt{N,\nu}} \in \mathbb{R}^{1 \times d}
        .
    \end{multline*} 
    Then, we may re-write $\Xi^r_{\mathcal{U}}\br{\tau(t),\Yt{N};\Boldomega}$ in~\eqref{eqn:lem:Reference:dV:Xi:Functions} as
    \begin{multline*}
        \begin{aligned}
            \Xi^r_{\mathcal{U}}\br{\tau(t),\Yt{N};\Boldomega}
            =&
            \int_0^{t} 
            \left( 
                \mathcal{U}^r_{\mu}\br{t,\nu,\Yt{N};\Boldomega}d\nu
                +
                \mathcal{U}^r_{\sigma}\br{t,\nu,\Yt{N};\Boldomega}d\Wt{\nu}
            \right)
            \\
            =&
            \sum_{i=1}^3 
            \int_0^{t}
                \psi_i^r(t,\nu,\Yt{N}) 
                \left[
                    \Lparamu{\nu, \Xrt{N,\nu}}d\nu
                    +
                    \Fparasigma{\nu,\Xrt{N,\nu}}d\Wt{\nu}
                \right]
        \end{aligned}
        \\
        + 
        \int_0^{t}
            \psi_{ad}^r(t,\nu,\Yt{N}) 
            \left[
                \Lparamu{\nu, \Xrt{N,\nu}}d\nu
                +
                \Fparasigma{\nu,\Xrt{N,\nu}}d\Wt{\nu}
            \right],
            \quad t \in [0,\tstar],
    \end{multline*} 
    where we have used $\tau(t) \in [0,\tstar]$, $\forall t \in \mathbb{R}_{\geq 0}$, since $\tau^\star = \tstar$.
    Hence, exploiting the linearity of the expectation operator, along with Minkowski's inequality, we obtain  
    \begin{multline}\label{eqn:lem:Appendix:ReferenceProcess:XirU:Bound:A:Initial}
        \begin{aligned}
            &\ELaw{y_0}{
                \expo{-(2\lambda+\Boldomega) \tau(t)}
                \Xi^r_{\mathcal{U}}\br{\tau(t),\Yt{N};\Boldomega}
            }
            \\
            &
            =
            \expo{-(2\lambda+\Boldomega) t}
            \sum_{i=1}^3 
            \ELaw{y_0}{
                \int_0^{t}
                    \psi_i^r(t,\nu,\Yt{N}) 
                    \left[
                        \Lparamu{\nu, \Xrt{N,\nu}}d\nu
                        +
                        \Fparasigma{\nu,\Xrt{N,\nu}}d\Wt{\nu}
                    \right]
            }
        \end{aligned}
        \\
        + 
        \expo{-(2\lambda+\Boldomega) t}
        \ELaw{y_0}{
            \int_0^{t}
                \psi_{ad}^r(t,\nu,\Yt{N}) 
                \left[
                    \Lparamu{\nu, \Xrt{N,\nu}}d\nu
                    +
                    \Fparasigma{\nu,\Xrt{N,\nu}}d\Wt{\nu}
                \right]
        },
            \quad \forall t \in [0,\tstar],
    \end{multline}
    and
    \begin{multline}\label{eqn:lem:Appendix:ReferenceProcess:XirU:Bound:B:Initial}
        \begin{aligned}
            &
            \pLpLaw{y_0}{
                \expo{-(2\lambda+\Boldomega) \tau(t)}
                \Xi^r_{\mathcal{U}}\br{\tau(t),\Yt{N};\Boldomega}
            }
            \\
            &
            \leq 
            \expo{-(2\lambda+\Boldomega) t}
            \sum_{i=1}^3 
            \pLpLaw{y_0}{
                \int_0^{t}
                    \psi_i^r(t,\nu,\Yt{N}) 
                    \left[
                        \Lparamu{\nu, \Xrt{N,\nu}}d\nu
                        +
                        \Fparasigma{\nu,\Xrt{N,\nu}}d\Wt{\nu}
                    \right]
            }
        \end{aligned}
        \\
        + 
        \expo{-(2\lambda+\Boldomega) t}
        \pLpLaw{y_0}{
            \int_0^{t}
                \psi_{ad}^r(t,\nu,\Yt{N}) 
                \left[
                    \Lparamu{\nu, \Xrt{N,\nu}}d\nu
                    +
                    \Fparasigma{\nu,\Xrt{N,\nu}}d\Wt{\nu}
                \right]
        }
        ,
    \end{multline}
    for all $(t,\sfp) \in [0,\tstar] \times \mathbb{N}_{\geq 2}$. 
    In the two expressions above, we have used the fact that since $\tau^\star = \tstar$, the definition of  $\tau(t)$ in~\eqref{eqn:Reference:StoppingTimes} leads to the conclusion that $\tau(t) = t$, for $t \in [0,\tstar]$.   
    Hence, the stopping time $\tau(t)$ is a deterministic function of $t$ since $\tstar$ is a constant, and thus 
    \begin{align*}
        \mathbb{E}\left[\expo{-(2\lambda+\Boldomega) \tau(t)}\right]
        =
        \expo{-(2\lambda+\Boldomega) \tau(t)}
        =
        \expo{-(2\lambda+\Boldomega) t},
        \quad \forall t \in [0,\tstar].
    \end{align*} 
    
    Next, using the definition of $\psi^r_1$ in~\eqref{eqn:lem:Appendix:ReferenceProcess:XirU:psi_r:Decomposition} produces
    \begin{multline}\label{eqn:lem:Appendix:ReferenceProcess:XirU:Psi1:Bound:1}
        \int_0^{t}
            \psi_1^r(t,\nu,\Yt{N}) 
            \left[
                \Lparamu{\nu, \Xrt{N,\nu}}d\nu
                +
                \Fparasigma{\nu,\Xrt{N,\nu}}d\Wt{\nu}
            \right]
        \\
        =
        \frac{\Boldomega  \expo{\Boldomega t}}{2\lambda - \Boldomega}
        \int_0^{t}
            \expo{\Boldomega \nu}
            \mathcal{P}^r_\circ\br{t,\nu}
            \left[
                \Lparamu{\nu, \Xrt{N,\nu}}d\nu
                +
                \Fparasigma{\nu,\Xrt{N,\nu}}d\Wt{\nu}
            \right]
        ,
        \quad t \in [0,\tstar].
    \end{multline} 
    Using the definition of $\mathcal{P}^r_\circ$ in~\eqref{eqn:lem:Appendix:ReferenceProcess:ddV:Expression:Pr}, one sees that  
    \begin{align*}
        \int_0^{t}
            \expo{\Boldomega \nu}
            \mathcal{P}^r_\circ\br{t,\nu}
            \left[
                \Lparamu{\nu, \Xrt{N,\nu}}d\nu
                +
                \Fparasigma{\nu,\Xrt{N,\nu}}d\Wt{\nu}
            \right]
        = \sum_{i=1}^4 N^r_i(t), \quad \forall t \in [0,T], 
    \end{align*}
    where the processes $N^r_i(t)$, $i \in \cbr{1,\cdots,4}$, are defined in~\eqref{eqn:prop:Appendix:ReferenceProcess:N:Definitions} in the statement of Proposition~\ref{prop:Appendix:ReferenceProcess:N:Bounds} as follows:
    \begin{align*}
            N^r_1(t)
            =
            \int_0^{t}
                \expo{\Boldomega \nu}
                M_\mu(t,\nu)\Lparamu{\nu, \Xrt{N,\nu}}d\nu,
            \quad 
            N^r_2(t)
            =
            \int_0^{t}
                \expo{\Boldomega \nu}
                M_\mu(t,\nu)\Fparasigma{\nu,\Xrt{N,\nu}}d\Wt{\nu},
            \\
            N^r_3(t)
            =
            \int_0^{t}
                \expo{\Boldomega \nu}
                M_\sigma(t,\nu)\Lparamu{\nu, \Xrt{N,\nu}}d\nu,
            \quad 
            N^r_4(t)
            =
            \int_0^{t}
                \expo{\Boldomega \nu}
                M_\sigma(t,\nu)\Fparasigma{\nu,\Xrt{N,\nu}}d\Wt{\nu},
    \end{align*}
    and
    \begin{align*}
        M_\mu(t,\nu)
        =
        \int_\nu^{t} 
            \expo{ (2\lambda - \Boldomega) \beta } \mathcal{P}^r_{\mu}(\beta)^\top
        d\beta, 
        \quad 
        M_\sigma(t,\nu)
        =
        \int_\nu^{t} 
            \expo{ (2\lambda - \Boldomega) \beta } 
            \left[
            \mathcal{P}^r_{\sigma}(\beta)d\Wt{\beta}
            +
            \mathcal{P}^r_{\sigma_\star}(\beta)d\Wstart{\beta}
            \right]^\top.
    \end{align*}
    Hence, we may write~\eqref{eqn:lem:Appendix:ReferenceProcess:XirU:Psi1:Bound:1} as 
    \begin{align*}
        \int_0^{t}
            \psi_1^r(t,\nu,\Yt{N}) 
            \left[
                \Lparamu{\nu, \Xrt{N,\nu}}d\nu
                +
                \Fparasigma{\nu,\Xrt{N,\nu}}d\Wt{\nu}
            \right]
        =
        \frac{\Boldomega  \expo{\Boldomega t}}{2\lambda - \Boldomega}
        \sum_{i=1}^4 N^r_i(t)
        ,
        \quad t \in [0,\tstar].
    \end{align*}
    It then follows from the linearity of the expectation operator and the Minkowski's inequality that 
    \begin{align*}
        \ELaw{y_0}{
            \int_0^{t}
                \psi_1^r(t,\nu,\Yt{N}) 
                \left[
                    \Lparamu{\nu, \Xrt{N,\nu}}d\nu
                    +
                    \Fparasigma{\nu,\Xrt{N,\nu}}d\Wt{\nu}
                \right]
        }
        \leq     
        \frac{\Boldomega  \expo{\Boldomega t}}{\absolute{2\lambda - \Boldomega}}
        \sum_{i=1}^4
        \absolute{\ELaw{y_0}{N^r_i(t)}}
        ,
        \forall t \in [0,\tstar],
    \end{align*}
    and 
    \begin{multline*}
        \pLpLaw{y_0}{
            \int_0^{t}
                \psi_1^r(t,\nu,\Yt{N}) 
                \left[
                    \Lparamu{\nu, \Xrt{N,\nu}}d\nu
                    +
                    \Fparasigma{\nu,\Xrt{N,\nu}}d\Wt{\nu}
                \right]
        }
        \\
        \leq     
        \frac{\Boldomega  \expo{\Boldomega t}}{\absolute{2\lambda - \Boldomega}}
        \sum_{i=1}^4
        \pLpLaw{y_0}{N^r_i(t)}
        ,
        \quad 
        \forall (t,\sfp) \in [0,\tstar] \times \mathbb{N}_{\geq 2}.
    \end{multline*}
    Hence, using Proposition~\ref{prop:Appendix:ReferenceProcess:N:Bounds} leads to 
    \begin{multline}\label{eqn:lem:Appendix:ReferenceProcess:XirU:Psi1:Bound:p=1:Final}
        \begin{aligned}
            &\ELaw{y_0}{
                \int_0^{t}
                    \psi_1^r(t,\nu,\Yt{N}) 
                    \left[
                        \Lparamu{\nu, \Xrt{N,\nu}}d\nu
                        +
                        \Fparasigma{\nu,\Xrt{N,\nu}}d\Wt{\nu}
                    \right]
            }
            \\
            &\leq     
            \frac{ \expo{ (2\lambda+\Boldomega) t }  }{\sqrt{\lambda}\absolute{2\lambda - \Boldomega}}
            \Delta^r_{\mathcal{P}_1}(1,\tstar)
            \sup_{\nu \in [0,\tstar]} 
                \LpLaw{2}{y_0}{\Lparamu{\nu, \Xrt{N,\nu}}}  
        \end{aligned}
        \\
        +
        \frac{\expo{ (2\lambda+\Boldomega) t } }{\absolute{2\lambda - \Boldomega}}
        \left( 
            \frac{\BoldomegaRoot }{\sqrt{\lambda}}
            \Delta^r_{\mathcal{P}_2}(1,\tstar)
            +
            \frac{\Boldomega }{\lambda}
            \Delta^r_{\mathcal{P}_3}(1,\tstar)
        \right) 
        \sup_{\nu \in [0,\tstar]}
                \LpLaw{2}{y_0}{\Fparasigma{\nu,\Xrt{N,\nu}}}    
        ,
    \end{multline}
    for all $t \in [0,\tstar]$ and 
    \begin{multline}\label{eqn:lem:Appendix:ReferenceProcess:XirU:Psi1:Bound:p>=2:Final}
        \begin{aligned}
            &\pLpLaw{y_0}{
                \int_0^{t}
                    \psi_1^r(t,\nu,\Yt{N}) 
                    \left[
                        \Lparamu{\nu, \Xrt{N,\nu}}d\nu
                        +
                        \Fparasigma{\nu,\Xrt{N,\nu}}d\Wt{\nu}
                    \right]
            }
            \\
            &\leq     
            \frac{ \expo{ (2\lambda+\Boldomega) t }  }{\sqrt{\lambda}\absolute{2\lambda - \Boldomega}}
            \Delta^r_{\mathcal{P}_1}(\sfp,\tstar)
            \sup_{\nu \in [0,\tstar]} 
                \LpLaw{2\sfp}{y_0}{\Lparamu{\nu, \Xrt{N,\nu}}} 
        \end{aligned}
        \\
        +
        \frac{\expo{ (2\lambda+\Boldomega) t } }{\absolute{2\lambda - \Boldomega}}
        \left( 
            \frac{\BoldomegaRoot }{\sqrt{\lambda}}
            \Delta^r_{\mathcal{P}_2}(\sfp,\tstar)
            +
            \frac{\Boldomega }{\lambda}
            \Delta^r_{\mathcal{P}_3}(\sfp,\tstar)
        \right)  
        \sup_{\nu \in [0,\tstar]}
                \LpLaw{2\sfp}{y_0}{\Fparasigma{\nu,\Xrt{N,\nu}}}
        ,
    \end{multline}
    for all $(t,\sfp) \in [0,\tstar] \times \mathbb{N}_{\geq 2}$.

    \begingroup
    \endgroup

    Next, using the definition of $\psi^r_2$ in~\eqref{eqn:lem:Appendix:ReferenceProcess:XirU:psi_r:Decomposition} we see that
    \begin{multline*}
        \int_0^{t}
            \psi_2^r(t,\nu,\Yt{N}) 
            \left[ 
                \Lparamu{\nu, \Xrt{N,\nu}}d\nu
                +
                \Fparasigma{\nu,\Xrt{N,\nu}}d\Wt{\nu}
            \right]
        \\
        =
        -
        \frac{\Boldomega \expo{ 2\lambda t } }{2\lambda - \Boldomega}
        \RefDiffNorm{\Yt{N,t}}^\top g(t) 
        \int_0^{t}
            \expo{ \Boldomega \nu }
            \left[ 
                \Lparamu{\nu, \Xrt{N,\nu}}d\nu
                +
                \Fparasigma{\nu,\Xrt{N,\nu}}d\Wt{\nu}
            \right]
        ,
        \quad t \in [0,\tstar].
    \end{multline*}
    Hence 
    \begin{multline*}
        \absolute{
            \int_0^{t}
                \psi_2^r(t,\nu,\Yt{N}) 
                \left[ 
                    \Lparamu{\nu, \Xrt{N,\nu}}d\nu
                    +
                    \Fparasigma{\nu,\Xrt{N,\nu}}d\Wt{\nu}
                \right]
        }
        \\
        \leq 
        \frac{\Boldomega \expo{ 2\lambda t } }{\absolute{2\lambda - \Boldomega}}
        \Delta_g 
        \norm{\RefDiffNorm{\Yt{N,t}}} 
        \norm{ 
            \int_0^{t}
                \expo{ \Boldomega \nu }
                \left[ 
                    \Lparamu{\nu, \Xrt{N,\nu}}d\nu
                    +
                    \Fparasigma{\nu,\Xrt{N,\nu}}d\Wt{\nu}
                \right]
        }
        ,
        \quad t \in [0,\tstar],
    \end{multline*}
    where we have used the bound on $g(t)$ from Assumption~\ref{assmp:KnownFunctions}.
    It then follows from the Cauchy-Schwarz inequality and the Minkowski's inequality that 
    \begin{multline*}
        \begin{aligned}
            &\LpLaw{\sfp}{y_0}{
                \int_0^{t}
                    \psi_2^r(t,\nu,\Yt{N}) 
                    \left[ 
                        \Lparamu{\nu, \Xrt{N,\nu}}d\nu
                        +
                        \Fparasigma{\nu,\Xrt{N,\nu}}d\Wt{\nu}
                    \right]
            }
            \\
            &\leq 
            \frac{\Boldomega \expo{ 2\lambda t } }{\absolute{2\lambda - \Boldomega}}
            \Delta_g 
            \LpLaw{2\sfp}{y_0}{\RefDiffNorm{\Yt{N,t}}} 
            \LpLaw{2\sfp}{y_0}{
                \int_0^{t}
                    \expo{ \Boldomega \nu }
                    \norm{\Lparamu{\nu, \Xrt{N,\nu}}}
                d\nu
            }
        \end{aligned}
        \\
        + 
        \frac{\Boldomega \expo{ 2\lambda t } }{\absolute{2\lambda - \Boldomega}}
        \Delta_g 
        \LpLaw{2\sfp}{y_0}{\RefDiffNorm{\Yt{N,t}}} 
        \LpLaw{2\sfp}{y_0}{
            \int_0^{t}
                \expo{ \Boldomega \nu }
                \Fparasigma{\nu,\Xrt{N,\nu}}
            d\Wt{\nu}
        }
        ,
        \forall (t,\sfp) \in [0,\tstar] \times \mathbb{N}_{\geq 1}.
    \end{multline*}
    Since the regularity assumptions, along with the $t$-continuity of the strong solution $\Xrt{N,t}$, implies that $\Lparamu{\cdot, \Xrt{N}} \in \mathcal{M}^{loc}_2\left(\mathbb{R}^{m}|\Wfilt{t} \right)$ and  $\Fparasigma{\cdot,\Xrt{N}} \in \mathcal{M}^{loc}_2\left(\mathbb{R}^{m \times d}|\Wfilt{t} \right)$.
    Hence, we may invoke Proposition~\ref{prop:TechnicalResults:LebesgueMoment} for $\cbr{\theta, S(t), \xi} = \cbr{\Boldomega, \Lparamu{\cdot, \Xrt{N}} (n_s=m), 0}$, and Lemma~\ref{lemma:TechnicalResults:MartingaleMoment} for $\cbr{\theta, \Qt{t}, L(t), \xi} = \cbr{\Boldomega, \Wt{t} (n_q = d), \Fparasigma{\cdot,\Xrt{N}} (n_l=m), 0}$, and obtain 
    \begin{multline}\label{eqn:lem:Appendix:ReferenceProcess:XirU:Psi2:Bound:Final}
        \begin{aligned}
            &\LpLaw{\sfp}{y_0}{
                \int_0^{t}
                    \psi_2^r(t,\nu,\Yt{N}) 
                    \left[ 
                        \Lparamu{\nu, \Xrt{N,\nu}}d\nu
                        +
                        \Fparasigma{\nu,\Xrt{N,\nu}}d\Wt{\nu}
                    \right]
            }
            \\
            &\leq 
            \frac{\expo{ (2\lambda+\Boldomega) t } }{\absolute{2\lambda - \Boldomega}}
            \Delta_g 
            \sup_{\nu \in [0,\tstar]}\LpLaw{2\sfp}{y_0}{\RefDiffNorm{\Yt{N,\nu}}}
            \sup_{\nu \in [0,\tstar]}\LpLaw{2\sfp}{y_0}{\Lparamu{\nu, \Xrt{N,\nu}}}
        \end{aligned}
        \\
        + 
        \frac{\BoldomegaRoot \expo{ (2\lambda + \Boldomega) t } }{\absolute{2\lambda - \Boldomega}}
        \Delta_g 
        \left(\sfp \frac{2\sfp-1}{2}\right)^\Half  
        \sup_{\nu \in [0,\tstar]}\LpLaw{2\sfp}{y_0}{\RefDiffNorm{\Yt{N,\nu}}}
        \sup_{\nu \in [0,\tstar]}
        \LpLaw{2\sfp}{}{\Fparasigma{\nu,\Xrt{N,\nu}}}
        ,
    \end{multline}
    for all $(t,\sfp) \in [0,\tstar] \times \mathbb{N}_{\geq 1}$, where we have bounded $\expo{\Boldomega t} - 1$ and $\expo{2\Boldomega t} - 1$ by $\expo{\Boldomega t}$ and $\expo{2\Boldomega t}$, respectively, and used the fact that  $\LpLaw{2\sfp}{y_0}{\RefDiffNorm{\Yt{N,t}}} \leq \sup_{\nu \in [0,\tstar]}\LpLaw{2\sfp}{y_0}{\RefDiffNorm{\Yt{N,\nu}}}$, $\forall t \in [0,\tstar]$. 

    \begingroup
    \endgroup

    Next, using the definition of $\psi^r_3$ in~\eqref{eqn:lem:Appendix:ReferenceProcess:XirU:psi_r:Decomposition} we see that
    \begin{multline*}
            \int_0^{t}
                \psi_3^r(t,\nu,\Yt{N}) 
                \left[ 
                    \Lparamu{\nu, \Xrt{N,\nu}}d\nu
                    +
                    \Fparasigma{\nu,\Xrt{N,\nu}}d\Wt{\nu}
                \right]
        \\
        =
        \frac{2\lambda \expo{\Boldomega t } }{2\lambda - \Boldomega}
        \int_0^{t}
            \expo{2 \lambda \nu} 
            \RefDiffNorm{\Yt{N,\nu}}^\top g(\nu) 
            \left[ 
                \Lparamu{\nu, \Xrt{N,\nu}}d\nu
                +
                \Fparasigma{\nu,\Xrt{N,\nu}}d\Wt{\nu}
            \right]
        , 
        \quad t \in [0,\tstar].
    \end{multline*}
    It then follows from the linearity of the expectation operator and the Minkowski's inequality that 
    \begin{multline}\label{eqn:lem:Appendix:ReferenceProcess:XirU:Psi3:Bound:p=1:1}
        \begin{aligned}
            &\ELaw{y_0}{
                \int_0^{t}
                    \psi_3^r(t,\nu,\Yt{N}) 
                    \left[
                        \Lparamu{\nu, \Xrt{N,\nu}}d\nu
                        +
                        \Fparasigma{\nu,\Xrt{N,\nu}}d\Wt{\nu}
                    \right]
            }
            \\
            &\leq
            \frac{2\lambda \expo{\Boldomega t } }{\absolute{2\lambda - \Boldomega}}
            \ELaw{y_0}{
                \int_0^{t}
                    \expo{2 \lambda \nu}
                    \absolute{ 
                        \RefDiffNorm{\Yt{N,\nu}}^\top g(\nu) 
                        \Lparamu{\nu, \Xrt{N,\nu}}
                    }
                d\nu
            }
        \end{aligned}
        \\
        +
        \frac{2\lambda \expo{\Boldomega t } }{\absolute{2\lambda - \Boldomega}}
        \absolute{
            \ELaw{y_0}{
                \int_0^{t}
                    \expo{2 \lambda \nu} 
                    \RefDiffNorm{\Yt{N,\nu}}^\top g(\nu) 
                    \Fparasigma{\nu,\Xrt{N,\nu}}
                d\Wt{\nu}
            }
        }
        ,
        \quad 
        \forall t \in [0,\tstar],
    \end{multline}
    and
    \begin{multline}\label{eqn:lem:Appendix:ReferenceProcess:XirU:Psi3:Bound:p>=2:1}
        \begin{aligned}
            &\LpLaw{\sfp}{y_0}{
                \int_0^{t}
                    \psi_3^r(t,\nu,\Yt{N}) 
                    \left[
                        \Lparamu{\nu, \Xrt{N,\nu}}d\nu
                        +
                        \Fparasigma{\nu,\Xrt{N,\nu}}d\Wt{\nu}
                    \right]
            }
            \\
            &\leq
            \frac{2\lambda \expo{\Boldomega t } }{\absolute{2\lambda - \Boldomega}}
            \LpLaw{\sfp}{y_0}{
                \int_0^{t}
                    \expo{2 \lambda \nu}
                    \absolute{ 
                        \RefDiffNorm{\Yt{N,\nu}}^\top g(\nu) 
                        \Lparamu{\nu, \Xrt{N,\nu}}
                    }
                d\nu
            }
        \end{aligned}
        \\
        +
        \frac{2\lambda \expo{\Boldomega t } }{\absolute{2\lambda - \Boldomega}}
        \LpLaw{\sfp}{y_0}{
            \int_0^{t}
                \expo{2 \lambda \nu} 
                \RefDiffNorm{\Yt{N,\nu}}^\top g(\nu) 
                \Fparasigma{\nu,\Xrt{N,\nu}}
            d\Wt{\nu}
        }
        ,
        \quad 
        \forall (t,\sfp) \in [0,\tstar] \times \mathbb{N}_{\geq 2}.
    \end{multline}
    Now, as before, Proposition~\ref{prop:TechnicalResults:LebesgueMoment} for $\cbr{\theta, S(t), \xi} = \cbr{2\lambda, \RefDiffNorm{\Yt{N,t}}^\top g(t)\Lparamu{t, \Xrt{N,t}} (n_s=1), 0}$ yields
    \begin{multline*}
        \LpLaw{\sfp}{y_0}{
            \int_0^{t}
                \expo{2 \lambda \nu}
                \absolute{ 
                    \RefDiffNorm{\Yt{N,\nu}}^\top g(\nu) 
                    \Lparamu{\nu, \Xrt{N,\nu}}
                }
            d\nu
        }
        \\
        \leq
        \frac{\expo{2\lambda t}}{2\lambda}      
        \sup_{\nu \in [0,\tstar]}\LpLaw{\sfp}{}{\RefDiffNorm{\Yt{N,\nu}}^\top g(\nu)  \Lparamu{\nu, \Xrt{N,\nu}}}
        ,
        \quad \forall (t,\sfp) \in [0,\tstar] \times \mathbb{N}_{\geq 1}
        ,
    \end{multline*}
    where we have bounded $\expo{2\lambda t} - 1$ by $\expo{2\lambda t}$.
    It then follows from the Cauchy-Schwarz inequality and the bound on $g(t)$ from Assumption~\ref{assmp:KnownFunctions} that 
    \begin{multline}\label{eqn:lem:Appendix:ReferenceProcess:XirU:Psi3:Mu}
        \LpLaw{\sfp}{y_0}{
            \int_0^{t}
                \expo{2 \lambda \nu}
                \absolute{ 
                    \RefDiffNorm{\Yt{N,\nu}}^\top g(\nu) 
                    \Lparamu{\nu, \Xrt{N,\nu}}
                }
            d\nu
        }
        \\
        \leq    
        \frac{\expo{2\lambda t}}{2\lambda}  
        \Delta_g
        \sup_{\nu \in [0,\tstar]}
            \LpLaw{2\sfp}{}{\RefDiffNorm{\Yt{N,\nu}}}
        \sup_{\nu \in [0,\tstar]}
            \LpLaw{2\sfp}{}{\Lparamu{\nu, \Xrt{N,\nu}}}
        ,
        \quad \forall (t,\sfp) \in [0,\tstar] \times \mathbb{N}_{\geq 1}
        .
    \end{multline}
    Next, we may write 
    \begin{align*}
        \int_0^{t}
            \expo{2 \lambda \nu} 
            \RefDiffNorm{\Yt{N,\nu}}^\top g(\nu) \Fparasigma{\nu,\Xrt{N,\nu}}
        d\Wt{\nu}
        =
        \int_0^{t}
            \expo{2 \lambda \nu} 
            \RefDiffNorm{\Yt{N,\nu}}^\top g(\nu) \begin{bmatrix}\Fparasigma{\nu,\Xrt{N,\nu}} & 0_{m,d} \end{bmatrix}
        d\Wrt{\nu}.
    \end{align*}
    Furthermore, the regularity assumptions and the $t$-continuity of the strong solutions $\Xrt{N,t}$ and $\Xstart{N,t}$ imply that $\left(\RefDiffNorm{\Yt{N,t}}^\top g(t) \begin{bmatrix}\Fparasigma{t,\Xrt{N,t}} & 0_{m,d} \end{bmatrix}\right) \in \mathcal{M}^{loc}_2\left(\mathbb{R}^{1 \times 2d}|\Wfilt{t} \times \Wstarfilt{t} \right)$.
    Hence, by~\cite[Thm.~1.5.21]{mao2007stochastic}
    \begin{equation}\label{eqn:lem:Appendix:ReferenceProcess:XirU:Psi3:Sigma:p=1}
        \ELaw{y_0}{
            \int_0^{t}
                \expo{2 \lambda \nu} 
                \RefDiffNorm{\Yt{N,\nu}}^\top g(\nu) 
                \Fparasigma{\nu,\Xrt{N,\nu}}
            d\Wt{\nu}
        }
        =
        0, \quad \forall t \in [0,\tstar].
    \end{equation}
    Moreover, Lemma~\ref{lemma:TechnicalResults:MartingaleMoment} for $\cbr{\theta, \Qt{t}, L(t), \xi} = \cbr{2\lambda, \Wrt{t} (n_d = 2d), \left(\RefDiffNorm{\Yt{N,t}}^\top g(t) \begin{bmatrix}\Fparasigma{t,\Xrt{N,t}} & 0_{m,d} \end{bmatrix}\right) (n_l=1), 0}$ produces 
    \begin{multline*}
        \LpLaw{\sfp}{y_0}{
            \int_0^{t}
                \expo{2 \lambda \nu} 
                \RefDiffNorm{\Yt{N,\nu}}^\top g(\nu) 
                \Fparasigma{\nu,\Xrt{N,\nu}}
            d\Wt{\nu}
        }
        \\
        \leq 
        \frac{\expo{ 2\lambda t }}{2\sqrt{\lambda}}
        \left(\sfp \frac{\sfp-1}{2}\right)^\Half  
        \sup_{\nu \in [0,\tstar]}
        \LpLaw{\sfp}{}{
            \RefDiffNorm{\Yt{N,\nu}}^\top g(\nu) 
            \Fparasigma{\nu,\Xrt{N,\nu}}
        }
        ,
        \quad \forall (t,\sfp) \in [0,\tstar] \times \mathbb{N}_{\geq 2}
        ,
    \end{multline*}
    where we have bounded $\left(\expo{4\lambda t} - 1\right)^\frac{1}{2}$ by $\expo{2\lambda t}$.
    It then follows from the Cauchy-Schwarz inequality and the bound on $g(t)$ from Assumption~\ref{assmp:KnownFunctions} that 
    \begin{multline}\label{eqn:lem:Appendix:ReferenceProcess:XirU:Psi3:Sigma:p>=2}
        \LpLaw{\sfp}{y_0}{
            \int_0^{t}
                \expo{2 \lambda \nu} 
                \RefDiffNorm{\Yt{N,\nu}}^\top g(\nu) 
                \Fparasigma{\nu,\Xrt{N,\nu}}
            d\Wt{\nu}
        }
        \\
        \leq 
        \frac{\expo{ 2\lambda t }}{2\sqrt{\lambda}}
        \Delta_g
        \left(\sfp \frac{\sfp-1}{2}\right)^\Half  
        \sup_{\nu \in [0,\tstar]}
        \LpLaw{2\sfp}{}{
            \RefDiffNorm{\Yt{N,\nu}} 
        }
        \sup_{\nu \in [0,\tstar]}
        \LpLaw{2\sfp}{}{
            \Fparasigma{\nu,\Xrt{N,\nu}}
        }
        ,
        \quad \forall (t,\sfp) \in [0,\tstar] \times \mathbb{N}_{\geq 2}
        .
    \end{multline}
    Substituting~\eqref{eqn:lem:Appendix:ReferenceProcess:XirU:Psi3:Mu} and~\eqref{eqn:lem:Appendix:ReferenceProcess:XirU:Psi3:Sigma:p=1} into~\eqref{eqn:lem:Appendix:ReferenceProcess:XirU:Psi3:Bound:p=1:1} yields 
    \begin{multline}\label{eqn:lem:Appendix:ReferenceProcess:XirU:Psi3:Bound:p=1:Final}
        \ELaw{y_0}{
            \int_0^{t}
                \psi_3^r(t,\nu,\Yt{N}) 
                \left[
                    \Lparamu{\nu, \Xrt{N,\nu}}d\nu
                    +
                    \Fparasigma{\nu,\Xrt{N,\nu}}d\Wt{\nu}
                \right]
        }
        \\
        \leq
        \frac{ \expo{(2\lambda+\Boldomega) t } }{\absolute{2\lambda - \Boldomega}}  
        \Delta_g
        \sup_{\nu \in [0,\tstar]}
            \LpLaw{2}{y_0}{\RefDiffNorm{\Yt{N,\nu}}}
        \sup_{\nu \in [0,\tstar]}
            \LpLaw{2}{y_0}{\Lparamu{\nu, \Xrt{N,\nu}}}
        \quad 
        \forall t \in [0,\tstar].
    \end{multline}
    Similarly, substituting~\eqref{eqn:lem:Appendix:ReferenceProcess:XirU:Psi3:Mu} and~\eqref{eqn:lem:Appendix:ReferenceProcess:XirU:Psi3:Sigma:p>=2} into~\eqref{eqn:lem:Appendix:ReferenceProcess:XirU:Psi3:Bound:p>=2:1} yields 
    \begin{multline}\label{eqn:lem:Appendix:ReferenceProcess:XirU:Psi3:Bound:p>=2:Final}
        \begin{aligned}
            &\LpLaw{\sfp}{y_0}{
                \int_0^{t}
                    \psi_3^r(t,\nu,\Yt{N}) 
                    \left[
                        \Lparamu{\nu, \Xrt{N,\nu}}d\nu
                        +
                        \Fparasigma{\nu,\Xrt{N,\nu}}d\Wt{\nu}
                    \right]
            }
            \\
            &\leq
            \frac{\expo{(2\lambda+\Boldomega) t } }{\absolute{2\lambda - \Boldomega}}
            \Delta_g
            \sup_{\nu \in [0,\tstar]}
                \LpLaw{2\sfp}{y_0}{\RefDiffNorm{\Yt{N,\nu}}}
            \sup_{\nu \in [0,\tstar]}
                \LpLaw{2\sfp}{y_0}{\Lparamu{\nu, \Xrt{N,\nu}}}
        \end{aligned}
        \\
        +
        \frac{\sqrt{\lambda} \expo{(2\lambda+\Boldomega) t } }{\absolute{2\lambda - \Boldomega}}
        \Delta_g
        \left(\sfp \frac{\sfp-1}{2}\right)^\Half  
        \sup_{\nu \in [0,\tstar]}
        \LpLaw{2\sfp}{ y_0}{
            {\RefDiffNorm{\Yt{N,\nu}}}
        }
        \sup_{\nu \in [0,\tstar]}
        \LpLaw{2\sfp}{ y_0}{
            \Fparasigma{\nu,\Xrt{N,\nu}}
        }
        ,
    \end{multline}
    for all $(t,\sfp) \in [0,\tstar] \times \mathbb{N}_{\geq 2}$.

    \begingroup
    \endgroup

    Finally, using the definition of $\psi^r_{ad}$ in~\eqref{eqn:lem:Appendix:ReferenceProcess:XirU:psi_r:Decomposition} we see that
    \begin{multline*}
        \int_0^{t}
            \psi_{ad}^r(t,\nu,\Yt{N}) 
            \left[ 
                \Lparamu{\nu, \Xrt{N,\nu}}d\nu
                +
                \Fparasigma{\nu,\Xrt{N,\nu}}d\Wt{\nu}
            \right]
        \\
        =
        \frac{\Boldomega \expo{\Boldomega t }}{2\lambda - \Boldomega}
            \int_0^{t}
                \expo{\Boldomega\nu}
                \mathcal{P}^r_{ad} \br{t,\nu} 
                \left[ 
                    \Lparamu{\nu, \Xrt{N,\nu}}d\nu
                    +
                    \Fparasigma{\nu,\Xrt{N,\nu}}d\Wt{\nu}
                \right]
        , \quad t \in [0,\tstar].
    \end{multline*}
    Using the definition of $\mathcal{P}^r_{ad}$ in~\eqref{eqn:lem:Appendix:ReferenceProcess:ddV:Expression:PrU}, Proposition~\ref{prop:Appendix:ReferenceProcess:ddV:Bound}, then leads to 
    \begin{multline*}
        \int_0^{t}
            \psi_{ad}^r(t,\nu,\Yt{N}) 
            \left[ 
                \Lparamu{\nu, \Xrt{N,\nu}}d\nu
                +
                \Fparasigma{\nu,\Xrt{N,\nu}}d\Wt{\nu}
            \right]
        \\
        =
        \frac{\Boldomega \expo{\Boldomega t }}{2\lambda - \Boldomega}
            \int_0^{t}
                \expo{\Boldomega\nu}
                \left( 
                    \int_\nu^{t} 
                        \expo{ (2\lambda - \Boldomega) \beta } 
                        \mathcal{P}^r_\mathcal{U}(\beta)^\top 
                    d\beta
                \right)
                \left[ 
                    \Lparamu{\nu, \Xrt{N,\nu}}d\nu
                    +
                    \Fparasigma{\nu,\Xrt{N,\nu}}d\Wt{\nu}
                \right]
            \\
            =
            \frac{\Boldomega \expo{\Boldomega t }}{2\lambda - \Boldomega}
            N^r_{\mathcal{U}}(t)
        , \quad t \in [0,\tstar],
    \end{multline*}
    where the process $N^r_{\mathcal{U}}$ is defined in~\eqref{eqn:prop:Appendix:ReferenceProcess:NU:Definitions}, Proposition~\ref{prop:Appendix:ReferenceProcess:NU:Bounds}.
    Thus, we invoke Proposition~\ref{prop:Appendix:ReferenceProcess:NU:Bounds} and obtain 
    \begin{multline}\label{eqn:lem:Appendix:ReferenceProcess:XirU:PsiAd:Bound:Final}
        \LpLaw{p}{y_0}{
            \int_0^{t}
                \psi_{ad}^r(t,\nu,\Yt{N}) 
                \left[ 
                    \Lparamu{\nu, \Xrt{N,\nu}}d\nu
                    +
                    \Fparasigma{\nu,\Xrt{N,\nu}}d\Wt{\nu}
                \right]
        }
        \\
        \leq  
        \frac{\expo{ (2\lambda+\Boldomega) t } }{\cancel{2}\lambda\absolute{2\lambda - \Boldomega}}
        \Delta_g^2 
        \left(
            \sup_{\nu \in [0,\tstar]}
            \LpLaw{2\sfp}{y_0}{\Lparamu{\nu,\Xrt{N,\nu}}}
            + 
            \BoldomegaRoot
            \left(\sfp \frac{2\sfp-1}{2}\right)^\Half  
            \sup_{\nu \in [0,\tstar]}
            \LpLaw{2\sfp}{y_0}{\Fparasigma{\nu,\Xrt{N,\nu}}}
        \right)^2
        ,
    \end{multline}
   for all $(t,\sfp) \in [0,\tstar] \times \mathbb{N}_{\geq 1}$.

    Substituting~\eqref{eqn:lem:Appendix:ReferenceProcess:XirU:Psi1:Bound:p=1:Final},~\eqref{eqn:lem:Appendix:ReferenceProcess:XirU:Psi2:Bound:Final},~\eqref{eqn:lem:Appendix:ReferenceProcess:XirU:Psi3:Bound:p=1:Final}, and~\eqref{eqn:lem:Appendix:ReferenceProcess:XirU:PsiAd:Bound:Final} into~\eqref{eqn:lem:Appendix:ReferenceProcess:XirU:Bound:A:Initial}, followed by grouping terms that are $\propto  \cbr{\Boldomega, \BoldomegaRoot, 1} / \absolute{2\lambda - \Boldomega}$ leads to~\eqref{eqn:lem:Appendix:ReferenceProcess:XirU:Bound:A:Final}.
    Similarly, substituting~\eqref{eqn:lem:Appendix:ReferenceProcess:XirU:Psi1:Bound:p>=2:Final},~\eqref{eqn:lem:Appendix:ReferenceProcess:XirU:Psi2:Bound:Final},~\eqref{eqn:lem:Appendix:ReferenceProcess:XirU:Psi3:Bound:p>=2:Final}, and~\eqref{eqn:lem:Appendix:ReferenceProcess:XirU:PsiAd:Bound:Final} into~\eqref{eqn:lem:Appendix:ReferenceProcess:XirU:Bound:B:Initial} leads to~\eqref{eqn:lem:Appendix:ReferenceProcess:XirU:Bound:B:Final}, thus concluding the proof.

\end{proof}

Next, we derive the expressions for the terms that constitute the bound in Proposition~\ref{prop:Appendix:ReferenceProcess:N:Bounds}.
\begin{proposition}\label{prop:Appendix:ReferenceProcess:BoundsRoundB}
    Consider the stopping times $\tau^\star$ and $\tstar$, as defined in~\eqref{eqn:Reference:StoppingTimes}, Lemma~\ref{lem:Reference:dV}, and assume that $\tau^\star = \tstar$. 
    If Assumptions~\ref{assmp:KnownFunctions}~-~\ref{assmp:knownDiffusion:Decomposition} hold, then with $\mathsf{p}^\star$ defined in Assumption~\ref{assmp:NominalSystem:FiniteMomentsWasserstein}, and 
    \begin{equation}\label{eqn:prop:Appendix:ReferenceProcess:Final:Bound:Condition2}
        \LpRefError{y_0}
        \doteq
        \sup_{\nu \in [0,\tstar]}
        \LpLaw{2 \sfp}{y_0}{\Xrt{N,\nu} - \Xstart{N,\nu}}, 
        \quad 
        \LpRef{y_0}
        \doteq
        \sup_{\nu \in [0,\tstar]}
        \LpLaw{2 \sfp}{y_0}{\Xrt{N,\nu}},
        \quad \sfp \in \cbr{1,\dots,\sfp^\star},
    \end{equation} 
    the following bounds hold for all $\sfp \in \cbr{1,\dots,\sfp^\star}$: 
    \begin{subequations}\label{eqn:prop:Appendix:ReferenceProcess:BoundsRoundB:Del:Final}
        \begin{align}
            \Delta^r_{\mathcal{P}_1}(\sfp,\tstar)
            \leq  
            \widehat{\Delta}^r_1(\sfp)
            +
            \widehat{\Delta}^r_2(\sfp)
            \sqrt{\LpRef{y_0}} 
                + 
                \widehat{\Delta}^r_3
                \LpRef{y_0}
                +
                \widehat{\Delta}^r_4
                \LpRefError{y_0}
            ,
            \\
            \Delta^r_{\mathcal{P}_3}(\sfp,\tstar)
            \leq
            \sqrt{m}
            \Delta_g
            \left(
                2\Delta_p + \Delta_\sigma
            \right)
            +
            \sqrt{m}
            \Delta_g
            \Delta_\sigma
            \sqrt{\LpRef{y_0}} 
            ,
        \end{align}
    \end{subequations}
    where $\Delta^r_{\mathcal{P}_1}(\sfp,\tstar)$ and $\Delta^r_{\mathcal{P}_3}(\sfp,\tstar)$  are defined in the statement of Proposition~\ref{prop:Appendix:ReferenceProcess:N:Bounds}.
    Furthermore, the constants $\mathfrak{p}(\sfp)$ and $\Lip{f}$ are defined in~\eqref{eqn:app:Global:Constants}, and the constants $\widehat{\Delta}^r_1(\sfp)$, $\widehat{\Delta}^r_2(\sfp)$, $\widehat{\Delta}^r_3$, and $\widehat{\Delta}^r_4$ are defined in~\eqref{eqn:app:Constants:Ref:DelHat}.


\end{proposition}
\begin{proof}
    We begin with the term $\Delta^r_{\mathcal{P}_\mu}$ defined in Proposition~\ref{prop:Appendix:ReferenceProcess:Pparts:Bound} and the definition $\bar{F}_\mu = f$ to get  
    \begin{subequations}
        \begin{align}
            \Delta^r_{\mathcal{P}_\mu}(\sfp)
            =& 
            \sup_{\nu \in [0,\tstar]}
            \left(
                \LpLaw{2\sfp}{y_0}{f(\nu,\Xrt{N,\nu})-f(\nu,\Xstart{N,\nu})}
                + 
                \LpLaw{2\sfp}{y_0}{\Lmu{\nu,\Xrt{N,\nu}}}  
            \right)
            \label{eqn:prop:Appendix:ReferenceProcess:BoundsRoundB:Delpmu:GlobalLip:Initial}
            \\
            \leq&
            \sup_{\nu \in [0,\tstar]}
            \left(
                \LpLaw{2\sfp}{y_0}{f(\nu,\Xrt{N,\nu})}
                + 
                \LpLaw{2\sfp}{y_0}{f(\nu,\Xstart{N,\nu})}
                + 
                \LpLaw{2\sfp}{y_0}{\Lmu{\nu,\Xrt{N,\nu}}}  
            \right),
            \label{eqn:prop:Appendix:ReferenceProcess:BoundsRoundB:Delpmu:Initial}
        \end{align}
    \end{subequations}
    where the second inequality is due to the Minkowski inequality. 
    Using Assumptions~\ref{assmp:KnownFunctions},~\ref{assmp:NominalSystem:FiniteMomentsWasserstein}, and~\ref{assmp:UnknownFunctions}, the bound in~\eqref{eqn:prop:Appendix:ReferenceProcess:BoundsRoundB:Delpmu:Initial} can be developed into
    \begin{align}\label{eqn:prop:Appendix:ReferenceProcess:BoundsRoundB:Delpmu}
        \Delta^r_{\mathcal{P}_\mu}(\sfp)
        \leq
        \Delta_f \left(2 + \Delta_\star \right)
        +
        \Delta_\mu
        + 
        (\Delta_f + \Delta_\mu)
        \LpRef{y_0}
        ,
    \end{align}
    where we have also used the subadditivity of the square root function.  
    If additionally Assumption~\ref{assmp:KnownFunctions:GlobalLip} holds, then we obtain a tighter bound   by developing~\eqref{eqn:prop:Appendix:ReferenceProcess:BoundsRoundB:Delpmu:GlobalLip:Initial} to get 
    \begin{align}\label{eqn:prop:Appendix:ReferenceProcess:BoundsRoundB:Delpmu:GlobalLip}
        \Delta^r_{\mathcal{P}_\mu}(\sfp)
        \leq 
        \Delta_\mu 
        +
        L_f
        \LpRefError{y_0}
        +  
        \Delta_\mu
        \LpRef{y_0}
        .
    \end{align}
    We may combine the bounds in~\eqref{eqn:prop:Appendix:ReferenceProcess:BoundsRoundB:Delpmu} and~\eqref{eqn:prop:Appendix:ReferenceProcess:BoundsRoundB:Delpmu:GlobalLip} by writing 
    \begin{align}\label{eqn:prop:Appendix:ReferenceProcess:BoundsRoundB:Delpmu:Combined}
        \Delta^r_{\mathcal{P}_\mu}(\sfp)
        \leq&
        \Delta_f \left(2 + \Delta_\star \right)
        \left(1 - \Lip{f}\right)
        +
        \Delta_\mu
        \notag 
        \\
        &+ 
        \left( 
            \Delta_f\left(1 - \Lip{f}\right) 
            + 
            \Delta_\mu 
        \right)
        \LpRef{y_0}
        +
        L_f
        \Lip{f}
        \LpRefError{y_0},
    \end{align}
    where $\Lip{f}$ is defined in~\eqref{eqn:app:Constants:GlobalLipschitz}.

    Next, we have the following from Proposition~\ref{prop:Appendix:ReferenceProcess:Pparts:Bound}: 
    \begin{align*}
        \Delta^r_{\mathcal{P}_\sigma}(\sfp)
        =
        \sup_{\nu \in [0,\tstar]}
        \left( 
            \LpLaw{2\sfp}{y_0}{p(\nu,\Xrt{N,\nu})}
            +
            \LpLaw{2\sfp}{y_0}{\Lsigma{\nu,\Xrt{N,\nu}}}
        \right)
        +
        \sup_{\nu \in [0,\tstar]}
        \LpLaw{2\sfp}{y_0}{p(\nu,\Xstart{N,\nu})}
        ,
    \end{align*}
    where we have used the definition $\bar{F}_\sigma = p$. 
    It then follows from Assumptions~\ref{assmp:KnownFunctions},~\ref{assmp:NominalSystem:FiniteMomentsWasserstein}, and~\ref{assmp:UnknownFunctions} that
    \begin{align}\label{eqn:prop:Appendix:ReferenceProcess:BoundsRoundB:Delpsigma}
        \Delta^r_{\mathcal{P}_\sigma}(\sfp)
        \leq     
        2\Delta_p + \Delta_\sigma
        +
        \Delta_\sigma
        \sqrt{\LpRef{y_0}}
        ,
    \end{align}
    where we have also used the following due to Cauchy-Schwarz inequality:
    \begin{align*}
        \LpLaw{2\sfp}{y_0}{
            \norm{\Xrt{N,\nu}}^\frac{1}{2}
        } 
        =
        \left(
            \LpLaw{\sfp}{y_0}{
                \Xrt{N,\nu}
            }
        \right)^\frac{1}{2}
        \leq 
        \left(
            \LpLaw{2\sfp}{y_0}{
                \Xrt{N,\nu}
            }
        \right)^\frac{1}{2},
    \end{align*}
    and thus 
    \begin{align}\label{eqn:lem:Appendix:ReferenceProcess:ComponentBounds:Manipulations}
        \sup_{\nu \in [0,\tstar]}
        \LpLaw{2\sfp}{y_0}{
            \norm{\Xrt{N,\nu}}^\frac{1}{2}
        } 
        \leq 
        \sup_{\nu \in [0,\tstar]}
        \left(
            \LpLaw{2\sfp}{y_0}{
                \Xrt{N,\nu}
            }
        \right)^\frac{1}{2}
        =
        \left(
            \sup_{\nu \in [0,\tstar]}
            \LpLaw{2\sfp}{y_0}{
                \Xrt{N,\nu}
            }
        \right)^\frac{1}{2}
        =
        \sqrt{\LpRef{y_0}}
        .
    \end{align}
    Using~\eqref{eqn:prop:Appendix:ReferenceProcess:BoundsRoundB:Delpmu:Combined} and~\eqref{eqn:prop:Appendix:ReferenceProcess:BoundsRoundB:Delpsigma}, we can now bound the terms in~\eqref{prop:Appendix:ReferenceProcess:Pparts:Bounds:Final} as follows: 
    \begin{subequations}\label{eqn:prop:Appendix:ReferenceProcess:BoundsRoundC:SumDelp}
        \begin{align}
            \begin{multlined}[b][0.9\linewidth]
                \sup_{\nu \in [0,\tstar]}
                \LpLaw{2\sfp}{y_0}{\mathcal{P}^r_{\mu}(\nu)}
                \leq
                2 \Delta_g
                \left[
                    \Delta_f \left(2 + \Delta_\star \right)
                    \left(1 - \Lip{f}\right)
                    +
                    \Delta_\mu
                \right] 
                \\
                +
                2 \Delta_g
                \left( 
                    \Delta_f\left(1 - \Lip{f}\right) 
                    + 
                    \Delta_\mu 
                \right)
                \LpRef{y_0}
                +
                2
                \left( 
                    \Delta_g
                    L_f
                    \Lip{f}
                    + 
                    \Delta_{\dot{g}}
                \right)
                \LpRefError{y_0},
            \end{multlined}
            \label{eqn:prop:Appendix:ReferenceProcess:BoundsRoundC:SumDelpmu}
            \\
            \sum_{i \in \cbr{\sigma,\sigma_\star}}
            \sup_{\nu \in [0,\tstar]}
            \LpLaw{2\sfp}{y_0}{\mathcal{P}^r_{i}(\nu)}
            \leq
            2
            \Delta_g
            \left(
                2\Delta_p + \Delta_\sigma
            \right)
            +
            2
            \Delta_g
                \Delta_\sigma
                \sqrt{\LpRef{y_0}}
            ,
            \label{eqn:prop:Appendix:ReferenceProcess:BoundsRoundC:SumDelpsigma}
        \end{align}
    \end{subequations}
    where we have used the definition $\RefDiffNorm{\Yt{N}} = 2 \left(\Xrt{N} - \Xstart{N}\right)$ from the statement of Lemma~\ref{lem:Reference:dV}.
    Now, recall the following definitions of $\Delta^r_{\mathcal{P}_1}(\sfp,\tstar)$ and $\Delta^r_{\mathcal{P}_3}(\sfp,\tstar)$ from the statement of Proposition~\ref{prop:Appendix:ReferenceProcess:N:Bounds}:
    \begin{align*}
        \Delta^r_{\mathcal{P}_1}(\sfp,\tstar)
        = 
        \frac{ 1 }{ 2\sqrt{\lambda} }
            \sup_{\nu \in [0,\tstar]}  
            \LpLaw{2\sfp}{y_0}{\mathcal{P}^r_{\mu}(\nu)}
        + 
        \frac{\mathfrak{p}(\sfp)}{2}
        \sup_{\nu \in [0,\tstar]}
        \left(
            \LpLaw{2\sfp}{y_0}{\mathcal{P}^r_{\sigma}(\nu)}
            +
            \LpLaw{2\sfp}{y_0}{\mathcal{P}^r_{\sigma_\star}(\nu)} 
        \right)
            ,
        \\
        \Delta^r_{\mathcal{P}_3}(\sfp,\tstar)
        = 
        \frac{\sqrt{m}}{2}  
        \sup_{\nu \in [0,\tstar]}
        \left(
            \LpLaw{2\sfp}{y_0}{\mathcal{P}^r_{\sigma}(\nu)}
            +
            \LpLaw{2\sfp}{y_0}{\mathcal{P}^r_{\sigma_\star}(\nu)} 
        \right)
        .
    \end{align*}
    Substituting~\eqref{eqn:prop:Appendix:ReferenceProcess:BoundsRoundC:SumDelp} into the above then leads to~\eqref{eqn:prop:Appendix:ReferenceProcess:BoundsRoundB:Del:Final}. 

\end{proof}

The final result of the section derives the bounds due to Lemmas~\ref{lem:Appendix:ReferenceProcess:Xir} and~\ref{lem:Appendix:ReferenceProcess:XirU}. 
\begin{lemma}\label{lem:Appendix:ReferenceProcess:Final:Bound}
    Consider the stopping times $\tau^\star$ and $\tstar$, as defined in~\eqref{eqn:Reference:StoppingTimes}, Lemma~\ref{lem:Reference:dV}, and assume that $\tau^\star = \tstar$. 
    If Assumptions~\ref{assmp:KnownFunctions}~-~\ref{assmp:knownDiffusion:Decomposition} hold, then with $\mathsf{p}^\star$ defined in Assumption~\ref{assmp:NominalSystem:FiniteMomentsWasserstein}, and 
    \begin{equation}\label{eqn:prop:ReferenceProcess:Final:Bound:Condition}
        \LpRefError{y_0}
        \doteq
        \sup_{\nu \in [0,\tstar]}
        \LpLaw{2 \sfp}{y_0}{\Xrt{N,\nu} - \Xstart{N,\nu}}, 
        \quad 
        \LpRef{y_0}
        \doteq
        \sup_{\nu \in [0,\tstar]}
        \LpLaw{2 \sfp}{y_0}{\Xrt{N,\nu}},
        \quad \sfp \in \cbr{1,\dots,\sfp^\star},
    \end{equation} 
    the following bound holds for all $(t,\sfp) \in \mathbb{R}_{\geq 0} \times \cbr{2,\dots,\sfp^\star}$:
    \begin{multline}\label{eqn:lem:Appendix:ReferenceProcess:Final:Bound:B:Final}
        \pLpLaw{y_0}{
            \expo{-2\lambda \tau(t)}
            \Xi^r\br{\tau(t),\Yt{N}}
        }
        +
        \pLpLaw{y_0}{
            \expo{-(2\lambda+\Boldomega) \tau(t)}
            \Xi^r_{\mathcal{U}}\br{\tau(t),\Yt{N};\Boldomega}
        }
        \\
        \leq   
        \Delta^r_\circ(\sfp, \Boldomega)
        +
        \widebreve{\Delta}^r_{\circledcirc}(\sfp, \Boldomega)
        \left(\LpRef{y_0}\right)^\frac{1}{2}
        +
        \widebreve{\Delta}^r_{\odot}(\sfp, \Boldomega)
        \LpRef{y_0}
        +
        \Delta^r_\odot(\sfp, \Boldomega)
        \LpRefError{y_0}
        \\
        +
        \left(
            \widebreve{\Delta}^r_{\otimes}(\sfp,\Boldomega)
            \LpRef{y_0}
            +
            \Delta^r_{\otimes}(\sfp,\Boldomega)
            \LpRefError{y_0}
        \right)
        \left(\LpRef{y_0}\right)^\frac{1}{2}
        \\
        +
        \left(
        \widebreve{\Delta}^r_{\circledast}(\sfp,\Boldomega)
            \LpRef{y_0}
            +
            \Delta^r_{\circledast}(\sfp,\Boldomega)
            \LpRefError{y_0}
        \right)
        \left(\LpRef{y_0}\right)
    ,
    \end{multline} 
    where 
    \begin{align*}
        \begin{aligned}
            \Delta^r_\circ(\sfp, \Boldomega)
            = 
            \frac{\Delta^r_{\circ_1}}{2\lambda}
            + 
            \frac{1}{\absolute{2\lambda - \Boldomega}}
            \left(  
                \Delta^r_{\circ_2}(\sfp)
                +
                \BoldomegaRoot 
                \Delta^r_{\circ_3}(\sfp)
                + 
                \Boldomega
                \Delta^r_{\circ_4}(\sfp)
            \right)
            ,
        \end{aligned}
        \\
        \begin{aligned}
            \widebreve{\Delta}^r_{\circledcirc}(\sfp, \Boldomega)
            = 
            \frac{\Delta^r_{\circledcirc_1}}{2\lambda}
            +
            \frac{1}{\absolute{2\lambda - \Boldomega}}
            \left(  
                \Delta^r_{\circledcirc_2}(\sfp) 
                +
                \BoldomegaRoot
                \Delta^r_{\circledcirc_3}(\sfp)
                +
                \Boldomega
                \Delta^r_{\circledcirc_4}(\sfp)
            \right)
            ,
        \end{aligned}
        \\
        \begin{aligned}
            \Delta^r_\odot(\sfp, \Boldomega)
            = 
            \frac{\Delta^r_{\odot_2}}{2\lambda}
            +
            \frac{\Delta^r_{\odot_3}(\sfp)}{2\sqrt{\lambda}}
            +
            \frac{1}{\absolute{2\lambda - \Boldomega}}
            \left(
                \Delta^r_{\odot_5}(\sfp)
                +
                \BoldomegaRoot
                \Delta^r_{\odot_7}(\sfp)
            \right),
        \end{aligned}
        \\
        \begin{aligned}
            \widebreve{\Delta}^r_{\odot}(\sfp, \Boldomega)
            = 
            \frac{\Delta^r_{\odot_1}}{2\lambda}
            +
            \frac{1}{\absolute{2\lambda - \Boldomega}}
            \left(
                \Delta^r_{\odot_4}(\sfp)
                +
                \BoldomegaRoot
                \Delta^r_{\odot_6}(\sfp)
                +
                \Boldomega
                \Delta^r_{\odot_8}(\sfp)  
            \right),
        \end{aligned}
        \\
        \begin{aligned}
            \Delta^r_\otimes(\sfp, \Boldomega)
            = 
            \frac{1}{\absolute{2\lambda - \Boldomega}}
            \left(
               \Delta^r_{\otimes_3}(\sfp)
                +    
                \BoldomegaRoot
               \Delta^r_{\otimes_5}(\sfp)
            \right),
            \quad
            \widebreve{\Delta}^r_{\otimes}(\sfp, \Boldomega)
            = 
            \frac{\Delta^r_{\otimes_1}(\sfp)}{2\sqrt{\lambda}}
            +
            \frac{1}{\absolute{2\lambda - \Boldomega}}
            \left(
               \Delta^r_{\otimes_2}(\sfp) 
               +
               \BoldomegaRoot
               \Delta^r_{\otimes_4}(\sfp)
            \right),
        \end{aligned}
        \\
        \begin{aligned}
            \Delta^r_\circledast(\sfp, \Boldomega)
            = 
            \frac{\Delta^r_{\circledast_1}}{2\lambda}
            +
            \frac{\Delta^r_{\circledast_3}(\sfp)}{\absolute{2\lambda - \Boldomega}}
            ,
            \quad
            \widebreve{\Delta}^r_{\circledast}(\sfp, \Boldomega)
            = 
            \frac{\Delta^r_{\circledast_2}(\sfp)}{\absolute{2\lambda - \Boldomega}}
            ,
        \end{aligned}
    \end{align*}
    and the constants $\Delta^r_i(\sfp)$, $i \in \cbr{ \circ_1,\dots,\circ_4, \circledcirc_1,\dots,\circledcirc_4, \odot_1,\dots,\odot_8, \otimes_1,\dots,\otimes_5, \circledast_1,\circledast_2 , \circledast_3   }$, are defined in~\eqref{eqn:app:Constants:Ref:DelCirc}~-~\eqref{eqn:app:Constants:Ref:DelCircledAst} in Section~\ref{subsec:app:Definitions:Reference}.
    Additionally, the term  
    \begin{align*}
        \ELaw{y_0}{
            \expo{-2\lambda \tau(t)}
            \Xi^r\br{\tau(t),\Yt{N}}
            }
        +
        \ELaw{y_0}{
            \expo{-(2\lambda+\Boldomega) \tau(t)}
            \Xi^r_{\mathcal{U}}\br{\tau(t),\Yt{N};\Boldomega}
            }
        ,
    \end{align*}
    can be bounded for all $t \in \mathbb{R}_{\geq 0}$ by the right hand side of~\eqref{eqn:lem:Appendix:ReferenceProcess:Final:Bound:B:Final} evaluated at $\sfp =1$.

\end{lemma}
\begin{proof}
    We begin by combining the results from Lemmas~\ref{lem:Appendix:ReferenceProcess:Xir} and~\ref{lem:Appendix:ReferenceProcess:XirU} to obtain he following bounds that hold for all $t \in [0,\tstar]$: 
    \begin{subequations}\label{eqn:lem:Appendix:ReferenceProcess:Final:Bounds:Initial}
        \begin{align}
            \begin{multlined}[b][0.9\linewidth]
                \ELaw{y_0}{
                    \expo{-2\lambda t}
                    \Xi^r\br{t,\Yt{N}}
                    }
                +
                \ELaw{y_0}{
                    \expo{-(2\lambda+\Boldomega) \tau(t)}
                    \Xi^r_{\mathcal{U}}\br{\tau(t),\Yt{N};\Boldomega}
                }
                \\
                \leq  
                \frac{\Delta_{\Xi_1}^r(1,\tstar)}{2\lambda} 
                +
                \frac{ 1  }{\absolute{2\lambda - \Boldomega}}
                \left(
                    \Delta^r_{\mathcal{U}_1}(1,\tstar)
                    +
                    \BoldomegaRoot 
                    \Delta^r_{\mathcal{U}_2}(1,\tstar)
                    +
                    \Boldomega
                    \Delta^r_{\mathcal{U}_3}(1,\tstar)
                \right) 
                ,
            \end{multlined}
            \label{eqn:lem:Appendix:ReferenceProcess:Final:Bound:2}
            \\
            \begin{multlined}[b][0.9\linewidth]
                \pLpLaw{y_0}{
                    \expo{-2\lambda t}
                    \Xi^r\br{t,\Yt{N}}
                }
                +
                \pLpLaw{y_0}{
                    \expo{-(2\lambda+\Boldomega) \tau(t)}
                    \Xi^r_{\mathcal{U}}\br{\tau(t),\Yt{N};\Boldomega}
                }
                \\
                \leq 
                \frac{\Delta_{\Xi_1}^r(\sfp)}{2\lambda} 
                +
                \frac{\Delta_{\Xi_2}^r(\sfp)}{2\sqrt{\lambda}} 
                +
                \frac{ 1  }{\absolute{2\lambda - \Boldomega}}
                \left(
                    \Delta^r_{\mathcal{U}_1}(\sfp,\tstar)
                    +
                    \BoldomegaRoot 
                    \Delta^r_{\mathcal{U}_2}(\sfp,\tstar)
                    +
                    \Boldomega
                    \Delta^r_{\mathcal{U}_3}(\sfp,\tstar)
                \right), 
                \quad \forall \sfp \in \mathbb{N}_{\geq 2}.
            \end{multlined}
            \label{eqn:lem:Appendix:ReferenceProcess:Final:Bound:1}
        \end{align}
    \end{subequations}

    Next, we use the definitions $\bar{F}_\sigma(\nu, \cdot) = p(\nu,\cdot)$, $\RefDiffNorm{\Yt{N}} = 2 \left(\Xrt{N} - \Xstart{N}\right)$, along with Assumptions~\ref{assmp:KnownFunctions},~\ref{assmp:UnknownFunctions}, and~\ref{assmp:knownDiffusion:Decomposition} to arrive at the following bounds: 
    \begin{subequations}\label{eqn:lem:Appendix:ReferenceProcess:ComponentBounds:A}
        \begin{align}
            \sup_{\nu \in [0,\tstar]}
            \LpLaw{2\sfp}{y_0}{
                \RefDiffNorm{\Yt{N,\nu}} 
            } 
            \leq     
            2\LpRefError{y_0},
            \quad
            \sup_{\nu \in [0,\tstar]}
            \LpLaw{2\sfp}{y_0}{
                \Lperpmu{\nu,\Xrt{N,\nu}}
            } 
            \leq     
            \Delta_\mu^\perp 
            +
            \Delta_\mu^\perp 
            \LpRef{y_0}
            , 
            \\
            \sup_{\nu \in [0,\tstar]}
            \LpLaw{2\sfp}{y_0}{
                \Fsigma{\nu,\Xrt{N,\nu}} 
            } 
            \leq     
            \Delta_p + \Delta_\sigma 
            + 
            \Delta_\sigma
            \sqrt{\LpRef{y_0}},
            \\
            \sup_{\nu \in [0,\tstar]}
            \LpLaw{2\sfp}{y_0}{
                \Fperpsigma{\nu,\Xrt{N,\nu}} 
            } 
            \leq     
            \Delta_p^{\perp} +  \Delta_\sigma^{\perp} 
            + 
            \Delta_\sigma^{\perp}
            \sqrt{\LpRef{y_0}},   
            \quad 
            \sup_{\nu \in [0,\tstar]}
            \LpLaw{2\sfp}{y_0}{
                \Fbarsigma{\nu,\Xstart{N,\nu}} 
            } 
            \leq     
            \Delta_p,
        \end{align}
    \end{subequations}
    where we have further used the subadditivity of the square root function. 
    We have also used the manipulations in~\eqref{eqn:lem:Appendix:ReferenceProcess:ComponentBounds:Manipulations}.

    Now, recall the definitions of $\Delta_{\Xi_1}^r(\sfp)$ and $\Delta_{\Xi_2}^r(\sfp)$ in Lemma~\ref{lem:Appendix:ReferenceProcess:Xir} for $\sfp \in \mathbb{N}_{\geq 1}$: 
    \begin{align*}
        \begin{multlined}[][0.8\linewidth]
            \Delta_{\Xi_1}^r(\sfp,\tstar)
            =
            \sup_{\nu \in [0,\tstar]}
            \left(
                \Delta_g^\perp
                \LpLaw{2\sfp}{y_0}{
                    \RefDiffNorm{\Yt{N,\nu}} 
                } 
                \LpLaw{2\sfp}{y_0}{
                    \Lperpmu{\nu,\Xrt{N,\nu}}
                }
            \right.
            \\
            \left.  
                +
                \ELaw{y_0}{
                    \Fsigma{\nu,\Xrt{N,\nu}}^{2\sfp}
                }^\frac{1}{\sfp}
                +
                \ELaw{y_0}{
                    \Fbarsigma{\nu,\Xstart{N,\nu}}^{2\sfp}
                }^\frac{1}{\sfp}     
            \right)
            ,
        \end{multlined}
        \\
        \Delta_{\Xi_2}^r(\sfp,\tstar)
        =
        \mathfrak{p}(\sfp)
        \sup_{\nu \in [0,\tstar]}
        \left(
            \Delta_g^\perp
            \LpLaw{2\sfp}{y_0}{
                \Fperpsigma{\nu,\Xrt{N,\nu}}
            }
            + 
            \LpLaw{2\sfp}{y_0}{
                \Fbarsigma{\nu,\Xstart{N,\nu}}
            }
        \right)
        \sup_{\nu \in [0,\tstar]}
        \LpLaw{2\sfp}{y_0}{
            \RefDiffNorm{\Yt{N,\nu}}
        } 
        .
    \end{align*}
    We can bound these terms by using~\eqref{eqn:lem:Appendix:ReferenceProcess:ComponentBounds:A} as follows:
    \begin{subequations}
        \begin{align*}
            \Delta^r_{\Xi_1}(\sfp,\tstar)
            \leq
            \Delta_p^2
            +
            2 
            \Delta_g^\perp
            \Delta_\mu^\perp
            \LpRefError{y_0}
            \left(
                1 
                + 
                \LpRef{y_0}
            \right) 
            +
            \left(
                \Delta_p + \Delta_\sigma 
                + 
                \Delta_\sigma
                \sqrt{\LpRef{y_0}}
            \right)^2
            ,
            \\
            \Delta_{\Xi_2}^r(\sfp,\tstar)
            \leq 
            2
            \mathfrak{p}(\sfp)  
            \left(
                \Delta_g^\perp
                \left(\Delta_p^{\perp} +  \Delta_\sigma^{\perp}\right) 
                + 
                \Delta_p
                + 
                \Delta_g^\perp
                \Delta_\sigma^{\perp}
                \sqrt{\LpRef{y_0}}
            \right)
            \LpRefError{y_0},
        \end{align*}
    \end{subequations}
    for all $\sfp \in \cbr{1,\sfp^\star}$.
    We can therefore write the above as 
    \begin{subequations}\label{eqn:lem:Appendix:ReferenceProcess:Final:Bound:DelXi}
        \begin{align}
            \Delta^r_{\Xi_1}(\sfp,\tstar)
            \leq
            \Delta^r_{\circ_1}
            + 
            \Delta^r_{\circledcirc_1}
                \left(\LpRef{y_0}\right)^\frac{1}{2}
            +
            \Delta^r_{\odot_1}
            \LpRef{y_0}
            +
            \Delta^r_{\odot_2}
            \LpRefError{y_0} 
            + 
            \Delta^r_{\circledast_1}
            \LpRef{y_0}
            \LpRefError{y_0}
            ,
            \\
            \Delta_{\Xi_2}^r(\sfp,\tstar)
            \leq   
            \Delta^r_{\odot_3}(\sfp)
            \LpRefError{y_0}
            + 
            \Delta^r_{\otimes_1}(\sfp)
            \left(\LpRef{y_0}\right)^\frac{1}{2}
            \LpRefError{y_0}
            ,
        \end{align}  
    \end{subequations}
    for all $\sfp \in \cbr{1,\sfp^\star}$.

    Next, recall the terms $\Delta^r_{\mathcal{U}_1}$, $\Delta^r_{\mathcal{U}_2}$, and $\Delta^r_{\mathcal{U}_3}$ in Lemma~\ref{lem:Appendix:ReferenceProcess:XirU}:
    \begin{align*}
        \begin{multlined}[][0.9\linewidth]
            \Delta^r_{\mathcal{U}_1}(\sfp,\tstar)
            =
            \sqrt{\lambda} 
            \Delta_g
            \mathfrak{p}(\sfp)  
            \sup_{\nu \in [0,\tstar]}
            \LpLaw{2\sfp}{}{
                \RefDiffNorm{\Yt{N,\nu}} 
            }
            \sup_{\nu \in [0,\tstar]}
            \LpLaw{2\sfp}{}{
                \Fparasigma{\nu,\Xrt{N,\nu}}
            }
            \\
            \qquad \qquad
            +
            \left(
                \frac{\Delta^r_{\mathcal{P}_1}(\sfp,\tstar)}{\sqrt{\lambda}}
                + 
                2
                \Delta_g
                \left[ 
                    \sup_{\nu \in [0,\tstar]}\LpLaw{2\sfp}{y_0}{\RefDiffNorm{\Yt{N,\nu}}}
                    + 
                    \frac{\Delta_g}{2\lambda}
                    \sup_{\nu \in [0,\tstar]}
                    \LpLaw{2\sfp}{y_0}{\Lparamu{\nu,\Xrt{N,\nu}}}
                \right]
            \right)
            \\
            \times 
            \sup_{\nu \in [0,\tstar]}
            \LpLaw{2\sfp}{y_0}{\Lparamu{\nu,\Xrt{N,\nu}}}
            ,
        \end{multlined}
        \\   
        \begin{multlined}[][0.9\linewidth] 
            \Delta^r_{\mathcal{U}_2}(\sfp,\tstar)
            =
            \mathfrak{p}'(\sfp)
            \left(
                \frac{\Delta^r_{\mathcal{P}_1}(\sfp,\tstar)}{\sqrt{\lambda}} 
                + 
                \Delta_g
                \left[   
                    \sup_{\nu \in [0,\tstar]}\LpLaw{2\sfp}{y_0}{\RefDiffNorm{\Yt{N,\nu}}}
                    +
                    \frac{ 2\Delta_g }{\lambda}  
                    \sup_{\nu \in [0,\tstar]}
                    \LpLaw{2\sfp}{y_0}{\Lparamu{\nu,\Xrt{N,\nu}}}
                \right]
            \right)
            \\
            \times
            \sup_{\nu \in [0,\tstar]}
                \LpLaw{2\sfp}{y_0}{\Fparasigma{\nu,\Xrt{N,\nu}}}
            ,
        \end{multlined}
        \\
        \begin{aligned}
            \Delta^r_{\mathcal{U}_3}(\sfp,\tstar)
            =
            \left(
                \frac{ \Delta^r_{\mathcal{P}_3}(\sfp,\tstar)}{\lambda}
                +
                \frac{ \Delta_g^2  \mathfrak{p}'(\sfp)^2 }{\lambda}
                    \sup_{\nu \in [0,\tstar]}
                    \LpLaw{2\sfp}{y_0}{\Fparasigma{\nu,\Xrt{N,\nu}}}
            \right)
            \sup_{\nu \in [0,\tstar]}
                \LpLaw{2\sfp}{y_0}{\Fparasigma{\nu,\Xrt{N,\nu}}}
            ,
        \end{aligned}
    \end{align*}
    where we have further used the definition $\Delta^r_{\mathcal{P}_2}(\sfp,\tstar)=
    \mathfrak{p}'(\sfp)
    \Delta^r_{\mathcal{P}_1}(\sfp,\tstar)$ from Proposition~\ref{prop:Appendix:ReferenceProcess:N:Bounds}.
    In order to bound these terms, in addition to~\eqref{eqn:prop:Appendix:ReferenceProcess:BoundsRoundB:Del:Final} in Proposition~\ref{prop:Appendix:ReferenceProcess:BoundsRoundB}, and the bounds in~\eqref{eqn:lem:Appendix:ReferenceProcess:ComponentBounds:A} above, we also appeal to the following due to Assumptions~\ref{assmp:KnownFunctions},~\ref{assmp:UnknownFunctions}, and~\ref{assmp:knownDiffusion:Decomposition}: 
    \begin{align}\label{eqn:lem:Appendix:ReferenceProcess:ComponentBounds:B}
        \begin{aligned}
            \sup_{\nu \in [0,\tstar]}
            \LpLaw{2\sfp}{y_0}{
                \Lparamu{\nu,\Xrt{N,\nu}}
            } 
            \leq     
            \Delta_\mu^{\paral} 
            +
            \Delta_\mu^{\paral}
            \LpRef{y_0}
            , 
            \\
            \sup_{\nu \in [0,\tstar]}
            \LpLaw{2\sfp}{y_0}{
                \Fparasigma{\nu,\Xrt{N,\nu}} 
            } 
            \leq     
            \Delta_p^{\paral} +  \Delta_\sigma^{\paral} 
            + 
            \Delta_\sigma^{\paral}
            \sqrt{\LpRef{y_0}}.
        \end{aligned}
    \end{align}
    Using the above then leads to
    \begin{subequations} \label{eqn:lem:Appendix:ReferenceProcess:Final:Bound:DelU}
        \begin{align}
            \begin{multlined}[b][0.9\linewidth]
                \Delta^r_{\mathcal{U}_1}(\sfp,\tstar)
                \leq 
                \Delta^r_{\circ_2}(\sfp)
                +
                \Delta^r_{\circledcirc_2}(\sfp)
                \left(\LpRef{y_0}\right)^\frac{1}{2} 
                +
                \Delta^r_{\odot_4}(\sfp)
                \LpRef{y_0}
                +
                \Delta^r_{\odot_5}(\sfp)
                \LpRefError{y_0}
                \\
                + 
                \left(
                    \Delta^r_{\otimes_2}(\sfp)
                    \LpRef{y_0} 
                    +
                    \Delta^r_{\otimes_3}(\sfp)   
                    \LpRefError{y_0}
                \right)
                \left(\LpRef{y_0}\right)^\frac{1}{2}
                \\
                +
                \left(
                    \Delta^r_{\circledast_2}
                    \LpRef{y_0}
                    +
                    \Delta^r_{\circledast_3}
                    \LpRefError{y_0}
                \right)
                \LpRef{y_0}
                ,
            \end{multlined}
            \\
            \begin{multlined}[b][0.9\linewidth] 
                \Delta^r_{\mathcal{U}_2}(\sfp,\tstar)
                \leq 
                \Delta^r_{\circ_3}(\sfp)
                +
                \Delta^r_{\circledcirc_3}(\sfp)
                \left(\LpRef{y_0}\right)^\frac{1}{2}
                +
                \Delta^r_{\odot_6}(\sfp)
                \LpRef{y_0}
                +
                \Delta^r_{\odot_7}(\sfp)
                \LpRefError{y_0}
                \\
                +
                \left(
                    \Delta^r_{\otimes_4}(\sfp)
                    \LpRef{y_0}
                    +
                    \Delta^r_{\otimes_5}(\sfp)
                    \LpRefError{y_0}
                \right)
                \left(\LpRef{y_0}\right)^\frac{1}{2}
                ,
            \end{multlined}
            \\
            \Delta^r_{\mathcal{U}_3}(\sfp,\tstar)
            \leq
            \Delta^r_{\circ_4}(\sfp)
            +
            \Delta^r_{\circledcirc_4}(\sfp)
            \left(\LpRef{y_0}\right)^\frac{1}{2}
            +
            \Delta^r_{\odot_8}(\sfp)
            \LpRef{y_0}.
        \end{align}
    \end{subequations}
    The proof is then concluded by substituting~\eqref{eqn:lem:Appendix:ReferenceProcess:Final:Bound:DelXi}~-~\eqref{eqn:lem:Appendix:ReferenceProcess:Final:Bound:DelU} into~\eqref{eqn:lem:Appendix:ReferenceProcess:Final:Bounds:Initial}.

\end{proof}


\setcounter{equation}{0}
\section{True (Uncertain) Process}\label{app:TrueProcess}


We first begin by deriving a couple of expressions regarding the \ellonedrac~feedback operator $\FL$ and the reference feedback operator $\ReferenceInput$.
\begin{proposition}\label{prop:Appendix:TrueProcess:FL:Expression}
    Consider the \ellonedrac~feedback operator $\FL$ and the reference feedback operator $\ReferenceInput$ defined in~\eqref{eqn:L1DRAC:Definition:FeedbackOperator} and~\eqref{eqn:ReferenceFeedbackOperatorProcess}, respectively.
    Assume that the closed-loop system~\eqref{eqn:True:SDE} admits $z \in \mathcal{M}_2\br{\mathbb{R}^n~|~\Wfilt{t}}$ as a unique strong solution, for all $t \in [0,\tau]$, where $\tau \in [0,T]$\footnote{For example, given any stopping time $\tau' \in \mathbb{R}_{\geq 0}$, we can choose $\tau \doteq \tau' \wedge T$.} is a stopping time.  
    Then, 
    \begin{equation}\label{eqn:prop:Appendix:TrueProcess:FL:Expression}
        \left(\FL - \ReferenceInput\right)(z)(t)
        =
        \Boldomega 
        \expo{-\Boldomega t}
        \ControlErrorTilde[z,t]
        ,
        \quad 
        t \in [0,\tau],
    \end{equation}
    where $\ControlErrorTilde[z,t]$ is defined as 
    \begin{align}\label{eqn:prop:Appendix:TrueProcess:FL:ControlError:Solution:Decomposed}
        \ControlErrorTilde[z,t]
        = 
        \int_0^t 
            \expo{\Boldomega \nu} 
            \left( \Sigma^{\paral}_\nu(z) - \Lparahat{\nu} d\nu \right)
        ,
    \end{align}
    for $t \in  [0,\tau]$, $z_0 = z(0)$, and
    \begin{align}\label{eqn:prop:Appendix:TrueProcess:FL:AdaptationLaw:Matched:Final}
        \Lparahat{t} 
        =& 
        0_m  \indicator{[0,\BoldTs) \cap [0,\tau]}{t}
        \notag 
        \\
        &
        -
        \sum_{i \in \mathbb{N}_{\geq 1}}
        \left( 
            \lambda_s \left(1 - e^{\lambda_s \BoldTs}\right)^{-1}
            \Theta_{ad}(i\BoldTs)
            \int_{(i-1)\BoldTs}^{i\BoldTs} \expo{-\lambda_s (i\BoldTs-\nu)} \Sigma_\nu(z)
        \right)
        \indicator{ [i\BoldTs,(i+1)\BoldTs) \cap [0,\tau] }{t}
        .
    \end{align}
    In the expressions above, we have (formally) defined
    \begin{align}\label{eqn:prop:Appendix:TrueProcess:FL:ControlError:Sigma}
        \Sigma^{\cbr{\cdot,\paral}}_t(z)
        \doteq 
        \Lambda^{\cbr{\cdot,\paral}}_\mu \left(t,z(t)\right)dt 
        +
        F^{\cbr{\cdot,\paral}}_\sigma \left(t,z(t)\right)d\Wt{t}
        ,
        \quad t \in [0,\tau].  
    \end{align}
\end{proposition}
\begin{proof}
    Since $z$ is assumed to be the unique strong solution of~\eqref{eqn:True:SDE}, we have that 
    \begin{align}\label{eqn:prop:Appendix:TrueProcess:FL:System:Differential}
        dz(t) =& \Fmu{t,z(t),\FL[z][t]}dt + \Fsigma{t,z(t)}d\Wt{t} 
        \notag 
        \\
        =&
        \left(f(t,z(t)) + g(t)\FL[z][t] + \Lmu{t,z(t)}\right)dt 
        +
        \Fsigma{t,z(t)}d\Wt{t} 
        , \quad z(0) = z_0 \sim \xi_0~(\text{\Pas}), 
    \end{align}
    where we have set $\ULt{t} = \FL[z][t]$.
    %
    Next, using the definition of $\FL$ from~\eqref{eqn:L1DRAC:Definition:FeedbackOperator:Alt} we set  
    \begin{align}\label{eqn:prop:Appendix:TrueProcess:FL:Input}
        \FL[z] 
        = 
        \Filter[\hat{\Lambda}^{\paral}],
        \quad
        \hat{\Lambda}^{\paral} = \AdaptationLawParal[\hat{\Lambda}],
        \quad 
        \hat{\Lambda} = \AdaptationLaw[\hat{z}][z],
        \quad 
        \hat{z} = \Predictor[z].
    \end{align} 
    Hence, using the definition of $\ReferenceInput$ from~\eqref{eqn:ReferenceFeedbackOperatorProcess}, we obtain 
    \begin{multline*}
        \left(\left(\FL - \ReferenceInput\right)(z)\right)(t)
        =
        - \Boldomega \int_0^t \expo{-\Boldomega (t-\nu)} \Lparahat{\nu} d\nu
        -
        \Filter[\Lparamu{\cdot,z}][t] - \FilterW[\Fparasigma{\cdot,z}, \Wt{}][t] 
        \\
        =
        \Boldomega
        \expo{-\Boldomega t }
        \int_0^t 
        \expo{ \Boldomega \nu}
            \left( 
            \Sigma^{\paral}_\nu(z)
            -
            \Lparahat{\nu} d\nu
            \right) 
            ,       
    \end{multline*}
    for $t \in [0,\tau]$, where we have used the definition of $\Sigma^{\paral}_t(z)$ in~\eqref{eqn:prop:Appendix:TrueProcess:FL:ControlError:Sigma}.
    Then, for any stopping time $\hat{\tau} \in [0,\tau]$
    one obtains the following: 
    \begin{multline*}
        \left(\left(\FL - \ReferenceInput\right)(z)\right)(t)
        =
        \Boldomega
        \expo{-\Boldomega t }
        \left( 
            \int_0^{\hat{\tau}} 
                \expo{ \Boldomega \nu}
                \left( 
                    \Sigma^{\paral}_\nu(z)
                    -
                    \Lparahat{\nu} d\nu
                \right)
            + 
            \int_{\hat{\tau}}^t 
                \expo{ \Boldomega \nu}
                \left( 
                    \Sigma^{\paral}_\nu(z)
                    -
                    \Lparahat{\nu} d\nu
                \right)
        \right)
        \\ 
        =
        \Boldomega
        \expo{-\Boldomega t }
        \left( 
            \ControlError[\hat{\tau}]
            + 
            \int_{\hat{\tau}}^t 
                \expo{ \Boldomega \nu}
                \left( 
                    \Sigma^{\paral}_\nu(z)
                    -
                    \Lparahat{\nu} d\nu
                \right)
        \right)
        ,       
    \end{multline*}
    for $t \in [\hat{\tau}, \tau]$, where we have defined 
    \begin{align*}
        \ControlError[t] 
        \doteq
        \int_0^t 
        \expo{ \Boldomega \nu}
        \left( 
            \Sigma^{\paral}_\nu(z)
            -
            \Lparahat{\nu} d\nu
        \right)
        .
    \end{align*}
    Since the temporal instances $i\BoldTs$, $i \in \mathbb{N}$, are constant, and hence stopping times, we can write
    \begin{align*}
        \left(\left(\FL - \ReferenceInput\right)(z)\right)(t)
        =
        \begin{cases}
            \Boldomega
            \expo{-\Boldomega t }
            \ControlError[t]
            = 
            \Boldomega
            \expo{-\Boldomega t}
            \int_0^t 
            \expo{\Boldomega \nu} 
            \left( \Sigma^{\paral}_\nu - \Lparahat{\nu} d\nu \right)
            , 
            & 
            \quad 
            t \in [0,\BoldTs) \cap [0,\tau]
            \\
            \Boldomega
            \expo{-\Boldomega t}
            \left( 
                \ControlError[i\BoldTs]
                +
                \int_{i\BoldTs}^t 
                \expo{ \Boldomega \nu} 
                \left( \Sigma^{\paral}_\nu(z) - \Lparahat{\nu} d\nu \right)
            \right) 
            , 
            & 
            \quad 
            t \in [i\BoldTs,(i+1)\BoldTs) \cap [0,\tau]
        \end{cases}
        ,
    \end{align*}
    for $i \in \mathbb{N}_{\geq 1}$, thus proving~\eqref{eqn:prop:Appendix:TrueProcess:FL:ControlError:Solution:Decomposed} since $\ControlError[t] = \ControlErrorTilde[z,t]$, over each interval $[i\BoldTs,(i+1)\BoldTs) \cap [0,\tau]$, for $i \in \mathbb{N}_{\geq 1}$.

    Next, we derive the expression for the adaptive estimate $\Lparahat{t}$, which is defined in~\eqref{eqn:prop:Appendix:TrueProcess:FL:Input} as 
    \begin{align*}
        \hat{\Lambda}^{\paral} = \AdaptationLawParal[\hat{\Lambda}],
        \quad 
        \hat{\Lambda} = \AdaptationLaw[\hat{z}][z].
    \end{align*}
    It then follows from the definitions of the adaptation law operators $\AdaptationLaw$ and $\AdaptationLawParal$ in~\eqref{eqn:L1DRAC:Definition:FeedbackOperator:AdaptationLaw} that 
    \begin{subequations}\label{eqn:prop:Appendix:TrueProcess:FL:AdaptationLaw}
        \begin{align}
            \Lparahat{t} 
            =&
            \sum_{i \in \mathbb{N}_{\geq 0}}  
            \Theta_{ad}(i\BoldTs) \Lhat{t}
            \indicator{[i\BoldTs,(i+1)\BoldTs) \cap [0,\tau]}{t}
            , 
            \label{eqn:prop:Appendix:TrueProcess:FL:AdaptationLaw:Matched}
            \\
            \Lhat{t} 
            =&
            0_n \indicator{[0,\BoldTs)}{t} 
            +
            \lambda_s \br{1 - e^{\lambda_s \BoldTs}}^{-1} 
            \sum_{i \in \mathbb{N}_{\geq 1}}    
            \tilde{z}\br{i\BoldTs}
            \indicator{[i\BoldTs,(i+1)\BoldTs) \cap [0,\tau]}{t}, 
            \label{eqn:prop:Appendix:TrueProcess:FL:AdaptationLaw:Total}
        \end{align}
    \end{subequations}
    where $\tilde{z} \doteq \hat{z} - z$.
    Moreover, $\hat{z} = \Predictor[z]$ is the prediction process defined via the operator $\Predictor$ in~\eqref{eqn:L1DRAC:Definition:FeedbackOperator:Predictor} as follows: 
    \begin{align*}
        \hat{z}(t)
        =& 
        z_0
        +
        \int_0^t \br{-\lambda_s \mathbb{I}_n \tilde{z}(\nu)+ f(\nu,z(\nu)) +  g(\nu)\FL[z][\nu] + \Lhat{\nu}} d\nu, 
        \quad t \in [0,\tau].
    \end{align*}
    We may write the integral equation above in its differential form as follows:
    \begin{align*}
        d\hat{z}(t)
        =& 
        \br{-\lambda_s \mathbb{I}_n \tilde{z}(t)+ f(t,z(t)) +  g(t)\FL[z][t] + \Lhat{t}} dt,
        \quad 
        \hat{z}(0) = z_0, 
        \quad  t \in [0,\tau].
    \end{align*}
    Subtracting~\eqref{eqn:prop:Appendix:TrueProcess:FL:System:Differential} from the above equation then leads to differential equation for the error process $\tilde{z} \doteq \hat{z} - z$ as follows:
    \begin{multline}\label{eqn:prop:Appendix:TrueProcess:FL:ProcessError:Differential}
        d\tilde{z}(t)
        =
        \left(-\lambda_s \mathbb{I}_n \tilde{z}(t) + \Lhat{t} - \Lmu{t,z(t)}\right) dt
        - \Fsigma{t,z(t)}d\Wt{t} 
        \\
        =
        \left(-\lambda_s \mathbb{I}_n \tilde{z}(t) + \Lhat{t} \right) dt
        - \Sigma_t(z),
        \quad 
        \tilde{z}(0) = 0_n, 
        \quad  t \in [0,\tau]
        ,
    \end{multline}
    where $\Sigma_t$ is given in~\eqref{eqn:prop:Appendix:TrueProcess:FL:ControlError:Sigma}.
    Therefore, we can write the integral form over the stopping times $i\BoldTs$, $i \in \mathbb{N}$, as follows  
    \begin{align*}
        \tilde{z}(t)
        =
        \int_0^{t} 
            \left(-\lambda_s \mathbb{I}_n \tilde{z}(\nu) + \Lhat{\nu} \right) 
        d\nu
        - 
        \int_0^{t} \Sigma_\nu(z)
        , 
        \quad 
        t \in [0,\BoldTs) \cap [0,\tau]
        ,
        \\
        \tilde{z}(t)
        =
        \tilde{z}(i\BoldTs)
        + 
        \int_{i\BoldTs}^{t} 
            \left(-\lambda_s \mathbb{I}_n \tilde{z}(\nu) + \Lhat{\nu} \right) 
        d\nu
        - 
        \int_{i\BoldTs}^{t} \Sigma_\nu(z)
        , 
        \quad 
        t \in [i\BoldTs,(i+1)\BoldTs) \cap [0,\tau]
        ,
    \end{align*}
    for $i \in \mathbb{N}_{\geq 1}$.
    Substituting the adaptive estimate $\Lhat{t}$ from~\eqref{eqn:prop:Appendix:TrueProcess:FL:AdaptationLaw} into the above equations, we obtain  
    \begin{align*}
        \tilde{z}(t)
        =
        -
        \int_0^{t} 
            \lambda_s \mathbb{I}_n \tilde{z}(\nu) 
        d\nu
        - 
        \int_0^{t} \Sigma_\nu(z)
        , 
        \quad 
        t \in [0,\BoldTs) \cap [0,\tau]
        ,
        \\
        \tilde{z}(t)
        =
        \tilde{z}(i\BoldTs)
        + 
        \int_{i\BoldTs}^{t} 
            \left(
                -\lambda_s \mathbb{I}_n \tilde{z}(\nu) 
                + 
                \lambda_s \br{1 - e^{\lambda_s \BoldTs}}^{-1}     
                \tilde{z}\br{i\BoldTs} 
            \right) 
        d\nu
        - 
        \int_{i\BoldTs}^{t} \Sigma_\nu(z)
        , 
        \quad 
        t \in [i\BoldTs,(i+1)\BoldTs) \cap [0,\tau]
        ,
    \end{align*}
    for $i \in \mathbb{N}_{\geq 1}$.
    Since the expressions above represent linear integral equations in the variable $\tilde{z}$, passing  through their linear differential forms, we obtain their solutions as follows (see e.g.~\cite[Sec.~5.4.2]{evans2012introduction}): 
    \begin{align*}
        \tilde{z}(t)
        =
        - 
        \int_0^{t} \expo{-\lambda_s (t-\nu)} \Sigma_\nu(z)
        , 
        \quad 
        t \in \left([0,\BoldTs) \cap [0,\tau]\right)
        ,
        \\
        \begin{multlined}[][0.9\linewidth]
            \tilde{z}(t)
            =
            \expo{-\lambda_s (t-i\BoldTs)}
            \tilde{z}(i\BoldTs)
            + 
            \lambda_s \br{1 - e^{\lambda_s \BoldTs}}^{-1}     
            \tilde{z}\br{i\BoldTs} 
            \int_{i\BoldTs}^{t} 
                \expo{-\lambda_s (t-\nu)}
            d\nu
            \\
            - 
            \int_{i\BoldTs}^{t} \expo{-\lambda_s (t-\nu)} \Sigma_\nu(z)
            , 
            \quad 
            t \in \left([i\BoldTs,(i+1)\BoldTs) \cap [0,\tau]\right) 
            .
        \end{multlined}
    \end{align*}
    Hence, at the temporal instances $t = i\BoldTs$, $i \in \mathbb{N}_{\geq 1}$, the error process $\tilde{z}$ is given by 
    \begin{align*}
        \tilde{z}( i \BoldTs)
        =
        - 
        \int_{(i-1)\BoldTs}^{i \BoldTs} 
            \expo{-\lambda_s (i\BoldTs-\nu)} \Sigma_\nu(z) 
        ,
        \quad 
        i \in \mathbb{N}_{\geq 1}.
    \end{align*} 
    Substituting the above into~\eqref{eqn:prop:Appendix:TrueProcess:FL:AdaptationLaw:Total} then leads to
    \begin{align*}
        \begin{aligned}
            \Lhat{t} = 0_n, \quad &t \in [0,\BoldTs) \cap [0,\tau],  
            \\
            \Lhat{t} 
            =
            -
            \lambda_s \br{1 - e^{\lambda_s \BoldTs}}^{-1}
                \int_{(i-1)\BoldTs}^{i\BoldTs} \expo{-\lambda_s (i\BoldTs-\nu)} \Sigma_\nu(\nu)
            ,
            \quad &t \in [i\BoldTs,(i+1)\BoldTs) \cap [0,\tau], \, i \in \mathbb{N}_{\geq 1}. 
        \end{aligned} 
    \end{align*}
    Thus,we obtain the expression for the matched adaptive estimate in~\eqref{eqn:prop:Appendix:TrueProcess:FL:AdaptationLaw:Matched:Final} by substituting the above into~\eqref{eqn:prop:Appendix:TrueProcess:FL:AdaptationLaw:Matched}.
\end{proof}

We will need the following trivial result.
\begin{proposition}\label{proposition:Appendix:Scratch:Technical}
    We have the following identities: 
    \begin{multline}\label{eqn:prop:Appendix:Scratch:Technical:Diffusion:Total:A}
        \Theta_{ad}(i\BoldTs)
        \int_{(i-1)\BoldTs}^{i\BoldTs} 
            \expo{ \lambda_s \nu }
            \Sigma_\nu(z) 
        =
        \Theta_{ad}(i\BoldTs)
        \int_{(i-1)\BoldTs}^{i\BoldTs}
            \expo{ \lambda_s \nu }
            \SigmaError{\nu}{i\BoldTs,z}  
        \\
        + 
        \int_{(i-1)\BoldTs}^{i\BoldTs}
            \expo{ \lambda_s \nu }
            \left[
                \Lparamu{i\BoldTs,z\br{i\BoldTs}}
                d\nu
                + 
                \Fparasigma{i\BoldTs,z\br{i\BoldTs}}
                d\Wt{\nu}       
            \right]
        , 
    \end{multline}
    and 
    \begin{align}\label{eqn:prop:Appendix:Scratch:Technical:Total:Identity:B}
        \int_{i\BoldTs}^{t} 
            \expo{\Boldomega \nu}
            \Sigma^{\paral}_\nu(z) 
        = 
        \int_{i\BoldTs}^{t} 
            \expo{\Boldomega \nu}
            \SigmaparaError{\nu}{i\BoldTs,z}
        + 
        \int_{i\BoldTs}^{t} 
            \expo{\Boldomega \nu} 
            \left[
                \Lparamu{i\BoldTs,z\br{i\BoldTs}}
                d\nu
                + 
                \Fparasigma{i\BoldTs,z\br{i\BoldTs}}
                d\Wt{\nu}       
            \right]
        ,
    \end{align}
    for all $t \in [i\BoldTs,\tau], \, i \in \mathbb{N}_{\geq 1}$, 
    where $\Sigma^{\cbr{\cdot,\paral}}_t(z)$ is defined in~\eqref{eqn:prop:Appendix:TrueProcess:FL:ControlError:Sigma}, and 
    \begin{subequations}\label{eqn:prop:Appendix:Scratch:Technical:Total:TildeFunctions}
        \begin{align}
            \SigmaErrorAll{t}{i\BoldTs,z}
            =
            \LmuErrorAll{}{t,i\BoldTs,z}
            dt 
            +
            \FsigmaErrorAll{}{t,i\BoldTs,z}
            d\Wt{t}
            , 
            \label{eqn:prop:Appendix:Scratch:Technical:Total:TildeFunctions:Error}
            \\
            \LmuErrorAll{}{t,i\BoldTs,z}
            =
            \LmuAll{t,z(t)} - \LmuAll{i\BoldTs,z\br{i\BoldTs}}
            , 
            \label{eqn:prop:Appendix:Scratch:Technical:Total:TildeFunctions:Drift}
            \\ 
            \FsigmaErrorAll{}{t,i\BoldTs,z}
            =
            \FsigmaAll{t,z(t)} - \FsigmaAll{i\BoldTs,z\br{i\BoldTs}}
            .
            \label{eqn:prop:Appendix:Scratch:Technical:Total:TildeFunctions:Diffusion}
        \end{align}
    \end{subequations}
\end{proposition}
\begin{proof}
    We begin by adding and subtracting the terms $\Lmu{i\BoldTs,z(i\BoldTs)}$ and $\Fsigma{i\BoldTs,z(i\BoldTs)}$ to obtain the following expression:
    \begin{multline}\label{eqn:prop:Appendix:Scratch:Technical:Drift}
        \Theta_{ad}(i\BoldTs)
        \int_{(i-1)\BoldTs}^{i\BoldTs} 
            \expo{\lambda_s \nu} 
            \Sigma_\nu(z)
        =
        \Theta_{ad}(i\BoldTs)
        \int_{(i-1)\BoldTs}^{i\BoldTs} 
            \expo{\lambda_s \nu}
            \SigmaError{\nu}{i\BoldTs,z}  
        \\ 
        +
        \int_{(i-1)\BoldTs}^{i\BoldTs}
        \expo{\lambda_s \nu}
        \left[
            \Lparamu{i\BoldTs,z\br{i\BoldTs}}
            d\nu
            + 
            \Fparasigma{i\BoldTs,z\br{i\BoldTs}}
            d\Wt{\nu}       
        \right]
        ,
    \end{multline}
    where we have used the definition of $\Theta_{ad}$ from~\eqref{eqn:L1DRAC:Definition:FeedbackOperator:AdaptationLaw} to observe that 
    \begin{multline*}
        \Theta_{ad}(i\BoldTs)\Lmu{i\BoldTs,z\br{i\BoldTs}}
        = 
        \begin{bmatrix}\mathbb{I}_m & 0_{m,n-m}  \end{bmatrix} \bar{g}\br{i\BoldTs}^{-1}
        \Lmu{i\BoldTs,z\br{i\BoldTs}}
        \\
        = 
        \begin{bmatrix}\mathbb{I}_m & 0_{m,n-m}  \end{bmatrix} 
        \begin{bmatrix} g\br{i\BoldTs} & g\br{i\BoldTs}^\perp  \end{bmatrix}^{-1}
        \Lmu{i\BoldTs,z\br{i\BoldTs}},
    \end{multline*}
    which further leads to  
    \begin{align*}
        \Theta_{ad}(i\BoldTs)\Lmu{i\BoldTs,z\br{i\BoldTs}}
        = 
        \begin{bmatrix}\mathbb{I}_m & 0_{m,n-m}  \end{bmatrix} 
        \begin{bmatrix} \Lparamu{i\BoldTs,z\br{i\BoldTs}} \\ \Lperpmu{i\BoldTs,z\br{i\BoldTs}} \end{bmatrix}
        = 
        \Lparamu{i\BoldTs,z\br{i\BoldTs}}
        .
    \end{align*}
    An identical line of reasoning leads to $\Theta_{ad}(i\BoldTs)\Fsigma{i\BoldTs,z\br{i\BoldTs}} = \Fparasigma{i\BoldTs,z\br{i\BoldTs}}$.
    The identity in~\eqref{eqn:prop:Appendix:Scratch:Technical:Diffusion:Total:A} is thus established.

    The expression in~\eqref{eqn:prop:Appendix:Scratch:Technical:Total:Identity:B} follows similarly by adding and subtracting $\Lparamu{i\BoldTs,z\br{i\BoldTs}}$ and $\Fparasigma{i\BoldTs,z\br{i\BoldTs}}$ to and from the integrand of $\int_{i\BoldTs}^{t} \expo{\Boldomega \nu}\Sigma^{\paral}_\nu(z)$.
\end{proof}

The following result is a further development of Proposition~\ref{prop:Appendix:TrueProcess:FL:Expression}.
\begin{proposition}\label{proposition:Appendix:TrueProcess:FL:ExpressionEvolved}
    Assume that the closed-loop system~\eqref{eqn:True:SDE} admits $z \in \mathcal{M}_2\br{\mathbb{R}^n~|~\Wfilt{t}}$ as a unique strong solution, for all $t \in [0,\tau]$, where $\tau \in [0,T]$ is a stopping time.
    Then, the process $\ControlErrorTilde[z,t]$ defined in the statement of Proposition~\ref{prop:Appendix:TrueProcess:FL:Expression} can be expressed as follows:
    \begin{multline}\label{eqn:prop:Appendix:TrueProcess:ControlError:General:Total:[0,tau]:MAIN}
        \begin{aligned}
            \ControlErrorTilde[z,t]
            =&
            \indicator{<\BoldTs}{t}
                \int_0^t 
                    \expo{ \Boldomega \nu} 
                    \Sigma^{\paral}_\nu(z)
            +
            \indicator{\geq \BoldTs}{t}
            \int_0^\BoldTs
                \expo{ \Boldomega \nu} 
                \Lparamu{\nu,z(\nu)}
            d\nu
            \\
            &+ 
            \indicator{\geq \BoldTs}{t}
            \int_{0}^{t}
                \expo{\Boldomega \nu}
                \left[
                    \left(
                        \widetilde{\mathcal{H}}_\mu^{\paral}(\nu,z)
                        + 
                        \expo{(\lambda_s-\Boldomega) \nu}
                        \widetilde{\mathcal{H}}_\mu(\nu,z)
                    \right)
                    d\nu
                    + 
                    \left(
                        \widetilde{\mathcal{H}}_\sigma^{\paral}(\nu,z)
                        + 
                        \expo{(\lambda_s-\Boldomega) \nu}
                        \widetilde{\mathcal{H}}_\sigma(\nu,z)
                    \right)
                    d\Wt{\nu}
                \right]
        \end{aligned}
        \\
        +
        \indicator{\geq \BoldTs}{t}
        \int_{0}^{t}
            \expo{\Boldomega \nu} 
            \left( 
                \widetilde{\mathcal{G}}_{\sigma_1}(\nu,z)
                + 
                \expo{(\lambda_s - \Boldomega) \nu} 
                \widetilde{\mathcal{G}}_{\sigma_2}(\nu,z)
            \right)
        d\Wt{\nu}
        \\
        +
        \indicator{\geq \BoldTs}{t}
        \int_0^t 
            \expo{\Boldomega \nu}
            \left[
                \mathcal{G}_\mu(\nu,z)
                d\nu
                + 
                \left( 
                    \mathcal{G}_{\sigma_1}(\nu,z)
                    + 
                    \mathcal{G}_{\sigma_2}(\nu,z)
                \right)
                d\Wt{\nu}
            \right]
        ,
    \end{multline}
    for $t \in [0,\tau]$, where $\Sigma^{\paral}_\nu(z)$ is defined in~\eqref{eqn:prop:Appendix:TrueProcess:FL:ControlError:Sigma}, and for $\nu \in [0,t]$, we have 
    \begin{align*}
        \widetilde{\mathcal{H}}_\mu^{\paral}(\nu,z)
        =&
        \indicator{<\BoldTs}{\nu} 0_m 
        +
        \sum_{i=1}^{\istar{t} - 1}
        \indicator{[i\BoldTs,(i+1)\BoldTs)}{\nu} 
            \LparamuError{}{\nu,i\BoldTs,z}
        + 
        \indicator{ \geq \istar{t}\BoldTs}{\nu}
            \LparamuError{}{\nu,\istar{t}\BoldTs,z}
        ,
        \\
        \widetilde{\mathcal{H}}_\mu(\nu,z)
        =&
        -
        \sum_{i=1}^{\istar{t} - 1}
        \indicator{[(i-1)\BoldTs,i\BoldTs)}{\nu}
        \widetilde{\gamma}_i(\BoldTs)
        \Theta_{ad}(i\BoldTs)
            \LmuError{}{\nu,i\BoldTs,z}
        \\
        &
        -
        \indicator{[(\istar{t}-1)\BoldTs,\istar{t}\BoldTs)}{\nu}
        \widetilde{\gamma}^\star(t,\BoldTs)
        \Theta_{ad}(\istar{t}\BoldTs)
            \LmuError{}{\nu,\istar{t}\BoldTs,z} 
        +
        \indicator{\geq \istar{t}\BoldTs}{\nu}
        0_m
        ,
        \\
        \widetilde{\mathcal{H}}_\sigma^{\paral}(\nu,z)
        =&
        \indicator{<\BoldTs}{\nu} 0_m 
        +
        \sum_{i=1}^{\istar{t} - 1}
        \indicator{[i\BoldTs,(i+1)\BoldTs)}{\nu} 
            \FparasigmaError{}{\nu,i\BoldTs,z}
        + 
        \indicator{ \geq \istar{t}\BoldTs}{\nu}
            \FparasigmaError{}{\nu,\istar{t}\BoldTs,z}
        ,
        \\
        \widetilde{\mathcal{H}}_\sigma(\nu,z)
        =&
        -
        \sum_{i=1}^{\istar{t} - 1}
        \indicator{[(i-1)\BoldTs,i\BoldTs)}{\nu}
        \widetilde{\gamma}_i(\BoldTs)
        \Theta_{ad}(i\BoldTs)
            \FsigmaError{}{\nu,i\BoldTs,z}
        \\
        &
        -
        \indicator{[(\istar{t}-1)\BoldTs,\istar{t}\BoldTs)}{\nu}
        \widetilde{\gamma}^\star(t,\BoldTs)
        \Theta_{ad}(\istar{t}\BoldTs)
            \FsigmaError{}{\nu,\istar{t}\BoldTs,z}
        +
        \indicator{\geq \istar{t}\BoldTs}{\nu}
        0_m
        .
    \end{align*}
    Furthermore, we have defined the following for $\nu \in [0,t]$:
    \begin{align*}
        \mathcal{G}_\mu(\nu,z)
        =&
        \indicator{<\BoldTs}{\nu}
        0_m
        +
        \sum_{i=1}^{\istar{t} - 1}
        \indicator{[i\BoldTs,(i+1)\BoldTs)}{\nu} 
        \left(1 - \expo{-\lambda_s \BoldTs}\right)
        \Lparamu{i\BoldTs,z\br{i\BoldTs}}
        \\
        &+
        \indicator{ \geq \istar{t}\BoldTs}{\nu} 
        \left(1 - \expo{-\lambda_s \BoldTs}\right)
        \Lparamu{\istar{t}\BoldTs,z\br{\istar{t}\BoldTs}}
        ,
        \\
        \widetilde{\mathcal{G}}_{\sigma_1}(\nu,z)
        =& 
        \indicator{<\BoldTs}{\nu}
        \FparasigmaError{}{\nu,0,z}
        +
        \indicator{\geq \BoldTs}{\nu} 
        0_{m,d}
        ,
        \\
        \widetilde{\mathcal{G}}_{\sigma_2}(\nu,z)
        =&
        -
        \sum_{i=1}^{\istar{t} - 1}
        \indicator{[(i-1)\BoldTs,i\BoldTs)}{\nu}       
        \FparasigmaError{}{i\BoldTs,(i-1)\BoldTs,z}
        \widetilde{\gamma}_i(\BoldTs)
        +
        \indicator{\geq (\istar{t} - 1)\BoldTs}{\nu}
        0_{m,d}
        ,
        \\
        \mathcal{G}_{\sigma_1}(\nu,z)
        =&
        \sum_{i=1}^{\istar{t} - 1}
        \indicator{[(i-1)\BoldTs,i\BoldTs)}{\nu}
        \Fparasigma{(i-1)\BoldTs,z\br{(i-1)\BoldTs}}
        \left( 
            1  
            -
            \widetilde{\gamma}_i(\BoldTs)
            \expo{ \left(\lambda_s - \Boldomega\right) \nu} 
        \right)
        \\
        &
        +
        \indicator{[(\istar{t} - 1)\BoldTs,\istar{t}\BoldTs)}{\nu}
        \Fparasigma{(\istar{t} - 1)\BoldTs,z\br{(\istar{t} - 1)\BoldTs}}
        +  
        \indicator{ \geq \istar{t}\BoldTs}{\nu}
        0_{m,d}
        ,
        \\
        \mathcal{G}_{\sigma_2}(\nu,z)
        =&
        \indicator{<(\istar{t}-1)\BoldTs}{\nu}
        0_{m,d}
        \\
        &
        -
        \Fparasigma{\istar{t}\BoldTs,z\br{\istar{t}\BoldTs}}
        \left(
            \indicator{[(\istar{t}-1)\BoldTs,\istar{t}\BoldTs)}{\nu}
            \widetilde{\gamma}^\star(t,\BoldTs)
            \expo{ (\lambda_s - \Boldomega) \nu} 
            -
            \indicator{ \geq \istar{t}\BoldTs}{\nu}
        \right)
        .
    \end{align*}
    The expressions above are defined using the entities in~\eqref{eqn:prop:Appendix:Scratch:Technical:Total:TildeFunctions} and the following:
    \begin{align*}
        \istar{t} \doteq \max\cbr{i \in \mathbb{N}_{\geq 1} \, : \, i\BoldTs \leq t} = \left\lfloor \frac{t}{\BoldTs} \right\rfloor, 
        \\
        \widetilde{\gamma}_i(\BoldTs)
        \doteq 
        \expo{\left(\Boldomega - \lambda_s\right)i\BoldTs}
        \frac{\lambda_s}{\Boldomega} 
        \frac{\expo{\Boldomega\BoldTs} - 1}{\expo{\lambda_s \BoldTs} - 1}
        =
        \int_{i\BoldTs}^{(i+1)\BoldTs}
            \expo{ \Boldomega \nu}
        d\nu
        \left( 
            \int_{i\BoldTs}^{(i+1)\BoldTs}
                \expo{\lambda_s \nu}
            d\nu
        \right)^{-1}
        ,
        \\
        \widetilde{\gamma}^\star(t,\BoldTs)
        \doteq 
        \expo{(\Boldomega - \lambda_s)\istar{t}\BoldTs}
        \frac{\lambda_s}{\Boldomega}
        \frac{\expo{\Boldomega (t - \istar{t}\BoldTs)} - 1}{\expo{\lambda_s \BoldTs} - 1}
        =
        \int_{\istar{t}\BoldTs}^{t}
            \expo{ \Boldomega \nu}
        d\nu
        \left( 
            \int_{\istar{t}\BoldTs}^{(\istar{t}+1)\BoldTs}
                \expo{\lambda_s \nu}
            d\nu
        \right)^{-1}
        .
    \end{align*}
\end{proposition}
\begin{proof}
    We have from~\eqref{eqn:prop:Appendix:TrueProcess:FL:AdaptationLaw:Matched:Final} that $\Lparahat{\nu} = 0_m$, for $\nu \in [0,\BoldTs)$, and hence
    \begin{align}\label{eqn:prop:Appendix:TrueProcess:ControlError:General:Total:[0,Ts)}
        \ControlErrorTilde[z,t]
        =&
        \int_0^t 
        \expo{ \Boldomega \nu}
        \left(  
            \Sigma^{\paral}_\nu(z)
            - 
            \Lparahat{\nu}d\nu
        \right)
        \notag 
        \\
        =&
        \int_0^t 
        \expo{ \Boldomega \nu} 
            \Sigma^{\paral}_\nu(z)
        =
        \int_{0}^{t}
            \expo{ \Boldomega \nu}
            \Lparamu{\nu,z(\nu)}
        d\nu
        +
        \int_{0}^{t}
            \expo{ \Boldomega \nu} 
            \Fparasigma{\nu,z(\nu)}
        d\Wt{\nu}
        , \quad \forall t \in [0,\BoldTs) \cap [0,\tau]
        .
    \end{align}
    Similarly, over the interval  $[i\BoldTs,(i+1)\BoldTs)$, $i \in \mathbb{N}_{\geq 1}$, the adaptive estimate is given by~\eqref{eqn:prop:Appendix:TrueProcess:FL:AdaptationLaw:Matched:Final} as 
    \begin{align*}
        \Lparahat{\nu} 
        =
        -
        \frac{\lambda_s}{1 - e^{\lambda_s \BoldTs}}
        \Theta_{ad}(i\BoldTs)
        \int_{(i-1)\BoldTs}^{i\BoldTs} \expo{-\lambda_s (i\BoldTs-\beta)} \Sigma_\beta(z)
        ,
    \end{align*}
    which implies that 
    \begin{align*}
        \int_{i\BoldTs}^{t}
            \expo{ \Boldomega \nu}
            \Lparahat{\nu}
        d\nu
        =
        -
        \frac{\lambda_s}{1 - e^{\lambda_s \BoldTs}}
        \Theta_{ad}(i\BoldTs)
        \left(
            \int_{i\BoldTs}^{t}
                \expo{ \Boldomega \nu}
            d\nu
        \right)
        \int_{(i-1)\BoldTs}^{i\BoldTs} \expo{-\lambda_s (i\BoldTs-\nu)} \Sigma_\nu(z)
        ,
    \end{align*}
    for all $(t,i) \in \left([i\BoldTs,(i+1)\BoldTs) \cap [0,\tau]\right) \times \mathbb{N}_{\geq 1}$, where note that we have replaced $\beta$ with $\nu$ as the integration variable in the last term. 
    Since 
    \begin{align*}
        -\frac{\lambda_s}{1 - e^{\lambda_s \BoldTs}}
        \int_{(i-1)\BoldTs}^{i\BoldTs}
            \expo{-\lambda_s \left(i \BoldTs - \nu \right)} 
            \Sigma_\nu(z)
        =&
        -\frac{ \lambda_s \expo{-\lambda_s i \BoldTs} }{1 - e^{\lambda_s \BoldTs}}
        \int_{(i-1)\BoldTs}^{i\BoldTs}
            \expo{ \lambda_s \nu } 
            \Sigma_\nu(z)
        \\
        =&
        \left( 
            \int_{i\BoldTs}^{(i+1)\BoldTs}
                \expo{\lambda_s \nu}
            d\nu
        \right)^{-1}
        \int_{(i-1)\BoldTs}^{i\BoldTs}
            \expo{ \lambda_s \nu } 
            \Sigma_\nu(z)
        ,
    \end{align*}
    the expression above the last can be written as 
    \begin{align*}
        \int_{i\BoldTs}^{t}
            \expo{ \Boldomega \nu}
            \Lparahat{\nu}
        d\nu
        =
        \frac{
            \int_{i\BoldTs}^{t}
                \expo{ \Boldomega \nu}
            d\nu
        }{
            \int_{i\BoldTs}^{(i+1)\BoldTs}
                \expo{\lambda_s \nu}
            d\nu
        }
        \Theta_{ad}(i\BoldTs)
        \int_{(i-1)\BoldTs}^{i\BoldTs} \expo{ \lambda_s \nu} \Sigma_\nu(z)
        ,
        \quad 
        \forall 
        (t,i) \in \left([i\BoldTs,(i+1)\BoldTs) \cap [0,\tau]\right) \times \mathbb{N}_{\geq 1}
        .
    \end{align*}
    It then follows from~\eqref{eqn:prop:Appendix:Scratch:Technical:Diffusion:Total:A}, Proposition~\ref{proposition:Appendix:Scratch:Technical}, that
    \begin{multline*}
        \int_{i\BoldTs}^{t}
        \expo{ \Boldomega \nu}
        \Lparahat{\nu}
        d\nu 
        =
        \frac{
            \int_{i\BoldTs}^{t}
                \expo{ \Boldomega \nu}
            d\nu
        }{
            \int_{i\BoldTs}^{(i+1)\BoldTs}
                \expo{\lambda_s \nu}
            d\nu
        }
        \Theta_{ad}(i\BoldTs)
        \int_{(i-1)\BoldTs}^{i\BoldTs}
            \expo{ \lambda_s \nu}
            \SigmaError{\nu}{i\BoldTs,z}   
        \\
        +
        \frac{
            \int_{i\BoldTs}^{t}
                \expo{ \Boldomega \nu}
            d\nu
        }{
            \int_{i\BoldTs}^{(i+1)\BoldTs}
                \expo{\lambda_s \nu}
            d\nu
        }
        \int_{(i-1)\BoldTs}^{i\BoldTs}
            \expo{ \lambda_s \nu}
            \Lparamu{i\BoldTs,z\br{i\BoldTs}}
        d\nu
        \\
        +
        \frac{
            \int_{i\BoldTs}^{t}
                \expo{ \Boldomega \nu}
            d\nu
        }{
            \int_{i\BoldTs}^{(i+1)\BoldTs}
                \expo{\lambda_s \nu}
            d\nu
        }
        \int_{(i-1)\BoldTs}^{i\BoldTs}
            \expo{ \lambda_s \nu} 
            \Fparasigma{i\BoldTs,z\br{i\BoldTs}}
        d\Wt{\nu}       
        ,
    \end{multline*}
    for all $(t,i) \in \left([i\BoldTs,(i+1)\BoldTs) \cap [0,\tau]\right) \times \mathbb{N}_{\geq 1}$.
    Additionally, we have from~\eqref{eqn:prop:Appendix:Scratch:Technical:Total:Identity:B}, Proposition~\ref{proposition:Appendix:Scratch:Technical}, that 
    \begin{align*}
        \int_{i\BoldTs}^{t} 
            \expo{\Boldomega \nu}
            \Sigma^{\paral}_\nu(z) 
        = 
        \int_{i\BoldTs}^{t} 
            \expo{\Boldomega \nu}
            \SigmaparaError{\nu}{i\BoldTs,z}
        + 
        \int_{i\BoldTs}^{t} 
            \expo{\Boldomega \nu} 
            \left[
                \Lparamu{i\BoldTs,z\br{i\BoldTs}}
                d\nu
                + 
                \Fparasigma{i\BoldTs,z\br{i\BoldTs}}
                d\Wt{\nu}       
            \right]
        ,
    \end{align*}
    for all $(t,i) \in \left([i\BoldTs,(i+1)\BoldTs) \cap [0,\tau]\right) \times \mathbb{N}_{\geq 1}$.
    Hence, we use the two expressions above and conclude that 
    \begin{multline}\label{eqn:prop:Appendix:TrueProcess:ControlError:General:Total:[iTs,(i+1))}
        \ControlErrorTilde[z,t]
        =
        \int_{i\BoldTs}^{t} 
            \expo{\Boldomega \nu}
            \SigmaparaError{\nu}{i\BoldTs,z}
        -
        \frac{
            \int_{i\BoldTs}^{t}
                \expo{ \Boldomega \nu}
            d\nu
        }{
            \int_{i\BoldTs}^{(i+1)\BoldTs}
                \expo{\lambda_s \nu}
            d\nu
        }
        \Theta_{ad}(i\BoldTs)
        \int_{(i-1)\BoldTs}^{i\BoldTs}
            \expo{ \lambda_s \nu}
            \SigmaError{\nu}{i\BoldTs,z}   
        \\
        + 
        \left(
            \int_{i\BoldTs}^{t} 
                \expo{\Boldomega \nu} 
            d\nu
            -
            \frac{
                \int_{i\BoldTs}^{t}
                    \expo{ \Boldomega \nu}
                d\nu
            }{
                \int_{i\BoldTs}^{(i+1)\BoldTs}
                    \expo{\lambda_s \nu}
                d\nu
            }
            \int_{(i-1)\BoldTs}^{i\BoldTs}
                \expo{ \lambda_s \nu}
            d\nu
        \right)
        \Lparamu{i\BoldTs,z\br{i\BoldTs}}
        \\
        + 
        \Fparasigma{i\BoldTs,z\br{i\BoldTs}}
        \left(
            \int_{i\BoldTs}^{t} 
                \expo{\Boldomega \nu}  
            d\Wt{\nu}       
            -
            \frac{
                \int_{i\BoldTs}^{t}
                    \expo{ \Boldomega \nu}
                d\nu
            }{
                \int_{i\BoldTs}^{(i+1)\BoldTs}
                    \expo{\lambda_s \nu}
                d\nu
            }
            \int_{(i-1)\BoldTs}^{i\BoldTs}
                \expo{ \lambda_s \nu} 
            d\Wt{\nu}
        \right)
        ,
    \end{multline}
    for all $(t,i) \in \left([i\BoldTs,(i+1)\BoldTs) \cap [0,\tau]\right) \times \mathbb{N}_{\geq 1}$.

    Using~\eqref{eqn:prop:Appendix:TrueProcess:ControlError:General:Total:[0,Ts)} and~\eqref{eqn:prop:Appendix:TrueProcess:ControlError:General:Total:[iTs,(i+1))}, we can write $\ControlErrorTilde[z,t]$ as
    \begin{multline}\label{eqn:prop:Appendix:TrueProcess:ControlError:General:Total:[0,tau]}
        \ControlErrorTilde[z,t]
        =
        \indicator{<\BoldTs}{t}
            \int_0^t 
                \expo{ \Boldomega \nu} 
                \Sigma^{\paral}_\nu(z)
        +
        \indicator{\geq \BoldTs}{t}
        \int_{0}^{\BoldTs}
            \expo{ \Boldomega \nu}
            \Lparamu{\nu,z(\nu)}
        d\nu
        \\
        + 
        \indicator{\geq \BoldTs}{t}
        \left(
            \widetilde{\mathcal{F}}_\circ\left(z,t\right)
            + 
            \ControlErrorHatDrift[z,t]
            +
            \ControlErrorHatDiffusion[z,t]
        \right)
        ,
        \quad 
        t \in [0,\tau],
    \end{multline}
    where we have defined the following for $t \in [\BoldTs,\infty) \cap [0,\tau]$:
    \begin{align*}
        \widetilde{\mathcal{F}}_\circ\left(z,t\right)
        =&
        \sum_{i=1}^{\istar{t} - 1}
        \left(
            \int_{i\BoldTs}^{(i+1)\BoldTs} 
                \expo{\Boldomega \nu}
                \SigmaparaError{\nu}{i\BoldTs,z}
            -
            \widetilde{\gamma}_i(\BoldTs)
            \Theta_{ad}(i\BoldTs)
            \int_{(i-1)\BoldTs}^{i\BoldTs}
                \expo{ \lambda_s \nu}
                \SigmaError{\nu}{i\BoldTs,z}   
        \right)
        \\
        &+ 
        \int_{\istar{t}\BoldTs}^{t} 
            \expo{\Boldomega \nu}
            \SigmaparaError{\nu}{\istar{t}\BoldTs,z}
        -
        \widetilde{\gamma}^\star(t,\BoldTs)
        \Theta_{ad}(\istar{t}\BoldTs)
        \int_{(\istar{t}-1)\BoldTs}^{\istar{t}\BoldTs}
            \expo{ \lambda_s \nu}
            \SigmaError{\nu}{\istar{t}\BoldTs,z}   
        ,
        \\
        \ControlErrorHatDrift[z,t]
        =&
        \sum_{i=1}^{\istar{t} - 1}
        \left(
            \int_{i\BoldTs}^{(i+1)\BoldTs} 
                \expo{\Boldomega \nu} 
            d\nu
            -
            \widetilde{\gamma}_i(\BoldTs)
            \int_{(i-1)\BoldTs}^{i\BoldTs}
                \expo{ \lambda_s \nu}
            d\nu
        \right)
        \Lparamu{i\BoldTs,z\br{i\BoldTs}}
        \\
        &+ 
        \left(
            \int_{\istar{t}\BoldTs}^{t} 
                \expo{\Boldomega \nu} 
            d\nu
            -
            \widetilde{\gamma}^\star(t,\BoldTs)
            \int_{(\istar{t}-1)\BoldTs}^{\istar{t}\BoldTs}
                \expo{ \lambda_s \nu}
            d\nu
        \right)
        \Lparamu{\istar{t}\BoldTs,z\br{\istar{t}\BoldTs}}
        , 
        \\
        \ControlErrorHatDiffusion[z,t]
        =&
        \int_{0}^{\BoldTs}
            \expo{ \Boldomega \nu} 
            \Fparasigma{\nu,z(\nu)}
        d\Wt{\nu}
        + 
        \sum_{i=1}^{\istar{t} - 1}
        \Fparasigma{i\BoldTs,z\br{i\BoldTs}}
        \left(
            \int_{i\BoldTs}^{(i+1)\BoldTs} 
                \expo{\Boldomega \nu}  
            d\Wt{\nu}       
            -
            \widetilde{\gamma}_i(\BoldTs)
            \int_{(i-1)\BoldTs}^{i\BoldTs}
                \expo{ \lambda_s \nu} 
            d\Wt{\nu}
        \right)
        \\
        &+ 
        \Fparasigma{\istar{t}\BoldTs,z\br{\istar{t}\BoldTs}}
        \left(
            \int_{\istar{t}\BoldTs}^{t} 
                \expo{\Boldomega \nu}  
            d\Wt{\nu}       
            -
            \widetilde{\gamma}^\star(t,\BoldTs)
            \int_{(\istar{t}-1)\BoldTs}^{\istar{t}\BoldTs}
                \expo{ \lambda_s \nu} 
            d\Wt{\nu}
        \right)
        .
    \end{align*}

    Now, we re-write $\widetilde{\mathcal{F}}_\circ$ as 
    \begin{multline*}
        \widetilde{\mathcal{F}}_\circ\left(z,t\right)
        =
        \int_{\BoldTs}^{\istar{t}\BoldTs}
            \expo{\Boldomega \nu}
            \left(
                \sum_{i=1}^{\istar{t} - 1}
                \indicator{[i\BoldTs,(i+1)\BoldTs)}{\nu} 
                \SigmaparaError{\nu}{i\BoldTs,z}
            \right)
        + 
        \int_{\istar{t}\BoldTs}^{t} 
            \expo{\Boldomega \nu}
            \SigmaparaError{\nu}{\istar{t}\BoldTs,z}   
        \\
        -
        \int_{0}^{(\istar{t}-1)\BoldTs}    
            \expo{\lambda_s \nu}
            \left(
                \sum_{i=1}^{\istar{t} - 1}
                \indicator{[(i-1)\BoldTs,i\BoldTs)}{\nu}
                \widetilde{\gamma}_i(\BoldTs)
                \Theta_{ad}(i\BoldTs)
                \SigmaError{\nu}{i\BoldTs,z}
            \right)
        \\
        -
        \int_{(\istar{t}-1)\BoldTs}^{\istar{t}\BoldTs}
        \expo{ \lambda_s \nu}
                \widetilde{\gamma}^\star(t,\BoldTs)
                \Theta_{ad}(\istar{t}\BoldTs)
                \SigmaError{\nu}{\istar{t}\BoldTs,z} 
            ,
    \end{multline*}
    for $t \in [\BoldTs,\infty) \cap [0,\tau]$.
    Thus, using the definition of $\SigmaErrorAll{t}{i\BoldTs,z}$ in~\eqref{eqn:prop:Appendix:Scratch:Technical:Total:TildeFunctions}, we can further express $\widetilde{\mathcal{F}}_\circ$ as
    \begin{equation}\label{eqn:prop:Appendix:TrueProcess:ControlError:Initial:[0,tau]:Final}
        \widetilde{\mathcal{F}}_\circ\left(z,t\right)
        =
        \int_{0}^{t}
            \expo{\Boldomega \nu}
            \left[
                \left(
                    \widetilde{\mathcal{H}}_\mu^{\paral}(\nu,z)
                    + 
                    \expo{(\lambda_s-\Boldomega) \nu}
                    \widetilde{\mathcal{H}}_\mu(\nu,z)
                \right)
                d\nu
                + 
                \left(
                    \widetilde{\mathcal{H}}_\sigma^{\paral}(\nu,z)
                    + 
                    \expo{(\lambda_s-\Boldomega) \nu}
                    \widetilde{\mathcal{H}}_\sigma(\nu,z)
                \right)
                d\Wt{\nu}
            \right]
        ,
    \end{equation}    
    for $t \in [\BoldTs,\infty) \cap [0,\tau]$.

    We next manipulate the terms $\ControlErrorHatDrift$ and $\ControlErrorHatDiffusion$, and start by observing that 
    \begin{align*}
        \int_{i\BoldTs}^{(i+1)\BoldTs} 
            \expo{\Boldomega \nu} 
        d\nu
        -
        \widetilde{\gamma}_i(\BoldTs)
        \int_{(i-1)\BoldTs}^{i\BoldTs}
            \expo{ \lambda_s \nu}
        d\nu
        =&
        \left(
        \int_{i\BoldTs}^{(i+1)\BoldTs} 
            \expo{\Boldomega \nu} 
        d\nu
        \right)
        \left(
            1
            -
            \left(
                \int_{i\BoldTs}^{(i+1)\BoldTs}
                    \expo{ \lambda_s \nu}
                d\nu
            \right)^{-1}
            \int_{(i-1)\BoldTs}^{i\BoldTs}
                \expo{ \lambda_s \nu}
            d\nu
        \right)
        \\
        =&
        \left(
            \int_{i\BoldTs}^{(i+1)\BoldTs} 
                \expo{\Boldomega \nu} 
            d\nu
        \right)
        \left(1 - \expo{-\lambda_s \BoldTs}\right)
        ,
    \end{align*} 
    and, similarly 
    \begin{align*}
        \int_{\istar{t}\BoldTs}^{t} 
            \expo{\Boldomega \nu} 
        d\nu
        -
        \widetilde{\gamma}^\star(t,\BoldTs)
        \int_{(\istar{t}-1)\BoldTs}^{\istar{t}\BoldTs}
            \expo{ \lambda_s \nu}
        d\nu
        =
        \left(
            \int_{\istar{t}\BoldTs}^{t} 
                \expo{\Boldomega \nu} 
            d\nu
        \right)
        \left(1 - \expo{-\lambda_s \BoldTs}\right)
        .
    \end{align*}
    Hence, $\ControlErrorHatDrift$, defined in~\eqref{eqn:prop:Appendix:TrueProcess:ControlError:General:Total:[0,tau]}, can be written as
    \begin{multline*}
        \ControlErrorHatDrift[z,t]
        =
        \sum_{i=1}^{\istar{t} - 1}
        \left(
            \int_{i\BoldTs}^{(i+1)\BoldTs} 
                \expo{\Boldomega \nu} 
            d\nu
        \right)
        \left(1 - \expo{-\lambda_s \BoldTs}\right)
        \Lparamu{i\BoldTs,z\br{i\BoldTs}}
        \\
        + 
        \left(
            \int_{\istar{t}\BoldTs}^{t} 
                \expo{\Boldomega \nu} 
            d\nu
        \right)
        \left(1 - \expo{-\lambda_s \BoldTs}\right)
        \Lparamu{\istar{t}\BoldTs,z\br{\istar{t}\BoldTs}}
        , 
    \end{multline*}
    for $t \in [\BoldTs,\infty) \cap [0,\tau]$, which we can further write as 
    \begin{equation}\label{eqn:prop:Appendix:TrueProcess:ControlError:Drift:[0,tau]:Final}
        \ControlErrorHatDrift[z,t]
        =
        \int_0^t 
            \expo{\Boldomega \nu}
            \mathcal{G}_\mu(\nu,z)
        d\nu
        ,
        \quad
        t \in [\BoldTs,\infty) \cap [0,\tau]
        .
    \end{equation}

    Next, by adding and subtracting $\Fparasigma{0,z(0)}$, we obtain 
    \begin{align}\label{eqn:prop:Appendix:TrueProcess:ControlError:Diffusion:Identity:A}
        \int_{0}^{\BoldTs}
            \expo{ \Boldomega \nu} 
            \Fparasigma{\nu,z(\nu)}
        d\Wt{\nu}
        =& 
        \int_{0}^{\BoldTs}
            \expo{ \Boldomega \nu} 
            \left(\Fparasigma{\nu,z(\nu)} - \Fparasigma{0,z(0)}\right)
        d\Wt{\nu}
        + 
        \Fparasigma{0,z(0)}
        \int_{0}^{\BoldTs}
            \expo{ \Boldomega \nu} 
        d\Wt{\nu}
        \notag 
        \\
        =& 
        \int_{0}^{\BoldTs}
            \expo{ \Boldomega \nu} 
            \FparasigmaError{}{\nu,0,z}
        d\Wt{\nu}
        + 
        \Fparasigma{0,z(0)}
        \int_{0}^{\BoldTs}
            \expo{ \Boldomega \nu} 
        d\Wt{\nu}
        .
    \end{align}
    Furthermore, we obtain the following by adding and subtracting $\Fparasigma{(i-1)\BoldTs,z((i-1)\BoldTs)}$: 
    \begin{align*}
        \Fparasigma{i\BoldTs,z\br{i\BoldTs}}
        =&
        \left(
            \Fparasigma{i\BoldTs,z\br{i\BoldTs}}
            - 
            \Fparasigma{(i-1)\BoldTs,z((i-1)\BoldTs)}
        \right)
        +
        \Fparasigma{(i-1)\BoldTs,z((i-1)\BoldTs)}
        \\
        =&
        \FparasigmaError{}{i\BoldTs,(i-1)\BoldTs,z}
        +
        \Fparasigma{(i-1)\BoldTs,z((i-1)\BoldTs)}
        .
    \end{align*}
    Using this expression, we may write
    \begin{multline*}
        \Fparasigma{i\BoldTs,z\br{i\BoldTs}}
        \widetilde{\gamma}_i(\BoldTs)
        \int_{(i-1)\BoldTs}^{i\BoldTs}
            \expo{ \lambda_s \nu} 
        d\Wt{\nu}
        =
        \FparasigmaError{}{i\BoldTs,(i-1)\BoldTs,z}
        \widetilde{\gamma}_i(\BoldTs)
        \int_{(i-1)\BoldTs}^{i\BoldTs}
            \expo{ \lambda_s \nu} 
        d\Wt{\nu}
        \\
        +
        \Fparasigma{(i-1)\BoldTs,z((i-1)\BoldTs)}
        \widetilde{\gamma}_i(\BoldTs)
        \int_{(i-1)\BoldTs}^{i\BoldTs}
            \expo{ \lambda_s \nu} 
        d\Wt{\nu}
        ,
    \end{multline*}
    which allows us to further write
    \begin{multline}\label{eqn:prop:Appendix:TrueProcess:ControlError:Diffusion:Identity:B}
        \begin{aligned}
            &\sum_{i=1}^{\istar{t} - 1}
            \Fparasigma{i\BoldTs,z\br{i\BoldTs}}
            \left(
                \int_{i\BoldTs}^{(i+1)\BoldTs} 
                    \expo{\Boldomega \nu}  
                d\Wt{\nu}       
                -
                \widetilde{\gamma}_i(\BoldTs)
                \int_{(i-1)\BoldTs}^{i\BoldTs}
                    \expo{ \lambda_s \nu} 
                d\Wt{\nu}
            \right)
            \\
            & =
            \sum_{i=1}^{\istar{t} - 1}
            \left(
                \Fparasigma{i\BoldTs,z\br{i\BoldTs}}
                \int_{i\BoldTs}^{(i+1)\BoldTs} 
                    \expo{\Boldomega \nu}  
                d\Wt{\nu}       
                -
                \Fparasigma{i\BoldTs,z\br{i\BoldTs}}
                \widetilde{\gamma}_i(\BoldTs)
                \int_{(i-1)\BoldTs}^{i\BoldTs}
                    \expo{ \lambda_s \nu} 
                d\Wt{\nu}
            \right)
            \\
            & = 
            \sum_{i=1}^{\istar{t} - 1}
            \left(
                \Fparasigma{i\BoldTs,z\br{i\BoldTs}}
                \int_{i\BoldTs}^{(i+1)\BoldTs} 
                    \expo{\Boldomega \nu}  
                d\Wt{\nu}       
                -
                \Fparasigma{(i-1)\BoldTs,z((i-1)\BoldTs)}
                \widetilde{\gamma}_i(\BoldTs)
                \int_{(i-1)\BoldTs}^{i\BoldTs}
                    \expo{ \lambda_s \nu} 
                d\Wt{\nu}
            \right)
        \end{aligned}
        \\
        -
        \sum_{i=1}^{\istar{t} - 1}       
        \FparasigmaError{}{i\BoldTs,(i-1)\BoldTs,z}
        \widetilde{\gamma}_i(\BoldTs)
        \int_{(i-1)\BoldTs}^{i\BoldTs}
            \expo{ \lambda_s \nu} 
        d\Wt{\nu}
        .
    \end{multline}
    Then, adding~\eqref{eqn:prop:Appendix:TrueProcess:ControlError:Diffusion:Identity:A} and~\eqref{eqn:prop:Appendix:TrueProcess:ControlError:Diffusion:Identity:B}, we obtain
    \begin{multline}\label{eqn:prop:Appendix:TrueProcess:ControlError:Diffusion:Identity:A+B}
        \begin{aligned}
            &\int_{0}^{\BoldTs}
                \expo{ \Boldomega \nu} 
                \Fparasigma{\nu,z(\nu)}
            d\Wt{\nu}
            +
            \sum_{i=1}^{\istar{t} - 1}
            \Fparasigma{i\BoldTs,z\br{i\BoldTs}}
            \left(
                \int_{i\BoldTs}^{(i+1)\BoldTs} 
                    \expo{\Boldomega \nu}  
                d\Wt{\nu}       
                -
                \widetilde{\gamma}_i(\BoldTs)
                \int_{(i-1)\BoldTs}^{i\BoldTs}
                    \expo{ \lambda_s \nu} 
                d\Wt{\nu}
            \right)
            \\
            & 
            =
            \int_{0}^{\BoldTs}
                \expo{ \Boldomega \nu} 
                \FparasigmaError{}{\nu,0,z}
            d\Wt{\nu}
            -
            \sum_{i=1}^{\istar{t} - 1}       
            \FparasigmaError{}{i\BoldTs,(i-1)\BoldTs,z}
            \widetilde{\gamma}_i(\BoldTs)
            \int_{(i-1)\BoldTs}^{i\BoldTs}
                \expo{ \lambda_s \nu} 
            d\Wt{\nu}
        \end{aligned}
        \\
        +
        \sum_{i=1}^{\istar{t} - 1}
        \Fparasigma{(i-1)\BoldTs,z\br{(i-1)\BoldTs}}
        \left(
            \int_{(i-1)\BoldTs}^{i\BoldTs}
                \expo{\Boldomega \nu} 
                \left[
                    1  
                    -
                    \widetilde{\gamma}_i(\BoldTs)
                    \expo{ \left(\lambda_s - \Boldomega\right) \nu}
                \right] 
            d\Wt{\nu}
        \right)
        \\
        +
        \Fparasigma{(\istar{t} - 1)\BoldTs,z\br{(\istar{t} - 1)\BoldTs}}
        \int_{(\istar{t} - 1)\BoldTs}^{\istar{t}\BoldTs} 
            \expo{\Boldomega \nu}  
        d\Wt{\nu}
        ,
    \end{multline}
    where we have used the manipulation 
    \begin{multline*}
        \Fparasigma{0,z(0)}
        \int_{0}^{\BoldTs}
            \expo{ \Boldomega \nu} 
        d\Wt{\nu}
        +
        \sum_{i=1}^{\istar{t} - 1}
        \Fparasigma{i\BoldTs,z\br{i\BoldTs}}
        \int_{i\BoldTs}^{(i+1)\BoldTs} 
            \expo{\Boldomega \nu}  
        d\Wt{\nu}
        \\
        =
        \sum_{i=1}^{\istar{t} - 1}
        \Fparasigma{(i-1)\BoldTs,z\br{(i-1)\BoldTs}}
        \int_{(i-1)\BoldTs}^{i\BoldTs} 
            \expo{\Boldomega \nu}  
        d\Wt{\nu}
        \\
        +
        \Fparasigma{(\istar{t} - 1)\BoldTs,z\br{(\istar{t} - 1)\BoldTs}}
        \int_{(\istar{t} - 1)\BoldTs}^{\istar{t}\BoldTs} 
            \expo{\Boldomega \nu}  
        d\Wt{\nu}
        ,  
    \end{multline*}
    which allows us to write
    \begin{align*}
        &
        \begin{multlined}[][0.75\linewidth]
            \Fparasigma{0,z(0)}
            \int_{0}^{\BoldTs}
                \expo{ \Boldomega \nu} 
            d\Wt{\nu}
            +
            \sum_{i=1}^{\istar{t} - 1}
            \Fparasigma{i\BoldTs,z\br{i\BoldTs}}
            \int_{i\BoldTs}^{(i+1)\BoldTs} 
                \expo{\Boldomega \nu}  
            d\Wt{\nu}       
            \\
            -
            \sum_{i=1}^{\istar{t} - 1}
            \Fparasigma{(i-1)\BoldTs,z((i-1)\BoldTs)}
            \widetilde{\gamma}_i(\BoldTs)
            \int_{(i-1)\BoldTs}^{i\BoldTs}
                \expo{ \lambda_s \nu} 
            d\Wt{\nu}
        \end{multlined}
        \\
        &=
        \begin{multlined}[t][0.75\linewidth]
            \sum_{i=1}^{\istar{t} - 1}
            \Fparasigma{(i-1)\BoldTs,z\br{(i-1)\BoldTs}}
            \left(
                \int_{(i-1)\BoldTs}^{i\BoldTs}
                    \expo{\Boldomega \nu} 
                    \left[
                        1  
                        -
                        \widetilde{\gamma}_i(\BoldTs)
                        \expo{ \left(\lambda_s - \Boldomega\right) \nu}
                    \right] 
                d\Wt{\nu}
            \right)
            \\
            +
            \Fparasigma{(\istar{t} - 1)\BoldTs,z\br{(\istar{t} - 1)\BoldTs}}
            \int_{(\istar{t} - 1)\BoldTs}^{\istar{t}\BoldTs} 
                \expo{\Boldomega \nu}  
            d\Wt{\nu}.
        \end{multlined}
    \end{align*}
    Substituting~\eqref{eqn:prop:Appendix:TrueProcess:ControlError:Diffusion:Identity:A+B} for the first two terms in the definition of $\ControlErrorHatDiffusion$ in~\eqref{eqn:prop:Appendix:TrueProcess:ControlError:General:Total:[0,tau]}, we obtain
    \begin{multline*}
        \begin{aligned}
            \ControlErrorHatDiffusion[z,t]
            =&
            \int_{0}^{\BoldTs}
                \expo{ \Boldomega \nu} 
                \FparasigmaError{}{\nu,0,z}
            d\Wt{\nu}
            -
            \sum_{i=1}^{\istar{t} - 1}       
            \FparasigmaError{}{i\BoldTs,(i-1)\BoldTs,z}
            \widetilde{\gamma}_i(\BoldTs)
            \int_{(i-1)\BoldTs}^{i\BoldTs}
                \expo{ \lambda_s \nu} 
            d\Wt{\nu} 
            \\
            &+
            \sum_{i=1}^{\istar{t} - 1}
            \Fparasigma{(i-1)\BoldTs,z\br{(i-1)\BoldTs}}
            \left(
                \int_{(i-1)\BoldTs}^{i\BoldTs}
                    \expo{\Boldomega \nu} 
                    \left[
                        1  
                        -
                        \widetilde{\gamma}_i(\BoldTs)
                        \expo{ \left(\lambda_s - \Boldomega\right) \nu}
                    \right] 
                d\Wt{\nu}
            \right) 
            \\
            &+
            \Fparasigma{(\istar{t} - 1)\BoldTs,z\br{(\istar{t} - 1)\BoldTs}}
            \int_{(\istar{t} - 1)\BoldTs}^{\istar{t}\BoldTs} 
                \expo{\Boldomega \nu}  
            d\Wt{\nu}
        \end{aligned} 
        \\
        + 
        \Fparasigma{\istar{t}\BoldTs,z\br{\istar{t}\BoldTs}}
        \left(
            \int_{\istar{t}\BoldTs}^{t} 
                \expo{\Boldomega \nu}  
            d\Wt{\nu}       
            -
            \widetilde{\gamma}^\star(t,\BoldTs)
            \int_{(\istar{t}-1)\BoldTs}^{\istar{t}\BoldTs}
                \expo{ \lambda_s \nu} 
            d\Wt{\nu}
        \right)
        , 
    \end{multline*}
    for $t \in [\BoldTs,\infty) \cap [0,\tau]$.
    We can therefore re-write the previous expression as
    \begin{multline}\label{eqn:prop:Appendix:TrueProcess:ControlError:Diffusion:[0,tau]:Final}
        \ControlErrorHatDiffusion[z,t]
        =
        \int_{0}^{t}
            \expo{\Boldomega \nu} 
            \left( 
                \widetilde{\mathcal{G}}_{\sigma_1}(\nu,z)
                + 
                \expo{(\lambda_s - \Boldomega) \nu} 
                \widetilde{\mathcal{G}}_{\sigma_2}(\nu,z)
            \right)
        d\Wt{\nu}
        \\
        +
        \int_0^t 
            \expo{\Boldomega \nu}
            \left( 
                \mathcal{G}_{\sigma_1}(\nu,z)
                + 
                \mathcal{G}_{\sigma_2}(\nu,z)
            \right)
        d\Wt{\nu}
        ,
        \quad 
        t \in [\BoldTs,\infty) \cap [0,\tau]
        .
    \end{multline}

    Adding the derived expressions for $\widetilde{\mathcal{F}}_\circ$, $\ControlErrorHatDrift$, and $\ControlErrorHatDiffusion$ in~\eqref{eqn:prop:Appendix:TrueProcess:ControlError:Initial:[0,tau]:Final},~\eqref{eqn:prop:Appendix:TrueProcess:ControlError:Drift:[0,tau]:Final} and~\eqref{eqn:prop:Appendix:TrueProcess:ControlError:Diffusion:[0,tau]:Final}, respectively, leads to   
    \begin{multline*}
        \widetilde{\mathcal{F}}_\circ\left(z,t\right)
        + 
        \ControlErrorHatDrift[z,t]
        +
        \ControlErrorHatDiffusion[z,t]
        \\
        =
        \int_{0}^{t}
            \expo{\Boldomega \nu}
            \left[
                \left(
                    \widetilde{\mathcal{H}}_\mu^{\paral}(\nu,z)
                    + 
                    \expo{(\lambda_s-\Boldomega) \nu}
                    \widetilde{\mathcal{H}}_\mu(\nu,z)
                \right)
                d\nu
                + 
                \left(
                    \widetilde{\mathcal{H}}_\sigma^{\paral}(\nu,z)
                    + 
                    \expo{(\lambda_s-\Boldomega) \nu}
                    \widetilde{\mathcal{H}}_\sigma(\nu,z)
                \right)
                d\Wt{\nu}
            \right]
        \\
        +
        \int_{0}^{t}
            \expo{\Boldomega \nu} 
            \left( 
                \widetilde{\mathcal{G}}_{\sigma_1}(\nu,z)
                + 
                \expo{(\lambda_s - \Boldomega) \nu} 
                \widetilde{\mathcal{G}}_{\sigma_2}(\nu,z)
            \right)
        d\Wt{\nu}
        \\
        +
        \int_0^t 
            \expo{\Boldomega \nu}
            \left[
            \mathcal{G}_\mu(\nu,z)
            d\nu
            + 
            \left( 
                \mathcal{G}_{\sigma_1}(\nu,z)
                + 
                \mathcal{G}_{\sigma_2}(\nu,z)
            \right)
            d\Wt{\nu}
            \right]
        ,
        \quad 
        t \in [\BoldTs,\infty) \cap [0,\tau]
        .
    \end{multline*}   
    Substituting the above into~\eqref{eqn:prop:Appendix:TrueProcess:ControlError:General:Total:[0,tau]} gives us the desired result in~\eqref{eqn:prop:Appendix:TrueProcess:ControlError:General:Total:[0,tau]:MAIN}.
\end{proof}

Next, we provide the proof of Proposition~\ref{prop:True:TruncatedWellPosedness}.
\begin{proof}[Proof of Proposition~\ref{prop:True:TruncatedWellPosedness}]
    It follows from Definitions~\ref{def:True:JointProcess} and~\ref{def:True:TruncatedJointProcess} that we need to show the well posedness of 
    \begin{align}\label{eqn:True:TruncatedWellPosedness:System:2}
        d\Zt{N,t}
        =&
        d\Xt{N,t} - d\Xrt{N,t} 
        \notag 
        \\
        =& 
        \left[\FNmu{t,\Xt{N,t},\ULt{t}} - \FNmu{t,\Xrt{N,t},\Urt{t}}\right]dt
        +
        \left[\FNsigma{t,\Xt{N,t}} - \FNsigma{t,\Xrt{N,t}}\right]d\Wt{t}
        \notag 
        \\
        =& 
        \left[\FNmu{t,\Xt{N,t},\ULt{t}} - \FNmu{t,\Xrt{N,t},\Urt{t}}\right]dt
        +
        \FNZsigma{t,\Zt{N,t}}d\Wt{t}
        , 
    \end{align}
    for $t \in [0,T]$, where $\Zt{N,0} = 0$, $U_{\mathcal{L}_1} = \FL[\Xt{N}]$, $\Urt{} = \ReferenceInput[\Xrt{N}]$, the functions $F_{N,\cbr{\mu,\sigma}}$ are defined analogously to $J_{N,\cbr{\mu,\sigma}}$ in~\eqref{eqn:True:TruncatedJointProcess}, and we have set 
    \begin{align*}
        \FNZsigma{t,\Zt{N,t}} \doteq
        \FNsigma{t,\Xt{N,t}} - \FNsigma{t,\Xrt{N,t}}.
    \end{align*}  
    Let us denote by $f_N$, $\Lambda_{N,\mu}$, $p_N$, and $\Lambda_{N,\sigma}$  the truncated versions of the functions $f$, $\Lambda_{\mu}$, $p$, and $\Lambda_{\sigma}$, respectively, and where the truncation is defined as in~\eqref{eqn:True:TruncatedJointProcess}.
    Then, by adding and subtracting $g(t)\ReferenceInput[\Xt{N}][t]$, along with Definition~\ref{def:VectorFields}, we get 
    \begin{align*}
        \FNmu{t,\Xt{N,t},\ULt{t}} - \FNmu{t,\Xrt{N,t},\Urt{t}}
        =& 
        \FNmu{t,\Xt{N,t},\FL[\Xt{N}][t]} - \FNmu{t,\Xrt{N,t},\ReferenceInput[\Xrt{N}][t]}
        \\
        =& 
        \FNZmu{t,\Zt{N,t}} + g(t)\left[\left( \FL - \ReferenceInput \right)\left(\Xt{N,t}\right)\right],
    \end{align*}  
    where we have defined 
    \begin{align*}
        \FNZmu{t,\Zt{N,t}} 
        \doteq 
        \FNmu{t,\Xt{N,t},\ReferenceInput[\Xt{N}][t]} 
        - \FNmu{t,\Xrt{N,t},\ReferenceInput[\Xrt{N}][t]}.
    \end{align*}
    Substituting the above into~\eqref{eqn:True:TruncatedWellPosedness:System:2}
    \begin{align*}
        d\Zt{N,t}
        = 
        \FNZmu{t,\Zt{N,t}}dt
        +
        \FNZsigma{t,\Zt{N,t}}d\Wt{t}
        +
        g(t)\left[\left( \FL - \ReferenceInput \right)\left(\Xt{N,t}\right)\right]dt
        , 
        \quad 
        \Zt{N,0} = 0,
    \end{align*}
    for $t \in [0,T]$.
    Now, consider any $z \in \mathcal{M}_2\br{[0,T],\mathbb{R}^n~|~\Wfilt{0,t}}$, for  any $t \in [0,T]$, and  define
    \begin{align*}
        M(z(t)) = M_1(z(t)) + M_2(z(t)), \quad t \in [0,T], 
    \end{align*}
    where 
    \begin{align*}
        M_1(z(t)) 
        =& 
        \int_0^t
            \FNZmu{\nu,\Zt{N,\nu}}
        d\nu
        +
        \int_0^\nu
            \FNZsigma{\nu,\Zt{N,\nu}}
        d\Wt{\nu}, 
        \\
        M_2(z(t)) 
        =&
        \int_0^t
            g(\nu)\left[\left( \FL - \ReferenceInput \right)\left(\Xt{N,\nu}\right)\right]
        d\nu.
    \end{align*}
    We can then conclude the well-posedness by following the proof of Proposition~\ref{prop:Reference:TruncatedWellPosedness} for $M_1(z(t))$, and deriving similar conclusions for $M_2(z(t))$ by invoking Proposition~\ref{prop:Appendix:TrueProcess:FL:Expression}.

\end{proof}

The next two results help us with the computation of the stochastic differential of $\ZNorm{\Zt{N,t}}$ in the proof of Lemma~\ref{lem:True:dV}. 
\begin{proposition}\label{prop:Appendix:TrueProcess:dV}
    Let $\Zt{N,t}$ be the strong solution of~\eqref{eqn:True:TruncatedJointProcess}, and let $\tau(t)$ be the stopping time defined in~\eqref{eqn:True:StoppingTimes}, Lemma~\ref{lem:True:dV}.
    Then,
    \begin{multline}\label{eqn:Appendix:TrueProcess:prop:dV:Bound:Total}
        \begin{aligned}
             &\int_0^{\tau(t)}  
                e^{2 \lambda \nu} 
                \nabla \ZNorm{\Zt{N,\nu}}^\top \Jsigma{\nu,\Zt{N,\nu}} 
            d\Wt{\nu}
            \\
            & \quad 
            +
            \int_0^{\tau(t)}  e^{2 \lambda \nu} 
                \left(  
                    \nabla \ZNorm{\Zt{N,\nu}}^\top
                    \Jmu{\nu, \Zt{N,\nu}}
                    +
                    \frac{1}{2} \Trace{ \Ksigma{\nu, \Zt{N,\nu}}  \nabla^2 \ZNorm{\Zt{N,\nu}}}
                \right)
            d\nu
        \end{aligned}
        \\
        \leq
        -2 \lambda
        \int_0^{\tau(t)}  e^{2 \lambda \nu} 
                \ZNorm{\Zt{N,\nu}}
        d\nu 
        +
        \int_0^{\tau(t)}  
            e^{2 \lambda \nu}
            \left( 
                \left[  
                    \phi_\mu\left(\nu,\Zt{N,\nu}\right)
                    +
                    \widetilde{\mathcal{U}}\left(\nu,\Zt{N,\nu}\right)
                \right]
                d\nu
                +
                \phi_\sigma\left(\nu,\Zt{N,\nu}\right)
                d\Wt{\nu}      
            \right)
            \\
            +
            \int_0^{\tau(t)}  
                e^{2 \lambda \nu}
                \left( 
                    \left[  
                        \phi_{\mu^{\paral}}\left(\nu,\Zt{N,\nu}\right)
                        +
                        \phi_{U}\left(\nu,\Zt{N,\nu}\right)
                    \right]
                    d\nu
                    +
                        \phi_{\sigma^{\paral}}\left(\nu,\Zt{N,\nu}\right)
                    d\Wt{\nu}      
                \right),          
    \end{multline} 
    for all $ t \in \mathbb{R}_{\geq 0}$, where $\Ksigma{\nu, \Zt{N,\nu}} = \Jsigma{\nu,\Zt{N,\nu}} \Jsigma{\nu,\Zt{N,\nu}}^\top$,$\Jmu{\nu,\Zt{N,\nu}}$ and $\Jsigma{\nu,\Zt{N,\nu}}$ are defined in~\eqref{eqn:True:JointProcess}, and $\phi_\mu$, $\phi_\mu$, and $\widetilde{\mathcal{U}}$ are defined in~\eqref{eqn:lem:True:dV:phi:Functions}. 
    Additionally, we have defined 
    \begin{align*}
        \phi_{\mu^{\paral}}\left(\nu,\Zt{N,\nu}\right)
        =
        \ZDiffNorm{\Zt{N,\nu}}^\top
        g(\nu)
        \LZparamu{\nu,\Zt{N,\nu}},
        \quad 
        \phi_{\sigma^{\paral}}\left(\nu,\Zt{N,\nu}\right)
        =
        \ZDiffNorm{\Zt{N,\nu}}^\top 
        g(\nu) 
        \FZparasigma{\nu,\Zt{N,\nu}},
        \\
        \phi_{U}\left(\nu,\Zt{N,\nu}\right)
        =
        \ZDiffNorm{\Zt{N,\nu}}^\top
        g(\nu) 
        \ReferenceInputZ[\Zt{N}][\nu],
    \end{align*}
    where 
    $\ZDiffNorm{\Zt{N,\nu}} = 2\left(\Xt{N,\nu} - \Xrt{N,\nu}\right)$, and 
    \begin{align*}
        \LZparamu{\nu,\Zt{N,\nu}}
        =
        \Lparamu{\nu,\Xt{N,\nu}}
        - 
        \Lparamu{\nu,\Xrt{N,\nu}}, 
        \quad 
        \FZparasigma{\nu,\Zt{N,\nu}}
        =
        \Fparasigma{\nu,\Xt{N,\nu}}
        - 
        \Fparasigma{\nu,\Xrt{N,\nu}}, 
        \\ 
        \ReferenceInputZ[\Zt{N}][\nu] 
        =
        \ReferenceInput[\Xt{N}][\nu] 
        -
        \ReferenceInput[\Xrt{N}][\nu].
    \end{align*}

\end{proposition}
\begin{proof}
    Using the definitions of $J_\mu$ in~\eqref{eqn:True:JointProcess} and the decomposition of $F_\mu$ in~\eqref{eqn:VectorFields:Decomposition}, we have that 
    \begin{align*}
        \nabla \ZNorm{\Zt{N,\nu}}^\top \Jmu{\nu,\Zt{N,\nu}}
        =& 
        2\left(\Xt{N,\nu} - \Xrt{N,\nu}\right)^\top
        \left(
            \Fmu{\nu,\Xt{N,\nu},\ULt{\nu}}
            - 
            \Fmu{\nu,\Xrt{N,\nu},\Urt{\nu}}
        \right)
        \\
        =& 
        2\left(\Xt{N,\nu} - \Xrt{N,\nu}\right)^\top
        \left(
            \Fbarmu{\nu,\Xt{N,\nu}}
            - 
            \Fbarmu{\nu,\Xrt{N,\nu}}
        \right)
        \\
        &+
        2\left(\Xt{N,\nu} - \Xrt{N,\nu}\right)^\top
        \left[
            g(\nu)\left(\ULt{\nu}-\Urt{\nu}\right)
            +
            \Lmu{\nu,\Xt{N,\nu}}
            - 
            \Lmu{\nu,\Xrt{N,\nu}}
        \right], 
    \end{align*}
    for all $\nu \in [0,\tau(t)]$, which, due to Lemma~\ref{lem:ILFConditions:Consequence} reduces to 
    \begin{multline}\label{eqn:Appendix:TrueProcess:prop:dV:Term1:A:Pre}
       \nabla \ZNorm{\Zt{N,\nu}}^\top \Jmu{\nu,\Zt{N,\nu}}
        \leq 
        -2 \lambda \ZNorm{\Zt{N,\nu}}
        +
        \ZDiffNorm{\Zt{N,\nu}}^\top
            g(\nu)\left(\ULt{\nu}-\Urt{\nu}\right) 
        \\
        +
        \ZDiffNorm{\Zt{N,\nu}}^\top
        \left(
            \Lmu{\nu,\Xt{N,\nu}}
            - 
            \Lmu{\nu,\Xrt{N,\nu}}
        \right),
    \end{multline}
    for all $\nu \in [0,\tau(t)]$, where we have used the definition $\ZDiffNorm{\Zt{N,\nu}} = 2\left(\Xt{N,\nu} - \Xrt{N,\nu}\right)$.
    We develop the expression further by using~\eqref{eqn:UnknownFunctions:Decomposition} in Assumption~\ref{assmp:UnknownFunctions} to conclude that
    \begin{align*}
        \Lmu{\nu,\cdot}
        =
        \begin{bmatrix} g(\nu) & g(\nu)^\perp \end{bmatrix}
        \begin{bmatrix} \Lparamu{\nu,\cdot} \\ \Lperpmu{\nu,\cdot} \end{bmatrix}
        = g(\nu)\Lparamu{\nu,\cdot} + g(\nu)^\perp\Lperpmu{\nu,\cdot}
        ,
    \end{align*}
    which allows us to write~\eqref{eqn:Appendix:TrueProcess:prop:dV:Term1:A:Pre} as 
    \begin{multline}\label{eqn:Appendix:TrueProcess:prop:dV:Term1:A:1}
       \nabla \ZNorm{\Zt{N,\nu}}^\top \Jmu{\nu,\Zt{N,\nu}}
        \leq 
        -2 \lambda \ZNorm{\Zt{N,\nu}}
        +
        \ZDiffNorm{\Zt{N,\nu}}^\top
            g(\nu)\left(\ULt{\nu}-\Urt{\nu}\right) 
        \\
        +
        \ZDiffNorm{\Zt{N,\nu}}^\top
        g(\nu)
        \left(
            \Lparamu{\nu,\Xt{N,\nu}}
            - 
            \Lparamu{\nu,\Xrt{N,\nu}}
        \right)
        \\
        +
        \ZDiffNorm{\Zt{N,\nu}}^\top
        g(\nu)^\perp
        \left(
            \Lperpmu{\nu,\Xt{N,\nu}}
            - 
            \Lperpmu{\nu,\Xrt{N,\nu}}
        \right),
        \quad \forall \nu \in [0,\tau(t)].
    \end{multline}

    For the truncated processes $\Xt{N,t}$ and $\Xrt{N,t}$, we have from~\eqref{eqn:ReferenceFeedbackOperatorProcess} and~\eqref{eqn:True:L1DRACOperator:Initial} that   
    \begin{align*}
        \Urt{t} = \ReferenceInput[\Xrt{N}][t]
        , \quad
        \ULt{t} = \FL[\Xt{N}][t].  
    \end{align*} 
    Thus, by adding and subtracting $\ReferenceInput[\Xt{N}][t]$, we obtain 
    \begin{align}\label{eqn:Appendix:TrueProcess:prop:dV:InputManipulation}
        \ULt{t} - \Urt{t} 
        = 
        \FL[\Xt{N}][t]
        -
        \ReferenceInput[\Xrt{N}][t]
        =&
        \left[\FL[\Xt{N}][t] - \ReferenceInput[\Xt{N}][t]\right]
        -
        \ReferenceInput[\Xrt{N}][t] 
        + 
        \ReferenceInput[\Xt{N}][t]
        \notag 
        \\ 
        =&
        \left(\FL - \ReferenceInput\right)\left(\Xt{N}\right)(t)
        +
        \ReferenceInputZ[\Zt{N}][\nu]
        .  
    \end{align}
    Substituting into~\eqref{eqn:Appendix:TrueProcess:prop:dV:Term1:A:1} yields 
    \begin{multline}\label{eqn:Appendix:TrueProcess:prop:dV:Term1:A:Final}
            \nabla \ZNorm{\Zt{N,\nu}}^\top \Jmu{\nu,\Zt{N,\nu}}
            \leq
            -2 \lambda \ZNorm{\Zt{N,\nu}}
            +
            \widetilde{\mathcal{U}}\left(\nu,\Zt{N,\nu}\right)
            +
            \phi_{U}\left(\nu,\Zt{N,\nu}\right)
            +
            \phi_{\mu^{\paral}}\left(\nu,\Zt{N,\nu}\right)
        \\
        +
        \ZDiffNorm{\Zt{N,\nu}}^\top
        g(\nu)^\perp
        \left(
            \Lperpmu{\nu,\Xt{N,\nu}}
            - 
            \Lperpmu{\nu,\Xrt{N,\nu}}
        \right),
    \end{multline}
    for all $\nu \in [0,\tau(t)]$, where 
    \begin{align*}
        \phi_{\mu^{\paral}}\left(\nu,\Zt{N,\nu}\right)
        =
        \ZDiffNorm{\Zt{N,\nu}}^\top
        g(\nu)
        \left[
            \Lparamu{\nu,\Xt{N,\nu}}
            - 
            \Lparamu{\nu,\Xrt{N,\nu}}
        \right],
        \\
        \phi_{U}\left(\nu,\Zt{N,\nu}\right)
        =
        \ZDiffNorm{\Zt{N,\nu}}^\top
        g(\nu) 
        \left[
            \ReferenceInput[\Xt{N}][\nu] 
            -
            \ReferenceInput[\Xrt{N}][\nu] 
        \right],
        \\
        \widetilde{\mathcal{U}}\left(\nu,\Zt{N,\nu}\right)
        =
        \ZDiffNorm{\Zt{N,\nu}}^\top
        g(\nu)
        \left(\FL - \ReferenceInput\right)(\Xt{N})(\nu)
        .
    \end{align*}

    Next, using the definition of $J_\sigma$ in~\eqref{eqn:True:JointProcess}, we have that 
    \begin{align}\label{eqn:Appendix:TrueProcess:prop:dV:Term1:B:Final}
        \frac{1}{2} \Trace{ \Ksigma{\nu, \Zt{N,\nu}}  \nabla^2 \ZNorm{\Zt{N,\nu}}}
        = 
        \Trace{ \Ksigma{\nu, \Zt{N,\nu}}}
        =& 
        \Frobenius{\Jsigma{\nu, \Zt{N,\nu}}}^2
        \notag 
        \\
        =& 
        \Frobenius{\Fsigma{\nu,\Xt{N,\nu}} - \Fsigma{\nu,\Xrt{N,\nu}}}^2,
    \end{align}
    and 
    \begin{align}\label{eqn:Appendix:TrueProcess:prop:dV:Term1:C:1}
        \nabla \ZNorm{\Zt{N,\nu}}^\top \Jsigma{\nu,\Zt{N,\nu}}
        = 
        \ZDiffNorm{\Zt{N,\nu}}^\top 
        \left( \Fsigma{\nu,\Xt{N,\nu}} - \Fsigma{\nu,\Xrt{N,\nu}} \right)
        . 
    \end{align}
    As before,~\eqref{eqn:UnknownFunctions:Decomposition} in Assumption~\ref{assmp:UnknownFunctions} allows us to write
    \begin{align*}
        \Fsigma{\nu,\cdot}
        = 
        g(\nu)^\perp \Fperpsigma{\nu,\cdot} + g(\nu) \Fparasigma{\nu,\cdot}
        ,
    \end{align*}
    which in turn lets us express ~\eqref{eqn:Appendix:TrueProcess:prop:dV:Term1:C:1} as
    \begin{align}\label{eqn:Appendix:TrueProcess:prop:dV:Term1:C:Final}
        \nabla \ZNorm{\Zt{N,\nu}}^\top \Jsigma{\nu,\Zt{N,\nu}} 
        =& 
        \ZDiffNorm{\Zt{N,\nu}}^\top 
        g(\nu) 
        \left( \Fparasigma{\nu,\Xt{N,\nu}} - \Fparasigma{\nu,\Xrt{N,\nu}} \right)
        \notag 
        \\
        & + 
        \ZDiffNorm{\Zt{N,\nu}}^\top
        g(\nu)^\perp  
        \left( \Fperpsigma{\nu,\Xt{N,\nu}}  - \Fperpsigma{\nu,\Xrt{N,\nu}}  \right)
        . 
    \end{align}
    The expressions in~\eqref{eqn:Appendix:TrueProcess:prop:dV:Term1:A:Final},~\eqref{eqn:Appendix:TrueProcess:prop:dV:Term1:B:Final}, and~\eqref{eqn:Appendix:TrueProcess:prop:dV:Term1:C:Final} then lead to the desired result in~\eqref{eqn:Appendix:TrueProcess:prop:dV:Bound:Total}.

\end{proof}

In the subsequent proposition, we derive the effect of reference feedback operator $\ReferenceInput$ on the truncated joint process $\Zt{N,t}$.
\begin{proposition}\label{prop:Appendix:TrueProcess:dV:U}
    Let $\Zt{N,t}$ be the strong solution of~\eqref{eqn:True:TruncatedJointProcess}, and let $\tau(t)$ be the stopping time defined in~\eqref{eqn:True:StoppingTimes}, Lemma~\ref{lem:True:dV}. 
    Then, for the term $\phi_U$ defined in the statement of Proposition~\ref{prop:Appendix:TrueProcess:dV}, we have that    
    \begin{multline}\label{eqn:Appendix:TrueProcess:prop:dV:U}
        \int_0^{\tau(t)}  \expo{ 2\lambda \nu } 
        \phi_{U}\br{\nu,\Zt{N,\nu}}
        d\nu
        = 
        \int_0^{\tau(t)}  
            \left(\hat{\mathcal{U}}_{\mu}\br{\tau(t),\nu,\Zt{N};\Boldomega}d\nu +  \hat{\mathcal{U}}_{\sigma}\br{\tau(t),\nu,\Zt{N};\Boldomega}d\Wt{\nu}\right)
        \\
        +
        \int_0^{\tau(t)}   
            e^{ 2\lambda \nu }  
            \left(\phi_{U_\mu}\br{\nu,\Zt{N,\nu};\Boldomega} d\nu + \phi_{U_\sigma}\br{\nu,\Zt{N,\nu};\Boldomega}  
            d\Wt{\nu}\right) 
        ,
        \quad 
        t \in \mathbb{R}_{\geq 0},
    \end{multline}
    where 
    \begin{align*}
        \begin{multlined}[b][0.9\linewidth]
            \hat{\mathcal{U}}_{\mu}\br{\tau(t),\nu,\Zt{N};\Boldomega}
            \\
            =
            \expo{-\Boldomega \tau(t)}  
            \frac{\Boldomega}{2\lambda - \Boldomega}
            \left( 
                \expo{\Boldomega \tau(t)}
                \mathcal{P}\br{\tau(t),\nu}
                - 
                e^{ 2\lambda  \tau(t) }  
                \ZDiffNorm{\Zt{N,\tau(t)}}^\top g(\tau(t))
            \right) 
            \expo{ \Boldomega \nu} 
            \LZparamu{\nu,\Zt{N,\nu}}
            ,
        \end{multlined}
        \\
        \begin{multlined}[b][0.9\linewidth]
            \hat{\mathcal{U}}_{\sigma}\br{\tau(t),\nu,\Zt{N};\Boldomega}
            \\
            =
            \expo{-\Boldomega \tau(t)}  
            \frac{\Boldomega}{2\lambda - \Boldomega}
            \left( 
                \expo{\Boldomega \tau(t)}
                \mathcal{P}\br{\tau(t),\nu}
                - 
                e^{ 2\lambda  \tau(t) } 
                \ZDiffNorm{\Zt{N,\tau(t)}}^\top g(\tau(t)) 
            \right)   
            \expo{ \Boldomega \nu}  
            \FZparasigma{\nu,\Zt{N,\nu}}
            ,
        \end{multlined}
        \\
        \phi_{U_\mu}\br{\nu,\Zt{N,\nu};\Boldomega}
        =
        \frac{\Boldomega}{2\lambda - \Boldomega}
        \ZDiffNorm{\Zt{N,\nu}}^\top g(\nu)
        \LZparamu{\nu,\Zt{N,\nu}}
        ,
        \\
        \phi_{U_\sigma}\br{\nu,\Zt{N,\nu};\Boldomega}
        = 
        \frac{\Boldomega}{2\lambda - \Boldomega}
        \ZDiffNorm{\Zt{N,\nu}}^\top g(\nu)
         \FZparasigma{\nu,\Zt{N,\nu}} 
        ,
    \end{align*}
    and where $\mathcal{P}\br{\tau(t),\nu}$ is defined in~\eqref{eqn:True:ddV}, Lemma~\ref{lem:True:dV}.
\end{proposition}
\begin{proof}
     Using the definition of $\phi_U$ in the statement of Proposition~\ref{prop:Appendix:TrueProcess:dV}, we have that 
    \begin{multline*}
        \begin{aligned}
            &\int_0^{\tau(t)}  \expo{ 2\lambda \nu} 
                \phi_U \left(\nu, \Zt{N,\nu} \right)
            d\nu
            \\
            &
            =
            \int_0^{\tau(t)}  
                \expo{ 2\lambda \nu } 
                \ZDiffNorm{\Zt{N,\nu}}^\top
                g(\nu) 
                \ReferenceInputZ[\Zt{N}][\nu]
            d\nu
            \\
            &
            =
            \int_0^{\tau(t)}  
                \expo{ 2\lambda \nu } 
                \ZDiffNorm{\Zt{N,\nu}}^\top
                g(\nu) 
                \left(
                     \ReferenceInput[\Xt{N}][\nu] 
                    -
                    \ReferenceInput[\Xrt{N}][\nu]
                \right)
            d\nu
            \\
            &= 
            \int_0^{\tau(t)}  
                \expo{ 2\lambda \nu } 
                \ZDiffNorm{\Zt{N,\nu}}^\top
                g(\nu) 
                \left(
                    \Filter[\Lparamu{\cdot,\Xt{N}}][\nu]
                    -
                    \Filter[\Lparamu{\cdot,\Xrt{N}}][\nu]
                \right)
            d\nu
        \end{aligned}
        \\
        +
        \int_0^{\tau(t)}  
            \expo{ 2\lambda \nu } 
            \ZDiffNorm{\Zt{N,\nu}}^\top
            g(\nu) 
            \left( 
                \FilterW[\Fparasigma{\cdot,\Xt{N}}, \Wt{}][\nu] 
                - 
                \FilterW[\Fparasigma{\cdot,\Xrt{N}}, \Wt{}][\nu]
            \right)
        d\nu,
    \end{multline*}
    for all $t \geq 0$, where we have used the definition of $\ReferenceInput$ in~\eqref{eqn:ReferenceFeedbackOperatorProcess}. 
     Next, using the definitions of $\Filter$ and $\FilterW[\cdot,\Wt{}]$ in~\eqref{eqn:L1DRAC:Definition:FeedbackOperator:Filter} and~\eqref{eqn:ReferenceFeedbackOperatorProcess}, respectively, we can re-write the previous expression as   
    \begin{multline*}
        \int_0^{\tau(t)}  \expo{ 2\lambda \nu } 
        \phi_{U}\br{\nu,\Zt{N,\nu}}
        d\nu
        = 
        \int_0^{\tau(t)} 
        \int_0^\nu  
            \left(-\Boldomega \expo{(2\lambda - \Boldomega)\nu}\right)  
            \ZDiffNorm{\Zt{N,\nu}}^\top
            g(\nu)
            \left(
                \expo{ \Boldomega \beta} 
                \LZparamu{\beta,\Zt{N,\beta}}
                d\beta 
            \right) 
        d\nu
        \\
        +
        \int_0^{\tau(t)} 
        \int_0^\nu  
            \left(-\Boldomega \expo{(2\lambda - \Boldomega)\nu}\right)  
            \ZDiffNorm{\Zt{N,\nu}}^\top 
            g(\nu) 
            \left(
                \expo{ \Boldomega \beta} 
                \FZparasigma{\beta,\Zt{N,\beta}}
                d\Wt{\beta} 
            \right) 
        d\nu
        ,
    \end{multline*}
    for all $t \in \mathbb{R}_{\geq 0}$.
    Changing the order of integration in the first integral on the right hand side, and applying Lemma~\ref{lem:TechnicalResults:Fubini-ish} to the second integral:
    \begin{multline}\label{eqn:Appendix:TrueProcess:prop:dV:U:1}
        \begin{aligned}
            &\int_0^{\tau(t)}  \expo{ 2\lambda \nu } 
            \phi_{U}\br{\nu,\Zt{N,\nu}}
            d\nu
            \\
            &= 
            \int_0^{\tau(t)}  
                \left( 
                    - 
                    \int_\nu^{\tau(t)} 
                        \Boldomega \expo{(2\lambda - \Boldomega)\beta} 
                        \ZDiffNorm{\Zt{N,\beta}}^\top
                        g(\beta)
                    d\beta
                \right)   
                \expo{ \Boldomega \nu} 
                \LZparamu{\nu,\Zt{N,\nu}}  
            d\nu 
        \end{aligned} 
        \\
        +
        \int_0^{\tau(t)}  
            \left( 
                - 
                \int_\nu^{\tau(t)} 
                    \Boldomega 
                    \expo{(2\lambda - \Boldomega)\beta} 
                    \ZDiffNorm{\Zt{N,\beta}}^\top
                    g(\beta)
                d\beta
            \right)   
            \expo{ \Boldomega \nu} 
            \FZparasigma{\nu,\Zt{N,\nu}}
        d\Wt{\nu}  
        ,
    \end{multline}
    for all $t \in \mathbb{R}_{\geq 0}$, where in the first integral, we switch between the variables $\beta$ and $\nu$ after changing the order of integration.
     We then complete the proof by using an identical line of reasoning in the proof of Proposition~\ref{prop:Appendix:ReferenceProcess:dV:U} from~\eqref{eqn:Appendix:ReferenceProcess:prop:dV:U:1} onwards.

\end{proof}

Similar to Proposition~\ref{prop:Appendix:ReferenceProcess:ddV:Bound}, we derive an alternative representation for the term $\mathcal{P}\br{\tau(t),\nu}$ next. 
\begin{proposition}\label{prop:Appendix:True:ddV:Bound}
    Recall the expression for $\mathcal{P}\br{\tau(t),\nu}$ in~\eqref{eqn:True:ddV} in the statement of Lemma~\ref{lem:True:dV} which we restate below:
    \begin{align}\label{eqn:lem:Appendix:TrueProcess:ddV:Expression:Main}
        \mathcal{P}\br{\tau(t),\nu}
        =& 
        \int_\nu^{\tau(t)}
            e^{ (2\lambda - \Boldomega) \beta }  
            d_\beta
            \left[
                \ZDiffNorm{\Zt{N,\beta}}^\top g(\beta)
            \right]
        \in \mathbb{R}^{1 \times m}
        , \quad 0 \leq \nu \leq \tau(t),
    \end{align}
    where the $\tau(t)$ is defined in~\eqref{eqn:True:StoppingTimes}.
    Then, $\mathcal{P}\br{\tau(t),\nu}$ admits the following representation:
    \begin{equation}\label{eqn:lem:Appendix:TrueProcess:ddV:Expression:Pr:Main}
        \mathcal{P}\br{\tau(t),\nu}
        =
        \mathcal{P}_\circ \br{\tau(t),\nu}
        +
        \mathcal{P}_{ad} \br{\tau(t),\nu}
        +
        \widetilde{\mathcal{P}} \br{\tau(t),\nu}
        \in \mathbb{R}^{1 \times m},
        \quad 
        0 \leq \nu \leq \tau(t), ~t \in \mathbb{R}_{\geq 0},
    \end{equation}
    where 
    \begin{subequations}
        \begin{align}
            \mathcal{P}_\circ \br{\tau(t),\nu}
            =
            \int_\nu^{\tau(t)}
                e^{ (2\lambda - \Boldomega) \beta }
                \left(  
                    \mathcal{P}_\mu(\beta)
                    d\beta
                    + 
                    \mathcal{P}_\sigma (\beta)
                    d\Wt{\beta}
                \right)^\top
            ,
            \label{eqn:lem:Appendix:TrueProcess:ddV:Expression:Pr}
            \\
            \mathcal{P}_{ad} \br{\tau(t),\nu}
            =
            \int_\nu^{\tau(t)}
                e^{ (2\lambda - \Boldomega) \beta }  
                    \mathcal{P}_{\mathcal{U}}(\beta)^\top 
            d\beta
            ,
            \quad 
            \widetilde{\mathcal{P}} \br{\tau(t),\nu}
            =
            \int_\nu^{\tau(t)}
                e^{ (2\lambda - \Boldomega) \beta }  
                \widetilde{\mathcal{P}}_{\mathcal{U}}(\beta)^\top 
            d\beta
            \label{eqn:lem:Appendix:TrueProcess:ddV:Expression:TildePrU}
            ,
        \end{align}
    \end{subequations}
    and where $\mathcal{P}_\mu (\beta),~\mathcal{P}_{\mathcal{U}} (\beta),~\widetilde{\mathcal{P}}_{\mathcal{U}} (\beta)\in \mathbb{R}^m$ and $\mathcal{P}_\sigma (\beta) \in \mathbb{R}^{m \times d}$ are defined using the expressions in~\eqref{def:True:ErrorFunctions} as    
    \begin{align*}
        \mathcal{P}_\mu (\beta)
        =
        \dot{g}(\beta)^\top
        \RefDiffNorm{\Zt{N,\beta}}
        +
        2
        g(\beta)^\top
        \left(
            \fZ{\beta,\Zt{N,\beta}}
            +
            \LZmu{\beta,\Zt{N,\beta}} 
        \right)
        ,
        \\
        \mathcal{P}_{\mathcal{U}} (\beta)
        =
        2
        g(\beta)^\top
        g(\beta)
        \ReferenceInputZ[\Zt{N}][\beta],
        \quad
        \widetilde{\mathcal{P}}_{\mathcal{U}} (\beta)
        =
        2
        g(\beta)^\top
        g(\beta)
        \left(\FL - \ReferenceInput\right)\left(\Xt{N}\right)(\beta), 
        \\
        \mathcal{P}_\sigma (\beta)
        =
        2
        g(\beta)^\top
        \FZsigma{\beta,\Zt{N,\beta}}.
    \end{align*}

\end{proposition}
\begin{proof}
    We closely follow the proof of Proposition~\ref{prop:Appendix:ReferenceProcess:ddV:Bound} and begin by writing
    \begin{align}\label{eqn:lem:Appendix:TrueProcess:ddV:Kernel:Decomposition}
        \RefDiffNorm{\Zt{N,\beta}}^\top g(\beta)
        =&
        \begin{bmatrix} 
            \RefDiffNorm{\Zt{N,\beta}}^\top g_{\cdot,1}(\beta) & \cdots &  
            \RefDiffNorm{\Zt{N,\beta}}^\top g_{\cdot,m}(\beta)   
        \end{bmatrix}
        \notag  
        \\
        =&
        \begin{bmatrix} 
            \sum_{i=1}^n \RefDiffNormPoint{\Zt{N,\beta}} g_{i,1}(\beta) & \cdots &  
            \sum_{i=1}^n \RefDiffNormPoint{\Zt{N,\beta}} g_{i,m}(\beta)   
        \end{bmatrix} \in \mathbb{R}^{1 \times m},
    \end{align} 
    where $g_{\cdot, j}(\beta) \in \mathbb{R}^n$ is the $j$-th column of $g(\beta)$, and 
    \begin{align*}
        \RefDiffNormPoint{\Zt{N,\beta}} \doteq \frac{\partial}{\partial \left(\Xt{N} - \Xrt{N}\right)_i}\RefNorm{\Zt{N,\beta}} = 2\left(\Xt{N,\beta} - \Xrt{N,\beta}\right)_i  \in \mathbb{R}, \quad i \in \cbr{1,\dots,n}.
    \end{align*} 
    Applying \ito's lemma to $\RefDiffNormPoint{\Zt{N,\beta}}g_{i,j}(\beta) \in \mathbb{R}$, $(i,j) \in \cbr{1,\dots,n} \times \cbr{1,\dots,m}$, and using the truncated dynamics in~\eqref{eqn:True:TruncatedJointProcess}  we get
    \begin{multline}\label{eqn:lem:Appendix:TrueProcess:ddV:Ito:Initial}
        d_\beta \sbr{\RefDiffNormPoint{\Zt{N,\beta}}g_{i,j}(\beta)} 
        =
        \left( 
            \RefDiffNormPoint{\Zt{N,\beta}}\dot{g}_{i,j}(\beta)
            +
            \nabla \RefDiffNormPoint{\Zt{N,\beta}}^\top  \Jmu{\beta, \Zt{N,\beta}} 
            g_{i,j}(\beta)
        \right)
        d\beta
        \\
        + 
        \nabla \RefDiffNormPoint{\Zt{N,\beta}}^\top  \Jsigma{\beta, \Zt{N,\beta}} g_{i,j}(\beta) d\Wt{\beta},
    \end{multline}
    where we have used the straightforward fact that $\nabla^2 \RefDiffNormPoint{\Zt{N,\beta}} = 0 \in \mathbb{S}^{n}$.
    We have further replaced $J_{N,\mu}$ and $J_{N,\sigma}$ with $J_{\mu}$ and $J_{\sigma}$ because from Proposition~\ref{prop:True:TruncatedWellPosedness}, $Z_{N,\beta}$ is also a strong solution of the joint process~\eqref{eqn:True:JointProcess} for all $\beta \in [\nu,\tau(t)] \subseteq [0,\tau^\star] \subseteq [0,\tau_N]$.
    Observe from~\eqref{eqn:lem:Appendix:TrueProcess:ddV:Kernel:Decomposition}  that 
    \begin{align*}
        d_\beta\left[ \RefDiffNorm{\Zt{N,\beta}}^\top g_{\cdot, j}(\beta)\right]
        =
        \sum_{i=1}^n d_\beta\left[ \RefDiffNormPoint{\Zt{N,\beta}}g_{i,j}(\beta)\right] \in \mathbb{R}, \quad j \in \cbr{1,\dots,m}.
    \end{align*} 
    It therefore follows from the expression in~\eqref{eqn:lem:Appendix:TrueProcess:ddV:Ito:Initial} 
    \begin{multline*}
        d_\beta\left[ \RefDiffNorm{\Zt{N,\beta}}^\top g_{\cdot, j}(\beta)\right]
        =
        \sum_{i=1}^n
        \left( 
            \RefDiffNormPoint{\Zt{N,\beta}}\dot{g}_{i,j}(\beta)
            +
            \nabla \RefDiffNormPoint{\Zt{N,\beta}}^\top  \Jmu{\beta, \Zt{N,\beta}} 
            g_{i,j}(\beta)
        \right)
        d\beta
        \\
        + 
        \sum_{i=1}^n
        \nabla \RefDiffNormPoint{\Zt{N,\beta}}^\top  \Jsigma{\beta, \Zt{N,\beta}} g_{i,j}(\beta) d\Wt{\beta} 
        ,
    \end{multline*}
    which can be further written as
    \begin{multline}\label{eqn:lem:Appendix:TrueProcess:ddV:Ito:Pre:1}
        d_\beta\left[ \RefDiffNorm{\Zt{N,\beta}}^\top g_{\cdot, j}(\beta)\right]
        =
        \left( 
            \RefDiffNorm{\Zt{N,\beta}}^\top
            \dot{g}_{\cdot,j}(\beta)
            +
            g_{\cdot,j}(\beta)^\top
            \nabla \RefDiffNorm{\Zt{N,\beta}}^\top  \Jmu{\beta, \Yt{N,\beta}} 
        \right)
        d\beta
        \\
        + 
        g_{\cdot,j}(\beta)^\top
        \nabla \RefDiffNorm{\Zt{N,\beta}}^\top  \Jsigma{\beta, \Zt{N,\beta}}  d\Wt{\beta} 
        \in \mathbb{R}
        ,
    \end{multline} 
    for $j \in \cbr{1,\dots,m}$.
    Once again,~\eqref{eqn:lem:Appendix:TrueProcess:ddV:Kernel:Decomposition} implies that 
    \begin{align*}
        d_\beta \left[\RefDiffNorm{\Zt{N,\beta}}^\top g(\beta)\right]
        =
        \begin{bmatrix} 
            d_\beta \left[\RefDiffNorm{\Zt{N,\beta}}^\top g_{\cdot,1}(\beta)\right]  
            & \cdots &  
            d_\beta \left[\RefDiffNorm{\Zt{N,\beta}}^\top g_{\cdot,m}(\beta)\right]   
        \end{bmatrix}
        \in \mathbb{R}^{1 \times m}.
    \end{align*} 
    Thus, we use the expression in~\eqref{eqn:lem:Appendix:TrueProcess:ddV:Ito:Pre:1} to write 
    \begin{multline}\label{eqn:lem:Appendix:TrueProcess:ddV:Ito:1}
        d_\beta \left[\RefDiffNorm{\Zt{N,\beta}}^\top g(\beta)\right]
        =
        \left( 
            \RefDiffNorm{\Zt{N,\beta}}^\top
            \dot{g}(\beta)
            +
            \left(
                g(\beta)^\top
                \nabla \RefDiffNorm{\Zt{N,\beta}}^\top  \Jmu{\beta, \Zt{N,\beta}} 
            \right)^\top
        \right)
        d\beta
        \\
        + 
        \left(
            g(\beta)^\top
            \nabla \RefDiffNorm{\Zt{N,\beta}}^\top  \Jsigma{\beta, \Zt{N,\beta}}  d\Wt{\beta}
        \right)^\top
        \in \mathbb{R}^{1 \times m}. 
    \end{multline}
    Now, since $\Zt{N} = \Xt{N} - \Xrt{N}$, using the definition of the gradient of vector valued functions in Sec.~\ref{sec:Notation}, we observe that
    \begin{align*}
        \nabla \RefDiffNorm{\Zt{N,\beta}}
        =
        2
        \begin{bmatrix}
            \nabla \left(\Xt{N,\beta} - \Xrt{N,\beta}\right)_1
            & \cdots & 
            \nabla \left(\Xt{N,\beta} - \Xrt{N,\beta}\right)_n
        \end{bmatrix}
        =&
        2
        \begin{bmatrix}
            \nabla \left(\Zt{N,\beta}\right)_1
            & \cdots & 
            \nabla \left(\Zt{N,\beta}\right)_n
        \end{bmatrix}
        \\
        =&
        2 
        \mathbb{I}_n 
        \in \mathbb{S}^{n}.    
    \end{align*}
     Thus, it follows from the definitions of the vector fields in~\eqref{eqn:True:JointProcess} that
    \begin{align*}
        \nabla \RefDiffNorm{\Zt{N,\beta}}^\top  
        \Jmu{\beta, \Zt{N,\beta}}
        =&
        2 
        \left(
            \Fmu{\beta,\Xt{N,\beta},\ULt{\beta}} - \Fmu{\beta,\Xrt{N,\beta},\Urt{\beta}}
        \right)
        \\
        \overset{(\star)}{=}
        &
        2 
        \fZ{\beta,\Zt{N,\beta}}
        +
        2
        g(\beta)
        \left(\FL - \ReferenceInput\right)\left(\Xt{N}\right)(\beta)
        +
        2
        g(\beta)
        \ReferenceInputZ[\Zt{N}][\beta]
        \notag 
        \\
        &
        \hspace{7cm}
        +
        2
        \LZmu{\beta,\Zt{N,\beta}}
        \in \mathbb{R}^n,
        \\
        \nabla \RefDiffNorm{\Zt{N,\beta}}^\top  \Jsigma{\beta, \Zt{N,\beta}}
        =& 
        2  
        \left(
            \Fsigma{\beta,\Xt{N,\beta}} - \Fsigma{\beta,\Xrt{N,\beta}}
        \right)
        =
        2  
        \FZsigma{\beta,\Zt{N,\beta}}
        \in \mathbb{R}^{n \times d},
    \end{align*}
    where we obtain $(\star)$ by using the decomposition from~\eqref{eqn:VectorFields:Decomposition} in Definition~\ref{def:VectorFields} to obtain  
    \begin{align*}
        &\Fmu{\beta,\Xt{N,\beta},\ULt{\beta}} - \Fmu{\beta,\Xrt{N,\beta},\Urt{\beta}}
        \\
        &=
        \Fbarmu{\beta,\Xt{N,\beta}}
        - 
        \Fbarmu{\beta,\Xrt{N,\beta}}
        +
        g(\beta)\left(\ULt{\beta}-\Urt{\beta}\right)
        +
        \Lmu{\beta,\Xt{N,\beta}}
        - 
        \Lmu{\beta,\Xrt{N,\beta}}
        \\ 
        &=
        \fZ{\beta,\Zt{N,\beta}}
        +
        g(\beta)
        \left(\FL - \ReferenceInput\right)\left(\Xt{N}\right)(\beta)
        +
        g(\beta)
        \ReferenceInputZ[\Zt{N}][\beta]
        +
        \LZmu{\beta,\Zt{N,\beta}},
    \end{align*}
    and where we have also used~\eqref{eqn:Appendix:TrueProcess:prop:dV:InputManipulation} in the proof of Proposition~\ref{prop:Appendix:TrueProcess:dV}.
    Substituting the above identities into~\eqref{eqn:lem:Appendix:TrueProcess:ddV:Ito:1} produces
    \begin{align*}
        d_\beta \left[\RefDiffNorm{\Zt{N,\beta}}^\top g(\beta)\right]
        =
        \left[
            \mathcal{P}_\mu (\beta) 
            +
            \mathcal{P}_{\mathcal{U}} (\beta)
            +
            \widetilde{\mathcal{P}}_{\mathcal{U}} (\beta)
        \right]^\top
        d\beta
        + 
        \left[
            \mathcal{P}_\sigma (\beta)
            d\Wt{\beta}
        \right]^\top
        \in \mathbb{R}^{1 \times m}. 
    \end{align*}
     Then,~\eqref{eqn:lem:Appendix:TrueProcess:ddV:Expression:Pr:Main} is established by substituting the above into~\eqref{eqn:lem:Appendix:TrueProcess:ddV:Expression:Main}. 

\end{proof}

The next result establishes the bounds for the pertinent entities derived in the last proposition. 
\begin{proposition}\label{prop:Appendix:TrueProcess:Pparts:Bound}
    Consider the functions $\mathcal{P}_{\mu}(t) \in \mathbb{R}^{m}$ and $\mathcal{P}_{\sigma}(t) \in \mathbb{R}^{m \times d}$  defined in the statement of Proposition~\ref{prop:Appendix:True:ddV:Bound}.
    If the stopping time $\tau^\star$, defined in~\eqref{eqn:True:StoppingTimes}, Lemma~\ref{lem:True:dV}, satisfies  $\tau^\star = \tstar$, then
    \begin{subequations}\label{prop:Appendix:TrueProcess:Pparts:Bound:Total:Final}
        \begin{align}
            \sup_{\nu \in [0,\tstar]} 
            \LpLaw{2\sfp}{}{\mathcal{P}_{\mu}(\nu)}
            \leq    
            2
            \Delta_{g}
            \Delta_{\mathcal{P}_\mu}(\sfp)
            +
            \Delta_{\dot{g}}
            \sup_{\nu \in [0,\tstar]}
                \LpLaw{2\sfp}{}{\RefDiffNorm{\Zt{N,\nu}}}
             ,
            \label{prop:Appendix:TrueProcess:Pparts:Bound:Mu:Final} 
            \\
            \sup_{\nu \in [0,\tstar]} 
            \LpLaw{\sfq}{}{\mathcal{P}_{\sigma}(\nu)}
            \leq 
            2
            \Delta_{g}
            \Delta_{\mathcal{P}_\sigma}(\sfp)
            ,
            \label{prop:Appendix:TrueProcess:Pparts:Bound:Sigma:Final}
        \end{align}
    \end{subequations}
    where 
    \begin{align*}
        \Delta_{\mathcal{P}_\mu}(\sfp)
        =
        \sup_{\nu \in [0,\tstar]}
        \left(
            \LpLaw{2\sfp}{}{\fZ{\nu,\Zt{N,\nu}}}
            +
            \LpLaw{2\sfp}{}{\LZmu{\nu,\Zt{N,\nu}}} 
        \right),
        \quad   
        \Delta_{\mathcal{P}_\sigma}(\sfp)
        =
        \sup_{\nu \in [0,\tstar]} 
                \LpLaw{2\sfp}{}{\FZsigma{\nu,\Zt{N,\nu}}}.
    \end{align*}
\end{proposition}
\begin{proof}
    We have the following bound for the term $\mathcal{P}_\mu$ defined in ~\eqref{eqn:lem:Appendix:TrueProcess:ddV:Expression:Pr}:
    \begin{align*}
        \norm{\mathcal{P}_\mu(\nu)}
        \leq& 
        \Frobenius{\dot{g}(\nu)}
        \norm{\RefDiffNorm{\Zt{N,\nu}}}
        +
        2
        \Frobenius{g(\nu)}
        \left(
            \norm{\fZ{\nu,\Zt{N,\nu}}}
            +
            \norm{\LZmu{\nu,\Zt{N,\nu}}} 
        \right) 
        \\
        \leq& 
        \Delta_{\dot{g}}
        \norm{\RefDiffNorm{\Zt{N,\nu}}}
        +
        2
        \Delta_{g}
        \left(
            \norm{\fZ{\nu,\Zt{N,\nu}}}
            +
            \norm{\LZmu{\nu,\Zt{N,\nu}}} 
        \right) 
        ,
        \quad \forall \nu \in [0,\tstar],
    \end{align*}
    where we have used the bounds on $g(t)$ and $\dot{g}(t)$ in Assumption~\ref{assmp:KnownFunctions}.
    It then follows from the Minkowski's inequality that 
    \begin{align*}
        \LpLaw{2\sfp}{}{\mathcal{P}_\mu(\nu)}
        \leq& 
        \Delta_{\dot{g}}
        \LpLaw{2\sfp}{}{\RefDiffNorm{\Zt{N,\nu}}}
        +
        2
        \Delta_{g}
        \left(
            \LpLaw{2\sfp}{}{\fZ{\nu,\Zt{N,\nu}}}
            +
            \LpLaw{2\sfp}{}{\LZmu{\nu,\Zt{N,\nu}}} 
        \right) 
        ,
        \quad \forall \nu \in [0,\tstar],
    \end{align*}
     which in turn implies~\eqref{prop:Appendix:TrueProcess:Pparts:Bound:Mu:Final}.
    The bound in~\eqref{prop:Appendix:TrueProcess:Pparts:Bound:Sigma:Final} follows trivially. 
\end{proof}

The next lemma establishes the bound on $\Xi$.
\begin{lemma}\label{lem:Appendix:TrueProcess:Xi}
    Consider the term $\Xi\br{\tau(t),\Zt{N}}$ defined in~\eqref{eqn:lem:True:dV:Xi:Functions:A}, Lemma~\ref{lem:True:dV}.
    If the stopping time $\tau^\star$, defined in~\eqref{eqn:True:StoppingTimes}, Lemma~\ref{lem:True:dV}, satisfies  $\tau^\star = \tstar$, then:
    \begin{subequations}\label{eqn:lem:Appendix:TrueProcess:Xi:Bound:Final}
        \begin{align}
            \ELaw{}{\expo{-2\lambda t}\Xi\left(t,\Zt{N} \right)}
            \leq
            \frac{\Delta_{\Xi_1}(1,\tstar)}{2\lambda}
            ,
            \\
            \LpLaw{\sfp}{}{
                \expo{-2\lambda t}\Xi\left(t,\Zt{N} \right)
            }
            \leq  
            \frac{\Delta_{\Xi_1}(\sfp,\tstar)}{2\lambda}  
            +
            \frac{\Delta_{\Xi_2}(\sfp,\tstar)}{2\sqrt{\lambda}}
            ,
        \end{align}
    \end{subequations}
    for all $(t,\sfp) \in [0,\tstar] \times \mathbb{N}_{\geq 2}$, where 
    \begin{align*}
        \Delta_{\Xi_1}(\sfp,\tstar) 
        =  
        \sup_{\nu \in [0,\tstar]}
        \left(
            \Delta_g^\perp
            \LpLaw{2\sfp}{}{\ZDiffNorm{\Zt{N,\nu}}}
            \LpLaw{2\sfp}{}{\LZperpmu{\nu,\Zt{N,\nu}}}
            + 
            \left(\LpLaw{2\sfp}{}{\FZsigma{\nu,\Zt{N,\nu}}}\right)^2
        \right),
        \\
        \Delta_{\Xi_2}(\sfp,\tstar) 
        =
        \Delta_g^\perp
        \mathfrak{p}(\sfp)
        \sup_{\nu \in [0,\tstar]}
        \LpLaw{2\sfp}{}{
            \ZDiffNorm{\Zt{N,\nu}}  
        }
        \LpLaw{2\sfp}{}{
                \FZperpsigma{\nu,\Zt{N,\nu}}
        }
        .
    \end{align*}

\end{lemma}
\begin{proof}
    Note that $\tau(t) \in [0,\tstar]$ since $\tau^\star = \tstar$.
    Thus, in the following, we replace $\tau(t)$ with $t$ with the condition that $t \in [0,\tstar]$.
    We then begin by writing 
    \begin{align}\label{eqn:lem:Appendix:TrueProcess:Xi:Bound:InitialSetup}
        \expo{-2\lambda t}
        \Xi\left(t,\Zt{N} \right)
        =
        \expo{-2\lambda t}
        \int_0^{t}   
            \expo{2 \lambda \nu} \phi_\mu\left(\nu,\Zt{N,\nu}\right) d\nu 
        +
        \expo{-2\lambda t}
        \int_0^{t} 
            \expo{2 \lambda \nu} \phi_\sigma\left(\nu,\Zt{N,\nu}\right)d\Wt{\nu},
        \quad 
        t \in [0,\tstar].
    \end{align}
    The regularity assumptions and the $t$-continuity of the strong solutions $\Xt{N,t}$ and $\Xrt{N,t}$ imply that $\phi_{\sigma}\br{\nu,\Zt{N,\nu}} \in \mathcal{M}^{loc}_2\left(\mathbb{R}^{1 \times d}|\Wfilt{t}\right)$.
    Hence, taking the expectation of both sides of~\eqref{eqn:lem:Appendix:TrueProcess:Xi:Bound:InitialSetup} and using~\cite[Thm.~1.5.21]{mao2007stochastic} produces 
    \begin{align}\label{eqn:lem:Appendix:TrueProcess:Xi:Bound:p=1:Initial}
        \ELaw{}{\expo{-2\lambda t}\Xi\left(t,\Zt{N} \right)}
        =&
        \expo{-2\lambda t}
        \ELaw{}{
            \int_0^{t}   
                \expo{2 \lambda \nu} \phi_\mu\left(\nu,\Zt{N,\nu}\right) d\nu 
        }
        \notag 
        \\
        \leq&     
        \expo{-2\lambda t}
        \ELaw{}{
            \int_0^{t}   
                \expo{2 \lambda \nu} \absolute{\phi_\mu\left(\nu,\Zt{N,\nu}\right)} d\nu 
        }
        ,
        \quad 
        \forall t \in [0,\tstar],
    \end{align}
    where we have used the fact that since $\tau^\star = \tstar$, and $\tstar$ is a constant, 
    \begin{align*} 
        \mathbb{E}\left[\expo{-2\lambda t}\right]
        =
        \expo{-2\lambda t}
        ,
        \quad  \forall t \in [0,\tstar].
    \end{align*}
    Using the fact above, we further apply the Minkowski's inequality to~\eqref{eqn:lem:Appendix:TrueProcess:Xi:Bound:InitialSetup} and obtain  
    \begin{multline}\label{eqn:lem:Appendix:TrueProcess:Xi:Bound:p>=2:Initial}
        \LpLaw{\sfp}{}{
            \expo{-2\lambda t}\Xi\left(t,\Zt{N} \right)
        }
        \leq 
        \expo{-2\lambda t}        
        \LpLaw{\sfp}{}{
            \int_0^{t}   
                \expo{2 \lambda \nu} \phi_\mu\left(\nu,\Zt{N,\nu}\right) d\nu
        } 
        \\
        +
        \expo{-2\lambda t}
        \LpLaw{\sfp}{}{
        \int_0^{t} 
            \expo{2 \lambda \nu} \phi_\sigma\left(\nu,\Zt{N,\nu}\right)d\Wt{\nu}
        }
        ,
        \quad 
        \forall (t,\sfp) \in [0,\tstar] \times \mathbb{N}_{\geq 2}.
    \end{multline}
    The regularity assumptions on the vector fields and the $t$-continuity of the strong solutions $\Xt{N,t}$ and $\Xrt{N,t}$ over $[0,\tstar] \subseteq [0,\tau_N]$ imply that $\phi_{\mu}\br{\nu,\Zt{N,\nu}} \in \mathcal{M}^{loc}_2\left(\mathbb{R}|\Wfilt{t}\right)$.
    Thus, we invoke Proposition~\ref{prop:TechnicalResults:LebesgueMoment} for $\cbr{\theta,\xi,n_s,S} = \cbr{2\lambda,0,1,\phi_{\mu}\br{\cdot,\Zt{N}}}$ to obtain
    \begin{align*}
        \LpLaw{\sfp}{}{
            \int_0^{t}   
                \expo{2 \lambda \nu} \absolute{\phi_\mu\left(\nu,\Zt{N,\nu}\right)} d\nu 
        }
        \leq& 
        \frac{\expo{2\lambda t}}{2\lambda} 
        \sup_{\nu \in [0,\tstar]}
        \pLpLaw{}{\phi_{\mu}\br{\nu,\Zt{N,\nu}}}
        , 
        \quad \forall (t,\sfp) \in [0,\tstar] \times \mathbb{N}_{\geq 1},
    \end{align*}
    where we have bounded $\expo{2\lambda t} -1$ by $\expo{2\lambda t}$. 
    Using the definition of $\phi_{\mu}$ in~\eqref{eqn:lem:True:dV:phi:Functions}, we can further bound the above by invoking the Minkowski's and Cauchy-Schwarz inequalities: 
    \begin{multline}\label{eqn:lem:Appendix:TrueProcess:Xi:Bound:Mu:Final}
        \LpLaw{\sfp}{}{
            \int_0^{t}   
                \expo{2 \lambda \nu} \absolute{\phi_\mu\left(\nu,\Zt{N,\nu}\right)} d\nu 
        }
        \\
        \leq 
        \frac{\expo{2\lambda t}}{2\lambda} 
        \sup_{\nu \in [0,\tstar]}
        \left(
            \Delta_g^\perp
            \LpLaw{2\sfp}{}{\ZDiffNorm{\Zt{N,\nu}}}
            \LpLaw{2\sfp}{}{\LZperpmu{\nu,\Zt{N,\nu}}}
            + 
            \left(\LpLaw{2\sfp}{}{\FZsigma{\nu,\Zt{N,\nu}}}\right)^2
        \right)
        , 
    \end{multline}
    for all $(t,\sfp) \in [0,\tstar] \times \mathbb{N}_{\geq 1}$, where we have also used the bound on $g(\nu)^\perp$ in Assumption~\ref{assmp:KnownFunctions}.

     Next, we have the following from the definition of $\phi_{\sigma}$ in~\eqref{eqn:lem:True:dV:phi:Functions}:
    \begin{align*}
        \int_0^{t}  
            \expo{ 2 \lambda  \nu }  
            \phi_{\sigma}\br{\nu,\Zt{N,\nu}}
        d\Wt{\nu}
        =
        \int_0^{t}  
            \expo{ 2 \lambda  \nu }  
            \left( 
                \ZDiffNorm{\Zt{N,\nu}}^\top g(\nu)^\perp \FZperpsigma{\nu,\Zt{N,\nu}}
            \right)
        d\Wt{\nu}.
    \end{align*}
    We bound the above by using Lemma~\ref{lemma:TechnicalResults:MartingaleMoment} with $\cbr{\theta,\xi,n_l,n_q,\Qt{}} = \cbr{2\lambda,0,1,d, \Wt{} }$, and 
    \begin{align*}
        L(\nu)
        =
        \ZDiffNorm{\Zt{N,\nu}}^\top g(\nu)^\perp \FZperpsigma{\nu,\Zt{N,\nu}}
        ,
    \end{align*}
    to obtain
    \begin{multline}\label{eqn:lem:Appendix:TrueProcess:Xi:Bound:Sigma:Final}
        \pLpLaw{}{
            \int_0^{t}  
                \expo{ 2 \lambda  \nu }  
                    \phi_{\sigma}\br{\nu,\Zt{N,\nu}}
            d\Wt{\nu}
        }
        \\
        \begin{aligned}[b]
            &\leq 
            \mathfrak{p}(\sfp)
            \frac{\expo{ 2\lambda t }}{2\sqrt{\lambda}}
            \sup_{\nu \in [0,\tstar]}
            \LpLaw{\sfp}{}{
                 \ZDiffNorm{\Zt{N,\nu}}^\top g(\nu)^\perp \FZperpsigma{\nu,\Zt{N,\nu}}
            }
            \\
            &\leq   
            \mathfrak{p}(\sfp)
            \Delta_g^\perp
            \frac{\expo{ 2\lambda t }}{2\sqrt{\lambda}}
            \sup_{\nu \in [0,\tstar]}
            \LpLaw{2\sfp}{}{
                \ZDiffNorm{\Zt{N,\nu}}  
            }
            \LpLaw{2\sfp}{}{
                    \FZperpsigma{\nu,\Zt{N,\nu}}
            }
            ,
            \quad 
            \forall (t,\sfp) \in [0,\tstar] \times \mathbb{N}_{\geq 2},
        \end{aligned}
    \end{multline}
     where we have bounded $\left(\expo{4\lambda t} -1\right)^\Half$ by $\expo{2\lambda t}$ and used the Cauchy-Schwarz inequality to obtain the last inequality.

    Substituting~\eqref{eqn:lem:Appendix:TrueProcess:Xi:Bound:Mu:Final} into~\eqref{eqn:lem:Appendix:TrueProcess:Xi:Bound:p=1:Initial} leads to
    \begin{multline}\label{eqn:lem:Appendix:TrueProcess:Xi:Bound:p=1:Final}
        \ELaw{}{\expo{-2\lambda t}\Xi\left(t,\Zt{N} \right)}
        \\
        \leq     
        \frac{1}{2\lambda} 
        \sup_{\nu \in [0,\tstar]}
       \left(
            \Delta_g^\perp
            \LpLaw{2}{}{\ZDiffNorm{\Zt{N,\nu}}}
            \LpLaw{2}{}{\LZperpmu{\nu,\Zt{N,\nu}}}
            + 
            \left(\LpLaw{2}{}{\FZsigma{\nu,\Zt{N,\nu}}}\right)^2
        \right)
        ,
        \quad 
        \forall t \in [0,\tstar].
    \end{multline}
    Similarly, substituting~\eqref{eqn:lem:Appendix:TrueProcess:Xi:Bound:Mu:Final} and~\eqref{eqn:lem:Appendix:TrueProcess:Xi:Bound:Sigma:Final} into~\eqref{eqn:lem:Appendix:TrueProcess:Xi:Bound:p>=2:Initial} produces 
    \begin{multline}\label{eqn:lem:Appendix:TrueProcess:Xi:Bound:p>=2:Final}
        \begin{aligned}
            &\LpLaw{\sfp}{}{
                \expo{-2\lambda t}\Xi\left(t,\Zt{N} \right)
            }
            \\
            &\leq     
            \Delta_g^\perp
            \frac{\mathfrak{p}(\sfp)}{2\sqrt{\lambda}}
            \sup_{\nu \in [0,\tstar]}
            \LpLaw{2\sfp}{}{
                \ZDiffNorm{\Zt{N,\nu}}  
            }
            \LpLaw{2\sfp}{}{
                    \FZperpsigma{\nu,\Zt{N,\nu}}
            }
        \end{aligned} 
        \\
        +       
        \frac{1}{2\lambda} 
        \sup_{\nu \in [0,\tstar]}
        \left(
            \Delta_g^\perp
            \LpLaw{2\sfp}{}{\ZDiffNorm{\Zt{N,\nu}}}
            \LpLaw{2\sfp}{}{\LZperpmu{\nu,\Zt{N,\nu}}}
            + 
            \left(\LpLaw{2\sfp}{}{\FZsigma{\nu,\Zt{N,\nu}}}\right)^2
        \right)
        ,
    \end{multline}
    for all $(t,\sfp) \in [0,\tstar] \times \mathbb{N}_{\geq 2}$.
    We have thus established the bounds in~\eqref{eqn:lem:Appendix:TrueProcess:Xi:Bound:Final}. 

    \begingroup
    \endgroup

    \begingroup

        \begingroup
        \endgroup

    \endgroup

\end{proof}

Next, we provide a result that is essential to the proof of the main results of the section.
\begin{proposition}\label{prop:Appendix:TrueProcess:N:Bounds}
    Consider the following scalar processes:
    \begin{align}\label{eqn:prop:Appendix:TrueProcess:N:Definitions}
        \begin{aligned}
            N_1(t)
            =
            \int_0^{t}
                \expo{\Boldomega \nu}
                M_\mu(t,\nu)
                \LZparamu{\nu,\Zt{N,\nu}}
            d\nu,
            \quad 
            N_2(t)
            =
            \int_0^{t}
                \expo{\Boldomega \nu}
                M_\mu(t,\nu)
                \FZparasigma{\nu,\Zt{N,\nu}}
            d\Wt{\nu},
            \\
            N_3(t)
            =
            \int_0^{t}
                \expo{\Boldomega \nu}
                M_\sigma(t,\nu)
                \LZparamu{\nu,\Zt{N,\nu}}
            d\nu,
            \quad 
            N_4(t)
            =
            \int_0^{t}
                \expo{\Boldomega \nu}
                M_\sigma(t,\nu)
                \FZparasigma{\nu,\Zt{N,\nu}}
            d\Wt{\nu},
        \end{aligned}
    \end{align}
    where 
    \begin{align*}
        M_\mu(t,\nu)
        =
        \int_\nu^{t} 
            \expo{ (2\lambda - \Boldomega) \beta } 
            \mathcal{P}_\mu(\beta)^\top
        d\beta, 
        \quad 
        M_\sigma(t,\nu)
        =
        \int_\nu^{t} 
            \expo{ (2\lambda - \Boldomega) \beta } 
            \left(
                \mathcal{P}_{\sigma}(\beta)d\Wt{\beta}
            \right)^\top,
    \end{align*}
    for $0 \leq \nu \leq t \leq T$, and where $\mathcal{P}_\mu(\beta) \in \mathbb{R}^{m}$ and $\mathcal{P}_{\sigma}(\beta) \in \mathbb{R}^{m \times d}$ are defined in the statement of Proposition~\ref{prop:Appendix:True:ddV:Bound}.

    If the stopping time $\tau^\star$, defined in~\eqref{eqn:Reference:StoppingTimes}, Lemma~\ref{lem:Reference:dV}, satisfies  $\tau^\star = \tstar$, then  we have the following bounds:
    \begin{multline}\label{eqn:prop:Appendix:TrueProcess:N:Bounds:p=1:Main}
        \sum_{i=1}^4 \absolute{\ELaw{}{N_i(t)}}
        \leq
        \frac{ \expo{ 2\lambda t } }{ \sqrt{\lambda} \Boldomega }  
        \Delta_{\mathcal{P}_1}(1,\tstar)
        \sup_{\nu \in [0,\tstar]} 
            \LpLaw{2}{}{\LZparamu{\nu, \Zt{N,\nu}}}
        \\
        + 
        \left(
            \frac{\expo{ 2\lambda t }}{\sqrt{\lambda \Boldomega}}
            \Delta_{\mathcal{P}_2}(1,\tstar)
            +
            \frac{\expo{2\lambda t}}{\lambda} 
            \Delta_{\mathcal{P}_3}(1,\tstar)
        \right)
        \sup_{\nu \in [0,\tstar]}
                \LpLaw{2}{}{\FZparasigma{\nu,\Zt{N,\nu}}}
        ,
    \end{multline}
    for all $t \in [0,\tstar]$, and
    \begin{multline}\label{eqn:prop:Appendix:TrueProcess:N:Bounds:p>=2:Main}
        \sum_{i=1}^4 \pLpLaw{}{N_i(t)}
        \leq 
        \frac{ \expo{ 2\lambda t } }{ \sqrt{\lambda} \Boldomega }
        \Delta_{\mathcal{P}_1}(\sfp,\tstar)
        \sup_{\nu \in [0,\tstar]} 
            \LpLaw{2\sfp}{}{\LZparamu{\nu, \Zt{N,\nu}}}
        \\
        +
        \left( 
            \frac{\expo{ 2\lambda t }}{\sqrt{\lambda \Boldomega }} 
            \Delta_{\mathcal{P}_2}(\sfp,\tstar)
            + 
            \frac{\expo{2\lambda t}}{\lambda}
            \Delta_{\mathcal{P}_3}(\sfp,\tstar)
        \right)
        \sup_{\nu \in [0,\tstar]}
        \LpLaw{2\sfp}{}{\FZparasigma{\nu,\Zt{N,\nu}}}
        ,
    \end{multline}
    for all $(t,\sfp) \in [0,\tstar] \times \mathbb{N}_{\geq 2}$, where 
    \begin{align*}
        \Delta_{\mathcal{P}_1}(\sfp,\tstar)
        = 
        \frac{ 1 }{ 2  \sqrt{\lambda} }
            \sup_{\nu \in [0,\tstar]}  
            \LpLaw{2\sfp}{}{\mathcal{P}_{\mu}(\nu)}
        + 
        \frac{\mathfrak{p}(\sfp)}{2 }
        \sup_{\nu \in [0,\tstar]}
            \LpLaw{2\sfp}{}{\mathcal{P}_{\sigma}(\nu)}
            ,
        \quad 
        \Delta_{\mathcal{P}_2}(\sfp,\tstar)
        =
        \mathfrak{p}'(\sfp)
        \Delta_{\mathcal{P}_1}(\sfp,\tstar)
        ,
        \\
        \Delta_{\mathcal{P}_3}(\sfp,\tstar)
        = 
        \frac{\sqrt{m}}{2}  
        \sup_{\nu \in [0,\tstar]}
            \LpLaw{2\sfp}{}{\mathcal{P}_{\sigma}(\nu)}
        ,
    \end{align*}
    and where the constants $\mathfrak{p}(\sfp)$ and $\mathfrak{p}'(\sfp)$ are defined in~\eqref{eqn:app:Constants:FrakP}. 

\end{proposition}
\begin{proof}
    We omit the details of the proof since it follows \emph{mutatis mutandis} the proof of Proposition~\ref{prop:Appendix:ReferenceProcess:N:Bounds}.
\end{proof}

The following result establishes the effect of the reference feedback operator $\ReferenceInput$ on the truncated joint process in Definition~\ref{def:True:TruncatedJointProcess}.  
\begin{proposition}\label{prop:Appendix:TrueProcess:NU:Bounds}
    Consider the following scalar process:
    \begin{align}\label{eqn:prop:Appendix:TrueProcess:NU:Definitions}
        N_{\mathcal{U}}(t)
        =  
        \int_0^{t}
            \expo{\Boldomega \nu}
            M_\mathcal{U}(t,\nu)
            \left[ \LZparamu{\nu,\Zt{N,\nu}}d\nu + \FZparasigma{\nu,\Zt{N,\nu}}d\Wt{\nu} \right]
        ,
    \end{align}
    where  
    \begin{align*}
        M_\mathcal{U}(t,\nu)
        =
        \int_\nu^{t} 
            \expo{ (2\lambda - \Boldomega) \beta } \mathcal{P}_\mathcal{U}(\beta)^\top
        d\beta, 
    \end{align*}
    for $0 \leq \nu \leq t \leq T$, and where where $\mathcal{P}_{\mathcal{U}}(\beta) \in \mathbb{R}^{m}$ is  defined in the statement of Proposition~\ref{prop:Appendix:True:ddV:Bound}.
     If the stopping time $\tau^\star$, defined in~\eqref{eqn:True:StoppingTimes}, Lemma~\ref{lem:True:dV}, satisfies  $\tau^\star = \tstar$, then 
    \begin{align}\label{eqn:prop:Appendix:TrueProcess:NU:Bounds:Main}
        \LpLaw{\sfp}{}{N_{\mathcal{U}}(t)}
        \leq
        \Delta_g^2
        \frac{\expo{ 2\lambda t }}{\lambda}
        \left( 
            \frac{1}{\sqrt{\Boldomega}}
            \sup_{\nu \in [0,\tstar]}
            \LpLaw{2\sfp}{}{\LZparamu{\nu,\Zt{N,\nu}}}
            + 
            \mathfrak{p}'(\sfp)
            \sup_{\nu \in [0,\tstar]}
            \LpLaw{2\sfp}{}{\FZparasigma{\nu,\Zt{N,\nu}}}
        \right)^2
        ,
    \end{align}
    for all $(t,\sfp) \in [0,\tstar] \times \mathbb{N}_{\geq 1}$.

    

\end{proposition}
\begin{proof}

    We follow the derivation of~\eqref{eqn:prop:Appendix:ReferenceProcess:NU:Redefined} in the proof of Proposition~\ref{prop:Appendix:ReferenceProcess:NU:Bounds} to obtain the following representation of the process $N_{\mathcal{U}}(t)$:
    \begin{align}\label{eqn:prop:Appendix:TrueProcess:NU:Redefined}
        N_{\mathcal{U}}(t)
        = 
        \int_0^{t}
            \expo{ (2\lambda - \Boldomega) \nu }
            \mathcal{P}_\mathcal{U}(\nu)^\top
            \left(
            \int_0^{\nu}
                \expo{\Boldomega \beta}
                \left[ \LZparamu{\beta,\Zt{N,\beta}}d\beta + \FZparasigma{\beta,\Zt{N,\beta}}d\Wt{\beta} \right]
            \right)
        d\nu
        .
    \end{align}
    Using the definition of $\mathcal{P}_{\mathcal{U}}$ in Proposition~\ref{prop:Appendix:True:ddV:Bound}, followed by the definition of $\ReferenceInputZ$ in~\eqref{def:True:ErrorFunctions}, we get:   
    \begin{align*}
        \mathcal{P}_\mathcal{U}(\nu)
        =&
        2
        g(\nu)^\top
        g(\nu)
        \ReferenceInputZ[\Zt{N}][\nu]
        =
        2
        g(\nu)^\top
        g(\nu)
        \left(\ReferenceInput[\Xt{N}][\nu] - \ReferenceInput[\Xrt{N}][\nu]\right)
        .
    \end{align*}
    We next incorporate the definition of $\ReferenceInput$ in~\eqref{eqn:ReferenceFeedbackOperatorProcess} for the truncated process~\eqref{eqn:True:TruncatedJointProcess} to obtain:
    \begin{align*}
        \mathcal{P}_\mathcal{U}(\nu)
        =&
        -2 
        \Boldomega
        \expo{-\Boldomega \nu}
        g(\nu)^\top
        g(\nu)
        \int_0^\nu 
            \expo{\Boldomega \beta}
            \left(\Lparamu{\beta,\Xt{N,\beta}}-\Lparamu{\beta,\Xrt{N,\beta}}\right)d\beta 
        \\
        &
        \hspace{2cm}
        -2 
        \Boldomega
        \expo{-\Boldomega \nu}
        g(\nu)^\top
        g(\nu)
        \int_0^\nu 
            \expo{\Boldomega \beta}
            \left(\Fparasigma{\beta,\Xt{N,\beta}} - \Fparasigma{\beta,\Xrt{N,\beta}}\right)d\Wt{\beta}
        \\
        =&
        -2 
        \Boldomega
        \expo{-\Boldomega \nu}
        g(\nu)^\top
        g(\nu)
        \int_0^\nu 
            \expo{\Boldomega \beta}
            \left[ \LZparamu{\beta,\Zt{N,\beta}} d\beta + \FZparasigma{\beta,\Zt{N,\beta}}d\Wt{\beta} \right], 
    \end{align*}
    where have used the definitions of $\LZparamu{\beta,\Zt{N,\beta}}$ and $\FZparasigma{\beta,\Zt{N,\beta}}$ in~\eqref{def:True:ErrorFunctions}.
    Substituting the above expression into~\eqref{eqn:prop:Appendix:TrueProcess:NU:Redefined} produces 
    \begin{multline}\label{eqn:prop:Appendix:TrueProcess:NU:Redefined:Final}
        N_{\mathcal{U}}(t)
        = 
        - 2 \Boldomega
        \int_0^{t}
            \expo{ 2(\lambda - \Boldomega) \nu }
            \left(
                \int_0^\nu 
                    \expo{\Boldomega \beta}
                    \left[ \LZparamu{\beta,\Zt{N,\beta}} d\beta + \FZparasigma{\beta,\Zt{N,\beta}}d\Wt{\beta} \right]
            \right)^\top
        g(\nu)^\top
        g(\nu)
        \\
        \times 
        \left(
            \int_0^\nu 
                \expo{\Boldomega \beta}
                \left[ \LZparamu{\beta,\Zt{N,\beta}} d\beta + \FZparasigma{\beta,\Zt{N,\beta}}d\Wt{\beta} \right]
        \right)
        d\nu
        .
    \end{multline}
     Observe that $N_\mathcal{U}(t)$ can be expressed as the process $N(t)$ in the statement of Lemma~\ref{cor:TechnicalResults:UTildeBound} by setting 
    \begin{align*}
        Q_t = \Wt{t} \in \mathbb{R}^{d}\,(n_q=d), 
        \quad 
        \mathfrak{F}_t = \Wfilt{t},
        \quad 
        \theta_1 = \lambda, 
        \quad 
        \theta_2 = \Boldomega,
        \quad 
        \xi =0
        ,
        \\
        R(t) = -2\Boldomega g(t)^\top g(t) \in \mathcal{M}_{2}^{loc}\br{\mathbb{S}^m|\Wfilt{t}},
        \\
        S_1(t) = S_2(t) = \LZparamu{t,\Zt{N,t}} \in \mathcal{M}_{2}^{loc}\br{\mathbb{R}^m|\Wfilt{t}}, 
        \\
        L_1(t) = L_2(t) = \FZparasigma{t,\Zt{N,t}} 
        \in     
        \mathcal{M}_{2}^{loc}\br{\mathbb{R}^{m \times d}|\Wfilt{t}}
        .   
    \end{align*}
    Furthermore, as a consequence of Assumption~\ref{assmp:KnownFunctions}, we may set $\Delta_R$ in the hypothesis of Lemma~\ref{cor:TechnicalResults:UTildeBound} as $\Delta_R = 2 \Delta_g^2  \Boldomega$.
    Hence, applying Lemma~\ref{cor:TechnicalResults:UTildeBound} to the process $N_{\mathcal{U}}(t)$ in~\eqref{eqn:prop:Appendix:TrueProcess:NU:Redefined:Final} leads to 
    \begin{multline*}
        \LpLaw{\sfp}{}{N_{\mathcal{U}}(t)}
        \leq
        \Delta_g^2  \Boldomega
        \frac{\expo{ 2\lambda t } - 1}{\lambda}
        \left( 
            \frac{1-\expo{-\Boldomega t}}{\Boldomega}
            \sup_{\nu \in [0,\tstar]}
            \LpLaw{2\sfp}{}{\LZparamu{\nu,\Zt{N,\nu}}}
        \right.
        \\
        \left.
            + 
            \left(\sfp \frac{2\sfp-1}{2}\right)^\Half  
            \left(\frac{1 - \expo{ -2\Boldomega t }}{\Boldomega}\right)^\Half
            \sup_{\nu \in [0,\tstar]}
            \LpLaw{2\sfp}{}{\FZparasigma{\nu,\Zt{N,\nu}}}
        \right)^2
        ,
    \end{multline*}
   for all $(t,\sfp) \in [0,\tstar] \times \mathbb{N}_{\geq 1}$.
   Then, bounding $\expo{ 2\lambda t } - 1$ by $\expo{ 2\lambda t }$, and  $1-\expo{-\Boldomega t}$ and $1 - \expo{ -2\Boldomega t }$  by $1$ produces desired result in~\eqref{eqn:prop:Appendix:TrueProcess:NU:Bounds:Main}.

    \begingroup
    \endgroup
    
\end{proof}

The subsequent results quantify the effect of the feedabck error $\FL -  \ReferenceInput$ on the truncated joint process $\Zt{N,t}$.
We begin by deriving the moment bounds over the sampling interval $\BoldTs$ for the truncated true (uncertain) \ellonedrac process $\Xt{N,t}$. 
\begin{proposition}\label{prop:Appendix:TrueProcess:TsBound}
    Consider the truncated true (uncertain) \ellonedrac process $\Xt{N,t}$ presented in Definition~\ref{def:True:TruncatedJointProcess}.
    Then, for any $k \in \mathbb{N}_{\geq 1}$, for which 
    $k\BoldTs < \tau^\star$, we have the following bound 
    \begin{align}\label{eqn:prop:Appendix:TrueProcess:TsBound:Main}
        \LpLaw{2\sfp}{}{\Xt{N,k\BoldTs} - \Xt{N,t}}
        \leq 
        \BoldTs
        \widetilde{\delta}_1(\BoldTs,\sfp,\tau^\star) 
        +
        \BoldTs^\frac{1}{2}
        \widetilde{\delta}_2(\sfp,\tau^\star)
        ,
        \quad 
        \forall 
        (t,\sfp) \in [(k-1)\BoldTs,k\BoldTs] \times \mathbb{N}_{\geq 1},
    \end{align} 
    where the stopping time $\tau^\star$ is defined in Proposition~\ref{prop:True:TruncatedWellPosedness}, and  
    \begin{align*}
        \widetilde{\delta}_1(\BoldTs,\sfp,\tau^\star)
        =
        2
        \sup_{\nu \in [0,\tau^\star]}
        \LpLaw{2\sfp}{}{\Fbarmu{\nu,\Xt{N,\nu}}}
        + 
        2
        \left(
            1
            + 
            \Delta_g 
            \Delta_\Theta
            e^{-\lambda_s \BoldTs}
        \right)
        \sup_{\nu \in [0,\tau^\star]}
        \LpLaw{2\sfp}{}{ 
            \Lmu{\nu,\Xt{N,\nu}}
        }
        \\
        +
        2 
        \mathfrak{p}'(\sfp)
        \Delta_g 
        \Delta_\Theta
        e^{-\lambda_s \BoldTs}
        \gamma_1\left(\BoldTs\right)
        \sup_{\nu \in [0,\tau^\star]}
        \LpLaw{2\sfp}{}{
            \Fsigma{\nu,\Xt{N,\nu}}
        } 
        ,
        \\
        \widetilde{\delta}_2(\sfp,\tau^\star)
        =
        2
        \sqrt{2}
        \mathfrak{p}'(\sfp)  
        \sup_{\nu \in [0,\tau^\star]}
        \LpLaw{2\sfp}{}{\Fsigma{\nu,\Xt{N,\nu}}}
        ,
        \\
        \gamma_1\left(\BoldTs\right)
        \doteq 
        \left(
            \lambda_s
            \frac{ \expo{ \lambda_s \BoldTs } + 1 }{ \expo{ \lambda_s \BoldTs } - 1 }
        \right)^\Half
        =
        \left(
            \lambda_s
            \frac{ 1 + \expo{ -\lambda_s \BoldTs } }{ 1 - \expo{ -\lambda_s \BoldTs } }
        \right)^\Half
        .
    \end{align*}
\end{proposition}
\begin{proof}
    From Proposition~\ref{prop:True:TruncatedWellPosedness}, $\Zt{N,t}$ is a unique strong solution of~\eqref{eqn:True:JointProcess}, for all $t \in [0,\tau_N]$. 
    Moreover, $\tau^\star \leq \tau_N$ by definition, therefore implying that $\Xt{N,t}$ is a unique strong solution of~\eqref{eqn:True:SDE} in Definition~\ref{def:True:TrueL1Process}, for all $t \in [0,\tau^\star]$.
    That is, 
    \begin{align*}
        d\Xt{N,t} = \Fmu{t,\Xt{N,t},\ULt{t}}dt + \Fsigma{t,\Xt{N,t}}d\Wt{t}, \quad t \in [0,\tau^\star],
        \quad 
        \ULt{t} \doteq \FL[\Xt{N}][t]
        , 
    \end{align*}
    with $\Xt{N,0} = x_0 \sim \xi_0$.
    Hence,
    \begin{align*}
        \Xt{N,t} 
        = 
        \Xt{N,0}
        + 
        \int_0^t
            \Fmu{\nu,\Xt{N,\nu},\ULt{\nu}}d\nu 
        +
        \int_0^t 
            \Fsigma{\nu,\Xt{N,\nu}}d\Wt{\nu}, \quad t \in [0,\tau^\star],
        \quad 
        \ULt{t} \doteq \FL[\Xt{N}][t]
        . 
    \end{align*}
    Since $(k-1)\BoldTs$ is a constant, and hence a stopping time, it then follows that for any $t \in [(k-1)\BoldTs, k\BoldTs]$, we have
    \begin{align*}
         \Xt{N,t} - \Xt{N,(k-1)\BoldTs} 
        = 
        \int_{(k-1)\BoldTs}^t
            \Fmu{\nu,\Xt{N,\nu},\ULt{\nu}}d\nu 
        +
        \int_{(k-1)\BoldTs}^t
            \Fsigma{\nu,\Xt{N,\nu}}d\Wt{\nu}, 
        \\
        \ULt{t} \doteq \FL[\Xt{N}][t]
        ,
    \end{align*}
    which yields the following due to the Minkowski's inequality:
    \begin{multline}\label{eqn:prop:Appendix:TrueProcess:TsBound:ErrorIntegral:1}
        \LpLaw{2\sfp}{}{
            \Xt{N,t} - \Xt{N,(k-1)\BoldTs}
        } 
        \leq  
        \LpLaw{2\sfp}{}{
            \int_{(k-1)\BoldTs}^t
                \norm{\Fmu{\nu,\Xt{N,\nu},\ULt{\nu}}} 
            d\nu
        } 
        \\
        +
        \LpLaw{2\sfp}{}{
            \int_{(k-1)\BoldTs}^t
                \Fsigma{\nu,\Xt{N,\nu}}
            d\Wt{\nu}
        }
        , 
    \end{multline} 
    for all $(t,\sfp) \in [(k-1)\BoldTs,k\BoldTs] \times \mathbb{N}_{\geq 1}$, where $\ULt{t} \doteq \FL[\Xt{N}][t]$.

    Now, using Corollary~\ref{cor:TechnicalResults:LebesgueMoment} with $\xi = (k-1)\BoldTs$ gives us 
    \begin{multline*}  
        \LpLaw{2\sfp}{}{
            \int_{(k-1)\BoldTs}^t
                \norm{\Fmu{\nu,\Xt{N,\nu},\ULt{\nu}}} 
            d\nu
        }
        \leq      
        (t-(k-1)\BoldTs)
        \sup_{\nu \in [(k-1)\BoldTs,t]}
        \LpLaw{2\sfp}{}{\Fmu{\nu,\Xt{N,\nu},\ULt{\nu}}}
        \\ 
        \leq      
        \BoldTs
        \sup_{\nu \in [(k-1)\BoldTs,k\BoldTs]}
        \LpLaw{2\sfp}{}{\Fmu{\nu,\Xt{N,\nu},\ULt{\nu}}}
        ,
    \end{multline*}
    for all $t \in [(k-1)\BoldTs,k\BoldTs]$.
    Using~\eqref{eqn:VectorFields:Decomposition} in Definition~\ref{def:VectorFields} and the Minkowski's inequality, we can further develop the above bound as
    \begin{multline}\label{eqn:prop:Appendix:TrueProcess:TsBound:ErrorIntegral:Drift:Initial}  
        \LpLaw{2\sfp}{}{
            \int_{(k-1)\BoldTs}^t
                \norm{\Fmu{\nu,\Xt{N,\nu},\ULt{\nu}}} 
            d\nu
        }
        \\ 
        \leq     
        \BoldTs
        \sup_{\nu \in [(k-1)\BoldTs,k\BoldTs]}
        \left(
            \LpLaw{2\sfp}{}{\Fbarmu{\nu,\Xt{N,\nu}}}
            + 
            \Delta_g 
            \LpLaw{2\sfp}{}{\ULt{\nu}}
            + 
            \LpLaw{2\sfp}{}{\Lmu{\nu,\Xt{N,\nu}}}
        \right)
        ,
    \end{multline}
    for all $ t \in [(k-1)\BoldTs,k\BoldTs]$, where we have used the uniform bound on $\Frobenius{g(t)}$ in Assumption~\ref{assmp:KnownFunctions}.   
    Next, from~\eqref{eqn:True:L1DRACInput} and~\eqref{eqn:True:Filter}, we have
    \begin{align*}
        \ULt{t} = \FL[\Xt{N}][t] = \Filter[\hat{\Lambda}^{\paral}][t]
        =  
        - \Boldomega \int_0^t \expo{-\Boldomega(t-\nu)}\Lparahat{\nu}d\nu,
        \quad 
        t \in [(k-1)\BoldTs,k\BoldTs]
        ,
    \end{align*}
    and thus  
    \begin{align*}
        \LpLaw{2\sfp}{}{\ULt{t} }
        \leq   
        \Boldomega
        \expo{-\Boldomega t } 
        \LpLaw{2\sfp}{}{
            \int_0^t \expo{ \Boldomega \nu } \norm{\Lparahat{\nu}}d\nu
        },
        \quad 
        \forall
        t \in [(k-1)\BoldTs,k\BoldTs]
        .
    \end{align*}
    It then follows from Proposition~\ref{prop:TechnicalResults:LebesgueMoment} that 
    \begin{align*}
        \LpLaw{2\sfp}{}{\ULt{t} }
        \leq 
        \left(1-\expo{-\Boldomega t} \right)
        \sup_{\nu \in [0,t]}
        \LpLaw{2\sfp}{}{ \Lparahat{\nu} }
        \leq&   
        \sup_{\nu \in [0,t]}
        \LpLaw{2\sfp}{}{ \Lparahat{\nu} }
        \leq   
        \sup_{\nu \in [0,k\BoldTs]}
        \LpLaw{2\sfp}{}{ \Lparahat{\nu} }
        ,
    \end{align*}
    for all $t \in [(k-1)\BoldTs,k\BoldTs]$.
    The above further implies
    \begin{align}\label{eqn:prop:Appendix:TrueProcess:TsBound:ErrorIntegral:ControlBound:Initial}
        \sup_{\nu \in [(k-1)\BoldTs,k\BoldTs]}
        \LpLaw{2\sfp}{}{\ULt{\nu}}
        \leq
        \sup_{\nu \in [0,k\BoldTs]}
        \LpLaw{2\sfp}{}{ \Lparahat{\nu} }
        .
    \end{align}  
    Now, we obtain the following definition of $\Lparahat{\nu}$ from~\eqref{eqn:prop:Appendix:TrueProcess:FL:AdaptationLaw:Matched:Final}, Propostion~\ref{prop:Appendix:TrueProcess:FL:Expression}, with $\tau = k\BoldTs$ and $z = \Xt{N}$: 
    \begin{multline*}
        \Lparahat{\nu} 
        = 
        0_m  \indicator{[0,\BoldTs)}{\nu}
        \\
        -
        \sum_{l=1}^k
        \left( 
            \lambda_s \left(1 - e^{\lambda_s \BoldTs}\right)^{-1}
            \Theta_{ad}(l\BoldTs)
            \int_{(l-1)\BoldTs}^{l\BoldTs} \expo{-\lambda_s (l\BoldTs-\beta)} \Sigma_\beta(\Xt{N})
        \right)
        \indicator{ [l\BoldTs,(l+1)\BoldTs)}{\nu},
        \quad  
        \nu \in [0,k\BoldTs]
        , 
    \end{multline*} 
    with $k \in \mathbb{N}_{\geq 1}$, where we have the following \emph{formal} definition:  
    \begin{align}\label{eqn:prop:Appendix:TrueProcess:TsBound:Sigma}
        \Sigma_t(\Xt{N})
        =
        \Lmu{t,\Xt{N,t}}dt
        +
        \Fsigma{t,\Xt{N,t}}d\Wt{t}
        , 
        \quad t \in [0,k\BoldTs].
    \end{align}
    Therefore, 
    \begin{multline*}
        \norm{\Lparahat{\nu}} 
        \leq 
        0 \cdot \indicator{[0,\BoldTs)}{\nu}
        +
        \sum_{l=1}^k
        \left( 
            \lambda_s \left(e^{\lambda_s \BoldTs}-1\right)^{-1}
            \Delta_\Theta
            \norm{\int_{(l-1)\BoldTs}^{l\BoldTs} \expo{-\lambda_s (l\BoldTs-\beta)} \Sigma_\beta(\Xt{N})}
        \right)
        \indicator{ [l\BoldTs,(l+1)\BoldTs)}{\nu}
        ,
    \end{multline*}
    for all $\nu \in [0,k\BoldTs]$, where we have used the uniform bound on $\Theta_{ad}(t)$ in Assumption~\ref{assmp:KnownFunctions}. 
    The last inequality allows us to bound the $2\sfp^{th}$ moment as follows:
    \begin{multline*}
        \LpLaw{2\sfp}{}{ \Lparahat{\nu} } 
        \leq 
        0 \cdot \indicator{[0,\BoldTs)}{\nu}
        \\
        +
        \sum_{l=1}^k
        \left( 
            \lambda_s \left(e^{\lambda_s \BoldTs}-1\right)^{-1}
            \Delta_\Theta
            \LpLaw{2\sfp}{}{\int_{(l-1)\BoldTs}^{l\BoldTs} \expo{-\lambda_s (l\BoldTs-\beta)} \Sigma_\beta(\Xt{N})}
        \right)
        \indicator{ [l\BoldTs,(l+1)\BoldTs)}{\nu}
        , 
    \end{multline*}
    for all $\nu \in [0,k\BoldTs]$.
    \begingroup
    \endgroup
    Thus, over each piecewise temporal interval $[l\BoldTs, (l+1)\BoldTs)$, $l \in \cbr{1,\dots,k}$, we have  
    \begin{align*}
        \LpLaw{2\sfp}{}{ \Lparahat{\nu} } 
        \leq&  
        \lambda_s \left(e^{\lambda_s \BoldTs}-1\right)^{-1}
        \Delta_\Theta
        \LpLaw{2\sfp}{}{\int_{(l-1)\BoldTs}^{l\BoldTs} \expo{-\lambda_s (l\BoldTs-\beta)} \Sigma_\beta(\Xt{N})}
        \notag 
        \\
        \leq&  
        \lambda_s \left(e^{\lambda_s \BoldTs}-1\right)^{-1}
        \Delta_\Theta
        \expo{-\lambda_s l\BoldTs }
        \LpLaw{2\sfp}{}{
            \int_{(l-1)\BoldTs}^{l\BoldTs} 
                \expo{\lambda_s \beta } 
                \Sigma_\beta(\Xt{N})},
        \quad 
        \forall 
        \nu \in [l\BoldTs, (l+1)\BoldTs),
    \end{align*}
    where the last inequality is due to both $l$ and $\BoldTs$ being constants.
    Substituing the expression in~\eqref{eqn:prop:Appendix:TrueProcess:TsBound:Sigma} into the above and using the Minkowski's inequality leads to 
    \begin{multline*}
        \LpLaw{2\sfp}{}{ \Lparahat{\nu} } 
        \leq  
        \lambda_s \left(e^{\lambda_s \BoldTs}-1\right)^{-1}
        \Delta_\Theta
        \expo{-\lambda_s l\BoldTs }
        \LpLaw{2\sfp}{}{
            \int_{(l-1)\BoldTs}^{l\BoldTs} 
                \expo{\lambda_s \beta } 
                \Lmu{\beta,\Xt{N,\beta}}
            d\beta
        }
        \\
        +
        \lambda_s \left(e^{\lambda_s \BoldTs}-1\right)^{-1}
        \Delta_\Theta
        \expo{-\lambda_s l\BoldTs }
        \LpLaw{2\sfp}{}{
            \int_{(l-1)\BoldTs}^{l\BoldTs} 
                \expo{\lambda_s \beta } 
                \Fsigma{\beta,\Xt{N,\beta}}
            d\Wt{\beta}
        }
        ,
    \end{multline*}
    for all $\nu \in [l\BoldTs, (l+1)\BoldTs)$, $l \in \cbr{1,\dots,k}$.
    We now use Proposition~\ref{prop:TechnicalResults:LebesgueMoment} and Lemma~\ref{lemma:TechnicalResults:MartingaleMoment} to bound the terms on the right-hand side of the above expression 
    \begin{multline*}
        \LpLaw{2\sfp}{}{ \Lparahat{\nu} }   
        \leq  
        \Delta_\Theta
        e^{-\lambda_s \BoldTs}
        \sup_{\nu \in [(l-1)\BoldTs,l \BoldTs]}
        \LpLaw{2\sfp}{}{ 
            \Lmu{\nu,\Xt{N,\nu}}
        }
        \\
        + 
        \Delta_\Theta
        \mathfrak{p}'(\sfp)
        e^{-\lambda_s \BoldTs}
        \left(
            \lambda_s
            \frac{ \expo{ \lambda_s \BoldTs } + 1 }{ \expo{ \lambda_s \BoldTs } - 1 }
        \right)^\Half 
        \sup_{\nu \in [(l-1)\BoldTs,l \BoldTs]}
        \LpLaw{2\sfp}{}{
            \Fsigma{\nu,\Xt{N,\nu}}
        }
        ,
    \end{multline*}
     for all $\nu \in [l\BoldTs, (l+1)\BoldTs)$, $l \in \cbr{1,\dots,k}$, where we have used the following: 
    \begin{align*}
        \left(e^{\lambda_s \BoldTs}-1\right)^{-1}
        \left( \expo{ 2\lambda_s \BoldTs } - 1  \right)^\Half 
        = 
        \left(e^{\lambda_s \BoldTs}-1\right)^{-1}
        \left( \expo{ \lambda_s \BoldTs } - 1  \right)^\Half
        \left( \expo{ \lambda_s \BoldTs } + 1  \right)^\Half
        = 
        \left(
            \frac{ \expo{ \lambda_s \BoldTs } + 1 }{ \expo{ \lambda_s \BoldTs } - 1 }
        \right)^\Half 
        .
    \end{align*}
    Since $k \in \mathbb{N}_{\geq 1}$ satisfies $k\BoldTs < \tau^\star$, we have that $[(l-1)\BoldTs,l \BoldTs]\subset [0,\tau^\star]$, for all $l \in \cbr{1,\dots,k}$.
    Thus, we can further develop the above inequality into
    \begin{multline*}
        \LpLaw{2\sfp}{}{ \Lparahat{\nu} }   
        \leq  
        \Delta_\Theta
        e^{-\lambda_s \BoldTs}
        \sup_{\nu \in [0,\tau^\star]}
        \LpLaw{2\sfp}{}{ 
            \Lmu{\nu,\Xt{N,\nu}}
        }
        \\
        + 
        \Delta_\Theta
        \mathfrak{p}'(\sfp)
        e^{-\lambda_s \BoldTs}
        \left(
            \lambda_s
            \frac{ \expo{ \lambda_s \BoldTs } + 1 }{ \expo{ \lambda_s \BoldTs } - 1 }
        \right)^\Half 
        \sup_{\nu \in [0,\tau^\star]}
        \LpLaw{2\sfp}{}{
            \Fsigma{\nu,\Xt{N,\nu}}
        }
        ,
    \end{multline*}
    for all $\nu \in [l\BoldTs, (l+1)\BoldTs)$, $l \in \cbr{1,\dots,k}$.
    Since 
    \begin{align*}
    \sup_{\nu \in [0,k\BoldTs]}
    \LpLaw{2\sfp}{}{ \Lparahat{\nu} }
    \leq& 
    \max_{l \in \cbr{1,\dots,k}}
    \sup_{\nu \in [(l-1)\BoldTs,l \BoldTs]}
    \LpLaw{2\sfp}{}{ \Lparahat{\nu} } ,
    \end{align*}
    the previous bound implies 
    \begin{multline*}
        \sup_{\nu \in [0,k\BoldTs]}
        \LpLaw{2\sfp}{}{ \Lparahat{\nu} }
        \leq 
        \Delta_\Theta
        e^{-\lambda_s \BoldTs}
        \sup_{\nu \in [0,\tau^\star]}
        \LpLaw{2\sfp}{}{ 
            \Lmu{\nu,\Xt{N,\nu}}
        }
        \\
        + 
        \Delta_\Theta
        \mathfrak{p}'(\sfp)
        e^{-\lambda_s \BoldTs}
        \left(
            \lambda_s
            \frac{ \expo{ \lambda_s \BoldTs } + 1 }{ \expo{ \lambda_s \BoldTs } - 1 }
        \right)^\Half 
        \sup_{\nu \in [0,\tau^\star]}
        \LpLaw{2\sfp}{}{
            \Fsigma{\nu,\Xt{N,\nu}}
        }
        .
    \end{multline*}
     Substiuting this bound into~\eqref{eqn:prop:Appendix:TrueProcess:TsBound:ErrorIntegral:ControlBound:Initial}:
    \begin{multline*}
        \sup_{\nu \in [(k-1)\BoldTs,k\BoldTs]}
        \LpLaw{2\sfp}{}{\ULt{\nu}}
        \leq 
        \Delta_\Theta
        e^{-\lambda_s \BoldTs}
        \sup_{\nu \in [0,\tau^\star]}
        \LpLaw{2\sfp}{}{ 
            \Lmu{\nu,\Xt{N,\nu}}
        }
        \\
        + 
        \Delta_\Theta
        \mathfrak{p}'(\sfp)
        e^{-\lambda_s \BoldTs}
        \left(
            \lambda_s
            \frac{ \expo{ \lambda_s \BoldTs } + 1 }{ \expo{ \lambda_s \BoldTs } - 1 }
        \right)^\Half 
        \sup_{\nu \in [0,\tau^\star]}
        \LpLaw{2\sfp}{}{
            \Fsigma{\nu,\Xt{N,\nu}}
        },
    \end{multline*}   
    which we further substitute into~\eqref{eqn:prop:Appendix:TrueProcess:TsBound:ErrorIntegral:Drift:Initial} and get
    \begin{multline}\label{eqn:prop:Appendix:TrueProcess:TsBound:ErrorIntegral:Drift:Final}  
        \LpLaw{2\sfp}{}{
            \int_{(k-1)\BoldTs}^t
                \norm{\Fmu{\nu,\Xt{N,\nu},\ULt{\nu}}} 
            d\nu
        }
        \\ 
        \leq     
        \BoldTs
        \sup_{\nu \in [0,\tau^\star]}
        \LpLaw{2\sfp}{}{\Fbarmu{\nu,\Xt{N,\nu}}}
        + 
        \BoldTs
        \left(
            1
            + 
            \Delta_g 
            \Delta_\Theta
            e^{-\lambda_s \BoldTs}
        \right)
        \sup_{\nu \in [0,\tau^\star]}
        \LpLaw{2\sfp}{}{ 
            \Lmu{\nu,\Xt{N,\nu}}
        }
        \\
        + 
        \mathfrak{p}'(\sfp)
        \BoldTs
        \Delta_g 
        \Delta_\Theta
        e^{-\lambda_s \BoldTs}
        \left(
            \lambda_s
            \frac{ \expo{ \lambda_s \BoldTs } + 1 }{ \expo{ \lambda_s \BoldTs } - 1 }
        \right)^\Half 
        \sup_{\nu \in [0,\tau^\star]}
        \LpLaw{2\sfp}{}{
            \Fsigma{\nu,\Xt{N,\nu}}
        }
        ,
    \end{multline}
    for all $ t \in [(k-1)\BoldTs,k\BoldTs]$, where, similar as above, we have generalized the upper bound to $[0,\tau^\star]$.

    Next, we invoke \cite[Thm.~1.7.1]{mao2007stochastic} (as used in~\cite[Thm.~2.4.3]{mao2007stochastic}) to obtain  
    \begin{multline*}
        \ELaw{}{
            \norm{
                \int_{(k-1)\BoldTs}^t
                    \Fsigma{\nu,\Xt{N,\nu}}
                d\Wt{\nu}
            }^{2\sfp}
        }
        \\
        \leq 
        \left(\sfp\left(2\sfp-1\right)\right)^\sfp 
        \left( t - (k-1)\BoldTs \right)^{\sfp-1}
        \ELaw{}{
            \int_{(k-1)\BoldTs}^t
                \norm{\Fsigma{\nu,\Xt{N,\nu}}}^{2\sfp}
            d\nu
        }
        ,
        \quad 
        \forall  t \in [(k-1)\BoldTs,k\BoldTs]
        .
    \end{multline*}
    Hence, it follows from the Fubini's theorem that 
    \begin{multline*}
        \ELaw{}{
            \norm{
                \int_{(k-1)\BoldTs}^t
                    \Fsigma{\nu,\Xt{N,\nu}}
                d\Wt{\nu}
            }^{2\sfp}
        }
        \\
        \leq 
        \left(\sfp\left(2\sfp-1\right)\right)^\sfp  
        \left( t - (k-1)\BoldTs \right)^{\sfp-1}
        \int_{(k-1)\BoldTs}^t
            \ELaw{}{\norm{\Fsigma{\nu,\Xt{N,\nu}}}^{2\sfp}}
        d\nu
        \\
        \leq 
        \left(\sfp\left(2\sfp-1\right)\right)^\sfp  
        \left( t - (k-1)\BoldTs \right)^\sfp
        \sup_{\nu \in [t,(k-1)\BoldTs]}
        \ELaw{}{\norm{\Fsigma{\nu,\Xt{N,\nu}}}^{2\sfp}}
        ,
    \end{multline*}
    for all $ t \in [(k-1)\BoldTs,k\BoldTs]$, which further implies
    \begin{multline*}
        \ELaw{}{
            \norm{
                \int_{(k-1)\BoldTs}^t
                    \Fsigma{\nu,\Xt{N,\nu}}
                d\Wt{\nu}
            }^{2\sfp}
        }
        \leq 
        \left(\sfp\left(2\sfp-1\right)\right)^\sfp  
        \BoldTs^\sfp
        \sup_{\nu \in [k\BoldTs,(k-1)\BoldTs]}
        \ELaw{}{\norm{\Fsigma{\nu,\Xt{N,\nu}}}^{2\sfp}}
        \\
        \leq 
        \left(\sfp\left(2\sfp-1\right)\right)^\sfp  
        \BoldTs^\sfp
        \sup_{\nu \in [0,\tau^\star]}
        \ELaw{}{\norm{\Fsigma{\nu,\Xt{N,\nu}}}^{2\sfp}}
        ,
    \end{multline*}
    for all $ t \in [(k-1)\BoldTs,k\BoldTs]$.
    Hence,  
    \begin{multline}\label{eqn:prop:Appendix:TrueProcess:TsBound:ErrorIntegral:Diffusion:Final}
        \LpLaw{2\sfp}{}{
            \int_{(k-1)\BoldTs}^t
                \Fsigma{\nu,\Xt{N,\nu}}
            d\Wt{\nu}
        }
        \leq 
        \left(\sfp\left(2\sfp-1\right)\right)^\frac{1}{2}  
        \BoldTs^\frac{1}{2}
        \sup_{\nu \in [0,\tau^\star]}
        \LpLaw{2\sfp}{}{\Fsigma{\nu,\Xt{N,\nu}}}
        \\
        =
        \sqrt{2}
        \mathfrak{p}'(\sfp)
        \BoldTs^\frac{1}{2}
        \sup_{\nu \in [0,\tau^\star]}
        \LpLaw{2\sfp}{}{\Fsigma{\nu,\Xt{N,\nu}}}
        ,
    \end{multline}
    for all $ t \in [(k-1)\BoldTs,k\BoldTs]$.
    Substituting~\eqref{eqn:prop:Appendix:TrueProcess:TsBound:ErrorIntegral:Drift:Final} and~\eqref{eqn:prop:Appendix:TrueProcess:TsBound:ErrorIntegral:Diffusion:Final} into~\eqref{eqn:prop:Appendix:TrueProcess:TsBound:ErrorIntegral:1} produces
     \begin{multline*}
        \LpLaw{2\sfp}{}{
            \Xt{N,t} - \Xt{N,(k-1)\BoldTs}
        } 
        \\
        \leq  
        \BoldTs
        \sup_{\nu \in [0,\tau^\star]}
        \LpLaw{2\sfp}{}{\Fbarmu{\nu,\Xt{N,\nu}}}
        + 
        \BoldTs
        \left(
            1
            + 
            \Delta_g 
            \Delta_\Theta
            e^{-\lambda_s \BoldTs}
        \right)
        \sup_{\nu \in [0,\tau^\star]}
        \LpLaw{2\sfp}{}{ 
            \Lmu{\nu,\Xt{N,\nu}}
        }
        \\
        + 
        \mathfrak{p}'(\sfp)
        \BoldTs
        \Delta_g 
        \Delta_\Theta
        e^{-\lambda_s \BoldTs}
        \left(
            \lambda_s
            \frac{ \expo{ \lambda_s \BoldTs } + 1 }{ \expo{ \lambda_s \BoldTs } - 1 }
        \right)^\Half 
        \sup_{\nu \in [0,\tau^\star]}
        \LpLaw{2\sfp}{}{
            \Fsigma{\nu,\Xt{N,\nu}}
        } 
        \\
        +
        \sqrt{2}
        \mathfrak{p}'(\sfp)  
        \BoldTs^\frac{1}{2}
        \sup_{\nu \in [0,\tau^\star]}
        \LpLaw{2\sfp}{}{\Fsigma{\nu,\Xt{N,\nu}}}
        , 
    \end{multline*} 
    for all $(t,\sfp) \in [(k-1)\BoldTs,k\BoldTs] \times \mathbb{N}_{\geq 1}$. 
    The proof is then concluded by using this bound into the following consequence of the triangle inequality: 
    \begin{align*}
        \LpLaw{2\sfp}{}{\Xt{N,k\BoldTs} - \Xt{N,t}}
        \leq 
        \LpLaw{2\sfp}{}{\Xt{N,t} - \Xt{N,(k-1)\BoldTs}}
        + 
        \LpLaw{2\sfp}{}{\Xt{N,k\BoldTs} - \Xt{N,(k-1)\BoldTs}} 
        ,
    \end{align*}
    for all $(t,\sfp) \in [(k-1)\BoldTs,k\BoldTs] \times \mathbb{N}_{\geq 1}$. 

    \begingroup
    \endgroup

    \begingroup
    \endgroup
   
\end{proof}

The next result establishes uniform bounds on pertinenet integrands that define the feedback error operator $\FL - \ReferenceInput$.
\begin{proposition}\label{prop:Appendix:TrueProcess:ControlErrorIntegrand:Bounds}
    Consider the terms $\widetilde{\mathcal{H}}^{\cbr{\cdot,\paral}}_{\cbr{\mu,\sigma}}$, $\mathcal{G}_\mu$, and $\widetilde{\mathcal{G}}_{\cbr{\sigma_1,\sigma_2}}$, defined in the statement of Proposition~\ref{proposition:Appendix:TrueProcess:FL:ExpressionEvolved}.  
    Then, if the stopping time $\tau^\star$, defined in~\eqref{eqn:True:StoppingTimes}, Lemma~\ref{lem:True:dV}, satisfies  $\tau^\star = \tstar$ and $\tstar \geq \BoldTs$, we have the following bounds for all $(t,\sfp) \in [0,\tstar] \times \mathbb{N}_{\geq 1}$:
    \begin{subequations}\label{eqn:lem:Appendix:TrueProcess:NUTilde:ErrorDrift:Bounds:All:Final}
        \begin{align}
            \LpLaw{2\sfp}{}{
                \widetilde{\mathcal{H}}_\mu^{\paral}(t,\Xt{N})
                + 
                \expo{(\lambda_s-\Boldomega)t}
                \widetilde{\mathcal{H}}_\mu(t,\Xt{N})
            }
            \leq
            \Delta_{\widetilde{\mathcal{H}}_\mu}(\BoldTs,\sfp,\tstar)
            ,
            \label{eqn:lem:Appendix:TrueProcess:NUTilde:ErrorDrift:Bounds:A:Final}
            \\
            \LpLaw{2\sfp}{}{
                \widetilde{\mathcal{H}}_\sigma^{\paral}(t,\Xt{N})
                + 
                \expo{(\lambda_s-\Boldomega)t}
                \widetilde{\mathcal{H}}_\sigma(t,\Xt{N})
            }
            \leq 
            \Delta_{\widetilde{\mathcal{H}}_\sigma}(\BoldTs,\sfp,\tstar)
            ,
            \label{eqn:lem:Appendix:TrueProcess:NUTilde:ErrorDrift:Bounds:B:Final}
            \\
            \LpLaw{2\sfp}{}{
                \widetilde{\mathcal{G}}_{\sigma_1}(t,\Xt{N})
                + 
                \expo{(\lambda_s-\Boldomega)t}
                \widetilde{\mathcal{G}}_{\sigma_2}(t,\Xt{N})
            }
            \leq 
            \Delta_{\widetilde{\mathcal{G}}_\sigma}(\BoldTs,\sfp,\tstar)
            ,
            \label{eqn:lem:Appendix:TrueProcess:NUTilde:ErrorDrift:Bounds:C:Final}
            \\ 
            \label{eqn:lem:Appendix:TrueProcess:NUTilde:ErrorDrift:Bounds:D:Final}
            \LpLaw{2\sfp}{}{
                \mathcal{G}_\mu(t,\Xt{N})
            }
            \leq 
            \Delta_{\mathcal{G}_\mu}(\BoldTs,\tstar)
            ,
        \end{align}
    \end{subequations}
    where
    \begin{align*}
        \Delta_{\widetilde{\mathcal{H}}_\mu}(\BoldTs,\sfp,\tstar)
        =&
        \BoldTsroot
        \gamma_\mu\left(\Boldomega,\BoldTs\right)
        \widetilde{\delta}_2(\sfp,\tstar)
        +
        \BoldTs
        \left(
            \widehat{\gamma}_\mu\left(\Boldomega,\BoldTs\right)
            + 
            \gamma_\mu\left(\Boldomega,\BoldTs\right)
            \widetilde{\delta}_1(\BoldTs,\sfp,\tstar)
        \right)
        ,
        \\
        \Delta_{\widetilde{\mathcal{H}}_\sigma}(\BoldTs,\sfp,\tstar)
        =&
        \gamma_{\sigma}\left(\Boldomega,\BoldTs\right)
        \left(
            \BoldTs
            \widetilde{\delta}_1(\BoldTs,\sfp,\tstar) 
            +
            \BoldTs^\frac{1}{2}
            \widetilde{\delta}_2(\sfp,\tstar)
        \right)^\frac{1}{2}
        +  
        \BoldTs
        \widehat{\gamma}_{\sigma}\left(\Boldomega,\BoldTs\right)
        , 
        \\
        \Delta_{\widetilde{\mathcal{G}}_\sigma}(\BoldTs,\sfp,\tstar)
        =&
        \left(1 + \gamma_2(\Boldomega,\BoldTs)\right)
        \\
        &\times 
        \begin{multlined}[t][0.4\linewidth]
            \left[
                \left(
                    L^{\paral}_p
                    + 
                    L^{\paral}_\sigma
                \right)
                \left(
                    \BoldTs
                    \widetilde{\delta}_1(\BoldTs,\sfp,\tstar) 
                    +
                    \BoldTs^\frac{1}{2}
                    \widetilde{\delta}_2(\sfp,\tstar)
                \right)^\frac{1}{2}
            +  
                \BoldTs
                \left( 
                    \hat{L}^{\paral}_p 
                    + 
                    \hat{L}^{\paral}_\sigma
                \right)
            \right],
        \end{multlined}
        \\
        \Delta_{\mathcal{G}_\mu}(\BoldTs,\tstar)
        =&
        \left(1 - \expo{-\lambda_s \BoldTs}\right)
        \sup_{t \in [0,\tstar]}
        \LpLaw{2\sfp}{}{\Lparamu{t,\Xt{N,t}}}
        ,
    \end{align*}
    and where $\widetilde{\delta}_1(\BoldTs,\sfp,\tstar)$, and $\widetilde{\delta}_2(\sfp,\tstar)$ are defined in the statement of Proposition~\ref{prop:Appendix:TrueProcess:TsBound}, $\gamma_2(\Boldomega,\BoldTs)$ is defined in~\eqref{eqn:app:Functions:True:gammaTs:Numbered}, and $\gamma_{\cbr{\mu,\sigma}}\left(\Boldomega,\BoldTs\right)$ and $\widehat{\gamma}_{\cbr{\mu,\sigma}}\left(\Boldomega,\BoldTs\right)$ are defined in~\eqref{eqn:app:Functions:True:gammaTs:MuSigma}.

\end{proposition}
\begin{proof}
    We first begin by recalling the definitions in~\eqref{eqn:prop:Appendix:Scratch:Technical:Total:TildeFunctions:Drift}-\eqref{eqn:prop:Appendix:Scratch:Technical:Total:TildeFunctions:Diffusion} with $z = \Xt{N}$ and for any $i \in \mathbb{N}_{\geq 1}$:
    \begin{align*}
        \LmuErrorAll{}{t,i\BoldTs,\Xt{N}}
        =&
        \LmuAll{t,\Xt{N,t}} - \LmuAll{i\BoldTs,\Xt{N,i\BoldTs}}
        , 
        \\ 
        \FsigmaErrorAll{}{t,i\BoldTs,\Xt{N}}
        =&
        \FsigmaAll{t,\Xt{N,t}} - \FsigmaAll{i\BoldTs,\Xt{N,i\BoldTs}}
        \\
        =&
        \left(\psigmaAll{t,\Xt{N,t}} - \psigmaAll{i\BoldTs,\Xt{N,i\BoldTs}}\right)
        + 
        \left(\LsigmaAll{t,\Xt{N,t}} - \LsigmaAll{i\BoldTs,\Xt{N,i\BoldTs}}\right)
        ,
    \end{align*}
    where we have used the decompositions $\Fsigma{\nu,\cdot} = p(\nu,\cdot) + \Lsigma{\nu,\cdot}$ and $\Fparasigma{\nu,\cdot} = \ppara{\nu,\cdot} + \Lparasigma{\nu,\cdot}$ from Definitions~\ref{def:VectorFields} and~\ref{def:Diffusion:Decomposed}, respectively.

    It then follows from Assumptions~\ref{assmp:KnownFunctions},~\ref{assmp:knownDiffusion:Decomposition},~\ref{assmp:LipschitzContinuity}, and~\ref{assmp:KnownFunctions:GlobalLip}, and the Minkowski's inequality that 
    \begin{align*}
        \LpLaw{2\sfp}{}{
            \LmuErrorAll{}{t,i\BoldTs,\Xt{N}}
        }
        \leq& 
        \hat{L}^{\cbr{\cdot,\paral}}_\mu 
        \BoldTs 
        + 
        L^{\cbr{\cdot,\paral}}_\mu
        \LpLaw{2\sfp}{}{
            \Xt{N,t} - \Xt{N,i\BoldTs}
        }
        , 
        \\ 
        \LpLaw{2\sfp}{}{
            \FsigmaErrorAll{}{t,i\BoldTs,\Xt{N}}
        }
        \leq&
        \left( 
            \hat{L}^{\cbr{\cdot,\paral}}_p 
            + 
            \hat{L}^{\cbr{\cdot,\paral}}_\sigma
        \right)
        \BoldTs
        +
        \left(
            L^{\cbr{\cdot,\paral}}_p
            + 
            L^{\cbr{\cdot,\paral}}_\sigma
        \right)
        \left(
            \LpLaw{2\sfp}{}{
                \Xt{N,t} - \Xt{N,i\BoldTs}
            }
        \right)^\frac{1}{2}
        ,
    \end{align*} 
    for all $(i,t,\sfp) \in \mathbb{N}_{\geq 1} \times [(i-1)\BoldTs,i\BoldTs] \times \mathbb{N}_{\geq 1}$.
    With $\tau^\star = \tstar$ and $k=i$, we use the bound for $\LpLaw{2\sfp}{}{\Xt{N,t} - \Xt{N,i\BoldTs}}$ from Proposition~\ref{prop:Appendix:TrueProcess:TsBound} and obtain 
     \begin{subequations}\label{eqn:lem:Appendix:TrueProcess:NUTilde:ErrorDrift:Bounds:General}
        \begin{align}
            \LpLaw{2\sfp}{}{
                \LmuErrorAll{}{t,i\BoldTs,\Xt{N}}
            }
            \leq& 
            \BoldTsroot
            L^{\cbr{\cdot,\paral}}_\mu
            \widetilde{\delta}_2(\sfp,\tstar)
            +
            \BoldTs
            \left(
                \hat{L}^{\cbr{\cdot,\paral}}_\mu  
                +  
                L^{\cbr{\cdot,\paral}}_\mu
                \widetilde{\delta}_1(\BoldTs,\sfp,\tstar)
            \right)
            , 
            \label{eqn:lem:Appendix:TrueProcess:NUTilde:ErrorDrift:Bounds:General:Drift}
            \\ 
            \LpLaw{2\sfp}{}{
                \FsigmaErrorAll{}{t,i\BoldTs,\Xt{N}}
            }
            \leq&
            \BoldTs
            \left( 
                \hat{L}^{\cbr{\cdot,\paral}}_p 
                + 
                \hat{L}^{\cbr{\cdot,\paral}}_\sigma
            \right)
            \notag 
            \\
            &
            +
            \left(
                L^{\cbr{\cdot,\paral}}_p
                + 
                L^{\cbr{\cdot,\paral}}_\sigma
            \right)
            \left(
                \BoldTs
                \widetilde{\delta}_1(\BoldTs,\sfp,\tstar) 
                +
                \BoldTs^\frac{1}{2}
                \widetilde{\delta}_2(\sfp,\tstar)
            \right)^\frac{1}{2}
            ,
            \label{eqn:lem:Appendix:TrueProcess:NUTilde:ErrorDrift:Bounds:General:Diffusion}
        \end{align} 
    \end{subequations}
    for all $(i,t,\sfp) \in \mathbb{N}_{\geq 1} \times [(i-1)\BoldTs,i\BoldTs] \times \mathbb{N}_{\geq 1}$.

    Now, with $z = \Xt{N}$ and $\tau = \tstar$, we use the definition of $\widetilde{\mathcal{H}}_\mu^{\paral}$ and the Minkowski's inequality to conclude that  
    \begin{align*}
        &\LpLaw{2\sfp}{}{\widetilde{\mathcal{H}}_\mu^{\paral}(t,\Xt{N})}
        \\
        &\leq 
        \indicator{<\BoldTs}{t} 0 
        +
        \sum_{i=1}^{\istar{\tstar} - 1}
        \indicator{[i\BoldTs,(i+1)\BoldTs)}{t} 
        \LpLaw{2\sfp}{}{\LparamuError{}{t,i\BoldTs, \Xt{N}}}
        + 
        \indicator{ \geq \istar{\tstar}\BoldTs}{t}
        \LpLaw{2\sfp}{}{\LparamuError{}{t,\istar{\tstar}\BoldTs, \Xt{N}}}
        \\
        &\leq
        \max_{i \in \cbr{1,\dots,\istar{\tstar} - 1}}  
        \left\{
            \sup_{t \in [i\BoldTs,(i+1)\BoldTs]} 
            \LpLaw{2\sfp}{}{\LparamuError{}{t,i\BoldTs, \Xt{N}}}
            , 
            \sup_{t \in [\istar{\tstar}\BoldTs,\tstar]}
            \LpLaw{2\sfp}{}{\LparamuError{}{t,\istar{\tstar}\BoldTs, \Xt{N}}}
        \right\}
        ,
    \end{align*}
    for all $(t,\sfp) \in [0,\tstar] \times \mathbb{N}_{\geq 1}$.
    Similarly, using the definition of $\widetilde{\mathcal{H}}_\mu$ from Proposition~\ref{proposition:Appendix:TrueProcess:FL:ExpressionEvolved}, with $z = \Xt{N}$ and $\tau = \tstar$, we get
    \begin{multline*}
        \LpLaw{2\sfp}{}{\expo{(\lambda_s-\Boldomega)t}\widetilde{\mathcal{H}}_\mu(t,\Xt{N})}
        \\
        \leq
        \Delta_\Theta
        \max_{i \in \cbr{1,\dots,\istar{\tstar} - 1}}  
        \left\{
            \sup_{t \in [(i-1)\BoldTs,i\BoldTs]}
            \left(
                \absolute{\expo{(\lambda_s-\Boldomega)t} \widetilde{\gamma}_i(\BoldTs)} 
                \LpLaw{2\sfp}{}{\LmuError{}{t,i\BoldTs,\Xt{N}}}
            \right)
            ,
            \hspace{4cm}
        \right. 
        \\ 
        \left. 
            \sup_{t \in [(\istar{\tstar}-1)\BoldTs,\istar{\tstar}\BoldTs]}
            \left(
                \absolute{
                    \expo{(\lambda_s-\Boldomega)t}
                    \widetilde{\gamma}^\star(\tstar,\BoldTs)
                }
                \LpLaw{2\sfp}{}{\LmuError{}{t,\istar{\tstar}\BoldTs,\Xt{N}}}
            \right)
        \right\}
        ,
    \end{multline*}
    for all $(t,\sfp) \in [0,\tstar] \times \mathbb{N}_{\geq 1}$, where we have used the uniform bound on $\Frobenius{\Theta_{ad}(t)}$ from Assumption~\ref{assmp:KnownFunctions}.
    It then follows from the bounds above and the Minkowski's inequality that   
    \begin{equation}\label{eqn:lem:Appendix:TrueProcess:NUTilde:ErrorDrift:Bounds:A:Initial}
        \LpLaw{2\sfp}{}{
            \widetilde{\mathcal{H}}_\mu^{\paral}(t,\Xt{N})
            + 
            \expo{(\lambda_s-\Boldomega)t}
            \widetilde{\mathcal{H}}_\mu(t,\Xt{N})
        }
        \leq 
        \Psi^{\paral}_{\mathcal{H}_\mu}
        + 
        \Psi_{\mathcal{H}_\mu}
        ,
        \quad 
        \forall 
        (t,\sfp) \in [0,\tstar] \times \mathbb{N}_{\geq 1}
        ,
    \end{equation}
    where 
    \begin{align*}
        \Psi^{\paral}_{\mathcal{H}_\mu}
        =
        \max_{i \in \cbr{1,\dots,\istar{\tstar} - 1}}  
        \left\{
            \sup_{t \in [i\BoldTs,(i+1)\BoldTs]} 
            \LpLaw{2\sfp}{}{\LparamuError{}{t,i\BoldTs, \Xt{N}}}
            , 
            \sup_{t \in [\istar{\tstar}\BoldTs,\tstar]}
            \LpLaw{2\sfp}{}{\LparamuError{}{t,\istar{\tstar}\BoldTs, \Xt{N}}}
        \right\}
        ,
        \\ 
        \begin{multlined}[b][0.9\linewidth]
            \Psi_{\mathcal{H}_\mu}
            =
            \Delta_\Theta
            \max_{i \in \cbr{1,\dots,\istar{\tstar} - 1}}  
            \left\{
                \sup_{t \in [(i-1)\BoldTs,i\BoldTs]}
                \left(
                    \absolute{\expo{(\lambda_s-\Boldomega)t} \widetilde{\gamma}_i(\BoldTs)} 
                    \LpLaw{2\sfp}{}{\LmuError{}{t,i\BoldTs,\Xt{N}}}
                \right)
                ,
            \right. 
            \\ 
            \left. 
                \sup_{t \in [(\istar{\tstar}-1)\BoldTs,\istar{\tstar}\BoldTs]}
                \left(
                    \absolute{
                        \expo{(\lambda_s-\Boldomega)t}
                        \widetilde{\gamma}^\star(\tstar,\BoldTs)
                    }
                    \LpLaw{2\sfp}{}{\LmuError{}{t,\istar{\tstar}\BoldTs,\Xt{N}}}
                \right)
            \right\}
            .
        \end{multlined}
    \end{align*}
    It is straightforward to see from~\eqref{eqn:lem:Appendix:TrueProcess:NUTilde:ErrorDrift:Bounds:General:Drift} that 
    \begin{equation}\label{eqn:lem:Appendix:TrueProcess:NUTilde:ErrorDrift:Bounds:A:Psi:1}
        \Psi^{\paral}_{\mathcal{H}_\mu}
        \leq    
        \BoldTsroot
        L^{\paral}_\mu
        \widetilde{\delta}_2(\sfp,\tstar)
        +
        \BoldTs
        \left(
            \hat{L}^{\paral}_\mu  
            + 
            L^{\paral}_\mu
            \widetilde{\delta}_1(\BoldTs,\sfp,\tstar)
        \right)
        .
    \end{equation}
    Next, using the definition of $\widetilde{\gamma}_i$ from Proposition~~\ref{proposition:Appendix:TrueProcess:FL:ExpressionEvolved}, one sees that 
    \begin{align}\label{eqn:lem:Appendix:TrueProcess:NUTilde:ErrorDrift:GammaIdent:1}
        \expo{(\lambda_s-\Boldomega)t} \widetilde{\gamma}_i(\BoldTs)
        = 
        \expo{(\lambda_s-\Boldomega)(t-i\BoldTs)} 
        \frac{\lambda_s}{\Boldomega} 
        \frac{\expo{\Boldomega\BoldTs} - 1}{\expo{\lambda_s \BoldTs} - 1}
        \leq     
        \gamma_2(\Boldomega,\BoldTs), 
    \end{align}
    for all $t \in [(i-1)\BoldTs,i\BoldTs]$.
    Similarly, using the definition of $\widetilde{\gamma}^\star$ from Proposition~~\ref{proposition:Appendix:TrueProcess:FL:ExpressionEvolved}, we have that
    \begin{align}\label{eqn:lem:Appendix:TrueProcess:NUTilde:ErrorDrift:GammaIdent:2}
        \expo{(\lambda_s-\Boldomega)t}
        \widetilde{\gamma}^\star(\tstar,\BoldTs)
        =
        \expo{(\lambda_s-\Boldomega)(t-\istar{\tstar}\BoldTs)}
        \frac{\lambda_s}{\Boldomega}
        \frac{\expo{\Boldomega (\tstar - \istar{\tstar}\BoldTs)} - 1}{\expo{\lambda_s \BoldTs} - 1}
        \leq     
        \gamma_2(\Boldomega,\BoldTs)
        ,
    \end{align}
    for all $t \in [(\istar{\tstar}-1)\BoldTs,\istar{\tstar}\BoldTs]$, where we have used the fact that $\istar{\tstar} \doteq \left\lfloor \tstar/\BoldTs \right\rfloor $, and thus $(\tstar - \istar{\tstar}\BoldTs) \leq \BoldTs$.
    Then,~\eqref{eqn:lem:Appendix:TrueProcess:NUTilde:ErrorDrift:Bounds:General:Drift}, along with~\eqref{eqn:lem:Appendix:TrueProcess:NUTilde:ErrorDrift:GammaIdent:1} and~\eqref{eqn:lem:Appendix:TrueProcess:NUTilde:ErrorDrift:GammaIdent:2}, allow us to obtain the following bound:
    \begin{equation}\label{eqn:lem:Appendix:TrueProcess:NUTilde:ErrorDrift:Bounds:A:Psi:2}
        \Psi_{\mathcal{H}_\mu}
        \leq    
        \BoldTsroot
        L_\mu
        \gamma_2(\Boldomega,\BoldTs)
        \Delta_\Theta
        \widetilde{\delta}_2(\sfp,\tstar)
        +
        \BoldTs
        \gamma_2(\Boldomega,\BoldTs)
        \Delta_\Theta
        \left(
            \hat{L}_\mu  
            + 
            L_\mu
            \widetilde{\delta}_1(\BoldTs,\sfp,\tstar)
        \right)
        .
    \end{equation}
    Substituting~\eqref{eqn:lem:Appendix:TrueProcess:NUTilde:ErrorDrift:Bounds:A:Psi:1} and~\eqref{eqn:lem:Appendix:TrueProcess:NUTilde:ErrorDrift:Bounds:A:Psi:2} into~\eqref{eqn:lem:Appendix:TrueProcess:NUTilde:ErrorDrift:Bounds:A:Initial} leads to~\eqref{eqn:lem:Appendix:TrueProcess:NUTilde:ErrorDrift:Bounds:A:Final}. 

    Next, with $z = \Xt{N}$ and $\tau = \tstar$, we use the definition of $\widetilde{\mathcal{H}}_\sigma^{\paral}$ and the Minkowski's inequality to conclude that  
    \begin{align*}
        &\LpLaw{2\sfp}{}{\widetilde{\mathcal{H}}_\mu^{\paral}(t,\Xt{N})}
        \\
        &\leq 
        \indicator{<\BoldTs}{t} 0 
        +
        \sum_{i=1}^{\istar{\tstar} - 1}
        \indicator{[i\BoldTs,(i+1)\BoldTs)}{t} 
        \LpLaw{2\sfp}{}{\FparasigmaError{}{t,i\BoldTs, \Xt{N}}}
        + 
        \indicator{ \geq \istar{\tstar}\BoldTs}{t}
        \LpLaw{2\sfp}{}{\FparasigmaError{}{t,\istar{\tstar}\BoldTs, \Xt{N}}}
        \\
        &\leq
        \max_{i \in \cbr{1,\dots,\istar{\tstar} - 1}}  
        \left\{
            \sup_{t \in [i\BoldTs,(i+1)\BoldTs]} 
            \LpLaw{2\sfp}{}{\FparasigmaError{}{t,i\BoldTs, \Xt{N}}}
            , 
            \sup_{t \in [\istar{\tstar}\BoldTs,\tstar]}
            \LpLaw{2\sfp}{}{\FparasigmaError{}{t,\istar{\tstar}\BoldTs, \Xt{N}}}
        \right\}
        ,
    \end{align*}
    for all $(t,\sfp) \in [0,\tstar] \times \mathbb{N}_{\geq 1}$.
    Similarly, using the definition of $\widetilde{\mathcal{H}}_\mu$ from Proposition~\ref{proposition:Appendix:TrueProcess:FL:ExpressionEvolved}, with $z = \Xt{N}$ and $\tau = \tstar$, we get
    \begin{multline*}
        \LpLaw{2\sfp}{}{\expo{(\lambda_s-\Boldomega)t}\widetilde{\mathcal{H}}_\mu(t,\Xt{N})}
        \\
        \leq
        \Delta_\Theta
        \max_{i \in \cbr{1,\dots,\istar{\tstar} - 1}}  
        \left\{
            \sup_{t \in [(i-1)\BoldTs,i\BoldTs]}
            \left(
                \absolute{\expo{(\lambda_s-\Boldomega)t} \widetilde{\gamma}_i(\BoldTs)} 
                \LpLaw{2\sfp}{}{\FsigmaError{}{t,i\BoldTs,\Xt{N}}}
            \right)
            ,
            \hspace{4cm}
        \right. 
        \\ 
        \left. 
            \sup_{t \in [(\istar{\tstar}-1)\BoldTs,\istar{\tstar}\BoldTs]}
            \left(
                \absolute{
                    \expo{(\lambda_s-\Boldomega)t}
                    \widetilde{\gamma}^\star(\tstar,\BoldTs)
                }
                \LpLaw{2\sfp}{}{\FsigmaError{}{t,\istar{\tstar}\BoldTs,\Xt{N}}}
            \right)
        \right\}
        ,
    \end{multline*}
    for all $(t,\sfp) \in [0,\tstar] \times \mathbb{N}_{\geq 1}$, where we have used the uniform bound on $\Frobenius{\Theta_{ad}(t)}$ from Assumption~\ref{assmp:KnownFunctions}.
    It then follows from the bounds above and the Minkowski's inequality that   
    \begin{equation}\label{eqn:lem:Appendix:TrueProcess:NUTilde:ErrorDiffusion:Bounds:A:Initial}
        \LpLaw{2\sfp}{}{
            \widetilde{\mathcal{H}}_\sigma^{\paral}(t,\Xt{N})
            + 
            \expo{(\lambda_s-\Boldomega)t}
            \widetilde{\mathcal{H}}_\sigma(t,\Xt{N})
        }
        \leq 
        \Psi^{\paral}_{\mathcal{H}_\sigma}
        + 
        \Psi_{\mathcal{H}_\sigma}
        ,
        \quad 
        \forall 
        (t,\sfp) \in [0,\tstar] \times \mathbb{N}_{\geq 1}
        ,
    \end{equation}
    where 
    \begin{align*}
        \Psi^{\paral}_{\mathcal{H}_\sigma}
        =
        \max_{i \in \cbr{1,\dots,\istar{\tstar} - 1}}  
        \left\{
            \sup_{t \in [i\BoldTs,(i+1)\BoldTs]} 
            \LpLaw{2\sfp}{}{\FparasigmaError{}{t,i\BoldTs, \Xt{N}}}
            , 
            \sup_{t \in [\istar{\tstar}\BoldTs,\tstar]}
            \LpLaw{2\sfp}{}{\FparasigmaError{}{t,\istar{\tstar}\BoldTs, \Xt{N}}}
        \right\}
        ,
        \\ 
        \begin{multlined}[b][0.9\linewidth]
            \Psi_{\mathcal{H}_\sigma}
            =
            \Delta_\Theta
            \max_{i \in \cbr{1,\dots,\istar{\tstar} - 1}}  
            \left\{
                \sup_{t \in [(i-1)\BoldTs,i\BoldTs]}
                \left(
                    \absolute{\expo{(\lambda_s-\Boldomega)t} \widetilde{\gamma}_i(\BoldTs)} 
                    \LpLaw{2\sfp}{}{\FsigmaError{}{t,i\BoldTs,\Xt{N}}}
                \right)
                ,
            \right. 
            \\ 
            \left. 
                \sup_{t \in [(\istar{\tstar}-1)\BoldTs,\istar{\tstar}\BoldTs]}
                \left(
                    \absolute{
                        \expo{(\lambda_s-\Boldomega)t}
                        \widetilde{\gamma}^\star(\tstar,\BoldTs)
                    }
                    \LpLaw{2\sfp}{}{\FsigmaError{}{t,\istar{\tstar}\BoldTs,\Xt{N}}}
                \right)
            \right\}
            .
        \end{multlined}
    \end{align*}
    Then, using~\eqref{eqn:lem:Appendix:TrueProcess:NUTilde:ErrorDrift:Bounds:General:Diffusion}, along with~\eqref{eqn:lem:Appendix:TrueProcess:NUTilde:ErrorDrift:GammaIdent:1} and~\eqref{eqn:lem:Appendix:TrueProcess:NUTilde:ErrorDrift:GammaIdent:2}, we obtain the following bounds: 
    \begin{align*}
        \Psi^{\paral}_{\mathcal{H}_\sigma} 
        \leq& 
        2
        \Delta^{\paral}_p \left(1 - \Lip{p} \right)
        +  
        \BoldTs
        \left( 
            \hat{L}^{\paral}_p 
            \Lip{p}
            + 
            \hat{L}^{\paral}_\sigma
        \right)
        \notag 
        \\
        &
        +
        \left(
            L^{\paral}_p
            \Lip{p}
            + 
            L^{\paral}_\sigma
        \right)
        \left(
            \BoldTs
            \widetilde{\delta}_1(\BoldTs,\sfp,\tstar) 
            +
            \BoldTs^\frac{1}{2}
            \widetilde{\delta}_2(\sfp,\tstar)
        \right)^\frac{1}{2},
        \\
        \Psi_{\mathcal{H}_\sigma}
        \leq &
        2
        \Delta_p 
         \gamma_2(\Boldomega,\BoldTs)
        \Delta_\Theta
        \left(1 - \Lip{p} \right)
        +  
        \BoldTs
         \gamma_2(\Boldomega,\BoldTs)
        \Delta_\Theta
        \left( 
            \hat{L}_p 
            \Lip{p}
            + 
            \hat{L}_\sigma
        \right)
        \notag 
        \\
        &
        +
         \gamma_2(\Boldomega,\BoldTs)
        \Delta_\Theta
        \left(
            L_p
            \Lip{p}
            + 
            L_\sigma
        \right)
        \left(
            \BoldTs
            \widetilde{\delta}_1(\BoldTs,\sfp,\tstar) 
            +
            \BoldTs^\frac{1}{2}
            \widetilde{\delta}_2(\sfp,\tstar)
        \right)^\frac{1}{2}.
    \end{align*} 
    Substituting the above bounds into~\eqref{eqn:lem:Appendix:TrueProcess:NUTilde:ErrorDiffusion:Bounds:A:Initial} leads to~\eqref{eqn:lem:Appendix:TrueProcess:NUTilde:ErrorDrift:Bounds:B:Final}.

     Next, with $z = \Xt{N}$ and $\tau = \tstar$, we use the definition of $\widetilde{\mathcal{G}}_{\sigma_1}$ and $\widetilde{\mathcal{G}}_{\sigma_2}$ and the Minkowski's inequality to conclude that  
    \begin{align*}
        \LpLaw{2\sfp}{}{\widetilde{\mathcal{G}}_{\sigma_1}(t,\Xt{N})}
        \leq& 
        \indicator{<\BoldTs}{t}
        \LpLaw{2\sfp}{}{\FparasigmaError{}{t,0,\Xt{N}}}
        +
        \indicator{\geq \BoldTs}{t} 
        0 
        \\
        \leq &
        \sup_{t \in [0,\BoldTs]}
        \LpLaw{2\sfp}{}{\FparasigmaError{}{t,0,\Xt{N}}}
        ,
        \\
        \LpLaw{2\sfp}{}{
            \expo{(\lambda_s-\Boldomega)t}
            \widetilde{\mathcal{G}}_{\sigma_2}(t,\Xt{N})
        }
        \leq&
        \sum_{i=1}^{\istar{\tstar} - 1}
        \indicator{[(i-1)\BoldTs,i\BoldTs)}{t}
        \absolute{
            \expo{(\lambda_s-\Boldomega)t}\widetilde{\gamma}_i(\BoldTs)
        }
        \LpLaw{2\sfp}{}{       
            \FparasigmaError{}{i\BoldTs,(i-1)\BoldTs,\Xt{N}}
        }
        \\
        &+
        \indicator{\geq (\istar{t} - 1)\BoldTs}{t}
        0
        \\
        \leq & 
        \max_{i \in \cbr{1,\dots,\istar{\tstar} - 1}}  
        \sup_{t \in [(i-1)\BoldTs,i\BoldTs]}
        \left(
            \absolute{
                \expo{(\lambda_s-\Boldomega)t}\widetilde{\gamma}_i(\BoldTs)
            }
            \LpLaw{2\sfp}{}{       
                \FparasigmaError{}{i\BoldTs,(i-1)\BoldTs,\Xt{N}}
            }
        \right)
    \end{align*}
    for all $(t,\sfp) \in [0,\tstar] \times \mathbb{N}_{\geq 1}$.
    Hence, it follows from the Minkowski's inequality that 
    \begin{align*}
        &\LpLaw{2\sfp}{}{
            \widetilde{\mathcal{G}}_{\sigma_1}(t,\Xt{N})
            + 
            \expo{(\lambda_s-\Boldomega)t}
            \widetilde{\mathcal{G}}_{\sigma_2}(t,\Xt{N})
        }
        \\
        &\leq
        \sup_{t \in [0,\BoldTs]}
        \LpLaw{2\sfp}{}{\FparasigmaError{}{t,0,\Xt{N}}}
        \\
        &
        \qquad  
        +
        \max_{i \in \cbr{1,\dots,\istar{\tstar} - 1}}  
        \sup_{t \in [(i-1)\BoldTs,i\BoldTs]}
        \left(
            \absolute{
                \expo{(\lambda_s-\Boldomega)t}\widetilde{\gamma}_i(\BoldTs)
            }
            \LpLaw{2\sfp}{}{       
                \FparasigmaError{}{i\BoldTs,(i-1)\BoldTs,\Xt{N}}
            }
        \right)
        ,
    \end{align*}
    for all $(t,\sfp) \in [0,\tstar] \times \mathbb{N}_{\geq 1}$.
    Using~\eqref{eqn:lem:Appendix:TrueProcess:NUTilde:ErrorDrift:Bounds:General:Diffusion} and~\eqref{eqn:lem:Appendix:TrueProcess:NUTilde:ErrorDrift:GammaIdent:1} leads to 
    \begin{multline*}
        \LpLaw{2\sfp}{}{
            \widetilde{\mathcal{G}}_{\sigma_1}(t,\Xt{N})
            + 
            \expo{(\lambda_s-\Boldomega)t}
            \widetilde{\mathcal{G}}_{\sigma_2}(t,\Xt{N})
        }
        \\
        \leq
        \left(1 + \gamma_2(\Boldomega,\BoldTs)\right)
        \left[
            2
            \Delta^{\paral}_p \left(1 - \Lip{p} \right)
             +
            \left(
                L^{\paral}_p
                \Lip{p}
                + 
                L^{\paral}_\sigma
            \right)
            \left(
                \BoldTs
                \widetilde{\delta}_1(\BoldTs,\sfp,\tstar) 
                +
                \BoldTs^\frac{1}{2}
                \widetilde{\delta}_2(\sfp,\tstar)
            \right)^\frac{1}{2}
        \right.
        \\
        \left.
           +  
            \BoldTs
            \left( 
                \hat{L}^{\paral}_p 
                \Lip{p}
                + 
                \hat{L}^{\paral}_\sigma
            \right)
        \right]
        ,
    \end{multline*}
    for all $(t,\sfp) \in [0,\tstar] \times \mathbb{N}_{\geq 1}$, thus establishing~\eqref{eqn:lem:Appendix:TrueProcess:NUTilde:ErrorDrift:Bounds:C:Final}.

    Finally, using the definition of $\mathcal{G}_\mu$, we deduce that  
    \begin{multline*}
        \LpLaw{2\sfp}{}{\mathcal{G}_\mu(t,\Xt{N})}
        \leq
        \left(1 - \expo{-\lambda_s \BoldTs}\right)
        \\ 
        \times 
        \max_{i \in \cbr{1,\dots,\istar{\tstar} - 1}} 
        \left\{ 
            \sup_{t \in [i\BoldTs,(i+1)\BoldTs]}
            \LpLaw{2\sfp}{}{\Lparamu{t,\Xt{N,t}}}
            ,
            \sup_{t \in [\istar{\tstar}\BoldTs,\tstar]}
            \LpLaw{2\sfp}{}{\Lparamu{t,\Xt{N,t}}}
        \right\},
    \end{multline*}
    for all $t \in [0,\tstar]$.
    Therefore,
    \begin{align*}
        \LpLaw{2\sfp}{}{\mathcal{G}_\mu(t,\Xt{N})}
        \leq
        \left(1 - \expo{-\lambda_s \BoldTs}\right)
        \sup_{t \in [0,\tstar]}
        \LpLaw{2\sfp}{}{\Lparamu{t,\Xt{N,t}}}
        ,
    \end{align*}
    for all $t \in [0,\tstar]$, thus establishing~\eqref{eqn:lem:Appendix:TrueProcess:NUTilde:ErrorDrift:Bounds:D:Final}.     
\end{proof}

The next result bounds the moments of a few pertinent processes that we need later. 

\begin{proposition}\label{prop:Appendix:NUProcess:ZProcess} 
    Let
    \begin{multline}\label{eqn:prop:Appendix:NUProcess:ZProcess:J:Definitions}
        \mathfrak{Z}_j(t)
        = 
        \int_{\BoldTs}^t  
            \expo{ 2(\lambda - \Boldomega) \nu }
            \left(
                \int_0^\nu 
                    \expo{\Boldomega \beta}
                        \mathcal{J}_{j}(\beta,\Xt{N})
                d\Wt{\beta}
            \right)^\top 
            g(\nu)^\top
            g(\nu)
            \\
            \times
            \left(
                \int_0^{\nu}
                \expo{\Boldomega \beta}
                \left[ 
                    \LZparamu{\beta,\Zt{N,\beta}}d\beta + \FZparasigma{\beta,\Zt{N,\beta}}d\Wt{\beta} 
                \right]
            \right)       
        d\nu
       ,
        \quad (j,t) \in \cbr{1,\dots,4} \times [\BoldTs,\tstar],
    \end{multline}
    where the piecewise functions $\mathcal{J}_{j}$ are defined as follows for $\beta \in [0,\nu]$, $\nu \in [\BoldTs,t]$, $t \geq \BoldTs$: 
    \begin{subequations}\label{eqn:prop:Appendix:NUProcess:PiecewiseFunctions:J:Definitions}
        \begin{align}
            \mathcal{J}_1(\beta,\Xt{N})
            =
            \sum_{i=1}^{\istar{\nu} - 1}
            \indicator{[(i-1)\BoldTs,i\BoldTs)}{\beta}
            \Fparasigma{(i-1)\BoldTs,\Xt{N,(i-1)\BoldTs}}
            \left( 
                1  
                -
                \widetilde{\gamma}_i(\BoldTs)
                \expo{ \left(\lambda_s - \Boldomega\right) \beta} 
            \right)
            , 
            \\
            \mathcal{J}_2(\beta,\Xt{N})
            =
            \indicator{[(\istar{\nu} - 1)\BoldTs,\istar{\nu}\BoldTs)}{\beta}
            \Fparasigma{(\istar{\nu} - 1)\BoldTs,\Xt{N,(\istar{\nu} - 1)\BoldTs}} 
            ,
            \\
            \mathcal{J}_3(\beta,\Xt{N})
            =
            -
            \indicator{[(\istar{\nu} - 1)\BoldTs,\istar{\nu}\BoldTs)}{\beta}
            \widetilde{\gamma}^\star(\nu,\BoldTs)
            \expo{ (\lambda_s - \Boldomega) \beta} 
            \Fparasigma{\istar{\nu}\BoldTs,\Xt{N,\istar{\nu}\BoldTs}}
            ,
            \\ 
            \mathcal{J}_4(\beta,\Xt{N})
            =
            \indicator{ \geq \istar{\nu}\BoldTs}{\beta}
            \Fparasigma{\istar{\nu}\BoldTs,\Xt{N,\istar{\nu}\BoldTs}}
            ,
        \end{align}
    \end{subequations}
    and the entities $\istar{\nu}$, $\widetilde{\gamma}_i(\BoldTs)$, and $ \widetilde{\gamma}^\star(\nu,\BoldTs)$ are defined in Proposition~\ref{proposition:Appendix:TrueProcess:FL:ExpressionEvolved} as: 
    \begin{align*}
        \istar{\nu} \doteq \max\cbr{i \in \mathbb{N}_{\geq 1} \, : \, i\BoldTs \leq \nu} 
        , 
        \\
        \widetilde{\gamma}_i(\BoldTs)
        \doteq 
        \expo{\left(\Boldomega - \lambda_s\right)i\BoldTs}
        \frac{\lambda_s}{\Boldomega} 
        \frac{\expo{\Boldomega\BoldTs} - 1}{\expo{\lambda_s \BoldTs} - 1}
        ,
        \quad 
        \widetilde{\gamma}^\star(\nu,\BoldTs)
        \doteq 
        \expo{(\Boldomega - \lambda_s)\istar{\nu}\BoldTs}
        \frac{\lambda_s}{\Boldomega}
        \frac{\expo{\Boldomega (\nu - \istar{\nu}\BoldTs)} - 1}{\expo{\lambda_s \BoldTs} - 1}
        .
    \end{align*}
    If the stopping time $\tau^\star$, defined in~\eqref{eqn:True:StoppingTimes}, Lemma~\ref{lem:True:dV}, satisfies  $\tau^\star = \tstar$ and $\tstar \geq \BoldTs$, then
    \begin{multline}
        \sum_{j=1}^4
        \LpLaw{\sfp}{}{
            \mathfrak{Z}_j(t)
        }
        \leq
        \Delta_g^2
        \frac{ \expo{ 2\lambda t } }{2\lambda \Boldomega}
        \gamma''\left(\sfp,\Boldomega,\BoldTs\right)
        \sup_{\nu \in [0,\tstar]}
            \LpLaw{4\sfp}{}{\Fparasigma{\nu,\Xt{N,\nu}}}
        \\
        \times
        \left( 
            \frac{1}{\sqrt{\Boldomega}}
            \sup_{\nu \in [0,\tstar]}
                \LpLaw{2\sfp}{}{\LZparamu{\nu,\Zt{N,\nu}}}
            + 
            \mathfrak{p}'(\sfp)
            \sup_{\nu \in [0,\tstar]}
                \LpLaw{2\sfp}{}{\FZparasigma{\nu,\Zt{N,\nu}}}
        \right)
        ,
    \end{multline}
    for all $(t,\sfp) \in [\BoldTs,\tstar] \times \mathbb{N}_{\geq 1}$, where $\gamma''\left(\sfp,\Boldomega,\BoldTs\right)$ is defined in~\eqref{eqn:app:Functions:True:gammaTs:Dblprime}.
      
\end{proposition}
\begin{proof}
    We begin with the bound for $\mathfrak{Z}_1(t)$.

    \ul{\emph{Bound for $\mathfrak{Z}_1(t)$:}}
    Invoking Lemma~\ref{cor:TechnicalResults:UTildeBound} by setting
    \begin{align*} 
        \xi = \BoldTs,
        \quad  
        \theta_1 = \lambda, 
        \quad
        \theta_2 = \Boldomega,
        \quad   
        Q_t = \Wt{t} (n_q =d),
        \\   
        R(t) 
        = 
        g(t)^\top g(t) 
        \in 
        \mathbb{R}^{m \times m}\,(n_1=n_2 = m)
        ,
        \\ 
        S_1(t) = 0_m,
        \quad 
        L_1(t) = \mathcal{J}_1(\beta,\Xt{N}),
        \quad  
        S_2(t) = \LZparamu{t,\Zt{N,t}},
        \quad 
        L_2(t) = \FZparasigma{t,\Zt{N,t}}
        ,
    \end{align*}
    we obtain 
    \begin{multline}\label{eqn:lem:Appendix:TrueProcess:SumProcessNew:Expectation:B:1:Initial}
        \LpLaw{\sfp}{}{
            \mathfrak{Z}_1(t)
        }
        \leq 
        \Delta_g^2
        \frac{ \expo{ 2\lambda t } }{2\lambda \Boldomega}
        \mathfrak{p}'(\sfp)
        \sup_{\nu \in [0,\tstar]}
        \LpLaw{2\sfp}{}{ 
            \mathcal{J}_1(\nu,\Xt{N})
        }
        \\
        \times
        \left( 
            \frac{1}{\sqrt{\Boldomega}}
            \sup_{\nu \in [0,\tstar]}
                \LpLaw{2\sfp}{}{\LZparamu{\nu,\Zt{N,\nu}}}
            + 
            \mathfrak{p}'(\sfp)
            \sup_{\nu \in [0,\tstar]}
                \LpLaw{2\sfp}{}{\FZparasigma{\nu,\Zt{N,\nu}}}
        \right)
        ,
    \end{multline}
    for all $(t,\sfp) \in [\BoldTs,\tstar] \times \mathbb{N}_{\geq 1}$, where we have used the bound $\Delta_R = \Delta_g^2 \geq  \Frobenius{g(t)}^2 = \Frobenius{R(t)}$ due to Assumption~\ref{assmp:KnownFunctions}. 
    Using the definition of $\mathcal{J}_1$ and the Minkowski's inequality leads to
    \begin{align*}
        \LpLaw{2\sfp}{}{\mathcal{J}_1(\nu,\Xt{N})}
        \leq& 
        \sum_{i=1}^{\istar{t} - 1}
        \indicator{[(i-1)\BoldTs,i\BoldTs)}{\nu}
        \absolute{
            1  
            -
            \widetilde{\gamma}_i(\BoldTs)
            \expo{ \left(\lambda_s - \Boldomega\right) \nu} 
        }
        \LpLaw{2\sfp}{}{\Fparasigma{(i-1)\BoldTs,\Xt{N,(i-1)\BoldTs}}}
        ,
    \end{align*} 
    for all $\nu \in [0,t]$ and $(t,\sfp) \in [\BoldTs,\tstar] \times \mathbb{N}_{\geq 1}$.
    Hence, 
    \begin{align}\label{eqn:prop:Appendix:TrueProcess:SumProcessNew:J1:Bound}
        \LpLaw{2\sfp}{}{\mathcal{J}_1(\nu,\Xt{N})}
        &
        \leq    
        \max_{i \in \cbr{1,\dots,\istar{\tstar} - 1}}
        \sup_{t \in [(i-1)\BoldTs,i\BoldTs]}
        \left(
            \absolute{ 
                1  
                -
                \widetilde{\gamma}_i(\BoldTs)
                \expo{ \left(\lambda_s - \Boldomega\right) t} 
            }
            \LpLaw{2\sfp}{}{\Fparasigma{(i-1)\BoldTs,\Xt{N,(i-1)\BoldTs}}}
        \right)
        \notag 
        \\
        &
        \leq  
        \left(
            \max_{i \in \cbr{1,\dots,\istar{\tstar} - 1}}
            \sup_{t \in [(i-1)\BoldTs,i\BoldTs]}
            \absolute{ 
                1  
                -
                \widetilde{\gamma}_i(\BoldTs)
                \expo{ \left(\lambda_s - \Boldomega\right) t} 
            }
        \right)
        \sup_{\nu \in [0,\tstar]}
        \LpLaw{2\sfp}{}{\Fparasigma{\nu,\Xt{N,\nu}}}
        \notag   
        \\
        &
        \leq     
        \gamma_2'(\Boldomega,\BoldTs)
        \sup_{\nu \in [0,\tstar]}
        \LpLaw{2\sfp}{}{\Fparasigma{\nu,\Xt{N,\nu}}}
        ,
    \end{align}
    for all $(t,\sfp) \in [\BoldTs,\tstar] \times \mathbb{N}_{\geq 1}$, where, we have used the definition of $\widetilde{\gamma}_i$ from Proposition~~\ref{proposition:Appendix:TrueProcess:FL:ExpressionEvolved} to conclude that
    \begin{align*}
        \sup_{t \in [(i-1)\BoldTs,i\BoldTs]}
        \absolute{
            1  
            -
            \widetilde{\gamma}_i(\BoldTs)
            \expo{ \left(\lambda_s - \Boldomega\right) t}
        } 
        =
        \sup_{t \in [(i-1)\BoldTs,i\BoldTs]}
        \absolute{
            1
            -
            \expo{\left(\Boldomega - \lambda_s\right)(i\BoldTs-t)}
            \frac{\lambda_s}{\Boldomega} 
            \frac{\expo{\Boldomega\BoldTs} - 1}{\expo{\lambda_s \BoldTs} - 1}
        }
        \leq
        \gamma_2'(\Boldomega,\BoldTs)
        ,  
    \end{align*} 
    for all $i \in \cbr{1,\dots,\istar{\tstar} - 1}$.
    Substituting~\eqref{eqn:prop:Appendix:TrueProcess:SumProcessNew:J1:Bound} into~\eqref{eqn:lem:Appendix:TrueProcess:SumProcessNew:Expectation:B:1:Initial} leads to
    \begin{multline}\label{eqn:lem:Appendix:TrueProcess:SumProcessNew:Expectation:B:1:Final}
        \LpLaw{\sfp}{}{
            \mathfrak{Z}_1(t)
        }
        \leq 
        \Delta_g^2
        \frac{ \expo{ 2\lambda t } }{2\lambda \Boldomega}
        \mathfrak{p}'(\sfp)
        \gamma_2'(\Boldomega,\BoldTs)
        \sup_{\nu \in [0,\tstar]}
            \LpLaw{4\sfp}{}{\Fparasigma{\nu,\Xt{N,\nu}}}
        \\
        \times
        \left( 
            \frac{1}{\sqrt{\Boldomega}}
            \sup_{\nu \in [0,\tstar]}
                \LpLaw{2\sfp}{}{\LZparamu{\nu,\Zt{N,\nu}}}
            + 
            \mathfrak{p}'(\sfp)
            \sup_{\nu \in [0,\tstar]}
                \LpLaw{2\sfp}{}{\FZparasigma{\nu,\Zt{N,\nu}}}
        \right)
        ,
    \end{multline}
    for all $(t,\sfp) \in [\BoldTs,\tstar] \times \mathbb{N}_{\geq 1}$.
    Note that we have also used the bound $\LpLaw{2\sfp}{}{\Fparasigma{\nu,\Xt{N,\nu}}} \leq \LpLaw{4\sfp}{}{\Fparasigma{\nu,\Xt{N,\nu}}}$ which is a consequence of the Jensen's inequality.

    \ul{\emph{Bound for $\mathfrak{Z}_2(t)$:}}
    The definition of $\mathcal{J}_2$ allows us to write 
    \begin{align*}
        \mathfrak{Z}_2(t)
        =& 
        \int_{\BoldTs}^t  
            \expo{ 2(\lambda - \Boldomega) \nu }
            \left(
                \int_0^\nu 
                    \expo{\Boldomega \beta}
                        \mathcal{J}_{2}(\beta,\Xt{N})
                d\Wt{\beta}
            \right)^\top 
            g(\nu)^\top
            g(\nu)
            \\
            &
            \hspace{2cm}
            \times
            \left(
                \int_0^{\nu}
                \expo{\Boldomega \beta}
                \left[ 
                    \LZparamu{\beta,\Zt{N,\beta}}d\beta + \FZparasigma{\beta,\Zt{N,\beta}}d\Wt{\beta} 
                \right]
            \right)       
        d\nu
        \\
        =& 
        \int_{\BoldTs}^t  
            \expo{ 2(\lambda - \Boldomega) \nu }
            \left(
                \Fparasigma{(\istar{\nu} - 1)\BoldTs,\Xt{N,(\istar{\nu} - 1)\BoldTs}}
                \int_{(\istar{\nu} - 1)\BoldTs}^{\istar{\nu}\BoldTs} 
                    \expo{\Boldomega \beta}
                    d\Wt{\beta}
            \right)^\top 
            g(\nu)^\top
            g(\nu)
            \\
            &
            \hspace{2cm}
            \times
            \left(
                \int_0^{\nu}
                \expo{\Boldomega \beta}
                \left[ 
                    \LZparamu{\beta,\Zt{N,\beta}}d\beta + \FZparasigma{\beta,\Zt{N,\beta}}d\Wt{\beta} 
                \right]
            \right)       
        d\nu
       ,
    \end{align*}
    for $t \in [\BoldTs,\tstar]$, This process can be cast as $N_a(t)$ in Corollary~\ref{cor:TechnicalResults:PCA} by setting 
    \begin{align*}
        \theta_1 = \lambda, 
        \quad
        \theta_2 = \Boldomega,
        \quad   
        Q_t = \Wt{t} (n_q =d)
        ,
        \\
        R(t) 
        = 
        g(t)^\top g(t) 
        \in 
        \mathbb{R}^{m \times m}\,(n_1=n_2 = m)
        , 
        \\ 
        L_a(\nu)  
        =
        \Fparasigma{(\istar{\nu} - 1)\BoldTs,\Xt{N,(\istar{\nu} - 1)\BoldTs}}
        \in \mathbb{R}^{m \times d},
        \quad 
        \varsigma_1(\nu) = \varsigma_2(\beta) = 1
        , 
        \\ 
        S(t) = \LZparamu{t,\Zt{N,t}},
        \quad 
        L(t) = \FZparasigma{t,\Zt{N,t}}
        .
    \end{align*}
    Hence, by~\eqref{cor:TechnicalResults:PCA:Na:Bound:Final} in Corollary~\ref{cor:TechnicalResults:PCA}, we have
    \begin{multline}\label{eqn:lem:Appendix:TrueProcess:SumProcessNew:Expectation:B:2:Final}
        \LpLaw{\sfp}{}{\mathfrak{Z}_2(t)} 
        \leq  
        \Delta_g^2
        \mathfrak{p}''(\sfp)   
        \frac{\expo{ 2\lambda  t }}{2\lambda \Boldomega} 
            \left(1 - \expo{ -2\Boldomega \BoldTs } \right)^\frac{1}{2}
        \sup_{\nu \in [0,\tstar]}
            \LpLaw{4\sfp}{}{ 
                \Fparasigma{\nu,\Xt{N,\nu}}
            }
        \\
        \times     
        \left( 
            \frac{1}{\sqrt{\Boldomega}}
            \sup_{\nu \in [0,\tstar]}
                \LpLaw{2\sfp}{}{\LZparamu{\nu,\Zt{N,\nu}}}
            + 
            \mathfrak{p}'(\sfp)
            \sup_{\nu \in [0,\tstar]}
                \LpLaw{2\sfp}{}{\FZparasigma{\nu,\Zt{N,\nu}}}
        \right)
        , 
    \end{multline}
    for all $(t,\sfp) \in [\BoldTs,\tstar] \times \mathbb{N}_{\geq 1}$, where we have used the fact that 
    \begin{align*}
        &\sup_{\nu \in [\BoldTs,t]}
        \LpLaw{4\sfp}{}{ 
            \Fparasigma{(\istar{\nu} - 1)\BoldTs,\Xt{N,(\istar{\nu} - 1)\BoldTs}}
        }
        \notag 
        \\
        &\leq
        \sup_{\nu \in [0,\tstar]}
        \LpLaw{4\sfp}{}{ 
            \Fparasigma{(\istar{\nu} - 1)\BoldTs,\Xt{N,(\istar{\nu} - 1)\BoldTs}}
        }
        \leq
        \sup_{\nu \in [0,\tstar]}
        \LpLaw{4\sfp}{}{ 
            \Fparasigma{\nu,\Xt{N,\nu}}
        }, \quad \forall t \in [\BoldTs,\tstar].
    \end{align*}

    \ul{\emph{Bound for $\mathfrak{Z}_3(t)$:}}
    Similar to the above, from the definition of $\mathcal{J}_3$, we have: 
    \begin{align*}
        \mathfrak{Z}_3(t)
        =& 
        \int_{\BoldTs}^t  
            \expo{ 2(\lambda - \Boldomega) \nu }
            \left(
                \int_0^\nu 
                    \expo{\Boldomega \beta}
                        \mathcal{J}_{3}(\beta,\Xt{N})
                d\Wt{\beta}
            \right)^\top 
            g(\nu)^\top
            g(\nu)
            \\
            &
            \hspace{2cm}
            \times
            \left(
                \int_0^{\nu}
                \expo{\Boldomega \beta}
                \left[ 
                    \LZparamu{\beta,\Zt{N,\beta}}d\beta + \FZparasigma{\beta,\Zt{N,\beta}}d\Wt{\beta} 
                \right]
            \right)       
        d\nu
        \\
        =& 
        \int_{\BoldTs}^t  
            \expo{ 2(\lambda - \Boldomega) \nu }
            \left(
                -
                \Fparasigma{\istar{\nu}\BoldTs,\Xt{N,\istar{\nu}\BoldTs}}
                \widetilde{\gamma}^\star(\nu,\BoldTs)
                \int_{(\istar{\nu} - 1)\BoldTs}^{\istar{\nu}\BoldTs} 
                    \expo{\Boldomega \beta}
                    \expo{ (\lambda_s - \Boldomega) \beta} 
                d\Wt{\beta}
            \right)^\top 
            g(\nu)^\top
            g(\nu)
            \\
            &
            \hspace{2cm}
            \times
            \left(
                \int_0^{\nu}
                \expo{\Boldomega \beta}
                \left[ 
                    \LZparamu{\beta,\Zt{N,\beta}}d\beta + \FZparasigma{\beta,\Zt{N,\beta}}d\Wt{\beta} 
                \right]
            \right)       
        d\nu,
    \end{align*}
    for all $t \in [\BoldTs,\tstar]$.
    We can cast the above process as $N_a(t)$ in Corollary~\ref{cor:TechnicalResults:PCA} by setting 
    \begin{align*}
        \theta_1 = \lambda, 
        \quad
        \theta_2 = \Boldomega,
        \quad   
        Q_t = \Wt{t} (n_q =d)
        ,
        \\
        R(t) 
        = 
        g(t)^\top g(t) 
        \in 
        \mathbb{R}^{m \times m}\,(n_1=n_2 = m)
        , 
        \\ 
        L_a(\nu)  
        =
        -
        \Fparasigma{\istar{\nu}\BoldTs,\Xt{N,\istar{\nu}\BoldTs}}
        \in \mathbb{R}^{m \times d},
        \quad 
        \varsigma_1(\nu) = \widetilde{\gamma}^\star(\nu,\BoldTs), 
        \quad 
        \varsigma_2(\beta) = \expo{ (\lambda_s - \Boldomega) \beta}
        , 
        \\ 
        S(t) = \LZparamu{t,\Zt{N,t}},
        \quad 
        L(t) = \FZparasigma{t,\Zt{N,t}}
        .
    \end{align*}
    Hence, we use~\eqref{cor:TechnicalResults:PCA:Na:Bound:Final} in Corollary~\ref{cor:TechnicalResults:PCA} to obtain
    \begin{multline}\label{eqn:lem:Appendix:TrueProcess:SumProcessNew:Expectation:B:3:Initial}
        \begin{aligned}
            \LpLaw{\sfp}{}{\mathfrak{Z}_3(t)} 
            \leq&  
            \Delta_g^2
            \mathfrak{p}''(\sfp)   
            \frac{\expo{ 2\lambda  t }}{2\lambda \Boldomega} 
                \left(1 - \expo{ -2\Boldomega \BoldTs } \right)^\frac{1}{2}
            \sup_{\nu \in [0,\tstar]}
                \LpLaw{4\sfp}{}{ 
                    \Fparasigma{\nu,\Xt{N,\nu}}
                }
            \\
            &
            \times
            \sup_{\nu \in [\BoldTs,t]}
            \left(
                \absolute{\widetilde{\gamma}^\star(\nu,\BoldTs)}
                \sup_{\beta \in [(\istar{\nu}-1)\BoldTs,\istar{\nu}\BoldTs]}
                \expo{ (\lambda_s - \Boldomega) \beta}
            \right)
        \end{aligned}
        \\
        \times     
        \left( 
            \frac{1}{\sqrt{\Boldomega}}
            \sup_{\nu \in [0,\tstar]}
                \LpLaw{2\sfp}{}{\LZparamu{\nu,\Zt{N,\nu}}}
            + 
            \mathfrak{p}'(\sfp)
            \sup_{\nu \in [0,\tstar]}
                \LpLaw{2\sfp}{}{\FZparasigma{\nu,\Zt{N,\nu}}}
        \right)
        , 
    \end{multline}
    for all $(t,\sfp) \in [\BoldTs,\tstar] \times \mathbb{N}_{\geq 1}$, where we have used the fact that 
    \begin{align}\label{eqn:lem:Appendix:TrueProcess:SumProcessNew:Expectation:Fact:1}
        &\sup_{\nu \in [\BoldTs,t]}
        \LpLaw{4\sfp}{}{ 
            \Fparasigma{\istar{\nu}\BoldTs,\Xt{N,\istar{\nu}\BoldTs}}
        }
        \notag 
        \\
        &\leq
        \sup_{\nu \in [0,\tstar]}
        \LpLaw{4\sfp}{}{ 
            \Fparasigma{\istar{\nu}\BoldTs,\Xt{N,\istar{\nu}\BoldTs}}
        }
        \leq
        \sup_{\nu \in [0,\tstar]}
        \LpLaw{4\sfp}{}{ 
            \Fparasigma{\nu,\Xt{N,\nu}}
        }, \quad \forall t \in [\BoldTs,\tstar].
    \end{align}
    We now use the defintion of $\widetilde{\gamma}^\star$ to obtain 
    \begin{multline*}
        \absolute{\widetilde{\gamma}^\star(\nu,\BoldTs)}
        \sup_{\beta \in [(\istar{\nu}-1)\BoldTs,\istar{\nu}\BoldTs]}
            \expo{ (\lambda_s - \Boldomega) \beta}
        \\
        = 
        \expo{(\Boldomega - \lambda_s)\istar{\nu}\BoldTs}
        \frac{\lambda_s}{\Boldomega}
        \frac{\expo{\Boldomega (\nu - \istar{\nu}\BoldTs)} - 1}{\expo{\lambda_s \BoldTs} - 1}
        \sup_{\beta \in [(\istar{\nu}-1)\BoldTs,\istar{\nu}\BoldTs]}
            \expo{ (\lambda_s - \Boldomega) \beta}
        \\
        \leq 
        \expo{(\Boldomega - \lambda_s)\istar{\nu}\BoldTs}
        \frac{\lambda_s}{\Boldomega}
        \frac{\expo{\Boldomega \BoldTs} - 1}{\expo{\lambda_s \BoldTs} - 1}
        \sup_{\beta \in [(\istar{\nu}-1)\BoldTs,\istar{\nu}\BoldTs]}
            \expo{ (\lambda_s - \Boldomega) \beta}
            ,
        \quad 
        \forall \nu \in [\BoldTs,\tstar]
        , 
    \end{multline*}
    where the last inequality is due to $\nu - \istar{\nu}\BoldTs \leq \BoldTs$, $\forall \nu \geq 0$.
    Hence, if $\Boldomega \geq \lambda_s$, then the inequality above can be further generalized into 
    \begin{multline*}
        \absolute{\widetilde{\gamma}^\star(\nu,\BoldTs)}
        \sup_{\beta \in [(\istar{\nu}-1)\BoldTs,\istar{\nu}\BoldTs]}
            \expo{ (\lambda_s - \Boldomega) \beta}
        \leq 
        \expo{(\Boldomega - \lambda_s)\istar{\nu}\BoldTs}
        \frac{\lambda_s}{\Boldomega}
        \frac{\expo{\Boldomega \BoldTs} - 1}{\expo{\lambda_s \BoldTs} - 1}
            \expo{ (\lambda_s - \Boldomega) (\istar{\nu}-1)\BoldTs}
        \\
        = 
        \expo{(\Boldomega - \lambda_s)\BoldTs}
        \frac{\lambda_s}{\Boldomega}
        \frac{\expo{\Boldomega \BoldTs} - 1}{\expo{\lambda_s \BoldTs} - 1}
        ,
        \quad 
        \forall \nu \in [\BoldTs,\tstar]
        . 
    \end{multline*}
    If instead $\Boldomega < \lambda_s$, then 
    \begin{multline*}
        \absolute{\widetilde{\gamma}^\star(\nu,\BoldTs)}
        \sup_{\beta \in [(\istar{\nu}-1)\BoldTs,\istar{\nu}\BoldTs]}
            \expo{ (\lambda_s - \Boldomega) \beta}
        \leq 
        \expo{(\Boldomega - \lambda_s)\istar{\nu}\BoldTs}
        \frac{\lambda_s}{\Boldomega}
        \frac{\expo{\Boldomega \BoldTs} - 1}{\expo{\lambda_s \BoldTs} - 1}
            \expo{ (\lambda_s - \Boldomega) \istar{\nu}\BoldTs}
        \\
        = 
        \frac{\lambda_s}{\Boldomega}
        \frac{\expo{\Boldomega \BoldTs} - 1}{\expo{\lambda_s \BoldTs} - 1}
        ,
        \quad 
        \forall \nu \in [\BoldTs,\tstar]
        . 
    \end{multline*}
    We therefore conclude that
    \begin{align}\label{eqn:lem:Appendix:TrueProcess:SumProcessNew:Expectation:B:3:Identity}
        \absolute{\widetilde{\gamma}^\star(\nu,\BoldTs)}
        \sup_{\beta \in [(\istar{\nu}-1)\BoldTs,\istar{\nu}\BoldTs]}
            \expo{ (\lambda_s - \Boldomega) \beta}
        \leq 
        \max\left\{ \expo{(\Boldomega - \lambda_s)\BoldTs},1 \right\} 
        \frac{\lambda_s}{\Boldomega} 
        \frac{\expo{\Boldomega\BoldTs} - 1}{\expo{\lambda_s \BoldTs} - 1}
        \doteq
        \gamma_2(\Boldomega,\BoldTs) 
        ,
    \end{align}
    for all $\nu \in [\BoldTs,\tstar]$.
    Substituting the above into~\eqref{eqn:lem:Appendix:TrueProcess:SumProcessNew:Expectation:B:3:Initial} then leads to 
    \begin{multline}\label{eqn:lem:Appendix:TrueProcess:SumProcessNew:Expectation:B:3:Final}
        \LpLaw{\sfp}{}{\mathfrak{Z}_3(t)} 
        \leq  
        \Delta_g^2
        \mathfrak{p}''(\sfp)   
        \frac{\expo{ 2\lambda  t }}{2\lambda \Boldomega}
            \gamma_2(\Boldomega,\BoldTs) 
            \left(1 - \expo{ -2\Boldomega \BoldTs } \right)^\frac{1}{2}
        \sup_{\nu \in [0,\tstar]}
            \LpLaw{4\sfp}{}{ 
                \Fparasigma{\nu,\Xt{N,\nu}}
            }
        \\
        \times     
        \left( 
            \frac{1}{\sqrt{\Boldomega}}
            \sup_{\nu \in [0,\tstar]}
                \LpLaw{2\sfp}{}{\LZparamu{\nu,\Zt{N,\nu}}}
            + 
            \mathfrak{p}'(\sfp)
            \sup_{\nu \in [0,\tstar]}
                \LpLaw{2\sfp}{}{\FZparasigma{\nu,\Zt{N,\nu}}}
        \right)
        , 
    \end{multline}
    for all $(t,\sfp) \in [\BoldTs,\tstar] \times \mathbb{N}_{\geq 1}$.
    
    \ul{\emph{Bound for $\mathfrak{Z}_4(t)$:}}
    Finally, we have  
    \begin{align*}
        \mathfrak{Z}_4(t)
        =& 
        \int_{\BoldTs}^t  
            \expo{ 2(\lambda - \Boldomega) \nu }
            \left(
                \int_0^\nu 
                    \expo{\Boldomega \beta}
                        \mathcal{J}_{4}(\beta,\Xt{N})
                d\Wt{\beta}
            \right)^\top 
            g(\nu)^\top
            g(\nu)
            \\
            &
            \hspace{2cm}
            \times
            \left(
                \int_0^{\nu}
                \expo{\Boldomega \beta}
                \left[ 
                    \LZparamu{\beta,\Zt{N,\beta}}d\beta + \FZparasigma{\beta,\Zt{N,\beta}}d\Wt{\beta} 
                \right]
            \right)       
        d\nu
        \\
        =& 
        \int_{\BoldTs}^t  
            \expo{ 2(\lambda - \Boldomega) \nu }
            \left(
                \Fparasigma{\istar{\nu}\BoldTs,\Xt{N,\istar{\nu}\BoldTs}}
                \int_{\istar{\nu}\BoldTs}^\nu 
                    \expo{\Boldomega \beta}
                d\Wt{\beta}
            \right)^\top 
            g(\nu)^\top
            g(\nu)
            \\
            &
            \hspace{2cm}
            \times
            \left(
                \int_0^{\nu}
                \expo{\Boldomega \beta}
                \left[ 
                    \LZparamu{\beta,\Zt{N,\beta}}d\beta + \FZparasigma{\beta,\Zt{N,\beta}}d\Wt{\beta} 
                \right]
            \right)       
        d\nu,
    \end{align*}
    for all $t \in [\BoldTs,\tstar]$, which we can use tocast the above process as $N_b(t)$ in Corollary~\ref{cor:TechnicalResults:PCA} by setting 
    \begin{align*}
        \theta_1 = \lambda, 
        \quad
        \theta_2 = \Boldomega,
        \quad   
        Q_t = \Wt{t} (n_q =d)
        ,
        \\
        R(t) 
        = 
        g(t)^\top g(t) 
        \in 
        \mathbb{R}^{m \times m}\,(n_1=n_2 = m)
        , 
        \\ 
        L_b(\nu)  
        =
        \Fparasigma{\istar{\nu}\BoldTs,\Xt{N,\istar{\nu}\BoldTs}}
        \in \mathbb{R}^{m \times d},
        \quad  
        S(t) = \LZparamu{t,\Zt{N,t}},
        \quad 
        L(t) = \FZparasigma{t,\Zt{N,t}}
        .
    \end{align*}
    Hence, we use~\eqref{cor:TechnicalResults:PCA:Nb:Bound:Final} in Corollary~\ref{cor:TechnicalResults:PCA}, and the bound in~\eqref{eqn:lem:Appendix:TrueProcess:SumProcessNew:Expectation:Fact:1} to obtain
    \begin{multline}\label{eqn:lem:Appendix:TrueProcess:SumProcessNew:Expectation:B:4:Final}
        \LpLaw{\sfp}{}{
            \mathfrak{Z}_4(t)
        }
        \leq   
        \Delta_g^2
        \mathfrak{p}''(\sfp)   
        \frac{\expo{ 2\lambda  t }}{2\lambda \Boldomega} 
            \left(1 - \expo{ -2\Boldomega \BoldTs } \right)^\frac{1}{2}
        \sup_{\nu \in [0,\tstar]}
            \LpLaw{4\sfp}{}{ 
                \Fparasigma{\nu,\Xt{N,\nu}}
            }
        \\
        \times     
        \left( 
            \frac{1}{\sqrt{\Boldomega}}
            \sup_{\nu \in [0,\tstar]}
                \LpLaw{2\sfp}{}{\LZparamu{\nu,\Zt{N,\nu}}}
            + 
            \mathfrak{p}'(\sfp)
            \sup_{\nu \in [0,\tstar]}
                \LpLaw{2\sfp}{}{\FZparasigma{\nu,\Zt{N,\nu}}}
        \right)
        , 
    \end{multline}
    for all $(t,\sfp) \in [\BoldTs,\tstar] \times \mathbb{N}_{\geq 1}$. 
    The proof is then concluded by adding together the bounds in~\eqref{eqn:lem:Appendix:TrueProcess:SumProcessNew:Expectation:B:1:Final}, ~\eqref{eqn:lem:Appendix:TrueProcess:SumProcessNew:Expectation:B:2:Final},~\eqref{eqn:lem:Appendix:TrueProcess:SumProcessNew:Expectation:B:3:Final}, and~\eqref{eqn:lem:Appendix:TrueProcess:SumProcessNew:Expectation:B:4:Final}.

\end{proof}

Using the bounds established in the last proposition, the following result establishes the effect of the feedback error operator $\FL - \ReferenceInput$ on the truncated joint process in Definition~\ref{def:True:TruncatedJointProcess}.  
\begin{lemma}\label{lem:Appendix:TrueProcess:NUTilde:Bounds}
    Consider the following scalar process:
    \begin{align}\label{eqn:lem:Appendix:TrueProcess:NUTilde:Definitions}
        N_{\mathcal{U}}(t)
        =  
        \int_0^{t}
            \expo{\Boldomega \nu}
            \widetilde{M}_\mathcal{U}(t,\nu)
            \left[ \LZparamu{\nu,\Zt{N,\nu}}d\nu + \FZparasigma{\nu,\Zt{N,\nu}}d\Wt{\nu} \right] 
            \in       
            \mathbb{R}
        ,
    \end{align}
    where  
    \begin{align*}
        \widetilde{M}_\mathcal{U}(t,\nu)
        =
        \int_\nu^{t} 
            \expo{ (2\lambda - \Boldomega) \beta } \widetilde{\mathcal{P}}_\mathcal{U}(\beta)^\top
        d\beta, 
    \end{align*}
    for $0 \leq \nu \leq t \leq T$, and where where $\widetilde{\mathcal{P}}_{\mathcal{U}}(\beta) \in \mathbb{R}^{m}$ is  defined in the statement of Proposition~\ref{prop:Appendix:True:ddV:Bound}.
    If the stopping time $\tau^\star$, defined in~\eqref{eqn:True:StoppingTimes}, Lemma~\ref{lem:True:dV}, satisfies  $\tau^\star = \tstar$, then
    \begin{align}\label{eqn:lem:Appendix:TrueProcess:NUTilde:Bounds:Main}
        \LpLaw{\sfp}{}{N_{\mathcal{U}}(t)}
        \leq
        \frac{\expo{ 2\lambda t }  }{\lambda}
        \Delta_{\mathcal{U}}(\sfp,\tstar,\Boldomega,\BoldTs),
        \quad   
        \forall
        (t,\sfp) \in [0,\tstar] \times \mathbb{N}_{\geq 1},
    \end{align}
    where 
    \begin{align*}
        \Delta_{\mathcal{U}}(\sfp,\tstar,\Boldomega,\BoldTs)
        \doteq 
        \Delta^{-}(\sfp,\tstar,\Boldomega,\BoldTs)
        + 
        \indicator{\geq \BoldTs}{\tstar}
        \sum_{k=1}^3
        \Delta_k^+(\sfp,\tstar,\Boldomega,\BoldTs),
    \end{align*} 
    and where 
    \begin{align*}
        \begin{multlined}[b][0.95\linewidth]
            \Delta^{-}(\sfp,\tstar,\Boldomega,\BoldTs)
            =
            \mathring{\Delta}(\sfp,\tstar,\Boldomega)
            \gamma_1'(\Boldomega,\BoldTs)
            \left( 
                \frac{ 1 }{\sqrt{\Boldomega}}  
                \sup_{\nu \in [0,\tstar]}
                    \LpLaw{2\sfp}{}{\Lparamu{\nu,\Xt{N,\nu}}}
                +
                2 
                \mathfrak{p}'(\sfp)
                \sup_{\nu \in [0,\tstar]}
                    \LpLaw{2\sfp}{}{\Fparasigma{\nu,\Xt{N,\nu}}}
            \right)
            ,
        \end{multlined}
        \\
        \begin{multlined}[b][0.95\linewidth]
            \Delta_1^+(\sfp,\tstar,\Boldomega,\BoldTs)
            \\
            =
            \frac{1}{\sqrt{\Boldomega}}
            \mathring{\Delta}(\sfp,\tstar,\Boldomega)
            \left( 
                \Delta_{\widetilde{\mathcal{H}}_\mu}(\BoldTs,\sfp,\tstar)
                +
                \Delta_{\mathcal{G}_\mu}(\BoldTs,\tstar)
                + 
                \sqrt{\Boldomega}
                \mathfrak{p}'(\sfp)
                \left[
                    \Delta_{\widetilde{\mathcal{H}}_\sigma}(\BoldTs,\sfp,\tstar)
                    + 
                    \Delta_{\widetilde{\mathcal{G}}_\sigma}(\BoldTs,\sfp,\tstar)
                \right]
            \right),
        \end{multlined}
        \\
        \Delta_2^+(\sfp,\tstar,\Boldomega,\BoldTs)
        =
        \mathring{\Delta}(\sfp,\tstar,\Boldomega)
        \gamma''\left(\sfp,\Boldomega,\BoldTs\right)
        \sup_{\nu \in [0,\tstar]}
            \LpLaw{4\sfp}{}{\Fparasigma{\nu,\Xt{N,\nu}}}
        ,
        \\
        \Delta_3^+(\sfp,\tstar,\Boldomega,\BoldTs)
        =
        \frac{\expo{\Boldomega \BoldTs}-1}{\BoldomegaRoot}
        \mathring{\Delta}(\sfp,\tstar,\Boldomega)
        \sup_{\nu \in [0,\tstar]}
            \LpLaw{2\sfp}{}{\Lparamu{\nu,\Xt{N,\nu}}}
        .
    \end{align*}
    In the above, the terms $\Delta_{\widetilde{\mathcal{H}}_\mu}$, $\Delta_{\widetilde{\mathcal{H}}_\sigma}$, $\Delta_{\widetilde{\mathcal{G}}_\sigma}$, and $\Delta_{\mathcal{G}_\mu}$ are defined in Proposition~\ref{prop:Appendix:TrueProcess:ControlErrorIntegrand:Bounds}, the terms $\mathfrak{p}'(\sfp)$, $\mathfrak{p}''(\sfp)$, $\gamma_1'$, $\gamma_2$, $\gamma'_2$, and $\gamma''$ are defined in Appendix~\ref{app:Definitions}, and 
    \begin{align*}
        \mathring{\Delta}(\sfp,\tstar,\Boldomega)
        =
        \Delta_g^2
        \left(
            \frac{1}{\BoldomegaRoot}
            \sup_{\nu \in [0,\tstar]}
            \LpLaw{2\sfp}{}{
                \LZparamu{\nu,\Zt{N,\nu}}
            }
            + 
            \mathfrak{p}'(\sfp)  
            \sup_{\nu \in [0,\tstar]}
            \LpLaw{2\sfp}{}{
                \FZparasigma{\nu,\Zt{N,\nu}}
            }
        \right).
    \end{align*}

\end{lemma}
\begin{proof}
    Assume w.l.o.g. that $\BoldTs < T$.
    Then, as in~\eqref{eqn:prop:Appendix:TrueProcess:NU:Redefined} in the proof of Proposition~\ref{prop:Appendix:TrueProcess:NU:Bounds}, we obtain 
    \begin{align*}
        N_{\mathcal{U}}(t)
        =
        \int_0^{t}
            \expo{ (2\lambda - \Boldomega) \nu }
            \widetilde{\mathcal{P}}_\mathcal{U}(\nu)^\top
            \left(
            \int_0^{\nu}
                \expo{\Boldomega \beta}
                \left[ \LZparamu{\beta,\Zt{N,\beta}}d\beta + \FZparasigma{\beta,\Zt{N,\beta}}d\Wt{\beta} \right]
            \right)
        d\nu
        , 
        \quad 
        t \in [0,T].
    \end{align*}
    Using the definition of $\widetilde{\mathcal{P}}_{\mathcal{U}}$ in Proposition~\ref{prop:Appendix:True:ddV:Bound}, we get 
    \begin{align}\label{eqn:lem:Appendix:TrueProcess:NUTilde:Redefined}
        &N_{\mathcal{U}}(t)
        \notag
        \\
        &=
        2
        \int_0^{t}
            \expo{ (2\lambda - \Boldomega) \nu }
            \left[\left(\FL - \ReferenceInput\right)\left(\Xt{N}\right)(\nu)\right]^\top
            g(\nu)^\top
            g(\nu)
            \left(
            \int_0^{\nu}
                \expo{\Boldomega \beta}
                \left[ \LZparamu{\beta,\Zt{N,\beta}}d\beta + \FZparasigma{\beta,\Zt{N,\beta}}d\Wt{\beta} \right]
            \right)
        d\nu
        \notag 
        \\
        &=
        2 \Boldomega
        \int_0^{t}
            \expo{ 2(\lambda - \Boldomega) \nu }
            \ControlErrorTilde[\Xt{N},\nu]^\top
            g(\nu)^\top
            g(\nu)
            \left(
            \int_0^{\nu}
                \expo{\Boldomega \beta}
                \left[ \LZparamu{\beta,\Zt{N,\beta}}d\beta + \FZparasigma{\beta,\Zt{N,\beta}}d\Wt{\beta} \right]
            \right)
        d\nu
        , 
    \end{align}
    for $t \in [0,T]$, where the last line is due to the conclusion in Proposition~\ref{prop:Appendix:TrueProcess:FL:Expression} that $\left(\FL - \ReferenceInput\right)\br{\Xt{N}}(\nu)=\Boldomega \expo{-\Boldomega \nu}\ControlErrorTilde[\Xt{N},\nu]$.

    To obtain the expression for $\ControlErrorTilde[\Xt{N},\nu]$, we invoke Proposition~\ref{proposition:Appendix:TrueProcess:FL:ExpressionEvolved} with $z = \Xt{N}$ to obtain: 
    \begin{align*}
         \ControlErrorTilde[\Xt{N},\nu]
        =
        \int_0^\nu 
            \expo{ \Boldomega \beta} 
            \Sigma^{\paral}_\beta(\Xt{N}), 
        \quad \nu < \BoldTs,
    \end{align*}
    and for $\nu \geq \BoldTs$, 
    \begin{multline*}
        \begin{aligned}
            \ControlErrorTilde[\Xt{N},\nu]
            =&
            \int_0^\BoldTs
                \expo{ \Boldomega \beta} 
                \Lparamu{\beta,\Xt{N,\beta}}
            d\beta
            \\
            &+ 
            \int_{0}^{\nu}
                \expo{\Boldomega \beta}
                \left[
                    \left(
                        \widetilde{\mathcal{H}}_\mu^{\paral}(\beta,\Xt{N})
                        + 
                        \expo{(\lambda_s-\Boldomega) \beta}
                        \widetilde{\mathcal{H}}_\mu(\beta,\Xt{N})
                    \right)
                    d\beta
                    + 
                    \left(
                        \widetilde{\mathcal{H}}_\sigma^{\paral}(\beta,\Xt{N})
                        + 
                        \expo{(\lambda_s-\Boldomega) \beta}
                        \widetilde{\mathcal{H}}_\sigma(\beta,\Xt{N})
                    \right)
                    d\Wt{\beta}
                \right]
        \end{aligned}
        \\
        +
        \int_{0}^{\nu}
            \expo{\Boldomega \beta} 
            \left( 
                \widetilde{\mathcal{G}}_{\sigma_1}(\beta,\Xt{N})
                + 
                \expo{(\lambda_s - \Boldomega) \beta} 
                \widetilde{\mathcal{G}}_{\sigma_2}(\beta,\Xt{N})
            \right)
        d\Wt{\beta}
        \\
        +
        \int_0^\nu 
            \expo{\Boldomega \beta}
            \left[
                \mathcal{G}_\mu(\beta,\Xt{N})
                d\beta
                + 
                \left( 
                    \mathcal{G}_{\sigma_1}(\beta,\Xt{N})
                    + 
                    \mathcal{G}_{\sigma_2}(\beta,\Xt{N})
                \right)
                d\Wt{\beta}
            \right]
            .
    \end{multline*}
    Susbtituting the above expression for $\ControlErrorTilde[\Xt{N},\nu]$ into~\eqref{eqn:lem:Appendix:TrueProcess:NUTilde:Redefined}:
    \begin{align}\label{eqn:lem:Appendix:TrueProcess:NUTilde:Representation:Evolved:Pre1}
        N_{\mathcal{U}}(t)
        =
        \begin{cases}
            2 \Boldomega
            \int_0^{t}
                \expo{ 2(\lambda - \Boldomega) \nu }
                \mathcal{N}_\star(\nu)^\top
                g(\nu)^\top
                g(\nu)
                \mathring{\mathcal{N}}(\nu)
            d\nu
            ,
            & \text{if } t < \BoldTs
            ,
            \\[10pt]
            \begin{aligned}
                &
                2 \Boldomega
                \int_0^{\BoldTs}
                    \expo{ 2(\lambda - \Boldomega) \nu }
                    \mathcal{N}_\star(\nu)^\top
                    g(\nu)^\top
                    g(\nu)
                    \mathring{\mathcal{N}}(\nu)
                d\nu
                \\
                &+ 
                2 \Boldomega
                \int_{\BoldTs}^t  
                    \expo{ 2(\lambda - \Boldomega) \nu }
                    \left(\mathcal{N}_1(\nu) + \mathcal{N}_2(\nu) + \mathcal{N}_3(\nu)\right)^\top 
                \\
                &\hspace{5cm} 
                \times
                    g(\nu)^\top
                    g(\nu)
                    \mathring{\mathcal{N}}(\nu)       
                d\nu
            \end{aligned}
            , 
            & \text{if otherwise } t \geq \BoldTs
            ,
        \end{cases}
    \end{align} 
    where
    \begin{align*}
        \mathcal{N}_\star(\nu)
        =
        \int_0^\nu 
            \expo{ \Boldomega \beta} 
            \Sigma^{\paral}_\beta(\Xt{N})
        =
        \int_0^\nu 
            \expo{ \Boldomega \beta}
            \left[
                \Lparamu{\beta,\Xt{N,\beta}}d\beta 
                +
                \Fparasigma{\beta,\Xt{N,\beta}}d\Wt{\beta}
            \right]
        ,
        \\
        \begin{multlined}[b][0.9\linewidth]
            \mathcal{N}_1(\nu)
            = 
            \int_{0}^{\nu}
                \expo{\Boldomega \beta}
                \left[
                    \left(
                        \widetilde{\mathcal{H}}_\mu^{\paral}(\beta,\Xt{N})
                        + 
                        \expo{(\lambda_s-\Boldomega) \beta}
                        \widetilde{\mathcal{H}}_\mu(\beta,\Xt{N})
                        + 
                        \mathcal{G}_\mu(\beta,\Xt{N})
                    \right)
                    d\beta
                \right. 
                \\
                \left.
                    + 
                    \left(
                        \widetilde{\mathcal{H}}_\sigma^{\paral}(\beta,\Xt{N})
                        + 
                        \expo{(\lambda_s-\Boldomega) \beta}
                        \widetilde{\mathcal{H}}_\sigma(\beta,\Xt{N})
                        + 
                        \widetilde{\mathcal{G}}_{\sigma_1}(\beta,\Xt{N})
                        + 
                        \expo{(\lambda_s - \Boldomega) \beta} 
                        \widetilde{\mathcal{G}}_{\sigma_2}(\beta,\Xt{N})
                    \right)
                    d\Wt{\beta}
                \right]
            ,
        \end{multlined}
        \\ 
        \mathcal{N}_2(\nu)
        =
        \int_0^\nu 
            \expo{\Boldomega \beta}
            \left( 
                \mathcal{G}_{\sigma_1}(\beta,\Xt{N})
                + 
                \mathcal{G}_{\sigma_2}(\beta,\Xt{N})
            \right)
        d\Wt{\beta}
        , 
        \\
        \mathcal{N}_3
        =
        \int_0^{\BoldTs} 
            \expo{\Boldomega \beta}
            \Lparamu{\beta,\Xt{N,\beta}}
        d\beta
        ,
        \\
        \mathring{\mathcal{N}}(\nu)
        =
        \int_0^{\nu}
            \expo{\Boldomega \beta}
            \left[ 
                \LZparamu{\beta,\Zt{N,\beta}}d\beta + \FZparasigma{\beta,\Zt{N,\beta}}d\Wt{\beta} 
            \right]
        .
    \end{align*}
    We can further write~\eqref{eqn:lem:Appendix:TrueProcess:NUTilde:Representation:Evolved:Pre1} as     
    \begin{align}\label{eqn:lem:Appendix:TrueProcess:NUTilde:Representation:Evolved:Pre}
        N_{\mathcal{U}}(t)
        =
        \begin{cases}
            2\Boldomega
            \mathfrak{N}_\star(t)
            ,
            & \text{if } t < \BoldTs
            ,
            \\[10pt]
            2\Boldomega
            \left(
                \mathfrak{N}_\star(\BoldTs)
                + 
                \sum_{k=1}^3
                \mathfrak{N}_k(t)
            \right)
            , 
            & \text{if otherwise } t \geq \BoldTs
            ,
        \end{cases}
    \end{align}
    where we have defined 
    \begin{align*}
        \mathfrak{N}_\star(t)
        =&
        \int_0^{t}
            \expo{ 2(\lambda - \Boldomega) \nu }
            \mathcal{N}_\star(\nu)^\top
            g(\nu)^\top
            g(\nu)
            \mathring{\mathcal{N}}(\nu)
        d\nu
        , 
        \\
        \mathfrak{N}_k(t)
        =& 
        \int_{\BoldTs}^t  
            \expo{ 2(\lambda - \Boldomega) \nu }
            \mathcal{N}_k(\nu)^\top 
            g(\nu)^\top
            g(\nu)
            \mathring{\mathcal{N}}(\nu)       
        d\nu
        , 
        \quad k \in \cbr{1,2,3}.
    \end{align*}
    Using the Minkowski's inequality along with the fact that $\tstar$ and $\BoldTs$ are constants leads to the following bounds for all $\sfp \in \mathbb{N}_{\geq 1}$:
    \begin{align}\label{eqn:lem:Appendix:TrueProcess:NUTilde:Bound:Initial}
        \LpLaw{\sfp}{}{
            N_{\mathcal{U}}(t)
        }
        \leq
        \begin{cases}
            2\Boldomega
            \LpLaw{\sfp}{}{\mathfrak{N}_\star(t)}
            ,
            & \forall t \in [0,\tstar] \cap [0,\BoldTs)
            ,
            \\[10pt]
            2\Boldomega
            \left(
                \LpLaw{\sfp}{}{\mathfrak{N}_\star(\BoldTs)}
                + 
                \sum_{k=1}^3
                \LpLaw{\sfp}{}{\mathfrak{N}_k(t)}
            \right)
            , 
            & \forall t \in [0,\tstar] \cap [\BoldTs,T]
            .
        \end{cases}
    \end{align} 

    We now bound the processes $\mathfrak{N}_\star(t)$ and $\mathfrak{N}_k(t)$, $k \in \cbr{1,2,3}$.

    \ul{\emph{Bound for $\mathfrak{N}_\star(t)$:}}
    Using the definitions in~\eqref{eqn:lem:Appendix:TrueProcess:NUTilde:Representation:Evolved:Pre1}-\eqref{eqn:lem:Appendix:TrueProcess:NUTilde:Representation:Evolved:Pre}, we have 
    \begin{multline*}
        \begin{aligned}
            \mathfrak{N}_\star(t)
            =&
            \int_0^{t}
                \expo{ 2(\lambda - \Boldomega) \nu }
                \mathcal{N}_\star(\nu)^\top
                g(\nu)^\top
                g(\nu)
                \mathring{\mathcal{N}}(\nu)
            d\nu
            \\
            =&
            \int_0^{t}
                \expo{ 2(\lambda - \Boldomega) \nu }
                \left(
                    \int_0^\nu 
                        \expo{ \Boldomega \beta}
                        \left[
                            \Lparamu{\beta,\Xt{N,\beta}}d\beta 
                            +
                            \Fparasigma{\beta,\Xt{N,\beta}}d\Wt{\beta}
                        \right]
                \right)^\top
                g(\nu)^\top
                g(\nu)
        \end{aligned}
        \\
        \times
            \left(
                \int_0^{\nu}
                    \expo{\Boldomega \beta}
                    \left[ 
                        \LZparamu{\beta,\Zt{N,\beta}}d\beta + \FZparasigma{\beta,\Zt{N,\beta}}d\Wt{\beta} 
                    \right]
            \right)
        d\nu
        ,
    \end{multline*}
    for all $(t,\sfp) \in [0,\tstar \wedge \BoldTs] \times \mathbb{N}_{\geq 1}$.
    Therefore, we can cast the process above as $N(t)$ in Lemma~\ref{cor:TechnicalResults:UTildeBound} by setting
    \begin{align*} 
        \xi = 0,
        \quad  
        \theta_1 = \lambda, 
        \quad
        \theta_2 = \Boldomega,
        \quad   
        Q_t = \Wt{t} (n_q =d),
        \\   
        R(t) 
        = 
        g(t)^\top g(t) 
        \in 
        \mathbb{R}^{m \times m}\,(n_1=n_2 = m)
        ,
        \\ 
        S_1(t) = \Lparamu{t,\Xt{N,t}},
        \quad 
        L_1(t) = \Fparasigma{t,\Xt{N,t}},
        \quad  
        S_2(t) = \LZparamu{t,\Zt{N,t}},
        \quad 
        L_2(t) = \FZparasigma{t,\Zt{N,t}}
        .
    \end{align*}
    Thus, we apply Lemma~\ref{cor:TechnicalResults:UTildeBound} and obtain the following upon re-arranging terms:
    \begin{multline*}
        \LpLaw{\sfp}{}{\mathfrak{N}_\star(t)}
        \leq 
        \Delta_g^2
        \frac{\expo{ 2\lambda t }  }{2\lambda \Boldomega}
        \left(1 - \expo{ -2\lambda (\tstar \wedge \BoldTs) } \right)
        \left(
            1 - \expo{-\Boldomega (\tstar \wedge \BoldTs)}   
        \right)
        \\
        \times
        \left( 
            \frac{ 1 }{\sqrt{\Boldomega}}  
            \sup_{\nu \in [0,\tstar \wedge \BoldTs]}
                \LpLaw{2\sfp}{}{\Lparamu{\nu,\Xt{N,\nu}}}
            +
            2 
            \mathfrak{p}'(\sfp)
            \sup_{\nu \in [0,\tstar \wedge \BoldTs]}
                \LpLaw{2\sfp}{}{\Fparasigma{\nu,\Xt{N,\nu}}}
        \right)
        \\
        \times     
        \left( 
            \frac{ 1 }{\sqrt{\Boldomega}}
            \sup_{\nu \in [0,\tstar \wedge \BoldTs]}
                \LpLaw{2\sfp}{}{\LZparamu{\nu,\Zt{N,\nu}}}
            + 
            2
            \mathfrak{p}'(\sfp)
            \sup_{\nu \in [0,\tstar \wedge \BoldTs]}
                \LpLaw{2\sfp}{}{\FZparasigma{\nu,\Xt{N,\nu}}}
        \right)
        ,
    \end{multline*}
    for all  $(t,\sfp) \in [0,\tstar \wedge \BoldTs] \times \mathbb{N}_{\geq 1}$, where we have used Assumption~\ref{assmp:KnownFunctions} to set $\Delta_R = \Delta_g^2$ in the hypothesis of Proposition~\ref{cor:TechnicalResults:UTildeBound}.
    We have further used the facts that $1 - \expo{-\Boldomega(\tstar \wedge \BoldTs)} \leq 1$ and $1 + \expo{-\Boldomega(\tstar \wedge \BoldTs)} \leq 2$, $\forall \tstar \wedge \BoldTs \geq 0$, after first using the following trivial manipulation: 
    \begin{multline}\label{eqn:lem:Appendix:TrueProcess:NUTilde:ExponentialManipulation}
        \left(
            \frac{1 - \expo{-2\Boldomega (\tstar \wedge \BoldTs)}}{\Boldomega}  
        \right)^\Half
        =
        \left(
            \frac{ \left(1 - \expo{-\Boldomega (\tstar \wedge \BoldTs)}\right)\left(1 + \expo{-\Boldomega (\tstar \wedge \BoldTs)}\right) }{\Boldomega}  
        \right)^\Half
        \\
        =
        \left(
            \frac{ 1 - \expo{-\Boldomega (\tstar \wedge \BoldTs)} }{\Boldomega}  
        \right)^\Half
        \left(1 + \expo{-\Boldomega (\tstar \wedge \BoldTs)}\right)^\Half
        .
    \end{multline}
    Note that the facts $\tstar \wedge \BoldTs \leq \tstar$ and $\tstar \wedge \BoldTs \leq \BoldTs$ imply that $\sbr{0,\tstar \wedge \BoldTs} \subseteq \sbr{0,\tstar}$ and
    \begin{align*}
        \left(1 - \expo{ -2\lambda (\tstar \wedge \BoldTs) } \right)
        \left(
            1 - \expo{-\Boldomega (\tstar \wedge \BoldTs)}   
        \right)
        \leq      
        \left(1 - \expo{ -2\lambda \BoldTs } \right)
        \left(
            1 - \expo{-\Boldomega \BoldTs}   
        \right)
        \doteq 
        \gamma_1'(\Boldomega,\BoldTs)
        .
    \end{align*}
    Hence, we can develop the previous bound into
    \begin{multline*}
        \LpLaw{\sfp}{}{\mathfrak{N}_\star(t)}
        \leq 
        \Delta_g^2
        \frac{\expo{ 2\lambda t }  }{2\lambda \Boldomega}
        \gamma_1'(\Boldomega,\BoldTs) 
        \left( 
            \frac{ 1 }{\sqrt{\Boldomega}}  
            \sup_{\nu \in [0,\tstar]}
                \LpLaw{2\sfp}{}{\Lparamu{\nu,\Xt{N,\nu}}}
            +
            2 
            \mathfrak{p}'(\sfp)
            \sup_{\nu \in [0,\tstar]}
                \LpLaw{2\sfp}{}{\Fparasigma{\nu,\Xt{N,\nu}}}
        \right)
        \\
        \times     
        \left( 
            \frac{ 1 }{\sqrt{\Boldomega}}
            \sup_{\nu \in [0,\tstar]}
                \LpLaw{2\sfp}{}{\LZparamu{\nu,\Zt{N,\nu}}}
            + 
            2
            \mathfrak{p}'(\sfp)
            \sup_{\nu \in [0,\tstar]}
                \LpLaw{2\sfp}{}{\FZparasigma{\nu,\Xt{N,\nu}}}
        \right)
        ,
    \end{multline*}
    which further implies
    \begin{align}\label{eqn:lem:Appendix:TrueProcess:NUTilde:Bound:A:Final}
        2\Boldomega
        \LpLaw{\sfp}{}{\mathfrak{N}_\star(t)}
        \leq 
        \frac{\expo{ 2\lambda t }  }{\lambda}
        \Delta^{-}(\sfp,\tstar,\Boldomega,\BoldTs),
        \quad 
        \forall (t,\sfp) \in [0,\tstar \wedge \BoldTs] \times \mathbb{N}_{\geq 1}.
    \end{align} 
    \begingroup
    \endgroup

    \ul{\emph{Bound for $\mathfrak{N}_1(t)$:}}
    The process $\mathfrak{N}_1$ in~\eqref{eqn:lem:Appendix:TrueProcess:NUTilde:Representation:Evolved:Pre} can be written as 
    \begin{multline*}
        \begin{aligned}
            \mathfrak{\widetilde{N}}_1(t)
            =& 
            \int_{\BoldTs}^t  
                \expo{ 2(\lambda - \Boldomega) \nu }
                \mathcal{N}_1(\nu)^\top 
                g(\nu)^\top
                g(\nu)
                \mathring{\mathcal{N}}(\nu)       
            d\nu
            \\
            =& 
            \int_{\BoldTs}^t  
                \expo{ 2(\lambda - \Boldomega) \nu }
                \left(
                    \int_{0}^{\nu}
                        \expo{\Boldomega \beta}
                        \left[
                            \mathcal{K}_\mu(\beta,\Xt{N})
                            d\beta
                            + 
                            \mathcal{K}_\sigma(\beta,\Xt{N})
                            d\Wt{\beta}
                        \right] 
                \right)^\top 
                g(\nu)^\top
                g(\nu)
        \end{aligned}
        \\
        \times    
            \left(
                \int_0^{\nu}
                    \expo{\Boldomega \beta}
                    \left[ 
                        \LZparamu{\beta,\Zt{N,\beta}}d\beta + \FZparasigma{\beta,\Zt{N,\beta}}d\Wt{\beta} 
                    \right]
            \right)
        d\nu      
        ,
    \end{multline*} 
    where we have defined
    \begin{subequations}\label{eqn:lem:Appendix:TrueProcess:NUTilde:KFunctions}
        \begin{align}
            \mathcal{K}_\mu(\beta,\Xt{N})
            =
            \widetilde{\mathcal{H}}_\mu^{\paral}(\beta,\Xt{N})
            + 
            \expo{(\lambda_s-\Boldomega) \beta}
            \widetilde{\mathcal{H}}_\mu(\beta,\Xt{N})
            + 
            \mathcal{G}_\mu(\beta,\Xt{N})
            , 
            \\ 
            \mathcal{K}_\sigma(\beta,\Xt{N})
            =
            \widetilde{\mathcal{H}}_\sigma^{\paral}(\beta,\Xt{N})
            + 
            \expo{(\lambda_s-\Boldomega) \beta}
            \widetilde{\mathcal{H}}_\sigma(\beta,\Xt{N})
            + 
            \widetilde{\mathcal{G}}_{\sigma_1}(\beta,\Xt{N})
            + 
            \expo{(\lambda_s - \Boldomega) \beta} 
            \widetilde{\mathcal{G}}_{\sigma_2}(\beta,\Xt{N})
            .
        \end{align}    
    \end{subequations}
    We apply Lemma~\ref{cor:TechnicalResults:UTildeBound} by setting
    \begin{equation}\label{eqn:lem:Appendix:TrueProcess:SumProcess:Idents} 
        \begin{aligned}
            \xi = \BoldTs,
            \quad  
            \theta_1 = \lambda, 
            \quad
            \theta_2 = \Boldomega,
            \quad   
            Q_t = \Wt{t} (n_q =d),
            \\   
            R(t) 
            = 
            g(t)^\top g(t) 
            \in 
            \mathbb{R}^{m \times m}\,(n_1=n_2 = m)
            ,
            \\ 
            S_1(t) 
            = 
            \mathcal{K}_\mu(t,\Xt{N})
            ,
            \quad 
            L_1(t) 
            = 
            \mathcal{K}_\sigma(t,\Xt{N})
            ,
            \quad 
            S_2(t) = \LZparamu{t,\Zt{N,t}},
            \quad 
            L_2(t) = \FZparasigma{t,\Zt{N,t}}
            ,
        \end{aligned}
    \end{equation}
    to obtain the following bound:
    \begin{multline}\label{eqn:lem:Appendix:TrueProcess:SumProcess:Expectation:A:1}
        \LpLaw{\sfp}{}{
            \mathfrak{N}_1(t)
        }
        \leq 
        \Delta_g^2
        \frac{ \expo{ 2\lambda t } }{2\lambda \Boldomega}
        \left( 
            \frac{1}{\sqrt{\Boldomega}}
            \sup_{\nu \in [0,\tstar]}
            \LpLaw{2\sfp}{}{ 
                \mathcal{K}_\mu(\nu,\Xt{N,\nu})
            }
            + 
            \mathfrak{p}'(\sfp)
            \sup_{\nu \in [0,\tstar]}
            \LpLaw{2\sfp}{}{ 
                \mathcal{K}_\sigma(\nu,\Xt{N,\nu})
            }
        \right)
        \\
        \times
        \left( 
            \frac{1}{\sqrt{\Boldomega}}
            \sup_{\nu \in [0,\tstar]}
                \LpLaw{2\sfp}{}{\LZparamu{\nu,\Zt{N,\nu}}}
            + 
            \mathfrak{p}'(\sfp)
            \sup_{\nu \in [0,\tstar]}
                \LpLaw{2\sfp}{}{\FZparasigma{\nu,\Zt{N,\nu}}}
        \right)
        ,
    \end{multline}
    for all $(t,\sfp) \in [\BoldTs,\tstar] \times \mathbb{N}_{\geq 1}$, where we have used the bound $\Delta_R = \Delta_g^2 \geq \Frobenius{R(t)}$ as above. 
    Using the definitions of $\mathcal{K}_\mu$ and $\mathcal{K}_\sigma$ in~\eqref{eqn:lem:Appendix:TrueProcess:NUTilde:KFunctions}, we use the Minkowski's inequality and the bounds derived in Proposition~\ref{prop:Appendix:TrueProcess:ControlErrorIntegrand:Bounds} to obtain the following:
    \begin{multline*}
        \LpLaw{2\sfp}{}{ 
            \mathcal{K}_\mu(\nu,\Xt{N})
        } 
        \leq
        \LpLaw{2\sfp}{}{ 
            \widetilde{\mathcal{H}}_\mu^{\paral}(\nu,\Xt{N})
            + 
            \expo{(\lambda_s-\Boldomega) \nu}
            \widetilde{\mathcal{H}}_\mu(\nu,\Xt{N})
        } 
        + 
        \LpLaw{2\sfp}{}{ 
            \mathcal{G}_\mu(\nu,\Xt{N})
        } 
        \\
        \leq      
        \Delta_{\widetilde{\mathcal{H}}_\mu}(\BoldTs,\sfp,\tstar)
        +
        \Delta_{\mathcal{G}_\mu}(\BoldTs,\tstar)
        ,
    \end{multline*}
    and 
    \begin{multline*}
        \LpLaw{2\sfp}{}{\mathcal{K}_\sigma(\nu,\Xt{N})}
        \leq 
        \LpLaw{2\sfp}{}{
            \widetilde{\mathcal{H}}_\sigma^{\paral}(\nu,\Xt{N})
            + 
            \expo{(\lambda_s-\Boldomega) \nu}
            \widetilde{\mathcal{H}}_\sigma(\nu,\Xt{N})
        }
        + 
        \LpLaw{2\sfp}{}{
            \widetilde{\mathcal{G}}_{\sigma_1}(\nu,\Xt{N})
            + 
            \expo{(\lambda_s - \Boldomega) \nu} 
            \widetilde{\mathcal{G}}_{\sigma_2}(\nu,\Xt{N})
        }
        \\
        \leq     
        \Delta_{\widetilde{\mathcal{H}}_\sigma}(\BoldTs,\sfp,\tstar)
        + 
        \Delta_{\widetilde{\mathcal{G}}_\sigma}(\BoldTs,\sfp,\tstar)
        ,
    \end{multline*}
    for all $(\nu,\sfp) \in [0,\tstar] \times \mathbb{N}_{\geq 1}$.
    Substituting the above bounds into~\eqref{eqn:lem:Appendix:TrueProcess:SumProcess:Expectation:A:1} then leads to
    \begin{align}\label{eqn:lem:Appendix:TrueProcess:NUTilde:Bound:B:1:Final}
        2\Boldomega
        \LpLaw{\sfp}{}{
            \mathfrak{N}_1(t)
        }
        \leq 
        \frac{ \expo{ 2\lambda t } }{\lambda}
        \Delta_1^+(\sfp,\tstar,\Boldomega,\BoldTs)
        ,
        \quad \forall (t,\sfp) \in [\BoldTs,\tstar] \times \mathbb{N}_{\geq 1}.
    \end{align}
    \begingroup
    \endgroup

    \ul{\emph{Bound for $\mathfrak{N}_2(t)$:}}
    Recall the definition of the process $\mathfrak{N}_2$ in~\eqref{eqn:lem:Appendix:TrueProcess:NUTilde:Representation:Evolved:Pre}: 
    \begin{multline*}
        \begin{aligned}
        \mathfrak{N}_2(t)
        =& 
        \int_{\BoldTs}^t  
            \expo{ 2(\lambda - \Boldomega) \nu }
            \mathcal{N}_2(\nu)^\top 
            g(\nu)^\top
            g(\nu)
            \mathring{\mathcal{N}}(\nu)       
        d\nu
        \\
        =& 
        \int_{\BoldTs}^t  
            \expo{ 2(\lambda - \Boldomega) \nu }
            \left(
                \int_0^\nu 
                    \expo{\Boldomega \beta}
                    \left( 
                        \mathcal{G}_{\sigma_1}(\beta,\Xt{N})
                        + 
                        \mathcal{G}_{\sigma_2}(\beta,\Xt{N})
                    \right)
                d\Wt{\beta}
            \right)^\top 
            g(\nu)^\top
            g(\nu)
        \end{aligned}
            \\
            \times
            \left(
                \int_0^{\nu}
                \expo{\Boldomega \beta}
                \left[ 
                    \LZparamu{\beta,\Zt{N,\beta}}d\beta + \FZparasigma{\beta,\Zt{N,\beta}}d\Wt{\beta} 
                \right]
            \right)       
        d\nu
        ,
        \quad t \in [\BoldTs,\tstar],
    \end{multline*}
    where, we have the following due to the definitions of $\mathcal{G}_{\sigma_1}$ and $\mathcal{G}_{\sigma_2}$ in Proposition~\ref{proposition:Appendix:TrueProcess:FL:ExpressionEvolved}:
    \begin{multline*}
        \int_0^\nu 
            \expo{\Boldomega \beta}
            \mathcal{G}_{\sigma_1}(\beta,\Xt{N})
        d\Wt{\beta}
        \\
        = 
        \int_0^\nu 
            \expo{\Boldomega \beta}
            \left(
                \sum_{i=1}^{\istar{\nu} - 1}
                \indicator{[(i-1)\BoldTs,i\BoldTs)}{\beta}
                \Fparasigma{(i-1)\BoldTs,\Xt{N,(i-1)\BoldTs}}
                \left( 
                    1  
                    -
                    \widetilde{\gamma}_i(\BoldTs)
                    \expo{ \left(\lambda_s - \Boldomega\right) \beta} 
                \right)
            \right)
        d\Wt{\beta}
        \\
        +
        \int_0^\nu 
            \expo{\Boldomega \beta}
            \indicator{[(\istar{\nu} - 1)\BoldTs,\istar{\nu}\BoldTs)}{\beta}
                \Fparasigma{(\istar{\nu} - 1)\BoldTs,\Xt{N,(\istar{\nu} - 1)\BoldTs}}      
        d\Wt{\beta},
    \end{multline*}
    and
    \begin{multline*}
        \int_0^\nu 
            \expo{\Boldomega \beta}
            \mathcal{G}_{\sigma_2}(\beta,\Xt{N})
        d\Wt{\beta}
        \\
        = 
        -
        \int_0^\nu 
            \expo{\Boldomega \beta}
            \Fparasigma{\istar{\nu}\BoldTs,\Xt{N,\istar{\nu}\BoldTs}}
                \indicator{[(\istar{\nu}-1)\BoldTs,\istar{\nu}\BoldTs)}{\beta}
                \widetilde{\gamma}^\star(\nu,\BoldTs)
                \expo{ (\lambda_s - \Boldomega) \beta} 
        d\Wt{\beta}
        \\
        +
        \int_0^\nu 
            \expo{\Boldomega \beta}
            \Fparasigma{\istar{\nu}\BoldTs,\Xt{N,\istar{\nu}\BoldTs}}
                \indicator{ \geq \istar{\nu}\BoldTs}{\beta}
        d\Wt{\beta}
        .
    \end{multline*}
    Using the above, it is straightforward to establish that 
    \begin{align*}
        \mathfrak{N}_2(t)
        =
        \sum_{j=1}^4
        \mathfrak{Z}_j(t)
        ,
        \quad t \in [\BoldTs,\tstar],
    \end{align*}
    where the processes $\mathfrak{Z}_j$ are defined in~\eqref{eqn:prop:Appendix:NUProcess:ZProcess:J:Definitions} in the statement of Proposition~\ref{prop:Appendix:NUProcess:ZProcess}.
    Hence, it follows from the Minkowski's inequality and Proposition~\ref{prop:Appendix:NUProcess:ZProcess}:
    \begin{multline*}
        \begin{aligned}
            \LpLaw{\sfp}{}{\mathfrak{N}_2(t)}
            \leq& 
            \sum_{j=1}^4
            \LpLaw{\sfp}{}{
                \mathfrak{Z}_j(t)
            }
            \\
            \leq&
            \Delta_g^2
            \frac{ \expo{ 2\lambda t } }{2\lambda \Boldomega}
            \left( 
                \mathfrak{p}'(\sfp)
                \gamma_2'(\Boldomega,\BoldTs)
                + 
                \mathfrak{p}''(\sfp)
                \left(1 - \expo{ -2\Boldomega \BoldTs } \right)^\frac{1}{2}
                \left(
                    2    
                    +   
                    \gamma_2(\Boldomega,\BoldTs) 
                \right)
            \right)
            \sup_{\nu \in [0,\tstar]}
                \LpLaw{4\sfp}{}{\Fparasigma{\nu,\Xt{N,\nu}}}
        \end{aligned}
        \\
        \times
        \left( 
            \frac{1}{\sqrt{\Boldomega}}
            \sup_{\nu \in [0,\tstar]}
                \LpLaw{2\sfp}{}{\LZparamu{\nu,\Zt{N,\nu}}}
            + 
            \mathfrak{p}'(\sfp)
            \sup_{\nu \in [0,\tstar]}
                \LpLaw{2\sfp}{}{\FZparasigma{\nu,\Zt{N,\nu}}}
        \right)
        ,
    \end{multline*}
    for all $(t,p) \in [\BoldTs,\tstar] \times \mathbb{N}_{\geq 1}$. 
    The above implies that
    \begin{align}\label{eqn:lem:Appendix:TrueProcess:NUTilde:Bound:B:2:Final}
        2\Boldomega
        \LpLaw{\sfp}{}{
            \mathfrak{N}_2(t)
        }
        \leq 
        \frac{ \expo{ 2\lambda t } }{\lambda}
        \Delta_2^+(\sfp,\tstar,\Boldomega,\BoldTs)
        ,
        \quad \forall (t,\sfp) \in [\BoldTs,\tstar] \times \mathbb{N}_{\geq 1}.
    \end{align}

    \ul{\emph{Bound for $\mathfrak{N}_3(t)$:}}
    Finally, using the definitions in~\eqref{eqn:lem:Appendix:TrueProcess:NUTilde:Representation:Evolved:Pre1}-\eqref{eqn:lem:Appendix:TrueProcess:NUTilde:Representation:Evolved:Pre}, we have 
    \begin{multline*}
            \mathfrak{N}_3(t)
            =
            \int_{\BoldTs}^t  
                \expo{ 2(\lambda - \Boldomega) \nu }
                \mathcal{N}_3(\nu)^\top 
                g(\nu)^\top
                g(\nu)
                \mathring{\mathcal{N}}(\nu)       
            d\nu
        \\
        =
        \int_{\BoldTs}^t  
            \expo{ 2(\lambda - \Boldomega) \nu }
            \left(
                \int_0^{\BoldTs} 
                    \expo{\Boldomega \beta}
                    \Lparamu{\beta,\Xt{N,\beta}}
                d\beta
            \right)^\top 
            g(\nu)^\top
            g(\nu)
        \\
        \times
            \left(
                \int_0^{\nu}
                    \expo{\Boldomega \beta}
                    \left[ 
                        \LZparamu{\beta,\Zt{N,\beta}}d\beta + \FZparasigma{\beta,\Zt{N,\beta}}d\Wt{\beta} 
                    \right]
            \right)       
        d\nu,
    \end{multline*}
    for $(t,\sfp) \in [0,\tstar \wedge \BoldTs] \times \mathbb{N}_{\geq 1}$, which we can cast as $N_a(t)$ in Corollary~\ref{cor:TechnicalResults:Alt} by setting 
    \begin{align*}
        \theta_1 = \lambda, 
        \quad
        \theta_2 = \Boldomega,
        \quad   
        \xi = \BoldTs, 
        \quad 
        Q_t = \Wt{t} (n_q =d)
        ,
        \quad 
        S_1(\beta)  
        = 
        \Lparamu{\beta,\Xt{N,\beta}}
        \in \mathbb{R}^{m} (n_1 = m)
        , 
        \\
        R(\nu) 
        = 
        g(\nu)^\top g(\nu)
        \in        
        \mathbb{R}^{m \times m} (n_1 = n_2 = m)
        , 
        \\
        S_2(\beta) = \LZparamu{\beta,\Zt{N,\beta}} \in \mathbb{R}^m,
        \quad 
        L_2(\beta) = \FZparasigma{\beta,\Zt{N,\beta}} \in \mathbb{R}^{m \times d}
        .
    \end{align*}
    Thus, we use~\eqref{cor:TechnicalResults:Alt:Na:Bound:Final} in Corollary~\ref{cor:TechnicalResults:Alt} and use the uniform bound on $g(t)$ from Assumption~\ref{assmp:KnownFunctions} to obtain
    \begin{multline*}
        \LpLaw{\sfp}{}{\mathfrak{N}_3(t)} 
        \leq
        \Delta_g^2 
        \frac{\expo{2\lambda t}}{2\lambda \Boldomega}
            \frac{\expo{\Boldomega \BoldTs}-1}{\BoldomegaRoot}
        \sup_{\nu \in [0,\tstar]}
            \LpLaw{2\sfp}{}{\Lparamu{\nu,\Xt{N,\nu}}}
        \\
        \times 
        \left(
            \frac{1}{\BoldomegaRoot}
            \sup_{\nu \in [0,\tstar]}
            \LpLaw{2\sfp}{}{
                \LZparamu{\nu,\Zt{N,\nu}}
            }
            + 
            \mathfrak{p}'(\sfp)  
            \sup_{\nu \in [0,\tstar]}
            \LpLaw{2\sfp}{}{
                \FZparasigma{\nu,\Zt{N,\nu}}
            }
        \right)
        , 
    \end{multline*}
    for all $(t,\sfp) \in [\BoldTs,\tstar] \times \mathbb{N}_{\geq 1}$, which allows us to write 
    \begin{align}\label{eqn:lem:Appendix:TrueProcess:NUTilde:Bound:B:3:Final}
        2\Boldomega
        \LpLaw{\sfp}{}{
            \mathfrak{N}_3(t)
        }
        \leq 
        \frac{ \expo{ 2\lambda t } }{\lambda}
        \Delta_3^+(\sfp,\tstar,\Boldomega,\BoldTs)
        ,
        \quad \forall (t,\sfp) \in [\BoldTs,\tstar] \times \mathbb{N}_{\geq 1}.
    \end{align}
    Substituting the bounds~\eqref{eqn:lem:Appendix:TrueProcess:NUTilde:Bound:A:Final},~\eqref{eqn:lem:Appendix:TrueProcess:NUTilde:Bound:B:1:Final},~\eqref{eqn:lem:Appendix:TrueProcess:NUTilde:Bound:B:2:Final} and~\eqref{eqn:lem:Appendix:TrueProcess:NUTilde:Bound:B:3:Final} into~\eqref{eqn:lem:Appendix:TrueProcess:NUTilde:Bound:Initial} yields:  
    \begin{align*}
        \LpLaw{\sfp}{}{
            N_{\mathcal{U}}(t)
        }
        \leq
        \frac{\expo{ 2\lambda t }  }{\lambda}
        \times 
        \begin{cases}
            \Delta^{-}(\sfp,\Boldomega,\BoldTs)
            ,
            & \forall t \in [0,\tstar] \cap [0,\BoldTs)
            ,
            \\[10pt]
            \Delta^{-}(\sfp,\Boldomega,\BoldTs)
            +
            \sum_{k=1}^3
            \Delta_k^+(\sfp,\tstar,\Boldomega,\BoldTs)
            , 
            & \forall t \in [0,\tstar] \cap [\BoldTs,T],
        \end{cases}
    \end{align*} 
    which completes the proof by observing that the bound on the right hand side can be expressed as 
    \begin{align*}
        \Delta_{\mathcal{U}}(\sfp,\tstar,\Boldomega,\BoldTs)
        = 
        \Delta^{-}(\sfp,\Boldomega,\BoldTs)
        + 
        \indicator{\geq \BoldTs}{\tstar}
        \sum_{k=1}^3
        \Delta_k^+(\sfp,\tstar,\Boldomega,\BoldTs).
    \end{align*} 


\end{proof}

We next derive the bound on the moments of $\Xi_{\mathcal{U}}$ in Lemma~\ref{lem:True:dV} using the results from Propostion~\ref{prop:Appendix:TrueProcess:TsBound} to Lemma~\ref{lem:Appendix:TrueProcess:NUTilde:Bounds} above. 
\begin{lemma}\label{lem:Appendix:TrueProcess:XiU}
    If the stopping time $\tau^\star$, defined in~\eqref{eqn:True:StoppingTimes}, Lemma~\ref{lem:True:dV}, satisfies  $\tau^\star = \tstar$, then the term $\Xi_{\mathcal{U}}\left(\tau(t),\Zt{N};\Boldomega \right)$ defined in~\eqref{eqn:lem:True:dV:Xi:Functions:B}, Lemma~\ref{lem:True:dV}, satisfies the following bounds: 
    \begin{multline}\label{eqn:lem:Appendix:TrueProcess:XirU:p=1:Bound:Final}
        \ELaw{}{
            \expo{-(2\lambda+\Boldomega) \tau(t)}
            \Xi_{\mathcal{U}}\left(\tau(t),\Zt{N};\Boldomega \right)
        }
        \\
        \leq  
        \frac{ 1 }{  \absolute{2\lambda - \Boldomega}}  
        \left(
            \Delta_{\mathcal{U}_1}(1,\tstar)
            +
            \sqrt{\Boldomega}
            \Delta_{\mathcal{U}_2}(1,\tstar)
            +
            \Boldomega
            \Delta_{\mathcal{U}_3}(1,\tstar)
            +
            \frac{\Boldomega}{\lambda }
            \Delta_{\mathcal{U}}(1,\tstar,\Boldomega,\BoldTs)
        \right)
        , 
        \quad \forall t \in [0,\tstar] 
        , 
    \end{multline}
    and
    \begin{multline}\label{eqn:lem:Appendix:TrueProcess:XirU:p>=2:Bound:Final}
        \LpLaw{\sfp}{}{
            \expo{-(2\lambda+\Boldomega) \tau(t)}
            \Xi_{\mathcal{U}}\left(\tau(t),\Zt{N};\Boldomega \right)
        }
        \\
        \leq  
        \frac{ 1 }{  \absolute{2\lambda - \Boldomega}}  
        \left(
            \Delta_{\mathcal{U}_1}(\sfp,\tstar)
            +
            \sqrt{\Boldomega}
            \Delta_{\mathcal{U}_2}(\sfp,\tstar)
            +
            \Boldomega
            \Delta_{\mathcal{U}_3}(\sfp,\tstar)
            +
            \frac{\Boldomega}{\lambda }
            \Delta_{\mathcal{U}}(\sfp,\tstar,\Boldomega,\BoldTs)
        \right)
        , 
    \end{multline}
    for all $(t,\sfp) \in [0,\tstar] \times \mathbb{N}_{\geq 2}$, where 
    \begin{align*}
        \begin{aligned}
            \Delta_{\mathcal{U}_1}(\sfp,\tstar)
            =&
            \sqrt{\lambda}
            \Delta_g
            \mathfrak{p}(\sfp)  
            \sup_{\nu \in [0,\tstar]}
            \LpLaw{2\sfp}{}{
                \RefDiffNorm{\Zt{N,\nu}}
            }
            \sup_{\nu \in [0,\tstar]}
                \LpLaw{2\sfp}{}{
                    \FZparasigma{\nu,\Zt{N,\nu}}
            }
            \\
            &+
            \left(
                    \frac{\Delta_{\mathcal{P}_1}(\sfp,\tstar)}{\sqrt{\lambda}} 
                    +
                    2
                    \Delta_g 
                    \sup_{\nu \in [0,\tstar]}\LpLaw{2\sfp}{}{\RefDiffNorm{\Zt{N,\nu}}}
            \right.
            \\
            & 
            \hspace{5cm}
            \left.
                +
                \frac{\Delta_g^2}{\lambda} 
                \sup_{\nu \in [0,\tstar]}\LpLaw{2\sfp}{}{\LZparamu{\nu,\Zt{N,\nu}}}
            \right)
            \sup_{\nu \in [0,\tstar]}\LpLaw{2\sfp}{}{\LZparamu{\nu, \Zt{N,\nu}}}
            ,
        \end{aligned}
        \\
        \begin{multlined}[b][0.95\linewidth]
            \Delta_{\mathcal{U}_2}(\sfp,\tstar)
            =
            \left(
                \frac{ \Delta_{\mathcal{P}_2}(\sfp,\tstar) }{\sqrt{\lambda }}
                + 
                \Delta_g 
                \mathfrak{p}'(\sfp)
                \sup_{\nu \in [0,\tstar]}
                    \LpLaw{2\sfp}{}{\RefDiffNorm{\Zt{N,\nu}}}
                + 
                2
                \frac{\Delta_g^2}{\lambda}
                \mathfrak{p}'(\sfp)  
                \sup_{\nu \in [0,\tstar]}
                    \LpLaw{2\sfp}{}{\LZparamu{\nu,\Zt{N,\nu}}}
            \right)
            \\
            \times
            \sup_{\nu \in [0,\tstar]}\LpLaw{2\sfp}{}{\FZparasigma{\nu, \Zt{N,\nu}}}
            ,
        \end{multlined}
        \\
        \Delta_{\mathcal{U}_3}(\sfp,\tstar)
        =
        \left(
            \frac{ \Delta_{\mathcal{P}_3}(\sfp,\tstar) }{\lambda}
            + 
            \frac{\Delta_g^2}{\lambda}
            \mathfrak{p}'(\sfp)^2  
            \sup_{\nu \in [0,\tstar]}\LpLaw{2\sfp}{}{\FZparasigma{\nu,\Zt{N,\nu}}}
        \right)
        \sup_{\nu \in [0,\tstar]}\LpLaw{2\sfp}{}{\FZparasigma{\nu, \Zt{N,\nu}}}
        ,
    \end{align*}
    and where $\cbr{\Delta_{\mathcal{P}_1}(\sfp,\tstar),\Delta_{\mathcal{P}_2}(\sfp,\tstar),\Delta_{\mathcal{P}_3}(\sfp,\tstar)}$, and $\Delta_{\mathcal{U}}(\sfp,\tstar,\Boldomega,\BoldTs)$  are defined in Proposition~\ref{prop:Appendix:TrueProcess:N:Bounds} and Lemma~\ref{lem:Appendix:TrueProcess:NUTilde:Bounds}, respectively.

\end{lemma}
\begin{proof}
    We begin by recalling the definition of $\psi(\tau(t),\nu,\Zt{N})$ from~\eqref{eqn:lem:True:dV:psi:Functions} 
    \begin{multline*}
        \psi(\tau(t),\nu,\Zt{N})
        =
        \frac{\Boldomega}{2\lambda - \Boldomega}
        \left(
            \expo{\Boldomega(\tau(t)+\nu)}
            \mathcal{P}\br{\tau(t),\nu}  
            -
            \expo{ (2\lambda \tau(t)+\Boldomega \nu)}
            \ZDiffNorm{\Zt{N,\tau(t)}}^\top g(\tau(t))
        \right)
        \\
        +
        \frac{2\lambda}{2\lambda - \Boldomega}
        \expo{(\Boldomega \tau(t) + 2 \lambda \nu)} 
        \ZDiffNorm{\Zt{N,\nu}}^\top g(\nu)  
        \in \mathbb{R}^{1 \times m},
    \end{multline*}
    which, upon using the decomposition~\eqref{eqn:lem:Appendix:TrueProcess:ddV:Expression:Pr:Main} in Proposition~\ref{prop:Appendix:True:ddV:Bound}, can be re-written as
    \begin{equation}\label{eqn:lem:Appendix:TrueProcess:XirU:psi_r:Decomposition}
        \psi(\tau(t),\nu,\Zt{N})
        = 
        \sum_{i=1}^3 \psi_i(\tau(t),\nu,\Zt{N})
        +
        \psi_{ad}(\tau(t),\nu,\Zt{N})
        +
        \widetilde{\psi}(\tau(t),\nu,\Zt{N})
        \in \mathbb{R}^{1 \times m},
    \end{equation}
    where 
    \begin{align*}
        \psi_1(\tau(t),\nu,\Zt{N})
        =  
        \frac{\Boldomega}{2\lambda - \Boldomega}
        \expo{\Boldomega(\tau(t)+\nu)}
        \mathcal{P}_\circ \br{\tau(t),\nu}
        ,
        \\
        \psi_2(\tau(t),\nu,\Zt{N})
        =
    -
        \frac{\Boldomega}{2\lambda - \Boldomega}
        \expo{ (2\lambda \tau(t)+\Boldomega \nu)}
        \ZDiffNorm{\Zt{N,\tau(t)}}^\top g(\tau(t))
        ,
        \\
        \psi_3(\tau(t),\nu,\Zt{N})
        =
        \frac{2\lambda}{2\lambda - \Boldomega}
        \expo{(\Boldomega \tau(t) + 2 \lambda \nu)} 
        \ZDiffNorm{\Zt{N,\nu}}^\top g(\nu)
        ,
    \end{align*}
    and 
    \begin{align*}
        \psi_{ad}(\tau(t),\nu,\Zt{N})
        =
        \frac{\Boldomega}{2\lambda - \Boldomega}
        \expo{\Boldomega(\tau(t)+\nu)}
        \mathcal{P}_{ad} \br{\tau(t),\nu}
        ,
        \quad 
        \widetilde{\psi}(\tau(t),\nu,\Zt{N})
        =
        \frac{\Boldomega}{2\lambda - \Boldomega}
        \expo{\Boldomega(\tau(t)+\nu)}
        \widetilde{\mathcal{P}} \br{\tau(t),\nu}
        .
    \end{align*}
    Using the decomposition in~\eqref{eqn:lem:Appendix:TrueProcess:XirU:psi_r:Decomposition}, we re-write the terms $\mathcal{U}_{\mu}\br{\tau(t),\nu,\Zt{N};\Boldomega}$ and $\mathcal{U}_{\sigma}\br{\tau(t),\nu,\Zt{N};\Boldomega}$ defined in~\eqref{eqn:lem:True:dV:phi:Functions:2}, as follows:
    \begin{align*}
        \mathcal{U}_{\mu}\br{\tau(t),\nu,\Zt{N};\Boldomega}
        =
        \left( \sum_{i=1}^3 \psi_i(\tau(t),\nu,\Zt{N})
        +
        \psi_{ad}(\tau(t),\nu,\Zt{N})
        +
        \widetilde{\psi}(\tau(t),\nu,\Zt{N}) \right)
        \LZparamu{\nu,\Zt{N,\nu}}
        \in \mathbb{R}
        ,
        \\
        \mathcal{U}_{\sigma}\br{\tau(t),\nu,\Zt{N};\Boldomega}
        =
        \left( \sum_{i=1}^3 \psi_i(\tau(t),\nu,\Zt{N})
        +
        \psi_{ad}(\tau(t),\nu,\Zt{N})
        +
        \widetilde{\psi}(\tau(t),\nu,\Zt{N}) \right)
        \FZparasigma{\nu,\Zt{N,\nu}}
        \in \mathbb{R}^{1 \times d}
        .
    \end{align*}
     Thus, using the definition of $\Xi_{\mathcal{U}}\br{\tau(t),\Zt{N};\Boldomega}$ in~\eqref{eqn:lem:True:dV:Xi:Functions}, we may write 
    \begin{multline*}
        \begin{aligned}
            \Xi_{\mathcal{U}}\left(\tau(t),\Zt{N};\Boldomega \right)
            =&
            \int_0^{\tau(t)}
            \left(  
                \mathcal{U}_{\mu}\br{\tau(t),\nu,\Zt{N};\Boldomega}d\nu 
                + 
                \mathcal{U}_{\sigma}\br{\tau(t),\nu,\Zt{N};\Boldomega}d\Wt{\nu}
            \right)
            \\
            =&
            \sum_{i=1}^3
            \int_0^{\tau(t)}
            \psi_i(\tau(t),\nu,\Zt{N})
            \left[ \LZparamu{\nu,\Zt{N,\nu}}d\nu + \FZparasigma{\nu,\Zt{N,\nu}}d\Wt{\nu} \right]
            \\
            &+
            \int_0^{\tau(t)} 
            \psi_{ad}(\tau(t),\nu,\Zt{N})
            \left[ \LZparamu{\nu,\Zt{N,\nu}}d\nu + \FZparasigma{\nu,\Zt{N,\nu}}d\Wt{\nu} \right]
        \end{aligned}
        \\
        +
        \int_0^{\tau(t)}
            \widetilde{\psi}(\tau(t),\nu,\Zt{N})
        \left[ \LZparamu{\nu,\Zt{N,\nu}}d\nu + \FZparasigma{\nu,\Zt{N,\nu}}d\Wt{\nu} \right]
        .
    \end{multline*}
    Hence, exploiting the linearity of the expectation operator, along with the Minkowski's inequality, we obtain 
    \begin{multline}\label{eqn:lem:Appendix:TrueProcess:XirU:p=1:Bound:Initial}
        \begin{aligned}
            &\ELaw{}{
                \expo{-(2\lambda+\Boldomega) \tau(t)}
                \Xi_{\mathcal{U}}\left(\tau(t),\Zt{N};\Boldomega \right)
            }
            \\
            &
            =
            \expo{-(2\lambda+\Boldomega) t}
            \sum_{i=1}^3
            \ELaw{}{ 
                \int_0^{t}
                    \psi_i(t,\nu,\Zt{N})
                    \left[ \LZparamu{\nu,\Zt{N,\nu}}d\nu + \FZparasigma{\nu,\Zt{N,\nu}}d\Wt{\nu} \right]
            }
        \end{aligned}
        \\
        +
        \expo{-(2\lambda+\Boldomega) t}
        \ELaw{}{
            \int_0^{t}
                \psi_{ad}(t,\nu,\Zt{N})
                \left[ \LZparamu{\nu,\Zt{N,\nu}}d\nu + \FZparasigma{\nu,\Zt{N,\nu}}d\Wt{\nu} \right]
        }
        \\
        +
        \expo{-(2\lambda+\Boldomega) t}
        \ELaw{}{
            \int_0^{t}
                \widetilde{\psi}(t,\nu,\Zt{N})
                \left[ \LZparamu{\nu,\Zt{N,\nu}}d\nu + \FZparasigma{\nu,\Zt{N,\nu}}d\Wt{\nu} \right]
        }
        , 
    \end{multline}
    for all $t \in [0,\tstar]$, and 
    \begin{multline}\label{eqn:lem:Appendix:TrueProcess:XirU:p>=2:Bound:Initial}
        \begin{aligned}
            &\LpLaw{\sfp}{}{
                \expo{-(2\lambda+\Boldomega) \tau(t)}
                \Xi_{\mathcal{U}}\left(\tau(t),\Zt{N};\Boldomega \right)
            }
            \\
            &
            \leq 
            \expo{-(2\lambda+\Boldomega) t}
            \sum_{i=1}^3
            \LpLaw{\sfp}{}{ 
                \int_0^{t}
                    \psi_i(t,\nu,\Zt{N})
                    \left[ \LZparamu{\nu,\Zt{N,\nu}}d\nu + \FZparasigma{\nu,\Zt{N,\nu}}d\Wt{\nu} \right]
            }
        \end{aligned}
        \\
        +
        \expo{-(2\lambda+\Boldomega) t}
        \LpLaw{\sfp}{}{
            \int_0^{t}
                \psi_{ad}(t,\nu,\Zt{N})
                \left[ \LZparamu{\nu,\Zt{N,\nu}}d\nu + \FZparasigma{\nu,\Zt{N,\nu}}d\Wt{\nu} \right]
        }
        \\
        +
        \expo{-(2\lambda+\Boldomega) t}
        \LpLaw{\sfp}{}{
            \int_0^{t}
                \widetilde{\psi}(t,\nu,\Zt{N})
                \left[ \LZparamu{\nu,\Zt{N,\nu}}d\nu + \FZparasigma{\nu,\Zt{N,\nu}}d\Wt{\nu} \right]
        }
        , 
    \end{multline}
    for all $(t,\sfp) \in [0,\tstar] \times \mathbb{N}_{\geq 2}$. 
    In the two expressions above, we have used the fact that since $\tau^\star = \tstar$, the definition of  $\tau(t)$ in~\eqref{eqn:Reference:StoppingTimes} leads to the conclusion that $\tau(t) = t$, for $t \in [0,\tstar]$.   
    Hence, the stopping time $\tau(t)$ is a deterministic function of $t$ since $\tstar$ is a constant, and thus 
    \begin{align*}
        \mathbb{E}\left[\expo{-(2\lambda+\Boldomega) \tau(t)}\right]
        =
        \expo{-(2\lambda+\Boldomega) \tau(t)}
        =
        \expo{-(2\lambda+\Boldomega) t},
        \quad \forall t \in [0,\tstar].
    \end{align*}

    Now, using the definition of $\psi_1$ in~\eqref{eqn:lem:Appendix:TrueProcess:XirU:psi_r:Decomposition}, we get 
    \begin{multline}\label{eqn:lem:Appendix:TrueProcess:XirU:Psi1:Bound:1}
        \int_0^{t}
        \psi_1(t,\nu,\Zt{N})
        \left[ \LZparamu{\nu,\Zt{N,\nu}}d\nu + \FZparasigma{\nu,\Zt{N,\nu}}d\Wt{\nu} \right]
        \\
        =
        \frac{\Boldomega \expo{\Boldomega t} }{2\lambda - \Boldomega} 
        \int_0^{t}
            \expo{\Boldomega \nu}
            \mathcal{P}_\circ \br{t,\nu} 
            \left[ \LZparamu{\nu,\Zt{N,\nu}}d\nu + \FZparasigma{\nu,\Zt{N,\nu}}d\Wt{\nu} \right]
        , 
        \quad  t \in [0,\tstar].
    \end{multline} 
    Using the definition of $\mathcal{P}_\circ$ in~\eqref{eqn:lem:Appendix:TrueProcess:ddV:Expression:Pr}, one sees that
    \begin{align*}
        \int_0^{t}
            \expo{\Boldomega \nu}
            \mathcal{P}_\circ \br{t,\nu} 
            \left[ \LZparamu{\nu,\Zt{N,\nu}}d\nu + \FZparasigma{\nu,\Zt{N,\nu}}d\Wt{\nu} \right]
        = 
        \sum_{i=1}^4 N_i(t), \quad  t \in [0,T], 
    \end{align*}
    where the processes $N_i(t)$, $i \in \cbr{1,\cdots,4}$, are defined in~\eqref{eqn:prop:Appendix:TrueProcess:N:Definitions} in the statement of Proposition~\ref{prop:Appendix:TrueProcess:N:Bounds}.
    Hence, we may write~\eqref{eqn:lem:Appendix:TrueProcess:XirU:Psi1:Bound:1} as 
    \begin{align*}
        \int_0^{t}
        \psi_1(t,\nu,\Zt{N})
        \left[ \LZparamu{\nu,\Zt{N,\nu}}d\nu + \FZparasigma{\nu,\Zt{N,\nu}}d\Wt{\nu} \right]
        =
        \frac{\Boldomega  \expo{\Boldomega t}}{2\lambda - \Boldomega}
        \sum_{i=1}^4 N_i(t)
        ,
        \quad t \in [0,\tstar].
    \end{align*}
    It then follows from the linearity of the expectation operator and the Minkowski's inequality that 
    \begin{align*}
        \ELaw{}{
            \int_0^{t}
                \psi_1(t,\nu,\Zt{N})
                \left[ \LZparamu{\nu,\Zt{N,\nu}}d\nu + \FZparasigma{\nu,\Zt{N,\nu}}d\Wt{\nu} \right]
        }
        \leq 
        \frac{\Boldomega \expo{\Boldomega t} }{\absolute{2\lambda - \Boldomega}} 
        \sum_{i=1}^4 \absolute{\ELaw{}{N_i(t)}}
        , 
        \quad \forall   t \in [0,\tstar]
        ,
    \end{align*}
    and  
    \begin{multline*}
        \LpLaw{\sfp}{}{
            \int_0^{t}
                \psi_1(t,\nu,\Zt{N})
                \left[ \LZparamu{\nu,\Zt{N,\nu}}d\nu + \FZparasigma{\nu,\Zt{N,\nu}}d\Wt{\nu} \right]
        }
        \\
        \leq 
        \frac{\Boldomega \expo{\Boldomega t} }{\absolute{2\lambda - \Boldomega}} 
        \sum_{i=1}^4 \LpLaw{\sfp}{}{N_i(t)}
        , 
        \quad \forall  (t,\sfp) \in [0,\tstar] \times \mathbb{N}_{\geq 2}
        .
    \end{multline*}
    Hence, using Proposition~\ref{prop:Appendix:TrueProcess:N:Bounds} leads to 
    \begin{multline}\label{eqn:lem:Appendix:TrueProcess:XirU:Psi1:Bound:p=1:Final}
        \begin{aligned}
            &\ELaw{}{
                \int_0^{t}
                    \psi_1(t,\nu,\Zt{N})
                    \left[ \LZparamu{\nu,\Zt{N,\nu}}d\nu + \FZparasigma{\nu,\Zt{N,\nu}}d\Wt{\nu} \right]
            }
            \\
            &\leq 
            \frac{\expo{(2\lambda + \Boldomega) t} }{\sqrt{\lambda} \absolute{2\lambda - \Boldomega}} 
            \Delta_{\mathcal{P}_1}(1,\tstar)
            \sup_{\nu \in [0,\tstar]} 
                \LpLaw{2}{}{\LZparamu{\nu, \Zt{N,\nu}}}
        \end{aligned}
        \\
        + 
        \frac{ \expo{(2\lambda + \Boldomega) t} }{\absolute{2\lambda - \Boldomega}} 
        \left(
            \frac{ \sqrt{\Boldomega} }{\sqrt{\lambda }}
            \Delta_{\mathcal{P}_2}(1,\tstar)
            +
            \frac{ \Boldomega}{\lambda} 
            \Delta_{\mathcal{P}_3}(1,\tstar)
        \right)
        \sup_{\nu \in [0,\tstar]}
                \LpLaw{2}{}{\FZparasigma{\nu,\Zt{N,\nu}}}
        ,
    \end{multline}
    for all $t \in [0,\tstar]$, and  
    \begin{multline}\label{eqn:lem:Appendix:TrueProcess:XirU:Psi1:Bound:p>=2:Final}
        \begin{aligned}
            &\LpLaw{\sfp}{}{
                \int_0^{t}
                    \psi_1(t,\nu,\Zt{N})
                    \left[ \LZparamu{\nu,\Zt{N,\nu}}d\nu + \FZparasigma{\nu,\Zt{N,\nu}}d\Wt{\nu} \right]
            }
            \\
            &\leq 
            \frac{ \expo{(2\lambda + \Boldomega) t} }{ \sqrt{\lambda} \absolute{2\lambda - \Boldomega}} 
            \Delta_{\mathcal{P}_1}(\sfp,\tstar) 
            \sup_{\nu \in [0,\tstar]} 
                \LpLaw{2\sfp}{}{\LZparamu{\nu, \Zt{N,\nu}}}
        \end{aligned}
        \\
        +
        \frac{ \expo{(2\lambda+\Boldomega) t} }{\absolute{2\lambda - \Boldomega}}
        \left(
            \frac{ \sqrt{\Boldomega} }{\sqrt{\lambda }}
            \Delta_{\mathcal{P}_2}(\sfp,\tstar)
            +
            \frac{ \Boldomega }{\lambda}
            \Delta_{\mathcal{P}_3}(\sfp,\tstar)
        \right)
        \sup_{\nu \in [0,\tstar]}
            \LpLaw{2\sfp}{}{\FZparasigma{\nu, \Zt{N,\nu}}}
        ,
    \end{multline}
    for all $ (t,\sfp) \in [0,\tstar] \times \mathbb{N}_{\geq 2}$.


    Next, using the definitions of $\psi_2$ and $\psi_3$  in~\eqref{eqn:lem:Appendix:TrueProcess:XirU:psi_r:Decomposition} we see that
    \begin{multline*}
        \int_0^{t}
            \psi_2(t,\nu,\Zt{N})
            \left[ \LZparamu{\nu,\Zt{N,\nu}}d\nu + \FZparasigma{\nu,\Zt{N,\nu}}d\Wt{\nu} \right]
        \\
        = 
        -\frac{\Boldomega \expo{ 2\lambda t}}{2\lambda - \Boldomega}
        \ZDiffNorm{\Zt{N,\tau(t)}}^\top g(\tau(t))
        \int_0^{t}
        \expo{\Boldomega \nu}
            \left[ \LZparamu{\nu,\Zt{N,\nu}}d\nu + \FZparasigma{\nu,\Zt{N,\nu}}d\Wt{\nu} \right]
        ,
    \end{multline*}
    and 
    \begin{multline*}
        \int_0^{t}
            \psi_3(t,\nu,\Zt{N})
            \left[ \LZparamu{\nu,\Zt{N,\nu}}d\nu + \FZparasigma{\nu,\Zt{N,\nu}}d\Wt{\nu} \right]
        \\
        = 
        \frac{2 \lambda  \expo{ \Boldomega t}}{2\lambda - \Boldomega}
        \int_0^{t}
            \expo{2\lambda \nu}
            \ZDiffNorm{\Zt{N,\nu}}^\top g(\nu)
            \left[ \LZparamu{\nu,\Zt{N,\nu}}d\nu + \FZparasigma{\nu,\Zt{N,\nu}}d\Wt{\nu} \right]
        ,
    \end{multline*}
    for $t \in [0,\tstar]$.
    Following the derivations of~\eqref{eqn:lem:Appendix:ReferenceProcess:XirU:Psi2:Bound:Final},~\eqref{eqn:lem:Appendix:ReferenceProcess:XirU:Psi3:Bound:p=1:Final}, and~\eqref{eqn:lem:Appendix:ReferenceProcess:XirU:Psi3:Bound:p>=2:Final} in the proof of Lemma~\ref{lem:Appendix:ReferenceProcess:XirU}, we obtain 
    \begin{multline}\label{eqn:lem:Appendix:TrueProcess:XirU:Psi2:Bound:Final}
        \begin{aligned}
            &\LpLaw{\sfp}{}{
                \int_0^{t}
                    \psi_2(t,\nu,\Zt{N})
                    \left[ \LZparamu{\nu,\Zt{N,\nu}}d\nu + \FZparasigma{\nu,\Zt{N,\nu}}d\Wt{\nu} \right]
            }
            \\
            &\leq 
            \frac{\expo{ (2\lambda+\Boldomega) t } }{\absolute{2\lambda - \Boldomega}}
            \Delta_g 
            \sup_{\nu \in [0,\tstar]}\LpLaw{2\sfp}{}{\RefDiffNorm{\Zt{N,\nu}}}
            \sup_{\nu \in [0,\tstar]}\LpLaw{2\sfp}{}{\LZparamu{\nu, \Zt{N,\nu}}}
        \end{aligned}
        \\
        + 
        \frac{\BoldomegaRoot \expo{ (2\lambda + \Boldomega) t } }{\absolute{2\lambda - \Boldomega}}
        \Delta_g 
        \mathfrak{p}'(\sfp)
        \sup_{\nu \in [0,\tstar]}
            \LpLaw{2\sfp}{}{\RefDiffNorm{\Zt{N,\nu}}}
        \sup_{\nu \in [0,\tstar]}
            \LpLaw{2\sfp}{}{\FZparasigma{\nu,\Zt{N,\nu}}}
        ,
    \end{multline}
    for all $(t,\sfp) \in [0,\tstar] \times \mathbb{N}_{\geq 1}$, along with 
    \begin{multline}\label{eqn:lem:Appendix:TrueProcess:XirU:Psi3:Bound:p=1:Final}
        \ELaw{}{
            \int_0^{t}
                \psi_3(t,\nu,\Zt{N})
                \left[ \LZparamu{\nu,\Zt{N,\nu}}d\nu + \FZparasigma{\nu,\Zt{N,\nu}}d\Wt{\nu} \right]
        }
        \\
        \leq
        \frac{ \expo{(2\lambda+\Boldomega) t } }{\absolute{2\lambda - \Boldomega}}  
        \Delta_g
        \sup_{\nu \in [0,\tstar]}
            \LpLaw{2}{}{\RefDiffNorm{\Zt{N,\nu}}}
        \sup_{\nu \in [0,\tstar]}
            \LpLaw{2}{}{\LZparamu{\nu, \Zt{N,\nu}}},
        \quad 
        \forall t \in [0,\tstar],
    \end{multline}
    and
    \begin{multline}\label{eqn:lem:Appendix:TrueProcess:XirU:Psi3:Bound:p>=2:Final}
        \begin{aligned}
            &\LpLaw{\sfp}{}{
                \int_0^{t}
                    \psi_3(t,\nu,\Zt{N})
                    \left[ \LZparamu{\nu,\Zt{N,\nu}}d\nu + \FZparasigma{\nu,\Zt{N,\nu}}d\Wt{\nu} \right]
            }
            \\
            &\leq
            \frac{\expo{(2\lambda+\Boldomega) t } }{\absolute{2\lambda - \Boldomega}}
            \Delta_g
            \sup_{\nu \in [0,\tstar]}
                \LpLaw{2\sfp}{}{\RefDiffNorm{\Zt{N,\nu}}}
            \sup_{\nu \in [0,\tstar]}
                \LpLaw{2\sfp}{}{\LZparamu{\nu, \Zt{N,\nu}}}
        \end{aligned}
        \\
        +
        \frac{\sqrt{\lambda} \expo{(2\lambda+\Boldomega) t } }{\absolute{2\lambda - \Boldomega}}
        \Delta_g
        \mathfrak{p}(\sfp)  
        \sup_{\nu \in [0,\tstar]}
        \LpLaw{2\sfp}{}{
            \RefDiffNorm{\Zt{N,\nu}}
        }
        \sup_{\nu \in [0,\tstar]}
            \LpLaw{2\sfp}{}{
                \FZparasigma{\nu,\Zt{N,\nu}}
        }
        ,
    \end{multline}
    for all $(t,\sfp) \in [0,\tstar] \times \mathbb{N}_{\geq 2}$.


    For the derivation of the penultimate component of the desired bound, we use the definition of $\psi_{ad}$ in~\eqref{eqn:lem:Appendix:TrueProcess:XirU:psi_r:Decomposition} to obtain 
    \begin{align*}
        &\int_0^{t}
            \psi_{ad}(t,\nu,\Zt{N})
            \left[ \LZparamu{\nu,\Zt{N,\nu}}d\nu + \FZparasigma{\nu,\Zt{N,\nu}}d\Wt{\nu} \right]
        \\
        &
        = 
        \frac{\Boldomega \expo{\Boldomega t}}{2\lambda - \Boldomega}
        \int_0^{t}
            \expo{\Boldomega \nu}
            \mathcal{P}_{ad} \br{t,\nu} 
            \left[ \LZparamu{\nu,\Zt{N,\nu}}d\nu + \FZparasigma{\nu,\Zt{N,\nu}}d\Wt{\nu} \right]
        \\
        &
        = 
        \frac{\Boldomega \expo{\Boldomega t } }{2\lambda - \Boldomega}
        \int_0^{t}
            \expo{\Boldomega\nu}
            \left(
                \int_\nu^{t}
                    e^{ (2\lambda - \Boldomega) \beta }  
                        \mathcal{P}_{\mathcal{U}}(\beta)^\top 
                d\beta
            \right) 
            \left[ \LZparamu{\nu,\Zt{N,\nu}}d\nu + \FZparasigma{\nu,\Zt{N,\nu}}d\Wt{\nu} \right]
        ,
        \quad t \in [0,\tstar] 
        ,
    \end{align*}
    where, in the last expression we have used the definition of $\mathcal{P}_{ad}$ in~\eqref{eqn:lem:Appendix:TrueProcess:ddV:Expression:TildePrU}, Proposition~\ref{prop:Appendix:True:ddV:Bound}.
    Observe that the expression above can be alternatively expressed as 
    \begin{align*}
       \int_0^{t}
            \psi_{ad}(t,\nu,\Zt{N})
            \left[ \LZparamu{\nu,\Zt{N,\nu}}d\nu + \FZparasigma{\nu,\Zt{N,\nu}}d\Wt{\nu} \right]
        =
        \frac{\Boldomega \expo{\Boldomega t} }{2\lambda - \Boldomega}
            N_{\mathcal{U}}(t)
        ,
        \quad t \in [0,\tstar]
        ,
    \end{align*}
    where the process $N_{\mathcal{U}}$ is defined in~\eqref{eqn:prop:Appendix:TrueProcess:NU:Definitions}, Proposition~\ref{prop:Appendix:TrueProcess:NU:Bounds}.
    Hence, 
    \begin{align*}
        \LpLaw{\sfp}{}{
            \int_0^{t}
            \psi_{ad}(t,\nu,\Zt{N})
            \left[ \LZparamu{\nu,\Zt{N,\nu}}d\nu + \FZparasigma{\nu,\Zt{N,\nu}}d\Wt{\nu} \right]
        }
        \leq 
        \frac{\Boldomega \expo{\Boldomega t} }{\absolute{2\lambda - \Boldomega}}
        \LpLaw{\sfp}{}{
            N_{\mathcal{U}}(t)
        }
        ,
    \end{align*}
    for all $(t,\sfp) \in [0,\tstar] \times \mathbb{N}_{\geq 1}$.
    Thus, we invoke Proposition~\ref{prop:Appendix:TrueProcess:NU:Bounds} and obtain
    \begin{multline}\label{eqn:lem:Appendix:TrueProcess:XirU:PsiAd:Bound:Final}
        \LpLaw{\sfp}{}{
            \int_0^{t}
            \psi_{ad}(t,\nu,\Zt{N})
            \left[ \LZparamu{\nu,\Zt{N,\nu}}d\nu + \FZparasigma{\nu,\Zt{N,\nu}}d\Wt{\nu} \right]
        }
        \\
        \leq 
        \frac{\expo{ (2\lambda + \Boldomega) t } }{ \lambda \absolute{2\lambda - \Boldomega}}
        \Delta_g^2
        \left( 
            \sup_{\nu \in [0,\tstar]}
            \LpLaw{2\sfp}{}{\LZparamu{\nu,\Zt{N,\nu}}}
            + 
            \sqrt{\Boldomega}
            \mathfrak{p}'(\sfp)
            \sup_{\nu \in [0,\tstar]}
            \LpLaw{2\sfp}{}{\FZparasigma{\nu,\Zt{N,\nu}}}
        \right)^2
        ,
    \end{multline}
    for all $(t,\sfp) \in [0,\tstar] \times \mathbb{N}_{\geq 1}$. 

    Finally, using the definition of $\widetilde{\psi}$ in~\eqref{eqn:lem:Appendix:TrueProcess:XirU:psi_r:Decomposition} we see that
    \begin{align*}
        & 
        \int_0^t
            \widetilde{\psi}(\tau(t),\nu,\Zt{N})
        \left[ \LZparamu{\nu,\Zt{N,\nu}}d\nu + \FZparasigma{\nu,\Zt{N,\nu}}d\Wt{\nu} \right] 
        \\
        &
        =
        \frac{\Boldomega \expo{\Boldomega t}}{2\lambda - \Boldomega} 
        \int_0^{t}
            \expo{\Boldomega \nu}
            \widetilde{\mathcal{P}}\br{t,\nu} 
            \left[ \LZparamu{\nu,\Zt{N,\nu}}d\nu + \FZparasigma{\nu,\Zt{N,\nu}}d\Wt{\nu} \right]
        \\
        &
        = 
        \frac{\Boldomega \expo{\Boldomega t}}{2\lambda - \Boldomega} 
        \int_0^{t}
            \expo{\Boldomega \nu}
            \left(
                \int_\nu^t
                    e^{ (2\lambda - \Boldomega) \beta }  
                    \widetilde{\mathcal{P}}_{\mathcal{U}}(\beta)^\top 
                d\beta
            \right) 
            \left[ \LZparamu{\nu,\Zt{N,\nu}}d\nu + \FZparasigma{\nu,\Zt{N,\nu}}d\Wt{\nu} \right]
        ,
    \end{align*}
    for $t \in [0,\tstar]$, where, in the last expression we have used the definition of $\widetilde{\mathcal{P}}$ in~\eqref{eqn:lem:Appendix:TrueProcess:ddV:Expression:TildePrU}, Proposition~\ref{prop:Appendix:True:ddV:Bound}.
    Since the process on the right hand side is identical to $N_{\mathcal{U}}(t)$ defined in the statement of Lemma~\ref{lem:Appendix:TrueProcess:NUTilde:Bounds}, we can write   
    \begin{align*}
        \int_0^t
            \widetilde{\psi}(\tau(t),\nu,\Zt{N})
        \left[ \LZparamu{\nu,\Zt{N,\nu}}d\nu + \FZparasigma{\nu,\Zt{N,\nu}}d\Wt{\nu} \right] 
        = 
        \frac{\Boldomega \expo{\Boldomega t}}{2\lambda - \Boldomega}
        N_{\mathcal{U}}(t)
        ,
        \quad  
        t \in [0,\tstar]
        ,
    \end{align*}
    which further implies that 
    \begin{align*}
        \LpLaw{\sfp}{}{
            \int_0^t
                \widetilde{\psi}(\tau(t),\nu,\Zt{N})
                \left[ \LZparamu{\nu,\Zt{N,\nu}}d\nu + \FZparasigma{\nu,\Zt{N,\nu}}d\Wt{\nu} \right]
        }
        \leq 
        \frac{\Boldomega \expo{\Boldomega t}}{\absolute{2\lambda - \Boldomega}}
        \LpLaw{\sfp}{}{N_{\mathcal{U}}(t)}
        ,
    \end{align*}
    for all $(t,\sfp) \in [0,\tstar] \times \mathbb{N}_{\geq 1}$.
    It then follows from Lemma~\ref{lem:Appendix:TrueProcess:NUTilde:Bounds} that
    \begin{multline}\label{eqn:lem:Appendix:TrueProcess:XirU:PsiTilde:Bound:Final}
        \LpLaw{p}{}{
            \int_0^t
                \widetilde{\psi}(\tau(t),\nu,\Zt{N})
                \left[ \LZparamu{\nu,\Zt{N,\nu}}d\nu + \FZparasigma{\nu,\Zt{N,\nu}}d\Wt{\nu} \right]
        }
        \\
        \leq  
        \frac{ \expo{(2\lambda+\Boldomega) t}}{\lambda \absolute{2\lambda - \Boldomega}}
        \Boldomega
        \Delta_{\mathcal{U}}(\sfp,\tstar,\Boldomega,\BoldTs)
        ,
        \quad  
        \forall (t,\sfp) \in [0,\tstar] \times \mathbb{N}_{\geq 1}.
    \end{multline}

    Substituting~\eqref{eqn:lem:Appendix:TrueProcess:XirU:Psi1:Bound:p=1:Final},~\eqref{eqn:lem:Appendix:TrueProcess:XirU:Psi2:Bound:Final},~\eqref{eqn:lem:Appendix:TrueProcess:XirU:Psi3:Bound:p=1:Final},~\eqref{eqn:lem:Appendix:TrueProcess:XirU:PsiAd:Bound:Final}, and~\eqref{eqn:lem:Appendix:TrueProcess:XirU:PsiTilde:Bound:Final} into~\eqref{eqn:lem:Appendix:TrueProcess:XirU:p=1:Bound:Initial}, followed by grouping terms that are $\propto  \cbr{\Boldomega, \BoldomegaRoot, 1} / \absolute{2\lambda - \Boldomega}$ leads to~\eqref{eqn:lem:Appendix:TrueProcess:XirU:p=1:Bound:Final}.
    Similarly, substituting~\eqref{eqn:lem:Appendix:TrueProcess:XirU:Psi1:Bound:p>=2:Final},~\eqref{eqn:lem:Appendix:TrueProcess:XirU:Psi2:Bound:Final},~\eqref{eqn:lem:Appendix:TrueProcess:XirU:Psi3:Bound:p>=2:Final},~\eqref{eqn:lem:Appendix:TrueProcess:XirU:PsiAd:Bound:Final}, and~\eqref{eqn:lem:Appendix:TrueProcess:XirU:PsiTilde:Bound:Final} into~\eqref{eqn:lem:Appendix:TrueProcess:XirU:p>=2:Bound:Initial} leads to~\eqref{eqn:lem:Appendix:TrueProcess:XirU:p>=2:Bound:Final}, thus concluding the proof.

\end{proof}

The next set of results below will allow us to bound the moments of $\widetilde{\Xi}_{\mathcal{U}}$ in Lemma~\ref{lem:True:dV}. 
We begin with a result which is similar to Proposition~\ref{prop:Appendix:NUProcess:ZProcess}.
\begin{proposition}\label{prop:Appendix:TrueProcess:ZProcess}
    Let
    \begin{align}\label{eqn:prop:Appendix:TrueProcess:ZProcess:J:Definitions}
        \mathfrak{J}_j(t)
        = 
        \int_{\BoldTs}^t  
            \expo{(2 \lambda -\Boldomega) \nu}
            \ZDiffNorm{\Zt{N,\nu}}^\top 
            g(\nu)
            \left(
                \int_0^\nu 
                    \expo{\Boldomega \beta}
                        \mathcal{J}_{j}(\beta,\Xt{N})
                d\Wt{\beta}
            \right)       
        d\nu
        ,
        \quad (j,t) \in \cbr{1,\dots,4} \times [\BoldTs,\tstar],
    \end{align}
     where the piecewise functions $\mathcal{J}_{j}$ are defined in~\eqref{eqn:prop:Appendix:NUProcess:PiecewiseFunctions:J:Definitions}, Proposition~\ref{prop:Appendix:NUProcess:ZProcess}, for $\beta \in [0,\nu]$, $\nu \in [\BoldTs,t]$, $t \geq \BoldTs$.
    If the stopping time $\tau^\star$, defined in~\eqref{eqn:True:StoppingTimes}, Lemma~\ref{lem:True:dV}, satisfies  $\tau^\star = \tstar$ and $\tstar \geq \BoldTs$, then
    \begin{align}
        \sum_{j=1}^4
        \LpLaw{\sfp}{}{
            \mathfrak{J}_j(t)
        }
        \leq 
        \frac{ \expo{2\lambda t} }{2\lambda \BoldomegaRoot  }
        \Delta_g
        \gamma''\left(\sfp,\Boldomega,\BoldTs\right)
        \sup_{\nu \in [0,\tstar]}  
            \LpLaw{2\sfp}{}{\ZDiffNorm{\Zt{N,\nu}}}
        \sup_{\nu \in [0,\tstar]}
        \LpLaw{4\sfp}{}{ 
            \Fparasigma{\nu,\Xt{N,\nu}}
        },
    \end{align}
    for all $(t,p) \in [\BoldTs,\tstar] \times \mathbb{N}_{\geq 1}$, where $\mathfrak{p}'(\sfp)$ and $\mathfrak{p}''(\sfp)$ are defined in~\eqref{eqn:app:Constants:FrakP}, and $\gamma_2(\Boldomega,\BoldTs)$, $\gamma_2'(\Boldomega,\BoldTs)$, and $\gamma''\left(\sfp,\Boldomega,\BoldTs\right)$ are defined in~\eqref{eqn:app:Functions:True:gammaTs:Numbered}.

\end{proposition}
\begin{proof}
    We begin by using the bound on $g(t)$ in Assumption~\ref{assmp:KnownFunctions} to  obtain
    \begin{align*}
        \LpLaw{\sfp}{}{
            \mathfrak{J}_j(t)
        }
        \leq 
        \Delta_g
        \LpLaw{\sfp}{}{
            \int_{\BoldTs}^t  
                \expo{(2 \lambda -\Boldomega) \nu} 
                \norm{\ZDiffNorm{\Zt{N,\nu}}} 
                \norm{
                    \int_0^\nu 
                        \expo{\Boldomega \beta}
                            \mathcal{J}_{j}(\beta,\Xt{N})
                    d\Wt{\beta}
                }       
            d\nu
        }
        ,
        \quad 
        j \in \cbr{1,\dots,4},
    \end{align*}
    for all $(t,p) \in [\BoldTs,\tstar] \times \mathbb{N}_{\geq 1}$.
    We now bound each of the processes above. 

    \ul{\emph{Bound for $\mathfrak{J}_1(t)$:}}
    Invoking Lemma~\ref{lemma:TechnicalResults:NestedSupMoment} by setting
    \begin{align*}
        \xi = \BoldTs, \quad \theta_1 = 2\lambda, \quad \theta_2 = \Boldomega,
        \quad 
        \cbr{\zeta_1,\zeta_2} = \cbr{0,t}, 
        \quad 
        S(\nu) =  \ZDiffNorm{\Zt{N,\nu}}, 
        \quad 
        L(\beta) =  
        \mathcal{J}_1(\beta,\Xt{N}),
    \end{align*}
    we obtain 
    \begin{multline}\label{eqn:prop:Appendix:TrueProcess:Zprocess:Z1:1}
         \LpLaw{\sfp}{}{
            \mathfrak{J}_1(t)
        }
        \leq 
        \Delta_g
        \LpLaw{\sfp}{}{
            \int_{\BoldTs}^t  
                \expo{(2 \lambda -\Boldomega) \nu} 
                \norm{\ZDiffNorm{\Zt{N,\nu}}} 
                \norm{
                    \int_0^\nu 
                        \expo{\Boldomega \beta}
                            \mathcal{J}_1(\beta,\Xt{N})
                    d\Wt{\beta}
                }       
            d\nu
        }
        \\
        \leq     
        \frac{ \expo{2\lambda t} }{2\lambda \BoldomegaRoot  }
        \mathfrak{p}'(\sfp)
        \Delta_g 
        \sup_{\nu \in [0,\tstar]}
            \LpLaw{2\sfp}{}{\ZDiffNorm{\Zt{N,\nu}}}
        \sup_{\nu \in [0,\tstar]} 
            \LpLaw{2\sfp}{}{\mathcal{J}_1(\nu,\Xt{N})}
        ,
    \end{multline}
    for all $(t,\sfp) \in [\BoldTs,\tstar] \times \mathbb{N}_{\geq 1}$.
    Recall from~\eqref{eqn:prop:Appendix:TrueProcess:SumProcessNew:J1:Bound} that
    \begin{align*}
        \LpLaw{2\sfp}{}{\mathcal{J}_1(\nu,\Xt{N})}
        &
        \leq     
        \gamma_2'(\Boldomega,\BoldTs)
        \sup_{\nu \in [0,\tstar]}
        \LpLaw{2\sfp}{}{\Fparasigma{\nu,\Xt{N,\nu}}}
        ,
        \quad \forall (t,\sfp) \in [\BoldTs,\tstar] \times \mathbb{N}_{\geq 1}.
    \end{align*}
    Substituting the above into~\eqref{eqn:prop:Appendix:TrueProcess:Zprocess:Z1:1} leads to
    \begin{align}\label{eqn:prop:Appendix:TrueProcess:Zprocess:J1:Final}
         \LpLaw{\sfp}{}{
            \mathfrak{J}_1(t)
        }
        \leq     
        \frac{ \expo{2\lambda t} }{2\lambda \BoldomegaRoot  }
        \mathfrak{p}'(\sfp)
        \Delta_g 
        \gamma_2'(\Boldomega,\BoldTs)
        \sup_{\nu \in [0,\tstar]}
            \LpLaw{2\sfp}{}{\ZDiffNorm{\Zt{N,\nu}}}
        \sup_{\nu \in [0,\tstar]}
            \LpLaw{4\sfp}{}{\Fparasigma{\nu,\Xt{N,\nu}}}
        ,
    \end{align}
    for all $(t,\sfp) \in [\BoldTs,\tstar] \times \mathbb{N}_{\geq 1}$.
    Note that we have also used the bound $\LpLaw{2\sfp}{}{\Fparasigma{\nu,\Xt{N,\nu}}} \leq \LpLaw{4\sfp}{}{\Fparasigma{\nu,\Xt{N,\nu}}}$, which is a consequence of the Jensen's inequality.
    
    \begingroup
    \endgroup

    \ul{\emph{Bound for $\mathfrak{J}_2(t)$:}}
    The definition of $\mathcal{J}_2$ allows us to write 
    \begin{align*}
        &\LpLaw{\sfp}{}{
            \mathfrak{J}_2(t)
        }
        \\
        &\leq 
        \Delta_g
        \LpLaw{\sfp}{}{
            \int_{\BoldTs}^t  
                \expo{(2 \lambda -\Boldomega) \nu} 
                \norm{\ZDiffNorm{\Zt{N,\nu}}} 
                \norm{
                    \int_0^\nu 
                        \expo{\Boldomega \beta}
                            \mathcal{J}_2(\beta,\Xt{N})
                    d\Wt{\beta}
                }       
            d\nu
        }
        \\
        &=
        \Delta_g
        \LpLaw{\sfp}{}{
            \int_{\BoldTs}^t  
                \expo{(2 \lambda -\Boldomega) \nu} 
                \norm{\ZDiffNorm{\Zt{N,\nu}}} 
                \norm{
                    \Fparasigma{(\istar{\nu} - 1)\BoldTs,\Xt{N,(\istar{\nu} - 1)\BoldTs}}
                    \int_{(\istar{\nu} - 1)\BoldTs}^{\istar{\nu}\BoldTs} 
                        \expo{\Boldomega \beta} 
                    d\Wt{\beta}
                }       
            d\nu
        }
        ,
    \end{align*}
    for all $(t,p) \in [\BoldTs,\tstar] \times \mathbb{N}_{\geq 1}$, which can be cast as the process $N_a(t)$ in Corollary~\ref{cor:TechnicalResults:NestedSupMoment} by setting 
    \begin{align*}
        \theta_1 = 2\lambda, 
        \quad
        \theta_2 = \Boldomega,
        \quad   
        Q_t = \Wt{t}
        ,
        \\
        S(\nu) =  \ZDiffNorm{\Zt{N,\nu}},
        \quad 
        L(\nu)  
        =
        \Fparasigma{(\istar{\nu} - 1)\BoldTs,\Xt{N,(\istar{\nu} - 1)\BoldTs}},
        \quad 
        \varsigma_1(\nu) = \varsigma_2(\beta) = 1
        .
    \end{align*}
    Hence, by~\eqref{eqn:cor:TechnicalResults:NestedSupMoment:Na:Bound:Final} in Corollary~\ref{cor:TechnicalResults:NestedSupMoment} we have
    \begin{align}\label{eqn:prop:Appendix:TrueProcess:Zprocess:J2:Final}
        \LpLaw{\sfp}{}{
            \mathfrak{J}_2(t)
        }
        \leq  
        \frac{ \expo{ 2\lambda t } }{ 2\lambda \sqrt{\Boldomega} }
        \mathfrak{p}''(\sfp)
        \Delta_g
        \left( 1 - \expo{ -2\Boldomega \BoldTs } \right)^\frac{1}{2}
        \sup_{\nu \in [0,\tstar]}  
            \LpLaw{2\sfp}{}{\ZDiffNorm{\Zt{N,\nu}}}
        \sup_{\nu \in [0,\tstar]}
        \LpLaw{4\sfp}{}{ 
            \Fparasigma{\nu,\Xt{N,\nu}}
        }
        , 
    \end{align}
    for all $(t,p) \in [\BoldTs,\tstar] \times \mathbb{N}_{\geq 1}$, where we have used the fact that 
    \begin{align*}
        &\sup_{\nu \in [\BoldTs,t]}
        \LpLaw{4\sfp}{}{ 
            \Fparasigma{(\istar{\nu} - 1)\BoldTs,\Xt{N,(\istar{\nu} - 1)\BoldTs}}
        }
        \notag 
        \\
        &\leq
        \sup_{\nu \in [0,\tstar]}
        \LpLaw{4\sfp}{}{ 
            \Fparasigma{(\istar{\nu} - 1)\BoldTs,\Xt{N,(\istar{\nu} - 1)\BoldTs}}
        }
        \leq
        \sup_{\nu \in [0,\tstar]}
        \LpLaw{4\sfp}{}{ 
            \Fparasigma{\nu,\Xt{N,\nu}}
        }, \quad \forall t \in [\BoldTs,\tstar].
    \end{align*}

    \ul{\emph{Bound for $\mathfrak{J}_3(t)$:}}
    Similar to the above, we use the definition of $\mathcal{J}_3$ to obtain the bound: 
    \begin{align*}
        &\LpLaw{\sfp}{}{
            \mathfrak{J}_3(t)
        }
        \\
        &\leq
        \Delta_g
        \LpLaw{\sfp}{}{
            \int_{\BoldTs}^t  
                \expo{(2 \lambda -\Boldomega) \nu} 
                \norm{\ZDiffNorm{\Zt{N,\nu}}} 
                \norm{
                    \int_0^\nu 
                        \expo{\Boldomega \beta}
                            \mathcal{J}_3(\beta,\Xt{N})
                    d\Wt{\beta}
                }       
            d\nu
        }
        \\
        &=
        \Delta_g
        \LpLaw{\sfp}{}{
            \int_{\BoldTs}^t  
                \expo{(2 \lambda -\Boldomega) \nu} 
                \norm{\ZDiffNorm{\Zt{N,\nu}}} 
                \norm{
                    \Fparasigma{\istar{\nu}\BoldTs,\Xt{N,\istar{\nu}\BoldTs}}
                    \widetilde{\gamma}^\star(\nu,\BoldTs)
                    \int_{(\istar{\nu} - 1)\BoldTs}^{\istar{\nu}\BoldTs} 
                        \expo{\Boldomega \beta}
                        \expo{ (\lambda_s - \Boldomega) \beta} 
                    d\Wt{\beta}
                }       
            d\nu
        }
        ,
    \end{align*}
    for all $(t,p) \in [\BoldTs,\tstar] \times \mathbb{N}_{\geq 1}$, which can be cast as the process $N_a(t)$ in Corollary~\ref{cor:TechnicalResults:NestedSupMoment} by setting 
    \begin{align*}
        \theta_1 = 2\lambda, 
        \quad
        \theta_2 = \Boldomega,
        \quad   
        Q_t = \Wt{t}
        ,
        \quad  
        \varsigma_1(\nu) = \widetilde{\gamma}^\star(\nu,\BoldTs), \quad 
        \varsigma_2(\beta) =  \expo{ (\lambda_s - \Boldomega) \beta} 
        ,
        \\
        S(\nu) =  \ZDiffNorm{\Zt{N,\nu}},
        \quad 
        L(\nu)  
        =
        \Fparasigma{\istar{\nu}\BoldTs,\Xt{N,\istar{\nu}\BoldTs}}
        .
    \end{align*}
    Hence, by~\eqref{eqn:cor:TechnicalResults:NestedSupMoment:Na:Bound:Final} in Corollary~\ref{cor:TechnicalResults:NestedSupMoment} we have
    \begin{multline}\label{eqn:prop:Appendix:TrueProcess:Zprocess:J3:Initial}
        \LpLaw{\sfp}{}{
            \mathfrak{J}_3(t)
        }
        \leq   
        \frac{ \expo{ 2\lambda t } }{ 2\lambda \sqrt{\Boldomega} }
        \mathfrak{p}''(\sfp)
        \Delta_g
        \left( 1 - \expo{ -2\Boldomega \BoldTs } \right)^\frac{1}{2}
        \sup_{\nu \in [\BoldTs,t]}
        \left(
            \absolute{\widetilde{\gamma}^\star(\nu,\BoldTs)}
            \sup_{\beta \in [(\istar{\nu}-1)\BoldTs,\istar{\nu}\BoldTs]}
            \expo{ (\lambda_s - \Boldomega) \beta}
        \right)
        \\   
        \times
        \sup_{\nu \in [0,\tstar]}  
            \LpLaw{2\sfp}{}{\ZDiffNorm{\Zt{N,\nu}}}
        \sup_{\nu \in [0,\tstar]}
        \LpLaw{4\sfp}{}{ 
            \Fparasigma{\nu,\Xt{N,\nu}}
        }
        , 
    \end{multline}
    for all $(t,p) \in [\BoldTs,\tstar] \times \mathbb{N}_{\geq 1}$, where we have used the fact that 
    \begin{align}\label{eqn:lem:Appendix:TrueProcess:XiTildeU:Fact:1}
        &\sup_{\nu \in [\BoldTs,t]}
        \LpLaw{4\sfp}{}{ 
            \Fparasigma{\istar{\nu}\BoldTs,\Xt{N,\istar{\nu}\BoldTs}}
        }
        \notag 
        \\
        &\leq
        \sup_{\nu \in [0,\tstar]}
        \LpLaw{4\sfp}{}{ 
            \Fparasigma{\istar{\nu}\BoldTs,\Xt{N,\istar{\nu}\BoldTs}}
        }
        \leq
        \sup_{\nu \in [0,\tstar]}
        \LpLaw{4\sfp}{}{ 
            \Fparasigma{\nu,\Xt{N,\nu}}
        }, \quad \forall t \in [\BoldTs,\tstar].
    \end{align}
    Once again we recall the following from~\eqref{eqn:lem:Appendix:TrueProcess:SumProcessNew:Expectation:B:3:Identity}:
    \begin{align*}
        \absolute{\widetilde{\gamma}^\star(\nu,\BoldTs)}
        \sup_{\beta \in [(\istar{\nu}-1)\BoldTs,\istar{\nu}\BoldTs]}
            \expo{ (\lambda_s - \Boldomega) \beta}
        \leq 
        \max\left\{ \expo{(\Boldomega - \lambda_s)\BoldTs},1 \right\} 
        \frac{\lambda_s}{\Boldomega} 
        \frac{\expo{\Boldomega\BoldTs} - 1}{\expo{\lambda_s \BoldTs} - 1}
        \doteq
        \gamma_2(\Boldomega,\BoldTs) 
        ,
    \end{align*}
    for all $\nu \in [\BoldTs,\tstar]$.
    Substituting the above into~\eqref{eqn:prop:Appendix:TrueProcess:Zprocess:J3:Initial} then leads to 
    \begin{multline}\label{eqn:prop:Appendix:TrueProcess:Zprocess:J3:Final}
        \LpLaw{\sfp}{}{
            \mathfrak{J}_3(t)
        }
        \leq   
        \frac{ \expo{ 2\lambda t } }{ 2\lambda \sqrt{\Boldomega} }
        \mathfrak{p}''(\sfp)
        \Delta_g
        \gamma_2(\Boldomega,\BoldTs)
        \left( 1 - \expo{ -2\Boldomega \BoldTs } \right)^\frac{1}{2} 
        \\   
        \times
        \sup_{\nu \in [0,\tstar]}  
            \LpLaw{2\sfp}{}{\ZDiffNorm{\Zt{N,\nu}}}
        \sup_{\nu \in [0,\tstar]}
        \LpLaw{4\sfp}{}{ 
            \Fparasigma{\nu,\Xt{N,\nu}}
        }
        , 
    \end{multline}
    for all $(t,\sfp) \in [\BoldTs,\tstar] \times \mathbb{N}_{\geq 1}$.    

    \begingroup
    \endgroup

    \ul{\emph{Bound for $\mathfrak{J}_4(t)$:}}
    Finally, we have 
    \begin{multline*}
        \LpLaw{\sfp}{}{
            \mathfrak{J}_4(t)
        }
        \leq        
        \Delta_g
        \LpLaw{\sfp}{}{
            \int_{\BoldTs}^t  
                \expo{(2 \lambda -\Boldomega) \nu} 
                \norm{\ZDiffNorm{\Zt{N,\nu}}} 
                \norm{
                    \int_0^\nu 
                        \expo{\Boldomega \beta}
                            \mathcal{J}_4(\beta,\Xt{N})
                    d\Wt{\beta}
                }       
            d\nu
        }
        \\
        =
        \Delta_g
        \LpLaw{\sfp}{}{
            \int_{\BoldTs}^t  
                \expo{(2 \lambda -\Boldomega) \nu} 
                \norm{\ZDiffNorm{\Zt{N,\nu}}} 
                \norm{
                    \Fparasigma{\istar{\nu}\BoldTs,\Xt{N,\istar{\nu}\BoldTs}}
                    \int_{\istar{\nu}\BoldTs}^\nu 
                        \expo{\Boldomega \beta}
                    d\Wt{\beta}
                }       
            d\nu
        }
        ,
    \end{multline*}
    for all $(t,p) \in [\BoldTs,\tstar] \times \mathbb{N}_{\geq 1}$. 
    The term inside the norm can be cast as the process $N_b(t)$ in Corollary~\ref{cor:TechnicalResults:NestedSupMoment} by setting 
    \begin{align*}
        \theta_1 = 2\lambda, 
        \quad
        \theta_2 = \Boldomega,
        \quad   
        Q_t = \Wt{t}
        ,
        \\
        S(\nu) =  \ZDiffNorm{\Zt{N,\nu}},
        \quad 
        L(\nu)  
        =
        \Fparasigma{\istar{\nu}\BoldTs,\Xt{N,\istar{\nu}\BoldTs}}
        .
    \end{align*}
    Hence, by~\eqref{eqn:cor:TechnicalResults:NestedSupMoment:Nb:Bound:Final} in Corollary~\ref{cor:TechnicalResults:NestedSupMoment}
    \begin{align}\label{eqn:prop:Appendix:TrueProcess:Zprocess:J4:Final}
        \LpLaw{\sfp}{}{
            \mathfrak{J}_4(t)
        }
        \leq   
        \frac{ \expo{ 2\lambda t }  }{ 2\lambda \sqrt{\Boldomega} }
        \mathfrak{p}''(\sfp)
        \Delta_g
        \left( 1 - \expo{ -2\Boldomega \BoldTs } \right)^\frac{1}{2}
        \sup_{\nu \in [0,\tstar]}  
            \LpLaw{2\sfp}{}{\ZDiffNorm{\Zt{N,\nu}}}
        \sup_{\nu \in [0,\tstar]}
        \LpLaw{4\sfp}{}{ 
            \Fparasigma{\nu,\Xt{N,\nu}}
        }
        , 
    \end{align}
    for all $(t,p) \in [\BoldTs,\tstar] \times \mathbb{N}_{\geq 1}$, where we have also used~\eqref{eqn:lem:Appendix:TrueProcess:XiTildeU:Fact:1}.
    The proof is then concluded by adding together the bounds in~\eqref{eqn:prop:Appendix:TrueProcess:Zprocess:J1:Final}, ~\eqref{eqn:prop:Appendix:TrueProcess:Zprocess:J2:Final},~\eqref{eqn:prop:Appendix:TrueProcess:Zprocess:J3:Final}, and~\eqref{eqn:prop:Appendix:TrueProcess:Zprocess:J4:Final}.


\end{proof}

We now derive a bound on the moments of $\widetilde{\Xi}_{\mathcal{U}}$. 
\begin{lemma}\label{lem:Appendix:TrueProcess:XiTildeU}
    If the stopping time $\tau^\star$, defined in~\eqref{eqn:True:StoppingTimes}, Lemma~\ref{lem:True:dV}, satisfies  $\tau^\star = \tstar$, then the term $\widetilde{\Xi}_{\mathcal{U}}\left(\tau(t),\Zt{N};\Boldomega \right)$ defined in~\eqref{eqn:lem:True:dV:Xi:Functions:C}, Lemma~\ref{lem:True:dV}, satisfies the following bound for all $(t,\sfp) \in [0,\tstar] \times \mathbb{N}_{\geq 1}$:
    \begin{equation}\label{eqn:lem:Appendix:TrueProcess:XiTildeU:Bound:Final}
        \pLpLaw{}{
            \expo{-2\lambda \tau(t)}\widetilde{\Xi}_{\mathcal{U}}\left(\tau(t),\Zt{N};\Boldomega \right)
        }
        \leq 
        \widetilde{\Delta}_{\mathcal{U}}(\sfp,\tstar,\Boldomega,\BoldTs)
        \doteq 
        \widetilde{\Delta}^-(\sfp,\tstar,\Boldomega,\BoldTs)
        + 
        \indicator{\geq \BoldTs}{\tstar}
        \sum_{k=1}^3
            \widetilde{\Delta}_k^+(\sfp,\tstar,\Boldomega,\BoldTs)
        ,
    \end{equation}
    where 
    \begin{multline*}
        \begin{aligned}
            &\widetilde{\Delta}^-(\sfp,\tstar,\Boldomega,\BoldTs)
            \\
            &=
             \frac{1 - \expo{-2\lambda \BoldTs}}{2\lambda}
            \Delta_g
            \sup_{\nu \in \sbr{0,\tstar \wedge \BoldTs}  }  
                    \LpLaw{2\sfp}{}{\ZDiffNorm{\Zt{N,\nu}}}
        \end{aligned}
            \\ 
            \times
            \left(
                \sup_{\nu \in \sbr{0,\tstar \wedge \BoldTs} } 
                    \LpLaw{2\sfp}{}{\Lparamu{\nu,\Xt{N,\nu}}}
                +
                \sqrt{\Boldomega}
                \mathfrak{p}'(\sfp)
                \left( 1 - \expo{-2\Boldomega \BoldTs}  \right)^\frac{1}{2} 
                \sup_{\nu \in \sbr{0,\tstar \wedge \BoldTs} } 
                    \LpLaw{2\sfp}{}{\Fparasigma{\nu,\Xt{N,\nu}}}
            \right),
    \end{multline*}
    and
    \begin{align*}    
        \widetilde{\Delta}_1^+(\sfp,\tstar,\Boldomega,\BoldTs)
        =&
        \begin{multlined}[t][0.8\linewidth]
            \frac{ 1 }{ 2\lambda  }
            \Delta_g
            \sup_{\nu \in [0,\tstar]}
            \LpLaw{2\sfp}{}{\ZDiffNorm{\Zt{N,\nu}}}
            \\
            \times 
            \left(
                \Delta_{\widetilde{\mathcal{H}}_\mu}(\BoldTs,\sfp,\tstar)
                +
                \Delta_{\mathcal{G}_\mu}(\BoldTs,\tstar)
                + 
                \sqrt{\Boldomega}
                \mathfrak{p}'(\sfp)
                \left[
                    \Delta_{\widetilde{\mathcal{H}}_\sigma}(\BoldTs,\sfp,\tstar)
                    + 
                    \Delta_{\widetilde{\mathcal{G}}_\sigma}(\BoldTs,\sfp,\tstar)
                \right]
            \right),
        \end{multlined} 
        \\
        \widetilde{\Delta}_2^+(\sfp,\tstar,\Boldomega,\BoldTs)
        =&
        \sqrt{\Boldomega}
        \frac{ 1 }{2\lambda }
        \Delta_g
        \gamma''\left(\sfp,\Boldomega,\BoldTs\right)
        \sup_{\nu \in [0,\tstar]}  
            \LpLaw{2\sfp}{}{\ZDiffNorm{\Zt{N,\nu}}}
        \sup_{\nu \in [0,\tstar]}
        \LpLaw{4\sfp}{}{ 
            \Fparasigma{\nu,\Xt{N,\nu}}
        },
        \\
        \widetilde{\Delta}_3^+(\sfp,\tstar,\Boldomega,\BoldTs)
        =&
        \frac{1}{\absolute{2 \lambda -\Boldomega}}
        \left(\expo{\Boldomega \BoldTs} - 1\right)
         \Delta_g
        \sup_{\nu \in [0,\tstar]}  
            \LpLaw{2\sfp}{}{\ZDiffNorm{\Zt{N,\nu}}}
        \sup_{\nu \in [0,\tstar]}  
            \LpLaw{2\sfp}{}{\Lparamu{\nu,\Xt{N,\nu}}},
    \end{align*}
    and where the terms $\Delta_{\widetilde{\mathcal{H}}_\mu}$, $\Delta_{\widetilde{\mathcal{H}}_\sigma}$, $\Delta_{\widetilde{\mathcal{G}}_\sigma}$, and $\Delta_{\mathcal{G}_\mu}$ are defined in Proposition~\ref{prop:Appendix:TrueProcess:ControlErrorIntegrand:Bounds}.

\end{lemma}
\begin{proof}
    We begin by using the respective definitions of $\widetilde{\Xi}_{\mathcal{U}}\left(\tau(t),\Zt{N};\Boldomega \right)$ and $\widetilde{\mathcal{U}}$ in~\eqref{eqn:lem:True:dV:Xi:Functions:C} and~\eqref{eqn:lem:True:dV:U:Functions} to write 
    \begin{align*}
        \widetilde{\Xi}_{\mathcal{U}}\left(\tau(t),\Zt{N} \right)
        =
        \int_0^{\tau(t)}  
            \expo{2 \lambda \nu} 
            \widetilde{\mathcal{U}}\left(\nu,\Zt{N,\nu}\right)d\nu
        =&
        \int_0^{\tau(t)}  
            \expo{2 \lambda \nu} 
            \ZDiffNorm{\Zt{N,\nu}}^\top
            g(\nu)
            \left(\FL - \ReferenceInput\right)(\Xt{N})(\nu)
        d\nu
        \\
        =& 
        \Boldomega 
        \int_0^{\tau(t)}  
            \expo{(2 \lambda -\Boldomega) \nu}
            \ZDiffNorm{\Zt{N,\nu}}^\top
            g(\nu)
            \ControlErrorTilde[\Xt{N},\nu]        
        d\nu
        ,
    \end{align*}
    where the last expression is due to the conclusion in Proposition~\ref{prop:Appendix:TrueProcess:FL:Expression} that $\left(\FL - \ReferenceInput\right)\br{\Xt{N}}(\nu)=\Boldomega \expo{-\Boldomega \nu}\ControlErrorTilde[\Xt{N},\nu]$.
    Note that $\tau(t) = t$, for all $t \in [0,\tstar]$, due to the hypothesis that $\tau^\star = \tstar$. 
    We can thus write the last expression as   
    \begin{align}\label{eqn:lem:Appendix:TrueProcess:XiTildeU:Representation:Initial}
        \widetilde{\Xi}_{\mathcal{U}}\left(\tau(t),\Zt{N} \right)
        = 
        \Boldomega 
            \int_0^{t}  
                \expo{(2 \lambda -\Boldomega) \nu}
                \ZDiffNorm{\Zt{N,\nu}}^\top
                g(\nu)
                \ControlErrorTilde[\Xt{N},\nu]        
            d\nu
        , 
        \quad t \in [0,\tstar].
    \end{align}
    To obtain the expression for $\ControlErrorTilde[\Xt{N},\nu]$, we invoke Proposition~\ref{proposition:Appendix:TrueProcess:FL:ExpressionEvolved} with $z = \Xt{N}$ to obtain: 
    \begin{align*}
         \ControlErrorTilde[\Xt{N},\nu]
        =
        \int_0^\nu 
            \expo{ \Boldomega \beta} 
            \Sigma^{\paral}_\beta(\Xt{N}), 
        \quad \nu < \BoldTs,
    \end{align*}
    and for $\nu \geq \BoldTs$, 
    \begin{multline*}
        \begin{aligned}
            \ControlErrorTilde[\Xt{N},\nu]
            =&
            \int_0^\BoldTs
                \expo{ \Boldomega \beta} 
                \Lparamu{\beta,\Xt{N,\beta}}
            d\beta
            \\
            &+ 
            \int_{0}^{\nu}
                \expo{\Boldomega \beta}
                \left[
                    \left(
                        \widetilde{\mathcal{H}}_\mu^{\paral}(\beta,\Xt{N})
                        + 
                        \expo{(\lambda_s-\Boldomega) \beta}
                        \widetilde{\mathcal{H}}_\mu(\beta,\Xt{N})
                    \right)
                    d\beta
                    + 
                    \left(
                        \widetilde{\mathcal{H}}_\sigma^{\paral}(\beta,\Xt{N})
                        + 
                        \expo{(\lambda_s-\Boldomega) \beta}
                        \widetilde{\mathcal{H}}_\sigma(\beta,\Xt{N})
                    \right)
                    d\Wt{\beta}
                \right]
        \end{aligned}
        \\
        +
        \int_{0}^{\nu}
            \expo{\Boldomega \beta} 
            \left( 
                \widetilde{\mathcal{G}}_{\sigma_1}(\beta,\Xt{N})
                + 
                \expo{(\lambda_s - \Boldomega) \beta} 
                \widetilde{\mathcal{G}}_{\sigma_2}(\beta,\Xt{N})
            \right)
        d\Wt{\beta}
        \\
        +
        \int_0^\nu 
            \expo{\Boldomega \beta}
            \left[
                \mathcal{G}_\mu(\beta,\Xt{N})
                d\beta
                + 
                \left( 
                    \mathcal{G}_{\sigma_1}(\beta,\Xt{N})
                    + 
                    \mathcal{G}_{\sigma_2}(\beta,\Xt{N})
                \right)
                d\Wt{\beta}
            \right]
            .
    \end{multline*}
    Susbtituting the above expression for $\ControlErrorTilde[\Xt{N},\nu]$ into~\eqref{eqn:lem:Appendix:TrueProcess:XiTildeU:Representation:Initial}: 
    \begin{align}\label{eqn:lem:Appendix:TrueProcess:XiTildeU:Representation:Evolved:Pre1}
        \widetilde{\Xi}_{\mathcal{U}}\left(\tau(t),\Zt{N} \right)
        =
        \begin{cases}
            \Boldomega 
            \int_0^{t}  
                \expo{(2 \lambda -\Boldomega) \nu} 
                \ZDiffNorm{\Zt{N,\nu}}^\top 
                g(\nu)
                \mathcal{N}_\star(\nu)       
            d\nu
            ,
            & \text{if } t < \BoldTs
            ,
            \\[10pt]
            \begin{aligned}
                &\Boldomega 
                \int_0^{\BoldTs}  
                    \expo{(2 \lambda -\Boldomega) \nu} 
                    \ZDiffNorm{\Zt{N,\nu}}^\top 
                    g(\nu)
                    \mathcal{N}_\star(\nu)       
                d\nu
                \\
                &+ 
                \Boldomega 
                \int_{\BoldTs}^t  
                    \expo{(2 \lambda -\Boldomega) \nu} 
                    \ZDiffNorm{\Zt{N,\nu}}^\top 
                    g(\nu)
                \\
                &\hspace{3cm} 
                \times
                    \left(\mathcal{N}_1(\nu) + \mathcal{N}_2(\nu) + \mathcal{N}_3(\nu)\right)       
                d\nu
            \end{aligned}
            , 
            & \text{if otherwise } t \geq \BoldTs
            ,
        \end{cases}
    \end{align}
     where $\mathcal{N}_\star$, $\mathcal{N}_1$, $\mathcal{N}_2$, and $\mathcal{N}_3$ are defined in~\eqref{eqn:lem:Appendix:TrueProcess:NUTilde:Representation:Evolved:Pre1} as follows:
    \begin{align*}
        \mathcal{N}_\star(\nu)
        =
        \int_0^\nu 
            \expo{ \Boldomega \beta} 
            \Sigma^{\paral}_\beta(\Xt{N})
        =
        \int_0^\nu 
            \expo{ \Boldomega \beta}
            \left[
                \Lparamu{\beta,\Xt{N,\beta}}d\beta 
                +
                \Fparasigma{\beta,\Xt{N,\beta}}d\Wt{\beta}
            \right]
        ,
        \\
        \begin{multlined}[b][0.9\linewidth]
            \mathcal{N}_1(\nu)
            = 
            \int_{0}^{\nu}
                \expo{\Boldomega \beta}
                \left[
                    \left(
                        \widetilde{\mathcal{H}}_\mu^{\paral}(\beta,\Xt{N})
                        + 
                        \expo{(\lambda_s-\Boldomega) \beta}
                        \widetilde{\mathcal{H}}_\mu(\beta,\Xt{N})
                        + 
                        \mathcal{G}_\mu(\beta,\Xt{N})
                    \right)
                    d\beta
                \right. 
                \\
                \left.
                    + 
                    \left(
                        \widetilde{\mathcal{H}}_\sigma^{\paral}(\beta,\Xt{N})
                        + 
                        \expo{(\lambda_s-\Boldomega) \beta}
                        \widetilde{\mathcal{H}}_\sigma(\beta,\Xt{N})
                        + 
                        \widetilde{\mathcal{G}}_{\sigma_1}(\beta,\Xt{N})
                        + 
                        \expo{(\lambda_s - \Boldomega) \beta} 
                        \widetilde{\mathcal{G}}_{\sigma_2}(\beta,\Xt{N})
                    \right)
                    d\Wt{\beta}
                \right]
            ,
        \end{multlined}
        \\ 
        \mathcal{N}_2(\nu)
        =
        \int_0^\nu 
            \expo{\Boldomega \beta}
            \left( 
                \mathcal{G}_{\sigma_1}(\beta,\Xt{N})
                + 
                \mathcal{G}_{\sigma_2}(\beta,\Xt{N})
            \right)
        d\Wt{\beta}
        , 
        \\
        \mathcal{N}_3
        =
        \int_0^{\BoldTs} 
            \expo{\Boldomega \beta}
            \Lparamu{\beta,\Xt{N,\beta}}
        d\beta
        .
    \end{align*}
    We can further write~\eqref{eqn:lem:Appendix:TrueProcess:XiTildeU:Representation:Evolved:Pre1} as     
    \begin{align}\label{eqn:lem:Appendix:TrueProcess:XiTildeU:Representation:Evolved:Pre}
        \widetilde{\Xi}_{\mathcal{U}}\left(\tau(t),\Zt{N} \right)
        =
        \begin{cases}
            \Boldomega
            \widetilde{\mathfrak{N}}_\star(t)
            ,
            & \text{if } t < \BoldTs
            ,
            \\[10pt]
            \Boldomega
            \left(
                \widetilde{\mathfrak{N}}_\star(\BoldTs)
                + 
                \sum_{k=1}^3
                \widetilde{\mathfrak{N}}_k(t)
            \right)
            , 
            & \text{if otherwise } t \geq \BoldTs
            ,
        \end{cases}
    \end{align}
    where we have defined 
    \begin{align*}
        \widetilde{\mathfrak{N}}_\star(t)
        =
        \int_0^{t}  
            \expo{(2 \lambda -\Boldomega) \nu} 
            \ZDiffNorm{\Zt{N,\nu}}^\top
            g(\nu)
            \mathcal{N}_\star(\nu)       
        d\nu
        , 
        \\
        \widetilde{\mathfrak{N}}_k(t)
        =
        \int_{\BoldTs}^t  
            \expo{(2 \lambda -\Boldomega) \nu} 
            \ZDiffNorm{\Zt{N,\nu}}^\top 
            g(\nu)
            \mathcal{N}_k(\nu)       
        d\nu
        , 
        \quad k \in \cbr{1,2,3}.
    \end{align*}
    Using the Minkowski's inequality along with the fact that $\tstar$ and $\BoldTs$ are constants leads to the following bounds for all $\sfp \in \mathbb{N}_{\geq 1}$:
    \begin{align}\label{eqn:lem:Appendix:TrueProcess:XiTildeU:Bound:Initial}
        \LpLaw{\sfp}{}{
            \expo{-2\lambda \tau(t)}\widetilde{\Xi}_{\mathcal{U}}\left(t,\Zt{N} \right)
        }
        \leq
        \begin{cases}
            \Boldomega
            \expo{-2\lambda t}
            \LpLaw{\sfp}{}{\widetilde{\mathfrak{N}}_\star(t)}
            ,
            & \forall t \in [0,\tstar] \cap [0,\BoldTs)
            ,
            \\[10pt]
            \Boldomega
            \expo{-2\lambda t}
            \left(
                \LpLaw{\sfp}{}{\widetilde{\mathfrak{N}}_\star(\BoldTs)}
                + 
                \sum_{k=1}^3
                \LpLaw{\sfp}{}{\widetilde{\mathfrak{N}}_k(t)}
            \right)
            , 
            & \forall t \in [0,\tstar] \cap [\BoldTs,T]
            .
        \end{cases}
    \end{align} 

    We now bound the processes $\widetilde{\mathfrak{N}}_\star(t)$ and $\widetilde{\mathfrak{N}}_k(t)$, $k \in \cbr{1,2,3}$.

    \ul{\emph{Bound for $\widetilde{\mathfrak{N}}_\star(t)$:}}
    Using the definitions in~\eqref{eqn:lem:Appendix:TrueProcess:XiTildeU:Representation:Evolved:Pre1}-\eqref{eqn:lem:Appendix:TrueProcess:XiTildeU:Representation:Evolved:Pre}, we have 
    \begin{align*}
        \widetilde{\mathfrak{N}}_\star(t)
        =&
        \int_0^{t}  
            \expo{(2 \lambda -\Boldomega) \nu} 
            \ZDiffNorm{\Zt{N,\nu}}^\top
            g(\nu)
            \mathcal{N}_\star(\nu)       
        d\nu
        \\
        =&
        \int_0^{t}  
            \expo{(2 \lambda -\Boldomega) \nu} 
            \ZDiffNorm{\Zt{N,\nu}}^\top
            g(\nu)
            \left(
                \int_0^\nu 
                    \expo{ \Boldomega \beta}
                    \left[
                        \Lparamu{\beta,\Xt{N,\beta}}d\beta 
                        +
                        \Fparasigma{\beta,\Xt{N,\beta}}d\Wt{\beta}
                    \right]
            \right)       
        d\nu
        ,
    \end{align*}
    which, upon using Minkowski's inequality, leads to
    \begin{multline*}
        \LpLaw{\sfp}{}{\widetilde{\mathfrak{N}}_\star(t)}
        \leq 
        \Delta_g
        \LpLaw{\sfp}{}{
            \int_0^{t}  
                \expo{(2 \lambda -\Boldomega) \nu} 
                \norm{\ZDiffNorm{\Zt{N,\nu}}} 
                \left(
                    \int_0^\nu 
                        \expo{ \Boldomega \beta}
                        \norm{\Lparamu{\beta,\Xt{N,\beta}}}
                    d\beta 
                \right)       
            d\nu
        }
        \\ 
        +
        \Delta_g 
        \LpLaw{\sfp}{}{
            \int_0^{t}  
                \expo{(2 \lambda -\Boldomega) \nu} 
                \norm{\ZDiffNorm{\Zt{N,\nu}}} 
                \norm{
                    \int_0^\nu 
                        \expo{ \Boldomega \beta}
                        \Fparasigma{\beta,\Xt{N,\beta}}
                    d\Wt{\beta}
                }       
            d\nu
        }
        ,
    \end{multline*}
    for all $(t,\sfp) \in [0,\tstar \wedge \BoldTs] \times \mathbb{N}_{\geq 1}$, where we have used the uniform bound on $g(t)$ in Assumption~\ref{assmp:KnownFunctions}. 
     We now invoke Proposition~\ref{prop:TechnicalResults:LebesgueNestedMoment} to bound the first term by setting 
    \begin{align*}
        \xi = 0, \quad \theta_1 = 2\lambda, \quad \theta_2 = \Boldomega, 
        \quad 
        S_1(\nu) = \ZDiffNorm{\Zt{N,\nu}}, 
        \quad 
        S_2(\beta) =  
        \Lparamu{\beta,\Xt{N,\beta}},
    \end{align*}
    and we invoke Lemma~\ref{lemma:TechnicalResults:NestedSupMoment} to bound the second term by setting
    \begin{align*}
        \xi = 0, \quad \theta_1 = 2\lambda, \quad \theta_2 = \Boldomega,
        \quad 
        \cbr{\zeta_1,\zeta_2} = \cbr{0,t}, 
        \quad 
        S(\nu) = \ZDiffNorm{\Zt{N,\nu}}, 
        \quad 
        L(\beta) =  
        \Fparasigma{\beta,\Xt{N,\beta}},
    \end{align*}
    and obtain 
    \begin{multline*}
        \LpLaw{\sfp}{}{\widetilde{\mathfrak{N}}_\star(t)}
        \leq 
        \frac{ \expo{2\lambda t} }{2\lambda \sqrt{\Boldomega} }
        \left(1 - \expo{-2\lambda t} \right)
        \Delta_g
        \sup_{\nu \in [0,t]}  
                \LpLaw{2\sfp}{}{\ZDiffNorm{\Zt{N,\nu}}}
        \\ 
        \times
        \left(
            \frac{1}{\sqrt{\Boldomega}}
            \sup_{\nu \in [0,t]} 
                \LpLaw{2\sfp}{}{\Lparamu{\nu,\Xt{N,\nu}}}
            +
            \mathfrak{p}'(\sfp)
            \left( 1 - \expo{-2\Boldomega t}  \right)^\frac{1}{2} 
            \sup_{\nu \in [0,t]} 
                \LpLaw{2\sfp}{}{\Fparasigma{\nu,\Xt{N,\nu}}}
        \right)
        ,
    \end{multline*}
    for all $(t,\sfp) \in [0,\tstar \wedge \BoldTs] \times \mathbb{N}_{\geq 1}$.
    Since $t \leq (\BoldTs \wedge \tstar) \leq \BoldTs$, we can develop the bound above into
    \begin{multline*}
        \LpLaw{\sfp}{}{\widetilde{\mathfrak{N}}_\star(t)}
        \leq 
        \frac{ \expo{2\lambda t} }{2\lambda \sqrt{\Boldomega} }
        \left(1 - \expo{-2\lambda \BoldTs} \right)
        \Delta_g
        \sup_{\nu \in \sbr{0,\tstar \wedge \BoldTs}  }  
                \LpLaw{2\sfp}{}{\ZDiffNorm{\Zt{N,\nu}}}
        \\ 
        \times
        \left(
            \frac{1}{\sqrt{\Boldomega}}
            \sup_{\nu \in \sbr{0,\tstar \wedge \BoldTs} } 
                \LpLaw{2\sfp}{}{\Lparamu{\nu,\Xt{N,\nu}}}
            +
            \mathfrak{p}'(\sfp)
            \left( 1 - \expo{-2\Boldomega \BoldTs}  \right)^\frac{1}{2} 
            \sup_{\nu \in \sbr{0,\tstar \wedge \BoldTs} } 
                \LpLaw{2\sfp}{}{\Fparasigma{\nu,\Xt{N,\nu}}}
        \right)
        ,
    \end{multline*}
    for all $(t,\sfp) \in [0,\tstar \wedge \BoldTs] \times \mathbb{N}_{\geq 1}$.
    Hence, we get 
    \begin{equation}\label{eqn:lem:Appendix:TrueProcess:XiTildeU:Bound:A:Final}
        \Boldomega
        \expo{-2\lambda t}
        \LpLaw{\sfp}{}{\widetilde{\mathfrak{N}}_\star(t)}
        \leq     
        \widetilde{\Delta}^-(\sfp,\tstar,\Boldomega,\BoldTs)
        , 
        \quad 
        \forall 
        (t,\sfp) \in [0,\tstar \wedge \BoldTs] \times \mathbb{N}_{\geq 1}.
    \end{equation} 
    \begingroup 
    \endgroup

    \ul{\emph{Bound for $\widetilde{\mathfrak{N}}_1(t)$:}}
    The process $\widetilde{\mathfrak{N}}_1$ in~\eqref{eqn:lem:Appendix:TrueProcess:XiTildeU:Representation:Evolved:Pre} can be written as 
    \begin{align*}
        \mathfrak{\widetilde{N}}_1(t)
        =& 
        \int_{\BoldTs}^t  
            \expo{(2 \lambda -\Boldomega) \nu} 
            \ZDiffNorm{\Zt{N,\nu}}^\top 
            g(\nu)
            \mathcal{N}_1(\nu)       
        d\nu
        \\
        =& 
        \int_{\BoldTs}^t  
            \expo{(2 \lambda -\Boldomega) \nu} 
            \ZDiffNorm{\Zt{N,\nu}}^\top 
            g(\nu)
            \left(
                \int_{0}^{\nu}
                \expo{\Boldomega \beta}
                \left[
                    \mathcal{K}_\mu(\beta,\Xt{N})
                    d\beta
                    + 
                    \mathcal{K}_\sigma(\beta,\Xt{N})
                    d\Wt{\beta}
                \right]
            \right)       
        d\nu
        ,
    \end{align*} 
    where we have defined
    \begin{subequations}\label{eqn:lem:Appendix:TrueProcess:XiTildeU:KFunctions}
        \begin{align}
            \mathcal{K}_\mu(\beta,\Xt{N})
            =
            \widetilde{\mathcal{H}}_\mu^{\paral}(\beta,\Xt{N})
            + 
            \expo{(\lambda_s-\Boldomega) \beta}
            \widetilde{\mathcal{H}}_\mu(\beta,\Xt{N})
            + 
            \mathcal{G}_\mu(\beta,\Xt{N})
            , 
            \\ 
            \mathcal{K}_\sigma(\beta,\Xt{N})
            =
            \widetilde{\mathcal{H}}_\sigma^{\paral}(\beta,\Xt{N})
            + 
            \expo{(\lambda_s-\Boldomega) \beta}
            \widetilde{\mathcal{H}}_\sigma(\beta,\Xt{N})
            + 
            \widetilde{\mathcal{G}}_{\sigma_1}(\beta,\Xt{N})
            + 
            \expo{(\lambda_s - \Boldomega) \beta} 
            \widetilde{\mathcal{G}}_{\sigma_2}(\beta,\Xt{N})
            .
        \end{align}    
    \end{subequations}
    Therefore, using  Minkowski's inequality and the bound on $g(t)$ in Assumption~\ref{assmp:KnownFunctions}, we obtain
    \begin{multline*}
        \LpLaw{\sfp}{}{\widetilde{\mathfrak{N}}_1(t)}
        \leq  
        \Delta_g
        \LpLaw{\sfp}{}{
            \int_{\BoldTs}^t  
                \expo{(2 \lambda -\Boldomega) \nu} 
                \norm{\ZDiffNorm{\Zt{N,\nu}}} 
                \left(
                    \int_{0}^{\nu}
                        \expo{\Boldomega \beta}
                        \norm{\mathcal{K}_\mu(\beta,\Xt{N})}
                        d\beta
                \right)       
            d\nu
        }
        \\ 
        + 
        \Delta_g
        \LpLaw{\sfp}{}{
            \int_{\BoldTs}^t  
                \expo{(2 \lambda -\Boldomega) \nu} 
                \norm{\ZDiffNorm{\Zt{N,\nu}}} 
                \norm{
                    \int_{0}^{\nu}
                        \expo{\Boldomega \beta}
                        \mathcal{K}_\sigma(\beta,\Xt{N})
                    d\Wt{\beta}
                }       
            d\nu
        }
        ,
    \end{multline*}
    for all $(t,\sfp) \in [\BoldTs,\tstar] \times \mathbb{N}_{\geq 1}$. 
    We now invoke Proposition~\ref{prop:TechnicalResults:LebesgueNestedMoment} to bound the first term by setting 
    \begin{align*}
        \xi = \BoldTs, \quad \theta_1 = 2\lambda, \quad \theta_2 = \Boldomega, 
        \quad 
        S_1(\nu) =  \ZDiffNorm{\Zt{N,\nu}}, 
        \quad 
        S_2(\beta) =  
        \mathcal{K}_\mu(\beta,\Xt{N}),
    \end{align*}
    and we invoke Lemma~\ref{lemma:TechnicalResults:NestedSupMoment} to bound the second term by setting
    \begin{align*}
        \xi = \BoldTs, \quad \theta_1 = 2\lambda, \quad \theta_2 = \Boldomega,
        \quad 
        \cbr{\zeta_1,\zeta_2} = \cbr{0,t}, 
        \quad 
        S(\nu) =  \ZDiffNorm{\Zt{N,\nu}}, 
        \quad 
        L(\beta) =  
        \mathcal{K}_\sigma(\beta,\Xt{N}),
    \end{align*}
    and obtain 
    \begin{multline*}
        \LpLaw{\sfp}{}{\widetilde{\mathfrak{N}}_1(t)}
        \leq  
        \frac{ \expo{ 2\lambda t } - \expo{ 2\lambda \BoldTs } }{ 2\lambda \sqrt{\Boldomega}  }
        \Delta_g
        \sup_{\nu \in [\BoldTs,t]}
            \LpLaw{2\sfp}{}{\ZDiffNorm{\Zt{N,\nu}}}
        \\
        \times 
        \left(
            \frac{1}{\sqrt{\Boldomega}}
            \sup_{\nu \in [\BoldTs,t]} 
                \LpLaw{2\sfp}{}{\mathcal{K}_\mu(\nu,\Xt{N})} 
            + 
            \mathfrak{p}'(\sfp)
            \left( 1 - \expo{-2\Boldomega t}  \right)^\frac{1}{2}
            \sup_{\nu \in [0,t]} 
            \LpLaw{2\sfp}{}{\mathcal{K}_\sigma(\beta,\Xt{N})}
        \right)
        ,
    \end{multline*}
    for all $(t,\sfp) \in [\BoldTs,\tstar] \times \mathbb{N}_{\geq 1}$.
    Since $t \in [\BoldTs,\tstar]$, we can further develop the bound above into 
    \begin{multline*}
        \LpLaw{\sfp}{}{\widetilde{\mathfrak{N}}_1(t)}
        \leq  
        \frac{ \expo{ 2\lambda t } }{ 2\lambda \sqrt{\Boldomega}  }
        \Delta_g
        \sup_{\nu \in [0,\tstar]}
        \LpLaw{2\sfp}{}{\ZDiffNorm{\Zt{N,\nu}}}
        \\
        \times 
        \left(
            \frac{1}{\sqrt{\Boldomega}}
            \sup_{\nu \in [0,\tstar]} 
                \LpLaw{2\sfp}{}{\mathcal{K}_\mu(\nu,\Xt{N})} 
            + 
            \mathfrak{p}'(\sfp)
            \sup_{\nu \in [0,\tstar]} 
            \LpLaw{2\sfp}{}{\mathcal{K}_\sigma(\beta,\Xt{N})}
        \right)
        ,
    \end{multline*}
    for all $(t,\sfp) \in [\BoldTs,\tstar] \times \mathbb{N}_{\geq 1}$.
    Therefore,
    \begin{multline}\label{eqn:lem:Appendix:TrueProcess:XiTildeU:Bound:B:1:Initial}
        \Boldomega
        \expo{-2\lambda t}
        \LpLaw{\sfp}{}{\widetilde{\mathfrak{N}}_1(t)}
        \leq  
        \frac{ 1 }{ 2\lambda  }
        \Delta_g
        \sup_{\nu \in [0,\tstar]}
        \LpLaw{2\sfp}{}{\ZDiffNorm{\Zt{N,\nu}}}
        \\
        \times 
        \left(
            \sup_{\nu \in [0,\tstar]} 
                \LpLaw{2\sfp}{}{\mathcal{K}_\mu(\nu,\Xt{N})} 
            + 
             \sqrt{\Boldomega}
            \mathfrak{p}'(\sfp)
            \sup_{\nu \in [0,\tstar]} 
            \LpLaw{2\sfp}{}{\mathcal{K}_\sigma(\beta,\Xt{N})}
        \right)
        ,
    \end{multline}
    for all $(t,\sfp) \in [\BoldTs,\tstar] \times \mathbb{N}_{\geq 1}$.
    Using the definitions of $\mathcal{K}_\mu$ and $\mathcal{K}_\sigma$ in~\eqref{eqn:lem:Appendix:TrueProcess:XiTildeU:KFunctions}, we use  Minkowski's inequality and the bounds derived in Proposition~\ref{prop:Appendix:TrueProcess:ControlErrorIntegrand:Bounds} to obtain the following:
    \begin{multline*}
        \LpLaw{2\sfp}{}{ 
            \mathcal{K}_\mu(\nu,\Xt{N})
        } 
        \leq
        \LpLaw{2\sfp}{}{ 
            \widetilde{\mathcal{H}}_\mu^{\paral}(\nu,\Xt{N})
            + 
            \expo{(\lambda_s-\Boldomega) \nu}
            \widetilde{\mathcal{H}}_\mu(\nu,\Xt{N})
        } 
        + 
        \LpLaw{2\sfp}{}{ 
            \mathcal{G}_\mu(\nu,\Xt{N})
        } 
        \leq      
        \Delta_{\widetilde{\mathcal{H}}_\mu}(\BoldTs,\sfp,\tstar)
        +
        \Delta_{\mathcal{G}_\mu}(\BoldTs,\tstar)
        ,
    \end{multline*}
    and 
    \begin{multline*}
        \LpLaw{2\sfp}{}{\mathcal{K}_\sigma(\nu,\Xt{N})}
        \leq 
        \LpLaw{2\sfp}{}{
            \widetilde{\mathcal{H}}_\sigma^{\paral}(\nu,\Xt{N})
            + 
            \expo{(\lambda_s-\Boldomega) \nu}
            \widetilde{\mathcal{H}}_\sigma(\nu,\Xt{N})
        }
        + 
        \LpLaw{2\sfp}{}{
            \widetilde{\mathcal{G}}_{\sigma_1}(\nu,\Xt{N})
            + 
            \expo{(\lambda_s - \Boldomega) \nu} 
            \widetilde{\mathcal{G}}_{\sigma_2}(\nu,\Xt{N})
        }
        \\
        \leq     
        \Delta_{\widetilde{\mathcal{H}}_\sigma}(\BoldTs,\sfp,\tstar)
        + 
        \Delta_{\widetilde{\mathcal{G}}_\sigma}(\BoldTs,\sfp,\tstar)
        ,
    \end{multline*}
    for all $(\nu,\sfp) \in [0,\tstar] \times \mathbb{N}_{\geq 1}$.
    Substituting the above bounds into~\eqref{eqn:lem:Appendix:TrueProcess:XiTildeU:Bound:B:1:Initial} then leads to
    \begin{align}\label{eqn:lem:Appendix:TrueProcess:XiTildeU:Bound:B:1:Final}
        \Boldomega
        \expo{-2\lambda t}
        \LpLaw{\sfp}{}{\widetilde{\mathfrak{N}}_1(t)}
        \leq  
        \widetilde{\Delta}_1^+(\sfp,\tstar,\Boldomega,\BoldTs)
        ,
        \quad \forall (t,\sfp) \in [\BoldTs,\tstar] \times \mathbb{N}_{\geq 1}.
    \end{align}
    \begingroup
    \endgroup

    \ul{\emph{Bound for $\widetilde{\mathfrak{N}}_2(t)$:}}
    Recall the definition of the process $\widetilde{\mathfrak{N}}_2$ in~\eqref{eqn:lem:Appendix:TrueProcess:XiTildeU:Representation:Evolved:Pre}: 
    \begin{align*}
        \widetilde{\mathfrak{N}}_2(t)
        =& 
        \int_{\BoldTs}^t  
            \expo{(2 \lambda -\Boldomega) \nu} 
            \ZDiffNorm{\Zt{N,\nu}}^\top 
            g(\nu)
            \mathcal{N}_2(\nu)       
        d\nu
        \\
        =& 
        \int_{\BoldTs}^t  
            \expo{(2 \lambda -\Boldomega) \nu} 
            \ZDiffNorm{\Zt{N,\nu}}^\top 
            g(\nu)
            \left(
                \int_0^\nu 
                    \expo{\Boldomega \beta}
                    \left( 
                        \mathcal{G}_{\sigma_1}(\beta,\Xt{N})
                        + 
                        \mathcal{G}_{\sigma_2}(\beta,\Xt{N})
                    \right)
                d\Wt{\beta}
            \right)       
        d\nu
        ,
        \quad t \in [\BoldTs,\tstar],
    \end{align*}
    where we have the following due to the definitions of $\mathcal{G}_{\sigma_1}$ and $\mathcal{G}_{\sigma_2}$ in Proposition~\ref{proposition:Appendix:TrueProcess:FL:ExpressionEvolved}:
    \begin{multline*}
        \int_0^\nu 
            \expo{\Boldomega \beta}
            \mathcal{G}_{\sigma_1}(\beta,\Xt{N})
        d\Wt{\beta}
        \\
        = 
        \int_0^\nu 
            \expo{\Boldomega \beta}
            \left(
                \sum_{i=1}^{\istar{\nu} - 1}
                \indicator{[(i-1)\BoldTs,i\BoldTs)}{\beta}
                \Fparasigma{(i-1)\BoldTs,\Xt{N,(i-1)\BoldTs}}
                \left( 
                    1  
                    -
                    \widetilde{\gamma}_i(\BoldTs)
                    \expo{ \left(\lambda_s - \Boldomega\right) \beta} 
                \right)
            \right)
        d\Wt{\beta}
        \\
        +
        \int_0^\nu 
            \expo{\Boldomega \beta}
            \indicator{[(\istar{\nu} - 1)\BoldTs,\istar{\nu}\BoldTs)}{\beta}
                \Fparasigma{(\istar{\nu} - 1)\BoldTs,\Xt{N,(\istar{\nu} - 1)\BoldTs}}      
        d\Wt{\beta},
    \end{multline*}
    and
    \begin{multline*}
        \int_0^\nu 
            \expo{\Boldomega \beta}
            \mathcal{G}_{\sigma_2}(\beta,\Xt{N})
        d\Wt{\beta}
        \\
        = 
        -
        \int_0^\nu 
            \expo{\Boldomega \beta}
            \Fparasigma{\istar{\nu}\BoldTs,\Xt{N,\istar{\nu}\BoldTs}}
                \indicator{[(\istar{\nu}-1)\BoldTs,\istar{\nu}\BoldTs)}{\beta}
                \widetilde{\gamma}^\star(\nu,\BoldTs)
                \expo{ (\lambda_s - \Boldomega) \beta} 
        d\Wt{\beta}
        \\
        +
        \int_0^\nu 
            \expo{\Boldomega \beta}
            \Fparasigma{\istar{\nu}\BoldTs,\Xt{N,\istar{\nu}\BoldTs}}
                \indicator{ \geq \istar{\nu}\BoldTs}{\beta}
        d\Wt{\beta}
        .
    \end{multline*}
    Using the above, it is straightforward to establish that 
    \begin{align*}
        \widetilde{\mathfrak{N}}_2(t)
        =
        \sum_{j=1}^4
        \mathfrak{J}_j(t)
        ,
        \quad t \in [\BoldTs,\tstar],
    \end{align*}
    where the processes $\mathfrak{J}_j$ are defined in~\eqref{eqn:prop:Appendix:TrueProcess:ZProcess:J:Definitions} in the statement of Proposition~\ref{prop:Appendix:TrueProcess:ZProcess}.
    Hence, we obtain the following by the Minkowski's inequality and Proposition~\ref{prop:Appendix:TrueProcess:ZProcess}:
    \begin{multline*}
        \LpLaw{\sfp}{}{\widetilde{\mathfrak{N}}_2(t)}
        \leq 
        \sum_{j=1}^4
        \LpLaw{\sfp}{}{
            \mathfrak{J}_j(t)
        }
        \leq      
        \frac{ \expo{2\lambda t} }{2\lambda \BoldomegaRoot  }
        \Delta_g
        \gamma''\left(\sfp,\Boldomega,\BoldTs\right) 
        \sup_{\nu \in [0,\tstar]}  
            \LpLaw{2\sfp}{}{\ZDiffNorm{\Zt{N,\nu}}}
        \sup_{\nu \in [0,\tstar]}
        \LpLaw{4\sfp}{}{ 
            \Fparasigma{\nu,\Xt{N,\nu}}
        }
    \end{multline*}
    for all $(t,p) \in [\BoldTs,\tstar] \times \mathbb{N}_{\geq 1}$, which further leads to  
    \begin{align}\label{eqn:lem:Appendix:TrueProcess:XiTildeU:Bound:B:2:Final}
        \Boldomega
        \expo{-2\lambda t}
        \LpLaw{\sfp}{}{\widetilde{\mathfrak{N}}_2(t)}
        \leq  
        \widetilde{\Delta}_2^+(\sfp,\tstar,\Boldomega,\BoldTs)
        ,
        \quad \forall (t,\sfp) \in [\BoldTs,\tstar] \times \mathbb{N}_{\geq 1}.
    \end{align}
    \begingroup
    \endgroup

    \ul{\emph{Bound for $\widetilde{\mathfrak{N}}_3(t)$:}}
    Using the definition of $\widetilde{\mathfrak{N}}_3$ from~\eqref{eqn:lem:Appendix:TrueProcess:XiTildeU:Representation:Evolved:Pre}, we obtain  
    \begin{align*}
        \LpLaw{\sfp}{}{\widetilde{\mathfrak{N}}_3(t)}
        \leq& 
        \Delta_g
        \LpLaw{\sfp}{}{
            \left(
                \int_{\BoldTs}^t  
                    \expo{(2 \lambda -\Boldomega) \nu} 
                    \norm{\ZDiffNorm{\Zt{N,\nu}}} 
                d\nu
            \right)
            \left(
                \int_0^{\BoldTs} 
                    \expo{\Boldomega \beta}
                    \norm{\Lparamu{\beta,\Xt{N,\beta}}}
                d\beta
            \right)       
        }
        \\
        \leq& 
        \Delta_g
        \LpLaw{2\sfp}{}{
            \int_{\BoldTs}^t  
                \expo{(2 \lambda -\Boldomega) \nu} 
                \norm{\ZDiffNorm{\Zt{N,\nu}}}     
            d\nu
        }      
        \LpLaw{2\sfp}{}{
            \int_0^{\BoldTs} 
                \expo{\Boldomega \beta}
                \norm{\Lparamu{\beta,\Xt{N,\beta}}}
            d\beta
        }
        ,
    \end{align*}
    for all $(t,\sfp) \in [\BoldTs,\tstar] \times \mathbb{N}_{\geq 1}$, where we have used the bound on $g(t)$ from Assumption~\ref{assmp:KnownFunctions} and applied the Cauchy-Schwarz inequality.
    Bounding both the integrals on the right hand side via Proposition~\ref{prop:TechnicalResults:LebesgueMoment} leads to 
    \begin{align*}
        \LpLaw{\sfp}{}{\widetilde{\mathfrak{N}}_3(t)}
        \leq&
        \Delta_g
        \frac{\expo{(2 \lambda -\Boldomega) t} - \expo{(2 \lambda -\Boldomega) \BoldTs}}{\absolute{2 \lambda -\Boldomega}}
        \frac{\expo{\Boldomega \BoldTs} - 1}{\Boldomega}
        \sup_{\nu \in [\BoldTs,t]}  
            \LpLaw{2\sfp}{}{\ZDiffNorm{\Zt{N,\nu}}}
        \sup_{\nu \in [0,\BoldTs]}  
            \LpLaw{2\sfp}{}{\Lparamu{\nu,\Xt{N,\nu}}}
        \\
        \leq&
        \Delta_g
        \frac{\expo{2 \lambda t}}{\absolute{2 \lambda -\Boldomega}}
        \frac{\expo{\Boldomega \BoldTs} - 1}{\Boldomega}
        \sup_{\nu \in [0,\tstar]}  
            \LpLaw{2\sfp}{}{\ZDiffNorm{\Zt{N,\nu}}}
        \sup_{\nu \in [0,\tstar]}  
            \LpLaw{2\sfp}{}{\Lparamu{\nu,\Xt{N,\nu}}}
        ,
    \end{align*}
    for all $(t,\sfp) \in [\BoldTs,\tstar] \times \mathbb{N}_{\geq 1}$.
    Therefore, 
    \begin{equation}\label{eqn:lem:Appendix:TrueProcess:XiTildeU:Bound:B:3:Final}
        \Boldomega
        \expo{-2\lambda t}
        \LpLaw{\sfp}{}{\widetilde{\mathfrak{N}}_3(t)}
        \leq 
        \widetilde{\Delta}_3^+(\sfp,\tstar,\Boldomega,\BoldTs)
        ,
        \quad 
        \forall (t,\sfp) \in [\BoldTs,\tstar] \times \mathbb{N}_{\geq 1},
    \end{equation}
    for all $(t,\sfp) \in [\BoldTs,\tstar] \times \mathbb{N}_{\geq 1}$.
    \begingroup
    \endgroup
    The desired bound in~\eqref{eqn:lem:Appendix:TrueProcess:XiTildeU:Bound:Final} is then established by substituting~\eqref{eqn:lem:Appendix:TrueProcess:XiTildeU:Bound:A:Final},~\eqref{eqn:lem:Appendix:TrueProcess:XiTildeU:Bound:B:1:Final},~\eqref{eqn:lem:Appendix:TrueProcess:XiTildeU:Bound:B:2:Final}, and~\eqref{eqn:lem:Appendix:TrueProcess:XiTildeU:Bound:B:3:Final} into~\eqref{eqn:lem:Appendix:TrueProcess:XiTildeU:Bound:Initial}. 


    \begingroup
    \endgroup
    
    \begingroup
    \endgroup

\end{proof}

We now further develop a few bounds that will eventually allow us to consolidate the results in Lemmas~\ref{lem:Appendix:TrueProcess:Xi},~\ref{lem:Appendix:TrueProcess:XiU}, and~\ref{lem:Appendix:TrueProcess:XiTildeU}. 
\begin{proposition}\label{prop:Appendix:TrueProcess:BoundsRoundB}
    Consider the stopping times $\tau^\star$ and $\tstar$, as defined in~\eqref{eqn:True:StoppingTimes}, Lemma~\ref{lem:True:dV}, and assume that $\tau^\star = \tstar$. 
    Define the following for $\sfp \in \cbr{1,\dots,\sfp^\star}$, where $\mathsf{p}^\star$ is defined in Assumption~\ref{assmp:NominalSystem:FiniteMomentsWasserstein}:
    \begin{equation}\label{eqn:prop:TrueProcess:Final:Bound:Condition}
        \LpTrueError
        \doteq
        \sup_{\nu \in [0,\tstar]}
        \LpLaw{2 \sfp}{}{\Xt{N,\nu} - \Xrt{N,\nu}}, 
        \quad 
        \LpTrue
        \doteq
        \sup_{\nu \in [0,\tstar]}
        \LpLaw{2 \sfp}{}{\Xt{N,\nu}},
        \quad 
        \LpTrueRef
        \doteq
        \sup_{\nu \in [0,\tstar]}
        \LpLaw{2 \sfp}{}{\Xrt{N,\nu}}.
    \end{equation} 
    If Assumptions~\ref{assmp:KnownFunctions} and~\ref{assmp:LipschitzContinuity} hold, then we have the following bounds for all $\sfp \in \cbr{1,\dots,\sfp^\star}$:
    \begin{subequations}\label{eqn:prop:Appendix:TrueProcess:BoundsRoundB:Del:Final}
        \begin{align}
            \Delta_{\mathcal{P}_1}(\sfp,\tstar)
            \leq  
            \widehat{\Delta}_1(\sfp)
            + 
            \widehat{\Delta}_2(\sfp)
            \sqrt{\LpTrueError}
            +
            \widehat{\Delta}_3
            \left( \LpTrue + \LpTrueRef \right)
            + 
            \widehat{\Delta}_4
            \LpTrueError
            ,
            \\
            \Delta_{\mathcal{P}_3}(\sfp,\tstar)
            \leq
            \widehat{\Delta}_5
            \sqrt{\LpTrueError}
            ,
        \end{align}
    \end{subequations}
    where $\Delta_{\mathcal{P}_1}(\sfp,\tstar)$ and $\Delta_{\mathcal{P}_3}(\sfp,\tstar)$  are defined in the statement of Proposition~\ref{prop:Appendix:TrueProcess:N:Bounds}, and the constants $\widehat{\Delta}_1(\sfp)$, $\widehat{\Delta}_2(\sfp)$, and $\widehat{\Delta}_{\cbr{3,\dots,6}}$ are defined in~\eqref{eqn:app:Constants:True:DelHat}.
    
\end{proposition}
\begin{proof}
    We begin with the term $\Delta_{\mathcal{P}_\mu}$ defined in Proposition~\ref{prop:Appendix:TrueProcess:Pparts:Bound} and the definitions $\fZ{\nu,\Zt{N,\nu}}= f\left(\nu,\Xt{N,\nu}\right) - f\left(\nu,\Xrt{N,\nu}\right)$ and $\LZmu{\nu,\Zt{N,\nu}}=\Lmu{\nu,\Xt{N,\nu}}- \Lmu{\nu,\Xrt{N,\nu}}$ from~\eqref{def:True:ErrorFunctions} to get  
    \begin{subequations}
        \begin{align}
            \Delta_{\mathcal{P}_\mu}(\sfp)
            =&
            \sup_{\nu \in [0,\tstar]}
            \left(
                \LpLaw{2\sfp}{}{\fZ{\nu,\Zt{N,\nu}}}
                +
                \LpLaw{2\sfp}{}{\LZmu{\nu,\Zt{N,\nu}}} 
            \right)
            \notag 
            \\
            =&
            \sup_{\nu \in [0,\tstar]}
            \left(
                \LpLaw{2\sfp}{}{f\left(\nu,\Xt{N,\nu}\right) - f\left(\nu,\Xrt{N,\nu}\right)}
                +
                \LpLaw{2\sfp}{}{\Lmu{\nu,\Xt{N,\nu}}- \Lmu{\nu,\Xrt{N,\nu}}} 
            \right)
            ,
            \label{eqn:prop:Appendix:TrueProcess:BoundsRoundB:Delpmu:GlobalLip:Initial}
            \\
            \leq&
            \sup_{\nu \in [0,\tstar]}
            \left(
                \LpLaw{2\sfp}{}{f\left(\nu,\Xt{N,\nu}\right)}
                +
                \LpLaw{2\sfp}{}{f\left(\nu,\Xrt{N,\nu}\right)}
                +
                \LpLaw{2\sfp}{}{\Lmu{\nu,\Xt{N,\nu}}- \Lmu{\nu,\Xrt{N,\nu}}} 
            \right)
            ,
            \label{eqn:prop:Appendix:TrueProcess:BoundsRoundB:Delpmu:Initial}
        \end{align}
    \end{subequations}
    where the second inequality is due to the Minkowski inequality. 
    Using Assumptions~\ref{assmp:KnownFunctions} and~\ref{assmp:LipschitzContinuity}, the bound in~\eqref{eqn:prop:Appendix:TrueProcess:BoundsRoundB:Delpmu:Initial} can be developed into
    \begin{align}\label{eqn:prop:Appendix:TrueProcess:BoundsRoundB:Delpmu}
        \Delta_{\mathcal{P}_\mu}(\sfp)
        \leq
        2 \Delta_f
        + 
        \Delta_f
        \left( \LpTrue + \LpTrueRef \right)
        +
        L_\mu
        \LpTrueError
        ,
    \end{align}
    where we have also used the subadditivity of the square root function.  
    If additionally Assumption~\ref{assmp:KnownFunctions:GlobalLip} holds, then we obtain a tighter bound   by developing~\eqref{eqn:prop:Appendix:TrueProcess:BoundsRoundB:Delpmu:GlobalLip:Initial} to get 
    \begin{align}\label{eqn:prop:Appendix:TrueProcess:BoundsRoundB:Delpmu:GlobalLip}
        \Delta_{\mathcal{P}_\mu}(\sfp)
        \leq 
        \left(L_f + L_\mu\right)
        \LpTrueError
        .
    \end{align}
    We thus combine the bounds in~\eqref{eqn:prop:Appendix:TrueProcess:BoundsRoundB:Delpmu} and~\eqref{eqn:prop:Appendix:TrueProcess:BoundsRoundB:Delpmu:GlobalLip} by writing 
    \begin{align}\label{eqn:prop:Appendix:TrueProcess:BoundsRoundB:Delpmu:Combined}
        \Delta_{\mathcal{P}_\mu}(\sfp)
        \leq&
        2\Delta_f\left(1-\Lip{f}\right)
        +
        \Delta_f\left(1-\Lip{f}\right)
        \left( \LpTrue + \LpTrueRef \right)
        + 
        \left(L_\mu + L_f \Lip{f}\right)
        \LpTrueError
        ,
    \end{align}
    where $\Lip{f}$ is defined in~\eqref{eqn:app:Constants:GlobalLipschitz}.

    Next, we have the following from Proposition~\ref{prop:Appendix:TrueProcess:Pparts:Bound}: 
    \begin{align*}
        \Delta_{\mathcal{P}_\sigma}(\sfp)
        =&
        \sup_{\nu \in [0,\tstar]} 
            \LpLaw{2\sfp}{}{\FZsigma{\nu,\Zt{N,\nu}}}
        \\
        =&
        \sup_{\nu \in [0,\tstar]}
             \LpLaw{2\sfp}{}{\Fsigma{\nu,\Xt{N,\nu}}-\Fsigma{\nu,\Xrt{N,\nu}}}
        \\
        \leq&
        \sup_{\nu \in [0,\tstar]}
        \left(
            \LpLaw{2\sfp}{}{p\left(\nu,\Xt{N,\nu}\right)-p\left(\nu,\Xrt{N,\nu}\right)}
            +
            \LpLaw{2\sfp}{}{\Lsigma{\nu,\Xt{N,\nu}}-\Lsigma{\nu,\Xrt{N,\nu}}}
        \right)
         \\
        \leq&
        \sup_{\nu \in [0,\tstar]}
        \left(
            \LpLaw{2\sfp}{}{p\left(\nu,\Xt{N,\nu}\right)}
            +
            \LpLaw{2\sfp}{}{p\left(\nu,\Xrt{N,\nu}\right)}
            +
            \LpLaw{2\sfp}{}{\Lsigma{\nu,\Xt{N,\nu}}-\Lsigma{\nu,\Xrt{N,\nu}}}
        \right)
        ,
    \end{align*}
    where we have used the definition $\FZsigma{\nu,\Zt{N,\nu}}=\Fsigma{\nu,\Xt{N,\nu}}-\Fsigma{\nu,\Xrt{N,\nu}}$ from~\eqref{def:True:ErrorFunctions}, and $F_\sigma=\bar{F}_\sigma+\Lambda_\sigma=p+\Lambda_\sigma$. 
    Then, as above, it follows from Assumptions~\ref{assmp:KnownFunctions} and~\ref{assmp:LipschitzContinuity} that 
    \begin{align}\label{eqn:prop:Appendix:TrueProcess:BoundsRoundB:Delpsigma}
        \Delta_{\mathcal{P}_\sigma}(\sfp)
        \leq
        \left(
            L_p 
            + 
            L_\sigma
        \right)
        \sqrt{\LpTrueError},     
    \end{align}
    where $\Lip{p}$ is defined in~\eqref{eqn:app:Constants:GlobalLipschitz} and we have used manipulations similar to~\eqref{eqn:lem:Appendix:ReferenceProcess:ComponentBounds:Manipulations}.
    Using~\eqref{eqn:prop:Appendix:TrueProcess:BoundsRoundB:Delpmu:Combined} and~\eqref{eqn:prop:Appendix:TrueProcess:BoundsRoundB:Delpsigma}, we can now bound the terms in~\eqref{prop:Appendix:TrueProcess:Pparts:Bound:Total:Final} as follows: 
    \begin{subequations}\label{eqn:prop:Appendix:TrueProcess:BoundsRoundC:SumDelp}
        \begin{align}
            \sup_{\nu \in [0,\tstar]} 
            \LpLaw{2\sfp}{}{\mathcal{P}_{\mu}(\nu)}
            \leq 
            4
            \Delta_{g}
            \Delta_f\left(1-\Lip{f}\right)
            +
            2
            \Delta_{g}
            \Delta_f
            \left(1-\Lip{f}\right)
            \left( \LpTrue + \LpTrueRef \right)
            \notag 
            \\
            + 
            2
            \left(
                \Delta_{g}
                \left(L_\mu + L_f \Lip{f}\right)
                +
                \Delta_{\dot{g}}
            \right)
            \LpTrueError
             ,
            \\
            \sup_{\nu \in [0,\tstar]} 
            \LpLaw{\sfq}{}{\mathcal{P}_{\sigma}(\nu)}
            \leq 
            2
            \Delta_{g}
            \left(
                L_p 
                + 
                L_\sigma
            \right)
            \sqrt{\LpTrueError}
            ,
        \end{align}
    \end{subequations}
    where have used the definition $\ZDiffNorm{\Zt{N,\nu}} = 2\left(\Xt{N,\nu} - \Xrt{N,\nu}\right)$ from the statement of Lemma~\ref{lem:True:dV}.
    Now, recall the following definitions of $\Delta_{\mathcal{P}_1}(\sfp,\tstar)$ and $\Delta_{\mathcal{P}_3}(\sfp,\tstar)$ from the statement of Proposition~\ref{prop:Appendix:TrueProcess:N:Bounds}:
    \begin{align*}
        \Delta_{\mathcal{P}_1}(\sfp,\tstar)
        = 
        \frac{ 1 }{ 2  \sqrt{\lambda} }
            \sup_{\nu \in [0,\tstar]}  
            \LpLaw{2\sfp}{}{\mathcal{P}_{\mu}(\nu)}
        + 
        \frac{\mathfrak{p}(\sfp)}{2 }
        \sup_{\nu \in [0,\tstar]}
            \LpLaw{2\sfp}{}{\mathcal{P}_{\sigma}(\nu)}
            ,
        \\
        \Delta_{\mathcal{P}_3}(\sfp,\tstar)
        = 
        \frac{\sqrt{m}}{2}  
        \sup_{\nu \in [0,\tstar]}
            \LpLaw{2\sfp}{}{\mathcal{P}_{\sigma}(\nu)}
        .
    \end{align*}
    Substituting~\eqref{eqn:prop:Appendix:TrueProcess:BoundsRoundC:SumDelp} into the above then leads to~\eqref{eqn:prop:Appendix:TrueProcess:BoundsRoundB:Del:Final}.

    \begingroup
    \endgroup

\end{proof}

The next result bounds the terms derived in Proposition~\ref{prop:Appendix:TrueProcess:ControlErrorIntegrand:Bounds}.
\begin{proposition}\label{prop:Appendix:TrueProcess:ControlErrorIntegrand:Evolved}
    Consider the stopping times $\tau^\star$ and $\tstar$, as defined in~\eqref{eqn:True:StoppingTimes}, Lemma~\ref{lem:True:dV}, and assume that $\tau^\star = \tstar$. 
    Define the following for $\sfp \in \cbr{1,\dots,\sfp^\star}$, where $\mathsf{p}^\star$ is defined in Assumption~\ref{assmp:NominalSystem:FiniteMomentsWasserstein}:
    \begin{equation}
        \LpTrue
        \doteq
        \sup_{\nu \in [0,\tstar]}
        \LpLaw{2 \sfp}{}{\Xt{N,\nu}}.
    \end{equation} 
    If Assumptions~\ref{assmp:KnownFunctions} and~\ref{assmp:UnknownFunctions} hold, then we have the following bounds for all $\sfp \in \cbr{1,\dots,\sfp^\star}$:
    \begin{multline}\label{eqn:prop:Appendix:TrueProcess:ControlErrorIntegrand:Bounds:Evolved}
        \Delta_{\widetilde{\mathcal{H}}_\mu}(\BoldTs,\sfp,\tstar)
        +
        \Delta_{\mathcal{G}_\mu}(\BoldTs,\tstar)
        + 
        \sqrt{\Boldomega}
        \mathfrak{p}'(\sfp)
        \left[
            \Delta_{\widetilde{\mathcal{H}}_\sigma}(\BoldTs,\sfp,\tstar)
            + 
            \Delta_{\widetilde{\mathcal{G}}_\sigma}(\BoldTs,\sfp,\tstar)
        \right]
        \\
        \leq
        \Delta_{\mu_1}\left(\sfp,\Boldomega,\BoldTs\right)
        + 
        \sqrt{\Boldomega}
        \mathfrak{p}'(\sfp)
        {\Delta}_{\sigma_1}\left(\sfp,\Boldomega,\BoldTs\right)
        + 
        \sqrt{\Boldomega}
        \mathfrak{p}'(\sfp)
        {\Delta}_{\sigma_2}\left(\sfp,\Boldomega,\BoldTs\right)
        \pwrfourth{\LpTrue}
        \\
        +
        \left(
            {\Delta}_{\mu_2}\left(\sfp,\Boldomega,\BoldTs\right)
            +   
            \sqrt{\Boldomega}
            \mathfrak{p}'(\sfp)
            {\Delta}_{\sigma_3}\left(\sfp,\Boldomega,\BoldTs\right) 
        \right)
        \sqrt{\LpTrue}
        + 
        {\Delta}_{\mu_3}\left(\sfp,\Boldomega,\BoldTs\right)
        \LpTrue,
    \end{multline}
    where $\Delta_{\widetilde{\mathcal{H}}_\mu}$, $\Delta_{\widetilde{\mathcal{H}}_\sigma}$, $\Delta_{\widetilde{\mathcal{G}}_\sigma}$, and $\Delta_{\mathcal{G}_\mu}$ are defined in Proposition~\ref{prop:Appendix:TrueProcess:ControlErrorIntegrand:Bounds}, and the constants $\Delta_{\mu_{\cbr{1,2,3}}}$ and $\Delta_{\sigma_{\cbr{1,2,3}}}$ are defined in~\eqref{eqn:app:Constants:True:DelMu} and~\eqref{eqn:app:Constants:True:DelSigma}, respectively.
\end{proposition}
\begin{proof}
    We begin by using $\Fbarmu{\nu,\cdot} =  f(\nu,\cdot)$ and $\Fsigma{\nu,\cdot} = p(\nu,\cdot) + \Lsigma{\nu,\cdot}$ from Definition~\ref{def:VectorFields}, along with Assumptions~\ref{assmp:KnownFunctions} and~\ref{assmp:UnknownFunctions} to arrive at the following bounds: 
    \begin{subequations}\label{eqn:lem:Appendix:TrueProcess:ComponentBounds:A}
        \begin{align}
            \sup_{\nu \in [0,\tstar]}
            \LpLaw{2\sfp}{}{
                \Fbarmu{\nu,\Xt{N,\nu}}
            } 
            \leq     
            \Delta_f 
            +
            \Delta_f 
            \LpTrue
            , 
            \quad
            \sup_{\nu \in [0,\tstar]}
            \LpLaw{2\sfp}{}{
                \Fsigma{\nu,\Xt{N,\nu}} 
            } 
            \leq     
            \Delta_p + \Delta_\sigma 
            + 
            \Delta_\sigma
            \sqrt{\LpTrue},
            \\
            \sup_{\nu \in [0,\tstar]}
            \LpLaw{2\sfp}{}{
                \Lmu{\nu,\Xt{N,\nu}} 
            } 
            \leq     
            \Delta_\mu 
            + 
            \Delta_\mu
            \LpTrue,   
        \end{align}
    \end{subequations}
    where we have further used the subadditivity of the square root function and performed manipulations similar to~\eqref{eqn:lem:Appendix:ReferenceProcess:ComponentBounds:Manipulations}. 
    Now, recall the following from Proposition~\ref{prop:Appendix:TrueProcess:TsBound} for $\tau^\star = \tstar$:
    \begin{align*}
        \widetilde{\delta}_1(\BoldTs,\sfp,\tstar)
        =
        2
        \sup_{\nu \in [0,\tstar]}
        \LpLaw{2\sfp}{}{\Fbarmu{\nu,\Xt{N,\nu}}}
        + 
        2
        \left(
            1
            + 
            \Delta_g 
            \Delta_\Theta
            e^{-\lambda_s \BoldTs}
        \right)
        \sup_{\nu \in [0,\tstar]}
        \LpLaw{2\sfp}{}{ 
            \Lmu{\nu,\Xt{N,\nu}}
        }
        \\
        +
        2 
        \mathfrak{p}'(\sfp)
        \Delta_g 
        \Delta_\Theta
        e^{-\lambda_s \BoldTs}
        \gamma_1\left(\BoldTs\right)
        \sup_{\nu \in [0,\tstar]}
        \LpLaw{2\sfp}{}{
            \Fsigma{\nu,\Xt{N,\nu}}
        } 
        ,
        \\
        \widetilde{\delta}_2(\sfp,\tstar)
        =
        2
        \sqrt{2}
        \mathfrak{p}'(\sfp)  
        \sup_{\nu \in [0,\tstar]}
        \LpLaw{2\sfp}{}{\Fsigma{\nu,\Xt{N,\nu}}}
        .
    \end{align*}
    Substituting~\eqref{eqn:lem:Appendix:TrueProcess:ComponentBounds:A} into the above leads to 
    \begin{align*}
        \begin{multlined}[b][0.9\linewidth]
            \widetilde{\delta}_1(\BoldTs,\sfp,\tstar)
            \leq
            2\Delta_f
            + 
            2
            \left(
                1
                + 
                \Delta_g 
                \Delta_\Theta
                e^{-\lambda_s \BoldTs}
            \right)
            \Delta_\mu 
            +
            2 
            \mathfrak{p}'(\sfp)
            \Delta_g 
            \Delta_\Theta
            e^{-\lambda_s \BoldTs}
            \gamma_1\left(\BoldTs\right)
            \left(\Delta_p + \Delta_\sigma\right) 
            \\
            +
            2 
            \mathfrak{p}'(\sfp)
            \Delta_g 
            \Delta_\Theta
            e^{-\lambda_s \BoldTs}
            \gamma_1\left(\BoldTs\right)
                \Delta_\sigma
                \sqrt{\LpTrue}
            \\
            +
            2
            \left(
                \Delta_f  
                + 
                \left(
                    1
                    + 
                    \Delta_g 
                    \Delta_\Theta
                    e^{-\lambda_s \BoldTs}
                \right)
                \Delta_\mu    
            \right)    
            \LpTrue
            ,
        \end{multlined}
        \\
        \widetilde{\delta}_2(\sfp,\tstar)
        \leq
        2
        \sqrt{2}
        \mathfrak{p}'(\sfp)  
        \left(\Delta_p + \Delta_\sigma\right) 
        + 
        2
        \sqrt{2}
        \mathfrak{p}'(\sfp)  
        \Delta_\sigma
        \sqrt{\LpTrue}
        .
    \end{align*}
    We further substitute these bounds into $\Delta_{\widetilde{\mathcal{H}}_\mu}$, $\Delta_{\widetilde{\mathcal{H}}_\sigma}$, $\Delta_{\widetilde{\mathcal{G}}_\sigma}$, and $\Delta_{\mathcal{G}_\mu}$ that are defined in Proposition~\ref{prop:Appendix:TrueProcess:ControlErrorIntegrand:Bounds} to obtain:
    \begin{subequations}\label{eqn:lem:Appendix:TrueProcess:ComponentBounds:TildeHG}
        \begin{align}
            &
            \begin{multlined}[b][0.9\linewidth]
                \Delta_{\widetilde{\mathcal{H}}_\mu}(\BoldTs,\sfp,\tstar)
                \leq 
                \BoldTs
                \widehat{\gamma}_\mu\left(\Boldomega,\BoldTs\right)
                +
                \bar{\Delta}_1\left(\BoldTs,\sfp\right)
                \gamma_\mu\left(\Boldomega,\BoldTs\right)
                +
                \bar{\Delta}_2\left(\BoldTs,\sfp\right)
                \gamma_\mu\left(\Boldomega,\BoldTs\right)
                \sqrt{\LpTrue} 
                \\
                + 
                \gamma_\mu\left(\Boldomega,\BoldTs\right)
                \bar{\Delta}_3\left(\BoldTs\right)
                \LpTrue
                ,
            \end{multlined}
            \\
            &
            \begin{multlined}[b][0.9\linewidth]
                \Delta_{\widetilde{\mathcal{H}}_\sigma}(\BoldTs,\sfp,\tstar)
                \leq 
                \gamma'_{\sigma}\left(\Boldomega,\BoldTs\right)
                +
                \gamma_{\sigma}\left(\Boldomega,\BoldTs\right)
                \left(\bar{\Delta}_1\left(\BoldTs,\sfp\right)\right)^\frac{1}{2}
                +  
                \BoldTs
                \widehat{\gamma}_{\sigma}\left(\Boldomega,\BoldTs\right) 
                \\
                + 
                \gamma_{\sigma}\left(\Boldomega,\BoldTs\right)
                \left( \bar{\Delta}_2\left(\BoldTs,\sfp\right) \right)^\frac{1}{2} 
                \pwrfourth{\LpTrue}
                +   
                \gamma_{\sigma}\left(\Boldomega,\BoldTs\right)
                (\bar{\Delta}_3\left(\BoldTs\right))^\frac{1}{2} 
                \sqrt{\LpTrue}
            \end{multlined}
            ,
            \\
            &
            \begin{multlined}[b][0.9\linewidth]
                \Delta_{\widetilde{\mathcal{G}}_\sigma}(\BoldTs,\sfp,\tstar)
                \leq 
                \left(1 + \gamma_2(\Boldomega,\BoldTs)\right)
                \left[
                    \left(
                        L^{\paral}_p
                        + 
                        L^{\paral}_\sigma
                    \right)
                    \left(\bar{\Delta}_1\left(\BoldTs,\sfp\right)\right)^\frac{1}{2} 
                    +  
                    \BoldTs
                    \left( 
                        \hat{L}^{\paral}_p 
                        + 
                        \hat{L}^{\paral}_\sigma
                    \right)
                \right]
                \\
                + 
                \left(1 + \gamma_2(\Boldomega,\BoldTs)\right)
                \left(
                    L^{\paral}_p
                    + 
                    L^{\paral}_\sigma
                \right)
                \left( \bar{\Delta}_2\left(\BoldTs,\sfp\right) \right)^\frac{1}{2} 
                \pwrfourth{\LpTrue}
                \\
                +  
                \left(1 + \gamma_2(\Boldomega,\BoldTs)\right)
                \left(
                    L^{\paral}_p
                    + 
                    L^{\paral}_\sigma
                \right) 
                \left(\bar{\Delta}_3\left(\BoldTs\right)\right)^\frac{1}{2} \sqrt{\LpTrue}
                ,
            \end{multlined}
            \\
            &\Delta_{\mathcal{G}_\mu}(\BoldTs,\tstar)
            \leq 
            \Delta^{\paral}_\mu
            \left(1 - \expo{-\lambda_s \BoldTs}\right) 
            + 
            \Delta^{\paral}_\mu
            \left(1 - \expo{-\lambda_s \BoldTs}\right)
            \LpTrue,
        \end{align}
    \end{subequations} 
    where the constants $\bar{\Delta}_{\cbr{1,2,3}}$ are defined in~\eqref{eqn:app:Constants:True:DelBar} and we have used the following due to Assumption~\ref{assmp:UnknownFunctions} and the subadditivity of the square root function: 
    \begin{align}\label{eqn:lem:Appendix:TrueProcess:ComponentBounds:B}
        \sup_{\nu \in [0,\tstar \wedge \BoldTs]}
        \LpLaw{2\sfp}{}{
            \Lparamu{\nu,\Xt{N,\nu}} 
        }
        \leq 
        \sup_{\nu \in [0,\tstar]}
        \LpLaw{2\sfp}{}{
            \Lparamu{\nu,\Xt{N,\nu}} 
        } 
        \leq     
        \Delta^{\paral}_\mu 
        + 
        \Delta^{\paral}_\mu
        \LpTrue.   
    \end{align}

    Using all of the derived bounds above, we obtain the following:
    \begin{align*}
        \Delta_{\widetilde{\mathcal{H}}_\mu}(\BoldTs,\sfp,\tstar)
        +
        \Delta_{\mathcal{G}_\mu}(\BoldTs,\tstar)
        \leq
        {\Delta}_{\mu_1}\left(\sfp,\Boldomega,\BoldTs\right)
        +
        {\Delta}_{\mu_2}\left(\sfp,\Boldomega,\BoldTs\right)
        \sqrt{\LpTrue} 
        + 
         {\Delta}_{\mu_3}\left(\sfp,\Boldomega,\BoldTs\right)
        \LpTrue
        ,        
    \end{align*}
    and 
    \begin{align*}
        \Delta_{\widetilde{\mathcal{H}}_\sigma}(\BoldTs,\sfp,\tstar)
        + 
        \Delta_{\widetilde{\mathcal{G}}_\sigma}(\BoldTs,\sfp,\tstar)
        \leq
        {\Delta}_{\sigma_1}\left(\sfp,\Boldomega,\BoldTs\right)
        + 
        {\Delta}_{\sigma_2}\left(\sfp,\Boldomega,\BoldTs\right)
        \pwrfourth{\LpTrue}
        +   
        {\Delta}_{\sigma_3}\left(\sfp,\Boldomega,\BoldTs\right)
        \sqrt{\LpTrue}
        .
    \end{align*}
    The expression in~\eqref{eqn:prop:Appendix:TrueProcess:ControlErrorIntegrand:Bounds:Evolved} then follows in a straightforward manner.

\end{proof}

We conclude the appendix by collecting and developing the bounds from Lemmas~\ref{lem:Appendix:TrueProcess:Xi},~\ref{lem:Appendix:TrueProcess:XiU}, and~\ref{lem:Appendix:TrueProcess:XiTildeU}.
\begin{lemma}\label{lem:Appendix:TrueProcess:Final:Bound}
    Consider the stopping times $\tau^\star$ and $\tstar$, as defined in~\eqref{eqn:True:StoppingTimes}, and assume that $\tau^\star = \tstar$.
    As in~\eqref{eqn:prop:TrueProcess:Final:Bound:Condition}, define the following for $\sfp \in \cbr{1,\dots,\sfp^\star}$, where $\mathsf{p}^\star$ is defined in Assumption~\ref{assmp:NominalSystem:FiniteMomentsWasserstein}:
    \begin{equation}\label{eqn:lemma:TrueProcess:Final:Bound:Condition}
        \LpTrueError
        \doteq
        \sup_{\nu \in [0,\tstar]}
        \LpLaw{2 \sfp}{}{\Xt{N,\nu} - \Xrt{N,\nu}}, 
        \quad 
        \LpTrue
        \doteq
        \sup_{\nu \in [0,\tstar]}
        \LpLaw{2 \sfp}{}{\Xt{N,\nu}},
        \quad 
        \LpTrueRef
        \doteq
        \sup_{\nu \in [0,\tstar]}
        \LpLaw{2 \sfp}{}{\Xrt{N,\nu}}.
    \end{equation} 
    If Assumptions~\ref{assmp:KnownFunctions},~\ref{assmp:UnknownFunctions},~\ref{assmp:knownDiffusion:Decomposition} and~\ref{assmp:LipschitzContinuity} hold, then we have the following for all $(t,\sfp) \in \mathbb{R}_{\geq 0} \times \cbr{2,\dots,\sfp^\star}$:
    \begin{align}\label{eqn:lem:Appendix:TrueProcess:Final:Bound:B:Final}
        &
        \LpLaw{\sfp}{}{
            \expo{-2\lambda \tau(t)}
            \Xi\left(\tau(t),\Zt{N} \right)
        }
        +
        \LpLaw{\sfp}{}{
            \expo{-(2\lambda+\Boldomega) \tau(t)}
            \Xi_{\mathcal{U}}\left(\tau(t),\Zt{N} \right)
        }
        +
        \LpLaw{\sfp}{}{
            \expo{-2\lambda \tau(t)}
            \widetilde{\Xi}_{\mathcal{U}}\left(\tau(t),\Zt{N} \right)
        }
        \notag 
        \\
        &
        \leq
        \Delta_{\circledcirc}(\sfp,\Boldomega)
        \left(\LpTrueError\right)^\frac{1}{2}
        +
        \Delta_{\odot}(\sfp,\Boldomega)
        \LpTrueError
        +
        \left(
            \widebreve{\Delta}_{\otimes}(\sfp,\Boldomega)
            \left( \LpTrue + \LpTrueRef \right)
            +
            \Delta_{\otimes}(\sfp,\Boldomega)
            \LpTrueError
        \right)
        \left(\LpTrueError\right)^\frac{1}{2}
        \notag 
        \\
        & \qquad 
        +
        \left(
            \widebreve{\Delta}_\circledast(\Boldomega)
            \left( \LpTrue + \LpTrueRef \right)
            +
            \Delta_\circledast(\Boldomega)
            \LpTrueError
        \right)
        \LpTrueError
        +
        \sum_{i=1}^3
        \Upsilon_{a_i}\left(\LpTrue,\LpTrueError; \sfp,\Boldomega,\BoldTs\right)
        ,
    \end{align}
    where 
    \begin{align*}
        \Delta_{\circledcirc}(\sfp,\Boldomega)
        =
        \frac{ 1 }{  \absolute{2\lambda - \Boldomega}}
            \sqrt{\Boldomega}
            \Delta_{\circledcirc_1}(\sfp)
        ,
        \\
        \Delta_{\odot}(\sfp,\Boldomega)
        =
        \frac{\Delta_{\odot_1}}{2 \lambda}  
        +  
        \frac{ 1 }{  \absolute{2\lambda - \Boldomega}}
        \left(
            \Delta_{\odot_2}(\sfp) 
            + 
            \sqrt{\Boldomega}
            \Delta_{\odot_3}(\sfp)
            +
            \Boldomega
            \Delta_{\odot_4}(\sfp)
        \right),
        \\
        \widebreve{\Delta}_{\otimes}(\sfp,\Boldomega)
        =
        \sqrt{\Boldomega} 
        \frac{ \Delta_{\otimes_3}(\sfp) }{  \absolute{2\lambda - \Boldomega}}
        , 
        \quad
        \Delta_{\otimes}(\sfp,\Boldomega)
        =
        \frac{\Delta_{\otimes_1}(\sfp)}{2\sqrt{\lambda}}
        +
        \frac{ 1 }{  \absolute{2\lambda - \Boldomega}}
        \left(
            \Delta_{\otimes_2}(\sfp)
            +
            \sqrt{\Boldomega} 
            \Delta_{\otimes_4}(\sfp)
        \right),
        \\
        \widebreve{\Delta}_\circledast(\Boldomega)
        = 
        \frac{ \Delta_{\circledast_2} }{  \absolute{2\lambda - \Boldomega}},
        \quad 
        \Delta_\circledast(\Boldomega)
        =
        \left(
            \frac{ \Delta_{\circledast_1} }{2\lambda}  
            + 
            \frac{ \Delta_{\circledast_3} }{  \absolute{2\lambda - \Boldomega}}
        \right),
    \end{align*}
    and where the constants $\Delta_i(\sfp)$, $i \in \cbr{ \circledcirc_1, \odot_1,\dots,\odot_4, \otimes_1,\dots,\otimes_4, \circledast_1,\dots, \circledast_3}$, are defined in~\eqref{eqn:app:Constants:True:DelCircledCirc}~-~\eqref{eqn:app:Constants:True:DelCircledAst} in Section~\ref{subsec:app:Definitions:True}, and $\Upsilon_{a_1 \dots a_3}$ are defined in~\eqref{eqn:Definitions:Total:SamplingPeriodCondition:MasterUpsilons} in Section~\ref{subsubsec:Design:SamplingPeriod}.

    Lastly, the right hand side of~\eqref{eqn:lem:Appendix:TrueProcess:Final:Bound:B:Final}, evaluated at $\sfp = 1$, is an upper bound for the following term, for all $t \in \mathbb{R}_{\geq 0}$:
    \begin{align*}
         \ELaw{}{
            e^{-2\lambda \tau(t)}  
            \Xi\left(\tau(t),\Zt{N} \right)
        }
        +
        \ELaw{}{
            \expo{-(2\lambda+\Boldomega) \tau(t)}
            \Xi_{\mathcal{U}}\left(\tau(t),\Zt{N} \right)
        }
        +
        \ELaw{}{
            e^{-2\lambda \tau(t)}
            \widetilde{\Xi}_{\mathcal{U}}\left(\tau(t),\Zt{N} \right)
        }.
    \end{align*}
\end{lemma}
\begin{proof}
    We begin by combining the results from Lemmas~\ref{lem:Appendix:TrueProcess:Xi},~\ref{lem:Appendix:TrueProcess:XiU}, and~\ref{lem:Appendix:TrueProcess:XiTildeU} to obtain he following bounds that hold for all $t \in [0,\tstar]$: 
    \begin{subequations}\label{eqn:lem:Appendix:TrueProcess:Final:Bounds:Initial}
        \begin{align}
            &
            \begin{aligned}[b]
                &\ELaw{}{
                    e^{-2\lambda \tau(t)}  
                    \Xi\left(\tau(t),\Zt{N} \right)
                }
                +
                \ELaw{}{
                    \expo{-(2\lambda+\Boldomega) \tau(t)}
                    \Xi_{\mathcal{U}}\left(\tau(t),\Zt{N} \right)
                }
                +
                \ELaw{}{
                    e^{-2\lambda \tau(t)}
                    \widetilde{\Xi}_{\mathcal{U}}\left(\tau(t),\Zt{N} \right)
                }
                \\
                &\leq  
                \frac{\Delta_{\Xi_1}(1,\tstar)}{2\lambda}
                 +
                \widetilde{\Delta}_{\mathcal{U}}(1,\tstar,\Boldomega,\BoldTs)
                \\
                &
                \quad +
                \frac{ 1 }{  \absolute{2\lambda - \Boldomega}}  
                \left(
                    \Delta_{\mathcal{U}_1}(1,\tstar)
                    +
                    \sqrt{\Boldomega}
                    \Delta_{\mathcal{U}_2}(1,\tstar)
                    +
                    \Boldomega
                    \Delta_{\mathcal{U}_3}(1,\tstar)
                    +
                    \frac{\Boldomega}{\lambda }
                    \Delta_{\mathcal{U}}(1,\tstar,\Boldomega,\BoldTs)
                \right)
                ,
            \end{aligned}
            \label{eqn:lem:Appendix:TrueProcess:Final:Bound:2}
            \\
            &
            \begin{aligned}[b]
                &\LpLaw{\sfp}{}{
                    \expo{-2\lambda \tau(t)}
                    \Xi\left(\tau(t),\Zt{N} \right)
                }
                +
                \LpLaw{\sfp}{}{
                    \expo{-(2\lambda+\Boldomega) \tau(t)}
                    \Xi_{\mathcal{U}}\left(\tau(t),\Zt{N} \right)
                }
                +
                \LpLaw{\sfp}{}{
                    \expo{-2\lambda \tau(t)}
                    \widetilde{\Xi}_{\mathcal{U}}\left(\tau(t),\Zt{N} \right)
                }
                \\
                & \leq
                \frac{\Delta_{\Xi_1}(\sfp,\tstar)}{2\lambda}  
                +
                \frac{\Delta_{\Xi_2}(\sfp,\tstar)}{2\sqrt{\lambda}}
                +
                \widetilde{\Delta}_{\mathcal{U}}(\sfp,\tstar,\Boldomega,\BoldTs)
                \\
                & \quad 
                +
                \frac{ 1 }{  \absolute{2\lambda - \Boldomega}}  
                \left(
                    \Delta_{\mathcal{U}_1}(\sfp,\tstar)
                    +
                    \sqrt{\Boldomega}
                    \Delta_{\mathcal{U}_2}(\sfp,\tstar)
                    +
                    \Boldomega
                    \Delta_{\mathcal{U}_3}(\sfp,\tstar)
                    +
                    \frac{\Boldomega}{\lambda }
                    \Delta_{\mathcal{U}}(\sfp,\tstar,\Boldomega,\BoldTs)
                \right)
                , 
                \quad \forall \sfp \in \mathbb{N}_{\geq 2}.
            \end{aligned}
            \label{eqn:lem:Appendix:TrueProcess:Final:Bound:1}
        \end{align}
    \end{subequations}
    We now obtain the expressions for each of the bounds.

    \ul{\emph{Bounds for $\Delta_{\Xi_1}$ and $\Delta_{\Xi_2}$:}} 
    We have from Definitions~\ref{def:VectorFields} and~\ref{def:Diffusion:Decomposed} that $\Fsigma{\nu,\cdot} = p(\nu,\cdot) + \Lsigma{\nu,\cdot}$ and $\Fperpsigma{\nu,\cdot} = \pperp{\nu,\cdot} + \Lperpsigma{\nu,\cdot}$, respectively.
    Furthermore, $\ZDiffNorm{\Zt{N,\nu}} = 2\left(\Xt{N,\nu} - \Xrt{N,\nu}\right)$ as defined in the statement of Lemma~\ref{lem:True:dV}.
    Hence, using Assumptions~\ref{assmp:KnownFunctions},~\ref{assmp:knownDiffusion:Decomposition}, and~\ref{assmp:LipschitzContinuity}, we obtain the following bounds: 
    \begin{subequations}\label{eqn:lem:Appendix:TrueProcess:ComponentBounds:C}
        \begin{align}
            &
            \sup_{\nu \in [0,\tstar \wedge \BoldTs]}
            \LpLaw{2\sfp}{}{
                \ZDiffNorm{\Zt{N,\nu}}
            }
            \leq
            \sup_{\nu \in [0,\tstar]}
            \LpLaw{2\sfp}{}{
                \ZDiffNorm{\Zt{N,\nu}}
            } 
            \leq     
            2 \LpTrueError, 
            \label{eqn:lem:Appendix:TrueProcess:ComponentBounds:dV}
            \\
            &
            \sup_{\nu \in [0,\tstar]}
                \LpLaw{2\sfp}{}{\LZperpmu{\nu,\Zt{N,\nu}}}
            =
            \sup_{\nu \in [0,\tstar]}
            \LpLaw{2\sfp}{}{
                \Lperpmu{\nu,\Xt{N,\nu}}
                - 
                \Lperpmu{\nu,\Xrt{N,\nu}}
            }
            \leq     
            L_\mu^\perp   
            \LpTrueError,
            \label{eqn:lem:Appendix:TrueProcess:ComponentBounds:LmuCircPerp}
            \\
            &
            \begin{aligned}[b]
                &\sup_{\nu \in [0,\tstar]}
                \LpLaw{2\sfp}{}{
                    \FZsigma{\nu,\Zt{N,\nu}}
                }
                \\ 
                & \hspace{1cm}
                = 
                \sup_{\nu \in [0,\tstar]}
                \LpLaw{2\sfp}{}{
                    \Fsigma{\nu,\Xt{N,\nu}}
                    - 
                    \Fsigma{\nu,\Xrt{N,\nu}}
                }  
                \\
                & \hspace{1cm}
                \leq
                \sup_{\nu \in [0,\tstar]}
                \left(
                    \LpLaw{2\sfp}{}{
                        p\left(\nu,\Xt{N,\nu}\right)
                        - 
                        p\left(\nu,\Xrt{N,\nu}\right)
                    }
                    +
                    \LpLaw{2\sfp}{}{
                        \Lsigma{\nu,\Xt{N,\nu}}
                        - 
                        \Lsigma{\nu,\Xrt{N,\nu}}
                    } 
                \right)
                \\
                & \hspace{1cm}
                \leq    
                \left( 
                    L_p  + L_\sigma 
                \right)
                \sqrt{\LpTrueError},
            \end{aligned}
            \label{eqn:lem:Appendix:TrueProcess:ComponentBounds:FsigmaCirc}
        \end{align} 
    \end{subequations}
    where we have also used the definitions in~\eqref{def:True:ErrorFunctions}.
    We similarly obtain 
    \begin{align}\label{eqn:lem:Appendix:TrueProcess:ComponentBounds:FsigmaCircPerp}
        \sup_{\nu \in [0,\tstar]}
        \LpLaw{2\sfp}{}{
            \FZperpsigma{\nu,\Zt{N,\nu}}
        }
        \leq    
        \left( 
            L_p^{\perp}  + L_\sigma^{\perp} 
        \right)
        \sqrt{\LpTrueError}.
    \end{align}
    Substituting the bounds above into the following from Lemma~\ref{lem:Appendix:TrueProcess:Xi}:
    \begin{align*}
        \Delta_{\Xi_1}(\sfp,\tstar) 
        =  
        \sup_{\nu \in [0,\tstar]}
        \left(
            \Delta_g^\perp
            \LpLaw{2\sfp}{}{\ZDiffNorm{\Zt{N,\nu}}}
            \LpLaw{2\sfp}{}{\LZperpmu{\nu,\Zt{N,\nu}}}
            + 
            \left(\LpLaw{2\sfp}{}{\FZsigma{\nu,\Zt{N,\nu}}}\right)^2
        \right),
        \\
        \Delta_{\Xi_2}(\sfp,\tstar) 
        =
        \Delta_g^\perp
        \mathfrak{p}(\sfp)
        \sup_{\nu \in [0,\tstar]}
        \LpLaw{2\sfp}{}{
            \ZDiffNorm{\Zt{N,\nu}}  
        }
        \LpLaw{2\sfp}{}{
                \FZperpsigma{\nu,\Zt{N,\nu}}
        }
        ,
    \end{align*}
    leads to the following for $\sfp \in \cbr{1,\dots,\sfp^\star}$: 
    \begin{align}\label{eqn:lem:Appendix:TrueProcess:Final:Bound:DelXi}
        \Delta_{\Xi_1}(\sfp,\tstar) 
        \leq
        \Delta_{\odot_1}
        \LpTrueError
        +
        \Delta_{\circledast_1}
        \LpTrueError^2  
        , 
        \quad
        \Delta_{\Xi_2}(\sfp,\tstar) 
        \leq
        \Delta_{\otimes_1}(\sfp)
        \left(\LpTrueError\right)^\frac{3}{2}
        .
    \end{align}

    \ul{\emph{Bounds for $\Delta_{\mathcal{U}_1}$, $\Delta_{\mathcal{U}_2}$, $\Delta_{\mathcal{U}_3}$:}} 
    Recall the following from Lemma~\ref{lem:Appendix:TrueProcess:XiU}:
    \begin{align*}
        \begin{aligned}
            \Delta_{\mathcal{U}_1}(\sfp,\tstar)
            =&
            \sqrt{\lambda}
            \Delta_g
            \mathfrak{p}(\sfp)  
            \sup_{\nu \in [0,\tstar]}
            \LpLaw{2\sfp}{}{
                \RefDiffNorm{\Zt{N,\nu}}
            }
            \sup_{\nu \in [0,\tstar]}
                \LpLaw{2\sfp}{}{
                    \FZparasigma{\nu,\Zt{N,\nu}}
            }
            \\
            &+
            \left(
                    \frac{\Delta_{\mathcal{P}_1}(\sfp,\tstar)}{\sqrt{\lambda}} 
                    +
                    2
                    \Delta_g 
                    \sup_{\nu \in [0,\tstar]}\LpLaw{2\sfp}{}{\RefDiffNorm{\Zt{N,\nu}}}
            \right.
            \\
            & 
            \hspace{5cm}
            \left.
                +
                \frac{\Delta_g^2}{\lambda} 
                \sup_{\nu \in [0,\tstar]}\LpLaw{2\sfp}{}{\LZparamu{\nu,\Zt{N,\nu}}}
            \right)
            \sup_{\nu \in [0,\tstar]}\LpLaw{2\sfp}{}{\LZparamu{\nu, \Zt{N,\nu}}}
            ,
        \end{aligned}
        \\
        \begin{multlined}[b][0.95\linewidth]
            \Delta_{\mathcal{U}_2}(\sfp,\tstar)
            =
            \mathfrak{p}'(\sfp)
            \left(
                \frac{  \Delta_{\mathcal{P}_1}(\sfp,\tstar) }{\sqrt{\lambda }}
                + 
                \Delta_g 
                \sup_{\nu \in [0,\tstar]}
                    \LpLaw{2\sfp}{}{\RefDiffNorm{\Zt{N,\nu}}}
                + 
                2
                \frac{\Delta_g^2}{\lambda}  
                \sup_{\nu \in [0,\tstar]}
                    \LpLaw{2\sfp}{}{\LZparamu{\nu,\Zt{N,\nu}}}
            \right)
            \\
            \times
            \sup_{\nu \in [0,\tstar]}\LpLaw{2\sfp}{}{\FZparasigma{\nu, \Zt{N,\nu}}}
            ,
        \end{multlined}
        \\
        \Delta_{\mathcal{U}_3}(\sfp,\tstar)
        =
        \left(
            \frac{ \Delta_{\mathcal{P}_3}(\sfp,\tstar) }{\lambda}
            + 
            \frac{\Delta_g^2}{\lambda}
            \mathfrak{p}'(\sfp)^2  
            \sup_{\nu \in [0,\tstar]}\LpLaw{2\sfp}{}{\FZparasigma{\nu,\Zt{N,\nu}}}
        \right)
        \sup_{\nu \in [0,\tstar]}\LpLaw{2\sfp}{}{\FZparasigma{\nu, \Zt{N,\nu}}}
        ,
    \end{align*}
    where we have further used the definition $\Delta_{\mathcal{P}_2}(\sfp,\tstar) = \mathfrak{p}'(\sfp) \Delta_{\mathcal{P}_1}(\sfp,\tstar)$ from Proposition~\ref{prop:Appendix:TrueProcess:N:Bounds}.
    To bound these terms, in addition to the results in Proposition~\ref{prop:Appendix:TrueProcess:BoundsRoundB}, we also use~\eqref{eqn:lem:Appendix:TrueProcess:ComponentBounds:dV}, and the following bounds that can be derived similarly to~\eqref{eqn:lem:Appendix:TrueProcess:ComponentBounds:LmuCircPerp} and~\eqref{eqn:lem:Appendix:TrueProcess:ComponentBounds:FsigmaCircPerp}:
    \begin{subequations}\label{eqn:lem:Appendix:TrueProcess:ComponentBounds:CircPara}
        \begin{align}
            \sup_{\nu \in [0,\tstar]}
                \LpLaw{2\sfp}{}{\LZparamu{\nu,\Zt{N,\nu}}}
            =
            \sup_{\nu \in [0,\tstar]}
            \LpLaw{2\sfp}{}{
                \Lparamu{\nu,\Xt{N,\nu}}
                - 
                \Lparamu{\nu,\Xrt{N,\nu}}
            }
            \leq     
            L_\mu^{\paral} \LpTrueError,
            \label{eqn:lem:Appendix:TrueProcess:ComponentBounds:LmuCircPara}
            \\
            \sup_{\nu \in [0,\tstar]}
            \LpLaw{2\sfp}{}{
                \FZparasigma{\nu,\Zt{N,\nu}}
            }
            \leq    
            \left( 
                L_p^{\paral}  + L_\sigma^{\paral} 
            \right)
            \sqrt{\LpTrueError}.
            \label{eqn:lem:Appendix:TrueProcess:ComponentBounds:FsigmaCircPara}
        \end{align}
    \end{subequations}
    Using the above then leads to the following for $\sfp \in \cbr{1,\dots,\sfp^\star}$:
    \begin{subequations}\label{eqn:lem:Appendix:TrueProcess:Final:Bound:NumberedDelU}
        \begin{align}
            \Delta_{\mathcal{U}_1}(\sfp,\tstar)
            \leq  
            \Delta_{\odot_2}(\sfp) 
            \LpTrueError
            +
            \Delta_{\otimes_2}(\sfp)
            \left(\LpTrueError\right)^\frac{3}{2}
            +
            \left(
                \Delta_{\circledast_2}
                \left( \LpTrue + \LpTrueRef \right)
                + 
                \Delta_{\circledast_3}
                \LpTrueError
            \right)
            \LpTrueError
            ,
            \\
            \begin{aligned}[b]
                \Delta_{\mathcal{U}_2}(\sfp,\tstar)
                \leq&
                \Delta_{\circledcirc_1}(\sfp)
                \sqrt{\LpTrueError}
                + 
                \Delta_{\odot_3}(\sfp)
                \LpTrueError
                \\
                &
                +
                \left(
                    \Delta_{\otimes_3}(\sfp) 
                    \left( \LpTrue + \LpTrueRef \right)
                    + 
                    \Delta_{\otimes_4}(\sfp)
                    \LpTrueError
                \right)
                \sqrt{\LpTrueError}
                ,
            \end{aligned}
            \\
            \Delta_{\mathcal{U}_3}(\sfp,\tstar)
            \leq 
            \Delta_{\odot_4}(\sfp)
            \LpTrueError
            .
        \end{align}
    \end{subequations}

    \ul{\emph{Bounds for $\widetilde{\Delta}_{\mathcal{U}}$:}} 
    Now, we have from Definitions~\ref{def:VectorFields} and~\ref{def:Diffusion:Decomposed} that $\Fsigma{\nu,\cdot} = p(\nu,\cdot) + \Lsigma{\nu,\cdot}$ and $\Fparasigma{\nu,\cdot} = \ppara{\nu,\cdot} + \Lparasigma{\nu,\cdot}$, respectively.  
    Thus, we obtain the following bound due to Assmptions~\ref{assmp:KnownFunctions} and~\ref{assmp:UnknownFunctions}:
    \begin{align}\label{eqn:lem:Appendix:TrueProcess:ComponentBounds:D}
        \sup_{\nu \in \sbr{0,\tstar \wedge \BoldTs} } 
        \LpLaw{2\sfp}{}{\Fparasigma{\nu,\Xt{N,\nu}}}
        \leq
        \sup_{\nu \in [0,\tstar]}
        \LpLaw{2\sfp}{}{
            \Fparasigma{\nu,\Xt{N,\nu}}
        } 
        \leq&      
        \sup_{\nu \in [0,\tstar]}
        \LpLaw{4\sfp}{}{
            \Fparasigma{\nu,\Xt{N,\nu}}
        } 
        \notag 
        \\
        \leq& 
        \Delta^{\paral}_p + \Delta^{\paral}_\sigma 
        + 
        \Delta^{\paral}_\sigma
        \sqrt{\LpTrue}.
    \end{align}
    Then, using the bounds~\eqref{eqn:lem:Appendix:TrueProcess:ComponentBounds:TildeHG} and~\eqref{eqn:lem:Appendix:TrueProcess:ComponentBounds:B} in the proof of Proposition~\ref{prop:Appendix:TrueProcess:ControlErrorIntegrand:Evolved}, and the bounds~\eqref{eqn:lem:Appendix:TrueProcess:ComponentBounds:dV} and~\eqref{eqn:lem:Appendix:TrueProcess:ComponentBounds:D} above, we derive the following expressions for the terms $\widetilde{\Delta}^-(\sfp,\tstar,\Boldomega,\BoldTs)$ and $\widetilde{\Delta}_k^+(\sfp,\tstar,\Boldomega,\BoldTs)$, $k=\cbr{1,\dots,3}$, defined in Lemma~\ref{lem:Appendix:TrueProcess:XiTildeU}:
    \begin{subequations}\label{eqn:lem:Appendix:TrueProcess:InterBounds:DeltaTilde}
        \begin{align}
            \widetilde{\Delta}^-(\sfp,\tstar,\Boldomega,\BoldTs)
            \leq 
            \widetilde{\Upsilon}^{-}\left(\LpTrue; \sfp,\Boldomega,\BoldTs\right)
            \LpTrueError
            ,
            \\
            \widetilde{\Delta}^+_1(\sfp,\tstar,\Boldomega,\BoldTs)
            \leq 
            \frac{ 1 }{ \lambda  }
            \Delta_g
            \Upsilon'_1\left(\LpTrue; \sfp,\Boldomega,\BoldTs\right)
            \LpTrueError,
            \\
            \widetilde{\Delta}^+_2(\sfp,\tstar,\Boldomega,\BoldTs)
            \leq 
            \sqrt{\Boldomega}
            \frac{ 1 }{\lambda }
            \Delta_g
            \Upsilon'_2\left(\LpTrue; \sfp,\Boldomega,\BoldTs\right)
            \LpTrueError,
            \\
            \widetilde{\Delta}^+_3(\sfp,\tstar,\Boldomega,\BoldTs)
            \leq 
            \frac{2}{\absolute{2 \lambda -\Boldomega}}
            \Delta_g
            \Upsilon'_3\left(\LpTrue; \sfp,\Boldomega,\BoldTs\right)
            \LpTrueError
            ,
        \end{align}
    \end{subequations}
    for all $\sfp \in \cbr{1,\dots,\sfp^\star}$:

    Reccall the definition of $\widetilde{\Delta}_{\mathcal{U}}(\sfp,\tstar,\Boldomega,\BoldTs)$ from~\eqref{eqn:lem:Appendix:TrueProcess:XiTildeU:Bound:Final} in Lemma~\ref{lem:Appendix:TrueProcess:XiTildeU}
    \begin{align*}
        \widetilde{\Delta}_{\mathcal{U}}(\sfp,\tstar,\Boldomega,\BoldTs)
        =
        \widetilde{\Delta}^-(\sfp,\tstar,\Boldomega,\BoldTs)
        + 
        \indicator{\geq \BoldTs}{\tstar}
        \sum_{k=1}^3
            \widetilde{\Delta}_k^+(\sfp,\tstar,\Boldomega,\BoldTs)
        ,
    \end{align*}
    which can be trivially bounded as  
    \begin{align*}
        \widetilde{\Delta}_{\mathcal{U}}(\sfp,\tstar,\Boldomega,\BoldTs)
        \leq
        \widetilde{\Delta}^-(\sfp,\tstar,\Boldomega,\BoldTs)
        + 
        \sum_{k=1}^3
            \widetilde{\Delta}_k^+(\sfp,\tstar,\Boldomega,\BoldTs)
        .
    \end{align*}
    
    Substituting the bounds from~\eqref{eqn:lem:Appendix:TrueProcess:InterBounds:DeltaTilde} then leads to
    \begin{align}\label{eqn:lem:Appendix:TrueProcess:TildeDeltaU:FinalBounds}
        \begin{aligned}
            \widetilde{\Delta}_{\mathcal{U}}(\sfp,\tstar,\Boldomega,\BoldTs)
            \leq&
            \widetilde{\Upsilon}^{-}\left(\LpTrue; \sfp,\Boldomega,\BoldTs\right)
            \LpTrueError
            \notag 
            \\
            &+
            \frac{1}{\lambda}
            \Delta_g
            \left(
                \Upsilon'_1\left(\LpTrue; \sfp,\Boldomega,\BoldTs\right)
                + 
                \sqrt{\Boldomega}
                \Upsilon'_2\left(\LpTrue; \sfp,\Boldomega,\BoldTs\right)
            \right)
            \LpTrueError
        \end{aligned}
        \notag 
        \\
        +
        \frac{2}{\absolute{2\lambda - \Boldomega}}
        \Delta_g
        \Upsilon'_3\left(\LpTrue; \sfp,\Boldomega,\BoldTs\right)
        \LpTrueError
        ,
    \end{align}
    for all $\sfp \in \cbr{1,\dots,\sfp^\star}$. 

    \ul{\emph{Bounds for $\Delta_{\mathcal{U}}$:}}
    We next use~\eqref{eqn:lem:Appendix:TrueProcess:ComponentBounds:B},~\eqref{eqn:lem:Appendix:TrueProcess:ComponentBounds:D}, and Proposition~\ref{prop:Appendix:TrueProcess:ControlErrorIntegrand:Evolved}, to derive the following expressions for the terms $\Delta^-(\sfp,\tstar,\Boldomega,\BoldTs)$ and $\Delta_k^+(\sfp,\tstar,\Boldomega,\BoldTs)$, $k=\cbr{1,\dots,3}$, defined in Lemma~\ref{lem:Appendix:TrueProcess:NUTilde:Bounds}:
    \begin{subequations}\label{eqn:lem:Appendix:TrueProcess:InterBounds:DeltaU}
        \begin{align}
            \Delta^{-}(\sfp,\tstar,\Boldomega,\BoldTs)
            \leq     
            \mathring{\Delta}(\sfp,\tstar,\Boldomega)
            \Upsilon^{-}\left(\LpTrue; \sfp,\Boldomega,\BoldTs\right)
            ,
            \\
            \Delta_1^+(\sfp,\tstar,\Boldomega,\BoldTs)
            \leq      
            \mathring{\Delta}(\sfp,\tstar,\Boldomega)
            \frac{1}{\sqrt{\Boldomega}}
            \Upsilon'_1\left(\LpTrue; \sfp,\Boldomega,\BoldTs\right),
            \\
            \Delta_2^+(\sfp,\tstar,\Boldomega,\BoldTs)
            \leq
            \mathring{\Delta}(\sfp,\tstar,\Boldomega)
            \Upsilon'_2\left(\LpTrue; \sfp,\Boldomega,\BoldTs\right)
            ,
            \\
            \Delta_3^+(\sfp,\tstar,\Boldomega,\BoldTs)
            \leq
            \mathring{\Delta}(\sfp,\tstar,\Boldomega)
            \frac{1}{\BoldomegaRoot}
            \Upsilon'_3\left(\LpTrue; \sfp,\Boldomega,\BoldTs\right).
        \end{align}
    \end{subequations}

    As before, the definition of $\Delta_{\mathcal{U}}(\sfp,\tstar,\Boldomega,\BoldTs)$ in Lemma~\ref{lem:Appendix:TrueProcess:NUTilde:Bounds} implies the following straightforward bound:  
    \begin{align*}
        \Delta_{\mathcal{U}}(\sfp,\tstar,\Boldomega,\BoldTs)
        \leq
        \Delta^{-}(\sfp,\tstar,\Boldomega,\BoldTs)
        + 
        \sum_{k=1}^3
        \Delta_k^+(\sfp,\tstar,\Boldomega,\BoldTs)
        .
    \end{align*}
    Substituting~\eqref{eqn:lem:Appendix:TrueProcess:InterBounds:DeltaU} then leads to: 
    \begin{multline}\label{eqn:lem:Appendix:TrueProcess:DeltaU:FinalBounds:Pre}
        \Delta_{\mathcal{U}}(\sfp,\tstar,\Boldomega,\BoldTs)
        \leq
        \mathring{\Delta}(\sfp,\tstar,\Boldomega)
        \left(
            \Upsilon^{-}\left(\LpTrue; \sfp,\Boldomega,\BoldTs\right)
            +
            \frac{1}{\sqrt{\Boldomega}}
            \Upsilon'_1\left(\LpTrue; \sfp,\Boldomega,\BoldTs\right)
        \right.
        \\
        \left.
            +
            \Upsilon'_2\left(\LpTrue; \sfp,\Boldomega,\BoldTs\right)
            +
            \frac{1}{\BoldomegaRoot}
            \Upsilon'_3\left(\LpTrue; \sfp,\Boldomega,\BoldTs\right)
        \right)
        .
    \end{multline}
    We next recall the definition of the term  $\mathring{\Delta}(\sfp,\tstar,\Boldomega)$ from the statement of Lemma~\ref{lem:Appendix:TrueProcess:NUTilde:Bounds} given by
    \begin{align*}
        \mathring{\Delta}(\sfp,\tstar,\Boldomega)
        =
        \Delta_g^2
        \left(
            \frac{1}{\BoldomegaRoot}
            \sup_{\nu \in [0,\tstar]}
            \LpLaw{2\sfp}{}{
                \LZparamu{\nu,\Zt{N,\nu}}
            }
            + 
            \mathfrak{p}'(\sfp)  
            \sup_{\nu \in [0,\tstar]}
            \LpLaw{2\sfp}{}{
                \FZparasigma{\nu,\Zt{N,\nu}}
            }
        \right),
    \end{align*}
    which can be bounded by using~\eqref{eqn:lem:Appendix:TrueProcess:ComponentBounds:CircPara} to obtain the following:
    \begin{align*}
        \mathring{\Delta}(\sfp,\tstar,\Boldomega)
        \leq&
        \mathring{\Upsilon}\left(\LpTrueError; \sfp,\Boldomega\right)
        \notag 
        \doteq
        \Delta_g^2
        \left(
            \mathfrak{p}'(\sfp)
            \left( 
                L_p^{\paral}  + L_\sigma^{\paral} 
            \right)
            \sqrt{\LpTrueError}
            + 
            \frac{1}{\BoldomegaRoot}
            L_\mu^{\paral} 
            \LpTrueError
        \right).
    \end{align*}
    Substituting the above into~\eqref{eqn:lem:Appendix:TrueProcess:DeltaU:FinalBounds:Pre} produces the following for $\sfp \in \cbr{1,\dots,\sfp^\star}$:
    \begin{multline}\label{eqn:lem:Appendix:TrueProcess:DeltaU:FinalBounds}
        \Delta_{\mathcal{U}}(\sfp,\tstar,\Boldomega,\BoldTs)
        \leq
        \mathring{\Upsilon}\left(\LpTrueError; \sfp,\Boldomega\right)
        \left(
            \Upsilon^{-}\left(\LpTrue; \sfp,\Boldomega,\BoldTs\right)
            +
            \frac{1}{\sqrt{\Boldomega}}
            \Upsilon'_1\left(\LpTrue; \sfp,\Boldomega,\BoldTs\right)
        \right.
        \\
        \left.
            +
            \Upsilon'_2\left(\LpTrue; \sfp,\Boldomega,\BoldTs\right)
            +
            \frac{1}{\BoldomegaRoot}
            \Upsilon'_3\left(\LpTrue; \sfp,\Boldomega,\BoldTs\right)
        \right)
        .
    \end{multline}
    The proof is then concluded by substituting~\eqref{eqn:lem:Appendix:TrueProcess:Final:Bound:DelXi},~\eqref{eqn:lem:Appendix:TrueProcess:Final:Bound:NumberedDelU},~\eqref{eqn:lem:Appendix:TrueProcess:TildeDeltaU:FinalBounds}, and~\eqref{eqn:lem:Appendix:TrueProcess:DeltaU:FinalBounds} into~\eqref{eqn:lem:Appendix:TrueProcess:Final:Bounds:Initial}.

\end{proof}

\end{document}